%% file: main.tex
\documentclass[10pt,fleqn]{report}
\usepackage{setspace}
\usepackage{booktabs}
\usepackage{amsmath}
\usepackage{txfonts}
\usepackage{natbib}
\usepackage[dvipdfmx]{graphicx}
\usepackage{mediabb}
\usepackage{fancyhdr}
\usepackage{bm}
\usepackage{caption}
\usepackage[toc,page]{appendix}
\renewcommand{\figurename}{Fig.}

\newcommand{\kmpers}{\mathrm{km} \, \mathrm{s}^{-1}}
\newcommand{\logt}{\log T \, [\mathrm{K}]}
\begin{document}
\abovedisplayskip=4pt
\belowdisplayskip=4pt
\abovecaptionskip=4pt

% --- Title
%\input{tex/title.tex}
\input{tex/title.tex}
\maketitle

\thispagestyle{empty}
\mbox{}

\vspace{11cm}
\begin{center}
\copyright \ Naomasa Kitagawa, 2014
\end{center}
\newpage
\pagenumbering{roman}
% --- List of figures
%\listoffigures
%\clearpage

% --- Abstract
\section*{Abstract}
%\input{tex/abstract.tex}
\input{tex/abstract.tex}
\clearpage

% --- Acknowledgement
\section*{Acknowledgement}
%\input{tex/abstract.tex}
\input{tex/acknowledgement.tex}
%\clearpage

% --- Contents
\tableofcontents

%   Chapter 1
%\input{tex/abstract.tex}
\input{tex/contents_itdn.tex}

%   Chapter 2
%\input{tex/contents_diag.tex}
\input{tex/contents_diag.tex}

%   Chapter 3
%\input{tex/contents_cal.tex}
\input{tex/contents_cal.tex}

%   Chapter 4
%\input{tex/contents_vel.tex}
\input{tex/contents_vel.tex}

%   chapter 5
%\input{tex/contents_dns.tex}
\input{tex/contents_dns.tex}

%   Chapter 6
%\input{tex/contents_ndv.tex}
\input{tex/contents_ndv.tex}

%   Chapter 7   
%\input{tex/contents_dis.tex}
\input{tex/contents_dis.tex}

%   Chapter 8
%\input{tex/contents_rmk.tex}
\input{tex/contents_rmk.tex}

%   Appendix
\appendix
\input{tex/contents_mph.tex}

% --- Bibliography
\bibliography{apj-jour,%
bib_nk_a,%
bib_nk_b,%
bib_nk_c,%
bib_nk_d,%
bib_nk_e,%
bib_nk_g,%
bib_nk_h,%
bib_nk_i,%
bib_nk_k,%
bib_nk_l,%
bib_nk_m,%
bib_nk_n,%
bib_nk_o,%
bib_nk_p,%
bib_nk_r,%
bib_nk_s,%
bib_nk_t,%
bib_nk_u,%
bib_nk_w,%
bib_nk_y}

% --- End of Document
\end{document}

%% file: tex/title.tex
% Title
\title{
  {\huge Thesis} \\
  \vspace{3mm}
  \begin{spacing}{1.3}
    {\Huge
    Coronal upflows from edges of an active region observed
    with EUV Imaging Spectrometer onboard \textit{Hinode}
    }
  \end{spacing}
  \vspace{20mm}
  }
\author{
  {\huge Naomasa Kitagawa} \\
  Department of Earth and Planetary Science, 
  The University of Tokyo
  }
\date{2012 December 18}

% --- End of Contents ---

%% file: tex/abstract.tex
% ==================================================================
%   Chapter:
%     Abstract.
%   Contents:
%     - What we have done.
%     - Introduction to AR outflows
%     - Method (velocity reference; Chapter 3)
%     - Analysis & results (Chapter 4) with fundamental info.
%     - Analysis & results (Chapter 5 & 6) with fundamental info.
%     - Conclusions.
%     - Appendix.
% ==================================================================

% What we have done
%{\color{dartmouthgreen} \textbf{
In order to better understand the plasma supply and leakage at active regions,
%}}
we investigated physical properties of the upflows from edges of active region NOAA AR10978 observed with the EUV Imaging Spectrometer (EIS) onboard \textit{Hinode}.  Our observational aim is to measure two quantities of the outflows: Doppler velocity and electron density.  
\vspace{0.25in}

% Introduction to AR outflows
These upflows in the corona, referred to as active region outflows (hereafter \textit{the outflows}), were discovered for the first time by EIS due to its unprecedented high sensitivity and spectral resolution.  Those outflows are emanated at the outer edge of a bright active region core, where the intensity is low (\textit{i.e.}, dark region).  
%{\color{dartmouthgreen} 
%\textbf{
It is well known by a number of EIS observations that the coronal emission lines at the outflow regions are composed of an enhanced component at the blue wing (EBW) corresponding to a speed of $\sim 100 \, \kmpers$, added to by the stronger major component almost at rest.
%}} 
This EBW can be seen in line profiles of Fe \textsc{xii}--\textsc{xv} whose formation temperatures are around $\logt=6.2\text{--}6.3$.  It has been suggested that the outflows are (1) an indication of upflows from the footpoints of coronal loops induced by impulsive heating in the corona, (2) induced by the sudden pressure change after the reconnection between closed active region loops and open extended loops located at the edge of an active region, (3) driven by the contraction occurring at the edge of an active region which is caused by horizontal expansion of the active region, and (4) the tips of chromospheric spicules heated up to coronal temperature.  
%{\color{red}
While a number of observations have been revealed such aspects of the outflows, however, their electron density has not been known until present, which is one of the important physical quantities to consider the nature of the outflows.  In addition, the Doppler velocity at the transition region temperature ($\logt \le 6.0$) has not been measured accurately in the outflow regions because of the difficulties in EIS spectroscopic analysis (\textit{e.g.}, the lack of onboard calibration 
%{\color{dartmouthgreen} 
%\textbf{
lamp for absolute wavelength,
%}}
and the temperature drift of line centroids according to the orbital motion of the satellite).
%}
In this thesis, we analyzed the outflow regions in NOAA AR10978 in order to measure Doppler velocity within wide temperature range ($\logt = 5.5\text{--}6.5$) and electron density by using an emission line pair Fe \textsc{xiv} $264.78${\AA}/$274.20${\AA}. 
\vspace{0.25in}

% Method (EIS, EUV emission lines, velocity reference)
Since EIS does not have an absolute wavelength reference onboard, we need another reference for the precise measurement of the Doppler velocity.  In this thesis, we exploited Doppler velocity of the quiet region as the reference, which was studied in Chapter \ref{chap:cal}.  EUV emission lines observed in the quiet region are known to indicate redshift corresponding to $v \simeq 10 \, \kmpers$ at $\logt \le 5.8$, while those above that temperature have not been established where a number of emission lines observed with EIS exist.  
%{\color{dartmouthgreen}
%\textbf{
Since the corona is optically thin, spectra outside the limb are superposed symmetrically along the line of sight, which leads to the reasonable idea that the limb spectra take a Doppler velocity of $v=0$.
%}} 
%{\color{red}
We derived the Doppler velocity of the quiet region at the disk center by studying the center-to-limb variation of line centroid shifts for eleven emission lines from the transition region and the corona.
%}
By analyzing the spectroscopic data which cover the meridional line of the Sun from the south pole to the north pole, we determined the Doppler velocity of the quiet region with $5.7 \le \logt \le 6.3$ in the accuracy of $\simeq 3 \, \kmpers$ for the first time.  It is shown that emission lines below $\logt = 6.0$ have Doppler velocity of almost zero with an error of $1\text{--}3 \, \kmpers$, while those above that temperature are blueshifted with gradually increasing magnitude: $v = - 6.3 \pm 2.1 \, \kmpers$ at $\logt=6.25$\footnote{Positive (Negative) velocity indicates a motion away from (toward) us.}.
\vspace{0.25in}

% Analysis & results (Chapter 4)
The Doppler velocity of the outflows was measured for twenty six emission lines which cover the temperature range of $5.5 \le \logt \le 6.5$ in Chapter \ref{chap:vel}.  Though it is well known that the outflows are prominent around $\logt=6.1\text{--}6.3$ and exhibit clear blueshift corresponding to several tens of $\kmpers$ due to the existence of EBW extending up to $\sim 100 \, \kmpers$, the behavior below that temperature has not been revealed.  Using the Doppler velocity of the quiet region (obtained in Chapter \ref{chap:cal}) for the temperature range of $5.7 \le \logt \le 6.3$ as a reference, we measured the Doppler velocity of several types of coronal structures in NOAA AR10978: active region core, fan loops, and the outflow regions.  Active region core, characterized by high temperature loops ($\logt \ge 6.3\text{--}6.4$), indicated almost the same centroid shifts as the quiet region selected in the field of view of the EIS scan.  Fan loops are extending structures from the periphery of active regions, which indicated $v \simeq 10 \, \kmpers$ at the transition region temperature, and the Doppler velocity decreased with increasing formation temperature: reaching $v = -20 \, \kmpers$ at $\logt=6.3$.  Different from fan loops, the outflow regions exhibited a blueshift corresponding to $v \simeq -20 \, \kmpers$ at all temperature range below $\logt = 6.3$, which implies that the plasma does not return to the solar surface.  
%{\color{red} 
The fact that the outflow region and fan loops are often located near each other has been made it complicated to understand the physical view of those structures.  By extracting the target regions with much carefulness, we revealed the definitive difference of the outflow regions and fan loops in the Doppler velocity at the transition region temperature.
%}
\vspace{0.25in}

% Analysis & results (Chapter 5)
In Chapter \ref{chap:dns}, the electron density of the outflows (EBW component in coronal emission line profiles) was derived for the first time by using a density-sensitive line pair Fe \textsc{xiv} $264.78${\AA}/$274.20${\AA}.  This line pair has a wide sensitivity for the electron density range of $n_{\mathrm{e}} = 10^{8\text{--}10} \, \mathrm{cm}^{-3}$, which includes the typical values in the solar corona.  We extracted EBW component from the line profiles of Fe \textsc{xiv} through double-Gaussian fitting.  Since those two emission lines are emitted from the same ionization degree of the same ion species, they should be shifted by the same amount of Doppler velocity and thermal width.
%{\color{red}  
We challenged the simultaneous fitting applied to those two Fe \textsc{xiv} lines with such physical restrictions on the fitting parameters.
%}
After the double-Gaussian fitting, we obtained the intensity ratio of Fe \textsc{xiv} $264.78${\AA}/$274.20${\AA} both for the major component and EBW component.  Electron density for both component ($n_{\mathrm{Major}}$ and $n_{\mathrm{EBW}}$) was calculated by referring to the theoretical intensity ratio as a function of electron density which is given by CHIANTI database.  We studied six locations in the outflow regions.  The average electron density in those six locations was $n_{\mathrm{Major}} = 10^{9.16 \pm 0.16} \, \mathrm{cm}^{-3}$ and $n_{\mathrm{EBW}} = 10^{8.74 \pm 0.29} \, \mathrm{cm}^{-3}$.  The magnitude relationship between $n_{\mathrm{Major}}$ and $n_{\mathrm{EBW}}$ was different in the eastern and western outflow regions, 
%{\color{red}
which was discussed in Section \ref{chap:dis} associated with the magnetic topology.  The column depth was also calculated by using the electron densities for each component in the line profiles, and it leads to the result that (1) the outflows possess only a small fraction ($\sim 0.1$) compared to the major rest component in the eastern outflow region, while (2) the outflows dominate over the rest plasma by a factor of around five in the western outflow region.
%}
\vspace{0.25in}

% Analysis & results (Chapter 6)
We developed a new method in line profile analysis to investigate the electron density of a minor component in Chapter \ref{chap:ndv}.  Instead of obtaining the electron density from the intensity ratio calculated by the double-Gaussian fitting, we derive the electron density for each spectral bin in Fe \textsc{xiv} line profiles, which we refer to as \textit{$\lambda$-$n_{\mathrm{e}}$ diagram}.  This method has an advantage that it does not depend on any fitting models.  By using the $\lambda$-$n_{\mathrm{e}}$ diagram, we confirmed that EBW component indeed has smaller electron density than that of the major component in the western outflow region while that was not the case in the eastern outflow region.  
\vspace{0.25in}

% Discussion and conclusions
Our implications are as follows. 

\hangindent=30pt
\hangafter=1
(1) The outflow regions and fan loops, which has been often discussed in the same context, exhibited different temperature dependence of Doppler velocity.  We concluded these structures are not identical.  

\hangindent=30pt
\hangafter=1
(2) We tried to interpret the outflows in terms of the siphon flow 
%{\color{dartmouthgreen} \textbf{
(\textit{i.e.}, steady and unidirectional)
%}}
along coronal loops, but it turned out to be unreasonable 
%{\color{dartmouthgreen}
%\textbf{
because both the mass flux and the gas pressure gradient were in the opposite sense to what should be expected theoretically.
%}}  

\hangindent=30pt
\hangafter=1
(3) The temperature dependence of the Doppler velocity in the outflow regions are different from that was predicted by a previous numerical simulation on impulsive heating with longer timescale than the cooling.  
%{\color{dartmouthgreen} 
%\textbf{
We observed upflow at the transition temperature, while the numerical simulation resulted in downflow at that temperature.
%}}

\hangindent=30pt
\hangafter=1
(4) As for the case if intermittent heating is responsible for the outflows,
%{\color{dartmouthgreen}
%\textbf{
we analytically considered a balance between heating and the enthalpy flux.
%}}
The duration of heating was crudely estimated to be longer than $\tau = 500 \, \mathrm{s}$ so that the density of upflows from the footpoints becomes compatible with that of the observed outflows.  

\hangindent=30pt
\hangafter=1
(5) Though EBW component contributes to the emission as a small fraction in a line profile, the volume amount is around five times as large as the major component in the western outflow region 
%{\color{dartmouthgreen}
%\textbf{
as calculated by using the electron density for each component in Fe \textsc{xiv} line profiles.
%}}

\hangindent=30pt
\hangafter=1
%{\color{dartmouthgreen}
%\textbf{
(6) Coronal loops rooted at the eastern outflow regions are connected to the opposite polarity region within the active region when taking into account the magnetic topology constructed from an MDI magnetogram, from which
%}}
%{\color{red}
we suggest a possibility that the outflows actually contribute to the mass supply to active region coronal loops at the eastern outflow region.
%}

% --- End of Contents ---

%% file: tex/acknowledgement.tex
% =====================
%   Acknowledgement
% =====================

First I would like to express my appreciation to Professor Takaaki Yokoyama for his great tolerance and a number of insightful comments.  This thesis would never see the light of day without his support for these six years.  In the first year, he told me to see observational data as they are, and to habitually evaluate physical quantities in the solar corona by using typical parameters.  I learned how to analyze the data in the second year.  From the third year, he gradually has been let me have my own way of thinking.  Even when I took a stupid mistake, he waited and saw how things go without excessive-teaching attitude more than is necessary.  He had listened my analysis on a solar flare with interest during the fourth year.  The first version of this thesis was written in the fifth year.  He made an enormous amount of helpful comments devoting much time to reading.  The sixth year was challenging period when almost all of the contents in this thesis have been greatly improved.  He is a major witness of my working toward improvement during these six years. 

\vspace{0.25in}
I extend my gratitude for all of the referees, Dr.\ Masaki Fujimoto, Dr.\ Hirohisa Hara, Dr.\ Takeshi Imamura, Dr.\ Toshifumi Shimizu, and Dr.\ Ichiro Yoshikawa, for providing me with extraordinary opportunity to improve this thesis as an education.  Dr.\ Fujimoto discussed with me on the background and incentive of my work, and helped me write an attractive abstract submitted along with this thesis.  Dr.\ Hara made a large number of scientific comments with expertise in EUV spectroscopy every time I went to National Astronomical Observatory.  Dr.\ Imamura checked foundation for an understanding of physical processes in the solar corona.  Dr.\ Shimizu pointed out implications of the results obtained in this thesis, which was highly suggestive.  Dr.\ Yoshikawa offered me encouraging words when I was in a daze.  This thesis has finally been completed thanks to all their help and support. 

\vspace{0.25in}
Members of the laboratory, Shin Toriumi, Hideyuki Hotta, Yuki Matsui, Haruhisa Iijima, Takafumi Kaneko, and Shuoyang Wang encouraged me so much for more than a year, especially when I was depressed because of my unsatisfactory situation.  A brisk hour of exercise with Shin and Hideyuki made me get refreshed.  I also thank to Yusuke Iida, who had been a member of the laboratory and is now working in JAXA, for concise advice about the way of thinking as a researcher.  

\vspace{0.25in}
Masaru Kitagawa, Keiko Kitagawa, Ami Kitagawa, and partner Shoko Sato were always beside me.  I never forget that they were waiting for the day this thesis would be approved.  

\vspace{0.25in}
As a token of my appreciation for a year of service and repayment, I will give my pledge to continue effort in years to come, built on the experience during one-year-lasting thesis defense.  

% --- End of Contents ---

%% file: tex/contents_itdn.tex
% !TEX root = main_dns.tex
%% ===============
\chapter{Introduction}
\pagenumbering{arabic}
\label{chap:itdn}
\section{The solar corona}
  \input{tex/itdn_solar_corona.tex}
  \input{tex/itdn_appearance_corona.tex}

\section{Active region outflows}
  \input{tex/itdn_flows.tex}
  \subsection{Observations of AR outflows by \textit{Hinode}/EIS}
    \input{tex/itdn_aroutflow.tex}
  \subsection{Driving mechanisms of AR outflows}
    \input{tex/itdn_aroutflows_mech.tex}
\section{Motivation}

\input{tex/itdn_motivation.tex}

%% file: tex/itdn_solar_corona.tex
% ====================================================
%   Chapter:
%     Introduction.
%   Section:
%     General introduction to the solar corona.
% ====================================================

The solar corona is an outer atmosphere of the Sun which has a temperature exceeding $10^6 \, \mathrm{K}$.  It is an outstanding issue how the corona can be heated up to so high temperature compared to the inner photosphere, where the temperature is around $6000 \, \mathrm{K}$.  The observations of the solar corona date back to ancient eclipse recorded by Indian, Babylonian and Chinese.  Routine coronal observations started when Beonard Lyot built the first coronagraph in 1930, which occults the brighter photosphere by using a disk.  Forbidden lines of highly ionized atoms (Fe \textsc{x}--\textsc{xiv}; Ni \textsc{xii}--\textsc{xvi}) were identified \citep{edlen1943,swings1943} and it was claimed for the first time that the coronal temperature exceeds million Kelvin ($\mathrm{MK}$).  As already mentioned, the physical explanation of the mechanism keeping this high temperature in the solar corona is still unknown.  The second law of thermodynamics seems to be violated in the point that the much cooler photosphere ($T \sim 6000 \, \mathrm{K}$) exists at inner atmosphere, closer to the energy source at the core of the star. 

\begin{figure}
  \centering
  \includegraphics[width=11cm,clip]{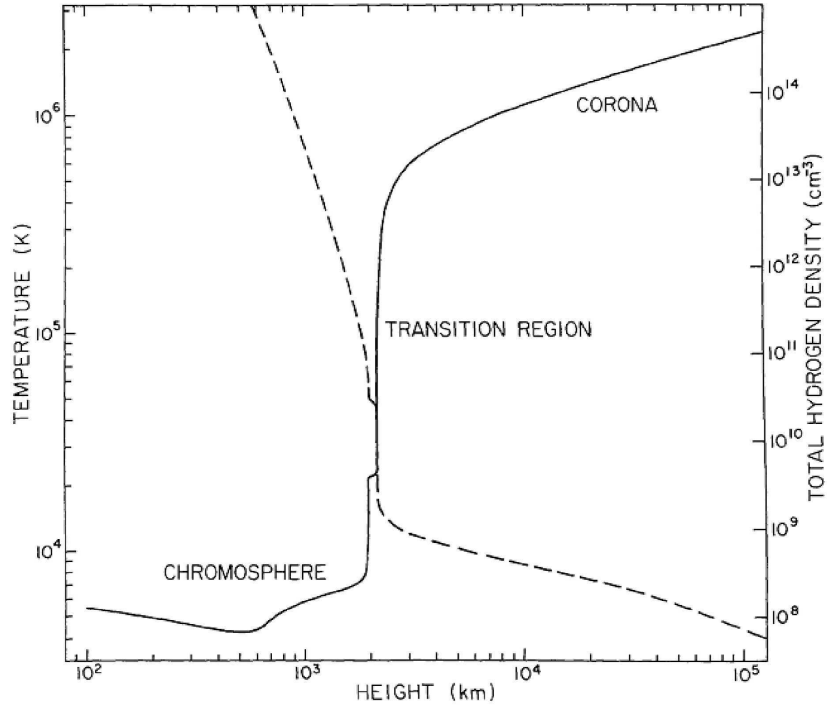}
  \caption{Distribution of temperature (\textit{solid} line) 
  and hydrogen density (\textit{dashed} line) in the solar atmosphere 
  as a function of height from the photosphere $\tau_{5000} = 1$. 
  Excerpted from \citet{withbroe1977}.}
  \label{itdn_temperature_profile_withbroe_1977}
\end{figure}

The temperature profile along the height is shown in Fig.~\ref{itdn_temperature_profile_withbroe_1977}, from which we can see the decreasing temperature in the photosphere and the increasing temperature from the bottom of the chromosphere where the temperature takes minimum value ($\sim 4200 \, \mathrm{K}$), to the upper atmosphere. There is a thin layer called the transition region between the cool chromosphere ($\sim 10^{4} \, \mathrm{K}$) and the hot corona ($\geq 10^{6} \, \mathrm{K}$). It is appropriate to think this layer as a temperature regime rather than a geometric layer because of the extremely spatial inhomogeneous structure in the solar atmosphere. The density profile also has steep gradient while the pressure must be continuous through the transition region. In the upper part of the transition region the temperature reaches up to $10^{6} \, \mathrm{K}$.  Due to the high temperature exceeding $10^{6} \, \mathrm{K}$, the corona consists of ions with high degree of ionization. These ions efficiently radiate the line emission in the EUV and X-ray wavelength range.  

% --- End of Contents ---

%% file: tex/itdn_appearance_corona.tex
% =====================================
%   Chapter:
%     Introduction.
%   Section:
%     Appearance of the corona.
% =====================================

\begin{figure}
  \centering
  \includegraphics[width=5.6cm]{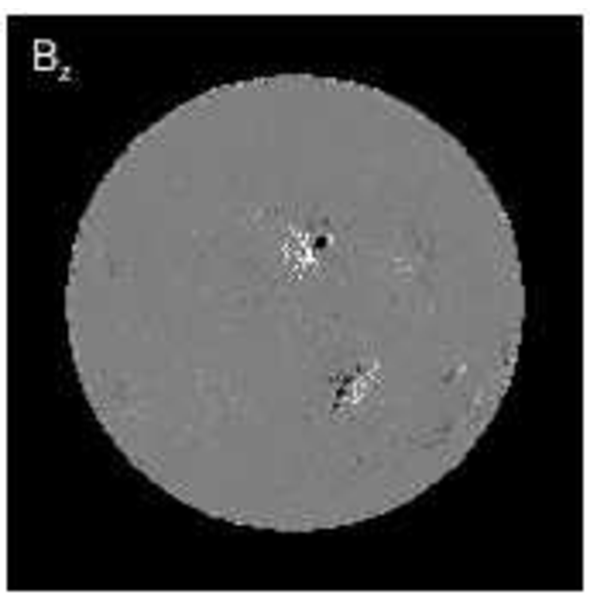}
  \includegraphics[width=5.6cm]{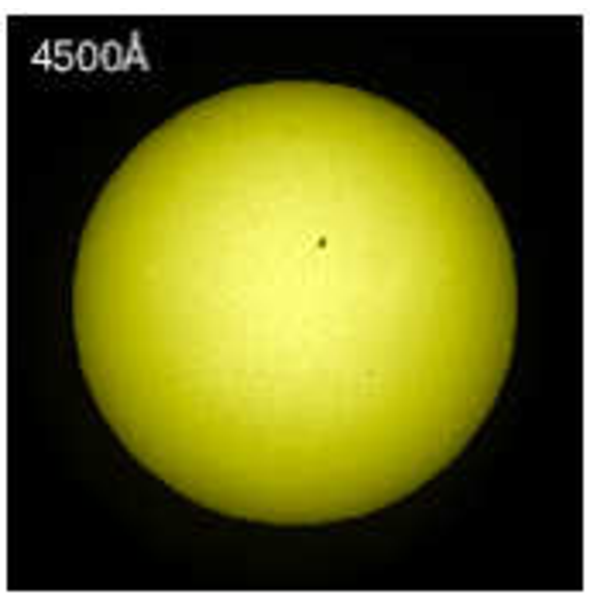}
  \includegraphics[width=5.6cm]{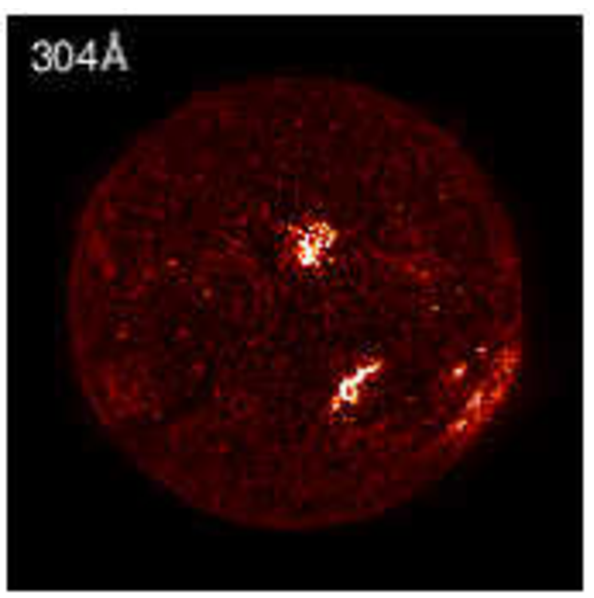}
  \includegraphics[width=5.6cm]{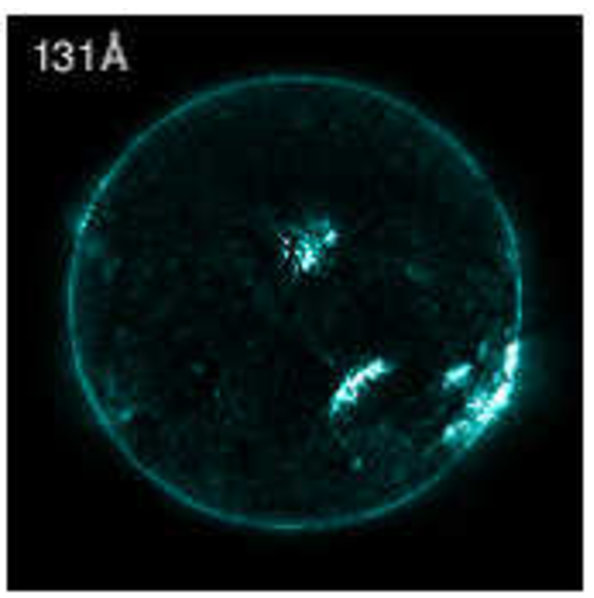}
  \includegraphics[width=5.6cm]{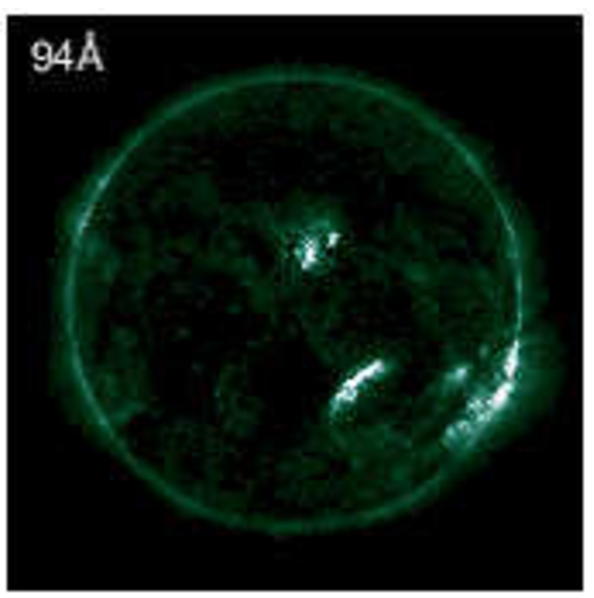}
  \includegraphics[width=5.6cm]{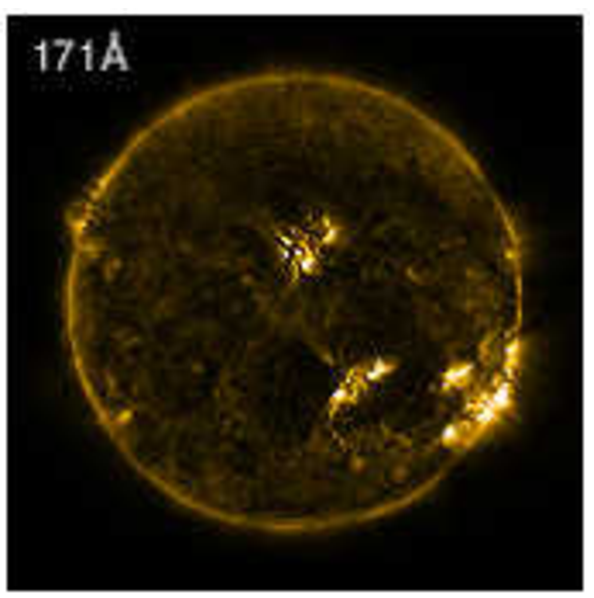}
  \includegraphics[width=5.6cm]{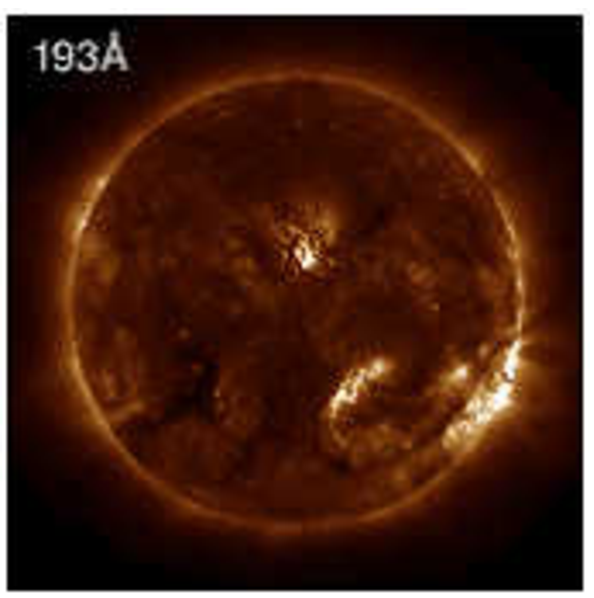}
  \includegraphics[width=5.6cm]{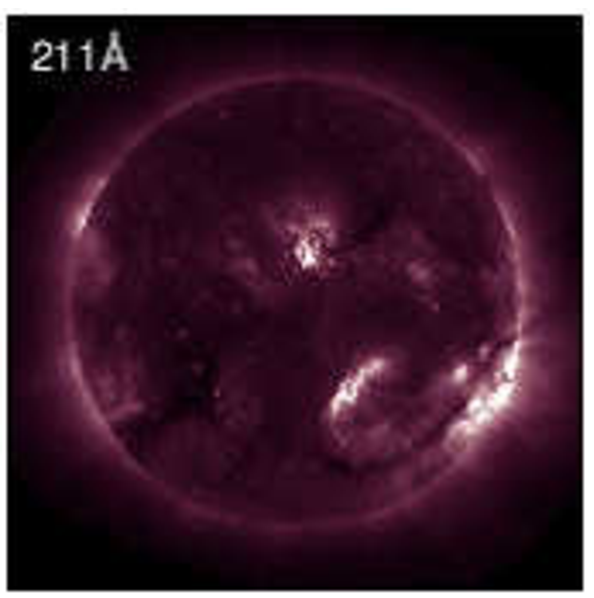}
  \includegraphics[width=5.6cm]{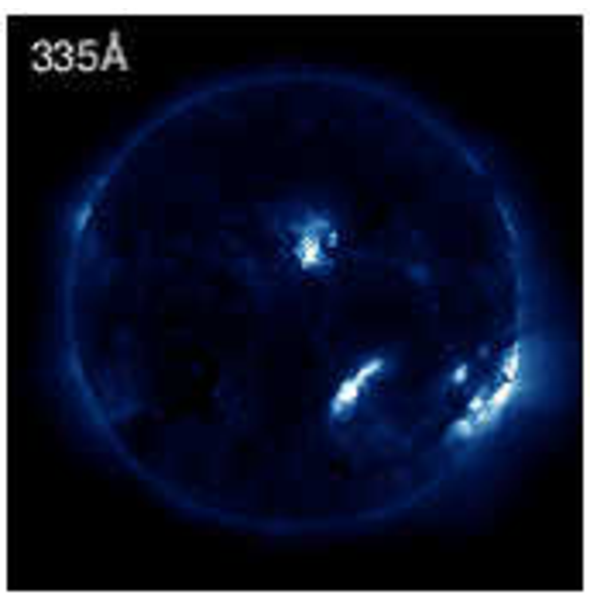}
  \caption{
    Images taken by \textit{SDO} on 2012 August 14.
    \textit{Right upper} panel shows a magnetogram taken by HMI.
    Other panels show EUV images taken by AIA. 
    Numbers at upper right in each panel represent the wavelength of passband.
    EUV images are displayed in an order of the formation temperature dominant in each filter.
  }
  \label{fig:itdn_aia_context}
\end{figure}

The solar corona has been categorized into three kinds historically by its X-ray brightness: active region, the quiet region, and coronal hole.  In Fig.~\ref{fig:itdn_aia_context}, various appearances of the solar corona are seen in the images taken by different filters of Atmospheric Imaging Assembly (AIA) onboard \textit{Solar Dynamics Observatory} (\textit{SDO}) launched by NASA.  There are several bright areas (active regions) slightly at the north of the center and in the south west\footnote{The east (west) is conventionally defined as left (right) in an image of the Sun. It is opposite to the Earth map.}. At the south east of the center, there is a dark region most clearly seen in the $193${\AA} passband image. Such a kind of region is called as a \textit{coronal hole}. The location other than active regions and a coronal hole is the \textit{quiet region}.  

% AR
Active regions are located in areas of strong magnetic field concentrations, visible as sunspot groups in optical wavelengths and magnetograms. Sunspot groups typically exhibit a strongly concentrated leading magnetic polarity, followed by a more fragmented trailing group of opposite polarity. Because of this bipolar nature active regions are mainly made up of closed magnetic field lines. Due to the persistent magnetic activity in terms of magnetic flux emergence, flux cancellation, magnetic reconfigurations, and magnetic reconnection processes, a number of dynamic phenomena such as transient brightenings, flares, and coronal mass ejections occur in active regions. We focus the persistent upflow seen at the edge of active regions in this thesis, which was discovered by EUV Imaging Spectrometer (EIS) onboard \textit{Hinode}.  
 
% LOOP (SDO AIA and HMI image)
Active regions are constructed by structures along the magnetic field, which has been called as ``coronal loops'', since X-ray observations from the space have enabled us to see the loop appearance along the coronal magnetic field \citep{rosner1978}.  Due to the nature of the solar corona that the plasma beta is much smaller than unity ($\beta \ll 1$), and that thermal conduction is strongly constrained in the direction parallel to magnetic field, the structures seen in EUV or X-ray images are basically configurated by the magnetic field.  A consequence of the plasma heating in the transition region and the chromosphere is the upflow into coronal part which makes the coronal loops filled with hotter and denser plasma than the background corona.  Those coronal loops produce bright emission. 

\begin{figure}
  \centering
  \includegraphics[width=5.6cm,clip]{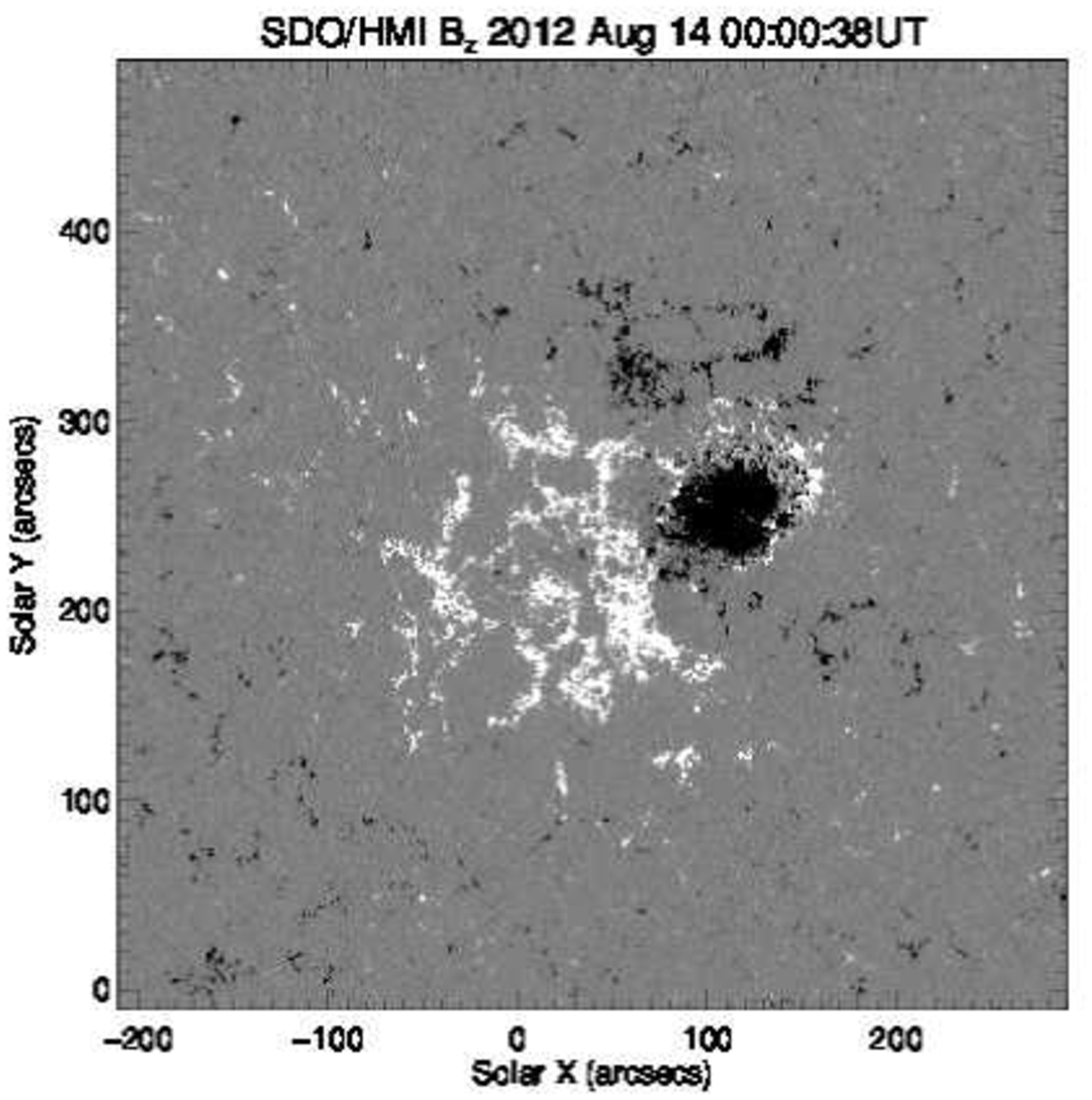}
  \includegraphics[width=5.6cm,clip]{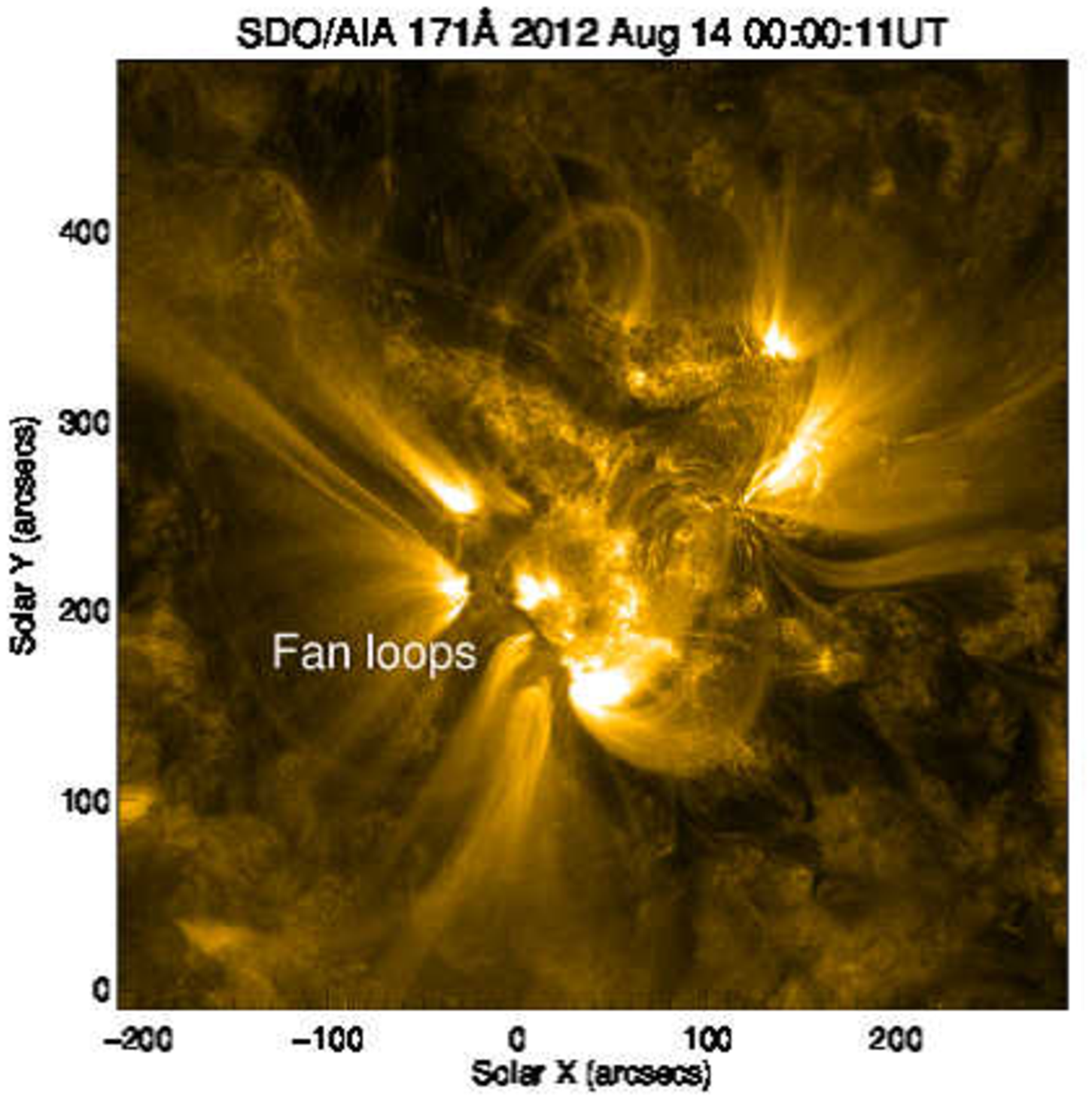}
  \includegraphics[width=5.6cm,clip]{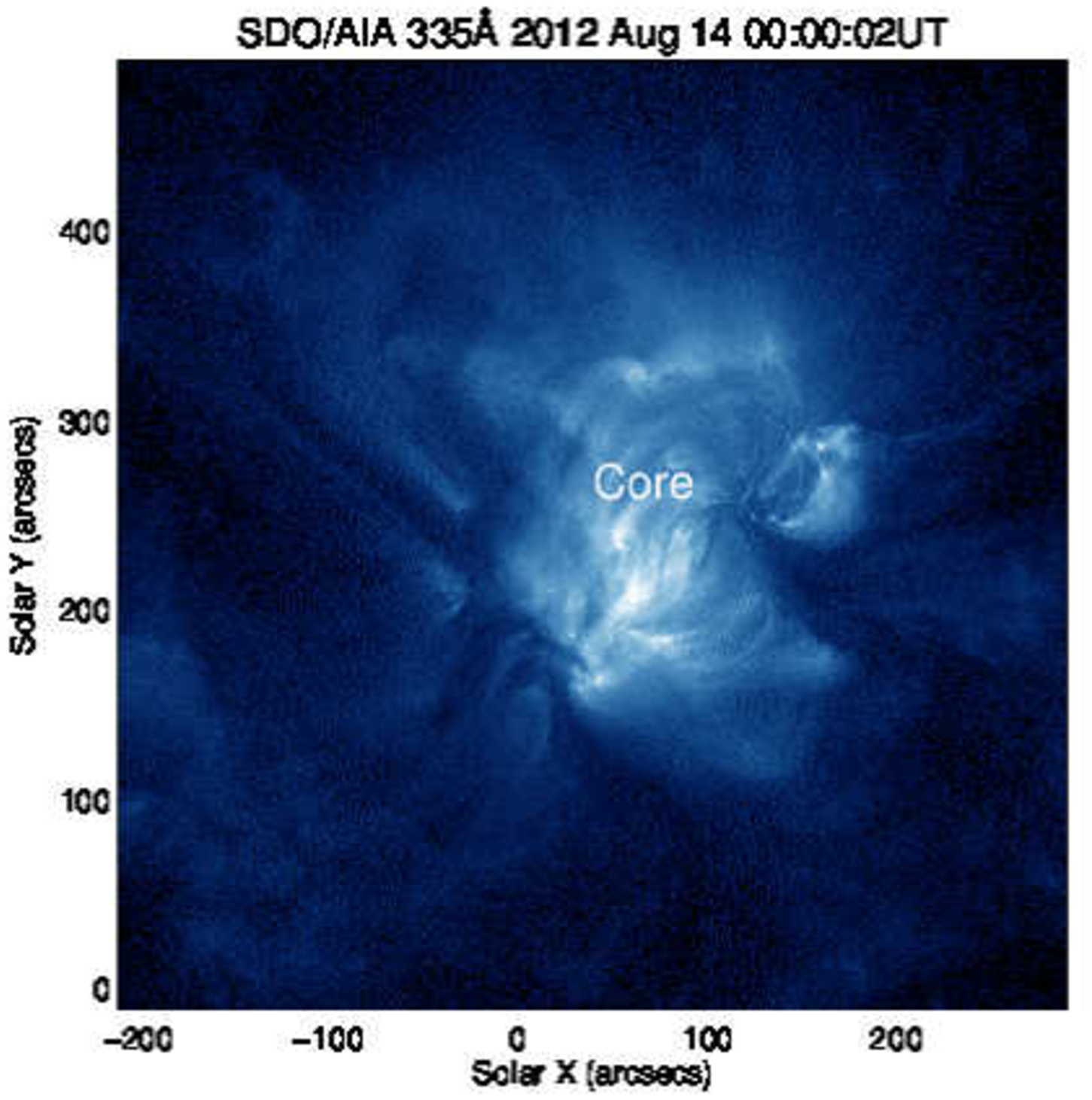}
  \includegraphics[width=5.6cm,clip]{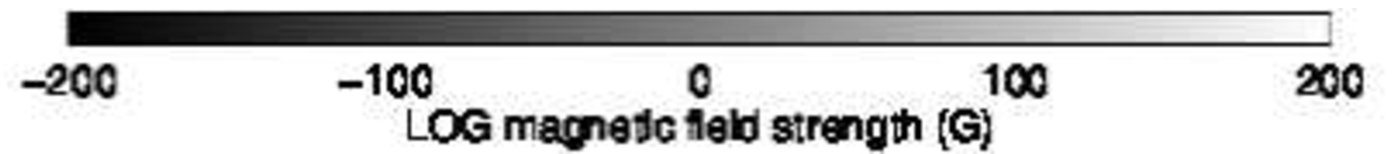}
  \includegraphics[width=5.6cm,clip]{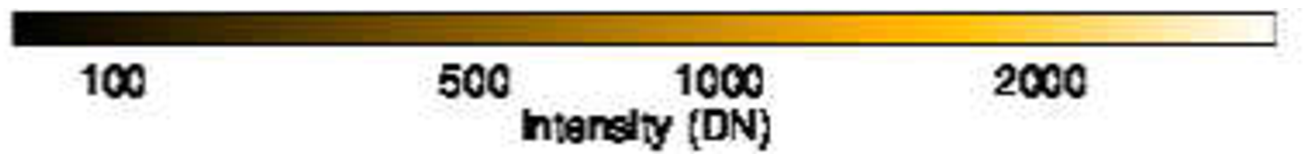}
  \includegraphics[width=5.6cm,clip]{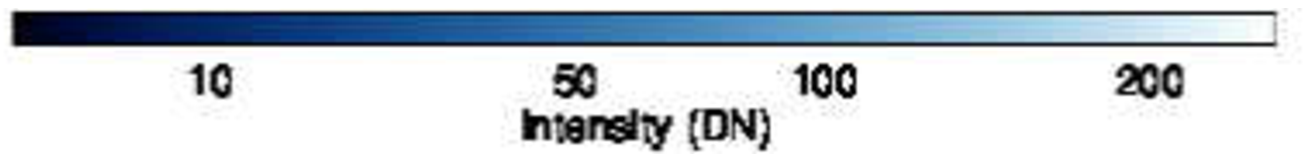}
  \caption{
    \textit{Left}:  Magnetogram taken by \textit{SDO}/HMI.
    \textit{Middle}: EUV image taken by \textit{SDO}/AIA $171${\AA} 
    passband ($\log T \, [\mathrm{K}] \simeq 5.9$). 
    \textit{Right}: EUV image taken by \textit{SDO}/AIA $335${\AA} 
    passband ($\log T \, [\mathrm{K}] \simeq 6.5$). 
  }
  \label{fig:itdn_aia_hmi_example}
\end{figure}

As an example of active region, a magnetogram taken by Heliospheric and Magnetic Imager (HMI) onboard \textit{SDO} and two EUV images taken by \textit{SDO}/AIA $171${\AA} ($\log T \, [\mathrm{K}] \simeq 5.9$) and $335${\AA} ($\log T \, [\mathrm{K}] \simeq 6.5$) passbands are shown in Fig.~\ref{fig:itdn_aia_hmi_example}.  The active region was extracted from near the center of the Sun shown in Fig.~\ref{fig:itdn_aia_context}.  An extended structure similar to a fan can be clearly seen at the east and west edge of the active region in the AIA $171${\AA} passband image (indicated by white characters), which is called ``fan loops''.  Fan loops are often clearly seen in EUV images corresponding to a temperature around $\log T \, [\mathrm{K}] \simeq 5.7$--$6.0$. In the AIA $335${\AA} passband image, there are multiple loop systems connecting the opposite polarities with strong magnetic field in the south east--north west direction as indicated by white characters. These loops form the dominant emission of the active region and are called as active region ``core''. 

% QS
The quiet region is the area outside active regions. It is relatively \textit{quiet}, however, various kinds of small dynamic phenomena have been observed all over the quiet region today by virtue of high resolution, for example, explosive events, bright points, jets, and giant arcade eruptions. The faint areas are called ``coronal holes'', where magnetic field lines are opened into the outer space. Thus, the coronal plasma can be ejected possibly as the solar wind. Frequently occurring soft X-ray jets have been observed by the previous Japanese satellite \textit{Yohkoh} and X-ray telescope (XRT) onboard \textit{Hinode}. 

% --- End of Contents ---

%% file: tex/itdn_flows.tex
% ============================
%   Chapter:
%    Flows.
%   Section:
%    Introduction to flows.
% ============================

The flows in the solar corona play a crucial role in dynamics and formation of variable structures, and observations of their properties constrain the coronal heating problem. Observations on the flows in the solar corona are described here. 

% Teriaca 1999
From observations by Solar Ultraviolet Measurements of Emitted Radiation (SUMER) onboard \textit{SoHO}, spectra of the transition region lines ($\le 10^6 \, \mathrm{K}$) are known to be redshifted both in the quiet region and in an active region core \citep{chae1998doppler,peterjudge1999,teriaca1999}.  On the other hand, coronal lines ($\simeq 10^6 \, \mathrm{K}$) indicate blueshift. 
%This result, redshift in transition region lines and blueshift in coronal lines, has been interpreted as upflows of hot plasma in coronal loops due to certain heating mechanism and downflows of radiatively cooling plasma after the heating ceases \citep{mcintosh2012}. The authors analyzed the downflow at the legs of coronal loops in cool emission lines (Si VII) and adopted that interpretation. 
\citet{hansteen2010} numerically showed that the redshift of transition region lines and slight blueshift of coronal lines are naturally produced by frequently occurring reconnections in the upper chromosphere as a response to the production of magnetic shear due to the braiding of magnetic fields by the photospheric convection. 
While those observations by SUMER focus on the low-temperature plasmas, the spectroscopic nature of the hotter component with $T \geq 10^{6} \, \mathrm{K}$ is still unclear, which is one of the main contents of this thesis (Chapter \ref{chap:cal}). 
%One of the unsolved topic is the temperature dependence of Doppler shift above the temperature of $\simeq 10^6 \, \mathrm{K}$, which now becomes feasible by analyzing the spectrum obtained with \textit{Hinode}/EIS, however, such analysis has not been reported yet. 
%It may be partly because of the difficulties in determining absolute Doppler velocities when we analyze the EIS data as described in Sect. \ref{chap:cal}.

% Apparent flows observed with TRACE
Flows have also been observed with imaging observations. \textit{Transition Region And Coronal Explorer} (TRACE) had enabled a discover of persistent, intermittent flow pattern in coronal loops \citep{winebarger2001}.  Since there was no any obvious periodicity, it was concluded that the flow is induced by magnetic reconnection instead of the waves coming from the photosphere.  The upward flows in coronal loops are also observed recently by AIA which has much higher temporal cadence than ever (full-Sun images at every $12 \, \mathrm{s}$ for seven EUV wavelength bands), which revealed the ubiquitous existence of such flows in coronal loops extending from the edge of active regions \citep{tian2011}. 
%In the next subsection, we describe a new type of flow discovered by \textit{Hinode}/EIS.
% --- End of Contents ---

%% file: tex/itdn_aroutflow.tex
% ================================
%   Chapter:
%     Introduction.
%   Section:
%     Discovery of AR outflows.
% ================================

Spectral coverage sensitive to the coronal temperature and unprecedented high signal-to-noise ratio of \textit{Hinode}/EIS enabled us to reveal the existence of upflows at the edge of active regions \citep{doschek2008,hara2008,harra2008}.  These upflows in the active region is called ``AR (active region) outflows'' and considered to be the upflows from the bottom of the corona.  It has previously been confirmed that these outflows persist for several days in the images taken by X-Ray Telescope (XRT) onboard \textit{Hinode} \citep{sakao2007}.  Some authors interpreted AR outflows as the source of the solar wind \citep{brooks2011}. 

% Location of AR outflows
\begin{figure}
  \centering
  %bb=0 0 810 359,
  \includegraphics[width=14cm,clip]{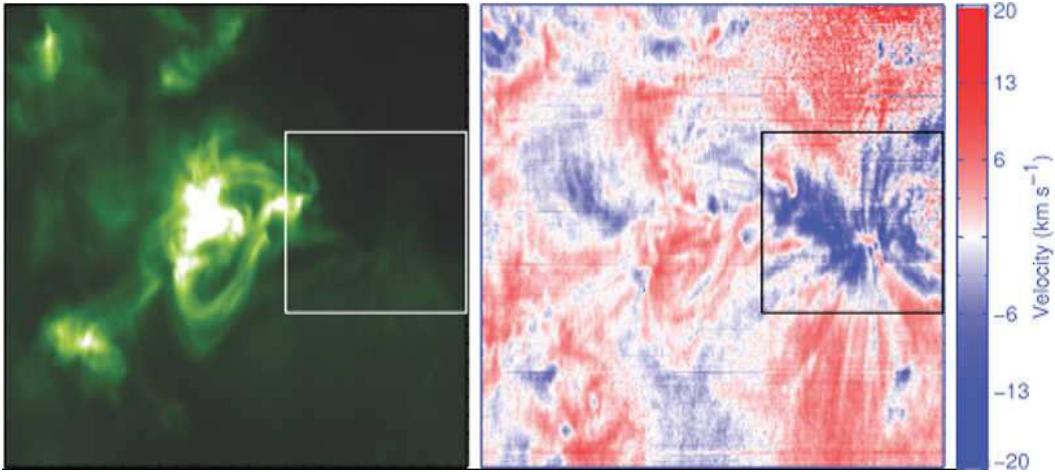}
  \vspace{2mm}
  \caption{Excerpts from \citet{doschek2008}. 
    \textit{Right}: Intensity map of Fe \textsc{xii} $195.12${\AA} for an active region taken on 2007 August 23. 
    \textit{Left}: Doppler velocity map. \textit{Blue} (\textit{Red}) indicates that the plasma moves toward (away) us.}
  \label{img:doschek_2008}
\end{figure}

%These AR outflows are apparently located at the less bright region in the AR edge rather than the core.  
\citet{doschek2008} analyzed emission line profiles of Fe \textsc{xii} $195.12${\AA} and revealed that the outflows are observed at the dark region outside an active region core as seen in Fig.~\ref{img:doschek_2008}.  A preliminary result from EIS has shown that there is a clear boundary between closed hot loops in the AR core ($\sim 3 \times 10^6 \, \mathrm{K}$) and extended cool loops ($\lesssim 1 \times 10^6 \, \mathrm{K}$) where the blueshift was observed \citep{delzanna2008}.  The upflows were seen in the low density and low radiance area.  Meanwhile, redshift was observed in the AR core for all emission lines (Fe \textsc{viii}--\textsc{xv}).  This apparent lack of signatures of any upflows at active region cores was explained as the situation that strong rest component in line profiles hinders the signal of upflows \citep{doschek2012}, but it has not been proved yet.  

% QSL
The magnetic configuration of the outflow region has been modeled by magnetic field extrapolation from the photospheric magnetogram \citep{harra2008,baker2009}, and it was revealed that AR outflows emanate from the footpoints of extremely long coronal loops in txhe edge of an active region \citep{harra2008}.  Close investigation revealed that AR outflows are located near the footpoints of quasi separatrix layers (QSLs), which forms the changes of the connectivity of the magnetic fields from closed coronal loops into open regions \citep{baker2009,delzanna2011}. 

% Velocity of AR outflows
The velocity of the outflow lies within the range of a few tens up to $\sim 100 \, \mathrm{km} \, \mathrm{s}^{-1}$.  These velocities were derived by subtracting the fitted single-Gaussian from raw line profiles \citep{hara2008}, and by double-Gaussian fitting \citep{bryans2010}.  By using extrapolated magnetic fields, the real velocity was derived from Doppler velocity and found to have a speed of $60$--$125 \, \mathrm{km} \, \mathrm{s}^{-1}$ \citep{harra2008}.  The upflow velocity of AR outflows increases with the formation temperature of which emission lines Si \textsc{vii}--Fe \textsc{xv} represent \citep{warren2011}.  The blueshift becomes larger in hotter emission line as $5\text{--}20 \, \mathrm{km} \, \mathrm{s}^{-1}$ for Fe \textsc{xii} (formed at $\sim 1 \times 10^6 \, \mathrm{K}$) and $10\text{--}30 \, \mathrm{km} \, \mathrm{s}^{-1}$ for Fe \textsc{xv} (formed at $\sim 3 \times 10^6 \, \mathrm{K}$) \citep{delzanna2008}.  The appearance of the blueshifted regions often seems to trace the loop-like structures, however, it is not completely understood whether the AR outflows are related to fan loop structures \citep{warren2011,tian2011,mcintosh2012}, which will be discussed later (Section \ref{sect:dis_outflow_fan}). 

% Line profiles of AR outflows
\begin{figure}
  \centering
  %bb=0 0 988 447,
  \includegraphics[width=14cm,clip]{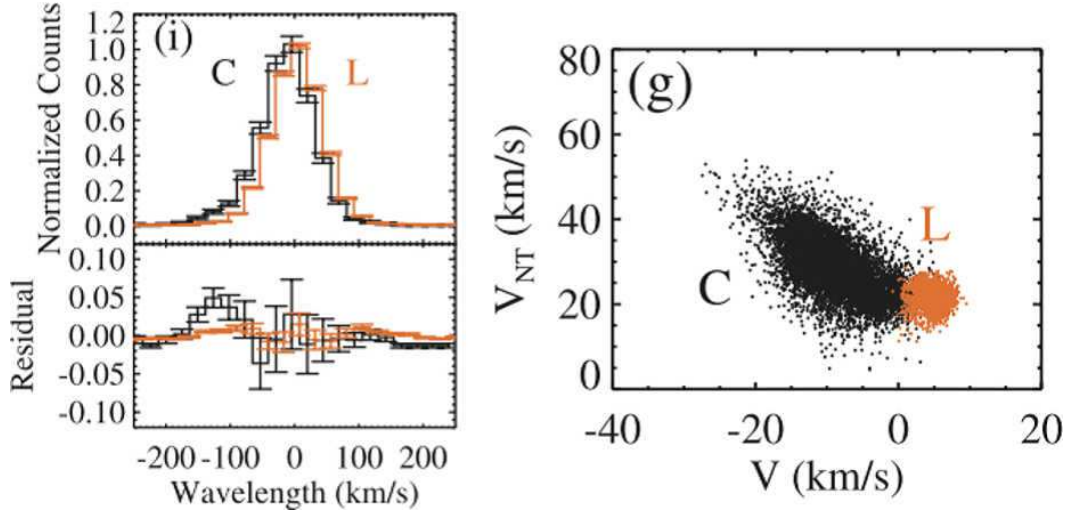}
  \vspace{2mm}
  \caption{Excerpts from \citet{hara2008}. 
    \textit{Left}: line profiles of Fe \textsc{xiv} $274.20${\AA} (\textit{top}) and residuals from a fitted single Gaussian (\textit{bottom}). 
    \textit{Right}: a scatter plot for Doppler and nonthermal velocities. 
    C (\textit{black}) and L (\textit{orange}) in both panels respectively indicate the observation at the disk center and at the limb.}
  \label{img:hara_2008}
\end{figure}

AR outflows are observed as an enhanced blue wing component (EBW) in the emission line profile of Fe \textsc{xii}--\textsc{xv}.  An example for Fe \textsc{xiv} $274.20${\AA} is shown in the \textit{left} panel of Fig.~\ref{img:hara_2008}.  By fitting the line profiles by a single Gaussian, it was revealed that there is a negative correlation between blueshifts and line widths \citep{doschek2008,hara2008} as seen in the \textit{right} panel of Fig.~\ref{img:hara_2008}, which indicates the existence of unresolved component in the blue wing emitted from the upflowing plasma.  \citet{hara2008} investigated the line profile of Fe \textsc{xiv} and Fe \textsc{xv} ($\sim 2 \times 10^{6} \, \mathrm{K}$) both at the disk center and at the limb (see \textit{black} and \textit{orange} plots in Fig.~\ref{img:hara_2008}), and revealed that EBWs were clearly observed only at the disk center, which means that the upflow is dominantly in the radial direction as to the solar surface.  This EBW does not exceed the major component at the rest by $\sim 25${\%} in terms of the intensity \citep{doschek2012}.

% Unresolved problem
Observations so far have revealed properties of the outflow from the edge of active region such as (1) location: less bright region outside the edge of active region core, (2) magnetic topology: boundary between open magnetic fields and closed loops, and (3) the velocity: reaching up to $\sim 100 \, \mathrm{km} \, \mathrm{s}^{-1}$ in the coronal temperature. 

% In this thesis,
The velocity of the outflows in the transition region has not been investigated which becomes important because it decides whether the plasma in all temperature range including the transition region and the corona flows out from the region, which will be described in Section \ref{chap:vel}.  In addition, the density of the outflow itself has not been investigated yet, which may be a crucial clue to reveal the driving mechanism (see the next section), and we will measure the density in Chapter \ref{chap:dns}.  

% --- End of TeX ---

%% file: tex/itdn_aroutflows_mech.tex
% =========================================
%   Chapter:
%     Introduction.
%   Section:
%     Driving mechanisms of AR outflows.
% =========================================

There are several types of driving mechanism of the AR outflow proposed so far. They are classified as (1) impulsive heating events concentrated at the footpoints of coronal loops \citep{hara2008,delzanna2008}, (2) the reconnection between open extended long loops located outside AR and inner loops \citep{harra2008,baker2009}, (3) horizontal expansion of active region \citep{murray2010}, and (4) chromospheric spicules \citep{mcintosh2009a,depontieu2011}. 

% Impulsive heating
\citet{delzanna2008} interpreted that the upflow observed outside the active region core as gentle chromospheric evaporation induced by the reconnection (\textit{i.e.}, nanoflare) as a consequence of the braiding of magnetic footpoints at the photosphere, though the clear evidence has not been shown yet.  While almost all of the observations on the outflow focus on the boundary where the magnetic topology changes (\textit{i.e.}, QSL), it was revealed that an upflow actually occurs at the footpoint of the closed active region loop \citep{hara2008}. The existence of upflows with a speed up to $\simeq 100 \, \mathrm{km} \, \mathrm{s}^{-1}$ in emission line profiles of Fe \textsc{xiv} $274.20${\AA} and \textsc{xv} $284.16${\AA} concentrated toward the footpoints was revealed, and that the line widths were also broadened. From those results, it was concluded that these observational results support the idea of impulsive heating of the lower corona \citep{serio1981,aschwanden2000c} instead of the uniform heating of corona loops \citep{rosner1978}. 

% Interchange reconnection
The outflows investigated so far tend to continue several days, from which some authors favor the interpretation of the driving mechanism in terms of interchange reconnection between the pre-existing long magnetic field in the quiet region and AR edge \citep{harra2008,baker2009}.  Once the reconnection between closed loops and open field occurs, the dense plasma in the closed loops are no longer trapped, and accerelated by a pressure gradient and magnetic tension into the reconnected longer structure \citep{baker2009}.  One-dimensional hydrodynamic simulation showed that the rarefaction wave develops at the reconnected point where the jump in the pressure exists between hot core loop and much longer cool loop, which could produce observed velocity up to $50 \, \mathrm{km} \, \mathrm{s}^{-1}$ and line width \citep{bradshaw2011}.  Their simulation also indicated the dependence of velocity on the temperature consistent with observations.  However, the emission line profiles synthesized in their study were all symmetric, which differ from those observed at the outflow region (\textit{i.e.}, asymmetric and have a enhanced tail in their blue wing).

% AR expansion
Among the driving mechanisms cited above, expansion of coronal loops \citep{murray2010} does not need the existence of magnetic reconnection.  In their simulation, homogeneous magnetic field vertical to the solar surface was imposed at the initial.  They set a strong magnetic flux tube with twist (which is much stronger than real AR) beneath the solar surface, which emerges from the interior of the Sun and expands until the magnetic pressure of the flux tube balances. The flux emergence is often observed in magnetograms, and it is regarded as the birth of an active region.  Expanding tube pushes the initially existing atmosphere where the gas pressure increases due to the compression by the expanding tube. As a result, increased pressure forces the plasma to be accelerated up to $45 \, \mathrm{km} \, \mathrm{s}^{-1}$ at the edge of expanding tube.  However, there is one difficulty when trying to explain the persist nature of outflows.  Their simulation only lasts for $25 \, \mathrm{min}$, which is much shorter than the whole lifetime of an active region ($\sim \mathrm{month}$). 

% Chromospheric spicules
Different from other three mechanisms, \citet{mcintosh2009a} and \citet{depontieu2011} suggested that the outflows are strongly coupled with chromospheric spicules.  A small fraction at the tip of spicules observed with AIA was heated to temperature above $10^{6} \, \mathrm{K}$, and the propagating features were detected which have a speed of $\sim 50\text{--}100 \, \mathrm{km} \, \mathrm{s}^{-1}$.  These were interpreted as a counterpart of the outflows observed with \textit{Hinode}/EIS.  However, the heating mechanism of the tips of spicules has not mentioned.  

% --- End of Contents ---

%% file: tex/itdn_motivation.tex
In this thesis, we focus on the outflow from the edge of active region which was discovered by \textit{Hinode}/EIS.  Main purpose is to clarify several aspects of the outflow region which have not been revealed yet: (1) Doppler velocity within a temperature range of $\log T \, [\mathrm{K}] = 5.5$--$6.5$, and (2) the electron density.  It has been already revealed that the fast upflow is seen in the coronal emission lines, but we do not precisely know the behavior of the transition region lines in the outflow region, which will be crucial information to understand the mass transport in the outflow region.  The electron density may be a clue to understand the origin of the outflow and also can be used to evaluate the gas pressure.  It helps us to consider how the outflow could be driven, combined with the magnetic field information. 

This thesis is structured as follows.  Two chapters following this introduction treat the preparation for the analysis.  Chapter \ref{chap:diag} is a brief introduction to the EIS instrument and the electron density diagnostics in emission line spectroscopy.  The measurement of the spatially averaged Doppler velocity in the quiet region by investigating the center-to-limb variation of Doppler shift will be described in Chapter \ref{chap:cal}.  After that measurement, the Doppler velocities of emission lines within a wide temperature range of $\log T \, [\mathrm{K}] = 5.5$--$6.5$ were measured in the outflow region by referring the quiet region as zero-point of the Doppler velocity which will described in Chapter \ref{chap:vel}.  The electron density of the AR outflow was derived in Chapter \ref{chap:dns} by using a density-sensitive line pair Fe \textsc{xiv} $264.78${\AA}/$274.20${\AA}.  Chapter \ref{chap:ndv} describes a new line profile analysis from a different point of view ($\lambda$-$n_\mathrm{e}$ diagram).  We discuss the nature of the outflow region in Chapter \ref{chap:dis}.  Finally, Chapter \ref{chap:rmk} will provide conclusions of this thesis.  The potential magnetic field will be calculated around the active region in Appendix \ref{chap:mph}, which helps us to consider the morphology of the outflow region.  

% --- End of Contents ---

%% file: tex/contents_diag.tex
\chapter{Diagnostics and instruments}
  \label{chap:diag}
\section{Emission line spectroscopy}
  \subsection{Spectral line profile}
    \input{tex/itdn_emission_corona.tex}
    \input{tex/itdn_line_profile.tex}
  \subsection{Density diagnostics}
    \label{itdn_dens_diag}
    \input{tex/itdn_dens_diag.tex}
\section{Instruments}
  \subsection{Hinode spacecraft}
    \input{tex/itdn_hinode.tex}
  \subsection{EUV Imaging Spectrometer onboard Hinode}
    \input{tex/itdn_eis.tex}

%% file: tex/itdn_emission_corona.tex
% =========================================
%   Chapter:
%     Introduction.
%   Section:
%     Emission from the solar corona.
% =========================================

\begin{figure}
  \centering
  \includegraphics[width=16.8cm,clip]{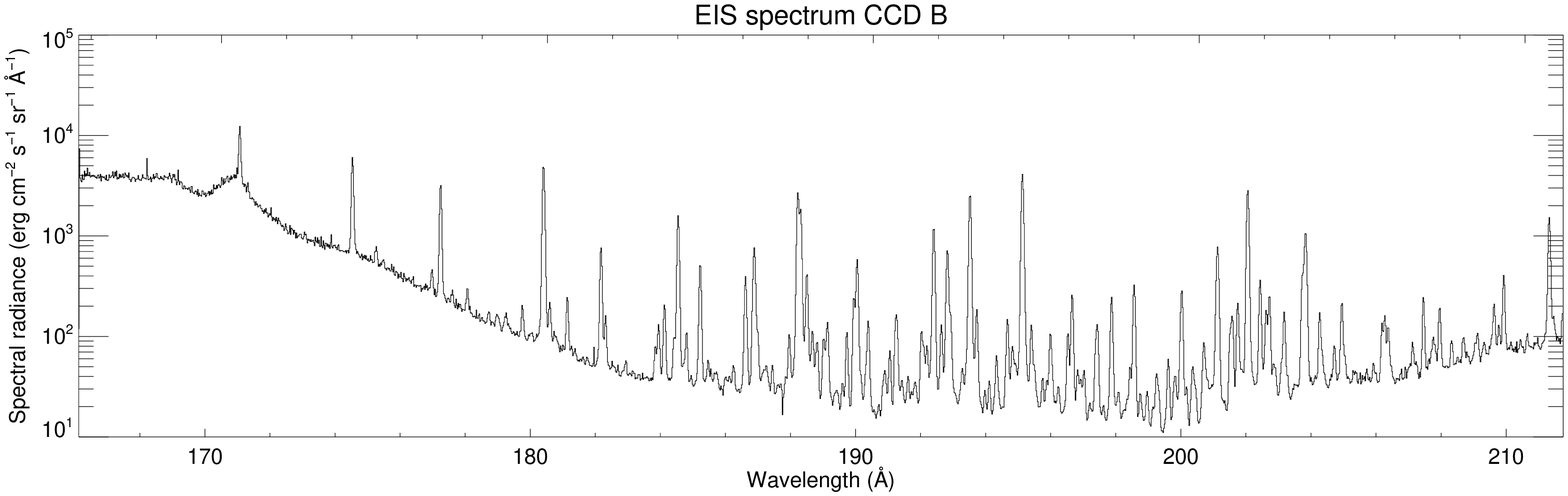}
  \includegraphics[width=16.8cm,clip]{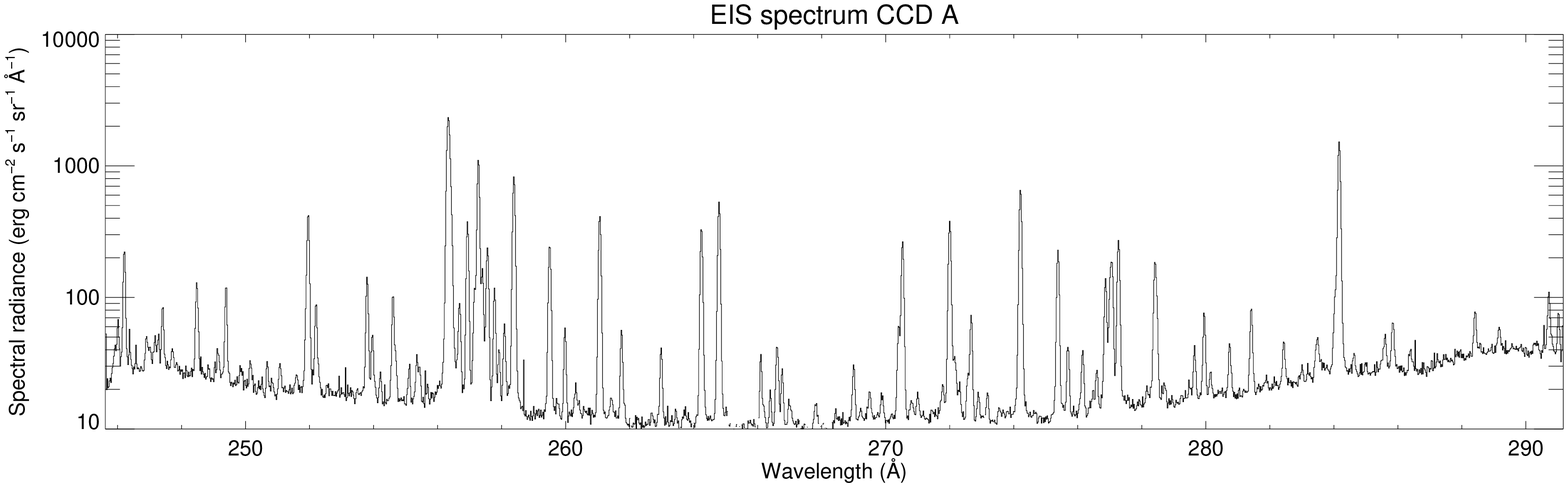}
  \caption{EUV spectra obtained by \textit{Hinode}/EIS on 2007 August 14.
    \textit{Upper}: $166$--$211${\AA} (CCD B).
    \textit{Lower}: $245$--$291${\AA} (CCD A).
  }
  \label{fig:eis_full_spec}
\end{figure}

The corona, filled with highly ionized ions, produces line emissions in the extreme ultraviolet (EUV) wavelength range.  There are several emission mechanisms such as bremsstrahlung, stimulated emission, spontaneous emission, radiative recombination, etc.  In the coronal condition, the spontaneous emission dominates which makes an ion decaying from an upper level into a lower level.  The most important process which causes the excitation from one energy level to upper level in the solar corona is inelastic collisions between ions and free electrons.  Inelastic collisions are involved with almost all of the emission lines whose wavelength is shorter than $2000${\AA}.  An electron-ion inelastic collision can be described as 
\begin{equation}
  X^{\, +m}_{i} (E_i) + e (E_1) \rightarrow X^{\, +m}_{j} (E_j) + e (E_2) \, \text{,}
\end{equation}
where $i$ and $j$ indicate the initial and final levels of the ion $X^{\, +m}$, $E_i$ and $E_j$ are the initial and final energies, and $E_1$ and $E_2$ are the initial and final energies of the free electron.  The ion in the final state can be de-excited spontaneously and emit one photon,  
\begin{equation}
  X^{\, +m}_{j} \rightarrow X^{\, +m}_{i} + h \nu_{ij} \, \text{,}
\end{equation}
where $\nu_{ij} = ( E_j - E_i ) / h$. Since the energy levels are discretized, a spectrum of an emission line has a sharp peak as a function of the frequency (\textit{i.e.}, also the wavelength).  Fig.~\ref{fig:eis_full_spec} shows an example of EUV spectra obtained by EUV Imaging Spectrometer (EIS) onboard \textit{Hinode}.  There are a number of peaks in the spectra emitted from highly ionized ions of He, O, Mg, Si, S, Ca, Fe, etc. 

% --- End of Contents ---

%% file: tex/itdn_line_profile.tex
% ==========================
%   Line profile analysis
% ==========================

Observed spectra actually have a broadened shape (\textit{i.e.}, not the delta function).  There are several reasons which make the spectra broadened: (1) \textit{natural broadening}, (2) \textit{pressure broadening}, (3) \textit{thermal Doppler broadening}, and (4) \textit{turbulence or superposition of flows}, etc. Each mechanism will be described shortly in the following.

% (1) Natural broadening
\paragraph{(1) Natural broadening}
Natural broadening essentially rises from the uncertainty in energy and time.  Here we deal with this mechanism from only a classical view point of damped oscillation.  Spontaneous emission coefficient $A_{ji}$ (Einstein's coefficient $A$; unit is $\mathrm{s}^{-1}$) is introduced.  The irradiance of transition $j \rightarrow i$ from an ion can be written as
\begin{equation}
  I(t) = I(0) \, \mathrm{e}^{-A_{ji} t} \, \text{.}
\end{equation}
Considering $I(t) \propto \lvert E (t) \rvert^2$, the electric field has the form
\begin{equation}
  E(t) = E(0) \, \mathrm{e}^{i \omega_{ij} t} \exp \left( -\frac{A_{ji}}{2} t \right) \, \text{,}
\end{equation}
where $\omega_{ij} = 2 \pi \nu_{ij}$. Taking the Fourier transform of this $E(t)$ results in
\begin{align}
  \hat{E} (\omega) & \propto 
  \int^{\infty}_{0} \mathrm{e}^{-i \omega t} 
  \exp \left( -\frac{A_{ji}}{2} t \right)
  \mathrm{e}^{i \omega_{ij} t} dt \nonumber \\
  & = \dfrac{1}{i(\omega_{ij} - \omega) - \dfrac{A_{ji}}{2}} \, \text{.}
\end{align}
The spectral line profile formed by natural broadening is then represented by the square absolute, 
\begin{equation}
  \phi_{\mathrm{Nat}} (\nu) 
  = \dfrac{1}{\pi} \dfrac{\dfrac{A_{ji}}{4 \pi}}{(\nu - \nu_{ij})^2 
    + \left( \dfrac{A_{ji}}{4 \pi} \right)^2} \, \text{,}
\end{equation}
where the coefficients are multiplied for the normalization in which the integration of line profile becomes unity. This profile is usually referred to as Lorentzian profile. The Full Width of Half Maximum (FWHM) of this profile is $\Delta \nu_{\mathrm{Nat}} = A_{ji} / (4 \pi)$, and in terms of the wavelength it is converted into 
\begin{equation}
  \Delta \lambda_{\mathrm{Nat}} = \dfrac{\lambda^2}{c} \dfrac{A_{ji}}{4 \pi} \, \text{.}
\end{equation}
For wavelength $\lambda = 200${\AA} and typical value $A_{ji} \sim 10^{9} \, \mathrm{s}^{-1}$, we can evaluate the natural broadening in the wavelength as $\Delta \lambda_{\mathrm{Nat}} \simeq 1.1 \times 10^{-6}${\AA}. 

% (2) Pressure broadening
\paragraph{(2) Pressure broadening}
Now we consider other mechanism which changes the phase of emission abruptly: collision with other electrons.  Using the mean free path $l_{\mathrm{mfp}}$ and thermal velocity $v_{\mathrm{th}}$, this process is characterized by the time scale $\tau_{\mathrm{pr}} = l_{\mathrm{mfp}} / v_{\mathrm{th}}$, inverse of which is a counterpart of the spontaneous emission coefficient $A_{ji}$ in natural broadening.  The mean free path is represented as
\begin{equation}
  l_{\mathrm{mfp}} = \dfrac{1}{n \sigma_{\mathrm{cs}}} \, \text{,}
\end{equation}
where $n$ is the particle density and $\sigma_{\mathrm{cs}}$ is the cross section for the collision between particles. For an ion in a ionized degree of $Z$, the cross section $\sigma_{\mathrm{cs}}$ can be evaluated by
\begin{equation}
  \dfrac{1}{2} k_{\mathrm{B}} T = \dfrac{1}{4 \pi} \dfrac{ Z e^2 }{\sqrt{\sigma_{\mathrm{cs}}}} \, \text{.}
  \label{eq:cross_section}
\end{equation}
Then, the FWHM for pressure broadening $\Delta \nu_{\mathrm{Pr}}$ can be written as
\begin{equation}
  \Delta \nu_{\mathrm{Pr}} = \dfrac{1}{4 \pi} \dfrac{v_{\mathrm{th}}}{l_{\mathrm{mfp}}}
  = \dfrac{1}{4 \pi} v_{\mathrm{th}} n \sigma_{\mathrm{cs}} 
  = \dfrac{1}{16 \pi^3} \frac{n Z^2 e^4}{m_e^{1/2} \left( k_{\mathrm{B}} T \right)^{3/2}} \, \text{.}
  \label{eq:pressure_broadening_fwhm}
\end{equation}
In terms of the wavelength, 
\begin{equation}
  \Delta \lambda_{\mathrm{Pr}} = \dfrac{\lambda^2}{c} \Delta \nu_{\mathrm{Pr}}
\end{equation}
as same as natural broadening. 
Using typical values in the corona $n = 10^{9} \, \mathrm{cm}^{-3}$, $T = 10^{6} \, \mathrm{K}$, and assuming $\lambda = 200${\AA} and $Z=11$ (\textit{e.g.}, Fe \textsc{XII}\footnote{In spectroscopic literature, we deal a neutral atom with denoting ``\textsc{I}''. First degree ion is represented by denoting ``\textsc{II}'', and so on.}), we obtain $\Delta \lambda_{\mathrm{Pr}} \simeq 3.6 \times 10^{-15}${\AA}. Obviously, pressure broadening is much smaller than natural broadening. Eq. (\ref{eq:pressure_broadening_fwhm}) shows that pressure broadening is proportional to the density, and in the solar corona where $n = 10^{9-11} \, \mathrm{cm}^{-3}$, pressure broadening is always negligible compared to other broadening mechanisms. 

% (3) Doppler broadening
\paragraph{(3) Thermal Doppler broadening}
The line-of-sight velocity of particles in plasma $u$ (we do not use $v$ to avoid the complexity with the frequency $\nu$) obeys to the Maxwell-Boltzmann distribution,
\begin{equation}
  f_{\mathrm{MB}} (u) = \left( \dfrac{M_{\mathrm{i}}}{2 \pi k_{\mathrm{B}} T_{\mathrm{i}}} \right)^{1/2}
  \exp \left[ - \dfrac{M_{\mathrm{i}} u^2}{2 \pi k_{\mathrm{B}} T_{\mathrm{i}}} \right] \, \text{,}
  \label{eq:mb_dist}
\end{equation}
after the velocity perpendicular to the line of sight is integrated.  The emission frequency from a moving particle increases (decreases) when the particle moves toward (away) from an observer.  This is the Doppler effect of the light, and in non-relativistic case the frequency and the wavelength are modified as
\begin{equation}
  \nu^{\prime} = \nu_{0} \left( 1 - \dfrac{u}{c} \right) \, \text{,} 
\end{equation}
\begin{equation}
  \lambda^{\prime} = \lambda_{0} \left( 1 + \dfrac{u}{c} \right) \, \text{.} 
\end{equation}
A suffix $_{0}$ indicates the original quantities, the prime $^{\prime}$ means the modified quantities, and positive (negative) velocity indicates the motion away (toward) from the observer.  The FWHM of thermal Doppler broadening becomes
\begin{equation}
  \Delta \nu_{\mathrm{Dop}} = 
  \dfrac{\nu_{0}}{c} \left( 4 \ln 2 \cdot 
  \dfrac{2 k_{\mathrm{B}} T_{\mathrm{i}}}{M_{\mathrm{i}}} \right)^{1/2} \, \text{,}
\end{equation}
which is written in terms of the wavelength as
\begin{equation}
  \Delta \lambda_{\mathrm{Dop}} = 
  \dfrac{\lambda_{0}}{c} \left( 4 \ln 2 \cdot 
  \dfrac{2 k_{\mathrm{B}} T_{\mathrm{i}}}{M_{\mathrm{i}}} \right)^{1/2} \, \text{.}
\end{equation}
For $\lambda_{0} = 200${\AA}, $T_{\mathrm{i}} = 10^{6} \, \mathrm{K}$, and assuming iron ions $M_{\mathrm{i}} = 56 m_{\mathrm{p}}$ ($m_{\mathrm{p}}=1.67 \times 10^{-24} \, \mathrm{g}$: proton mass), this takes a value of $\Delta \lambda_{\mathrm{Dop}} \simeq 1.9 \times 10^{-2} \, \text{\AA}$. Comparing three broadening mechanisms, it is clear that thermal Doppler broadening dominates in the solar corona. 
% Voigt function
Now we derive the spectral line profile taking into account the Doppler-shifted natural broadening,
\begin{equation}
  \phi_{\mathrm{Nat}} (\nu, u) = \dfrac{1}{\pi} 
  \dfrac{\dfrac{A}{4 \pi}}
        {\left( \nu - \nu_{0} + \nu_{0} \dfrac{u}{c} \right)^2 
          + \left( \dfrac{A}{4 \pi} \right)^2} \, \text{.}
  \label{eq:ds_nb}
\end{equation}
The term $A$ in the right-hand side denotes $A_{ji}$, where the suffix is omitted for the simplicity. The spectral line profile can be calculated as a convolution of Eq. (\ref{eq:ds_nb}) and the Maxwell-Boltzmann distribution Eq. (\ref{eq:mb_dist}), 
\begin{align}
  \psi (\nu) 
  & = \int_{-\infty}^{\infty} f_{\mathrm{MB}} (u) \phi_{\mathrm{Nat}} (\nu, u) du \nonumber \\
  & = \dfrac{1}{\pi} \left( \dfrac{M_{\mathrm{i}}}{2 \pi k_{\mathrm{B}} T_{\mathrm{i}}} \right)^{1/2}
  \int_{-\infty}^{\infty}
  \dfrac{\dfrac{A}{4 \pi}}
       {\left( \nu - \nu_{0} + \nu_{0} \dfrac{u}{c} \right)^2 
         + \left( \dfrac{A}{4 \pi} \right)^2}
  \exp \left[ - \dfrac{M_{\mathrm{i}} u^2}{2 \pi k_{\mathrm{B}} T_{\mathrm{i}}} \right] du \nonumber \\
  & = \dfrac{1}{\sqrt{2 \pi} \sigma_{\nu}}
  \dfrac{a}{\pi} 
  \int_{-\infty}^{\infty} 
  \dfrac{\mathrm{e}^{-\mu^2}}{\left( \mu + x_{\nu} \right)^2 + a^2} d\mu \nonumber \\
  & = \dfrac{1}{\sqrt{2 \pi} \sigma_{\nu}} V (x_{\nu}; \sigma_{\nu}, a) \, \text{,} 
\end{align}
where $\mu = \left[ M_{\mathrm{i}} / \left(2 \pi k_{\mathrm{B}} T_{\mathrm{i}} \right) \right]^{1/2} u$, $\sigma_{\nu} = \left( \nu_{0} / c \right) \left( k_{\mathrm{B}} T_{\mathrm{i}} / M_{\mathrm{i}} \right)^{1/2}$, $a = A / ( 4 \pi \cdot \sqrt{2} \sigma_{\nu} )$, and $x_{\nu} = \left( \nu - \nu_{0} \right) / ( \sqrt{2} \sigma_{\nu} )$. The function $V (x_{\nu}; \sigma_{\nu}, a)$ is called Voigt profile
\begin{align}
  V (x_{\nu}; \sigma_{\nu}, a) 
  & = \dfrac{a}{\pi} 
  \int_{-\infty}^{\infty} 
  \dfrac{\mathrm{e}^{-\mu^2}}{\left( \mu + x_{\nu} \right)^2 + a^2} d\mu \nonumber \\
  & \simeq 
  \begin{cases}
    \mathrm{e}^{- x_{\nu}^2} & (x_{\nu} \ll 1) \\
    \dfrac{a}{\sqrt{\pi}} \dfrac{1}{x_{\nu}^2} & (x_{\nu} \gg 1) \, \text{.}
  \end{cases}
\end{align}
Thus, the center of line profile is approximately Gaussian profile, and the wing of line profile is dominated by Lorentzian profile.  
%{\color{red} 
However, observed EUV line profiles are well fitted by a Gaussian profile because the line wings are weak compared to the sensitivity of spectrometers. 
%}
Therefore, spectral line profile is represented as
\begin{equation}
  \psi (\nu) = \dfrac{1}{\sqrt{2 \pi} \sigma_{\nu}} 
  \exp \left[ - \dfrac{\left( \nu - \nu_{0} \right)^2}{2 \sigma_{\nu}^2} \right] \, \text{,} 
\end{equation}
or, if writing it in terms of wavelength, it becomes
\begin{equation}
  \tilde{\psi} (\lambda) = \dfrac{1}{\sqrt{2 \pi} \sigma_{\lambda}} 
  \exp \left[ - \dfrac{\left( \lambda - \lambda_{0} \right)^2}{2 \sigma_{\lambda}^2} \right] \, \text{,} 
\end{equation}
where $\lambda_{0} = c / \nu_{0}$, and $\sigma_{\lambda} = \lambda_{0}^2 \sigma_{\nu} / c$. 

% (4) Other broadening mechanisms
\paragraph{(4) Other broadening mechanisms}
The emission from plasma in isothermal (\textit{i.e.}, homogeneous temperature) and without incoherent bulk motion forms spectral line profile represented by a Gaussian as described above.  If several plasma blobs exist along the line of sight, the spectral line profile should be the superposition of each Gaussian formed by each blob since the plasma in the solar corona is optically thin.  In addition, the spectral line profile would be broadened by instrumental effects. Thus, observed line profile are formed through these factors, and the width of observed Gaussian $\sigma_{\mathrm{obs}}$ can be represented as
\begin{equation}
  \sigma_{\mathrm{obs}} = \left( \sigma_{\lambda}^2 
  + \sigma_{\mathrm{NT}}^2 
  + \sigma_{\mathrm{Inst}}^2 \right)^{1/2} \, \text{,} 
\end{equation}
where $\sigma_{\mathrm{NT}}$ is nonthermal width (it does not mean those of high energy particles beyond the Maxwellian distribution, but it does mean excess broadening which cannot be attributed to thermal Doppler motion), and $\sigma_{\mathrm{Inst}}$ is the broadening caused by the instrument. 

% Example of line profiles
\begin{figure}
  \centering
  \includegraphics[width=14cm,clip]{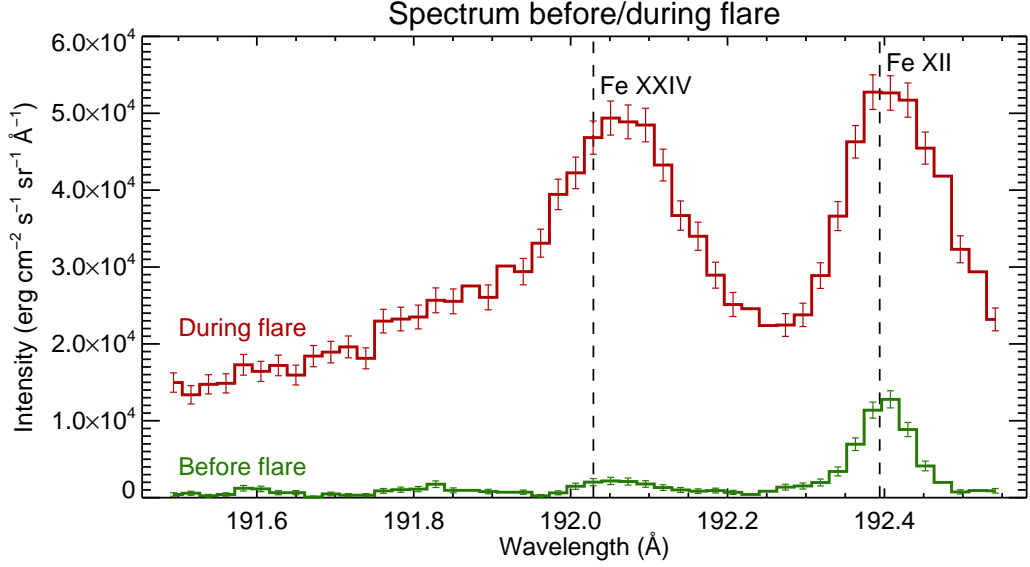}
  \caption{Line profiles of Fe \textsc{xxiv} $192.03${\AA} and Fe \textsc{xii} $192.39${\AA} observed by \textit{Hinode}/EIS on 2011 September 9.  \textit{Green} (\textit{Red}) profile shows the line profile at the flare site before (during) an M1.2-class flare.  \textit{Vertical dashed} lines indicate the rest wavelengths given by CHIANTI ver.~7 \citep{dere1997,landi2012}.}
  \label{itdn_eis_lp_examples}
\end{figure}

An example of line profiles is shown in Fig.~\ref{itdn_eis_lp_examples}.  \textit{Green} (\textit{Red}) profile shows the line profile at the flare site before (during) an M-class flare on 2011 September 9.  In the wavelength range plotted in the figure, there are two strong lines: Fe \textsc{xxiv} $192.03${\AA} ($\sim 10^7 \, \mathrm{K}$) and Fe \textsc{xii} $192.39${\AA} ($\sim 10^6 \, \mathrm{K}$).  Before the onset of the flare, Fe \textsc{xii} $192.39${\AA} was the only one prominent emission line.  During the flare, Fe \textsc{xxiv} $192.03${\AA} increases in its strength comparable to the neighbor Fe \textsc{xii} $192.39${\AA} because the high-temperature plasma (up to $\sim 10^7 \, \mathrm{K}$) is produced by the flare.  In addition, the spectrum of Fe \textsc{xxiv} $192.03${\AA} during the flare shows the shape far different from a single Gaussian profile.  It has a long tail in the shorter side in the wavelength direction, which is often referred to as enhanced blue wing (EBW).  This line profile is considered to be composed of the rest component and a broadened blueshifted component.  As this example shows, line profiles change their shapes depending on cases. 

% --- End of Contents ---

%% file: tex/itdn_dens_diag.tex
% ============
%   Density
% ============

Density of the solar corona has been often derived by two methods: EM method and the line ratio method. 
% EM method
In EM method, we use the intensity of an emission line.  When the assumption that the observed plasma has uniform temperature (\textit{i.e.}, isothermal) along the line of sight, the intensity of an emission line is expressed by
\begin{equation}
  I = n_{\mathrm{e}}^2 \, G(n_{\mathrm{e}}, T) \, h \, \text{,} 
  \label{eq:int_isothermal}
\end{equation}
where $n_{\mathrm{e}}$ is electron density ($\mathrm{cm}^{-3}$), $G(n_{\mathrm{e}}, T)$ is so-called the contribution function of emission line ($\mathrm{erg} \, \mathrm{cm}^3 \, \mathrm{s}^{-1} \, \mathrm{sr}^{-1} \, \text{\AA}^{-1}$), and $h$ is the column depth of the observed plasma ($\mathrm{cm}$). Note that the unit of intensity is $\mathrm{erg} \, \mathrm{cm}^{-2} \, \mathrm{s}^{-1} \, \mathrm{sr}^{-1} \, \text{\AA}^{-1}$.  The dependence of the contribution function on electron density is usually much weaker than that on temperature, but some emission lines have strong dependence on the electron density, which can be exploited to line ratio method as described after. 
%Example of the contribution function is shown in Fig. xx, from which we see the steep behavior of the contribution function as a function of temperature. 
Using Eq.~(\ref{eq:int_isothermal}), the electron density can be estimated as
\begin{equation}
  n_{\mathrm{e}} = 
  \left[
  \frac{I}{G(n_{\mathrm{e}}, T) h} 
  \right]^{1/2} \, \text{,}
\end{equation}
where the temperature is often assumed to be the formation temperature of the emission line.  Practically, it is difficult to know the precise column depth in observation because the 3D morphology of the solar corona cannot be obtained in most cases.  One way to determine the column depth $h$ is that we assume the circle cross section of coronal loops or the semi-spherical shape of bright points (small scale loops).  Then the column depth is obtained by using the coronal imaging observation.  However, the assumption of isothermal plasma is often violated in the solar corona because (1) the overlapping of several structures along the line of sight, and (2) sub-spatial-resolution fine structure of the solar corona. In addition, we obtain only the lower limit of electron density considering that the contribution function is convex upward as a function of temperature. 

% Line ratio method
Second way for the density diagnostics of the solar corona is the line ratio method.  We use an emission line pair whose intensity ratio has significant dependence on electron density.  If the line ratio is monotonically increases or decreases as a function of the electron density, we can use the line ratio as a tool for the density diagnostics. 

\begin{figure}
  \centering
  \includegraphics[width=12cm,clip]{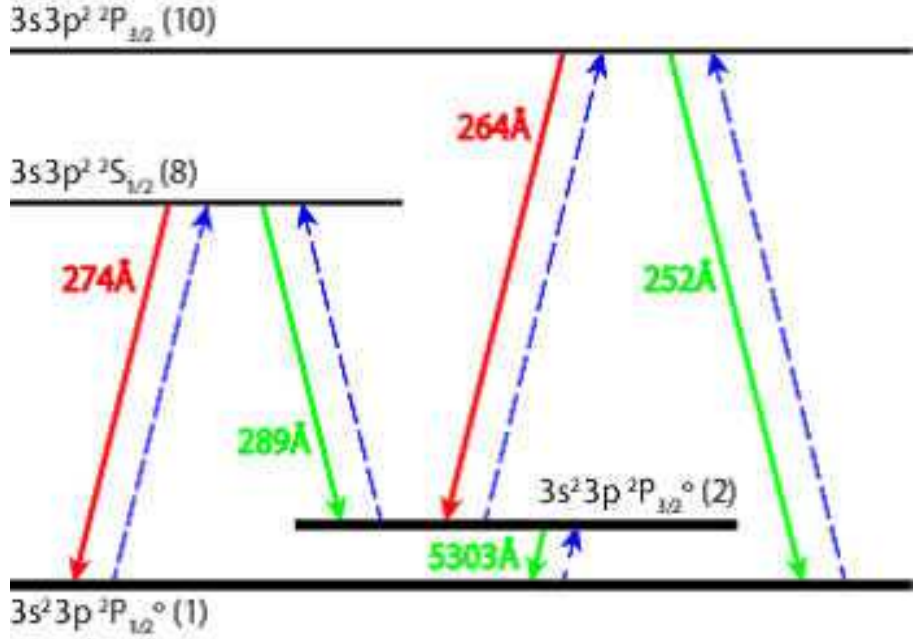}
  \caption{The energy levels and transitions of Fe \textsc{xiv}.  \textit{Horizontal} lines indicate energy levels.  The characters beside the lines indicate configurations of the electrons in Fe \textsc{xiv} ion and terms.  Numbers in the parenthesis after the characters are the index named for each energy level, which are commonly used in the atomic physics.  \textit{Full} (\textit{Dashed}) arrows indicate transitions for collisional excitation (emission).  Two \textit{red} arrows are the emission line pair ($8 \rightarrow 1$; $\lambda = 274.20${\AA}, $10 \rightarrow 2$; $\lambda = 264.78${\AA}) used for density diagnostics in this thesis.}
  \label{itdn_trans_level_fexiv}
\end{figure}

The theory about the dependence of intensity on the electron density is described below.  An example of the emission line pair Fe \textsc{xiv} $264.78${\AA}/$274.20${\AA} is given, both of which involve allowed transitions. 
% Formulation
The energy levels of Fe \textsc{xiv} and related emissions are shown in Fig.~\ref{itdn_trans_level_fexiv}.  In the figure, only transitions having significant influences on the balance of the number of electrons in each level are shown.  Numbers in the parenthesis after the characters are the index named for each energy level, which are commonly used in the atomic physics.  The transition $10 \leftrightarrow 8$ (which produces the emission $\lambda 3143${\AA}) is omitted because it has negligible contribution on the number equation in this system, considering that almost all of the ions are at the configuration $3s^2 3p$ (energy level index: $1$ or $2$). 

We assume the equilibrium between the numbers of ions in each levels, where the timescales of collisional excitation and radiative decay ($\sim 10^{-9} \, \mathrm{s}$) are much shorter than that of the change of temperature.  It almost always holds in the solar corona. Representing the number of ions in the level $i$ by $N_i$, the collisional excitation coefficient by $C^{\mathrm{ex}}_{i, j}$ (transition from lower level $i$ to upper level $j$) and the radiative decay by $A_{j, i}$ (transition from upper level $j$ to lower level $i$), the equilibrium between the numbers in each energy level can be written as
\begin{align}
  A_{2, 1} N_2 + A_{8, 1} N_8 + A_{10, 1} N_{10} 
  & = C^{\mathrm{ex}}_{1, 2} N_1 N_{\mathrm{e}} 
  + C^{\mathrm{ex}}_{1, 8} N_1 N_{\mathrm{e}}
  + C^{\mathrm{ex}}_{1 ,10} N_1 N_{\mathrm{e}} \, \text{,} \\
  C^{\mathrm{ex}}_{1, 2} N_1 N_{\mathrm{e}}
  + A_{8, 2} N_8 
  + A_{10, 2} N_{10} 
  & = C^{\mathrm{ex}}_{2, 8} N_2 N_{\mathrm{e}}
  + C^{\mathrm{ex}}_{2, 10} N_2 N_{\mathrm{e}}
  + A_{2, 1} N_2 \, \text{,} \label{eq:number_trans_level_2} \\
  C^{\mathrm{ex}}_{1, 8} N_1 N_{\mathrm{e}}
  + C^{\mathrm{ex}}_{2, 8} N_2 N_{\mathrm{e}} 
  & = A_{8, 1} N_{8} + A_{8, 2} N_{8} \, \text{,} \label{eq:number_trans_level_3} \\
  C^{\mathrm{ex}}_{1, 10} N_1 N_{\mathrm{e}} + C^{\mathrm{ex}}_{2, 10} N_2 N_{\mathrm{e}}
  & = A_{10, 1} N_{10} + A_{10, 2} N_{10} \, \text{.} \label{eq:number_trans_level_4}
\end{align}
Only three of these equations are independent. If the relative fraction of energy levels as to the ground state would be defined like
\begin{equation}
  \alpha_{2} = \frac{N_2}{N_1} \, \text{,} 
  \, \alpha_{8} = \frac{N_8}{N_1} \, \text{,} 
  \, \alpha_{10} = \frac{N_{10}}{N_1} \, \text{,} 
\end{equation}
Eqs.~(\ref{eq:number_trans_level_2})\text{--}(\ref{eq:number_trans_level_4}) are reduced to
\begin{align}
  \alpha_{2} & = \frac{X_{2}}{A_{2, 1} + X_{1} N_{\mathrm{e}}} N_{\mathrm{e}} \, \text{,} \\
  \alpha_{8} & = \frac{C^{\mathrm{ex}}_{1, 8} 
  + C^{\mathrm{ex}}_{2, 8} \alpha_{2}}{A_{8, \mathrm{tot}}} N_{\mathrm{e}} \, \text{,} \\
  \alpha_{10} & = \frac{C^{\mathrm{ex}}_{1, 10} 
  + C^{\mathrm{ex}}_{2, 10} \alpha_{2}}{A_{10, \mathrm{tot}}} N_{\mathrm{e}} \, \text{,}
\end{align}
where 
\begin{align}
  A_{8, \mathrm{tot}} & = A_{8, 1} + A_{8, 2} \, \mathrm{s}^{-1} \, \text{,} \\
  A_{10, \mathrm{tot}} & = A_{10, 1} + A_{10, 2} \, \mathrm{s}^{-1} \, \text{,} \\
  X_1 & = C^{\mathrm{ex}}_{2, \mathrm{tot}} 
        - \frac{A_{8, 2}}{A_{8, \mathrm{tot}}} C^{\mathrm{ex}}_{2, 8}
        - \frac{A_{10, 2}}{A_{10, \mathrm{tot}}} C^{\mathrm{ex}}_{2, 10} \, \mathrm{cm}^3 \,\mathrm{s}^{-1} \, \text{,}\\
  X_2 & = C^{\mathrm{ex}}_{1, 2} 
        + \frac{A_{8, 2}}{A_{8, \mathrm{tot}}} C^{\mathrm{ex}}_{1, 8}
        + \frac{A_{10, 2}}{A_{10, \mathrm{tot}}} C^{\mathrm{ex}}_{1, 10} \,\mathrm{cm}^3\,\mathrm{s}^{-1} \, \text{,}\\
  C^{\mathrm{ex}}_{2, \mathrm{tot}} & = C^{\mathrm{ex}}_{2, 8} + C^{\mathrm{ex}}_{2, 10} 
  \, \mathrm{cm}^3 \,\mathrm{s}^{-1} \, \text{.}
\end{align}
The radiance of each emission line ($I_{j, i}$; intensity involving the transition from upper level $j$ to lower level $i$) is represented as
\begin{equation}
  I_{j, i} = A_{j, i} N_j \, \text{,} 
\end{equation}
therefore, the ratio between two emission lines $264.78${\AA} (transition: $10 \rightarrow 2$)/$274.20${\AA} (transition: $8 \rightarrow 1$) can be calculated as 
\begin{align}
   \frac{I_{10, 2} (264.78\text{\AA})}{I_{8, 1} (274.20\text{\AA})} 
   & = \frac{A_{10, 2} N_{10}}{A_{8, 1} N_{8}}
     = \frac{A_{10, 2} \alpha_{10}}{A_{8, 1} \alpha_{8}} \\
   & = \frac{A_{10, 2}}{A_{10, \mathrm{tot}}} \frac{A_{8, \mathrm{tot}}}{A_{8, 1}}
       \frac{C^{\mathrm{ex}}_{1, 10} + C^{\mathrm{ex}}_{2, 10} \alpha_{2}}
            {C^{\mathrm{ex}}_{1, 8} + C^{\mathrm{ex}}_{2, 8} \alpha_{2}} \\
   & = \frac{A_{10, 2}}{A_{10, \mathrm{tot}}} \frac{A_{8, \mathrm{tot}}}{A_{8, 1}}
       \frac{X_3 N_{\mathrm{e}} + X_4}{X_5 N_{\mathrm{e}} + X_6} \, \text{,} 
       \label{eq:line_ratio_fexiv}
\end{align}
where
\begin{align}
   X_3 & = C^{\mathrm{ex}}_{1, 10} X_1 + C^{\mathrm{ex}}_{2, 10} X_2 \,\mathrm{cm}^6\,\mathrm{s}^{-2} \, \text{,} \\
   X_4 & = C^{\mathrm{ex}}_{1, 10} A_{2, 1} \,\mathrm{cm}^3\,\mathrm{s}^{-2} \, \text{,} \\
   X_5 & = C^{\mathrm{ex}}_{1, 8} X_1 + C^{\mathrm{ex}}_{2, 8} X_2 \,\mathrm{cm}^6\,\mathrm{s}^{-2} \, \text{,} \\
   X_6 & = C^{\mathrm{ex}}_{1, 8} A_{2, 1} \,\mathrm{cm}^3\,\mathrm{s}^{-2} \, \text{.}
\end{align}
Now it is clear that the line ratio Fe \textsc{xiv} $264.78${\AA}/$274.20${\AA} has a dependence on electron density in terms of a fractional function.  The coefficients $X_{1\text{--}6}$ can be calculated by using the atomic data given by \citet{storey2000} and \citet{tayal2008}, which is listed in Table \ref{tab:atomic_data_fexiv}.  Calculated coefficients $X_{1-6}$ are tabulated in Table \ref{tab:X_1-6}. 

\input{tex/table_atomic_data_fexiv.tex}

\input{tex/table_X_1-6.tex}

\begin{figure}
  \centering
  \includegraphics[width=12.5cm]{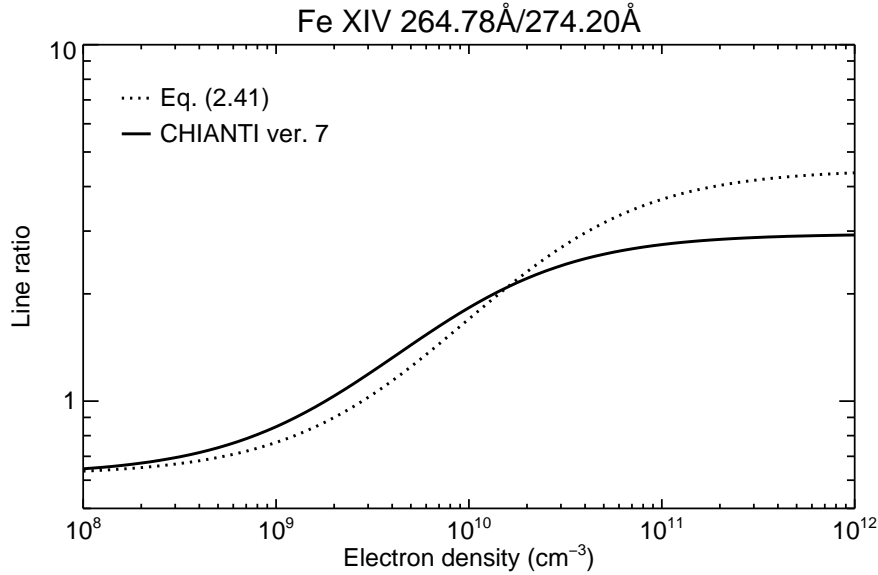}
  \caption{Line ratio of Fe \textsc{xiv} $264.78${\AA}/$274.20${\AA} 
           as a function of electron density calculated by using CHIANTI database. 
           While contribution function of Fe \textsc{xiv} $264.78${\AA} increases 
           with electron density, that of Fe \textsc{xiv} $274.20${\AA} decreases 
           with electron density. This makes the line ratio of these two lines 
           behaves monotonically as a function of electron density.}
  \label{itdn_line_ratio_example}
\end{figure}

% Graph of intensity ratio Fe \textsc{xiv} 264/274
The ratio of intensity from two emission lines Fe \textsc{xiv} $264.78${\AA}/$274.20${\AA} as a function of electron density is shown in Fig.~\ref{itdn_line_ratio_example}. 

The ratio was calculated by using CHIANTI database version 7 \citep{dere1997,landi2012}.  Note that calculating the ratio by using Eq.~(\ref{eq:line_ratio_fexiv}) results in slightly larger value than CHIANTI, however, the behavior is fundamentally the same.  This discrepancy may come from the difference in the atomic data.  In this thesis, we adopt the CHIANTI database because its data is the newest one available at present. 

% --- End of Contents ---

%% file: tex/table_atomic_data_fexiv.tex
\begin{table}
  \centering
  \caption{Transitions involving energy levels $1$, $2$, $8$, and $10$.  \textit{First column}: transitions indicated by indexes of lower and upper level.  \textit{Second column}: wavelength of emission line involving the transition.  \textit{Third column}: collisional excitation coefficients calculated by using effective collision strength in \citet{tayal2008}.  The temperature was assumed to be $2 \times 10^{6} \, \mathrm{K}$. \textit{Fourth column}: spontaneous emission coefficients (Einstein's coefficient $A$) from \citet{storey2000}.}
  %\vspace{1mm}
  \begin{tabular}{lrll} 
    \toprule
    Transition 
    & Wavelength ({\AA}) 
    & $C^{\mathrm{ex}}_{i, j}$ ($\mathrm{cm}^3 \, \mathrm{s}^{-1}$) 
    & $A_{j, i}$ ($\mathrm{s}^{-1}$) \\
    \midrule
    $1-2$ & $5304.49$ & $1.55 \times 10^{-9}$ & $6.023 \times 10^{1}$ \\
    $1-8$ & $274.20$ & $2.21 \times 10^{-9}$ & $1.777 \times 10^{10}$ \\
    $1-10$ & $252.20$ & $1.59 \times 10^{-9}$ & $7.598 \times 10^{9}$ \\
    $2-8$ & $289.15$ & $3.70 \times 10^{-10}$ & $1.147 \times 10^{9}$ \\
    $2-10$ & $264.79$ & $8.03 \times 10^{-9}$ & $3.291 \times 10^{10}$ \\
    \bottomrule
  \end{tabular}
  \label{tab:atomic_data_fexiv}
\end{table} 
% --- End of Contents ---

%% file: tex/table_X_1-6.tex
\begin{table}
  \centering
  \caption{Values of coefficients $X_{1\text{--}6}$.}
  %\vspace{1mm}
  \begin{tabular}{lc} 
    \toprule
    $X_1$ & $1.85 \times 10^{-9}$ ($\mathrm{cm}^3 \, \mathrm{s}^{-1}$) \\
    $X_2$ & $2.97 \times 10^{-9}$ ($\mathrm{cm}^3 \, \mathrm{s}^{-1}$) \\
    $X_3$ & $2.68 \times 10^{-17}$ ($\mathrm{cm}^6 \, \mathrm{s}^{-2}$) \\
    $X_4$ & $9.55 \times 10^{-8}$ ($\mathrm{cm}^3 \, \mathrm{s}^{-2}$) \\
    $X_5$ & $5.19 \times 10^{-18}$ ($\mathrm{cm}^6 \, \mathrm{s}^{-2}$) \\
    $X_6$ & $1.33 \times 10^{-7}$ ($\mathrm{cm}^3 \, \mathrm{s}^{-2}$) \\
    \bottomrule
  \end{tabular}
  \label{tab:X_1-6}
\end{table} 
% --- End of Contents ---

%% file: tex/itdn_hinode.tex
% ======================
%   HINODE spacecraft
% ======================

\textit{Hinode} \citep{kosugi2007} is a Japanese satellite of Institute of Space and Astronomical Science of the Japan Aerospace Exploration Agency (ISAS/JAXA), launched on 2006 September 23 6:36 JST. \textit{Hinode} has three instruments onboard: the Solar Optical Telescope (SOT), the X-ray Telescope (XRT), the EUV Imaging Spectrometer (EIS). The scientific aims are: (1) to understand the processes of magnetic field generation and transport including the magnetic modulation of the Sun's luminosity, (2) to investigate the processes responsible for energy transfer from the photosphere up to the chromosphere and the corona, and (3) to determine the mechanisms which induce eruptive phenomena, such as flares and coronal mass ejections (CMEs), and understand these phenomena in the context of the space weather. %Based upon this scientific motivation, \textit{Hinode} mission was constructed by an international collaboration including institutes in Japan, the United States, and the United Kingdom. ISAS/JAXA has responsibility for the design, development, test and integration of the \textit{Hinode} spacecraft with the National Astronomical Observatory of Japan as a domestic partner and the Mitsubishi Electronic Corporation as a leading contractor. The \textit{Hinode} spacecraft had been called by its development name Solar-B and the name \textit{Hinode} was given after the launch. 

% SOT
The Solar Optical Telescope (SOT) \citep{tsuneta2008} consists of the Optical Telescope and the Focal Plane Package (FPP). The SOT consists of a $50$-cm diffraction limit Gregorian telescope, and the FPP includes the narrowband imager (NFI) and the broadband imager (BFI), and the Stokes Spectropolarimeter (SP). The SOT provides unprecedented high spatial and temporal resolution image of the photosphere and the chromosphere by filtergram of NFI/BFI and vector magnetograms calculated through inversion of SP data. The SOT has revealed many kinds of magnetic activity of the Sun such as magnetic flux emergence, submergence, cancellation, and related response of the photosphere and chromosphere.

% XRT
The X-ray Telescope (XRT) \citep{golub2007} has a grazing-incidence optic and a CCD array. Kinds of filters are utilized: entrance aperture prefilters and focal plane analysis filters. The entrance aperture prefilters have two main purposes: (1) to reduce the visible light entering the instrument and (2) to reduce the heat load in the instrument. The focal plane analysis filters have two purposes: (1) to reduce the visible light reaching the focal plane and (2) to provide varying X-ray passbands for plasma temperature diagnostics. The science objectives of the XRT are chromospheric evaporations, reconnection dynamics, polar jets, and coronal holes. 

% --- End of Contents ---

%% file: tex/itdn_eis.tex
% ===============================
%   Introduction to Hinode/EIS
% ===============================

The EUV Imaging Spectrometer (EIS) \citep{culhane2007} onboard \textit{Hinode} observes the solar corona and the upper transition region emission lines in the wavelength range of $170\text{--}210${\AA} and $250\text{--}290${\AA}.  The emission line centroid position and the emission line width allow us to know the Doppler velocity and the nonthermal velocity of the observed plasma.  The plasma temperature and density can be measured by using the intensity ratio of temperature or density sensitive line pair (detail was given in Section \ref{itdn_dens_diag}).  The science aims of EIS is to investigate the coronal/photospheric velocity field comparison in active regions and understand the dynamics of flares 
%{\color{red}
(\textit{e.g.}, by coordination with SOT),
%} 
and to detect the heating signatures in the corona. 

\begin{figure}
   \centering
   \includegraphics[width=16.8cm,clip]{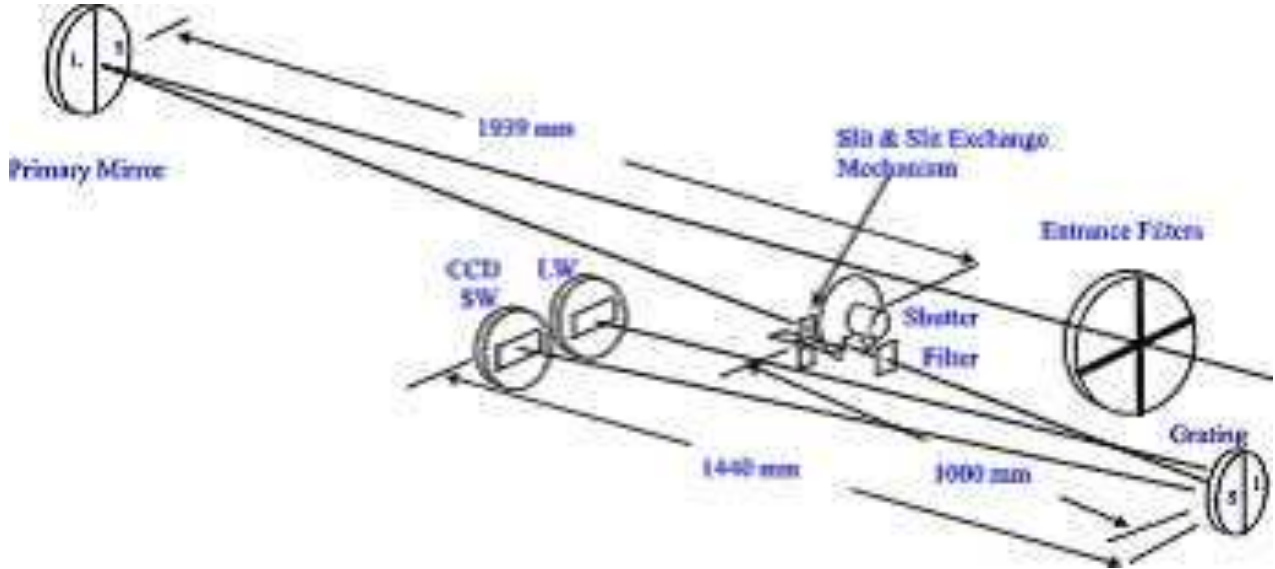}
   \caption{Mechanical design of the EUV Imaging Spectrometer (EIS) onboard \textit{Hinode}. 
            Excerpted from \citet{culhane2007}.}
   \label{itdn_eis_mech_design}
\end{figure}

% INSTRUMENTAL OVERVIEW
Previous spectrometers designed to operate in the wavelength range of $50\text{--}500${\AA} have employed grazing incidence optical systems, since the normal incidence reflectivity within this wavelength range is quite small for the usual optical materials.  The microchannel plate array detectors are commonly used, which provide high spatial resolution, however, they have low quantum efficiencies ($\mathrm{QE} \le 20${\%}).  EIS adopted normal incidence operation through the use of multilayer coatings applied to both mirror and gratings.  Furthermore, the use of thinned back-illuminated CCDs to register the diffracted photons make QE values be $2\text{--}3$ times larger than those for micro channel plate systems.  EIS has a large effective area in two EUV wavelength ranges, $170\text{--}210${\AA} and $250\text{--}290${\AA}.  The optical design of EIS is displayed in Fig.~\ref{itdn_eis_mech_design}. 

The solar radiation enters EIS through a thin $1500${\AA} Al filter which interrupts the transmission of visible radiation (\textit{i.e.}, the brightest wavelength range in the solar spectrum).  Incident photons are focused by the primary mirror onto a slit/slot and then on a toroidal concave grating.  Two differently optimized Mo/Si multilayer coatings are used to matching halves of both mirror and grating.  Then, diffracted photons are registered by a pair of thinned back-illuminated CCDs.  Exposure times are controlled by a rotating shutter.  Two spectroscopic slits ($1''$ and $2''$) and two spectroscopic imaging slots ($40''$ and $266''$) are used in a slit exchange mechanism, which allows the selection of four different apertures corresponding to each scientific purpose.  Raster scan observations are made by a piezoelectric drive system which rotates the primary mirror.  A raster scan has a field of view of $600''$ in the dispersion direction (the east-west direction on the solar surface) and $512''$ in the slit height direction (the north-south direction on the solar surface)\footnote{The solar radius roughly corresponds to $1000''$.}.  There is a coarse mechanism that can offset the mirror by $\pm 15'$ from the spacecraft pointing in the east-west direction.  The overall instrumental properties are given in Table \ref{EIS properties}. 

\input{tex/table_eis_design.tex}

% OBSERVATION MODE
EIS carries out observations in two modes: raster scan mode and sit-and-stare mode. In raster scan mode, the slit/slot scans the objective region on the solar surface.  The obtained data will be three dimensional ($x,y$, and wavelength), which is suitable for studying spatial variation of the spectra.  In the sit-and-stare mode, the slit/slot is fixed at a target position on the solar surface and is tracking the target by compensating for the solar rotation during the observation.  The data obtained in the sit-and-stare mode provide the temporal variability of the spectra, which is suitable for investigating phenomena like oscillations, jets, or microflares. 

% --- End of Contents ---

%% file: tex/table_eis_design.tex
\begin{table}
  \centering
  \caption{EIS properties \citep{korendyke2006,culhane2007}.}
  \begin{tabular}{ll} 
    \toprule
    Wavelength bands     & $170\text{--}210${\AA} and $250\text{--}290${\AA} \\
    Peak effective areas & $0.30\,\mathrm{cm}^2$ and $0.11\,\mathrm{cm}^2$ \\
    Primary mirror       & $15 \, \mathrm{cm}$ diameter; two Mo/Si multilayer coatings \\
    Grating              & Toroidal/laminar, $4200\,\mathrm{lines \, mm^{-1}}$, two Mo/Si multilayers \\
    CCD cameras          & Two back-thinned e2v CCDs, $2048 \times 1024 \times 13.5 \, \mathrm{\mu \, m\, pixels}$ \\
    Plate scales         & $13.53 \, \mathrm{\mu \, m \, arcsec^{-1}}$ (at CCD); 
                           $9.40 \, \mathrm{\mu \,m \, arcsec^{-1}}$ (at slit) \\
    Spatial resolution   & $\sim 2''$ \\
    Field of view        & $600'' \times 512''$, offset center: $\pm 825''$ in E-W direction \\
    Raster               & $1''$ in $0.7 \, \mathrm{s}$ (Minimum step size: $0''.123$) \\
    Slit/slot widths     & $1''$, $2''$ (slit), $40''$ and $266''$ (slot) \\
    Instrumental broadening & $\sim 2.5 \,\mathrm{pixels}$ \\
    Spectral resolution  & $47 \, \mathrm{m}${\AA} (FWHM) at $185${\AA}; 
                           $1\,\mathrm{pixel} = 22 \,\mathrm{m}${\AA} or \\
                         & approx.\ $25\,\mathrm{km \, s^{-1}}$ \\
    Temperature coverage & $\log T = 4.7, \, 5.6, \, 5.8, \, 5.9, \, 6.0\text{--}7.3$ \\
    CCD frame read time  & $0.8 \,\mathrm{s}$ \\
    Line observations    & Simultaneous observation of up to $25$ lines \\
    \bottomrule 
  \end{tabular}
  \label{EIS properties}
\end{table}

%% file: tex/contents_cal.tex
\chapter{Average Doppler shifts of the quiet region}
\label{chap:cal}
\section{Introduction}
  \input{tex/cal_itdn.tex}
  \label{sect:cal_itdn}
\section{Observations}
  \input{tex/cal_data_hop79.tex}
\section{Data reduction and analysis}
  \input{tex/cal_ana_pro.tex}
  \label{sect:cal_data_reduc}
  \subsection{Line profiles}
    \input{tex/cal_ns_lp.tex}
    \subsubsection{The transition lines}
      \input{tex/cal_lp_tr.tex}
      \label{sect:cal_lp_tr}
    \subsubsection{Coronal lines}
      \input{tex/cal_lp_co.tex}
      \label{sect:cal_lp_co}
  \subsection{Fitting}
    \input{tex/cal_lp_fit.tex}
  \subsection{Calibration of the spectrum tilt}
    \input{tex/cal_tilt.tex}
    \label{sect:cal_spec_tilt}
  \subsection{Alignment of data between exposures}
    \input{tex/cal_align.tex}
\section{Center-to-limb variation}
  \input{tex/cal_results_limb2limb.tex}
  \label{sect:cal_limb2limb}
\section{Summary}
  \input{tex/cal_sum.tex}
\clearpage
\begin{subappendices}
\section{Calibration of the spectrum tilt}
  \input{tex/cal_tilt_corr_itdn.tex}
  \label{sect:cal_append_itdn}
  \subsection{Observation and data analysis}
    \input{tex/cal_tilt_corr_obs.tex}
    \label{sect:cal_append_spec_tilt}
  \subsection{Results}
    \subsubsection{Spectrum tilt in the short-wavelength CCD}
      \input{tex/cal_tilt_corr.tex}
      \label{sect:cal_tilt_sw}
    \subsubsection{Spectrum tilt in the long-wavelength CCD}
      \input{tex/cal_tilt_diff.tex}
      \label{sect:cal_tilt_diff}
  \subsection{Summary}
    \input{tex/cal_tilt_sum.tex}
\section{Oxygen lines ($\log T \, [\mathrm{K}] = 5.4$)}
  \input{tex/cal_results_ov.tex}
  \label{sect:cal_results_ov}
\end{subappendices}

%% file: tex/cal_itdn.tex
% =======================
%   Chapter:
%     Calibration.
% =======================

Measurement of the Doppler shift of an emission line is an important method to investigate the velocity of a target in astrophysics.  EIS onboard \textit{Hinode} aims to measure a Doppler shift with an accuracy in the order of $10^{-3}${\AA} which corresponds to several $\kmpers$ in the EUV wavelength range.  The flow speed in the corona is typically the order of $\kmpers$ in the quiet region and up to several tens of $\kmpers$ in active regions.  The method is simple in the sense that we measure a line centroid and calculate the difference from the rest wavelength known in advance, however, the practical analysis is far more complex. 

Since EIS does not have the absolute calibration mechanism for wavelength, we often refer the average line centroid of the quiet region included in the field of view as zero.  This is based on the idea that the quiet region has smaller velocity than that of active regions.  The Doppler velocity derived through this procedure should be actually regarded as just a difference of the Doppler shift from that of the quiet region.  At present, the Doppler velocities of coronal emission lines in the quiet region have not been investigated with an precision better than $\simeq 10 \, \kmpers$, considering the several uncertainties described below.  We need another way to deduce the Doppler velocity in the quiet region different from previous data analysis in the literature.  For the precise measurements of the Doppler shift, we need to take care of several points below.

First point is a lack of our knowledge about precise rest wavelengths of some emission lines.  The database of emission lines provided by NIST\footnote{\texttt{www.nist.gov/pml/data/asd.cfm}} shows that the rest wavelengths are determined in the order of $10^{-3}${\AA} in most cases. We actually sometimes observe an emission line in different wavelength predicted by the theoretical calculation. This means that it is not possible to measure the Doppler shift more accurately than that deviation even if we can obtain the precise line centroid with small statistical error by long exposure time in an observation.

\begin{figure}
  \centering
  \includegraphics[width=12cm,clip]{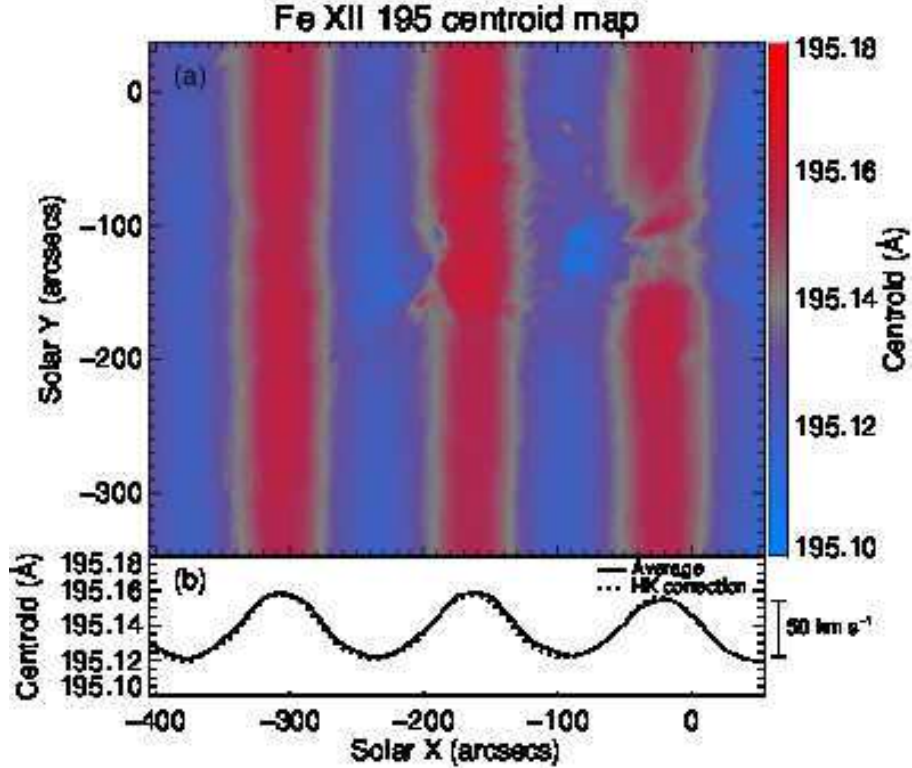}
  \caption{(a) Doppler shift map without any wavelength corrections for orbital variation. Only the spectrum tilt was removed by using the SSW package. (b) The line centroid averaged in $y$ direction (\textit{solid}) and the orbital variation estimated by \citet{kamio2010}'s model (\textit{dotted}). Note that the \textit{dotted} curve was shifted so that it has the same value as the \textit{solid} curve at the left edge of the FOV ($x=-405''$).}
  \label{fig:orbit_var_ex_dop_map}
\end{figure}

Second point is the drift of the spectrum signals on the CCDs which mainly comes from the displacement of the grating component in accordance with the thermal environment of the EIS instrument \citep{brown2007}.  The \textit{Hinode} spacecraft flies in the Sun synchronous orbit and the angle of which the spacecraft faces to the Earth changes periodically in $\simeq 98 \, \mathrm{min}$.  The temperatures of the components in the EIS instrument change due to the variation of the Earth radiation with that period. This causes the quasi-periodic drift of the spectrum of $\sim 0.05${\AA} which corresponds to $50$--$75 \, \kmpers$ at wavelength of $200$--$300${\AA}.  Fig.~\ref{fig:orbit_var_ex_dop_map} shows an example of this effect.  The displacement of the spectrum is obviously larger than the variation in the solar corona, therefore it becomes much important to remove the instrumental effect before the measurement of the Doppler shift.  Although the temperature changes in the components of EIS are roughly periodic, the calibration is never simple because the temporal behavior also changes with the seasonal variation of the orbit of the spacecraft, and there are also phase difference in the temperature variation in each component.  The package developed by \citet{kamio2010} has been widely used to correct the wavelength scale which varies in accordance with the temperature of the instrument.  They modeled the orbital variation of the wavelength scale by a linear relationship with the temperatures of components in the EIS instrument and some other parameters (\textit{e.g.}, pointing coordinates).  The developed model basically reproduces the observed spectrum drift, but there are residuals of $4$--$5 \, \kmpers$ in the standard deviation from the observed data. 

Third point is that the model above assumes that the Doppler shift of Fe \textsc{xii} $195.12${\AA} averaged in each exposure equals zero, which may not be the actual case precisely.  The SUMER observations revealed that an emission line Ne \textsc{viii} $770${\AA} ($\logt=5.8$) are blueshifted at the disk center corresponding to the Doppler velocity\footnote{We follow the convention that positive velocity indicates redshift (downward as to the solar surface) and negative velocity indicates blueshift (upward as to the solar surface).} of $-2$ -- $-3 \, \kmpers$ in the quiet region \citep{peter1999,teriaca1999}.  The Doppler shift of Fe \textsc{xii} $1242${\AA} was measured by \citet{teriaca1999}, which reported that the emission line is blueshifted by $-10 \, \kmpers$ in an active region \citep{teriaca1999}.  That Fe \textsc{xii} emission line was too weak for the measurement in the quiet region, and we cannot exclude the possibility that emission lines in the coronal temperature are shifted from the rest wavelength in the quiet region.  Although the model of \citet{kamio2010} is an useful tool to compensate the orbital variation of the wavelength scale and it is now included in the standard EIS software in SSW, taking into account all three factors described above, we can not discuss the Doppler velocity smaller than $\sim 5 \, \kmpers$ by using only their model. 

In this chapter, we exploited the data which covers the meridional line of the Sun in order to deduce the Doppler shifts at the center of the solar disk compared to those at the solar limb where the Doppler velocities are considered to be the best zero point at present. Such observations enable us to study the center-to-limb variation of the Doppler shift of emission lines, which solve the first point in the measurement of the Doppler shift described above (\textit{i.e.}, a lack of the knowledge about the precise rest wavelength). This analysis is based on the idea that the corona flows 
%{\color{red} 
statistically
%} 
in the radial direction from the global view.  
%{\color{red} 
We analyzed the spectra by integrating with a spatial scale of $50''$ in order to compensate the non-radial motions in a statistical sense.
%}

Since there are only few coronal emission lines in the spectra obtained by SUMER and its predecessors with strength enough for their line centroids to be measured in the quiet region, previous observations could only measure the center-to-limb variation of the transition region lines \citep{roussel-dupre1982,peter1999} whose formation temperature was $\logt \leq 5.8$. Our analysis challenges the center-to-limb variation of the Doppler shifts of several coronal emission lines ($\logt \geq 6.0$), and determine the average Doppler shifts in the quiet region at the disk center. The results in this chapter will be used as a reference of Doppler velocities for the analysis of the outflow in an active region (Chapter \ref{chap:vel}). 

% --- End of TeX ---

%% file: tex/cal_data_hop79.tex
%%%%%%%%%%%%%%%%%%%%%%%%%%%%%%%%%%%%%%%%%%%%%%%%%%%%
%  Chapter:
%    Doppler shifts in the quiet region.
%  Contents:
%    Description on data taken in HOP79.
%%%%%%%%%%%%%%%%%%%%%%%%%%%%%%%%%%%%%%%%%%%%%%%%%%%%

Based on the idea that emission lines are not shifted at the limb because our line of sight passes symmetrically the solar corona \citep{peter1999}, we can set the Doppler velocity at the limb as zero then derive the Doppler velocities at the disk center.  In order to investigate the center-to-limb variation of the Doppler shifts and measure those at the disk center, we exploited data taken during the coordinated observations between three instruments onboard \textit{Hinode} (usually referred to as Hinode Observing Plan 79; hereafter we use the term HOP79).  While these observations were originally intended to investigate the variation of solar irradiance along the 11-year solar cycle mainly by SOT observations, EIS is requested to take spectra with long exposures.  During the observations, the pointings of the satellite are gradually moved from the south pole to the north pole (north--south scan), or from the east limb to the west limb (east--west scan), so that the data cover the solar surface from one limb to the other limb. 

\begin{figure}
  \centering
  %bb=0 0 1471 536,
  \includegraphics[width=16.8cm,clip]{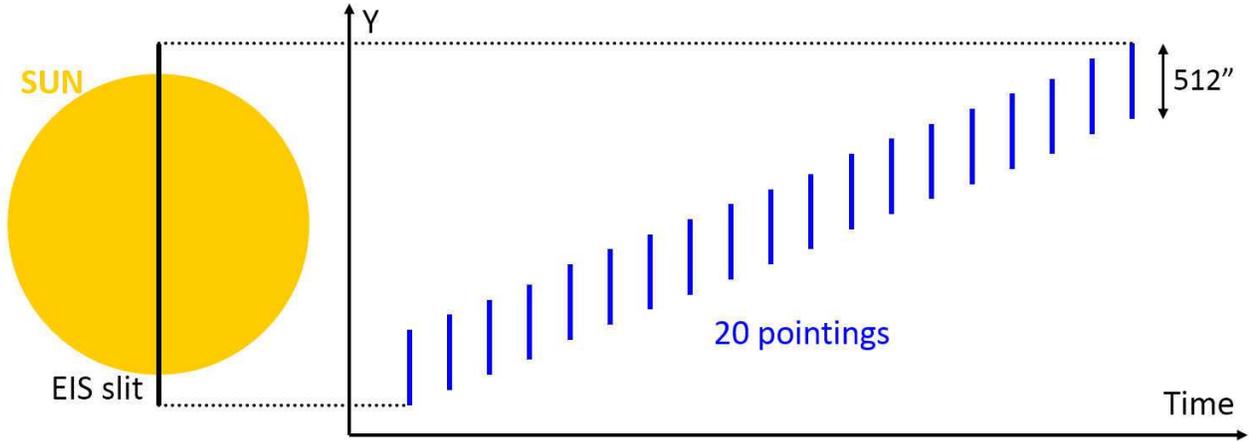}
  \caption{Schematic picture of the north--south scan in HOP79.}
  \label{fig:ns_scan_schem}
\end{figure}

In this analysis, we used the north--south scans which obtain the solar spectra without spatial gap between pointing changes (\textit{i.e.,} overlapping FOVs in each pointing by $300''$).  The schematic picture of the scan is shown in Fig.~\ref{fig:ns_scan_schem}.  This observation enables us to investigate the center-to-limb variation of Doppler shifts of emission lines by aligning the line centroids between the overlapping locations.  Note that in the east--west scans, EIS FOVs do not overlap with each pointing so that it is not possible to investigate the center-to-limb variation of the Doppler shifts.  Thus, we concentrate on the north--south scans in this analysis here. 

In the observations, $1''$ slit was used and the FOV in each pointing was $5'' \times 512''$ (\textit{i.e.,} five exposures at each pointing).  The exposure time was $120 \, \mathrm{s}$, which is enough to obtain good signal-to-noise (S/N) ratio for many coronal emission lines even in the quiet region.  The EIS usually records the spectra with a finite width in the wavelength direction (called as spectral window).  The EIS study analyzed here consists of 16 spectral windows with the spectral widths of $24$--$48$ pixels ($\simeq 0.5$--$1.0${\AA}), which were wide enough to include whole of emission lines (\textit{cf.} typical Gaussian width of emission lines in the quiet region does not exceed $\simeq 0.05${\AA}). 

Some recent studies have challenged the precise measurement of Doppler velocities of coronal structures by referring the centroids of emission lines determined from EIS spectra at the limb \citep{warren2011,young2012,dadashi2012}, however, there is a remaining factor for the uncertainty in the measurement.  They all used a calibration of wavelength developed by \citet{kamio2010} when comparing the centroids measured at the limb with those measured at their target coronal structures.  Therefore, their results are thought to have the error of $\simeq 5 \, \mathrm{km} \, \mathrm{s}^{-1}$.  The analysis here is free from that problem with overlapped scans from the south limb to the north limb, and will help us to determine the Doppler velocities with much carefulness. 

%\begin{figure}
%  \centering
%  \includegraphics[width=7cm,clip]{images/wvl_cal/eit/eit_0195_2010-Jan_00.eps}
%  \includegraphics[width=7cm,clip]{images/wvl_cal/eit/eit_0195_2010-Feb_00.eps}
%  \includegraphics[width=7cm,clip]{images/wvl_cal/eit/eit_0195_2010-Mar_00.eps}
%  \includegraphics[width=7cm,clip]{images/wvl_cal/aia/aia_193_2010-May_00.eps}
%  \includegraphics[width=7cm,clip]{images/wvl_cal/aia/aia_193_2010-Jul_00.eps}
%  \includegraphics[width=7cm,clip]{images/wvl_cal/aia/aia_193_2010-Aug_00.eps}
%  \caption{EIT or AIA images at the start of each HOP79. EIT $195${\AA} filter images are shown for January to March, and AIA $193${\AA} filter images are shown for May to December. White vertical line in each image indicates the location where EIS took spectral data.}
%  \label{fig:data_hop79}
%\end{figure}

\begin{figure}
  \centering
  \includegraphics[width=11.3cm,clip]{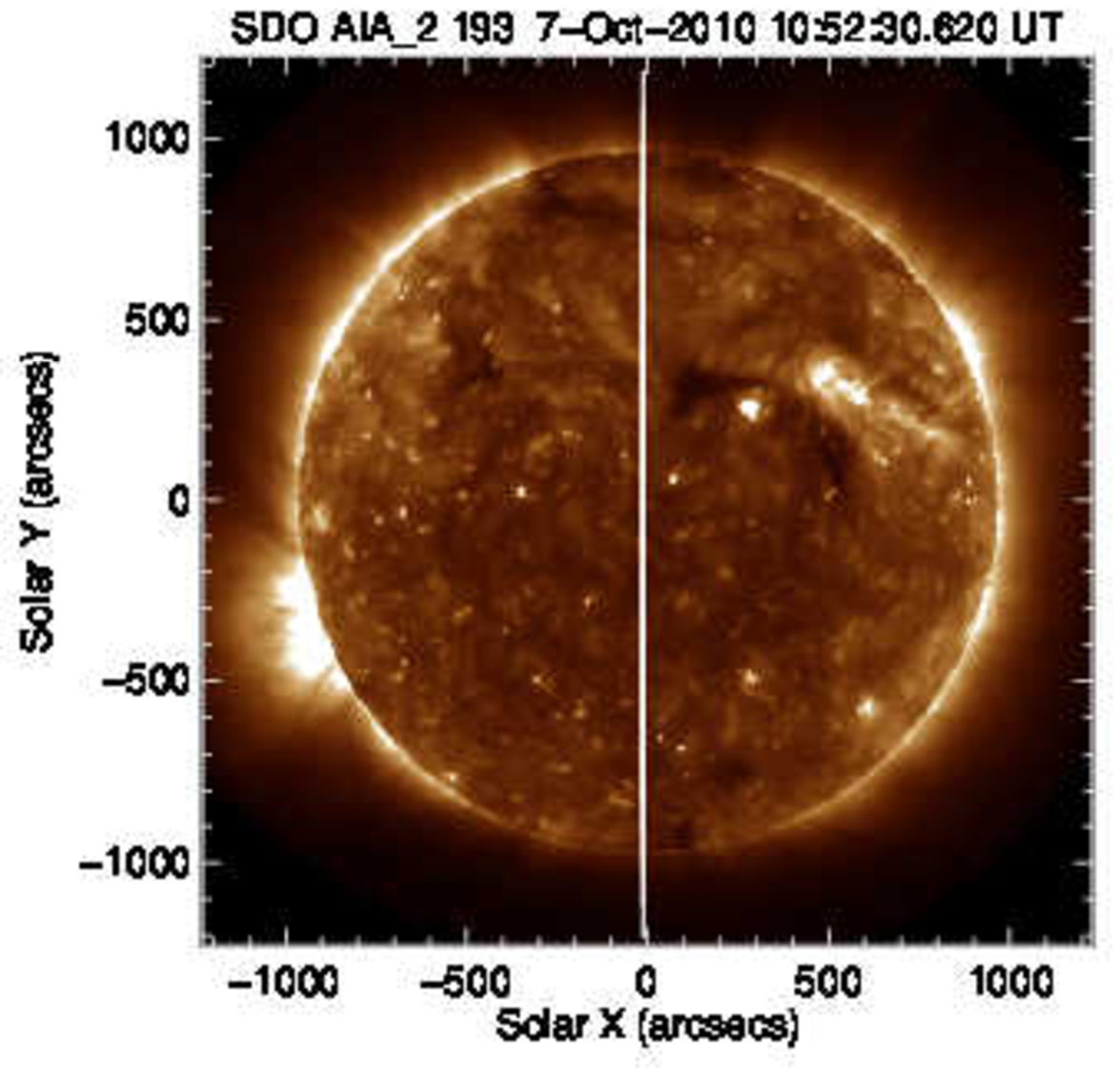}
  \includegraphics[width=11.3cm,clip]{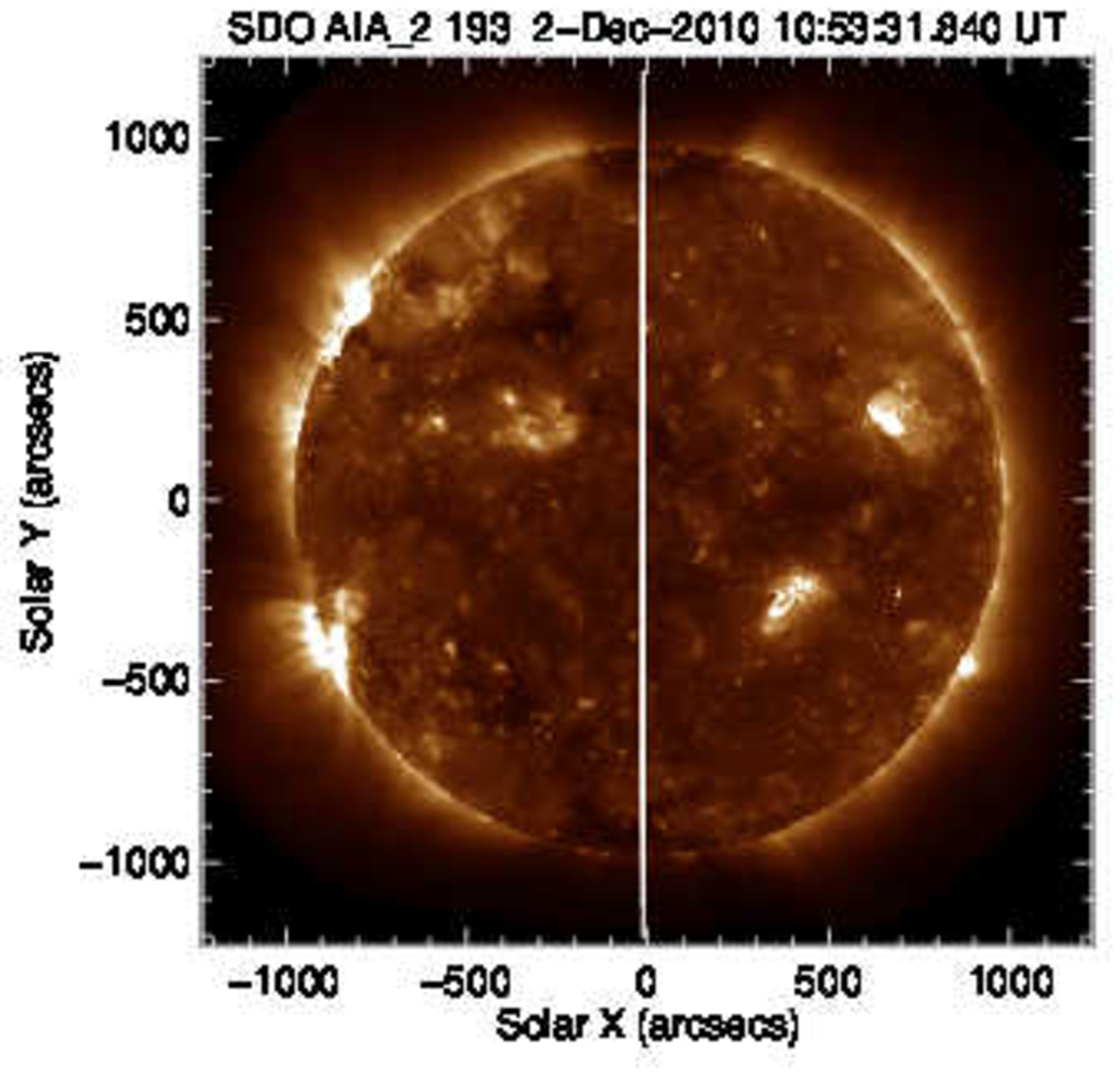}
  \caption{\textit{SDO}/AIA $193${\AA} passband images at the start of each HOP79. 
  \textit{Upper}: October. 
  \textit{Lower}: December. 
  A \textit{white vertical} line in each image indicates the location where EIS took spectral data.}
  \label{fig:data_hop79}
\end{figure}

The analyzed HOP79 data were taken during 2010 close to the bottom of the solar cycle in order to avoid the influence of active regions with relatively larger systematic flows than the quiet region.  When a spectral scan of EIS includes some active regions, there are several possibilities which cause the Doppler shift of an emission line.  Firstly, there can be seen many active phenomena like microflares, which induce plasma flows up to several ten $\mathrm{km} \, \mathrm{s}^{-1}$ in the corona.  Secondly, the corona in active regions generally has higher electron density ($10^{9-10} \, \mathrm{cm}^{-3}$; not flare condition) than in the quiet region ($10^{8-9} \, \mathrm{cm}^{-3}$), which often produces a fake shift of an emission line when another emission line exists in the neighbor whose emissivity strongly depends on the electron density.  Obviously, this is not the indication of the real flow.  Thirdly, a persistent upflow up to several tens of $\mathrm{km} \, \mathrm{s}^{-1}$ is often observed at the edge of active regions \citep{doschek2008,harra2008} which causes the blueshift of coronal emission lines. 

Many of the data taken in HOP79 after 2011 include active region(s) and its remnant in high latitude, and since the presence of active regions in spectral scan may affect our analysis, we analyzed the spectral data during 2010.  The data including a large coronal hole or active region(s) were not used in this analysis: January, February, March, May, July, August, September, and November.  In the analysis below, we concentrated the data in October and December.  The context images taken by \textit{SDO}/AIA are shown in Fig.~\ref{fig:data_hop79}.  A \textit{white vertical} line in each image indicates the location where EIS took spectral data. 

% --- End of Tex ---

%% file: tex/cal_ana_pro.tex
% ==================================================
%   Chapter:
%     Average Doppler shifts of the quiet region.
%   Description:
%     Analysis procedure.
% ==================================================
In this section, the procedure of analysis is described. First, we look over line profiles in order to check whether the single-Gaussian fitting is suitable for each emission line or not. Emission lines within the EUV range observed with EIS are often blended by neighboring ones, and this effect might cause a fake shift of the target emission line. %First we see a difference of line profiles between inside the solar disk and above the limb, and then describe properties of each emission line.

% --- End of TeX ---

%% file: tex/cal_ns_lp.tex
% ====================================
%   Chap. Doppler shift of the QS
% ====================================

\begin{figure}
  \centering
  \includegraphics[width=7.9cm,clip]{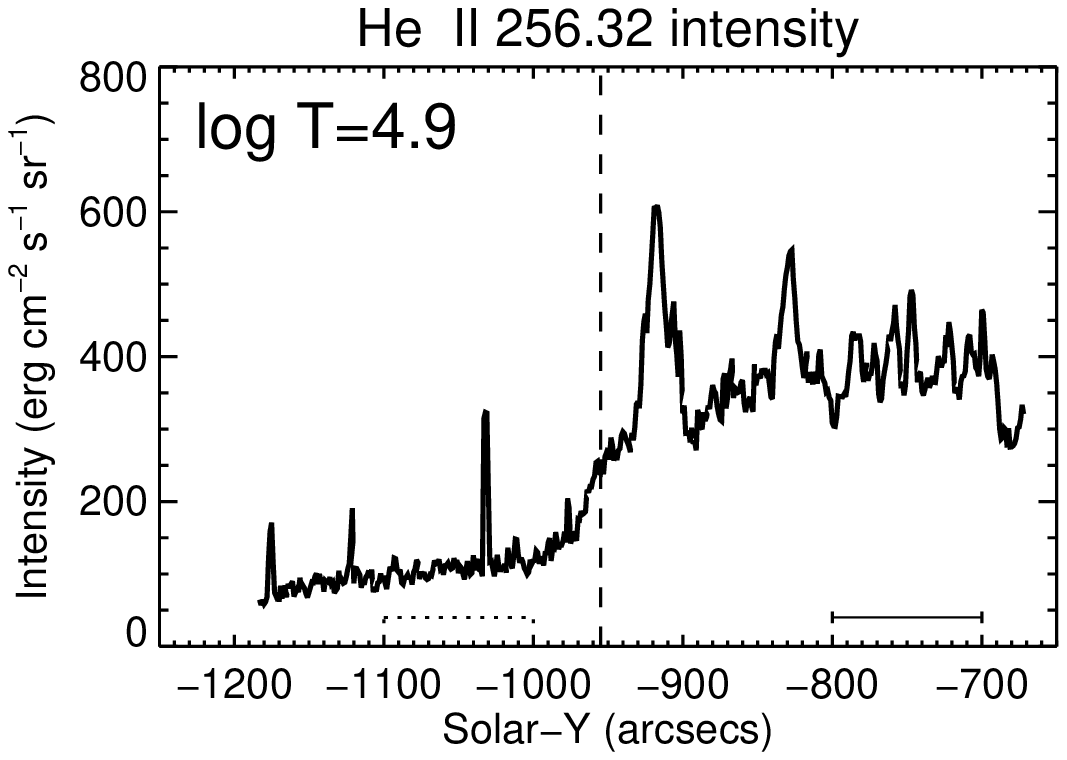}
  \includegraphics[width=7.9cm,clip]{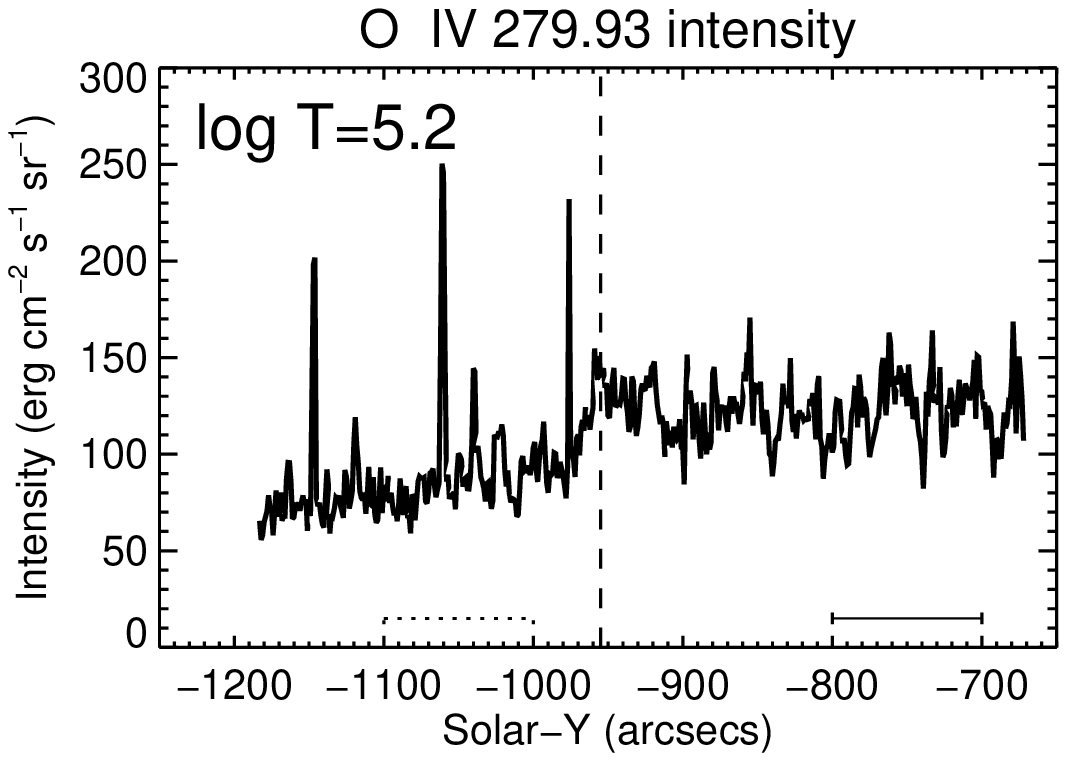}
  \includegraphics[width=7.9cm,clip]{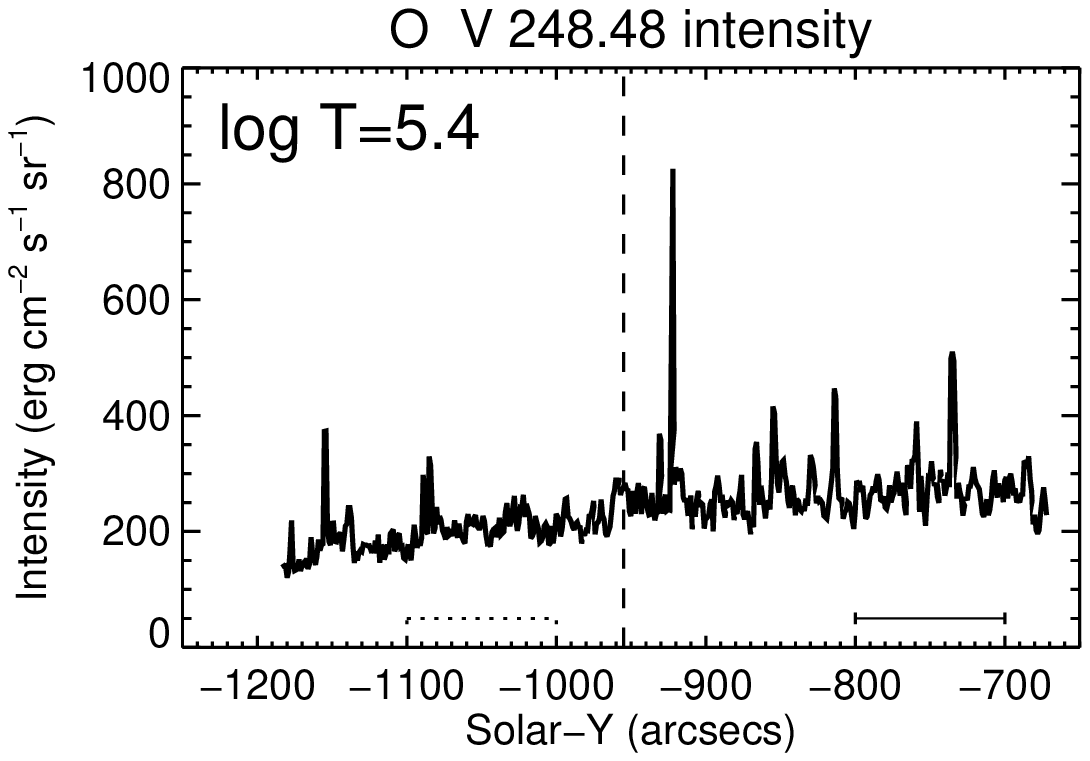}
  \includegraphics[width=7.9cm,clip]{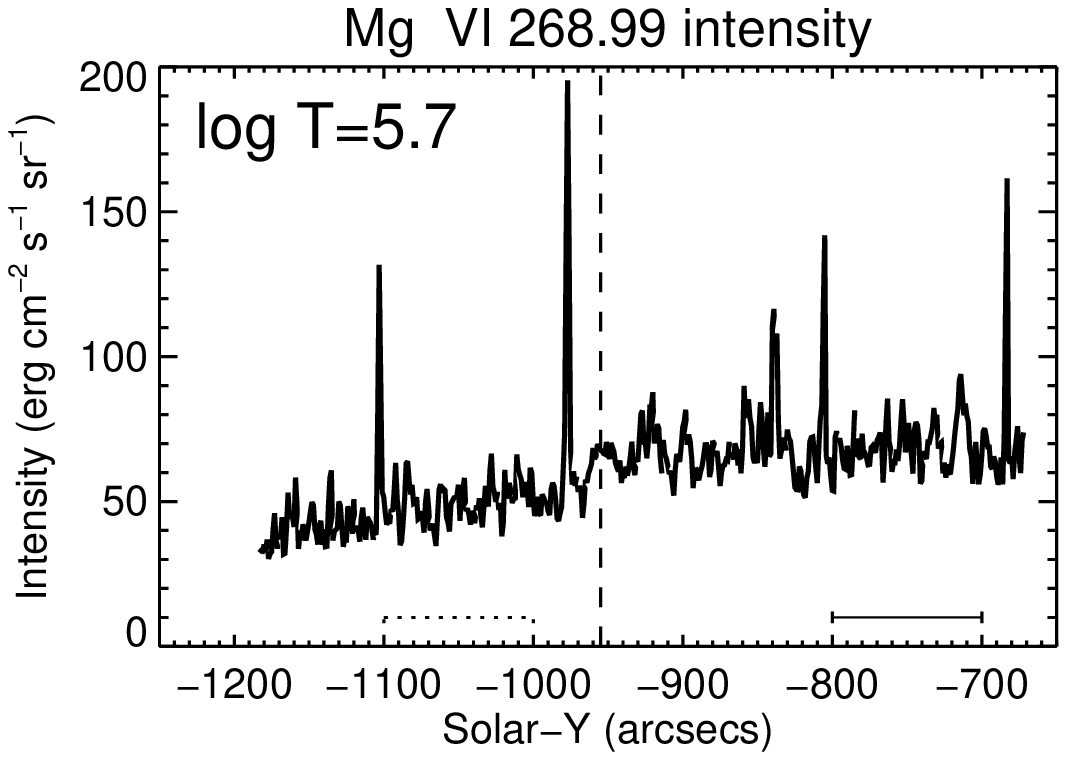}
  \includegraphics[width=7.9cm,clip]{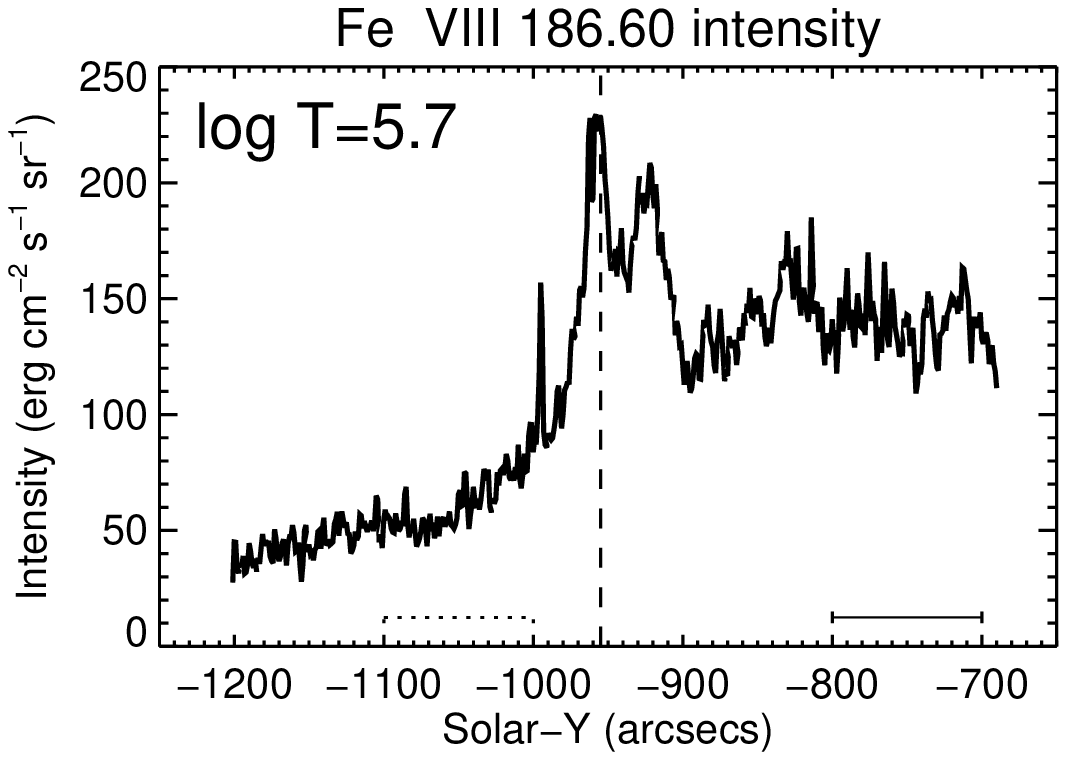}
  \includegraphics[width=7.9cm,clip]{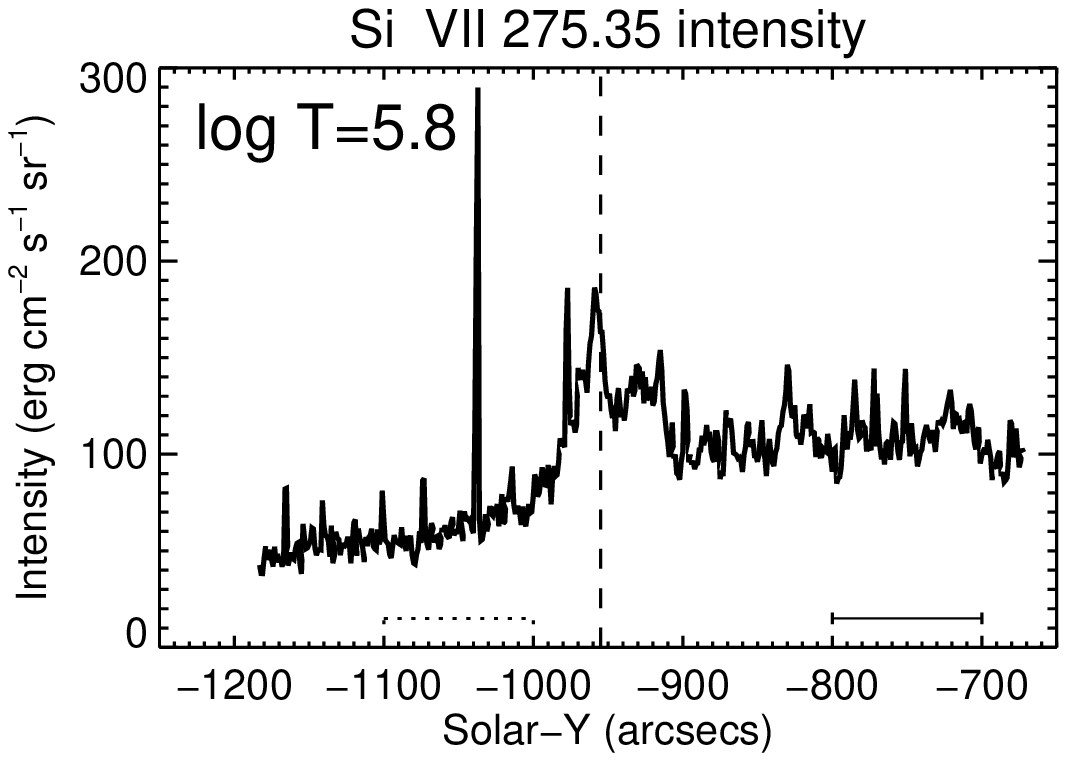}
  \includegraphics[width=7.9cm,clip]{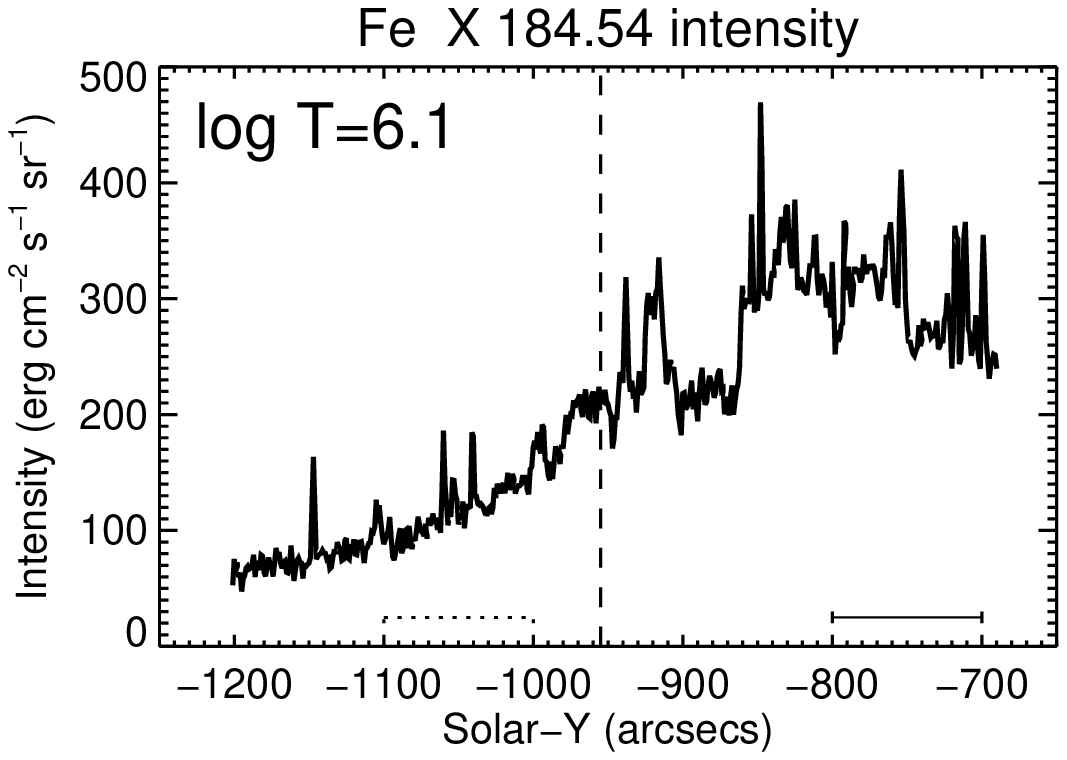}
  \includegraphics[width=7.9cm,clip]{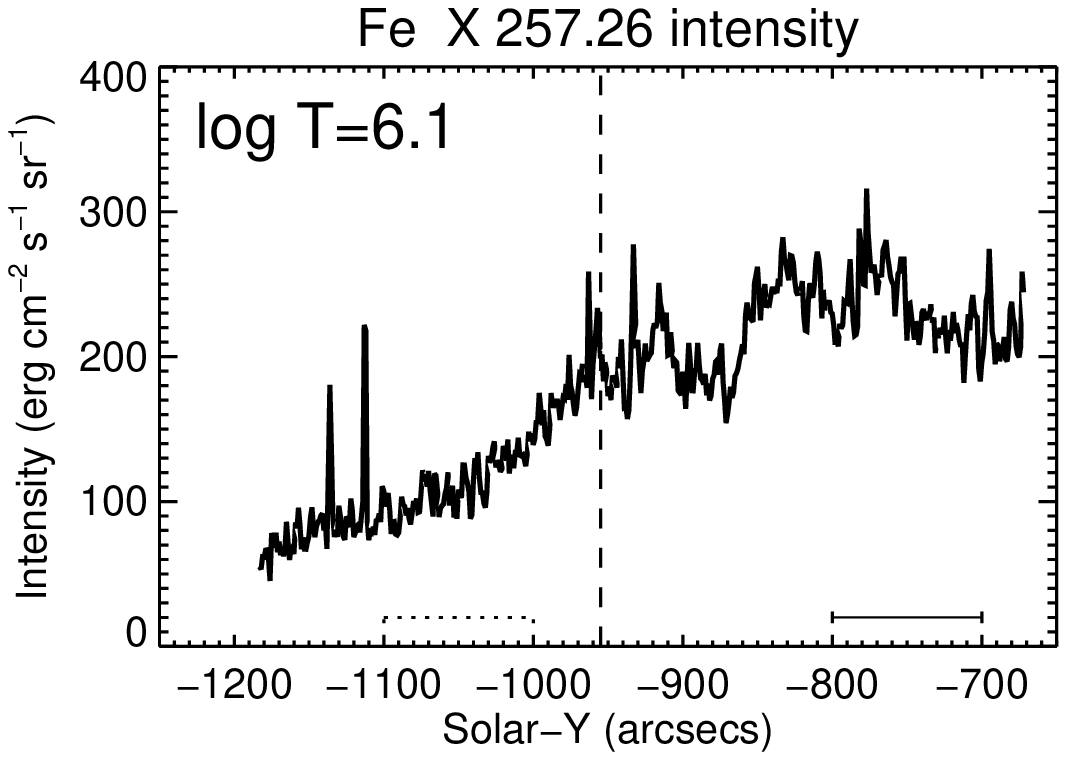}
  \caption{Intensity (\textit{ordinate}) as a function of the solar $Y$ (\textit{abscissa}). 
    A number in the left upper corner in each panel is the logarithmic formation temperature for the emission line.  A \textit{Vertical dashed} line indicates the limb location determined from the Fe \textsc{viii} intensity.  \textit{Horizontal} bars at the bottom in each panel shows the location where the spectrum was averaged and shown in Fig.~\ref{fig:cal_lp_ex}.
  }
  \label{fig:cal_int_limb}
\end{figure}
\addtocounter{figure}{-1}

\begin{figure}
  \centering
  \includegraphics[width=7.9cm,clip]{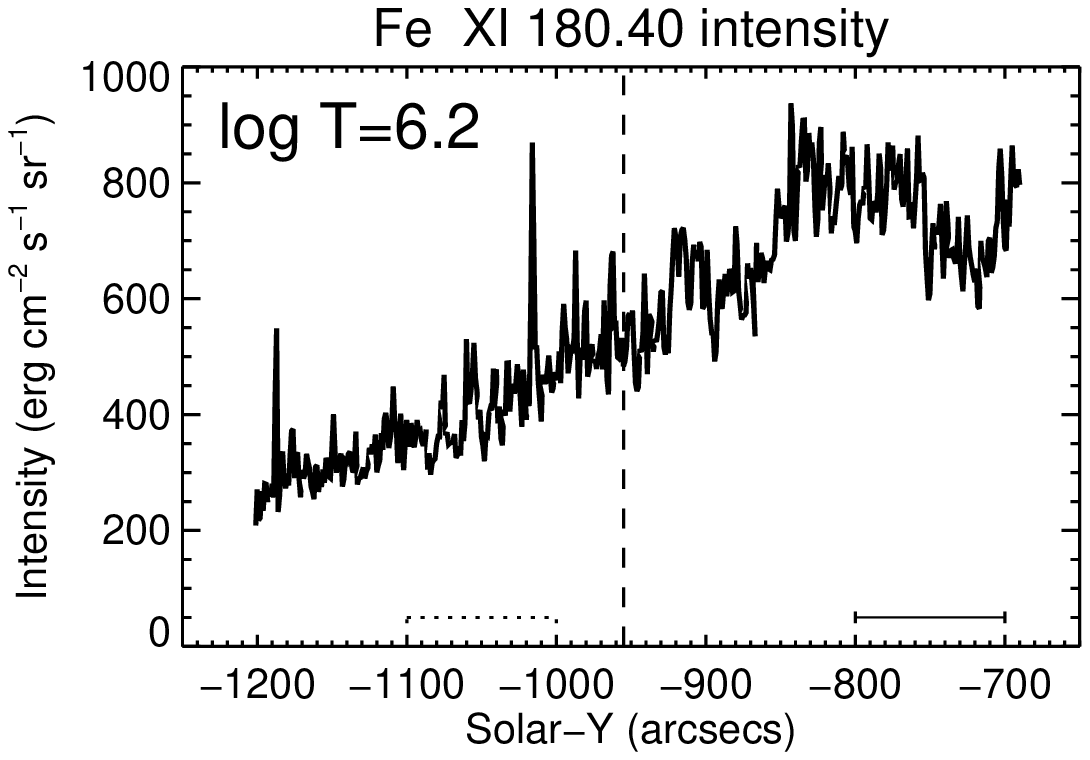}
  \includegraphics[width=7.9cm,clip]{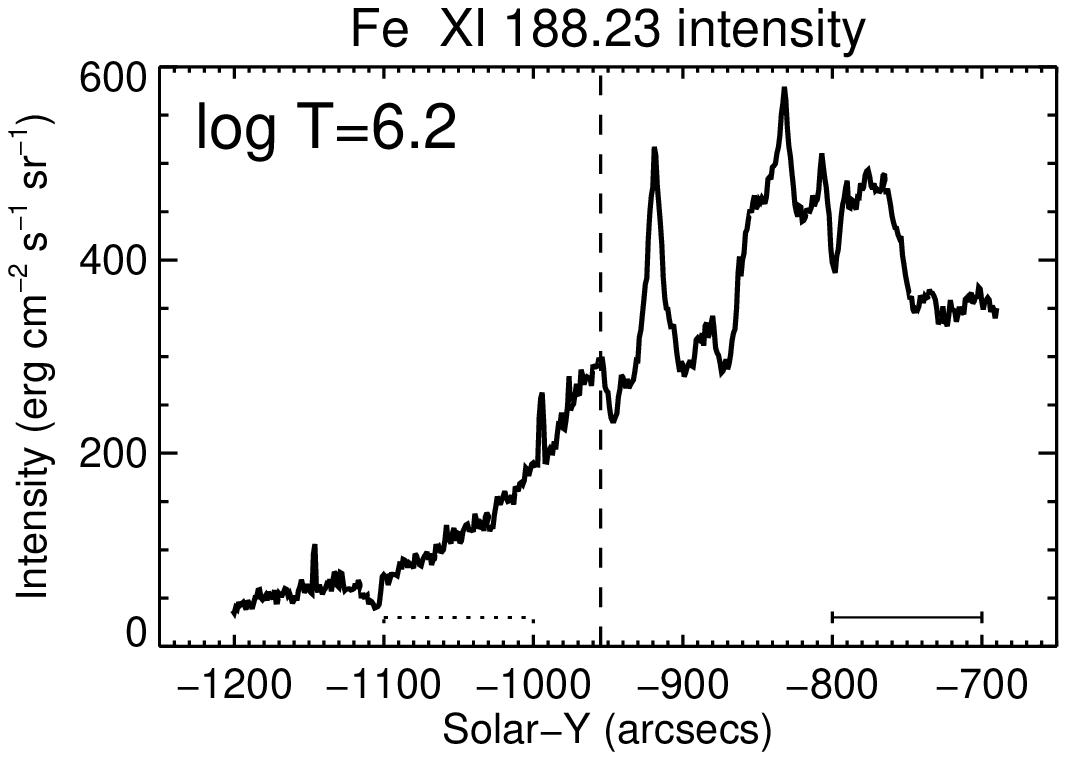}
  \includegraphics[width=7.9cm,clip]{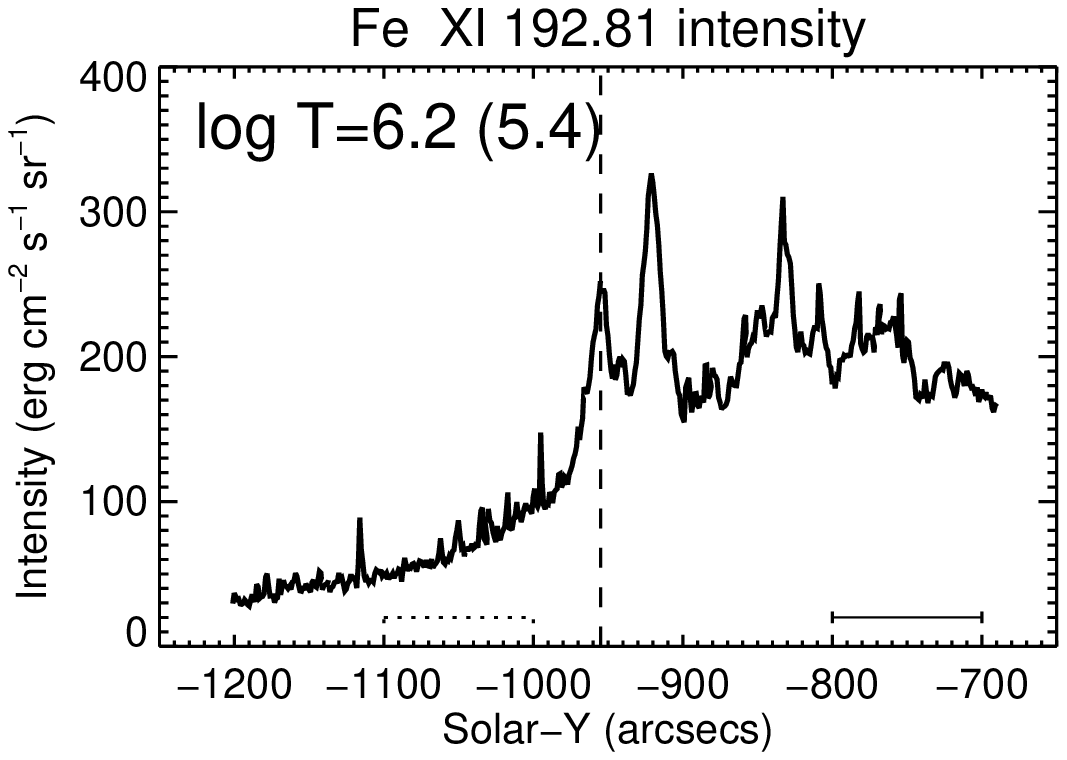}
  \includegraphics[width=7.9cm,clip]{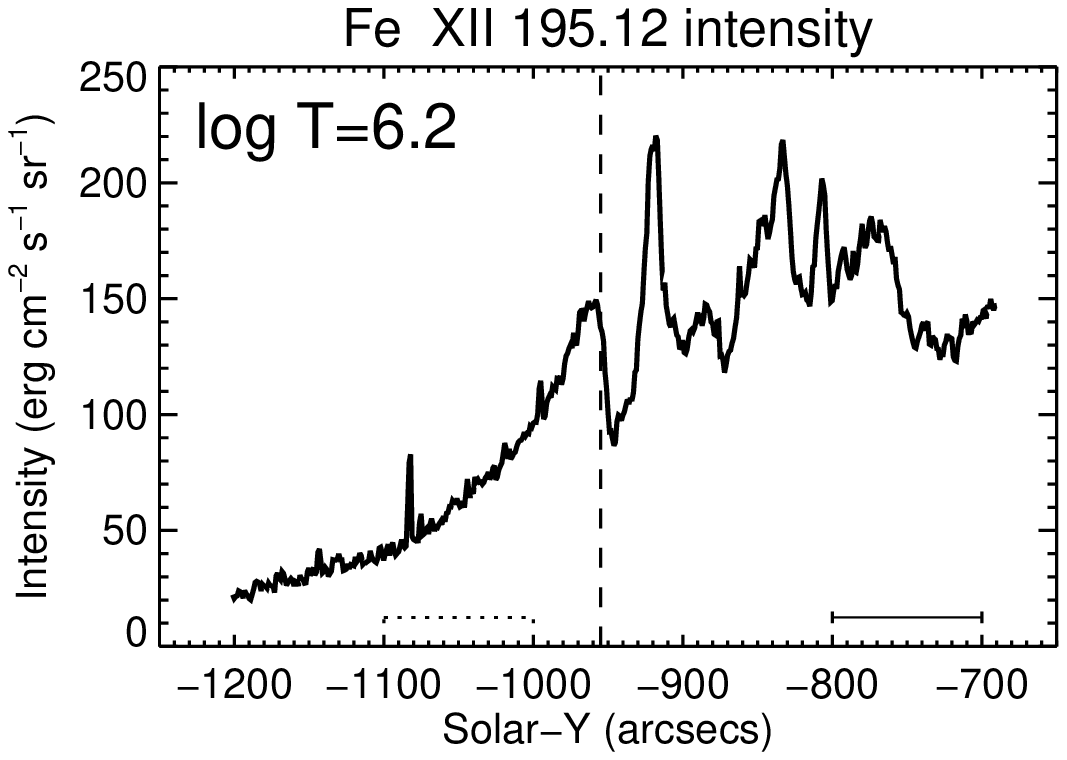}
  \includegraphics[width=7.9cm,clip]{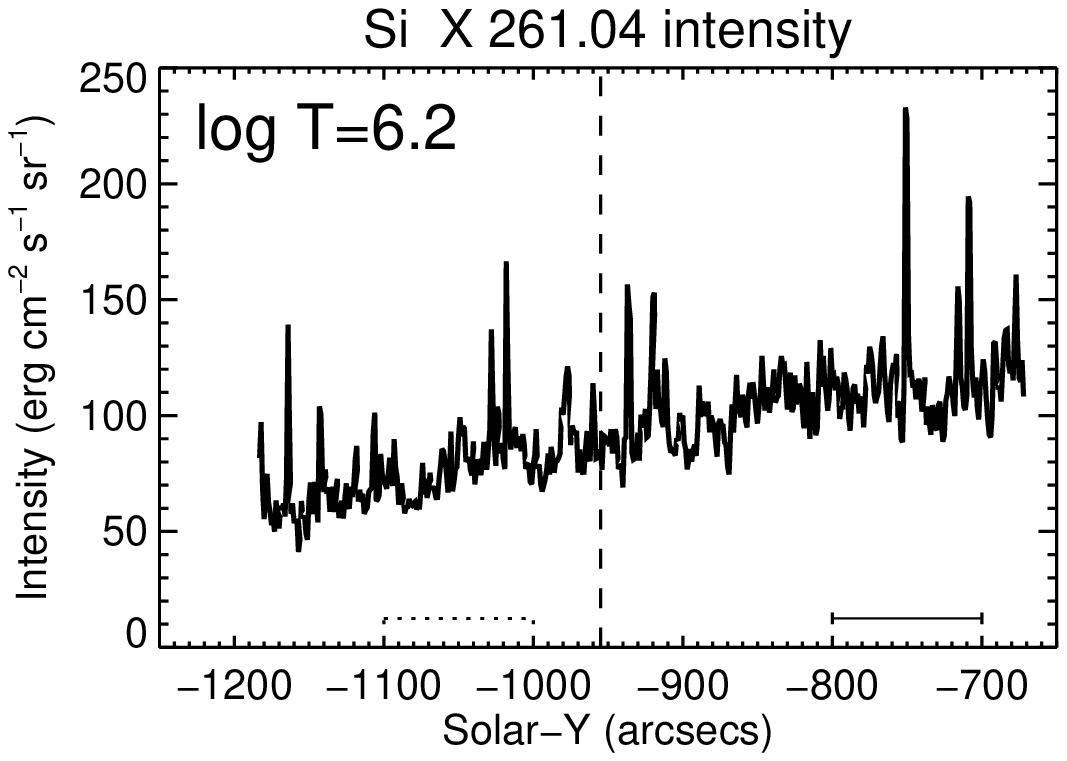}
  \includegraphics[width=7.9cm,clip]{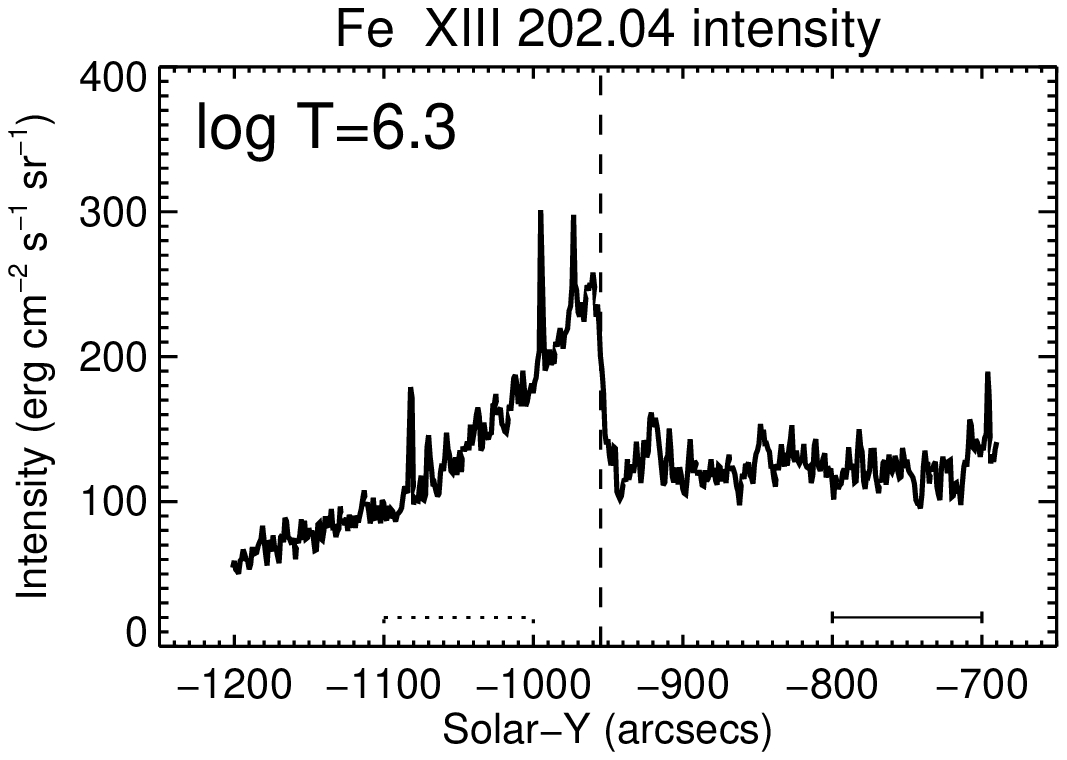}
  \includegraphics[width=7.9cm,clip]{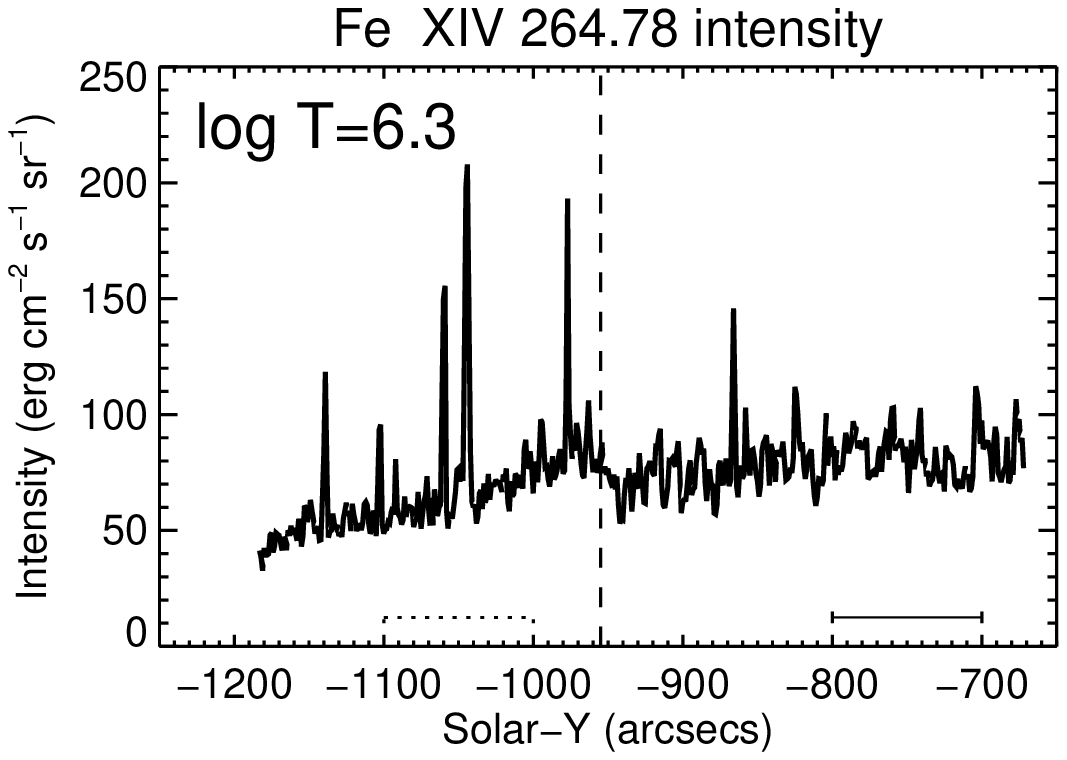}
  \includegraphics[width=7.9cm,clip]{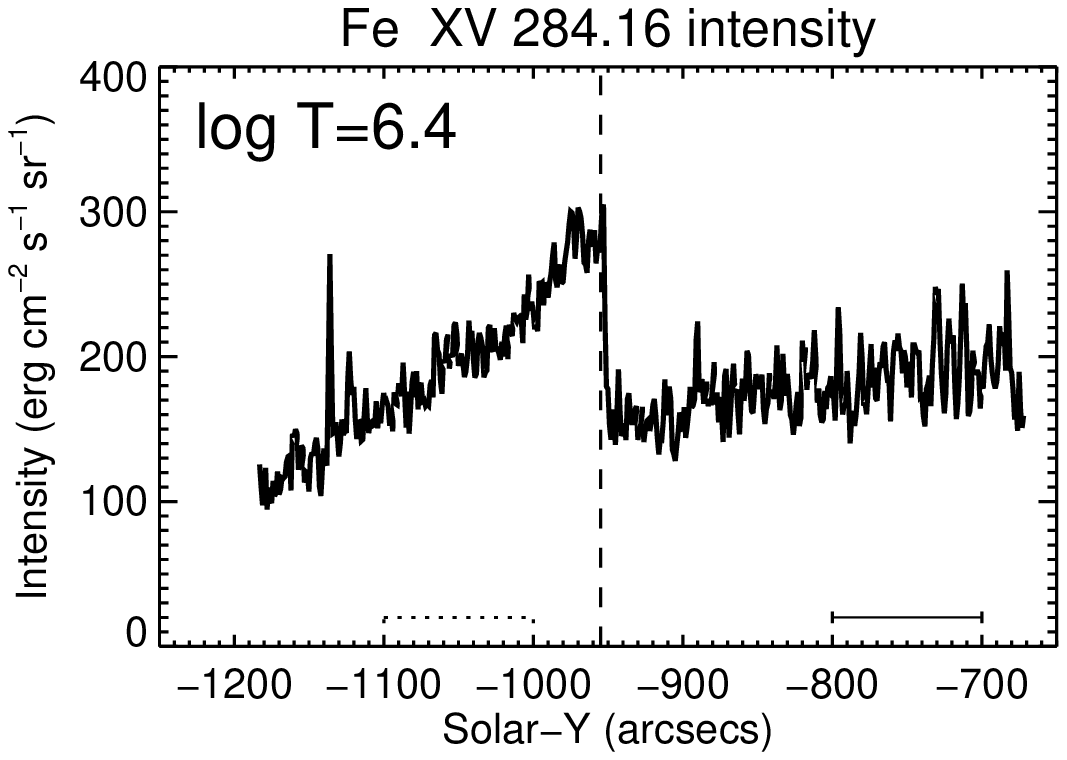}
  \caption{\textit{Continued.}}
\end{figure}

In order to decide emission lines to be analyzed, we first started from looking line profiles taken by EIS.  Fig.~\ref{fig:cal_int_limb} shows the spatial distribution of intensities near the south limb during HOP79 in 2010 October 7--8.  The intensities shown here are calculated by integrating over each spectral window.  Panels in the figure are in the order of the formation temperature.  The \textit{vertical dashed} line at $y=-960''$ in each panel indicates the limb location which was determined from the maximum point of Fe \textsc{viii} intensity since the limb was clearly seen in the transition region lines (O \textsc{iv}--\textsc{v}, Fe \textsc{viii} and Si \textsc{vii}) due to the well-known ``limb brightening'' effect which arises from the fact that the solar corona is optically thin.  In that situation, when we move our line of sight from the solar disk inside the limb to above the limb, the length of our line of sight becomes twice because there are no occulting structures above the limb.  For coronal lines from Fe \textsc{x}--\textsc{xii}, the intensity is stronger inside the limb than that off the limb, while for coronal lines like Fe \textsc{xiii} and Fe \textsc{xv} the intensity is stronger outside the limb compared to the disk (\textit{i.e.,} inside the limb). 

Line profiles on the solar disk ($y=-750''$; \textit{solid line}) and above the limb ($y=-1050''$; \textit{dashed line}) for all spectral windows taken by EIS during HOP79 on 2010 October 7--8 are shown in Fig.~\ref{fig:cal_lp_ex}.  Panels are in the order of the formation temperature as same as in Fig.~\ref{fig:cal_int_limb}.  The line profiles were integrated and averaged by the span of $100''$ in the $y$ direction and the integrated ranges are indicated by horizontal bar in Fig.~\ref{fig:cal_int_limb}.  We note characteristics of the emission lines seen in each spectral window below.

\begin{figure}
  \centering
  \includegraphics[width=7.9cm,clip]{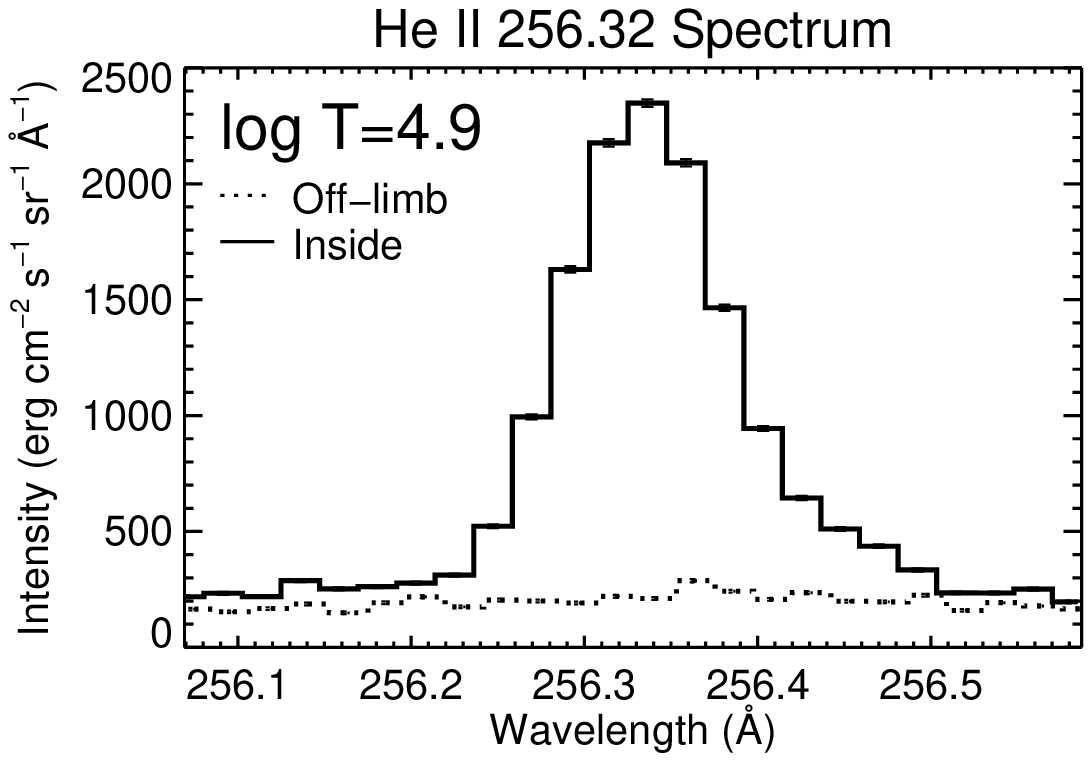}
  \includegraphics[width=7.9cm,clip]{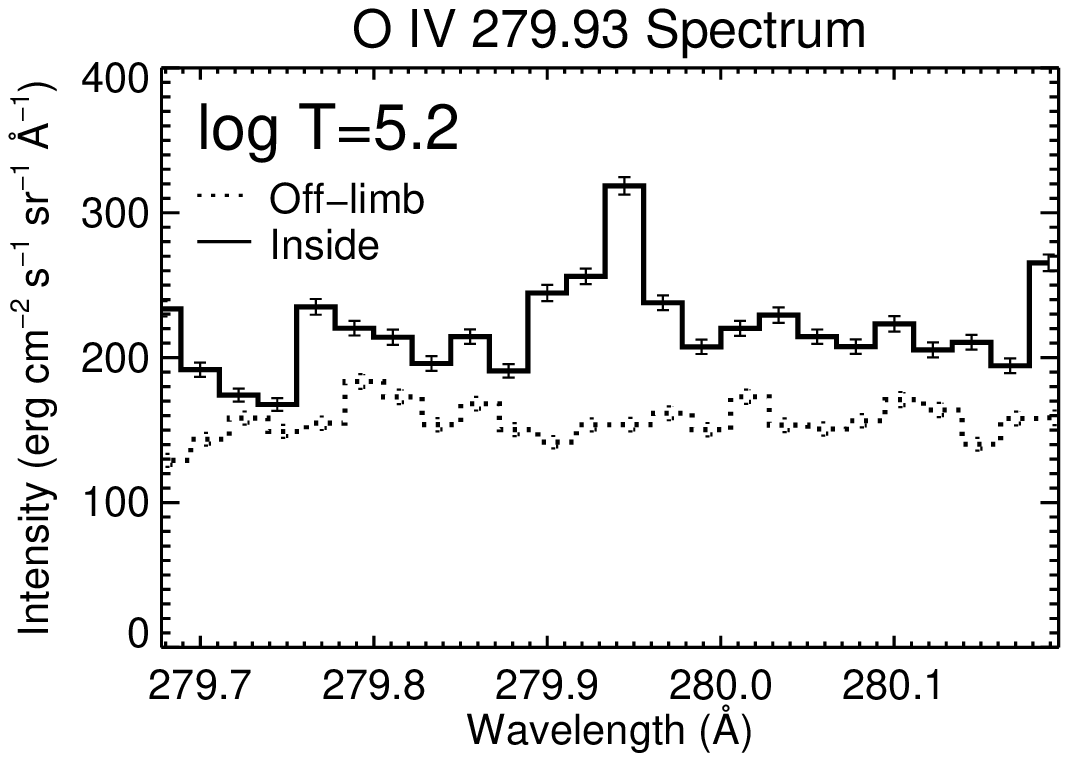}
  \includegraphics[width=7.9cm,clip]{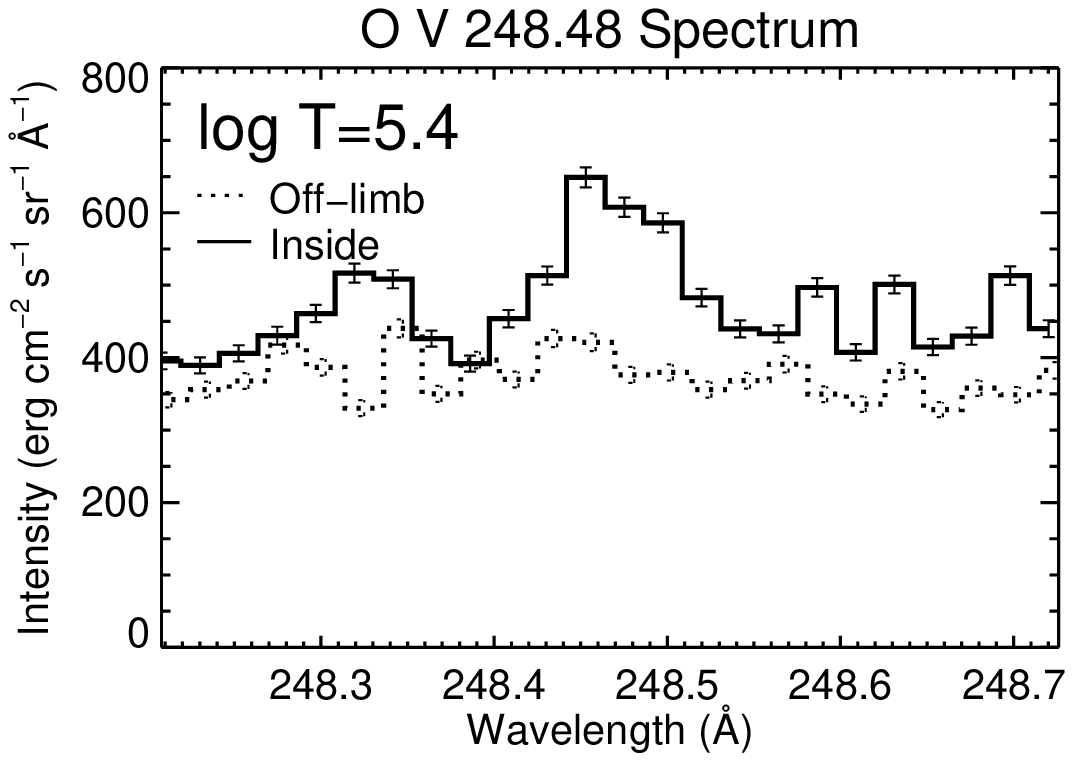}
  \includegraphics[width=7.9cm,clip]{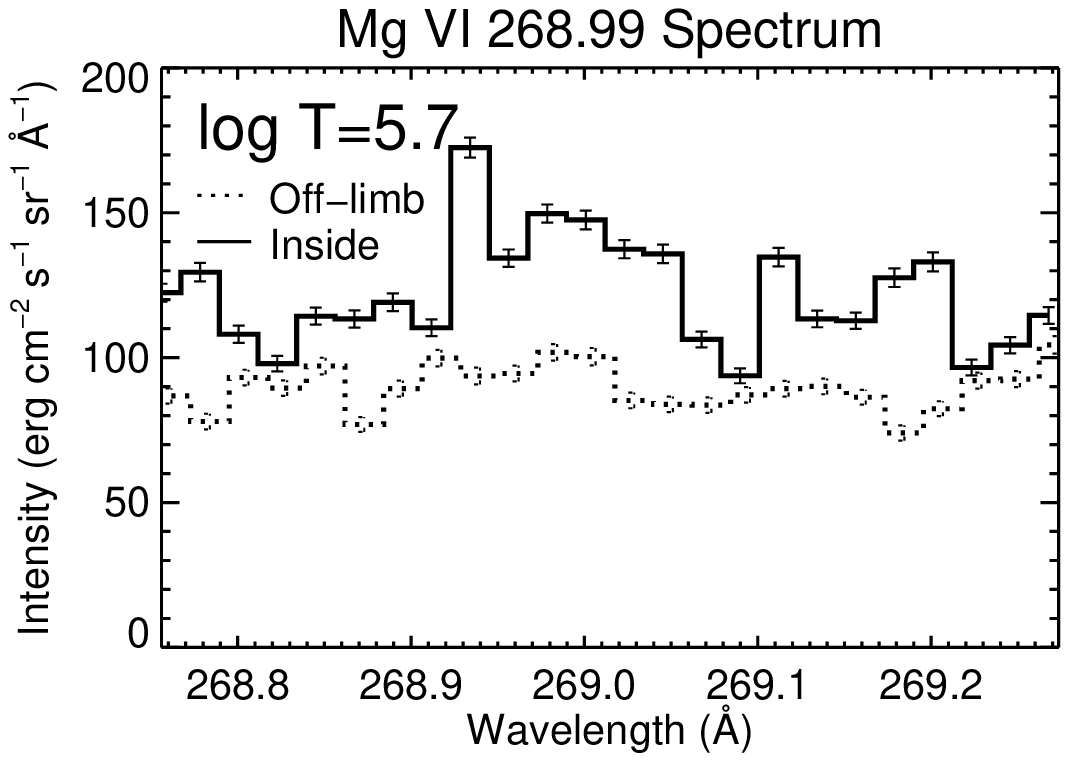}
  \includegraphics[width=7.9cm,clip]{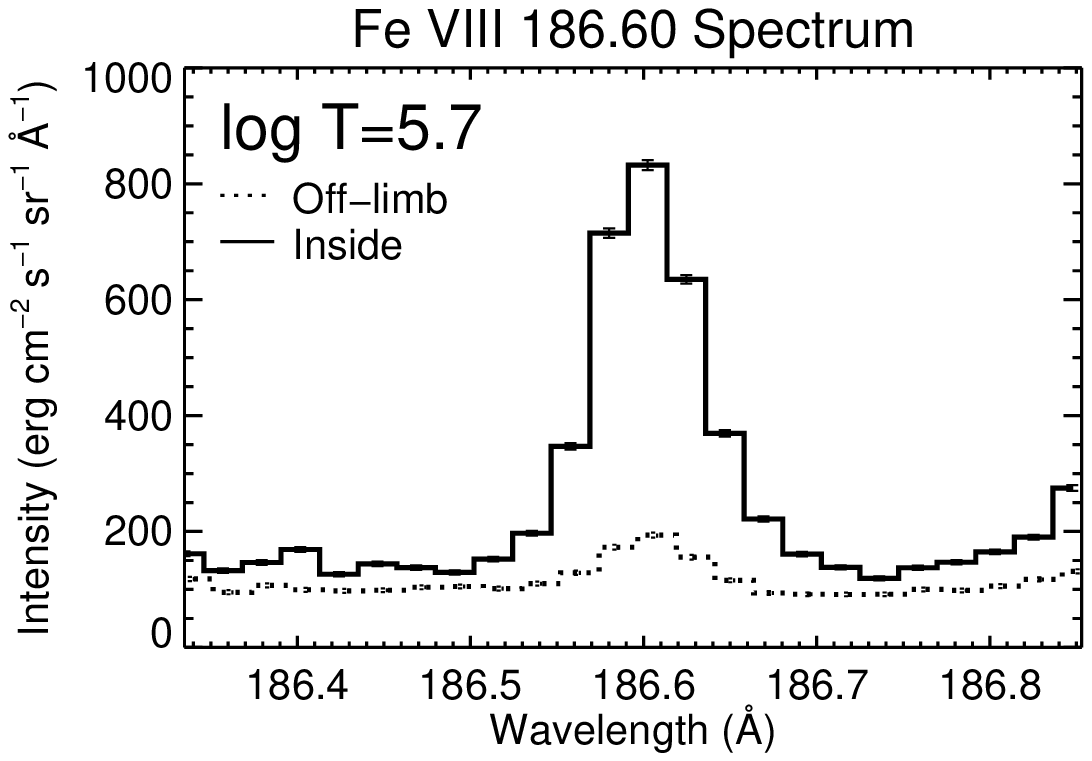}
  \includegraphics[width=7.9cm,clip]{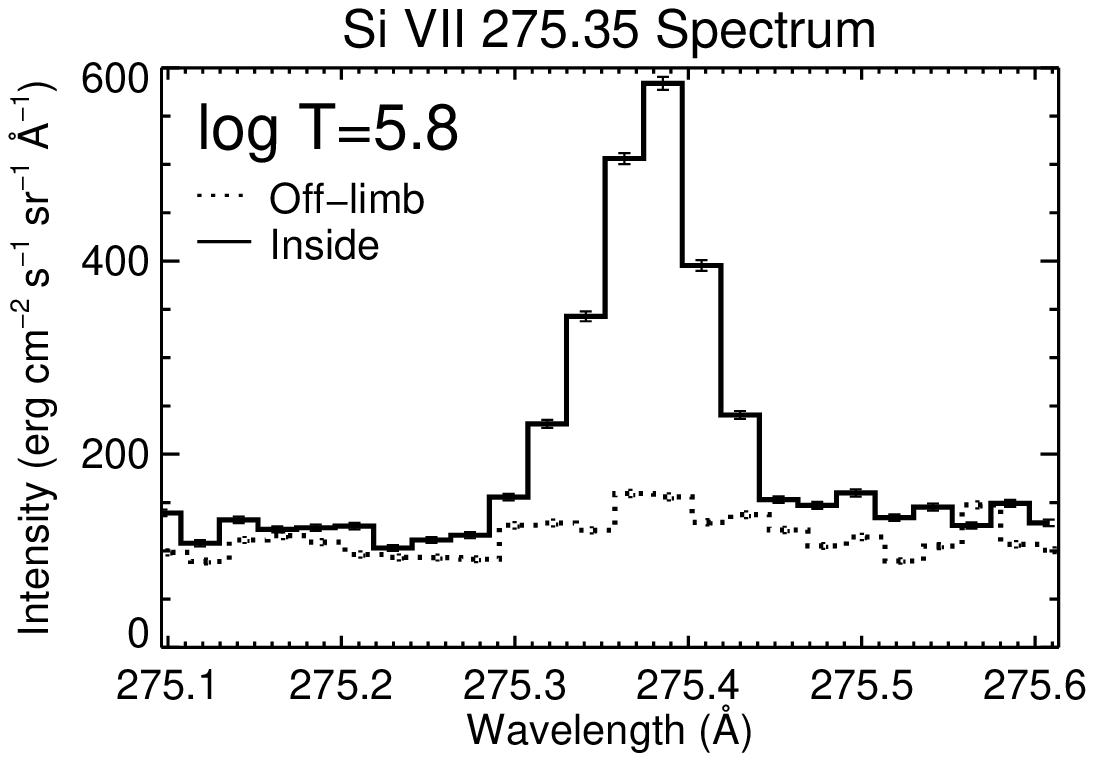}
  \includegraphics[width=7.9cm,clip]{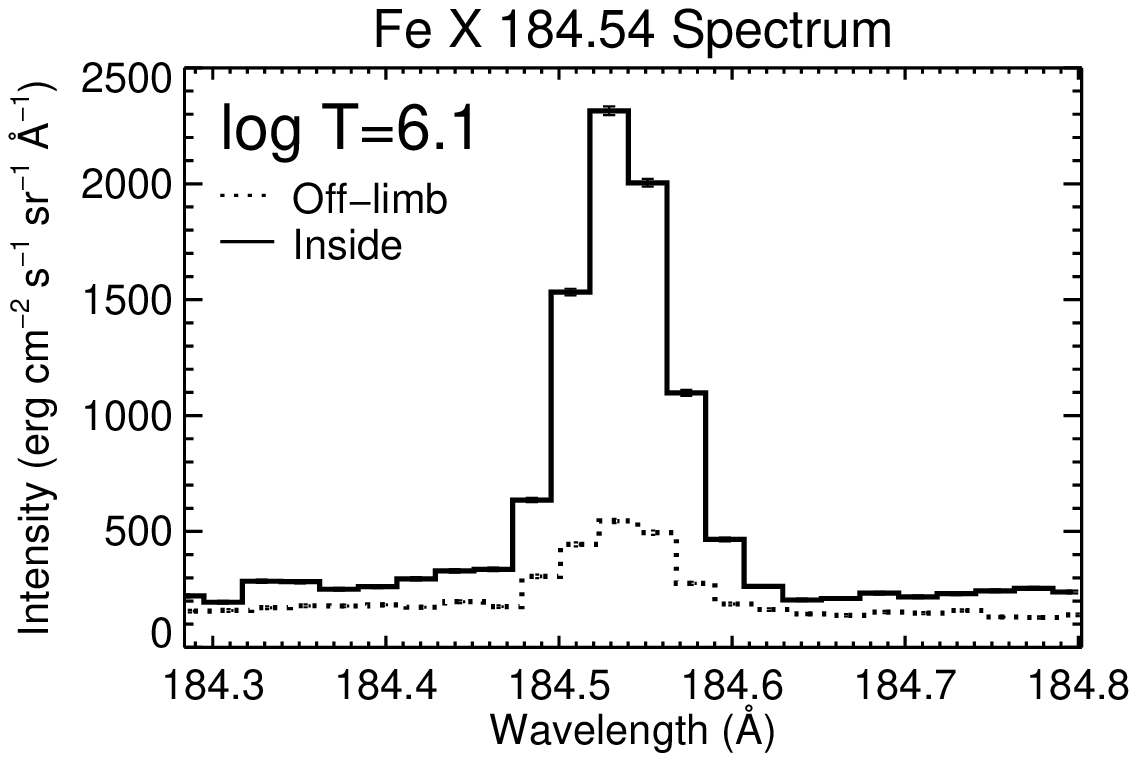}
  \includegraphics[width=7.9cm,clip]{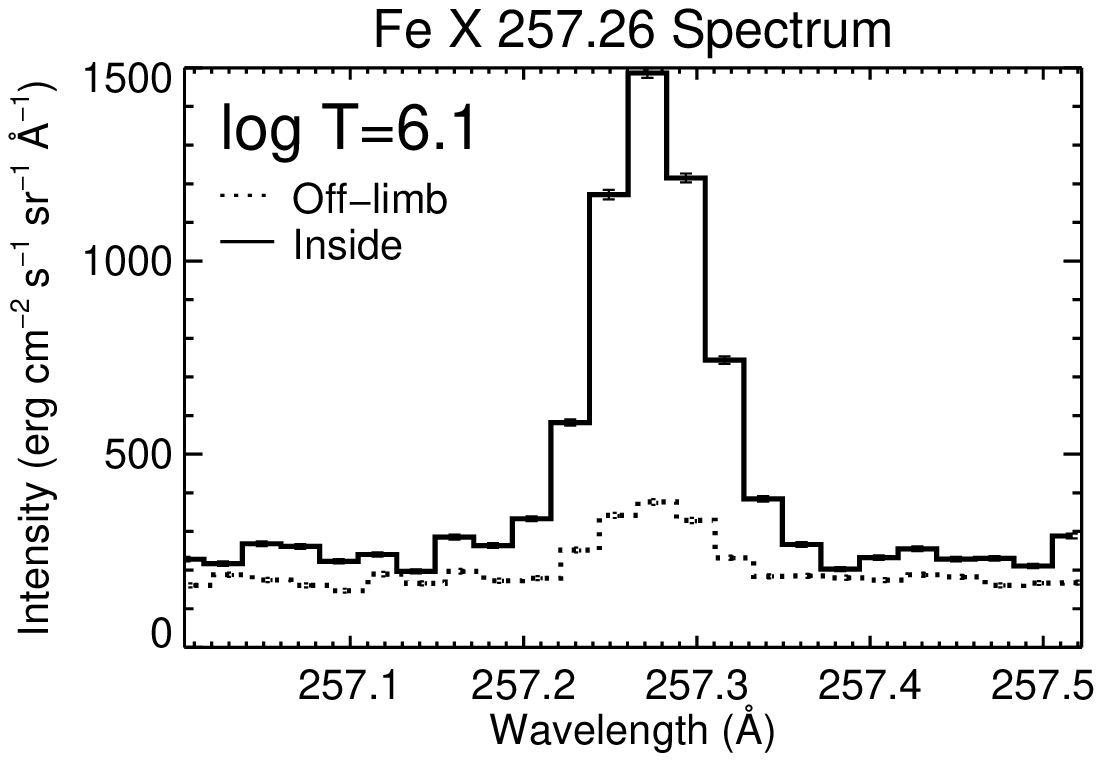}
  \caption{Line profiles on the disk (\textit{solid}) and off the limb (\textit{dotted}).
    A number in the left upper corner in each panel is the logarithmic formation temperature for the emission line.  Error bars include photon noise and uncertainty in CCD pedestal and dark current.
  }
  \label{fig:cal_lp_ex}  
\end{figure}
\addtocounter{figure}{-1}

\begin{figure}
  \centering
  \includegraphics[width=7.9cm,clip]{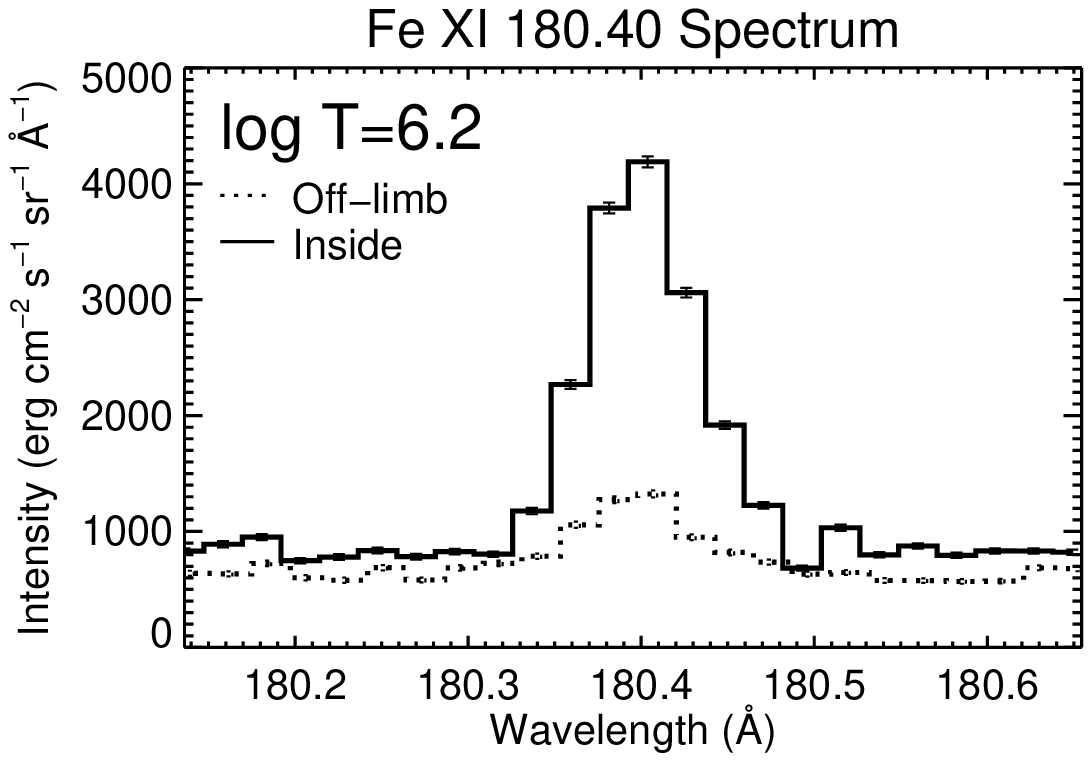}
  \includegraphics[width=7.9cm,clip]{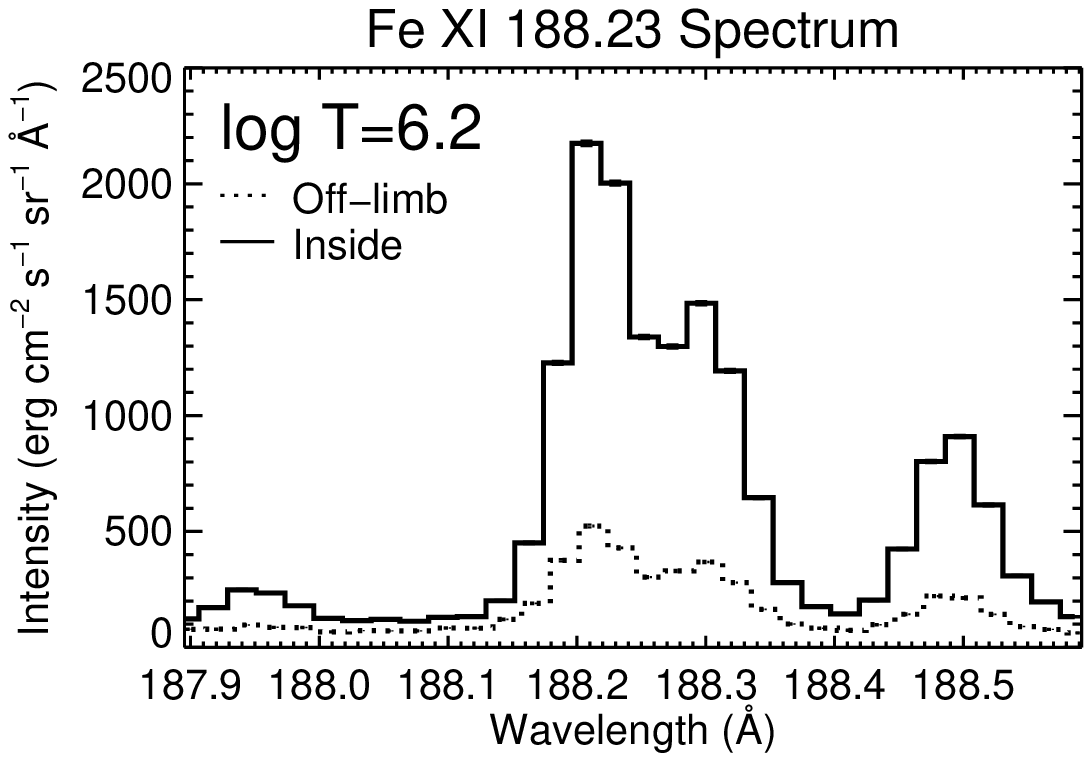}
  \includegraphics[width=7.9cm,clip]{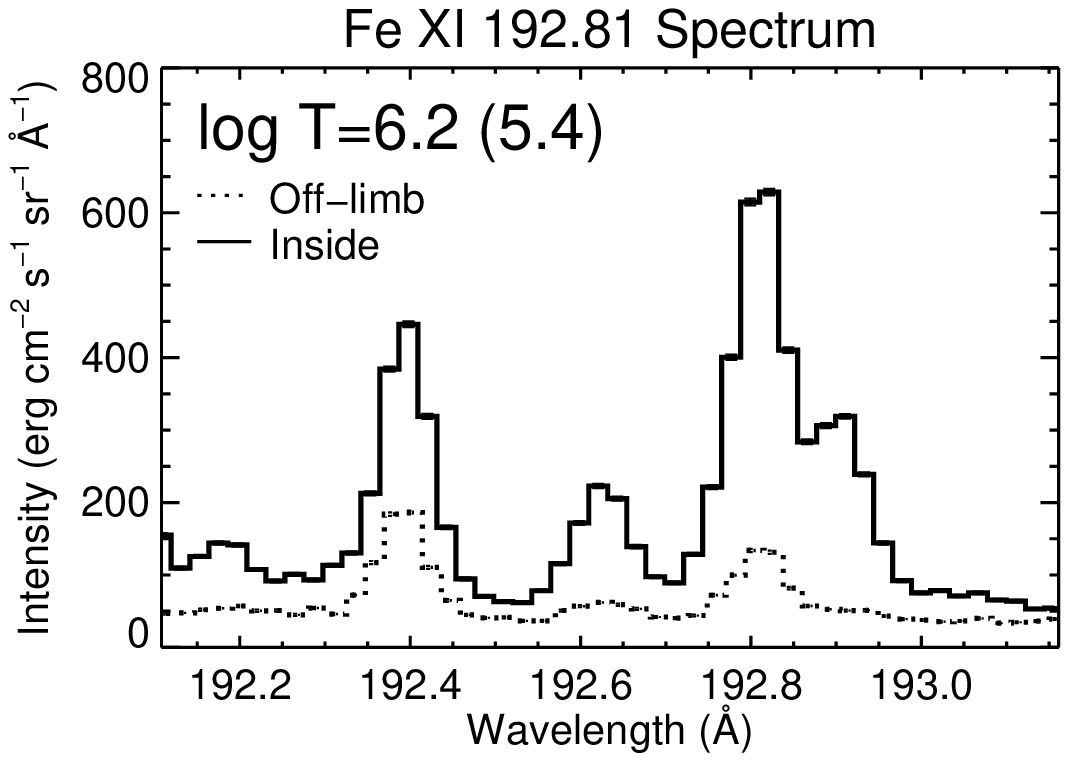}
  \includegraphics[width=7.9cm,clip]{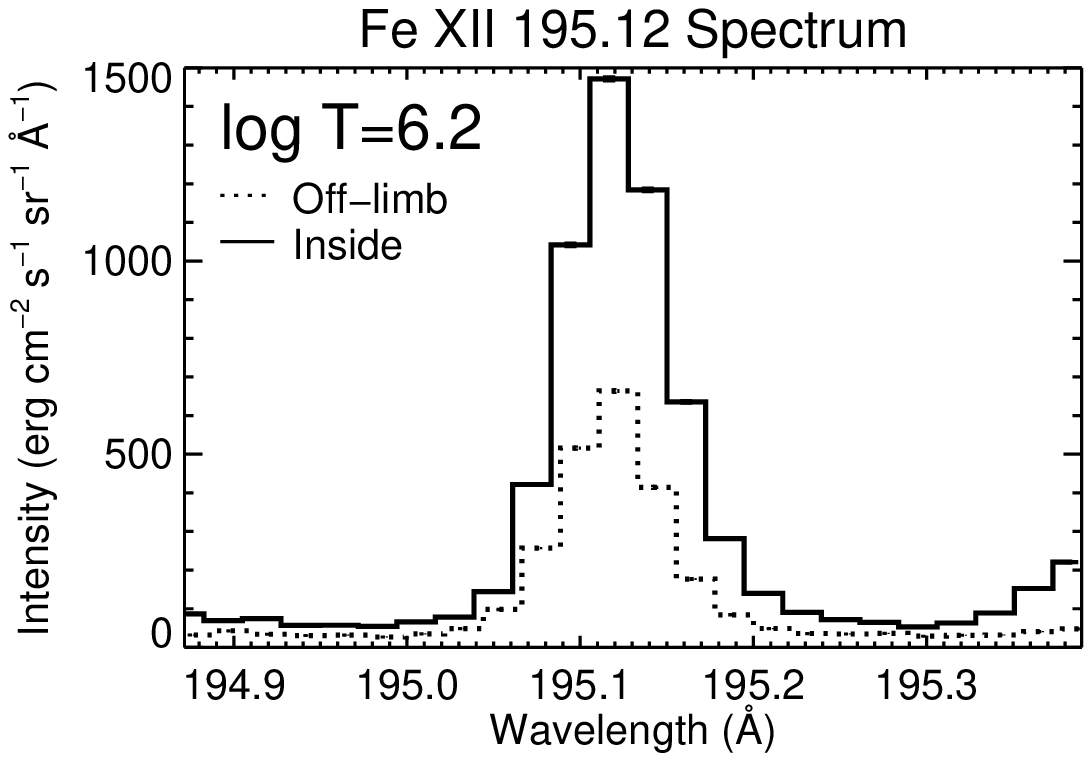}
  \includegraphics[width=7.9cm,clip]{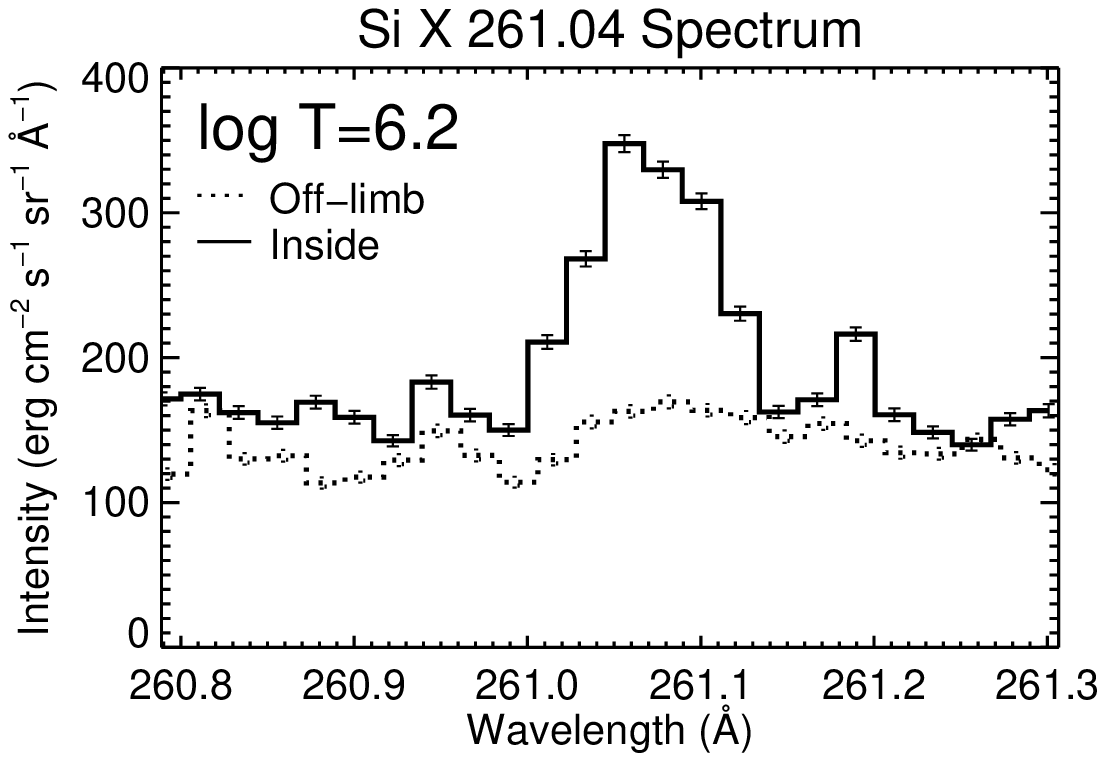}
  \includegraphics[width=7.9cm,clip]{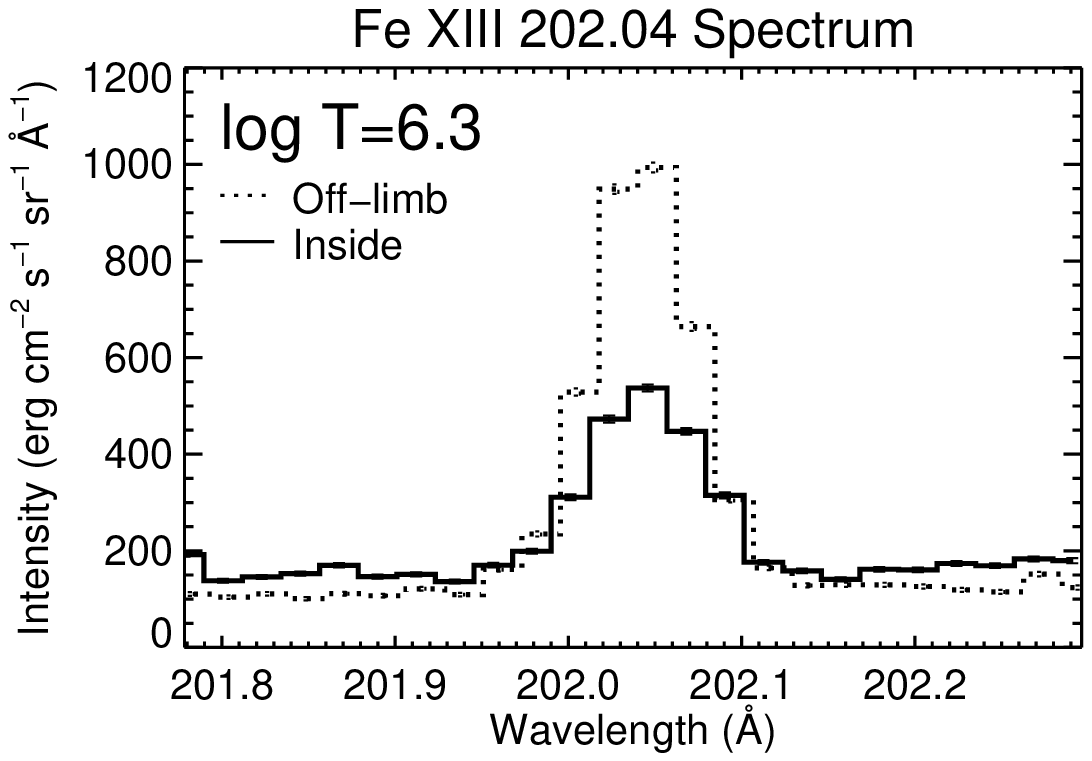}
  \includegraphics[width=7.9cm,clip]{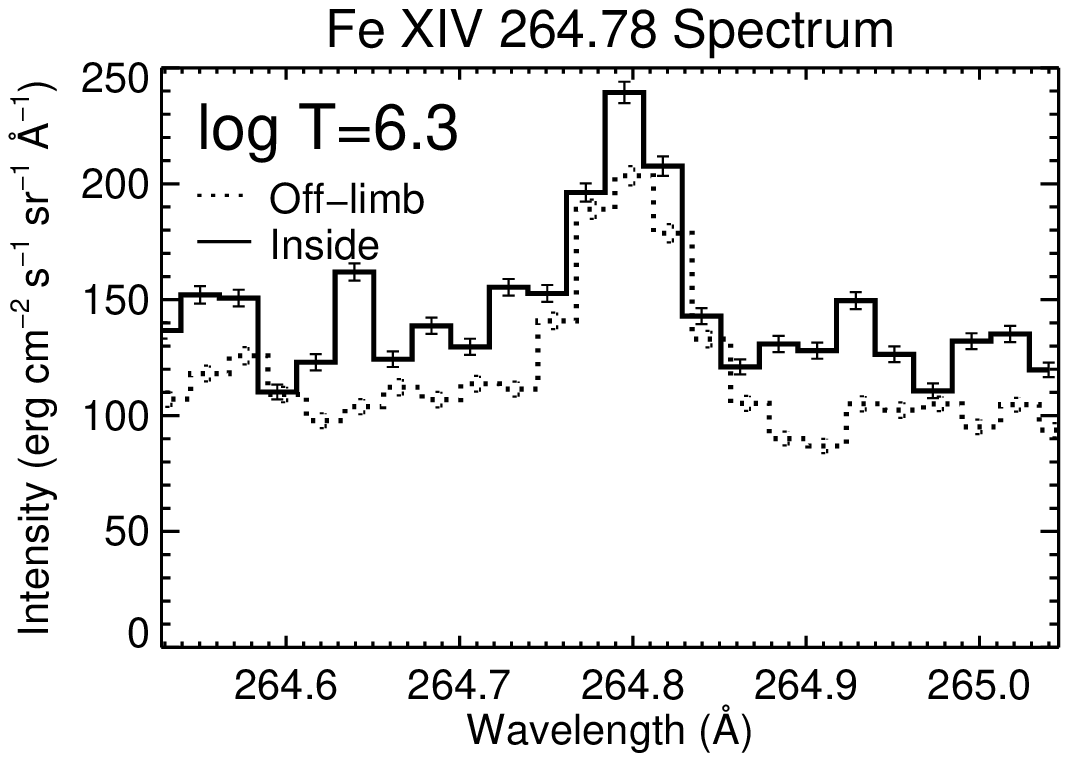}
  \includegraphics[width=7.9cm,clip]{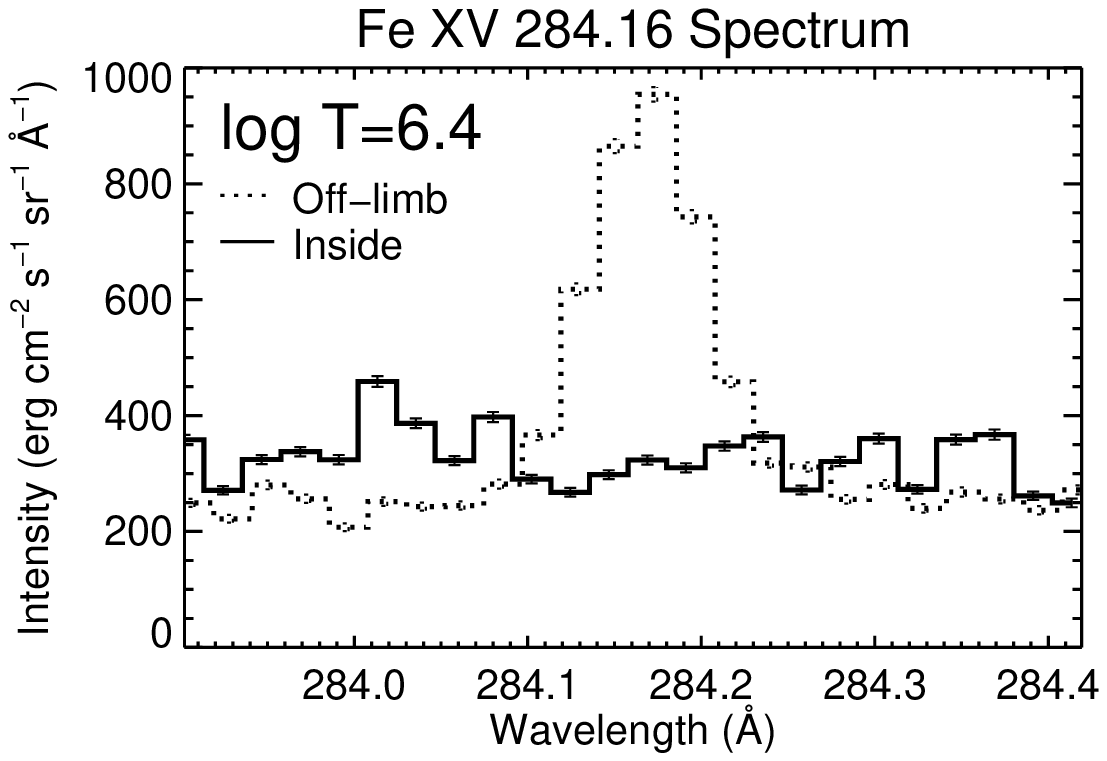}
  \caption{\textit{Continued.}}
\end{figure}

% --- End of TeX ---

%% file: tex/cal_lp_tr.tex
% ====================================
%   Chap. Doppler shift of the QS
% ====================================

\paragraph{He \textsc{ii}}
He \textsc{ii} $256.32${\AA} is known to be one of the strongest emission lines in EIS spectra and the only one with the formation temperature below $\log T \, [\mathrm{K}] = 5.0$.  The emission is very weak above the limb which indicates that it comes from the bottom of the corona or lower.  As seen in the solid line profile, He \textsc{ii} $256.32${\AA} has a long enhanced red wing.  This is the contribution from Si \textsc{x} $256.37${\AA}, and this blend makes the analysis of He \textsc{ii} $256.32${\AA} much complex.  Ideally, we can remove this Si \textsc{x} by referring Si \textsc{x} $261.04${\AA} since these line pair has constant intensity ratio of $I_{256.37} / I_{261.04} = 1.25$ (CHIANTI ver.\ 7; \citeauthor{landi2012} \citeyear{landi2012}) because their upper level of the transition are the same.  The intensity ratio might be possibly measured above the limb where He \textsc{ii} becomes much more weaker than inside the solar disk.  However, as seen from the off-limb spectrum (\textit{dotted histogram}) of Si \textsc{x} $261.04${\AA} in Fig.~\ref{fig:cal_lp_ex}, it was not strong enough to be used as a reference emission line (\textit{i.e.,} noisy).  Therefore, we did not use He \textsc{ii} $256.32${\AA} in this analysis.

\paragraph{O \textsc{iv}--\textsc{v}}
The EIS data analyzed here includes two oxygen emission lines: O \textsc{iv} $279.93${\AA} ($\log T \, [\mathrm{K}] = 5.2$) and O \textsc{v} $248.48${\AA} ($\log T \, [\mathrm{K}] = 5.4$).  Previous observations have reported that the transition region lines around $\log T \, [\mathrm{K}] \simeq 5.0$--$5.5$ are redshifted by up to $\sim 10 \, \mathrm{km} \, \mathrm{s}^{-1}$ at the disk center \citep{chae1998doppler,peterjudge1999,teriaca1999}, and that it is meaningful to analyze those oxygen lines to confirm the consistency between the previous observations and our results.  Since the spectra of these emission lines are much weaker compared to other emission lines observed (\textit{e.g.,} Fe emission lines), we integrated the spectra almost all along the slit ($500''$) at the expense of spatial resolution in the analysis as described in Section \ref{sect:cal_results_ov}.  As seen from the spectra in Fig.~\ref{fig:cal_lp_ex}, the integration of $100$ pixels obviously does not look enough to measure the precise line centroid.

\paragraph{Fe \textsc{viii} and Si \textsc{vii}}
Two emission lines Fe \textsc{viii} $186.60${\AA} and Si \textsc{vii} $275.35${\AA} are strong and well-isolated from other strong ones, in addition, the formation temperatures are similar.  There is a Ca \textsc{xiv} emission line near the line centroid of Fe \textsc{viii} $186.60${\AA}, but its influence is thought to be very weak in the quiet region due to its high formation temperature of Ca \textsc{xiv} ($\log T \, [\mathrm{K}]=6.7$).  Mg \textsc{vi} $268.99${\AA} has formation temperature similar to that of Fe \textsc{viii} and Si \textsc{vii}, but it was too noisy to achieve the precision of several $\mathrm{km} \, \mathrm{s}^{-1}$ so we did not use this line. 

% --- End of TeX ---

%% file: tex/cal_lp_co.tex
% ====================================
%   Chap. Doppler shift of the QS
% ====================================

\paragraph{Fe \textsc{ix}}
At the longer wavelength side in the spectral window of Fe \textsc{xi} $188.21${\AA}/$188.30${\AA}, there is a Fe \textsc{ix} $188.49${\AA}, which is isolated and relatively strong. We can fill the wide temperature gap between Fe \textsc{viii} ($\log T \, [\mathrm{K}] = 5.69$) and Fe \textsc{x} ($\log T \, [\mathrm{K}] = 6.04$) by using this line. 

\paragraph{Fe \textsc{x}}
There are two emission lines from Fe \textsc{x}: $184.54${\AA}/$257.26${\AA} in the analyzed EIS data.  Both are free from any significant blend by other lines near the line center.  Note that at the red wing of Fe \textsc{x} $184.54${\AA}, a weak line Fe \textsc{xi} $184.41${\AA} exists.  But this line is much weaker than Fe \textsc{x} $184.54${\AA} in the quiet region. 

\paragraph{Fe \textsc{xi}}
For Fe \textsc{xi} emission lines, there are three spectral windows including them: $180.40${\AA}, $188.21${\AA}, and $192.81${\AA}.  All these three lines unfortunately suffer from a significant blending.  Near the line center of Fe \textsc{xi} $180.40${\AA}, there is Fe \textsc{x} $180.44${\AA} ($\sim 2$ pixels apart each other in the EIS CCD).  This emission line is density sensitive and becomes stronger at a location where the electron density is higher.  This may cause a systematic redshift compared to the result from other Fe \textsc{xi} emission lines.  As seen in the Fe \textsc{xi} $188.21$ spectrum in Fig.~\ref{fig:cal_lp_ex}, two emission lines with comparative strength are blending each other: Fe \textsc{xi} $188.21${\AA}/$188.30${\AA}.  We fitted Fe \textsc{xi} $188.21${\AA}/$188.30${\AA} by double Gaussians, which is considered to be robust because these two emission lines are both strong and their line profiles usually have two distinct peaks.  The third emission line Fe \textsc{xi} $192.81${\AA} is significantly blended by the transition region lines O \textsc{v} $192.90${\AA} in the quiet region, so we did not use that line.  

\paragraph{Fe \textsc{xii}}
Two emission lines Fe \textsc{xii} $192.39${\AA} and $195.12${\AA} are both strong and suitable for the analysis of the quiet region.  One problem in the analysis of Fe \textsc{xii} $195.12${\AA} is that there exists a blend by Fe \textsc{xii} $195.18${\AA} and the line ratio $195.18${\AA}/$195.12${\AA} has a sensitivity for the electron density.  This will cause an apparent shift of the emission line toward the longer wavelength (\textit{i.e.,} redshift) especially in active regions and at bright points where the electron density typically becomes higher by an order of magnitude than that in the quiet region.

\paragraph{Fe \textsc{xiii}}
Fe \textsc{xiii} $202.04${\AA} is only one strong emission line from Fe \textsc{xiii} in this EIS study and known to be a clean line without any significant blend.  Different from emission lines with lower formation temperature, the spectra of Fe \textsc{xiii} $202.04${\AA} above the limb and inside the solar disk shown in Fig.~\ref{fig:cal_lp_ex} indicate that the off-limb spectrum is stronger than the disk spectrum by approximately twice. This value is what can be expected from the limb brightening effect. 

\paragraph{Fe \textsc{xiv}--\textsc{xv}}
Emission lines Fe \textsc{xiv} $264.78${\AA} and Fe \textsc{xv} $284.16${\AA} were very weak in the quiet region even with the exposure time of $120 \, \mathrm{s}$.  The off-limb spectrum and the disk spectrum of Fe \textsc{xv} indicate the same behavior as those of Fe \textsc{xiii}.  However, the spectra of Fe \textsc{xiv} behave differently from them.  This is considered to be the influence of an emission line Fe \textsc{xi} $264.77${\AA} existing near the line centroid of Fe \textsc{xiv}.  In the quiet region, contribution from Fe \textsc{xi} could become relatively strong compared to Fe \textsc{xiv} because the average temperature is slightly lower than that in active regions.  It is possible that Fe \textsc{xv} has the similar problem in the quiet region.  Therefore, we did not analyze Fe \textsc{xiv} and Fe \textsc{xv} emission lines here in order not to derive improper results. 

% --- End of TeX ---

%% file: tex/cal_lp_fit.tex
% ==============================
%   Fitting of line profiles.
% ==============================

\begin{figure}
  \centering
  \includegraphics[width=8.4cm,clip]{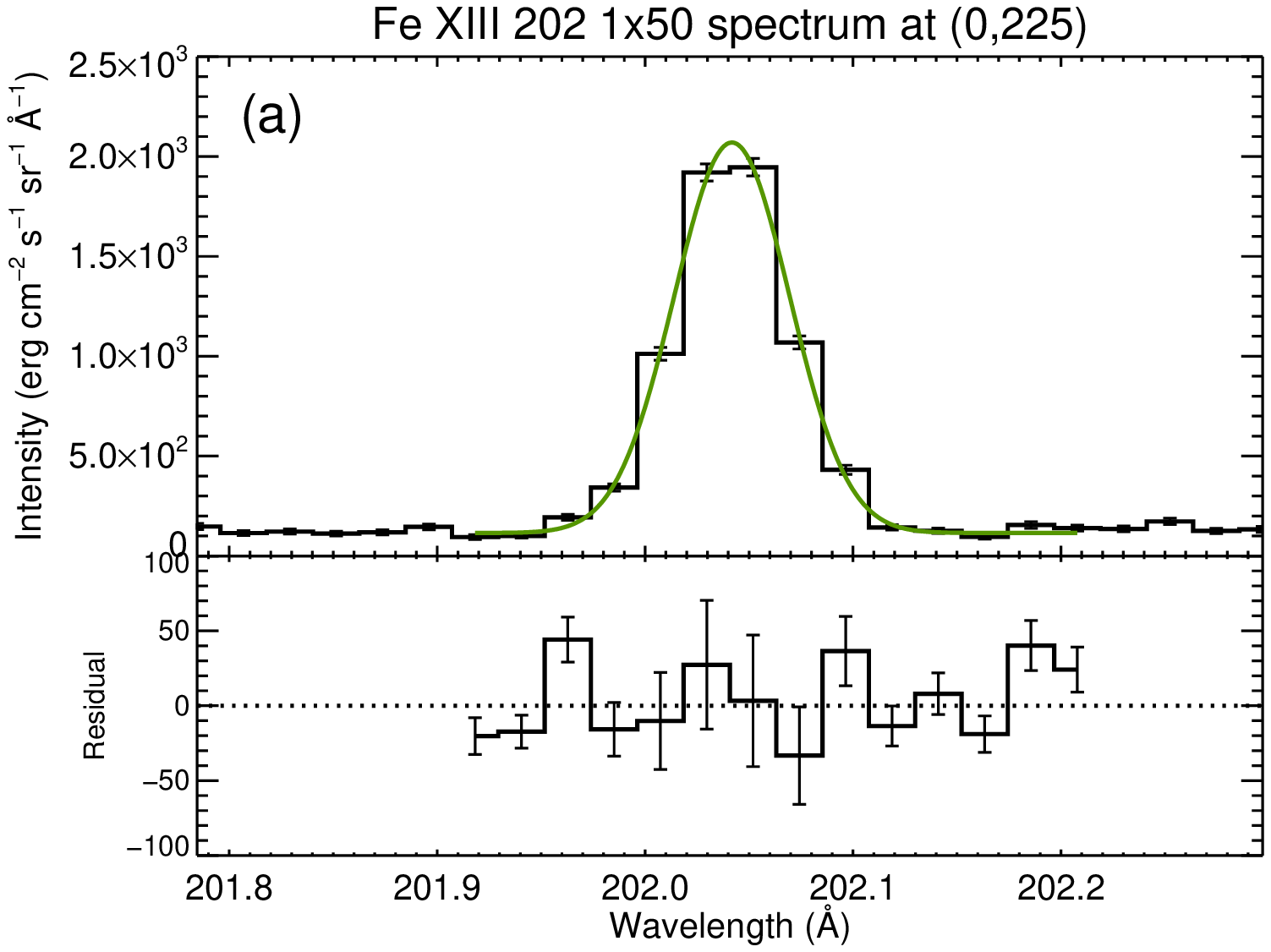}
  \includegraphics[width=8.4cm,clip]{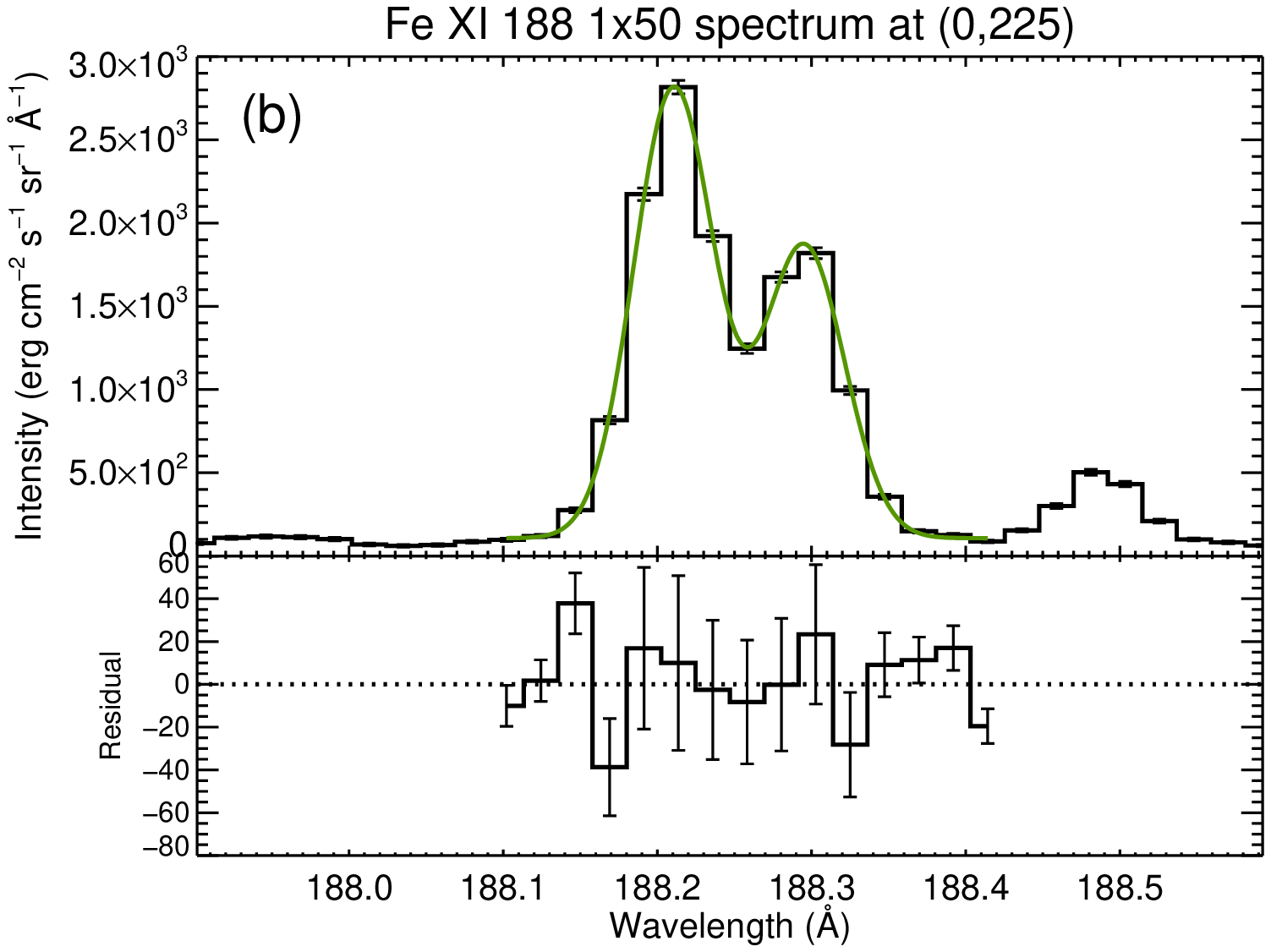}
  \caption{Line profile at the disk center taken during HOP79 in 2010 October. 
    (a) Fe \textsc{xi} $188.21${\AA}/$188.30${\AA}, and 
    (b) Fe \textsc{xiii} $202.04${\AA}. 
    Error bars include photon noise and uncertainty in CCD pedestal and dark current.}
  \label{fig:lp_ex_dc}
\end{figure}

Most of the emission lines used in this analysis can be considered to be well fitted by a single Gaussian because they are isolated and strong.  
%{\color{red} 
In order to reduce the fluctuations caused by the existence of coronal structures (\textit{e.g.}, bright points) and non-radial motions, we spatially integrated the spectra by $50''$ in the $y$ direction.
%}
One example of fitting Fe \textsc{xiii} $202.04${\AA} by a single Gaussian is shown in panel (a) of Fig.~\ref{fig:lp_ex_dc}.  Upper part of the panel shows the line profile at the disk center taken during HOP79 in 2010 October and a green line is a fitted single Gaussian.  We used a single Gaussian in the form of
\begin{equation}
  f (\lambda) = a_0 \exp \left[ - \dfrac{\left( \lambda - a_1 \right)^2}{2 a_2^2} \right] + a_3 
  \, \mathrm{.}
\end{equation}
Coefficients $a_i$ ($i=0,1,2,3$) respectively represent peak, line centroid, line width (Gaussian width), and constant background.  Lower part shows residuals from the fitted Gaussian, which do not exceed $\sim 2${\%} of the peak in the spectrum and are comparative to the errors.  Only the wavelength range used for the fitting is plotted in the lower panel.  The single Gaussian fitting was applied to the emission lines Fe \textsc{viii} $186.60${\AA}, Si \textsc{vii} $275.35${\AA}, Fe \textsc{ix} $188.49${\AA}, Fe \textsc{x} $184.54${\AA}/$257.26${\AA}, Fe \textsc{xi} $180.40${\AA}, Fe \textsc{xii} $192.39${\AA}/$195.18${\AA}, and Fe \textsc{xiii} $202.04${\AA}.  Each spectra were fitted by using $8$--$14$ pixels which include each emission line.

As an exception, Fe \textsc{xi} $188.21${\AA} and $188.30${\AA} were fitted by double Gaussians because they clearly overlap with each wing.  In this case we used double Gaussians with constant background.  Upper part of panel (b) shows a line profile of the Fe \textsc{xi} emission lines at the disk center as same as Fe \textsc{xiii} $202.04${\AA}, and a green line indicates the result of fitting. %Lower part of the panel again shows residuals from the double Gaussian from which we see that the fitting worked well. 

\begin{figure}
  \centering
  \includegraphics[width=8.4cm,clip]{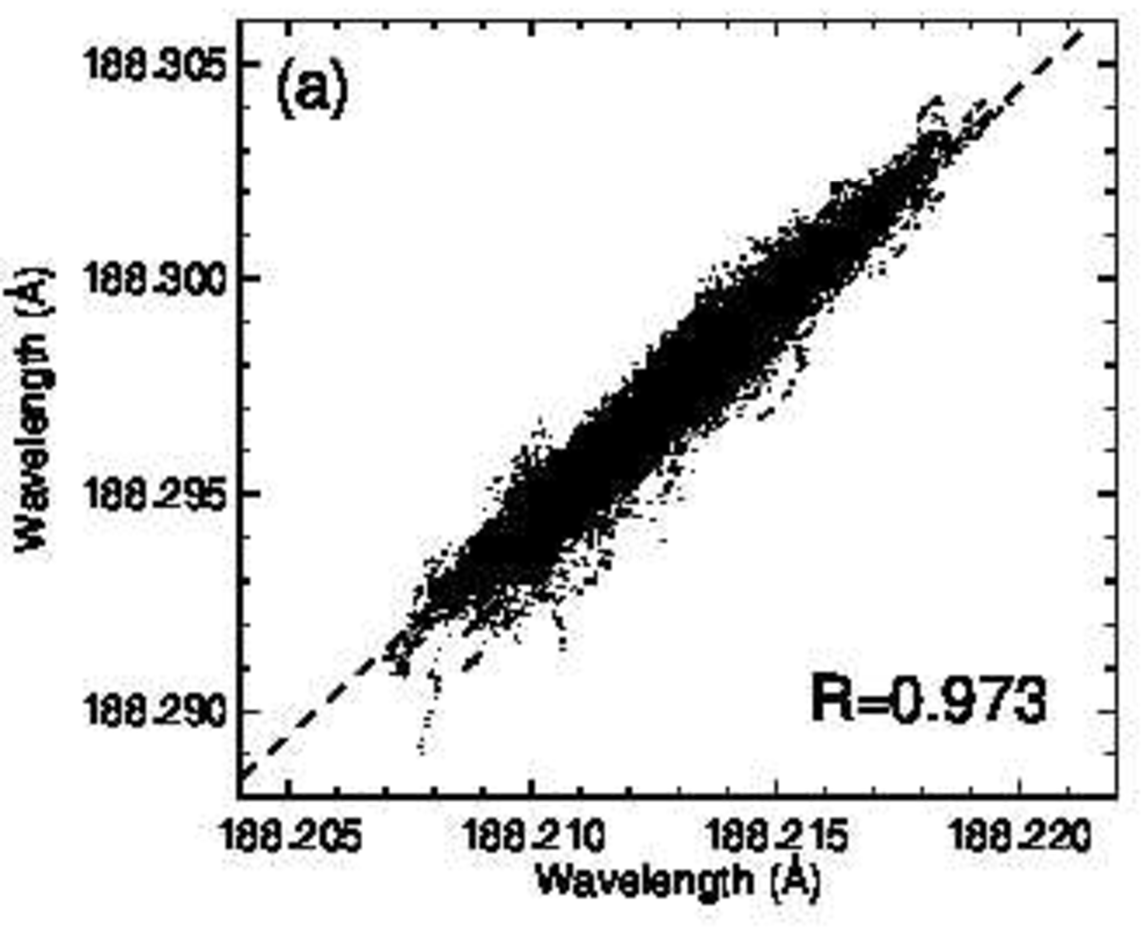}
  \includegraphics[width=8.4cm,clip]{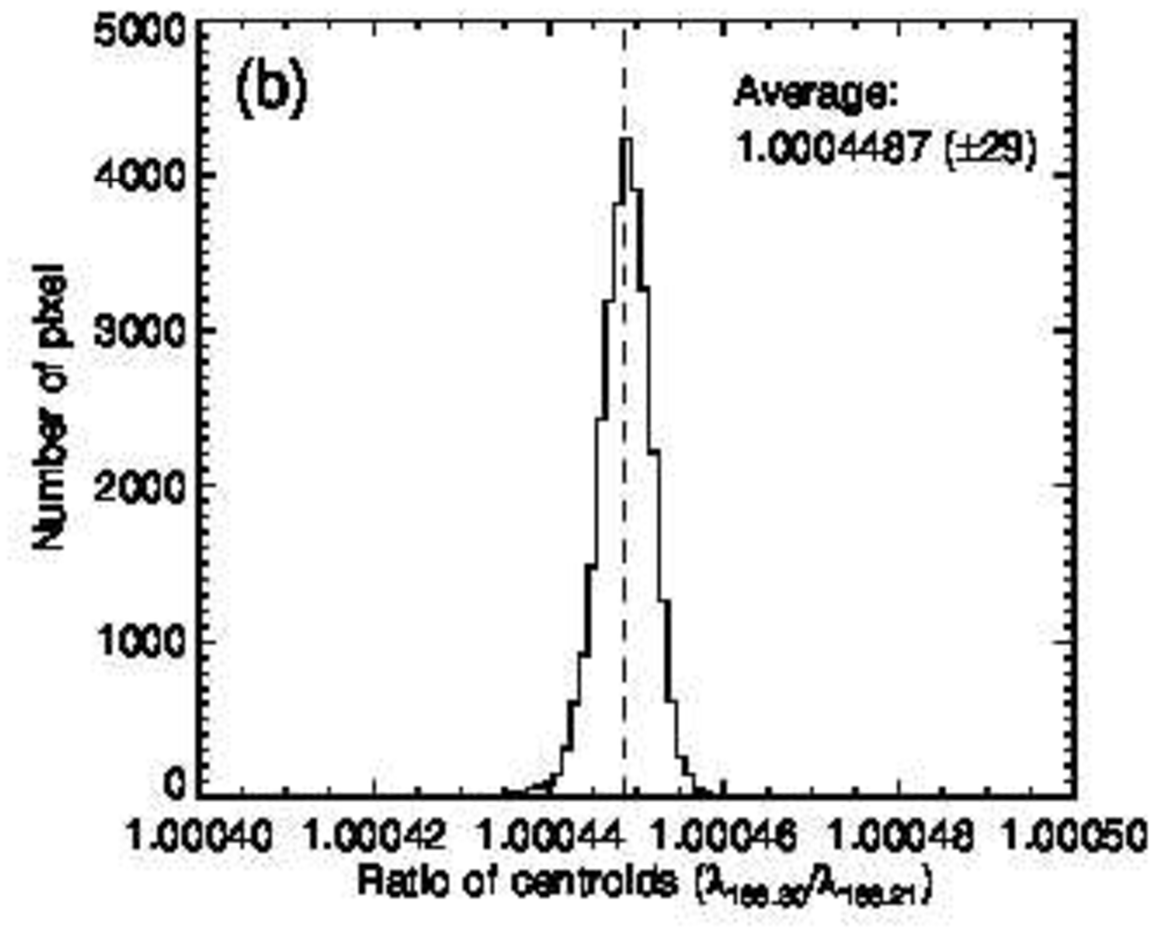}
  \caption{(a) Scatter plot for line centroids of Fe \textsc{xi} $188.21${\AA}/$188.30${\AA}.  Dashed line indicates a regression line based on a model function $y=ax$.  The coefficient $a$ was $1.0004487 (\pm 1 \times 10^{-7})$.  The correlation coefficient was $R=0.973$.  (b) Histogram for the ratio of line centroids ($\lambda_{188.30}/\lambda_{188.21}$).  The average value was $1.0004487$ and the standard deviation was $2.9 \times 10^{-6}$.}
  \label{fig:sct_188}
\end{figure}

In order to check the robustness of our double Gaussian fitting for Fe \textsc{xi} $188.21${\AA}/$188.30${\AA}, the scatter plot for fitted line centroids of two emission lines is made as shown in panel (a) of Fig.~\ref{fig:sct_188}.  Theoretically, the line centroids from the same ion have the relationship $\lambda_2 / \lambda_1 = \mathrm{const.}$ ($\lambda_1$ and $\lambda_2$ are the line centroid of two emission lines) considering the Doppler effect cancels out because the factor $1 + v / c$ is common between the emission lines from the same ion.  The two line centroids clearly have the positive correlation with the correlation coefficient of $R=0.973$. In addition, the ratio of two line centroids was $\lambda_{188.30} / \lambda_{188.21} = 1.0004487 \pm 2.9 \times 10^{-6}$ in the average as shown in panel (b), which is almost identical to the theoretical value ($1.0004405$).  Thus, we conclude that the double Gaussian is reliable.

% --- End of TeX ---

%% file: tex/cal_tilt.tex
% ==========================================
%   Chapter:
%     Doppler shifts in the quiet region.
%   Description:
%     Spectrum tilt.
% ==========================================

The spectra taken by EIS are known to be slightly tilted from $y$ axis when projected onto the CCDs (hereafter we call this effect as the spectrum tilt).  This arises from the subtle misalignment of the spectroscopic slits, the grating component and the CCDs.  The tilts of the slits and the grating component should cause the same degree of the tilts in the observed spectra.  The two CCDs are known to be displaced a little each other, which makes the spectrum of the two CCDs different.  The spectrum tilt is crudely $1$ pixel in the wavelength direction along the full height of the CCDs ($1024$ pixels), which corresponds to $\simeq 20$--$30 \, \kmpers$.

The current standard EIS software calibrates this effect by referring the spectrum tilt obtained at the off limb.  It is fixed and has no dependence on wavelength.  However, the spectrum tilt may have different characteristics on the two CCDs and even with wavelength. Since we aim to deduce Doppler velocities in the order of a few $\kmpers$, we carefully investigated the spectrum tilt by analyzing the scan taken in the quiet region at the solar disk as described in Section \ref{sect:cal_append_itdn}.  The obtained tilts were used to calibrate the north--south scan data from HOP79. 

% --- End of TeX ---

%% file: tex/cal_align.tex
% ==================================================
%   Chapter:
%     Average Doppler shifts of the quiet region.
%   Description:
%     Alignment of data between exposures.
% ==================================================

\begin{figure}
  \centering
  \includegraphics[width=8.4cm]{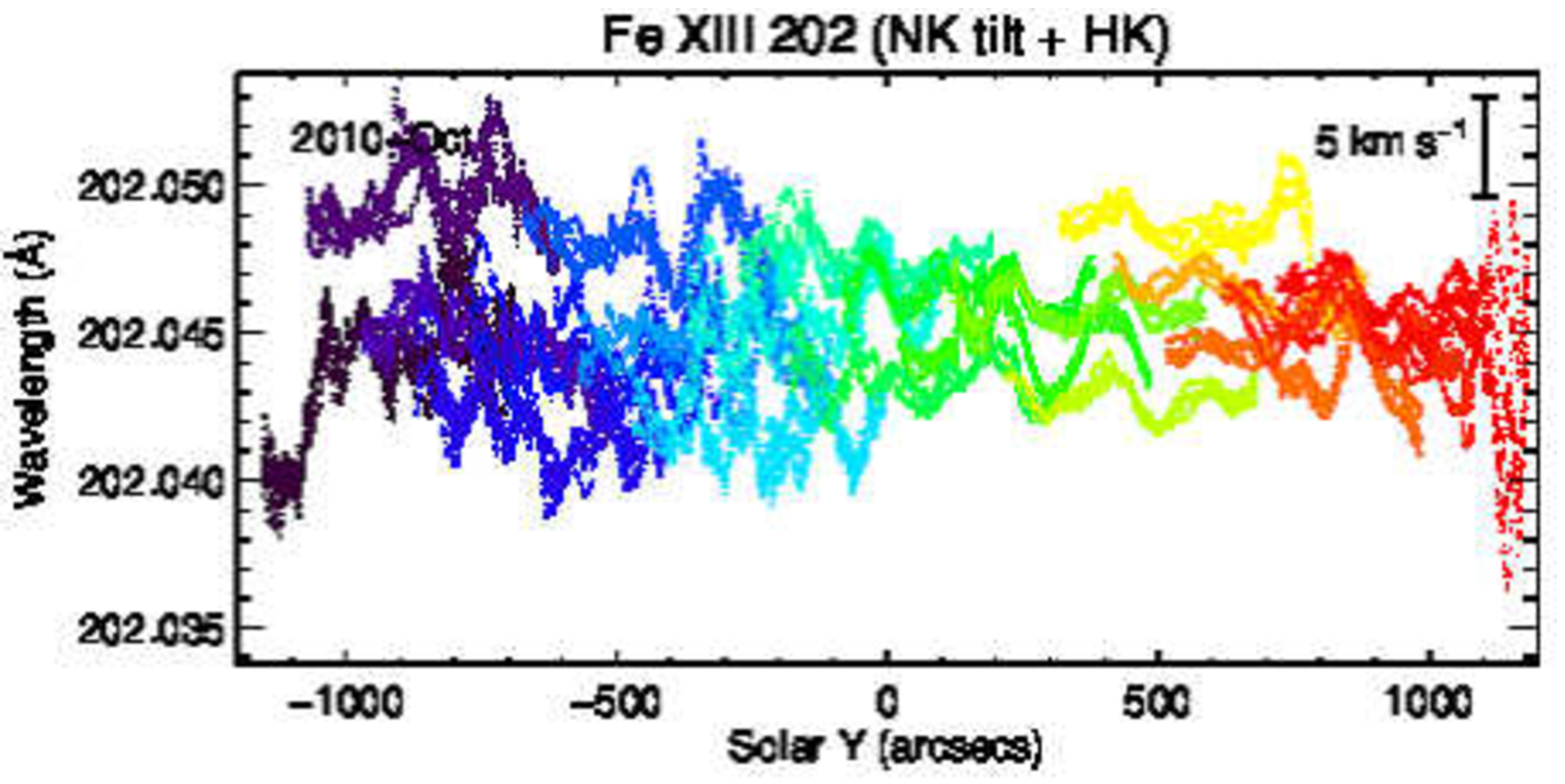}
  \includegraphics[width=8.4cm]{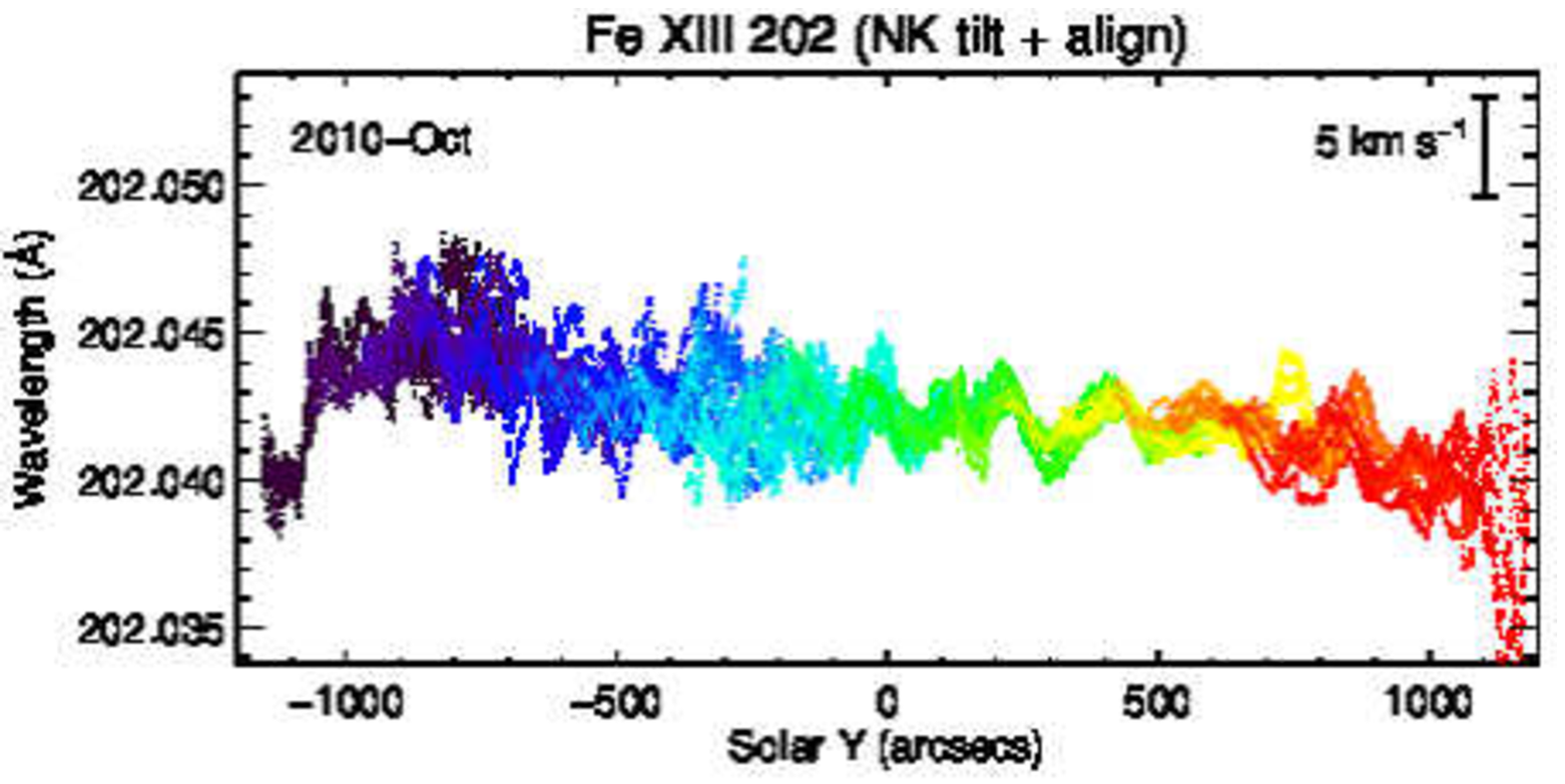}
  \caption{\textit{Left}: line centroids of Fe \textsc{xiii} $202.04${\AA} during the north--south scans derived through the standard EIS analysis software provided by the SSW package.  %{\color{red}
Different colors respectively indicate data points obtained at each pointing of the satellite ($20$ locations from the south pole to the north pole).  \textit{Right}: line centroids aligned between neighboring pointings.  Data arrays were shifted in \textit{vertical} axis of the panel so that the difference of line centroids between overlapping region becomes smallest.%}
}
  \label{fig:cal_align}
\end{figure}

The analysis in this chapter requires carefulness by the order of $1 \, \mathrm{km} \, \mathrm{s}^{-1}$, which means we need a little more finer reduction than the procedures provided by the standard EIS software in the SSW package.  The current SSW package includes a robust wavelength calibration developed by \citet{kamio2010}, but there remains an uncertainty of $\lesssim 10 \, \mathrm{km} \, \mathrm{s}^{-1}$ due to two reasons described in Section \ref{sect:cal_itdn}.  In order to reduce this uncertainty, we (1) aligned five exposures in one pointing, and (2) aligned them between the neighboring pointings so that the sum of squared difference in overlapped region becomes smallest.  Here we assumed that the average Doppler shift in the quiet region is independent of time.  An example of alignment is shown in Fig.~\ref{fig:cal_align}.  In order to compare our reduction to the standard package, the line centroid calibrated by SSW is plotted in panel (a).  The data after aligned are shown in panel (b), which is much less dispersed than the data in panel (a).  From this result, we confirmed that the fluctuation up to $0.005${\AA} ($\simeq 7 \, \mathrm{km} \, \mathrm{s}^{-1}$) is still remained after analyzed through the standard SSW package.

Note that a systematic linear behavior remains for some emission lines even after those careful analysis above.  This may come from the residual of the spectrum tilt removal, but we do not specify the exact reason at present.  In order to compensate the linear component appropriately as long as possible, we adopted the idea that the Doppler shifts must be symmetric about the disk center, which should be safe in the global FOV. 

% --- End of TeX ---

%% file: tex/cal_results_limb2limb.tex
% ==========================================
%   Chapter:
%     Doppler shifts in the quiet region.
% ==========================================

\begin{figure}
  \centering
  \includegraphics[width=15.7cm,clip]{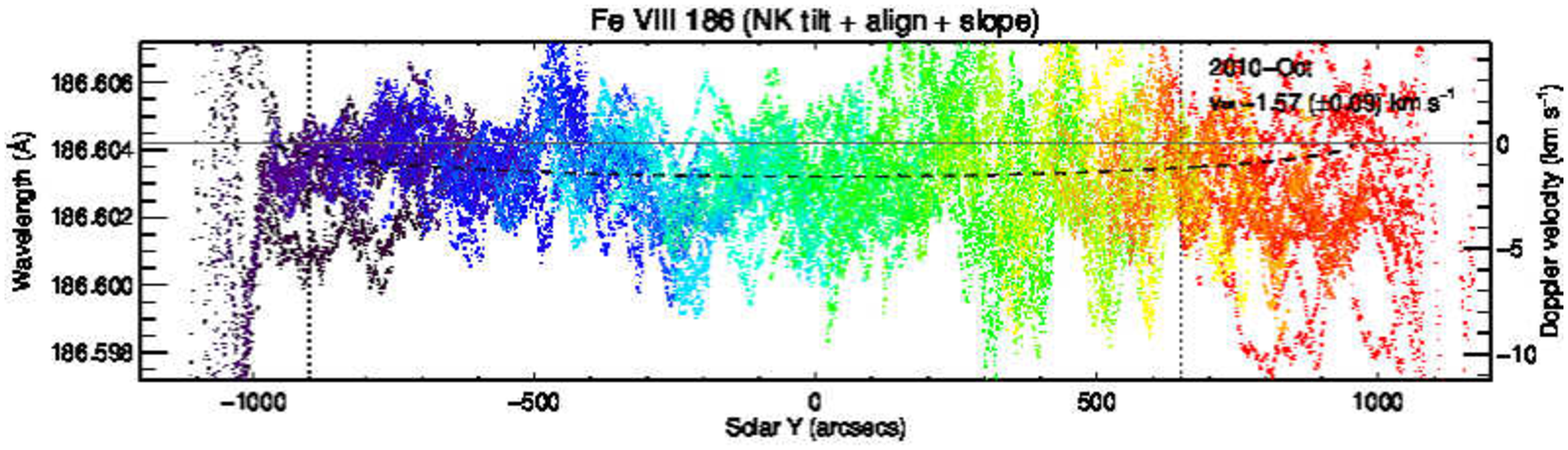}
  \includegraphics[width=15.7cm,clip]{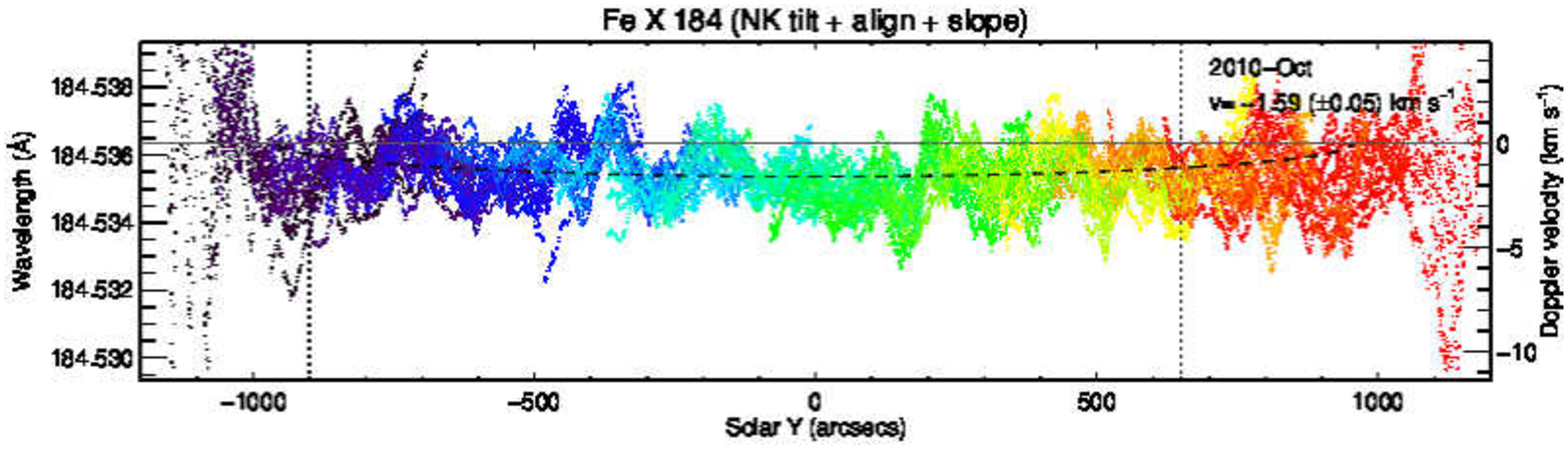}
  \includegraphics[width=15.7cm,clip]{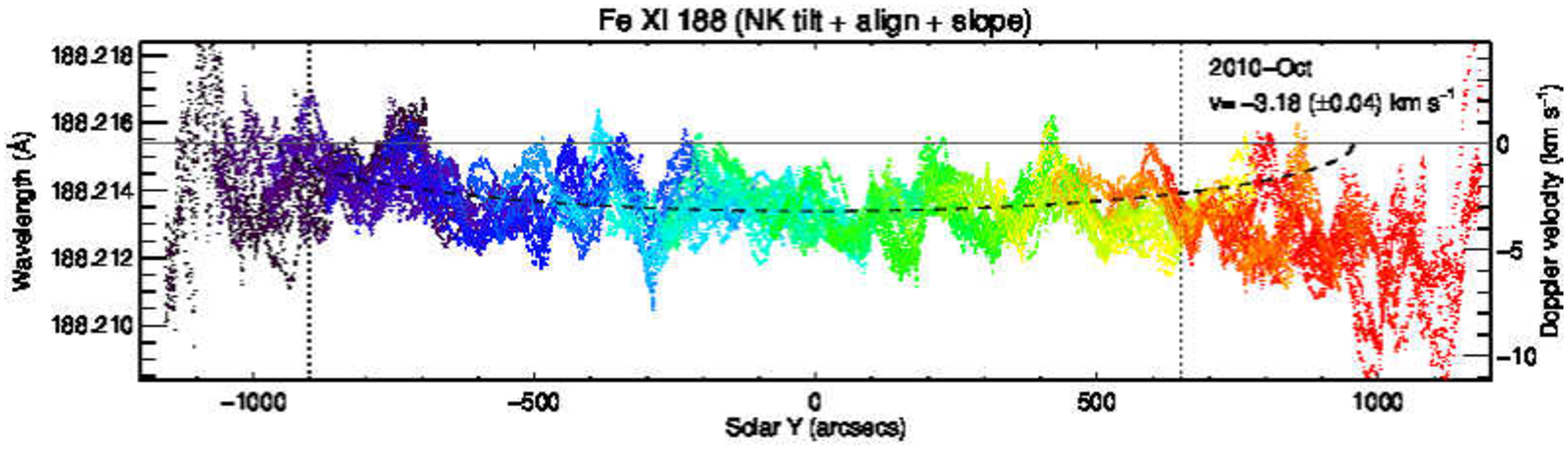}
  \includegraphics[width=15.7cm,clip]{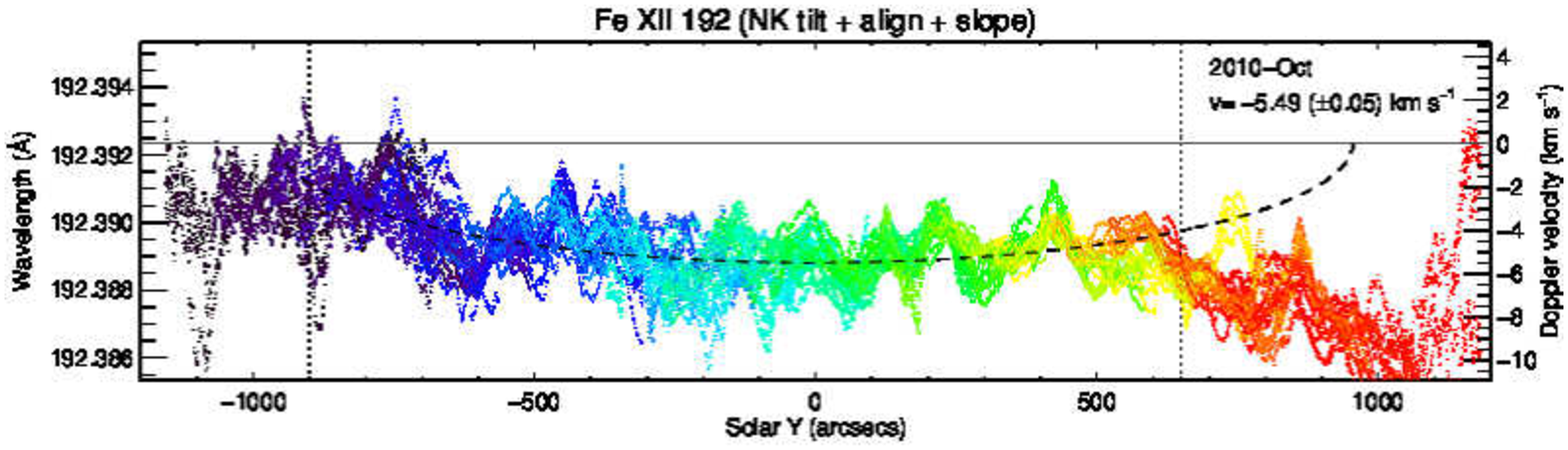}
  \includegraphics[width=15.7cm,clip]{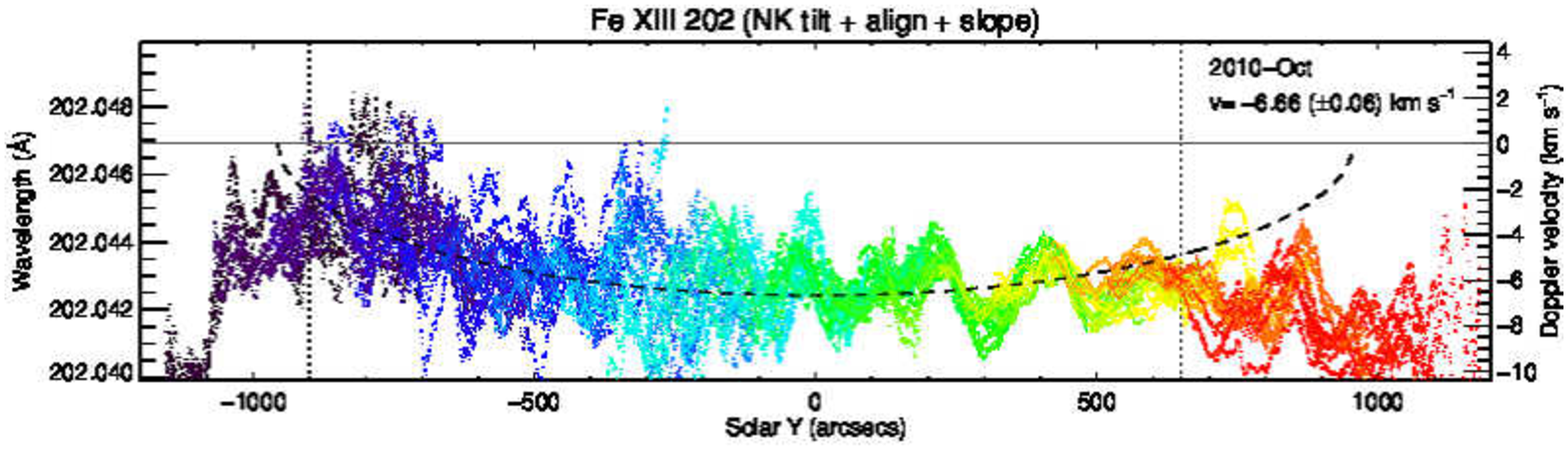}
  \caption{Center-to-limb variations of emission line centroid obtained 
           from the north--south scan during HOP79 in 2010 October.  Each panel 
           shows the variation of different emission line: 
    (a) Fe \textsc{viii} $186.60${\AA},
    (b) Fe \textsc{x} $184.54${\AA},
    (c) Fe \textsc{xi} $188.21${\AA}, 
    (d) Fe \textsc{xii} $192.39${\AA}, and 
    (e) Fe \textsc{xiii} $202.04${\AA}.
    Two vertical dotted line indicate that the data were fitted between those.
    \textit{Dashed} lines are the fitted line which has a form of 
    $v(y)=v_0 \cos \theta$ ($y=R_{\odot}\sin \theta$).
  }
  \label{fig:limb2limb_var_Oct}
\end{figure}

\begin{figure}
  \centering
  \includegraphics[width=15.7cm,clip]{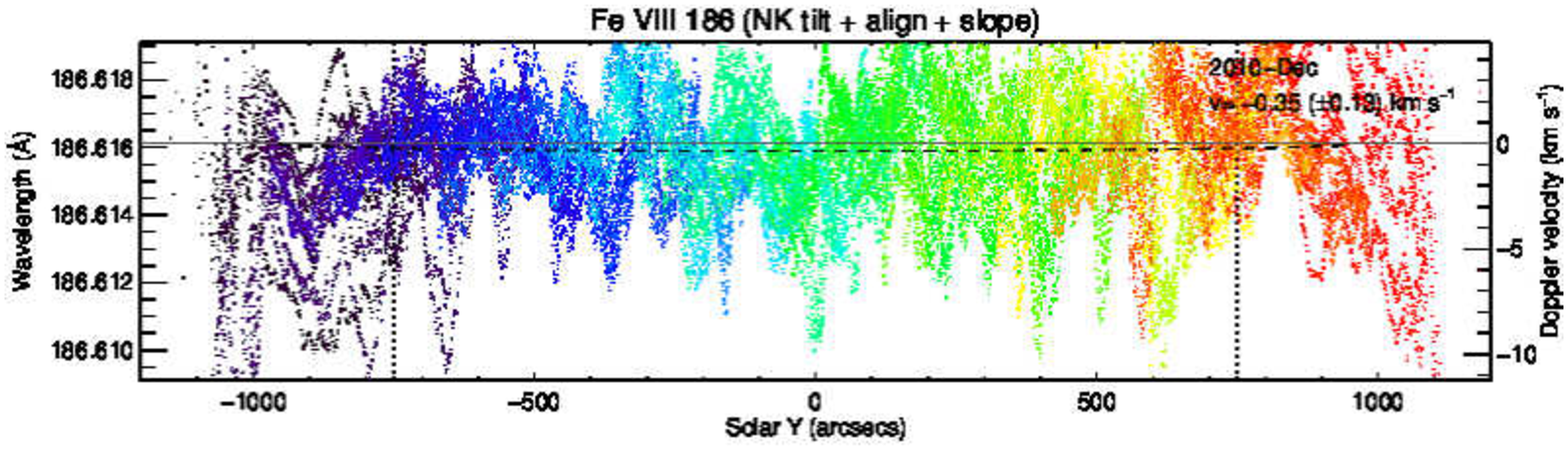}
  \includegraphics[width=15.7cm,clip]{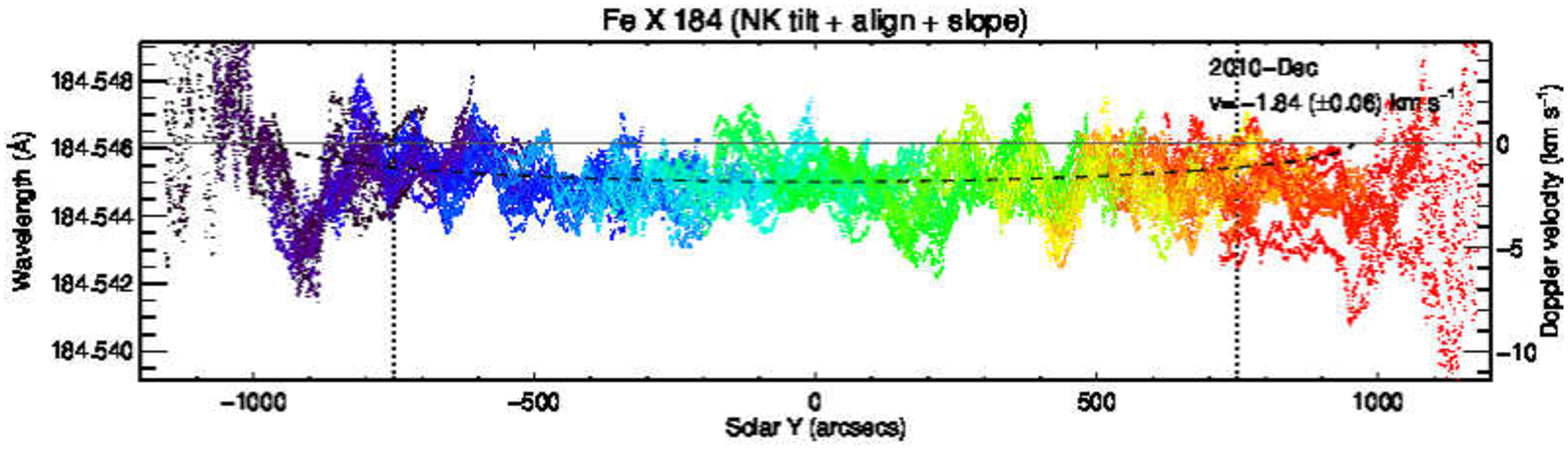}
  \includegraphics[width=15.7cm,clip]{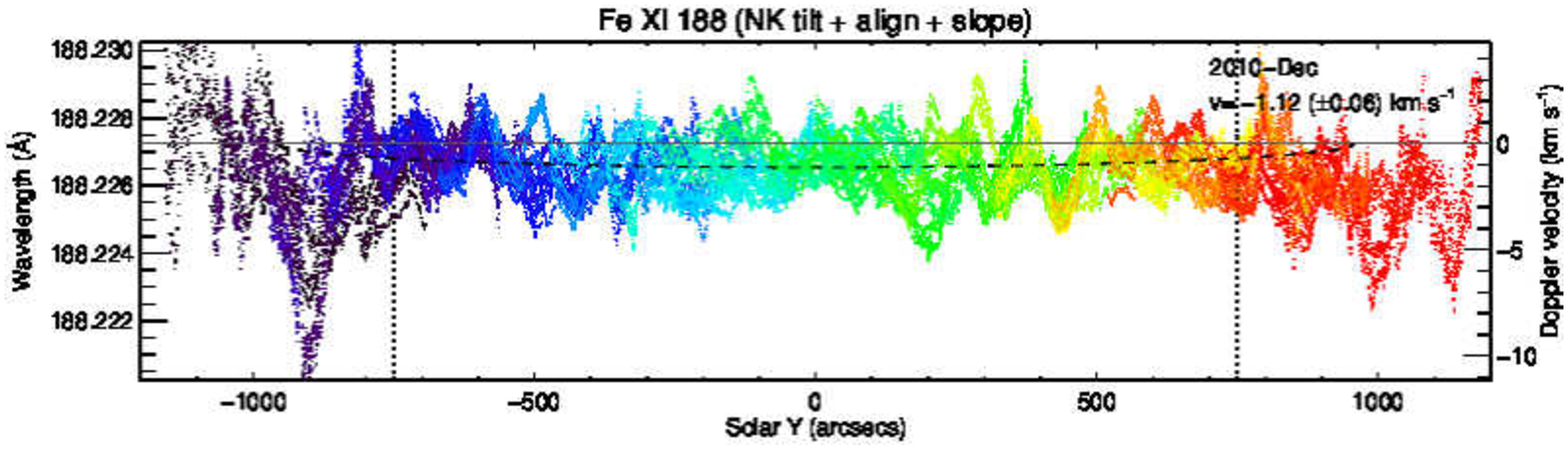}
  \includegraphics[width=15.7cm,clip]{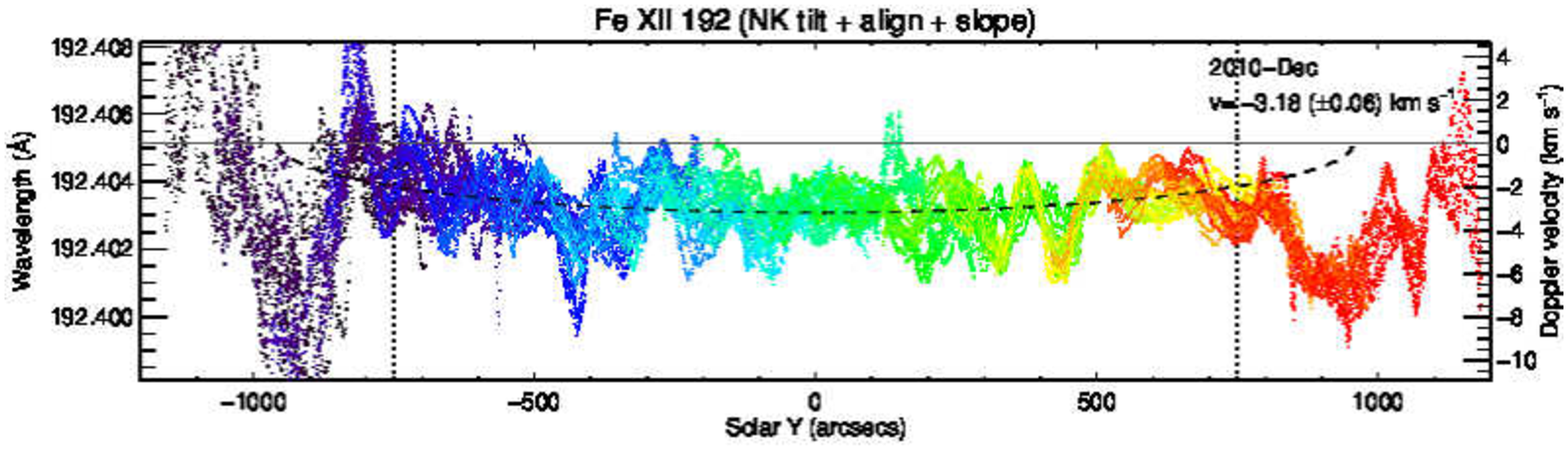}
  \includegraphics[width=15.7cm,clip]{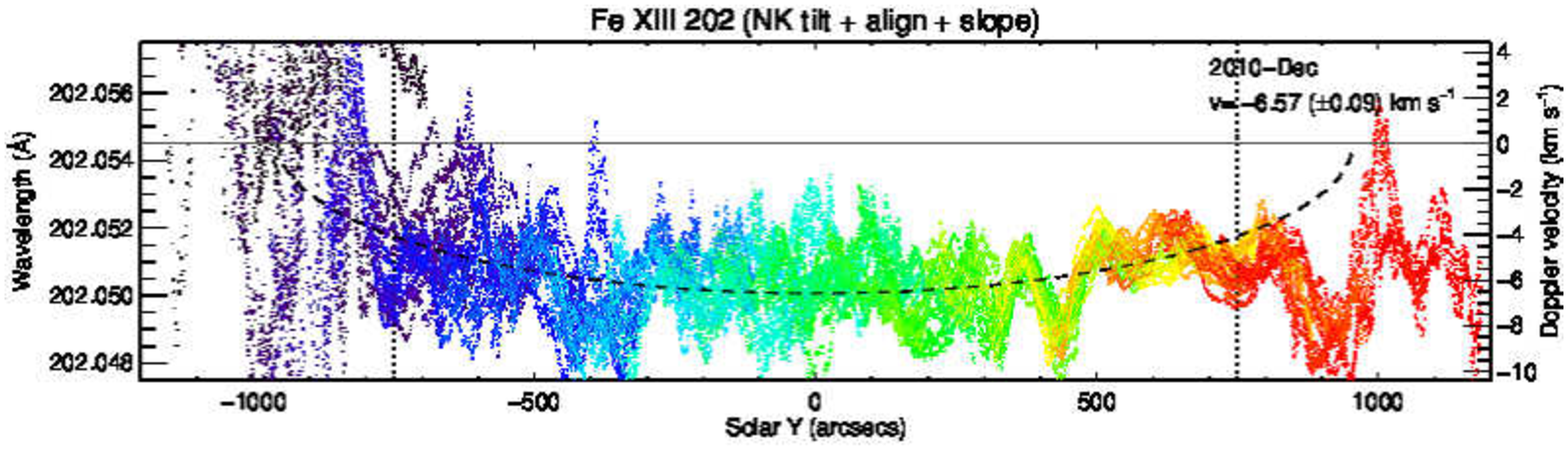}
  \caption{Center-to-limb variations of emission line centroid obtained 
           from the north--south scan during HOP79 in 2010 December. 
           Panels are in the same format as in Fig.~\ref{fig:limb2limb_var_Oct}.
  }
  \label{fig:limb2limb_var_Dec}
\end{figure}

We obtained the center-to-limb variations from the north--south scan during HOP79 after analyzing the data as described above.  Here results for five emission lines are respectively shown in Fig.~\ref{fig:limb2limb_var_Oct} and Fig.~\ref{fig:limb2limb_var_Dec} with an order of the formation temperature from \textit{upper} panel to \textit{lower} one: Fe \textsc{viii} $186.60${\AA}, Fe \textsc{x} $184.54${\AA}, Fe \textsc{xi} $188.21${\AA}, Fe \textsc{xii} $192.39${\AA}, and Fe \textsc{xiii} $202.04${\AA}.  In those observations, there were no active regions along the meridional line on the solar disk.  

Dashed lines in each panel represent the fitted curve when the center-to-limb variation of the Doppler shift is considered to be caused by the radial flow in the solar corona.  When the flow is in the radial direction only, the dependence of the Doppler velocity should be $v \, (\theta) = v_0 \cos \theta$ where $v_0$ is the radial velocity and $\theta$ is the angle between line of sight and normal vector as to the solar surface.  The solar $Y$ is represented as $y=R_{\odot} \sin \theta$.  We fitted the data by converting the abscissa into $\cos \theta$ and applied the linear function.  Note that the results were fitted within the range indicated by the region between two vertical dotted lines which indicates the quiet region.  There is a small coronal hole at the north pole on 2010 October 7--8 and emission lines are clearly blueshifted at $y \geq 700''$ which may be the indication of an outflow.  The radial velocity at the disk center is written in the right upper corner of each panel from which we see that the velocity decreases (\textit{i.e.,} upflow becomes stronger) with increasing formation temperature. The indicated errors are those calculated in the fitting procedure from the variance of the data points.  We hereafter denote these errors as $\sigma_{\mathrm{fit}}$. 

\begin{figure}
  \centering
  \includegraphics[width=6.2cm,clip]{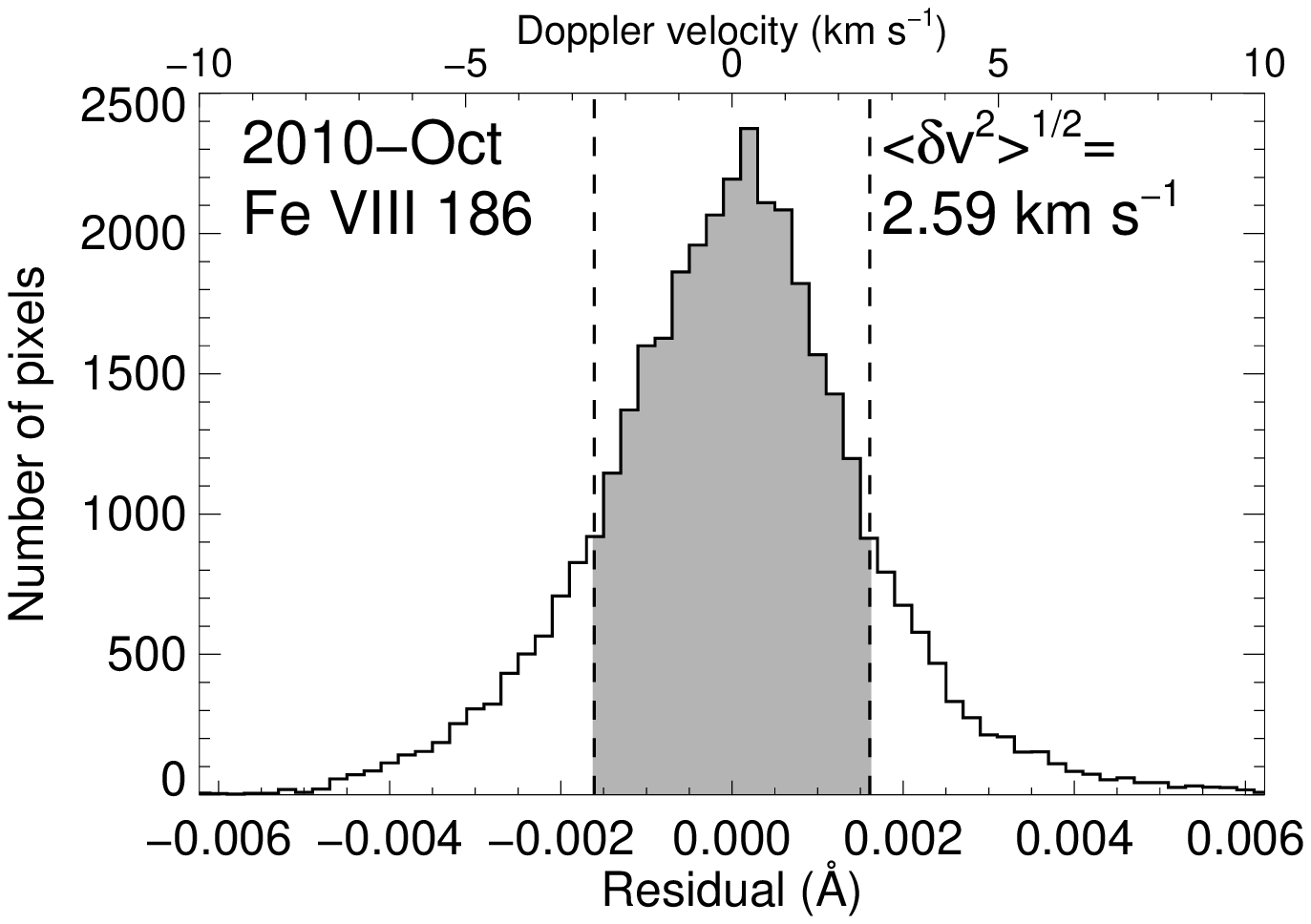}
  \includegraphics[width=6.2cm,clip]{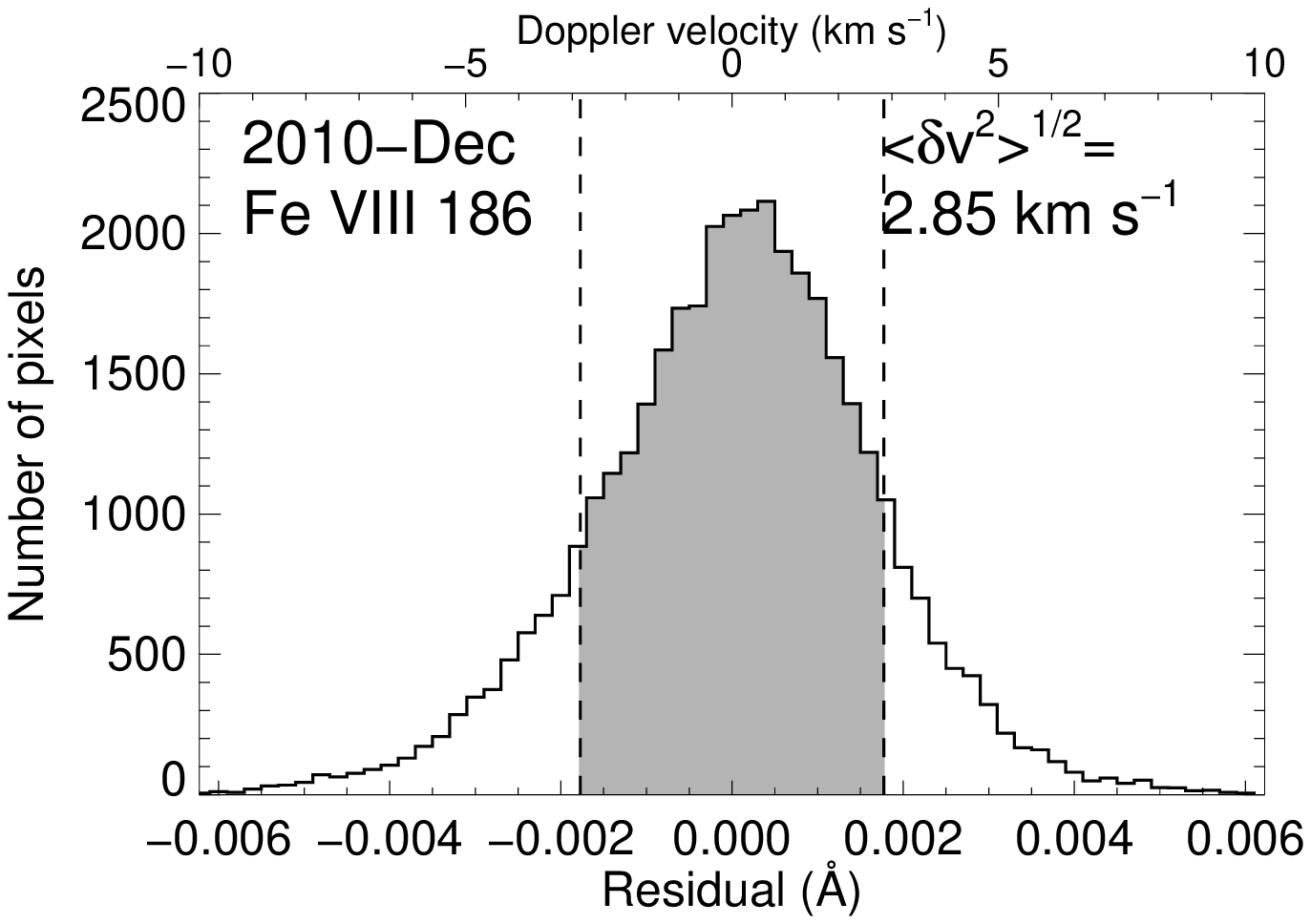} \\
  \includegraphics[width=6.2cm,clip]{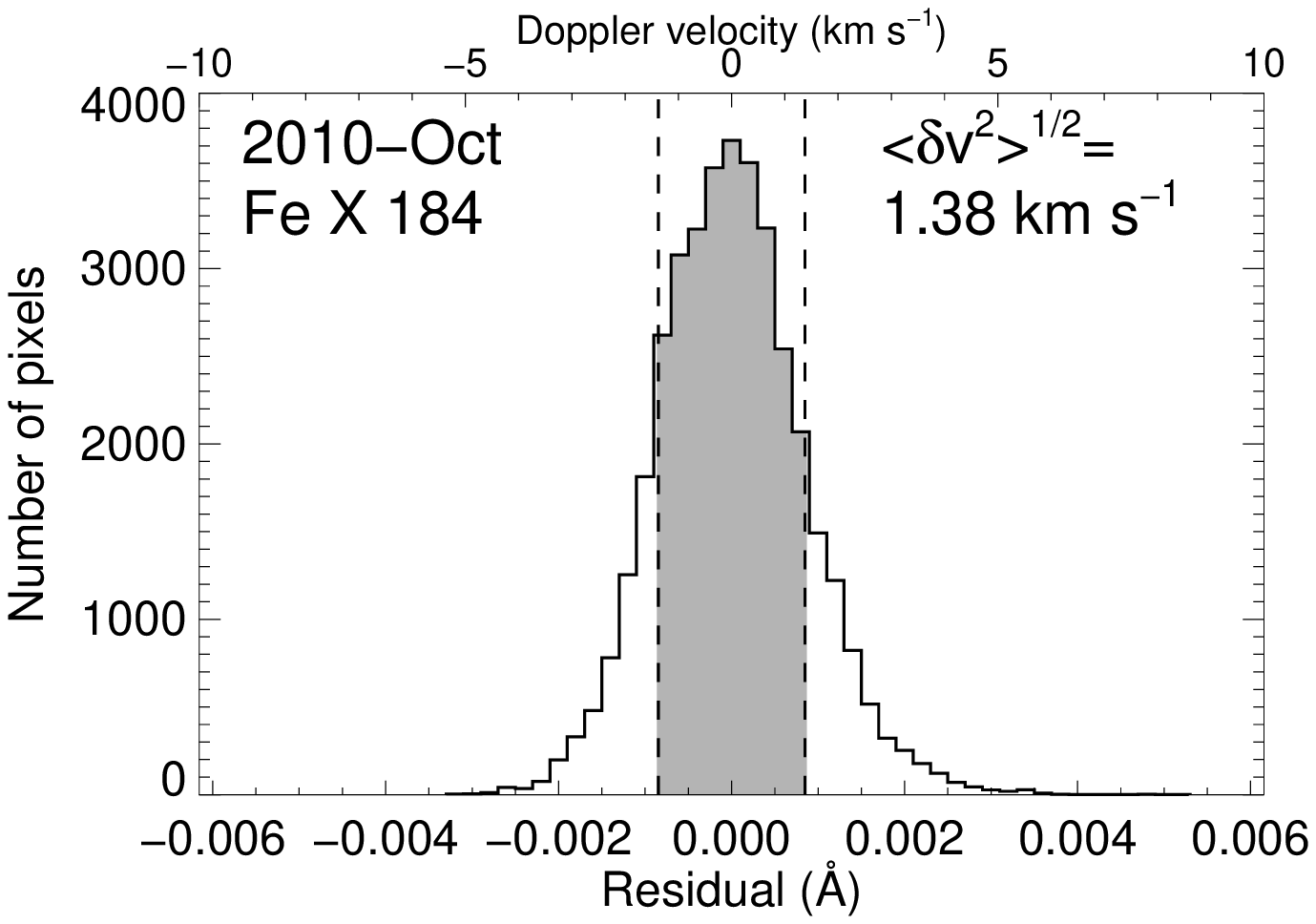}
  \includegraphics[width=6.2cm,clip]{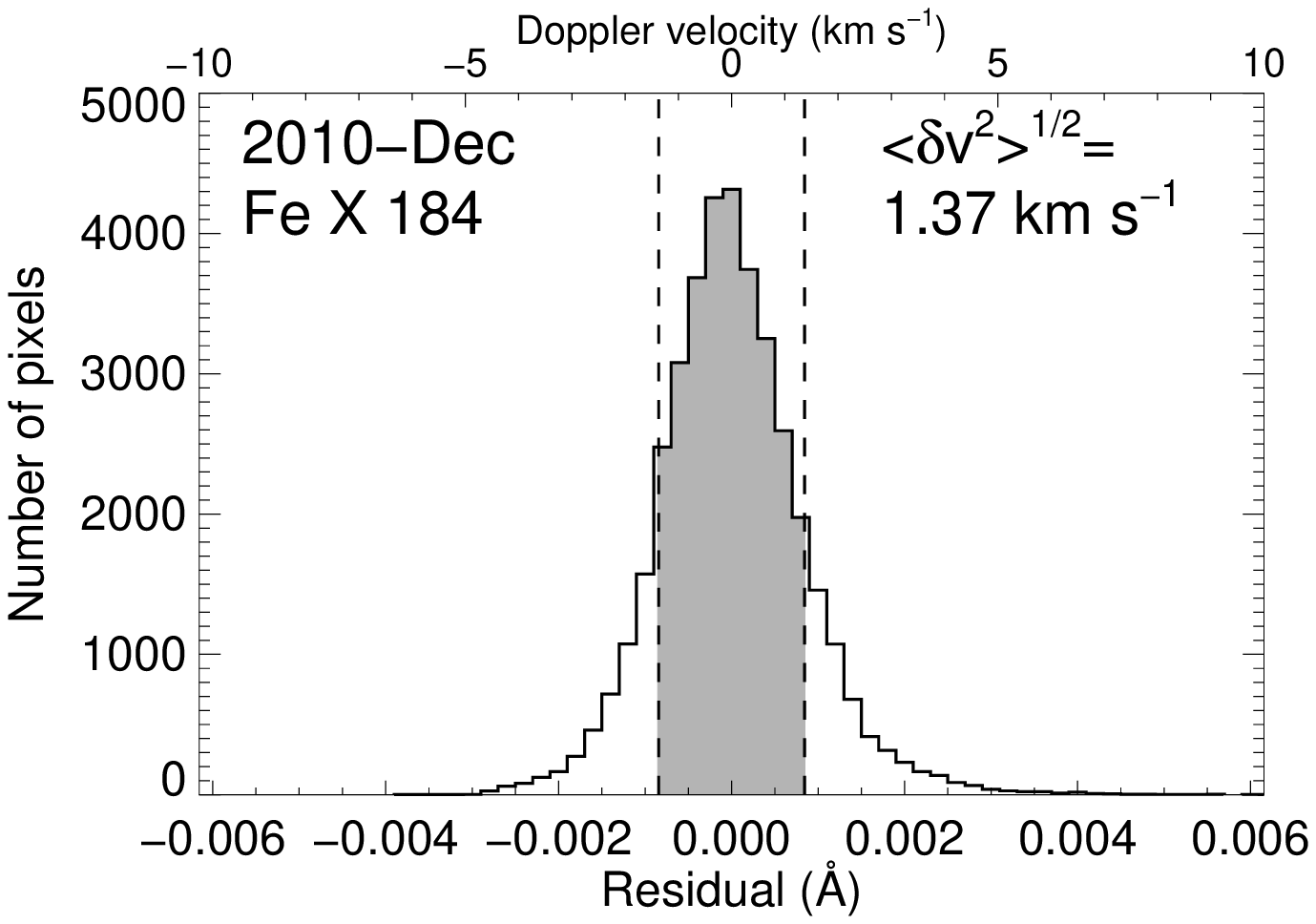} \\
  \includegraphics[width=6.2cm,clip]{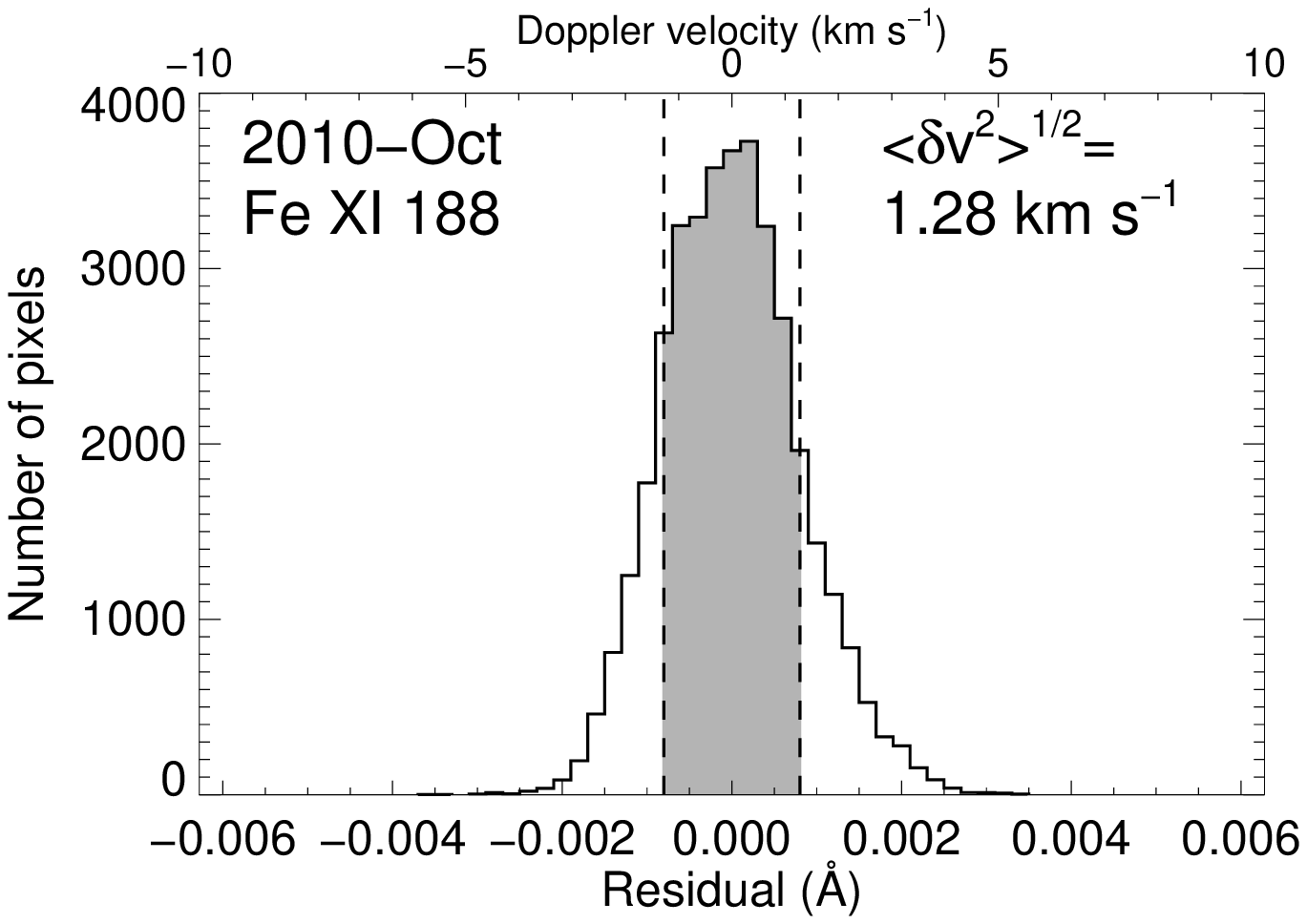}
  \includegraphics[width=6.2cm,clip]{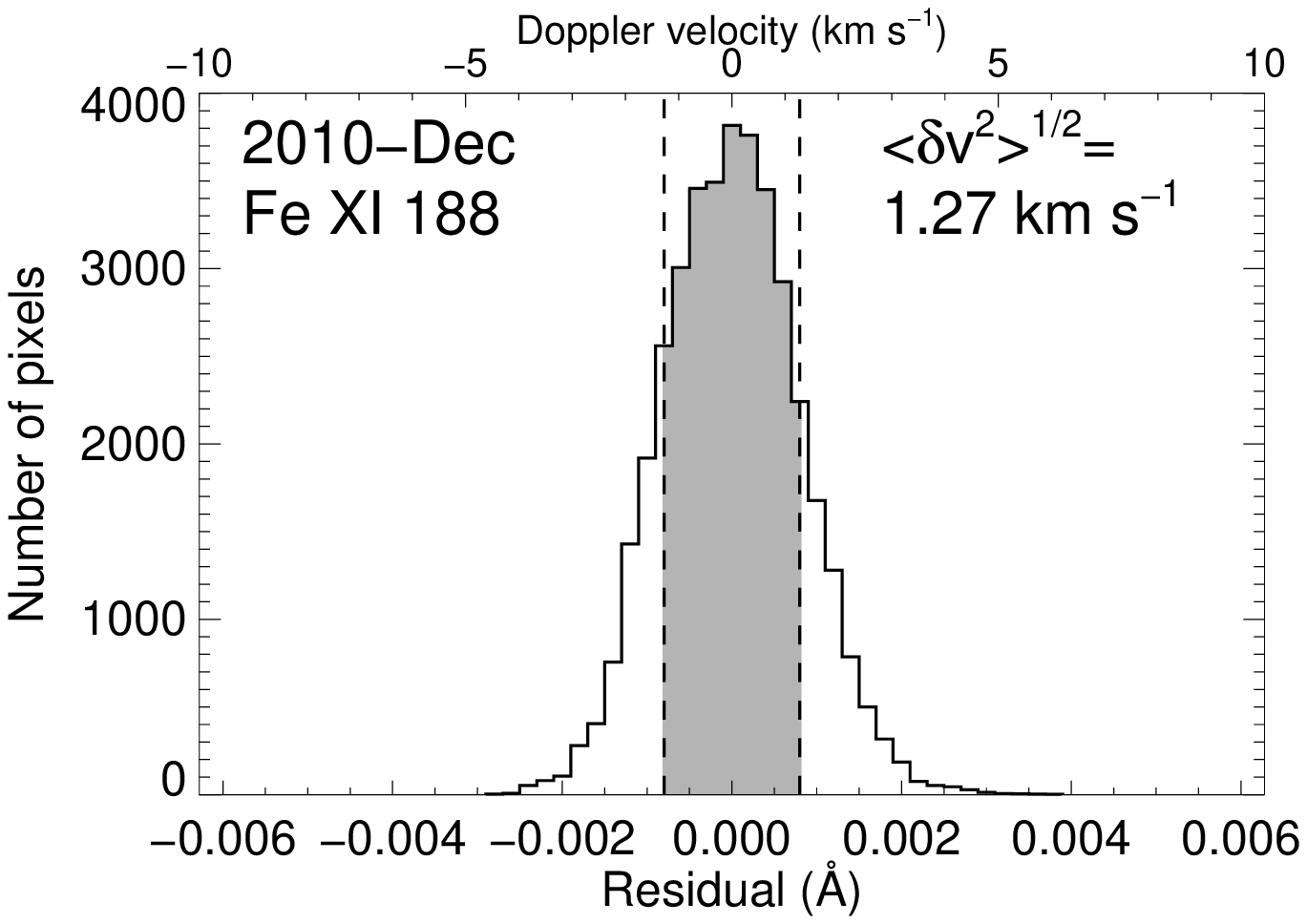} \\
  \includegraphics[width=6.2cm,clip]{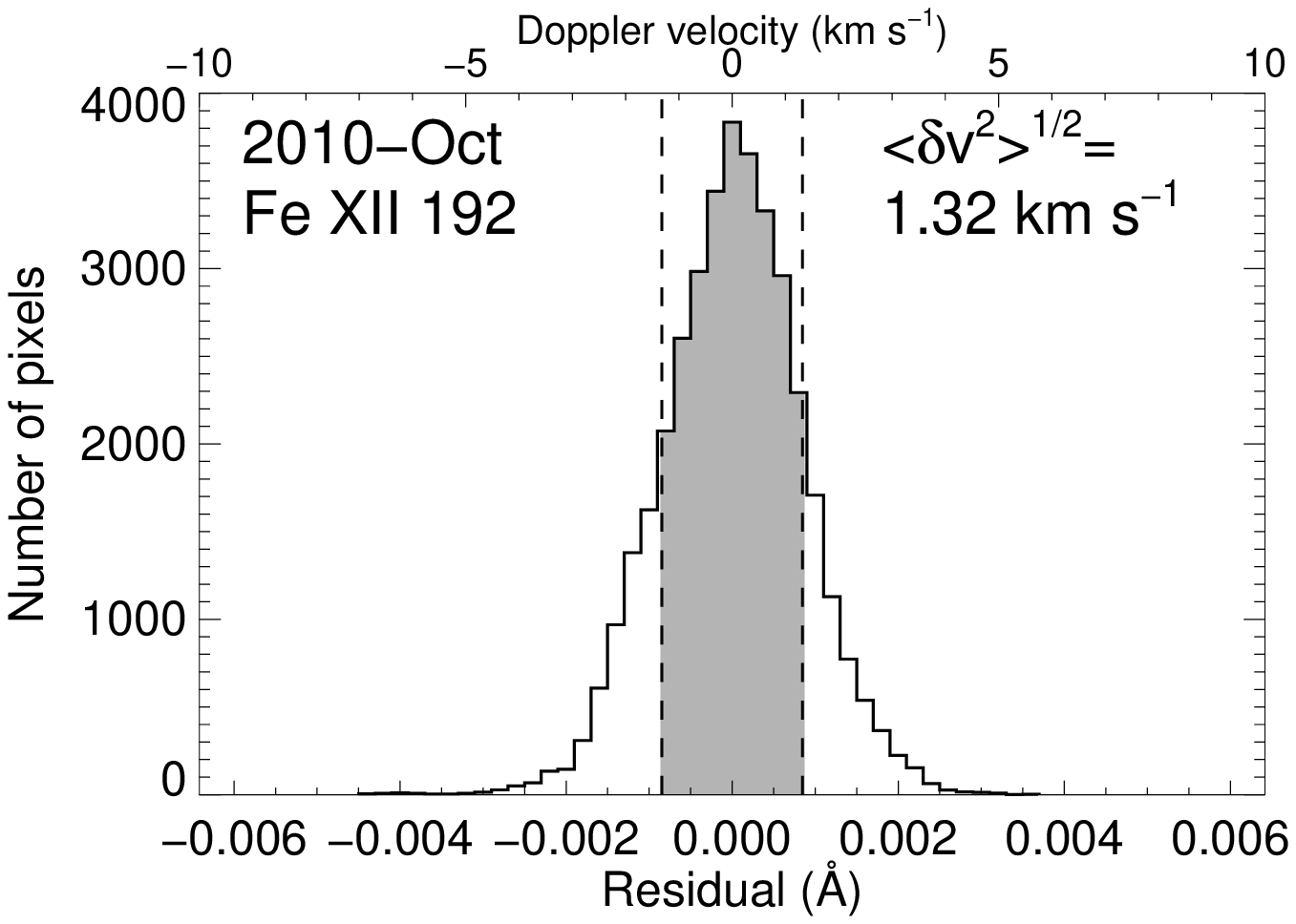}
  \includegraphics[width=6.2cm,clip]{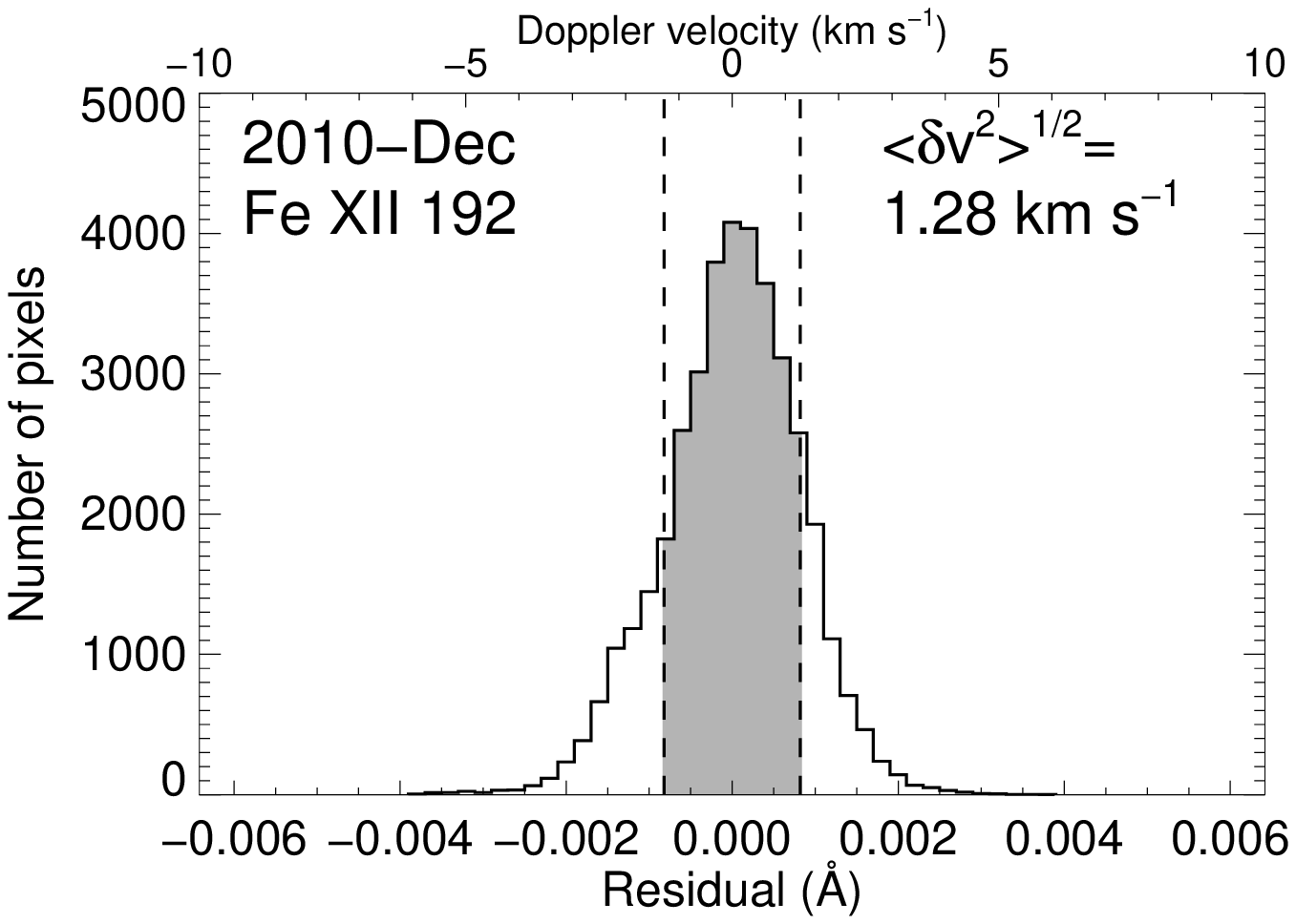} \\
  \includegraphics[width=6.2cm,clip]{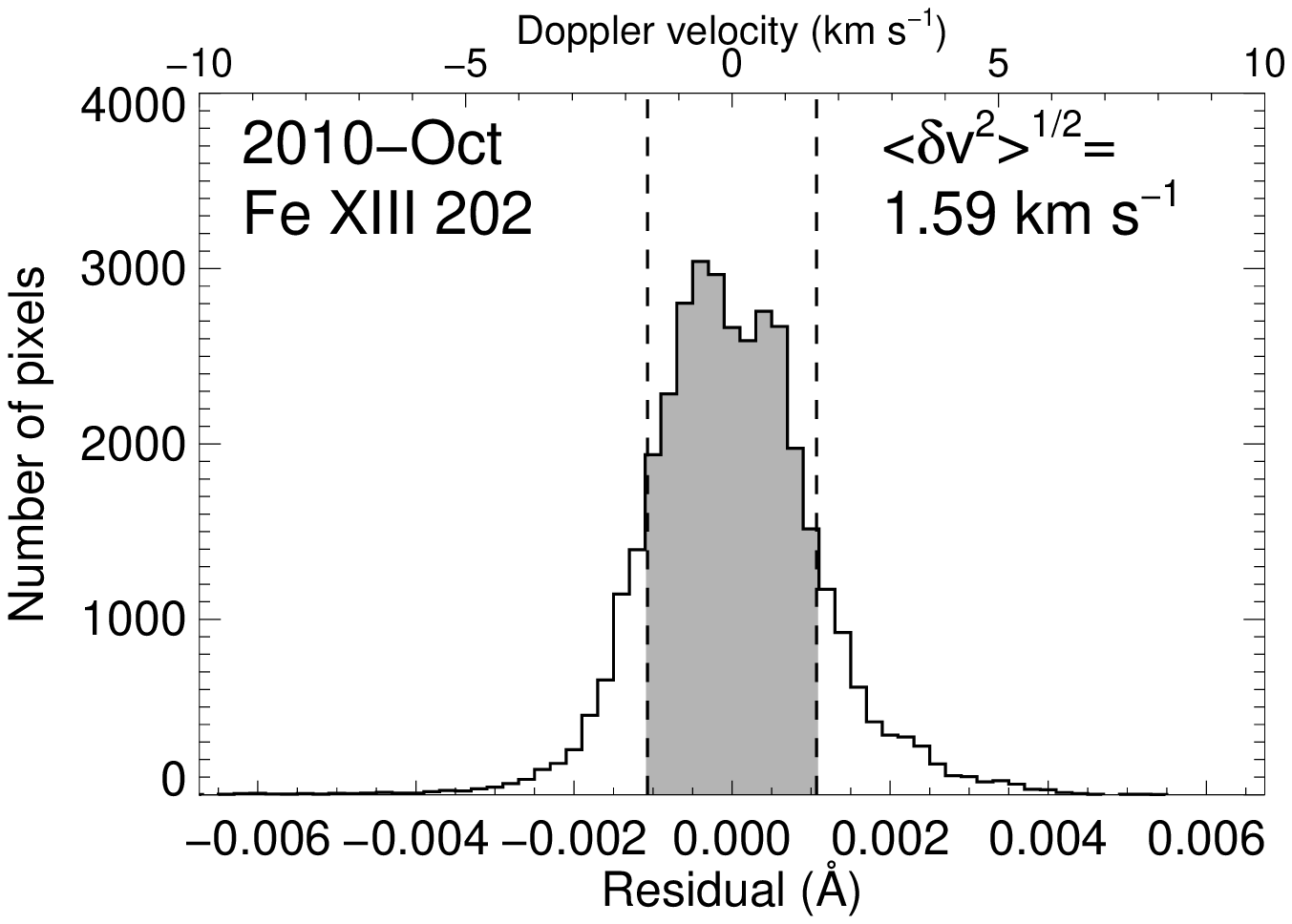}
  \includegraphics[width=6.2cm,clip]{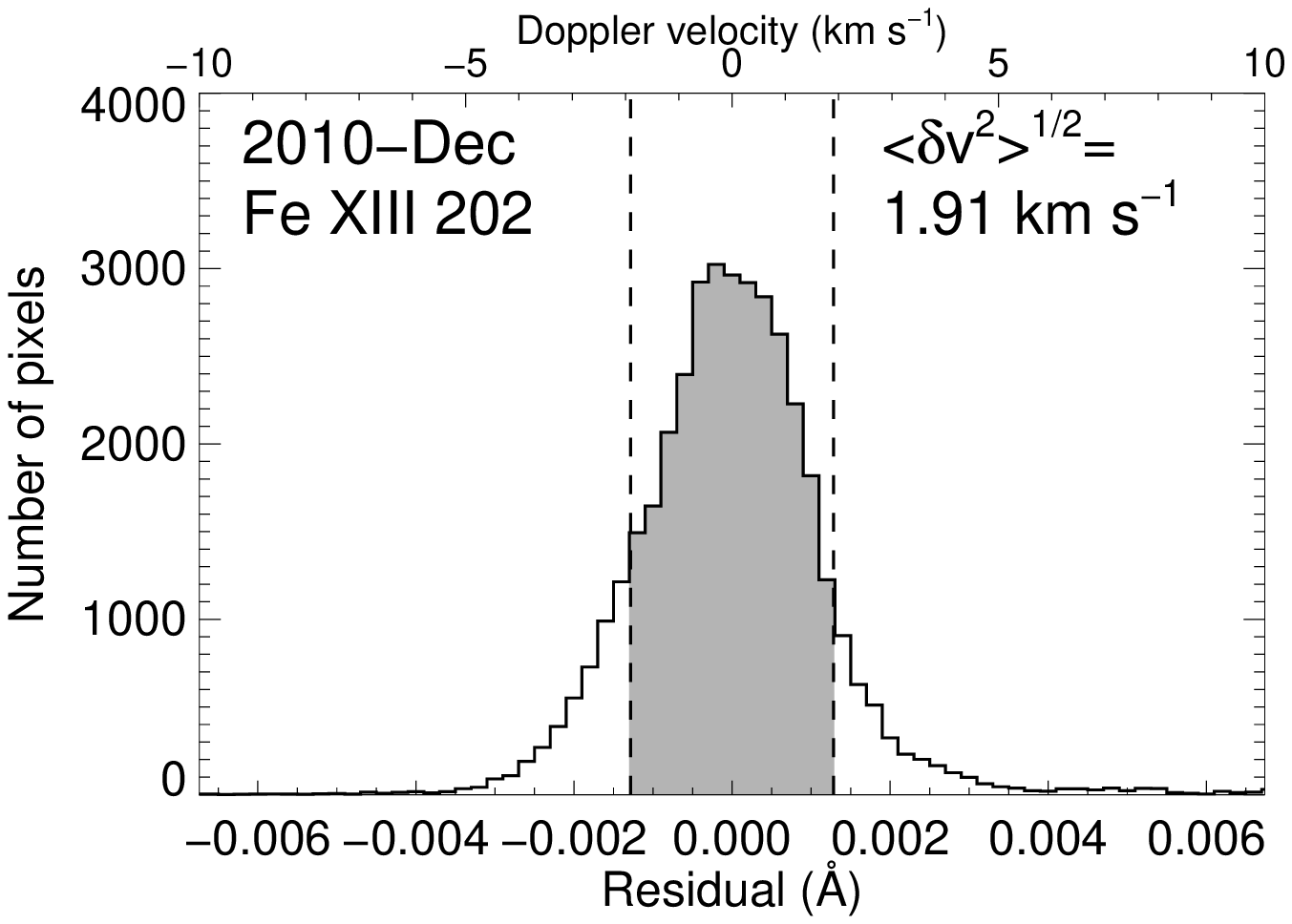} \\
  \caption{Histograms of residual from the fitted curves in Fig.~\ref{fig:limb2limb_var_Oct} (\textit{left column}) and Fig.~\ref{fig:limb2limb_var_Dec} (\textit{right column}).  Residuals of Fe \textsc{viii}, and \textsc{x}--\textsc{xiii} are shown in each row.  A number $\langle \delta v^2 \rangle^{1/2}$ in each panel indicates the standard deviation of residual. Two \textit{vertical dashed lines} indicate $\pm \langle \delta v^2\rangle^{1/2}$ and the area between those is painted by \textit{gray}.  Note that data points were extracted from only the fitted region. }
  \label{fig:cal_limb2limb_res}
\end{figure}

Fig.~\ref{fig:cal_limb2limb_res} shows histograms of the residual from the fitted curves in Fig.~\ref{fig:limb2limb_var_Oct} (October; \textit{left column}) and Fig.~\ref{fig:limb2limb_var_Dec} (December; \textit{right column}).  Residuals of Fe \textsc{viii}--\textsc{xiii} are shown from \textit{upper} to \textit{bottom} panels.  A number $\langle \delta v^2 \rangle^{1/2}$ in each panel indicates the standard deviation of the residual in the unit of $\kmpers$.  Two \textit{vertical dashed lines} indicate $\pm \langle \delta v^2\rangle^{1/2}$ and the area between those is painted by \textit{gray}.  The values of $\langle \delta v^2 \rangle^{1/2}$ are around $1\text{--}3 \, \kmpers$, which is much larger than the estimated error of the fitted velocity ($\sigma_{\mathrm{fit}}$) indicated in Fig.~\ref{fig:limb2limb_var_Oct} and Fig.~\ref{fig:limb2limb_var_Dec}.  Considering that $\langle \delta v^2 \rangle^{1/2}$ includes the real fluctuation of the quiet region, we regard $\sigma_{\mathrm{tot}} = \left( \sigma_{\mathrm{fit}}^{2} + \langle \delta v^2 \rangle \right)^{1/2}$ as an error of Doppler velocities at the disk center.  

\input{tex/tab_qr_dop_cmpl.tex}

\begin{figure}
  \centering
  \includegraphics[width=17.0cm,clip]{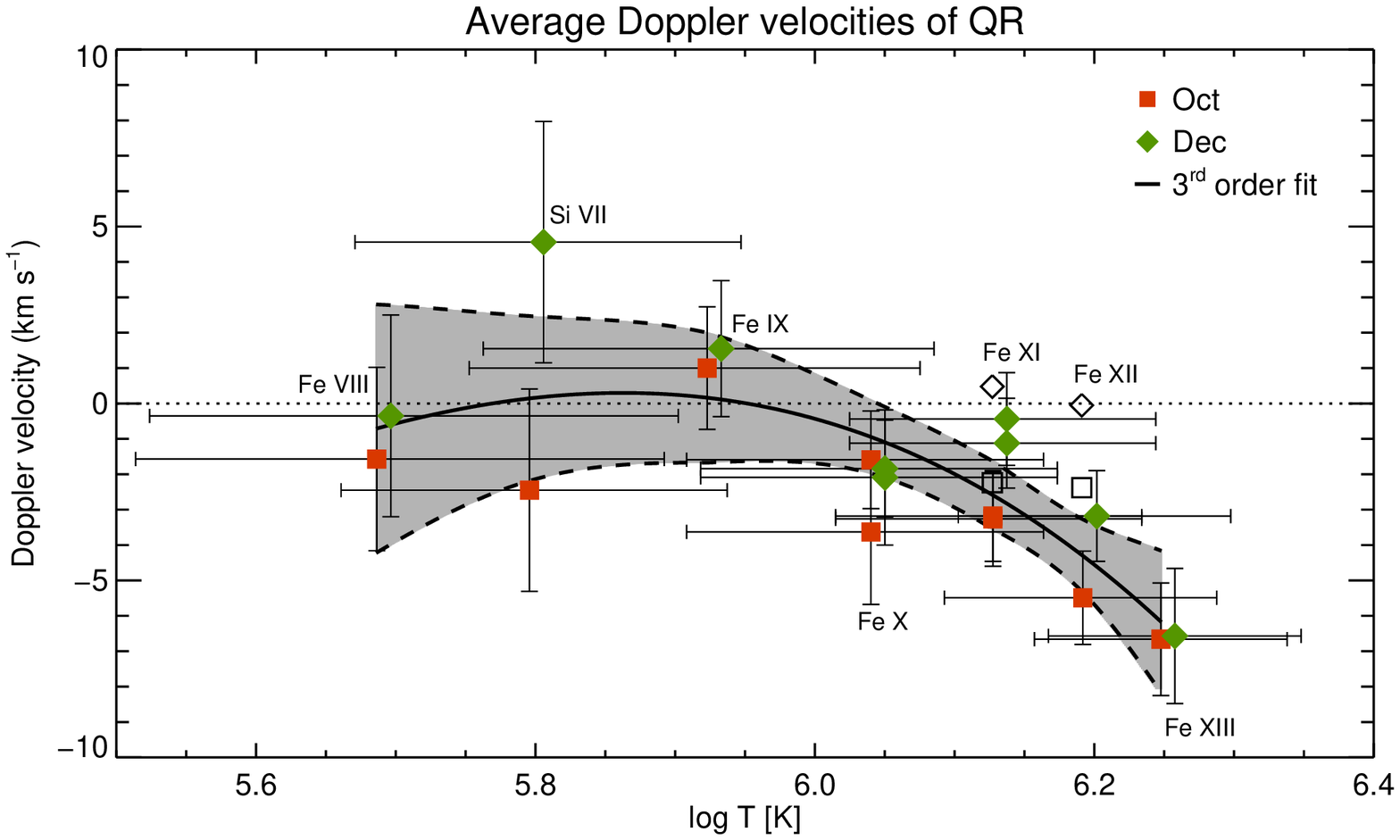}
  \caption{Doppler velocities at the disk center as a function of the formation temperature.  \textit{Red squares (Green diamonds)} indicate October (December) data.  \textit{Black symbols} show the results for potentially-blended emission lines Fe \textsc{xi} $180.40${\AA} and Fe \textsc{xii} $195.12${\AA}.  The \textit{Solid line} indicates a third order polynomial function fitted to all data points except for Fe \textsc{xi} $180.40${\AA} and Fe \textsc{xii} $195.12${\AA}.  The \textit{gray} region between two \textit{dashed lines} shows the standard deviation in the fitted curve.  }
  \label{fig:qr_dop_cmpl}
\end{figure}

Obtained Doppler velocities from eleven emission lines for the temperature range $\logt = 5.7$--$6.3$ are listed in Table \ref{tab:qr_dop} and these results were plotted in Fig.~\ref{fig:qr_dop_cmpl}.  The October and December results are respectively indicated by \textit{red squares} and \textit{green diamonds}.  December results are shifted by $+0.01$ in abscissa ($\logt$) to facilitate visualization.  The \textit{vertical} error bars indicate $\sigma_{\mathrm{tot}}$.  The \textit{horizontal} error bars indicate the full width of half maximum of contribution function.  The Doppler velocities of two potentially blended emission lines Fe \textsc{xi} $180.40${\AA} and Fe \textsc{xii} $195.12${\AA} are indicated by \textit{black symbols}.  As being considered in Section \ref{sect:cal_lp_co}, those emission lines are redshifted by several $\kmpers$ compared to the isolated emission line from the same ion.  The \textit{Solid line} indicates a third order polynomial function fitted to the all data points except for Fe \textsc{xi} $180.40${\AA} and Fe \textsc{xii} $195.12${\AA}.  The \textit{gray} region between two \textit{dashed lines} shows the standard deviation in the fitted curves.  The important conclusion here is that the Doppler velocities are almost zero or slightly positive (\textit{i.e.,} downward) at the temperature below $\logt = 6.0$, and above that temperature the emission lines are blueshifted with increasing temperature, and the Doppler velocity reaches ${-7} \, \text{--} \, {-6} \, \kmpers$ at $\logt = 6.25$ (Fe \textsc{xiii}).

% --- End of TeX ---

%% file: tex/tab_qr_dop_cmpl.tex
\begin{table}
  %\captionsetup{width=13cm}
  \centering
  \caption{Doppler velocities at the disk center obtained from north--south scan with \textit{Hinode}/EIS during 2010.  A symbol ${}^{\mathrm{b}}$ after wavelength (denoted by Wvl.\ in the \textit{second} column) means that the emission line is potentially blended by another emission line.  The unit of velocity is $\mathrm{km} \, \mathrm{s}^{-1}$.  $\sigma_{\mathrm{fit}}$ is the error in the fitted velocity and $\langle \delta v^2 \rangle^{1/2}$ is the standard deviation of the residual from the fitted curve.  The second column from right shows weighted average which were calculated by using the sum of those two errors squared ($\sigma_{\mathrm{tot}}$ in the text).  The most \textit{right} column shows the error in the weighted average.}
  \begin{tabular}{llrrrrrrrrr} %11 rows
    \toprule
    & 
    & 
    & \multicolumn{6}{c}{\small Doppler velocity at the disk center 
      ($\mathrm{km} \, \mathrm{s}^{-1}$)} 
    & 
    & \\
    \cmidrule(lr){4-9}
    &  
    & 
    %\rule[-1pt]{0pt}{12pt}
    & \multicolumn{3}{c}{\small October} 
    & \multicolumn{3}{c}{\small December} 
    & 
    & \\
    \cmidrule(lr){4-6}
    \cmidrule(lr){7-9}
    \multicolumn{1}{l}{\footnotesize Ion}
    & \multicolumn{1}{l}{\footnotesize Wvl.~({\AA})} 
    & \multicolumn{1}{c}{\footnotesize $\log T \, [\mathrm{K}]$} 
    %\rule[-1pt]{0pt}{12pt}
    & \multicolumn{1}{c}{\footnotesize $v_0$}
    & \multicolumn{1}{c}{\footnotesize $\sigma_{\mathrm{fit}}$}
    & \multicolumn{1}{c}{\footnotesize $\langle \delta v^2 \rangle^{1/2}$}
    & \multicolumn{1}{c}{\footnotesize $v_0$}
    & \multicolumn{1}{c}{\footnotesize $\sigma_{\mathrm{fit}}$}
    & \multicolumn{1}{c}{\footnotesize $\langle \delta v^2 \rangle^{1/2}$}
    & \multicolumn{1}{c}{\footnotesize Average}
    & \multicolumn{1}{c}{\footnotesize Error} \\
    \midrule
    Fe \textsc{viii} & $186.60$ & $5.69$ & $-1.57$ & $0.09$ & $2.59$ & $-0.35$ & $0.13$ & $2.85$ & $-1.02$ & $1.92$ \\
    Si \textsc{vii}  & $275.35$ & $5.80$ & $-2.45$ & $0.10$ & $2.86$ & $ 4.56$ & $0.16$ & $3.41$ & $ 0.44$ & $2.19$ \\
    Fe \textsc{ix}   & $188.49$ & $5.92$ & $ 1.00$ & $0.06$ & $1.73$ & $ 1.55$ & $0.09$ & $1.92$ & $ 1.25$ & $1.29$ \\
    Fe \textsc{x}    & $184.54$ & $6.04$ & $-1.59$ & $0.05$ & $1.38$ & $-1.84$ & $0.06$ & $1.37$ & $-1.72$ & $0.97$ \\
                     & $257.26$ &        & $-3.63$ & $0.08$ & $2.05$ & $-2.09$ & $0.09$ & $1.91$ & $-2.81$ & $1.40$ \\
    Fe \textsc{xi}   
    & $180.40^{\mathrm{b}}$ & $6.12$ & $-2.24$ & $0.07$ & $2.07$ & $ 0.48$ & $0.09$ & $2.00$ & $-0.83$ & $1.44$ \\
                     & $188.21$ &        & $-3.18$ & $0.04$ & $1.28$ & $-1.12$ & $0.06$ & $1.27$ & $-2.14$ & $0.90$ \\
                     & $188.30$ &        & $-3.26$ & $0.06$ & $1.34$ & $-0.44$ & $0.06$ & $1.31$ & $-1.82$ & $0.94$ \\
    Fe \textsc{xii}  & $192.39$ & $6.19$ & $-5.49$ & $0.05$ & $1.32$ & $-3.18$ & $0.06$ & $1.28$ & $-4.30$ & $0.92$ \\
    & $195.12^{\mathrm{b}}$ &        & $-2.38$ & $0.05$ & $1.29$ & $-0.05$ & $0.07$ & $1.44$ & $-1.34$ & $0.96$ \\
    Fe \textsc{xiii} & $202.04$ & $6.25$ & $-6.66$ & $0.06$ & $1.59$ & $-6.57$ & $0.09$ & $1.91$ & $-6.62$ & $1.22$ \\
    \bottomrule
  \end{tabular}
  \label{tab:qr_dop}
\end{table}

%% file: tex/cal_sum.tex
% ==========================
%   Doppler shifts of QR.
%   Section:
%     Summary.
% ==========================

In order to determine the reference velocities for emission lines in the quiet region, we analyzed the data taken during HOP79.  The consecutive scans on the meridional line enable us to investigate the center-to-limb variations of spectra.  We derived the center-to-limb variations of the Doppler velocities for the emission lines whose formation temperature is above $\log T \, [\mathrm{K}] \geq 6.0$ for the first time. It is concluded that below the temperature of $\log T \, [\mathrm{K}] = 6.0$ the Doppler velocities are almost zero or slightly positive (\textit{i.e.,} downward), while the Doppler velocity clearly becomes negative up to $-6 \, \mathrm{km} \, \mathrm{s}^{-1}$ above that temperature.  Previous observations have shown that the Doppler velocity in the quiet region at $\log T \, [\mathrm{K}] = 5.8$ measured by using Ne \textsc{viii} $770.43${\AA} is $-2.6 \pm 2.2 \, \mathrm{km} \, \mathrm{s}^{-1}$ \citep{peter1999}, $-2.4 \pm 1.5 \, \mathrm{km} \, \mathrm{s}^{-1}$ \citep{peterjudge1999}, and $-1.9 \pm 2.0 \, \mathrm{km} \, \mathrm{s}^{-1}$ \citep{teriaca1999}.  Our results were in good agreement with those studies within the error.

The results obtained in this chapter will be used as a reference for the Doppler velocities of the outflow region at the edge of an active region measured in the next chapter.  Although the results themselves would have much importance on the coronal dynamics as discussed in the literature \citep{peterjudge1999}, we do not discuss our results further since that is not the main purpose of this thesis. 

% --- End of TeX ---

%% file: tex/cal_tilt_corr_itdn.tex
%%%%%%%%%%%%%%%%%%%%%%%%%%%%%%%%%%%%%%%%%%%%%%%%%%%%
%  Chapter:
%    T vs. V
%  Contents:
%    Correction of spectrum tilt.
%%%%%%%%%%%%%%%%%%%%%%%%%%%%%%%%%%%%%%%%%%%%%%%%%%%%

EIS slits and CCDs are known to be slightly tilted from the solar $Y$ direction, which creates a systematic shift in observed spectra (hereafter referred to as the spectrum tilt) in CCD $Y$ direction. Since the projection of the slit is tilted clockwise on the CCDs, the spectra is redshifted in the northern part and blueshifted in the southern part. The standard SSW package corrects this effect by referring the tilt derived by the study \texttt{CALIB\_SLOT\_SLIT} (EIS study No. 352) which was designed to calibrate the the slit tilt and the instrumental width from the off-limb spectra \citep{young2010}.

There are two remained problems in that calibration: one is that it is unclear whether the spectrum tilt has a dependence on wavelength, and the other is that the spectrum tilt in the long-wavelength (LW) CCD was not investigated. The current standard EIS software uses a single set of fitted parameters derived from Fe \textsc{xii} $193.51${\AA}/$195.12${\AA} in the short-wavelength (SW) CCD to calibrate the spectrum tilt. Note that \citet{kamio2010} investigated the tilts of both CCDs and it was concluded that the SW CCD is more tilted than the LW CCD. In our analysis, we analyze the same data as \citet{kamio2010} taken at the quiet region near the solar disk and investigate the spectrum tilt using ten emission lines from the two CCDs. The results obtained in this appendix are used to calibrate the spectrum tilt when we deduce the center-to-limb variation of the line centroids (Section \ref{sect:cal_data_reduc}). 

% --- End of Tex ---

%% file: tex/cal_tilt_corr_obs.tex
% ============================================
%   Chapter:
%     Doppler shifts in the quiet region.
%   Section:
%     Observation for the tilt calibration.
% ============================================

\begin{figure}
  \centering
  \includegraphics[width=10cm,clip]{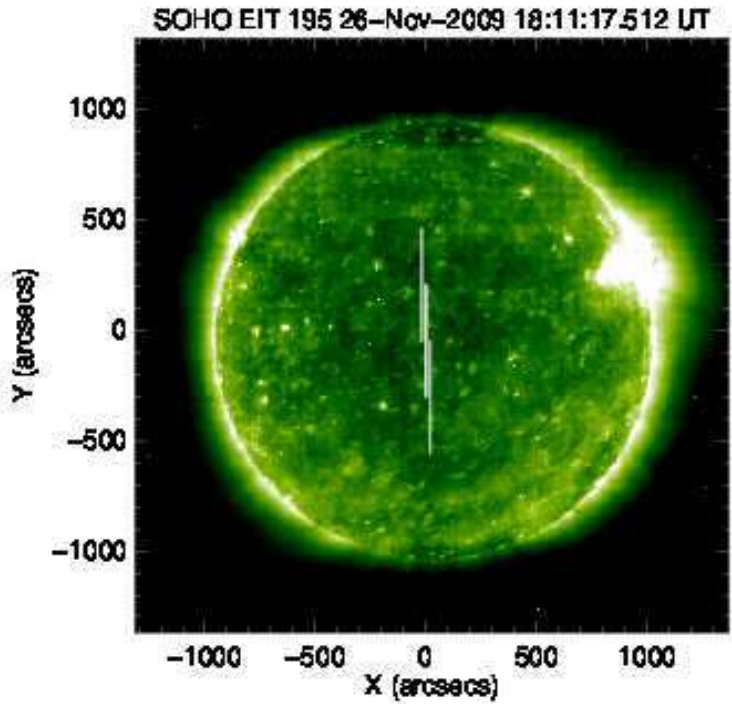}
  \caption{EIT image taken on 2009 November 26 18:11:17UT. Vertical lines indicate the location of EIS pointing during the observation. Three lines are drawn which indicate the pointings. Note that the slight shift of the three lines is imposed for visibility. Those are overlapped completely in the actual observation.}
  \label{fig:cal_tilt_eit}
\end{figure}

\begin{figure}
  \centering
  \includegraphics[width=16.8cm,clip]{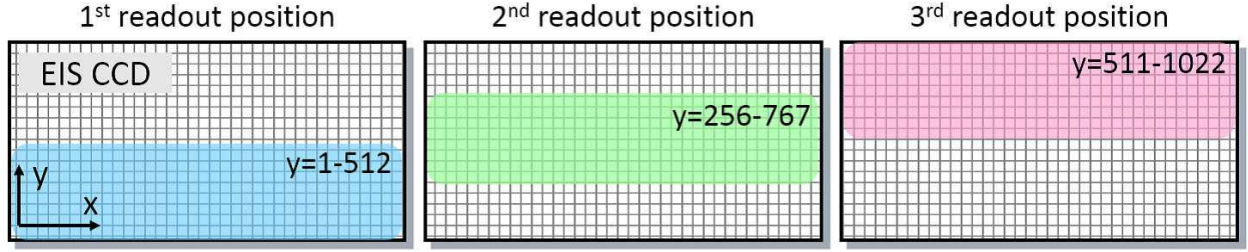}
  \caption{Schematic picture of the EIS observation used to calibrate the spectrum tilt.}
  \label{fig:tilt_study_schem}
\end{figure}

In order to calibrate the spectrum tilt, we used the data taken by the same EIS study as that analyzed to investigate the Doppler shifts of the quiet region (north--south scans in HOP79). The data was taken on 2009 November 26. While data taken during HOP79 recorded the spectra with the spatial span of $512$ pixels in $y$ direction ($y=306$--$817$ on the CCDs), this data contains spectra at all $y$ position on the CCDs. This enables us to investigate the spectrum tilt along the EIS slit more precisely. The analysis assumes that the variation of line centroids in the data comes from the instrumental effect, since the variation of the Doppler shift near the disk center is small even if the plasma has a constant radial velocity. The region studied is within approximately $500''$ from the disk center, so the curvature produced by the radial velocity does not exceed $1.5 \, \kmpers$ when the velocity is $\lesssim 10 \, \kmpers$. A context image of the Sun is shown in Fig.~\ref{fig:cal_tilt_eit} in which the location of the EIS slit is shown by white line. There were no active regions on the solar disk during the observation. 

Since EIS can obtain spectra of $512''$ height in the maximum at one exposure, the exposures are divided into three times and take the spectra on pixels $1$--$512$, $256$--$767$, and $511$--$1022$ from the bottom of the CCDs as shown in Fig.~\ref{fig:tilt_study_schem}. The spectra were integrated by $50$ pixels in order to enhance S/N ratio when they are fitted by a single Gaussian. At each $y$ position, EIS took the spectra in neighboring five location within $5''$ in the east-west direction. The fitted line centroids in these five exposures show almost the same behavior as seen in Fig.~\ref{fig:tilt_sk}. In order to compensate the influence of the orbital variation, we aligned those five arrays of the line centroid so that the sum of squared difference between each array becomes the minimum. In the next, assuming that the velocity in the quiet region is steady, we connected the three parts of the data and obtained the line centroid as a function of all $y$ position on the CCDs. 

% --- End of TeX ---

%% file: tex/cal_tilt_corr.tex
%%%%%%%%%%%%%%%%%%%%%%%%%%%%%%%%%%%%%%%%%%%%%%%%%%%%
%  Chapter:
%    T vs. V
%  Contents:
%    Correction of spectrum tilt.
%%%%%%%%%%%%%%%%%%%%%%%%%%%%%%%%%%%%%%%%%%%%%%%%%%%%

\begin{figure}
  \centering
  \includegraphics[width=8.1cm,clip]{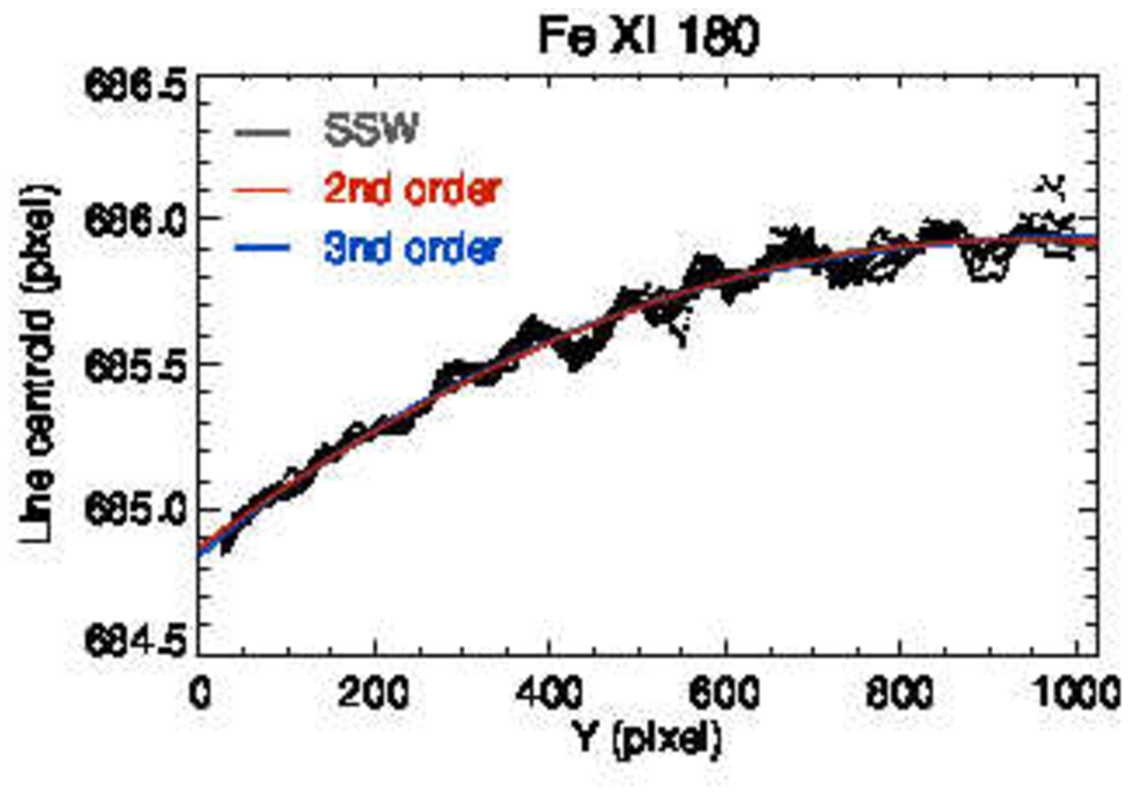}
  \includegraphics[width=8.1cm,clip]{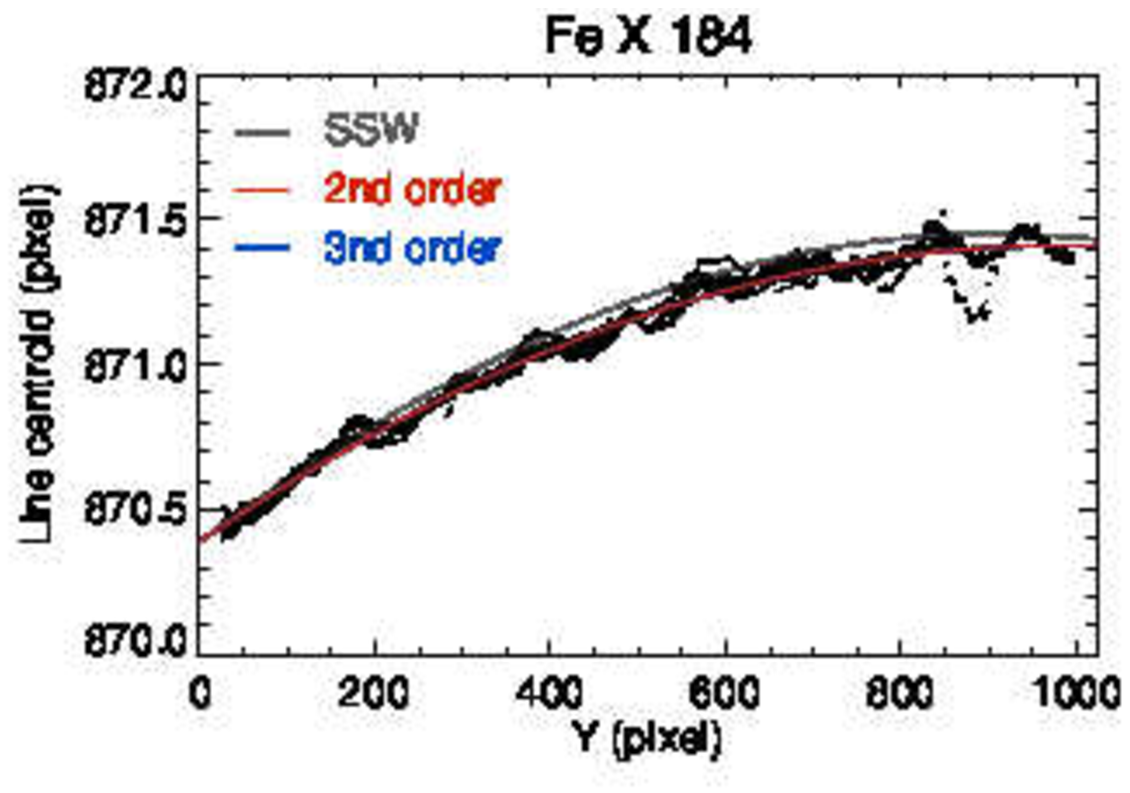}
  \includegraphics[width=8.1cm,clip]{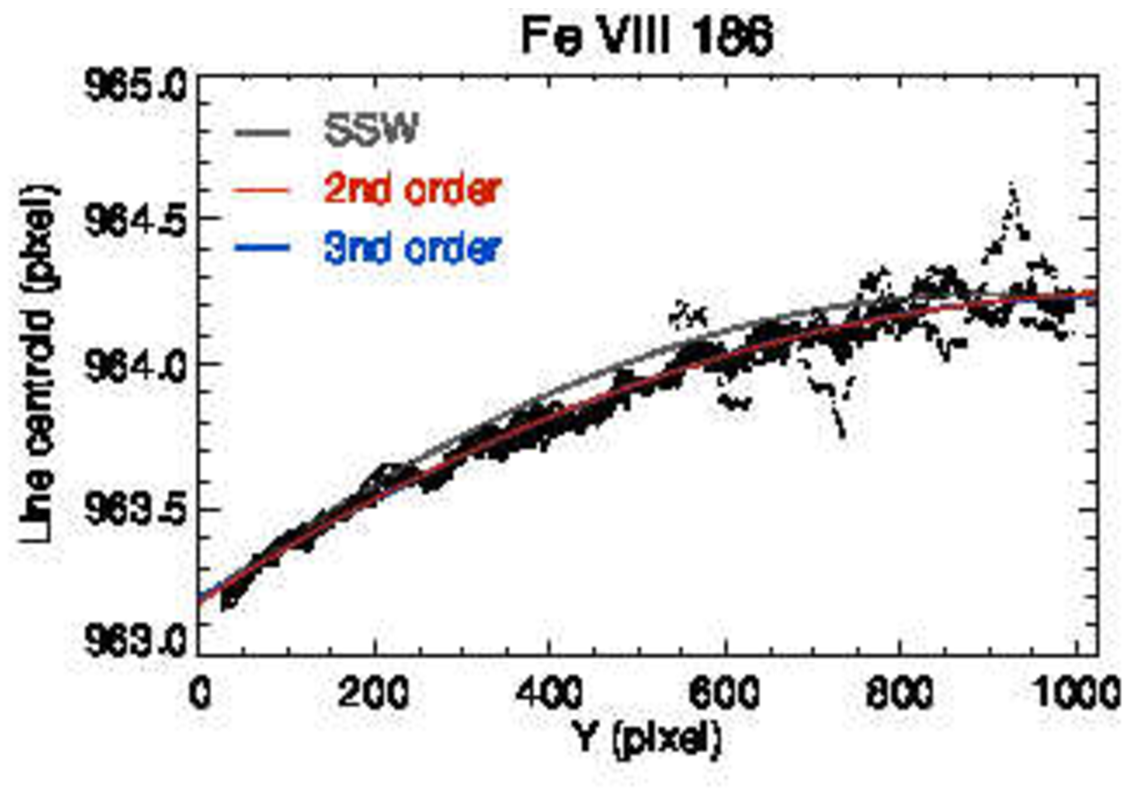}
  \includegraphics[width=8.1cm,clip]{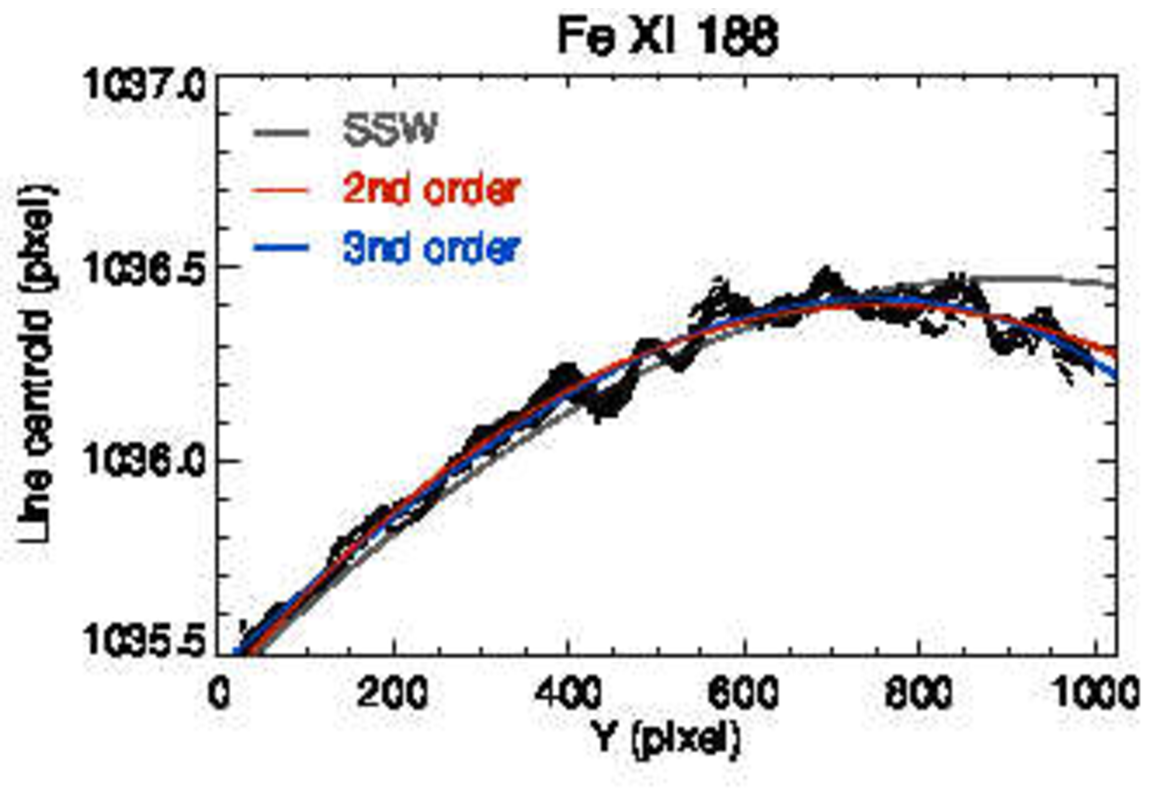}
  \includegraphics[width=8.1cm,clip]{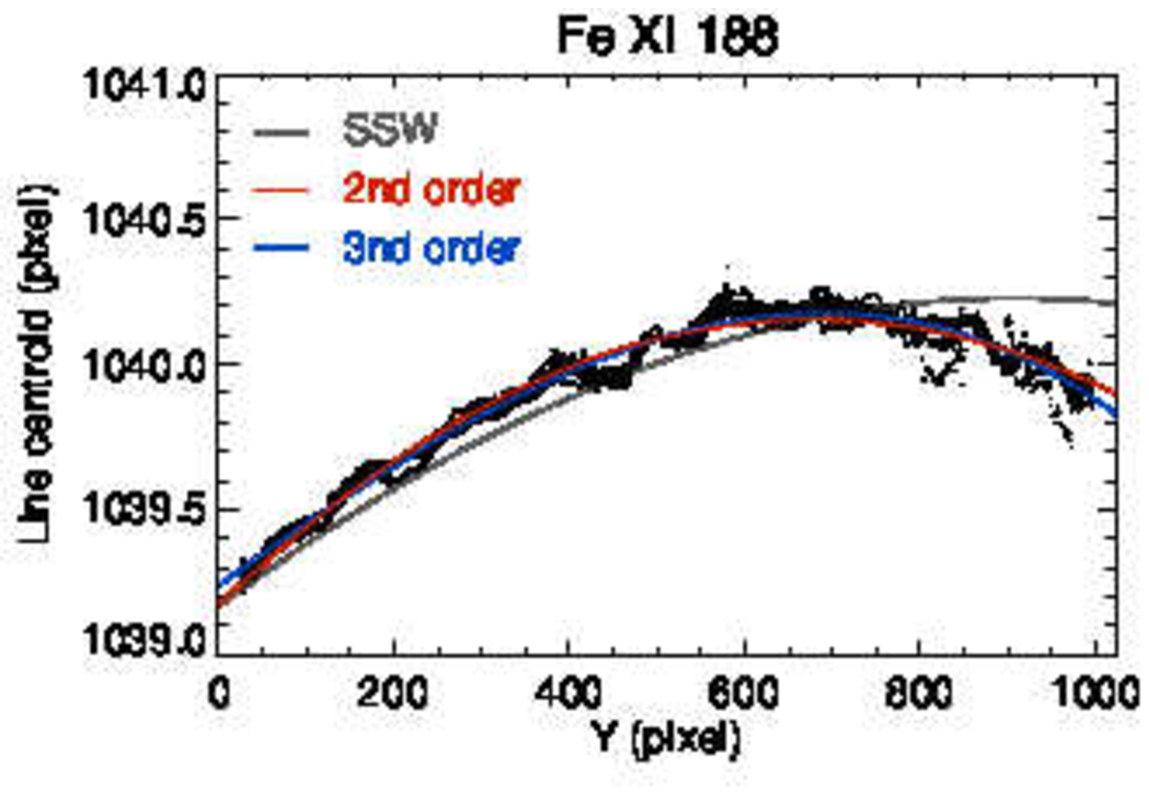}
  \includegraphics[width=8.1cm,clip]{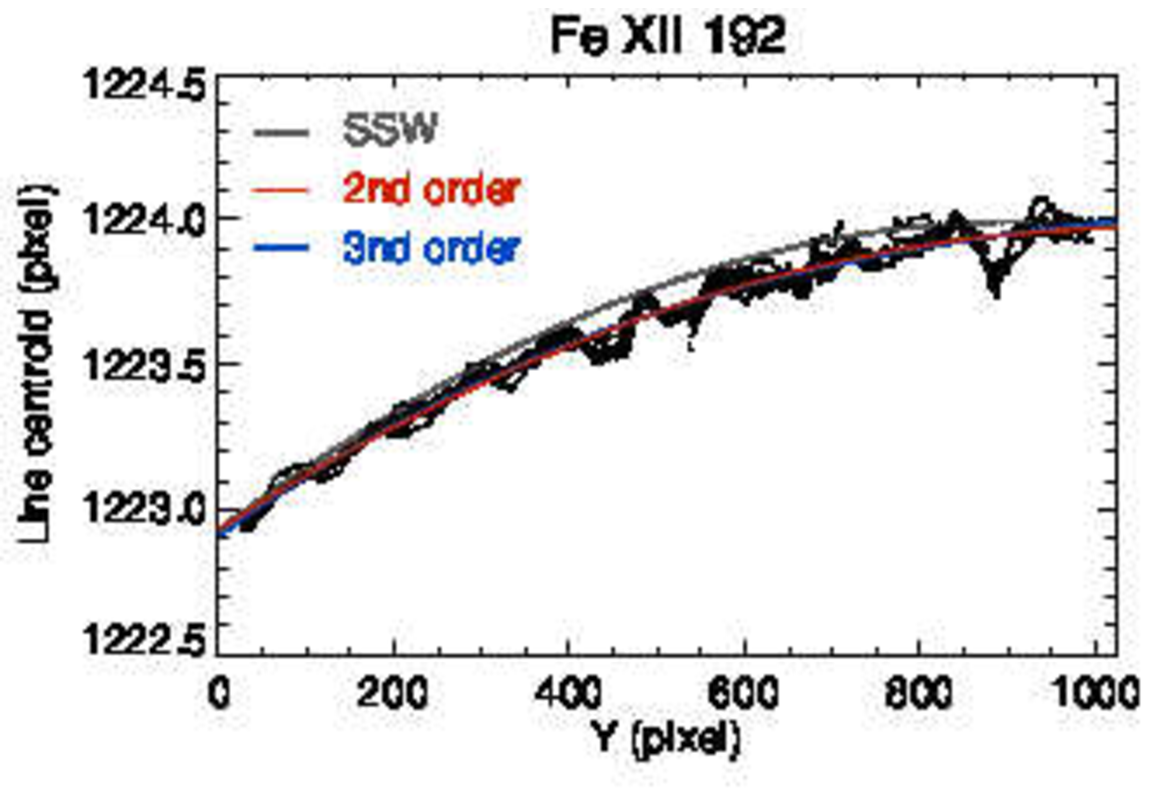}
  \includegraphics[width=8.1cm,clip]{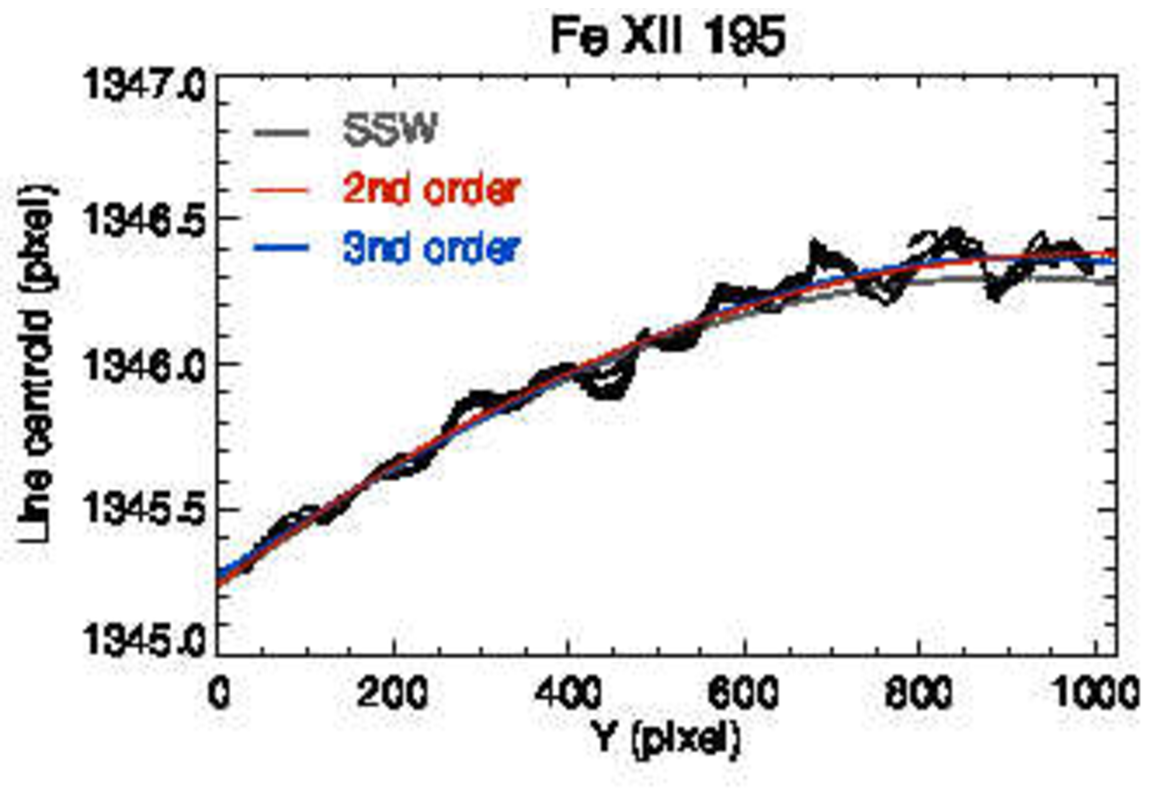}
  \includegraphics[width=8.1cm,clip]{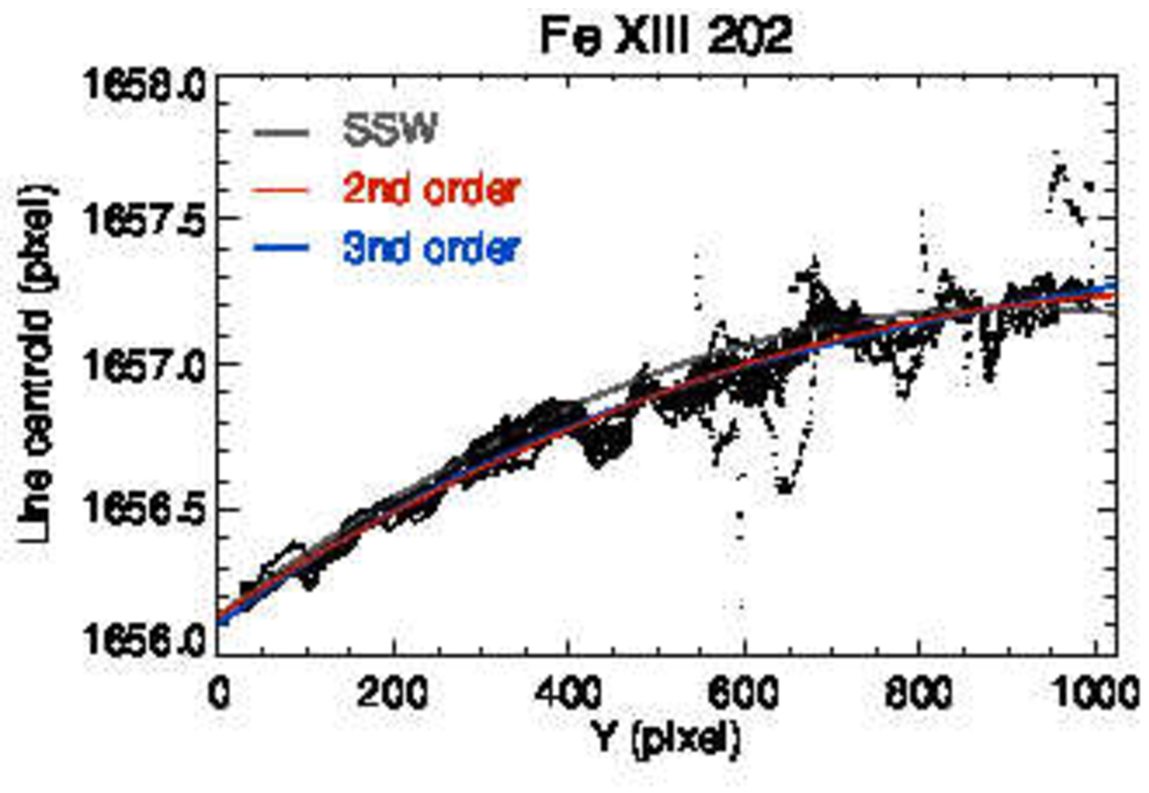}
  \caption{Line centroid as a function of $y$ position on the SW CCD. In each panel, data points are indicated by \textit{black dots.} \textit{Gray, blue,} and \textit{red lines} respectively show the spectrum tilt returned by the standard EIS software, second order polynomial fitting, and third order polynomial fitting.}
  \label{fig:tilt_sk}
\end{figure}

Line centroids of the emission lines in the SW CCD are shown in Fig.~\ref{fig:tilt_sk}. Panels show the line centroid as a function of $y$ position on the CCD. Data points are indicated by black dots. Gray, blue, and red lines respectively show the spectrum tilt returned by the standard EIS software, second order polynomial fitting, and third order polynomial fitting. Note that the line centroids are plotted in the unit of spectral pixel which corresponds to $\sim 0.0223${\AA} $\mathrm{pix}^{-1}$. 

All emission lines analyzed here indicate the similar behavior as clearly seen: the line centroids increases from the bottom of the CCD to the top, and they have the curvature with upwardly convex shape. Among them, the line centroid of Fe \textsc{xi} $188.21${\AA}/$188.30${\AA} shows different behavior at the top of the CCD above the pixel number around $800$. The curvature is larger than that of other emission lines and only the line centroid of Fe \textsc{xi} $188.21${\AA} decreases significantly at that location. Another emission line from the same ion Fe \textsc{xi} $180.40${\AA} did not show such behavior, from which we concluded that this abnormal curvature does not arise from the solar feature but the instrumental effect. We concluded that third order polynomial fitting is better than second order polynomial fitting because the spectrum curvatures seen in Fig.~\ref{fig:tilt_sk} may be asymmetric in the $y$ direction. 

\begin{figure}
  \centering
  \includegraphics[width=8.4cm,clip]{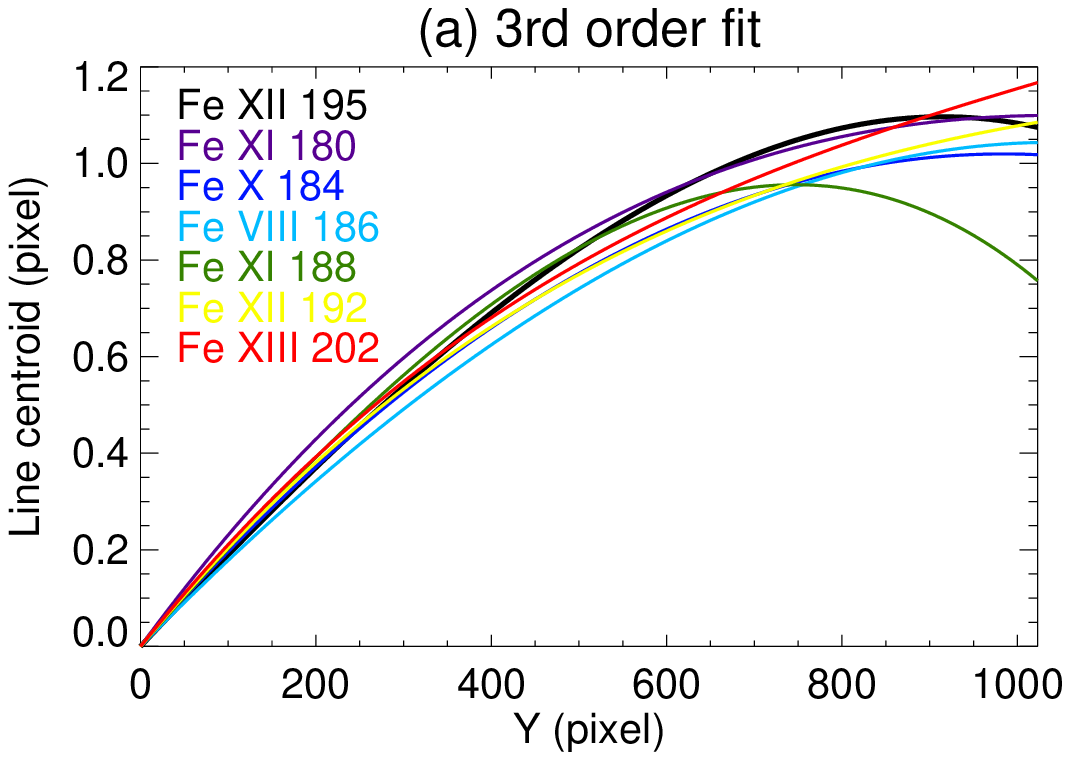}
  \includegraphics[width=8.4cm,clip]{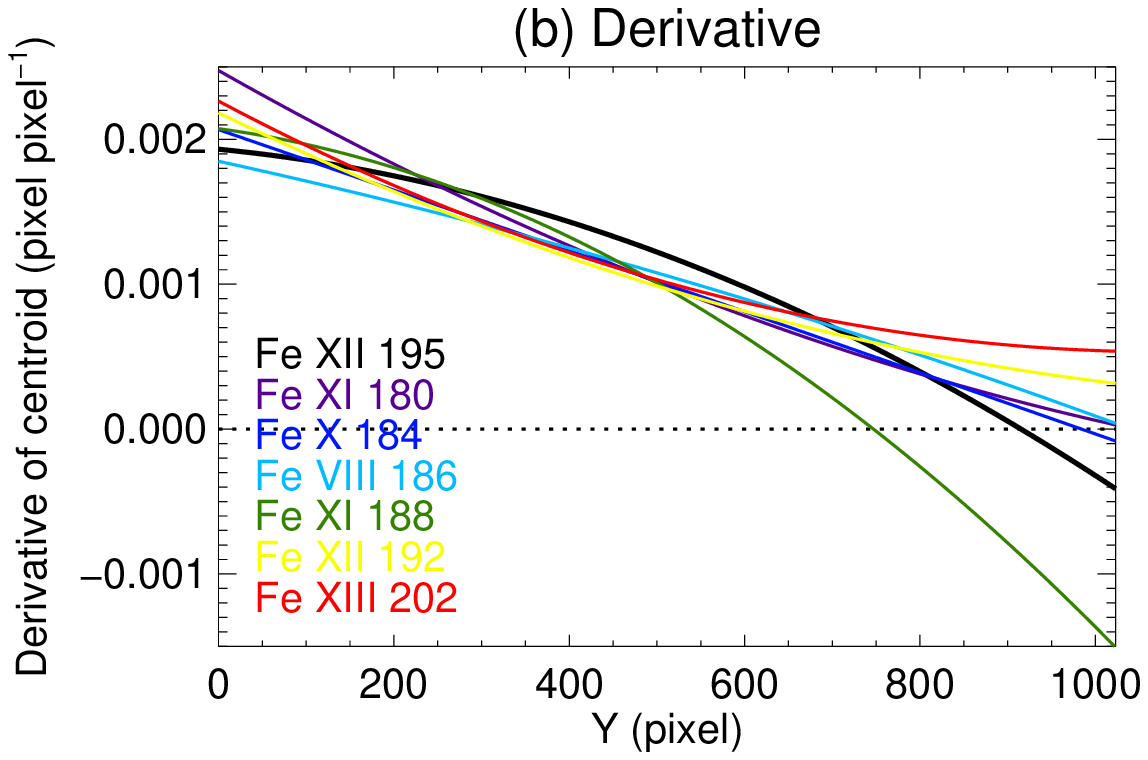}
  \caption{Derivatives of line centroid as a function of $y$ position on the SW CCD.}
  \label{fig:tilt_cmpl_3rd}
\end{figure}

The third order polynomial functions fitted to each spectrum tilt in the SW CCD are plotted together in Fig.~\ref{fig:tilt_cmpl_3rd}. Colors indicate the tilt obtained from each emission line. Panel (a) shows the fitted functions and panel (b) shows their derivative. As described above, it is easily seen that the behavior of Fe \textsc{xi} $188.21${\AA} differs from other emission lines at the top of the CCD. The graph for Fe \textsc{xii} $195.12${\AA} (indicated by black line) has steeper gradient in the middle of the CCD as seen from panel (b). The standard EIS software calibrates the spectrum tilt by using the result from Fe \textsc{xii} $193.51${\AA}/$195.12${\AA}. It means that the spectrum tilt is usually corrected too much for emission lines other than Fe \textsc{xii} $195.12${\AA} (though the deviation could be reduced by using not only $195.12${\AA} but also $193.51${\AA}). From these results, we concluded that it is better to use each fitted spectrum tilt to calibrate each emission line. We used the fitted parameters obtained by third polynomial fitting for each emission line to calibrate the data in HOP79.

% --- End of Tex ---

%% file: tex/cal_tilt_diff.tex
%%%%%%%%%%%%%%%%%%%%%%%%%%%%%%%%%%%%%%%%%%%%%%%%%%%%
%  Chapter:
%    T vs. V
%  Contents:
%    Difference of the spectrum tilt in two CCDs.
%%%%%%%%%%%%%%%%%%%%%%%%%%%%%%%%%%%%%%%%%%%%%%%%%%%%

\begin{figure}
  \centering
  \includegraphics[width=8.1cm,clip]{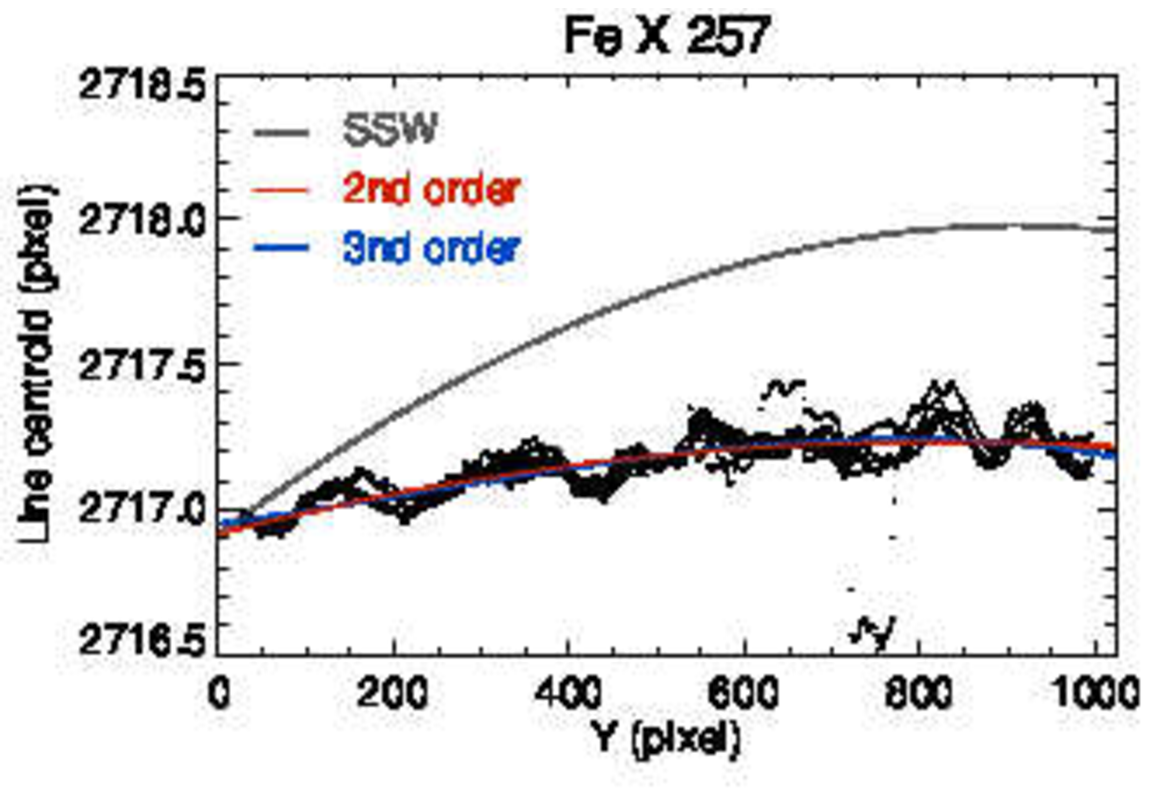}
  \includegraphics[width=8.1cm,clip]{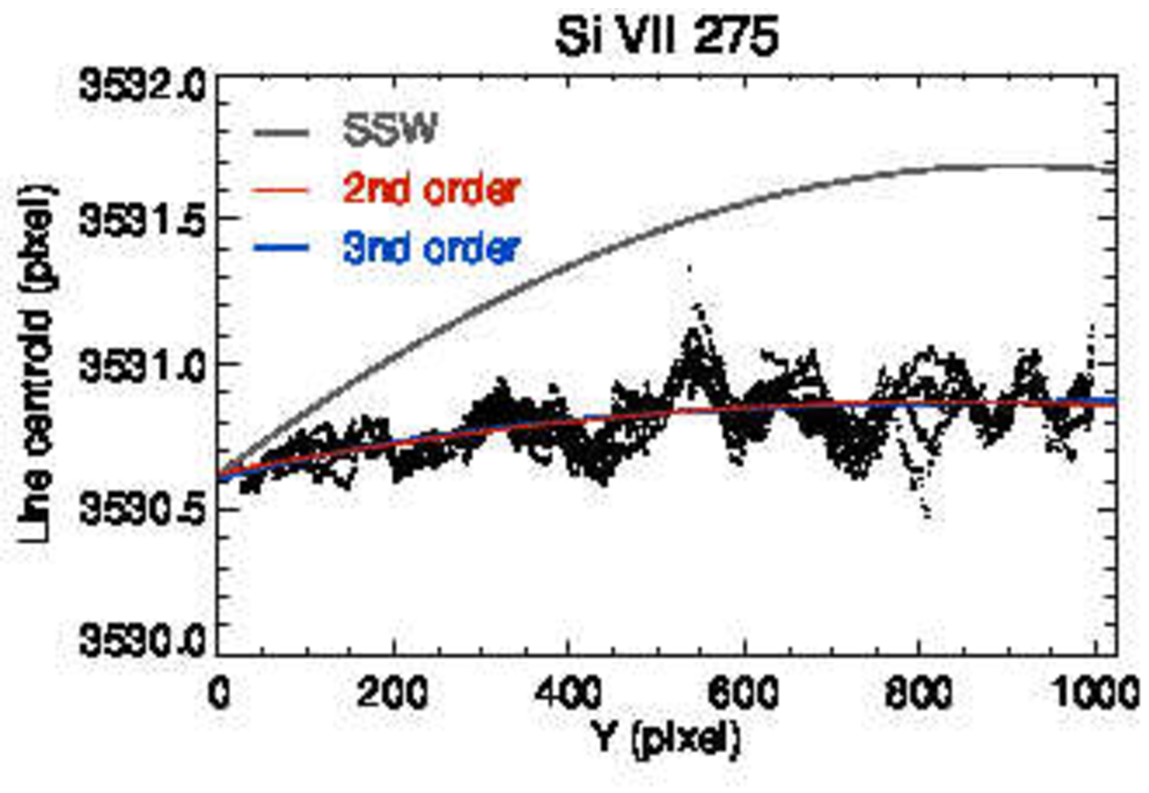}
  \caption{Line centroid as a function of $y$ position on the LW CCD.}
  \label{fig:tilt_lw}
\end{figure}

The spectrum tilt of the LW CCD is known to have a different property from that of the SW CCD \citep{kamio2010}, though it is not reflected in the standard EIS software. We investigated two line centroids of Fe \textsc{x} $257.26${\AA} and Si \textsc{vii} $275.35${\AA} which are shown in Fig.~\ref{fig:tilt_lw}. Both panels show the line centroid as a function of $y$ position on the CCD. Data points are indicated by black dots. Gray, blue, and red lines respectively show the spectrum tilt returned by the standard EIS software, second order polynomial fitting, and third order polynomial fitting. Comparing the data to gray lines, it is clear that the spectrum tilt in the LW CCD is much smaller than that in the SW CCD. The results shown here are consistent with those by \citet{kamio2010}. As same as for results in the SW CCD, we used the fitted parameters for each emission line to calibrate the data in HOP79.

% --- End of Tex ---

%% file: tex/cal_tilt_sum.tex
%%%%%%%%%%%%%%%%%%%%%%%%%%%%%%%%%%%%%%%%%%%%%%%%%%%%
%  Chapter:
%    Doppler sfhits in the quiet region.
%  Contents:
%    Summary.
%%%%%%%%%%%%%%%%%%%%%%%%%%%%%%%%%%%%%%%%%%%%%%%%%%%%

We investigated the spectrum tilts from ten emission lines (eight from the SW CCD and two from the LW CCD) using the data taken at the quiet region near the disk center. The spectra were obtained covering all $y$ position on the CCDs, from which we could analyze the behavior of line centroid along the slit. Our analysis revealed that the spectrum tilt has some characteristic dependence on wavelength. The difference of the tilt in the SW CCD and the LW CCD was also derived which is consistent with the result of \citet{kamio2010}. We use the spectrum tilt obtained here to calibrate the data since the analysis in this chapter dealt the Doppler shifts in the quiet region which is more or less several $\mathrm{km} \, \mathrm{s}^{-1}$ and requires much carefulness for the line centroid compared to other studies. The other reason why we must be such careful is that we investigate the center-to-limb variation of line centroids, which means that even subtle deviation which does not matter in the data with small field of view would become large enough to produce systematic variation in the long distance up to $2000''$. 

% --- End of Tex ---

%% file: tex/cal_results_ov.tex
% ========================================
%   Doppler shifts of the quiet region.
% ========================================

\begin{figure}
  \centering
  \includegraphics[width=12cm,clip]{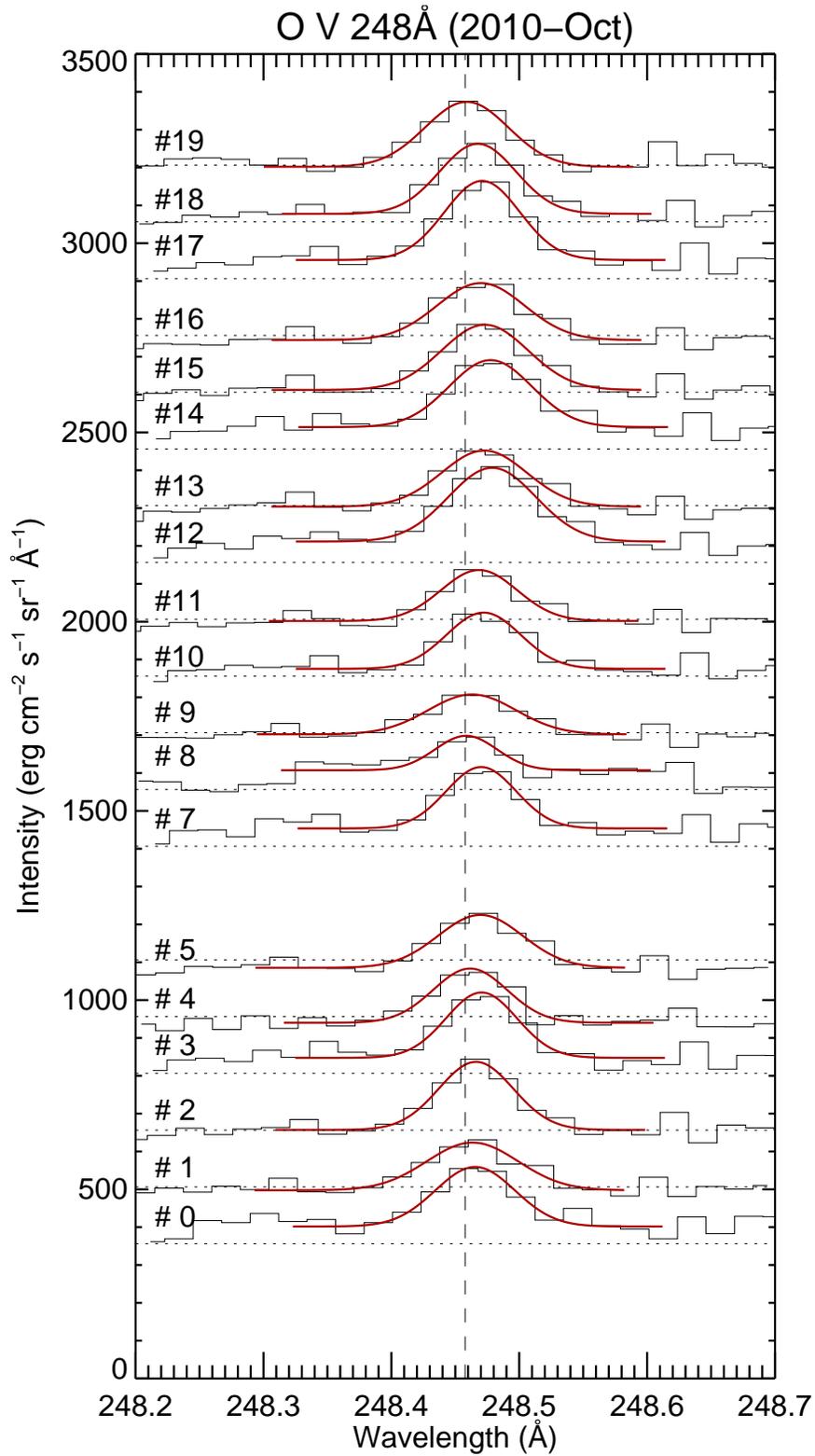}
  \caption{The spectra of O \textsc{v} $248.48${\AA} integrated by $500''$ in the solar $Y$ direction.  Twenty spectra were obtained in each pointing from south (\#0) to north (\#19).  \textit{Red} curves respectively indicate a fitted Gaussian.  A vertical dashed line is located at the average wavelength between the spectra \#0 and \#19.}
  \label{fig:ov_spectra}
\end{figure}

As introduced in Section \ref{sect:cal_lp_tr}, there are two transition region lines from oxygen ion in the EIS data analyzed in this chapter.  In order to check the consistency of our results and those from previous observations, here we investigated a center-to-limb variation of O \textsc{v} $248.48${\AA}.  This emission line is stronger than O \textsc{iv} $279.93${\AA}, but in order to obtain the spectra with high S/N ratio which leads to the precision of $\leq 5 \, \mathrm{km} \, \mathrm{s}$, we needed spatial integration with $500''$ in the solar-Y direction (\textit{i.e.}, along the EIS spectroscopic slit).  After the integration, five spectra at each pointing were obtained, which means we obtained $100$ spectra from twenty pointing locations in total.  The integrated spectra of O \textsc{v} $248.48${\AA} at first exposure in each pointing from October data are shown in Fig.~\ref{fig:ov_spectra}.  Twenty spectra were obtained in each pointing from south (\#0) to north (\#19).  Note that the spectrum \#6 was absent in this data due to data loss.  Red curves plotted over the spectra are a fitted Gaussian.  A vertical dashed line is located at the wavelength at the limb as calculated by fitting the result as same as for coronal lines.  The spectra at the solar disk are redshifted relative to those near both limbs.

\begin{figure}
  \centering
  \includegraphics[width=10cm,clip]{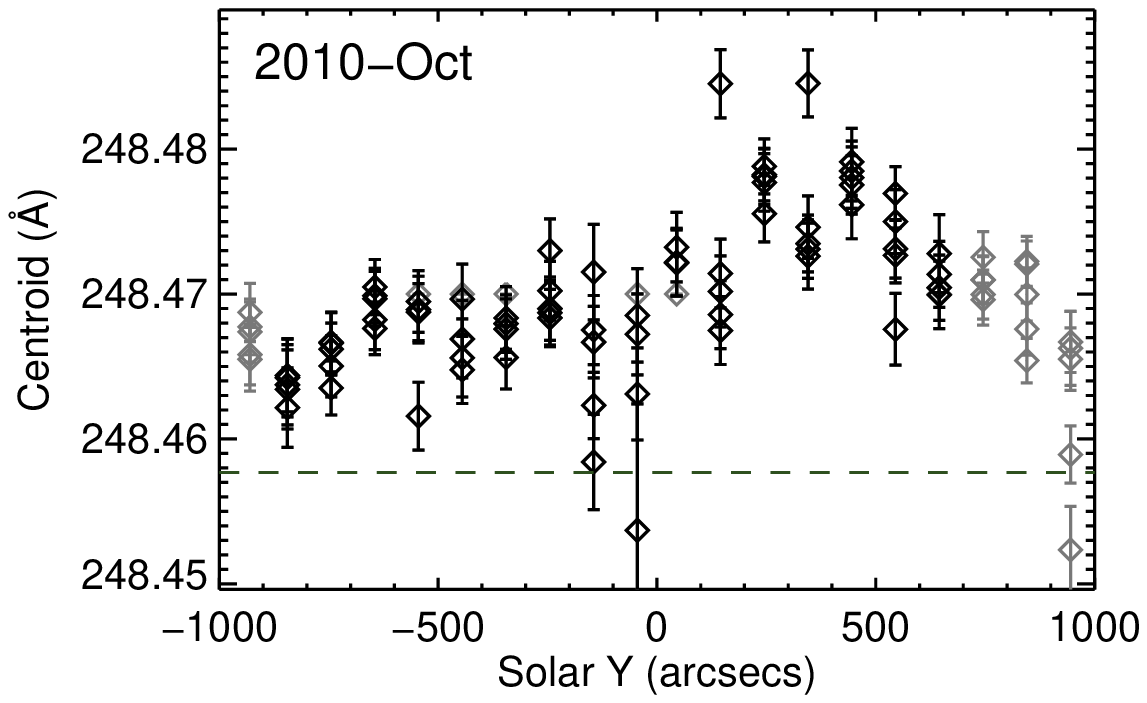}
  \includegraphics[width=10cm,clip]{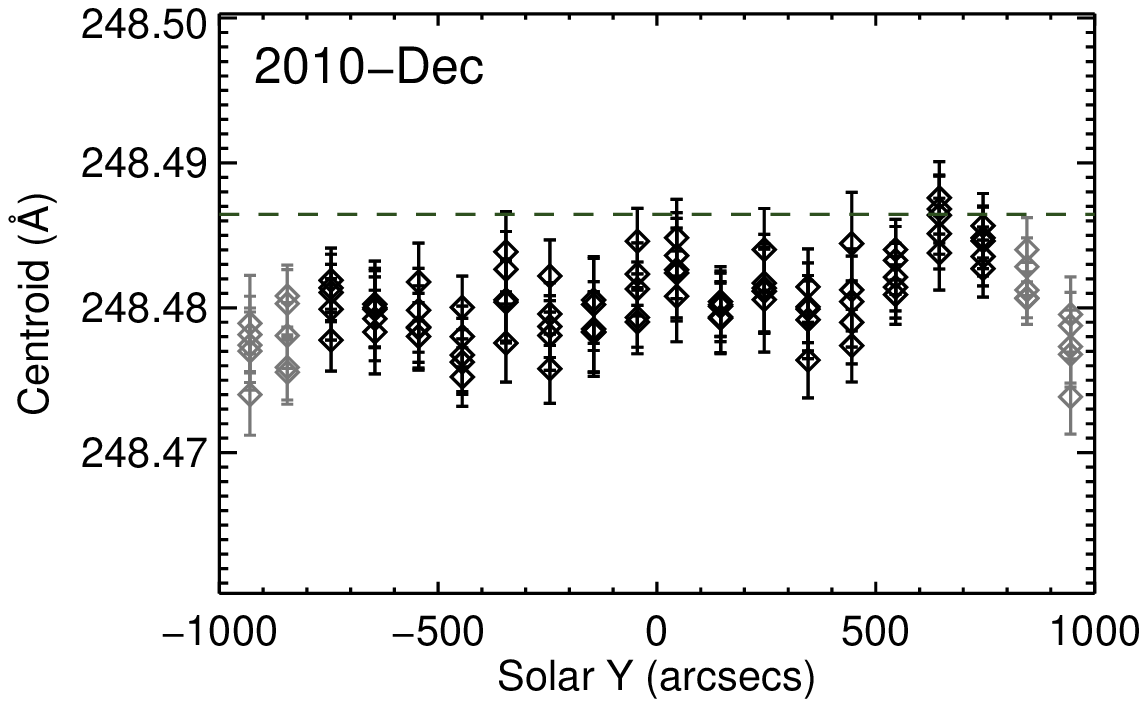}
  \caption{Centroid of O \textsc{v} $248.48${\AA} as a function of the solar 
    $Y$ coordinate for 2010 October (\textit{left}) and December (\textit{right}).
    \textit{Black} (\textit{Gray}) \textit{diamonds} indicate 
    that those data points are in (out) the quiet region. 
    \textit{Green dashed} lines are the centroid at the limb derived from the fitting by $v (\theta) = v_0 \cos \theta$ (\textit{cf.} Section \ref{sect:cal_limb2limb}).
  }
  \label{fig:cal_cen_mu}
\end{figure}

The centroid of the fitted Gaussian as a function of the solar $Y$ coordinate is shown in Fig.~\ref{fig:cal_cen_mu}.  The line centroids are basically more redshifted at the solar disk than at the limb.  
%In order to evaluate this redshift at the solar disk, the centroid as a function of $\cos \theta$ ($\theta$ is an angle between line of sight and normal vector as to the solar surface) is plotted in panel (b).  The dashed line in the panel is a fitted linear function and its slope indicates the radial velocity of $10 \, \mathrm{km} \, \mathrm{s}^{-1}$ downward to the solar surface.  
Previous observations by SUMER reported the redshift of the transition region lines whose formation temperatures are similar to that of O \textsc{v} \citep{chae1998doppler,peterjudge1999}.  However, we notice that there is a peculiar tendency in the center-to-limb variation of the Doppler shift.  The line centroid indeed increases (\textit{i.e.}, redshifted) from the limb to inside the disk as expected from the previous observations, but at some point it decreases in reversal toward the disk center. 

% --- End of TeX ---

%% file: tex/contents_vel.tex
\chapter{Doppler velocity measurement for AR outflows}
  \label{chap:vel}
\section{Introduction}
  \input{tex/vel_itdn.tex}
\section{Observations and data reduction}
  \subsection{EIS scan}
    \input{tex/vel_ar10978_obs.tex}
  \subsection{Data reduction}
    \input{tex/vel_ar10978_obs_lines.tex}
\section{Line profiles}
  \input{tex/vel_ar10978_lp.tex}
  \label{sect:vel_lp}
  \subsection{$\log T \, [\mathrm{K}] = 5.7$--$5.8$}
    \input{tex/vel_ar10978_lp_tr.tex}
    \label{sect:lp_tr}
  \subsection{$\log T \, [\mathrm{K}] = 6.1$--$6.2$}
    \input{tex/vel_ar10978_lp_ct.tex}
  \subsection{$\log T \, [\mathrm{K}] = 6.3$}
    \input{tex/vel_ar10978_lp_im.tex}
  \subsection{$\log T \, [\mathrm{K}] = 6.4$--$6.5$}
    \input{tex/vel_ar10978_lp_ht.tex}
    \label{sect:lp_ht}
\section{Measurement of the Doppler velocities of AR outflows}
  \subsection{Doppler velocity maps}
    \label{sect:vel_map}
    \input{tex/vel_ar10978_vel_map.tex}
  \subsection{Histogram of Doppler velocities}
    \input{tex/vel_ar10978_hist_vel.tex}
    \subsubsection{$\log T \, [\mathrm{K}] = 5.7$--$5.8$}
      \input{tex/vel_ar10978_hist_vel_tr.tex}
    \subsubsection{$\log T \, [\mathrm{K}] = 6.1$--$6.2$}
      \input{tex/vel_ar10978_hist_vel_ct.tex}
    \subsubsection{$\log T \, [\mathrm{K}] = 6.3$}
      \input{tex/vel_ar10978_hist_vel_im.tex}
    \subsubsection{$\log T \, [\mathrm{K}] = 6.4$--$6.5$}
      \input{tex/vel_ar10978_hist_vel_ht.tex}
\section{Temperature dependence of the Doppler velocities}
  \input{tex/vel_ar10978_tvsv.tex}
\section{Summary and discussion}
  \label{sect:vel_sum}
  \input{tex/vel_sum.tex}
\clearpage
\begin{subappendices}
  \section{Calibration of the spectrum tilt}
    \input{tex/vel_app_spec_tilt.tex}
    \label{sect:vel_app_spec_tilt}
  \section{Mg emission lines}
    \input{tex/vel_app_mg_lines.tex}
    \label{sect:vel_app_mg_lines}
  \section{Temperature dependence of the Doppler velocities (all samples)}
    \input{tex/vel_app_tvsv.tex}
  \section{Residual from single Gaussian fitting in the outflow region}
    \label{sect:vel_res}
    \input{tex/vel_ar10978_lp_res.tex}
%  \subsection{Spatial distribution of blueshifted component}
%    \input{tex/vel_localization.tex}
%  \subsection{Correlation between intensity, Doppler velocity, and line width}
%    \label{sect:vel_sct}
%    \input{tex/vel_sct.tex}
%\section{Doppler velocity for Si \textsc{vii} and Fe \textsc{viii}}
%  \input{tex/vel_app_feviii.tex}
\end{subappendices}

%% file: tex/vel_itdn.tex
% =============================================
%   Chapter:
%     Temperature dependence of AR outflows.
%   Section:
%     Introduction.
% =============================================

The solar corona shows different appearance when we observe it in the different temperatures.  Focusing on an active region, patchy structures are seen for the temperature of $\log T \, [\mathrm{K}] \leq 5.6$ (\textit{i.e.}, the transition region).  For the temperature range of $\log T \, [\mathrm{K}] = 5.6$--$6.0$, there are prominent elongated loops structures from the edge of the active region to the surrounding region, named ``fan loops'' after their appearance.  They are often observed at east or west edge of active regions.  Going up to higher temperature around $\log T \, [\mathrm{K}] = 6.0$--$6.3$, both fan loops and the core of the active region are bright, and loop structures at the core region can be seen which connect positive and negative magnetic polarity.  Above the temperature $\log T \, [\mathrm{K}] \simeq 6.4$, radiation from the core region dominates.  

The temperature dependence of Doppler velocities gives us a clue to understand the process occurring at the coronal structures.  Fan loops are known to indicate characteristic velocity dependence on the temperature \citep{warren2011}.  In the transition region temperature ($\log T \, [\mathrm{K}] \leq 6.0$), they show a redshift of $\sim 10 \, \mathrm{km} \, \mathrm{s}^{-1}$ which has been interpreted that the plasma in fan loops around the transition region temperature flows down to the footpoints.  On the other hand, a blueshift of $\gtrsim 10 \, \mathrm{km} \, \mathrm{s}^{-1}$ is observed above the temperature around $\log T \, [\mathrm{K}] = 6.0$ (\textit{e.g.}, Fe \textsc{x}--\textsc{xv}).  \citet{mcintosh2012} think this dependence as the indication of the coronal heating followed by cooling.  \citet{bradshaw2008} has analytically estimated the speed of cooling downflow by considering the balance of enthalpy flux, which could account for the reported value of several tens of $\mathrm{km} \, \mathrm{s}^{-1}$. 

Since outflow regions often takes its place adjacent to fan loops, and the blueshift of fan loops in corona lines has been reported, outflow regions and fan loops had been often regarded as the identical structure \citep{tian2011,mcintosh2012}.  \citet{young2012} recently mentioned that the location of outflows and fan loops are slightly displaced with each other.  It was reported that there was no Si \textsc{vii} emission in the outflow region \citep{warren2011}, 
%{\color{red} 
which implies that the outflow region could be separated from fan loops.  The physical properties of fan loops and the outflows are noted in Table \ref{tab:vel_outflow_fan}.  We analyzed the data taken at relatively initial phase of \textit{Hinode} when the sensitivity of EIS was high, and tried to measure the Doppler velocity at the transition region temperature in the outflow region.
%}

\input{tex/tab_vel_outflow_fan.tex}

In this chapter, we measure the Doppler velocities of a fan loop and the outflow region in NOAA AR10978 using a reference velocity of the quiet region obtained in Chapter \ref{chap:cal} and derive the temperature dependence.  Previous observations include the error up to $10 \, \mathrm{km} \, \mathrm{s}^{-1}$ at most, which might produce even the reversal of the sign of Doppler velocity because the observed values are also the order of $\sim 10 \, \mathrm{km} \, \mathrm{s}^{-1}$.  Our analysis in Chapter \ref{chap:cal} enables us to deduce the Doppler velocity within an error of $\sim 3 \, \mathrm{km} \, \mathrm{s}^{-1}$.  By measuring the Doppler velocity, we aim to reveal the nature of the outflow region and clarify differences between the outflow region and fan loops.  
%{\color{red} 
In this thesis, fan loops are defined as the loop structure extending from the periphery of an active region which is bright in the transition region lines, while the outflow region is defined as the region where the emission line width increases in the coronal lines and no fan loops exist.
%}

% --- End of TeX ---

%% file: tex/tab_vel_outflow_fan.tex
\begin{table}
  \centering
  \begin{minipage}{17cm}
    \caption{%
%\color{red} 
Obtained physical properties and inferences on fan loops and outflows.  The results obtained in specific literature are followed by a reference.  References are abbreviated as B09 \citep{baker2009}, NH11 \citep{nishizuka2011}, U11 \citep{ugarte-urra2011}, M12 \citep{mcintosh2012}, and Y12 \citep{young2012}.  \textit{TR} represents the transition region.}
    \label{tab:vel_outflow_fan}
    \begin{tabular}{lll}
      \toprule
                             & Fan loop          & Outflow \\
      \midrule
      %Location               & Periphery of active regions & Periphery of active regions \\
      Emission lines         & Si \textsc{vii} and Fe \textsc{viii} & Fe \textsc{xi}--\textsc{xv} \\
      Temperature ($\log T \, [\mathrm{K}]$) & $5.7\text{--}6.0$ & $6.2\text{--}6.3$ \\
      Intensity              & Bright in TR lines & Dark. Extending structures in Doppler maps of \\
                             &                    & coronal lines. \\
      Doppler velocity       &                   & \\
      \multicolumn{1}{r}{\small The transition region}  & Downflow of $v \simeq 20 \, \kmpers$ 
                                                 & Upflow of several $\kmpers$ (B09)\footnote{Accuracy of the Doppler velocity measurement was the order of $10 \, \kmpers$ at worst.  To be measured in this chapter.} \\
      \multicolumn{1}{r}{\small Corona}                 & Upflow or no net velocity (Y12)   
                                                        & Upflow of several tens to a hundred $\kmpers$ \\
      Line width             & Not studied\footnote{Obviously less than that in the outflow region by $\Delta W \simeq 10${m\AA}.} & Enhanced \\
      Electron density 
      ($\mathrm{cm}^{-3}$)    & $5.0 \times 10^{8}$ (Y12)
                             & Not studied\footnote{Electron density of total emission (the major component $+$ EBW) has been already studied and known to be smaller than that of active regions \citep{doschek2008,brooks2012}, but that of the upflow component has not been measured yet, which is to be studied in Chapter \ref{chap:dns}.} \\
      Emission line profiles & Symmetric at leg (NH11) & Major component $+$ EBW \\
                             & Major component $+$ EBW at footpoint & \\
                             & \multicolumn{1}{r}{(M12)} & \\
      Inferred origin        & Mass draining during the cooling phase & 1) Impulsive heating \\
                             &                   & 2) Reconnection between short and long loops \\
                             &                   & 3) Horizontal expansion of active regions \\
                             &                   & 4) Plasma at the tips of chromospheric spicules \\
      \bottomrule
    \end{tabular}
  \end{minipage}
\end{table}

%% file: tex/vel_ar10978_obs.tex
% =======================================
%   Chapter:
%     Doppler velocity of outflow.
%   Section:
%     Observations and data reduction.
% =======================================

\begin{figure}
  \centering
  \includegraphics[width=16cm,clip]{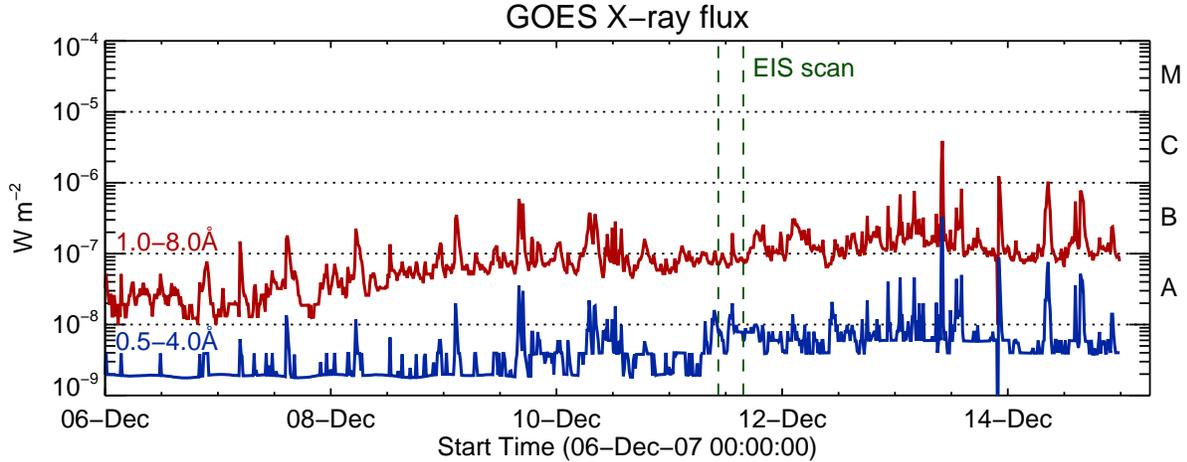}
  \caption{\textit{GOES} X-ray plot during the disk passage of NOAA AR10978 during 2007 December 6--15.  Alphabets written in the right side of the plot denote the conventional classification of flares.  \textit{Red} (\textit{Blue}) data represents low (high) energy channel of \textit{GOES} satellite.  The interval between \textit{green} lines indicate the duration of the EIS observation near at the disk center, which was used in this chapter. }
  \label{fig:goes_plot_ar10978}
\end{figure}

\begin{figure}
  \centering
  \includegraphics[width=8.4cm,clip]{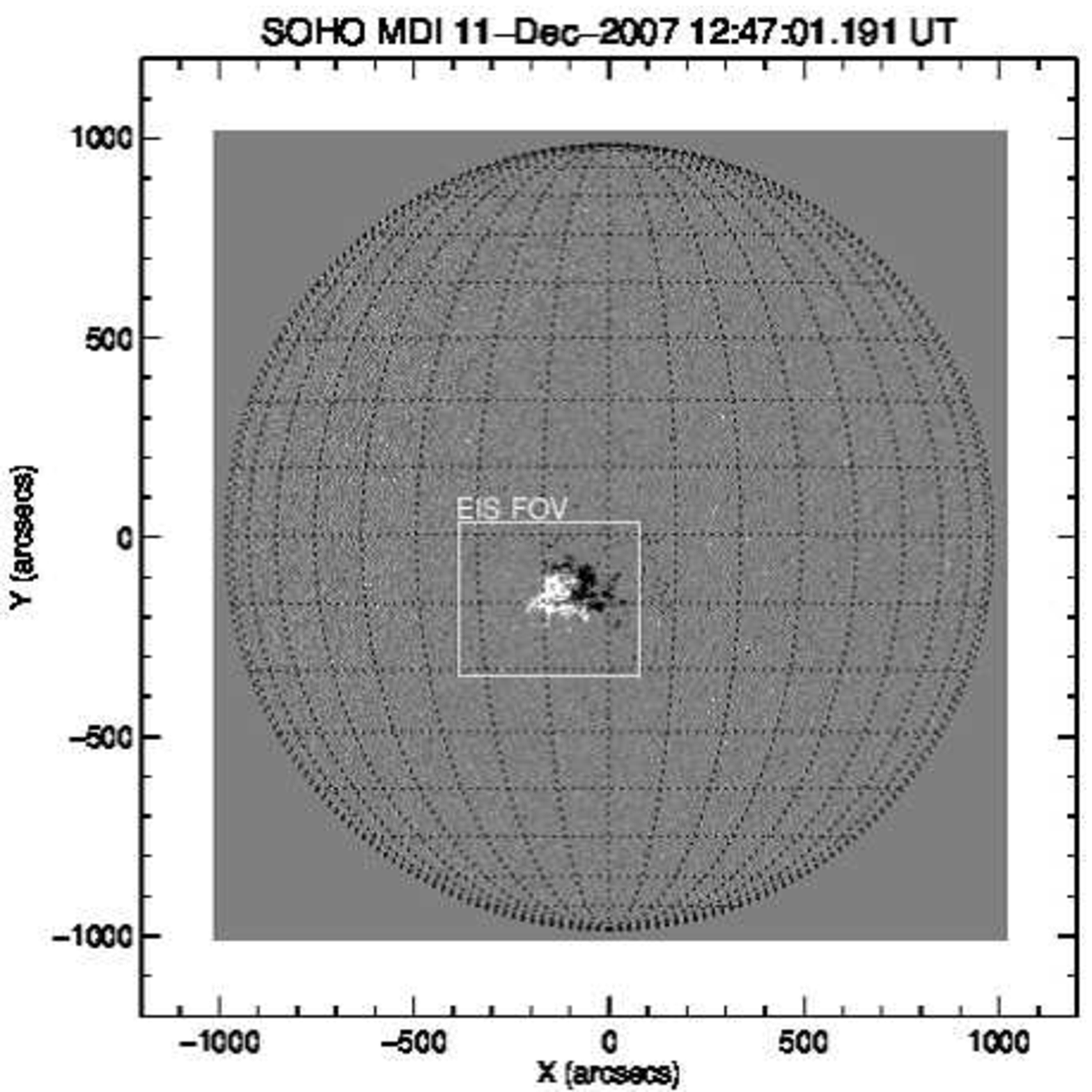}
  \includegraphics[width=8.4cm,clip]{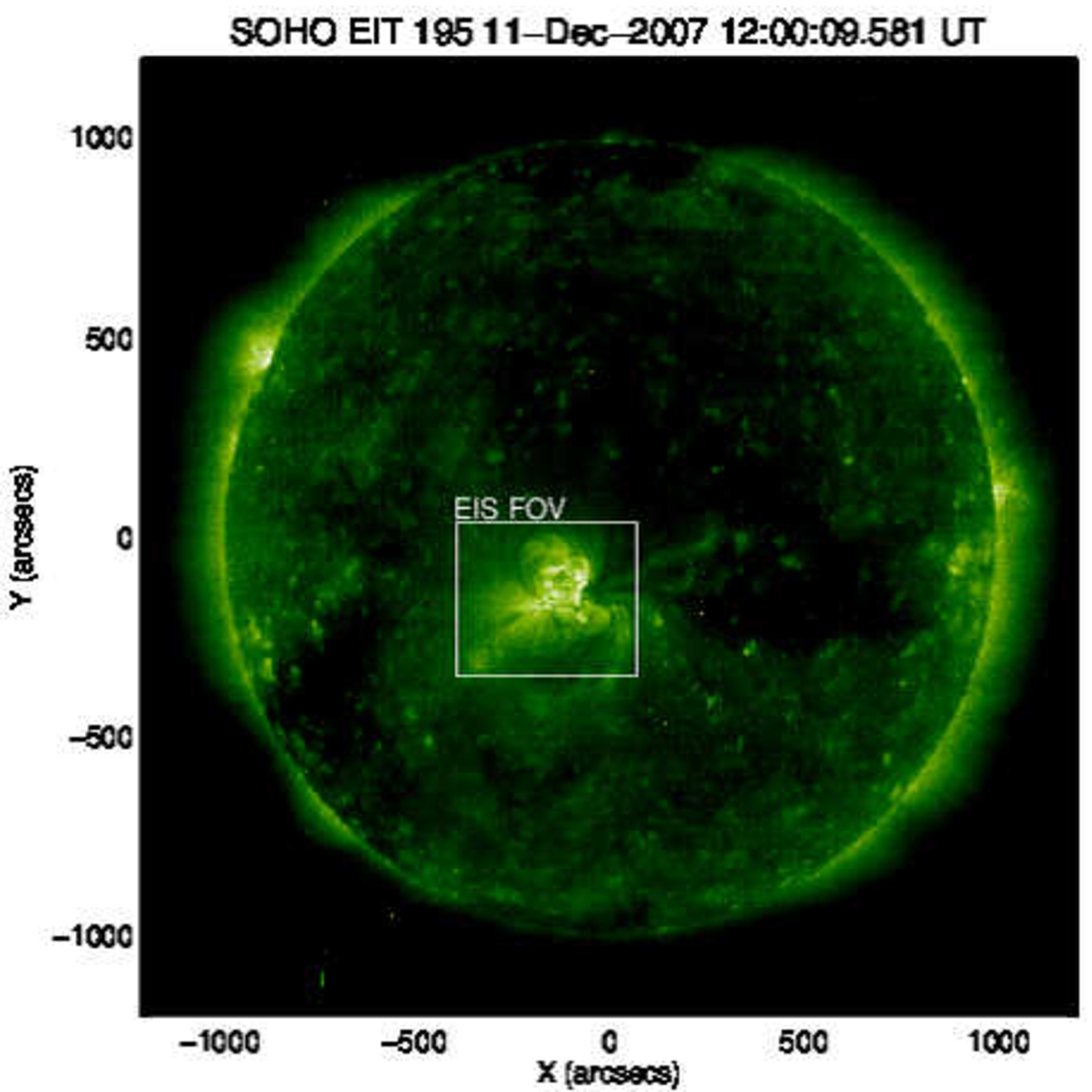}
  \caption{Images of AR10978 taken by \textit{SoHO}/MDI and EIT. 
    \textit{Left}: MDI magnetogram taken on 2007 December 11 12:47:01UT.
    \textit{Right}: EIT image taken on 2007 December 11 12:00:09UT.  
    The region surrounded by a white dotted line in each panel indicates 
    the field of view (FOV) of an EIS scan analyzed in this chapter. 
  }
  \label{fig:whole_context}
\end{figure}

% AR10978
A raster scan of NOAA AR10978 obtained with \textit{Hinode}/EIS was analyzed to study the temperature dependence of the Doppler velocity in the outflow region.  The active region appeared on the solar disk from the east limb on 2007 December 6, and disappeared into the west limb on 2007 December 18.  During this period, no other active regions existed, and the observation by Hinode were focused on AR 10978 except for a few synoptic observations.  AR10978 showed overall low activities especially in the former half of its disk passage as shown in Fig. \ref{fig:goes_plot_ar10978} (one B-class flare during the scan).  There is a negative-polarity leading sunspot and following positive-polarity strong magnetic field regions.  A magnetogram and an EUV image of the whole Sun taken by \textit{SoHO} on 2007 December 11 are shown in Fig.~\ref{fig:whole_context}. 

%Details on the EIS observation are shown Table \ref{tab:vel_data}.  
The EIS data analyzed in this chapter was taken on 2007 December 11 10:25:42--15:44:33UT.  The scan has a large-area FOV ($460'' \times 384''$) which includes entire AR10978.  The $1''$ spectroscopic slit was used with an exposure time of $40 \, \mathrm{sec}$.  The center of the field of view was $(-175'', -155'')$, which is suitable to derive the Doppler velocities with less superposition of coronal structures. %We also checked a scan at the west limb, from which the spectrum tilts were checked (Section \ref{sect:vel_app_spec_tilt}). 

%\input{../tex/tab_vel_data.tex}
% Overview of raster image as a context
\begin{figure}
  \centering
  \includegraphics[width=8.3cm,clip]{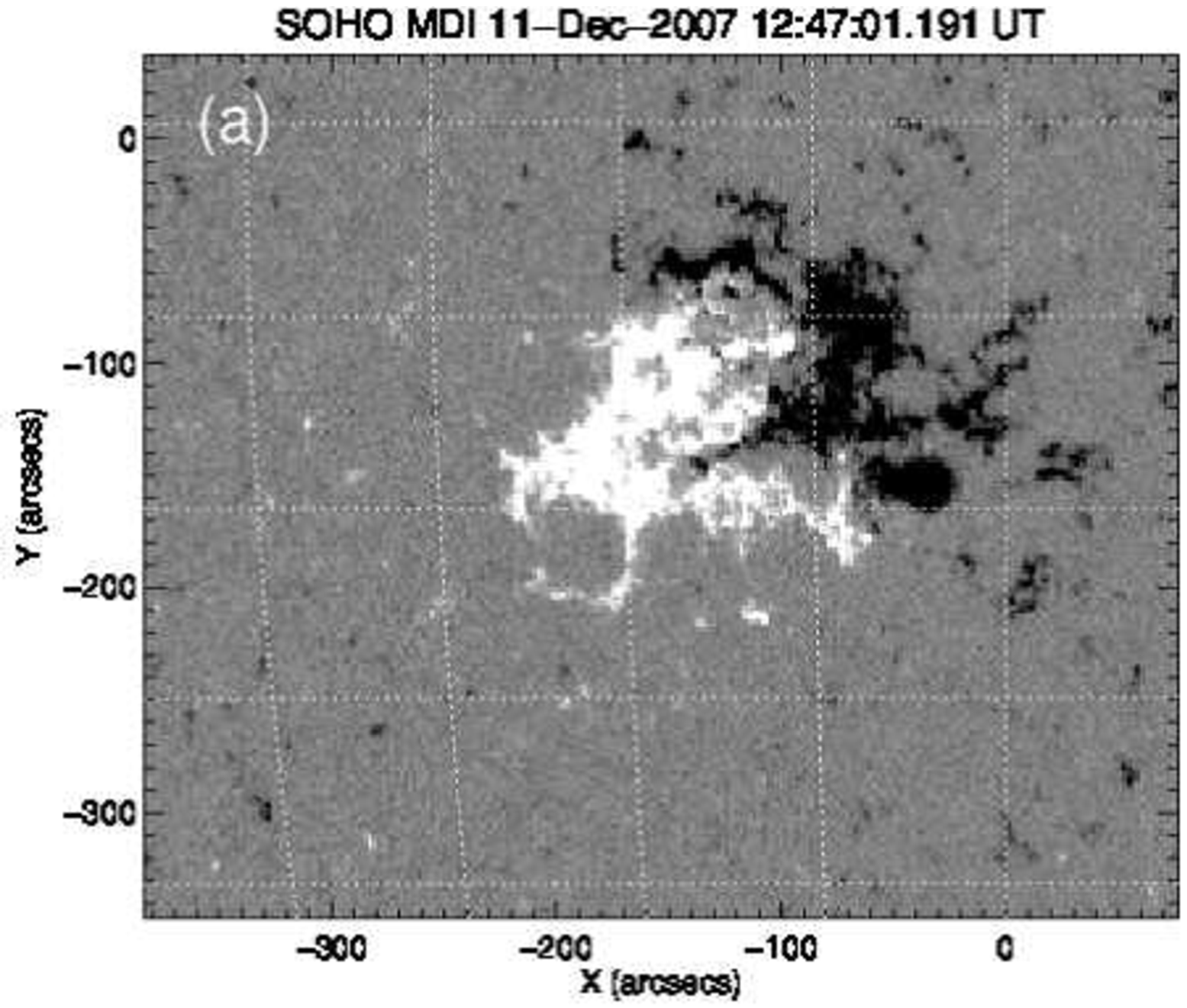}
  \includegraphics[width=8.3cm,clip]{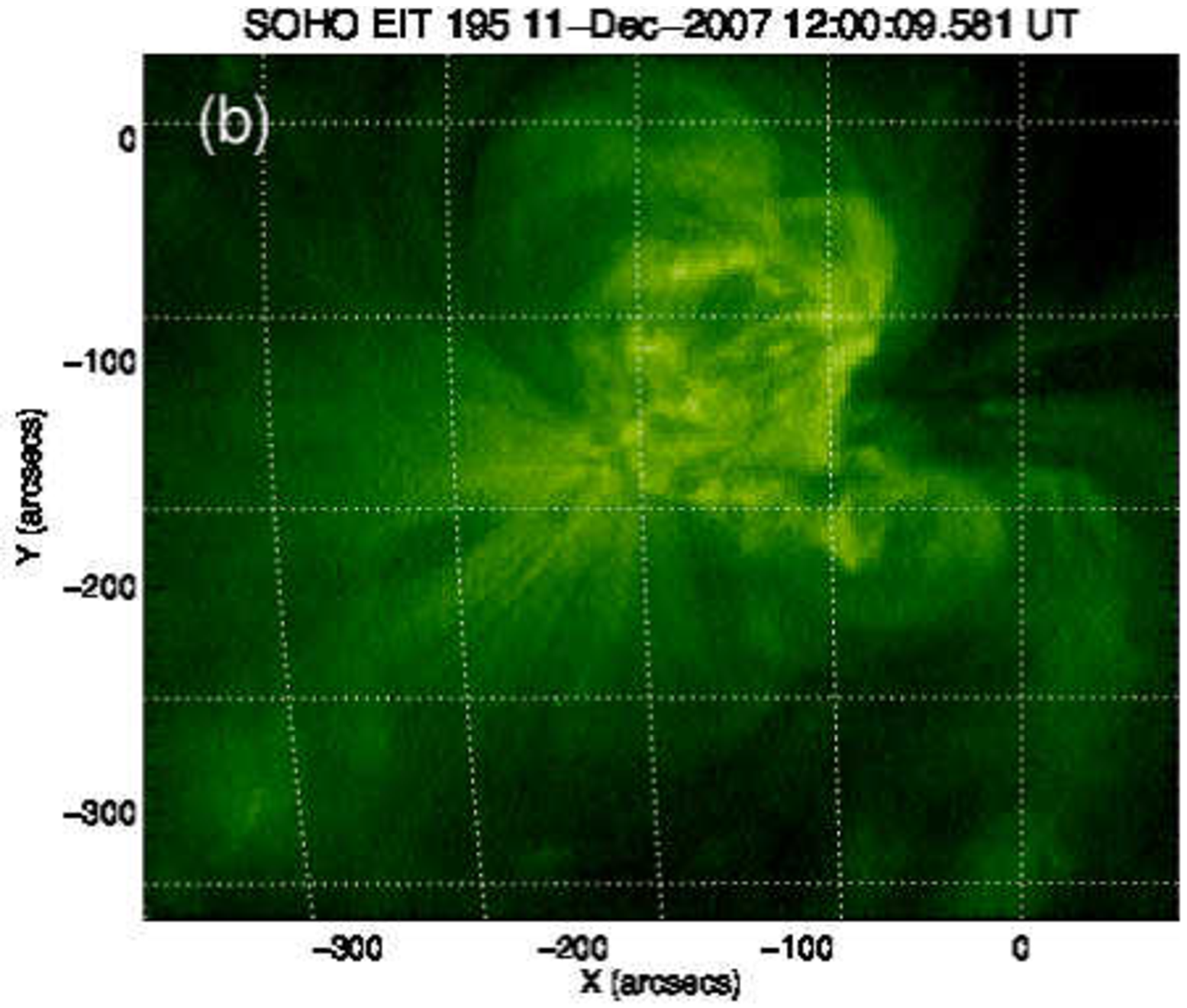}
  \includegraphics[width=8.3cm,clip]{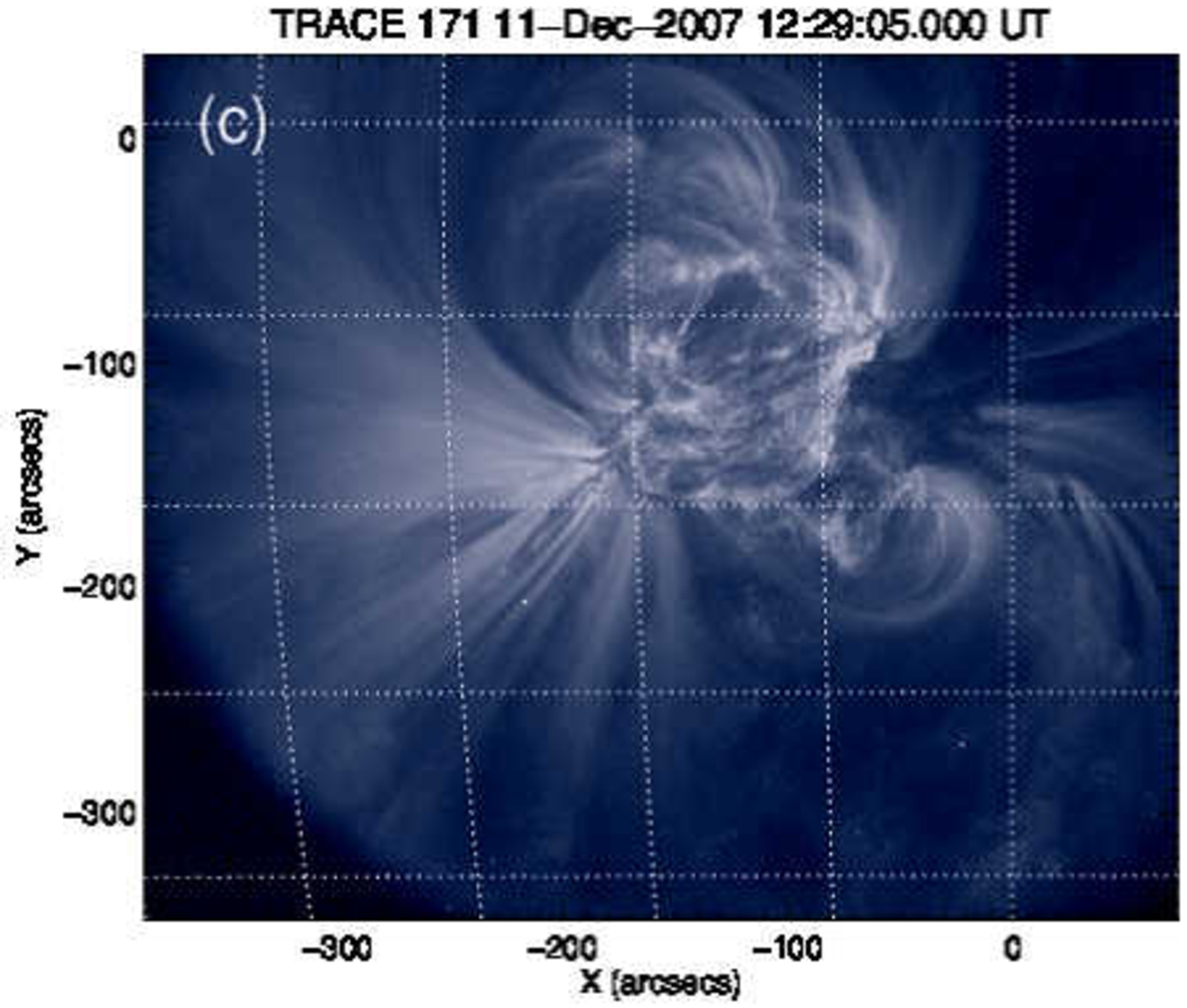}
  \includegraphics[width=8.3cm,clip]{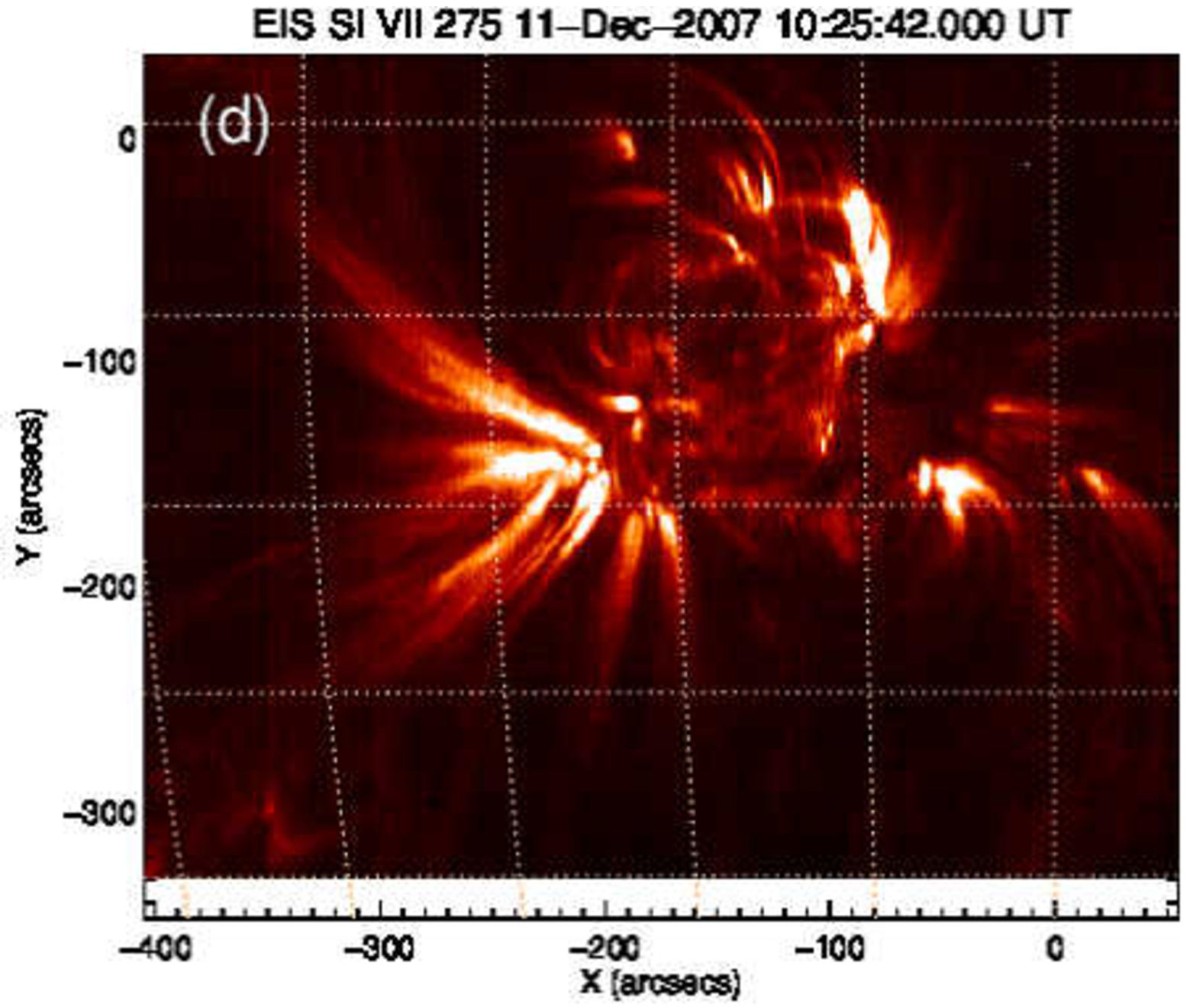}
  \includegraphics[width=8.3cm,clip]{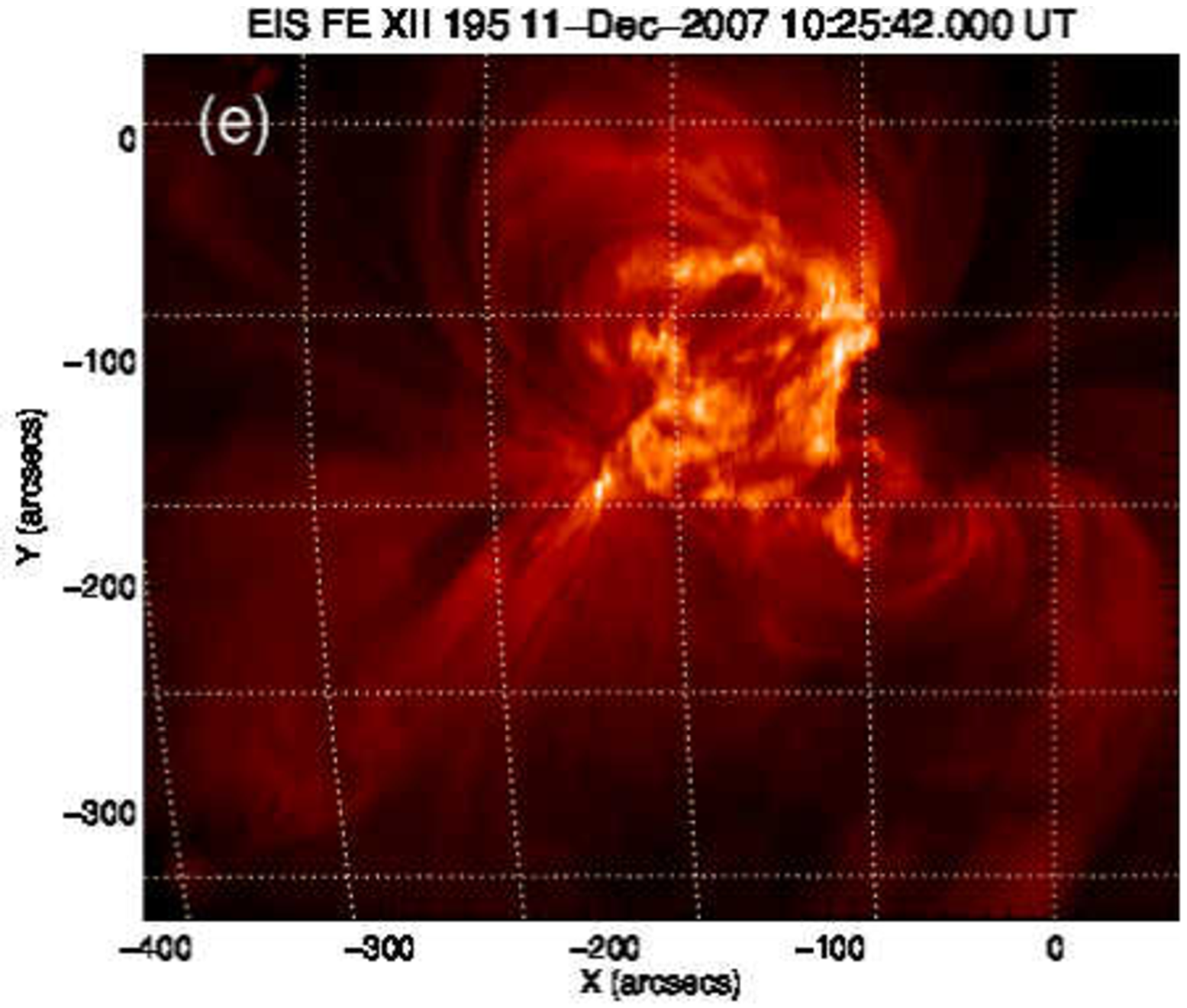}
  \includegraphics[width=8.3cm,clip]{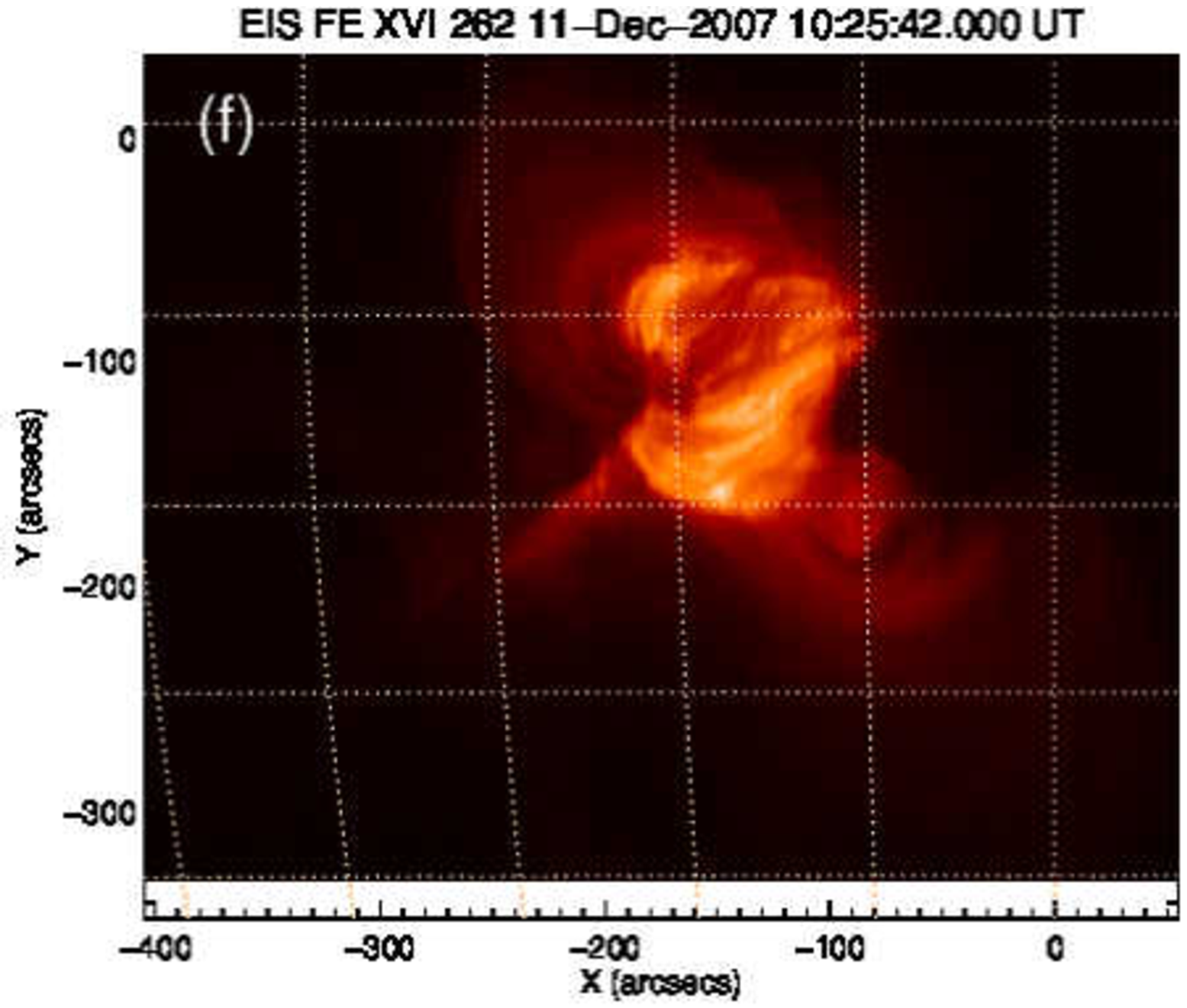}
  \caption{(a) MDI magnetogram, (b) EIT $195${\AA} passband image, (c) TRACE $171${\AA} passband image (mainly contributed by Fe \textsc{ix}--\textsc{x}), and intensity maps of three emission lines whose formation temperature is different: (d) Si \textsc{vii} $275.35${\AA} ($\log T \, [\mathrm{K}]=5.8$), (e) Fe \textsc{xii} $195.12${\AA} ($\log T \, [\mathrm{K}]=6.2$), and (f) Fe \textsc{xvi} $262.98${\AA} ($\log T \, [\mathrm{K}]=6.5$).}
  \label{fig:vel_context_ar10978}
\end{figure}

Fig.~\ref{fig:vel_context_ar10978} shows (a) an MDI magnetogram, (b) an EIT $195${\AA} passband image, (c) a TRACE $171${\AA} passband image, and (d)--(f) intensity maps of three emission lines whose formation temperature is different: Si \textsc{vii} $275.35${\AA} ($\log T \, [\mathrm{K}]=5.8$), Fe \textsc{xii} $195.12${\AA} ($\log T \, [\mathrm{K}]=6.2$), and Fe \textsc{xvi} $262.98${\AA} ($\log T \, [\mathrm{K}]=6.5$).  In the Si \textsc{vii} map, several elongated structures from east and west edge of the active region are seen, which is called as ``fan loops''.  This structure is typical around this temperature range ($\log T \, [\mathrm{K}] \lesssim 6.0$).  The emission from Si \textsc{vii} is relatively dark at the core of the active region.  The morphology becomes more complex for the intensity map of Fe \textsc{xii} than that of Si \textsc{vii}.  Not only the elongated structure, which is less clearly discernible than fan loops, but also loop-like structure (coronal loops) connecting east and west part of the active region can be seen.  Above the temperature around $\log T \, [\mathrm{K}]=6.4$, the emission from the core region dominates over the surroundings as seen in the intensity map of Fe \textsc{xvi}.  Emission outside the core region from this temperature range was very weak. 

% --- End of TeX ---

%% file: tex/vel_ar10978_obs_lines.tex
% =======================================
%   Chapter:
%     Doppler velocity of outflow.
%   Section:
%     Observations and data reduction.
% =======================================

\input{tex/tab_vel_emission_lines.tex}

\begin{figure}
  \centering
  \includegraphics[width=16.8cm,clip]{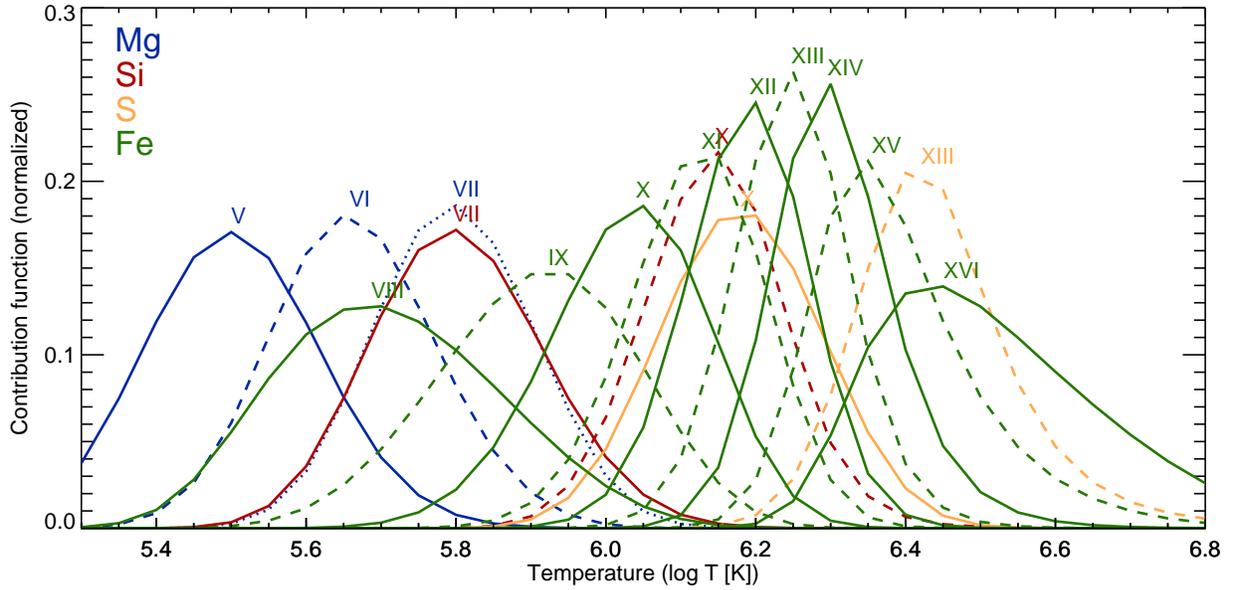}
  \caption{Contribution functions $G (T)$ given by CHIANTI database ver.~7 \citep{dere1997,landi2012}.  $G (T)$ for all emission lines used in this chapter are plotted. Colors indicate ion species (\textit{blue}: Mg, \textit{red}: Si, \textit{yellow}: S, and \textit{green}: Fe).  Each contribution function was normalized so that the area under the curve becomes unity.}
  \label{fig:vel_cnt_fnct}
\end{figure}

We used twenty six emission lines in total whose formation temperature ranges widely within $\log T \, [\mathrm{K}]=5.6\text{--}6.5$ in order to investigate temperature dependence of the Doppler velocity in coronal structures without gaps in temperature as long as possible.  The emission lines are listed in Table \ref{tab:vel_emission_lines}.  We selected emission lines which are relatively strong, isolated and free from significant blend by other unidentified lines.  The contribution functions of the emission lines used in this study are shown in Fig.~\ref{fig:vel_cnt_fnct}.  \textit{Blue}, \textit{red}, \textit{yellow}, and \textit{green} lines respectively indicate those of Mg, Si, S, and Fe ion. 

The data was calibrated through the standard EIS software to remove hot/warm pixels, subtract dark current and CCD pedestal.  Note that some programs in the current standard software calibrate the spectrum tilt by using same parameters in both SW and LW CCDs, however, we used different parameters for each CCD given by \citet{kamio2010}.  The validity of them was checked as described in Section \ref{sect:vel_app_spec_tilt}.  

\begin{figure}
  \centering
  \includegraphics[width=8.4cm,clip]{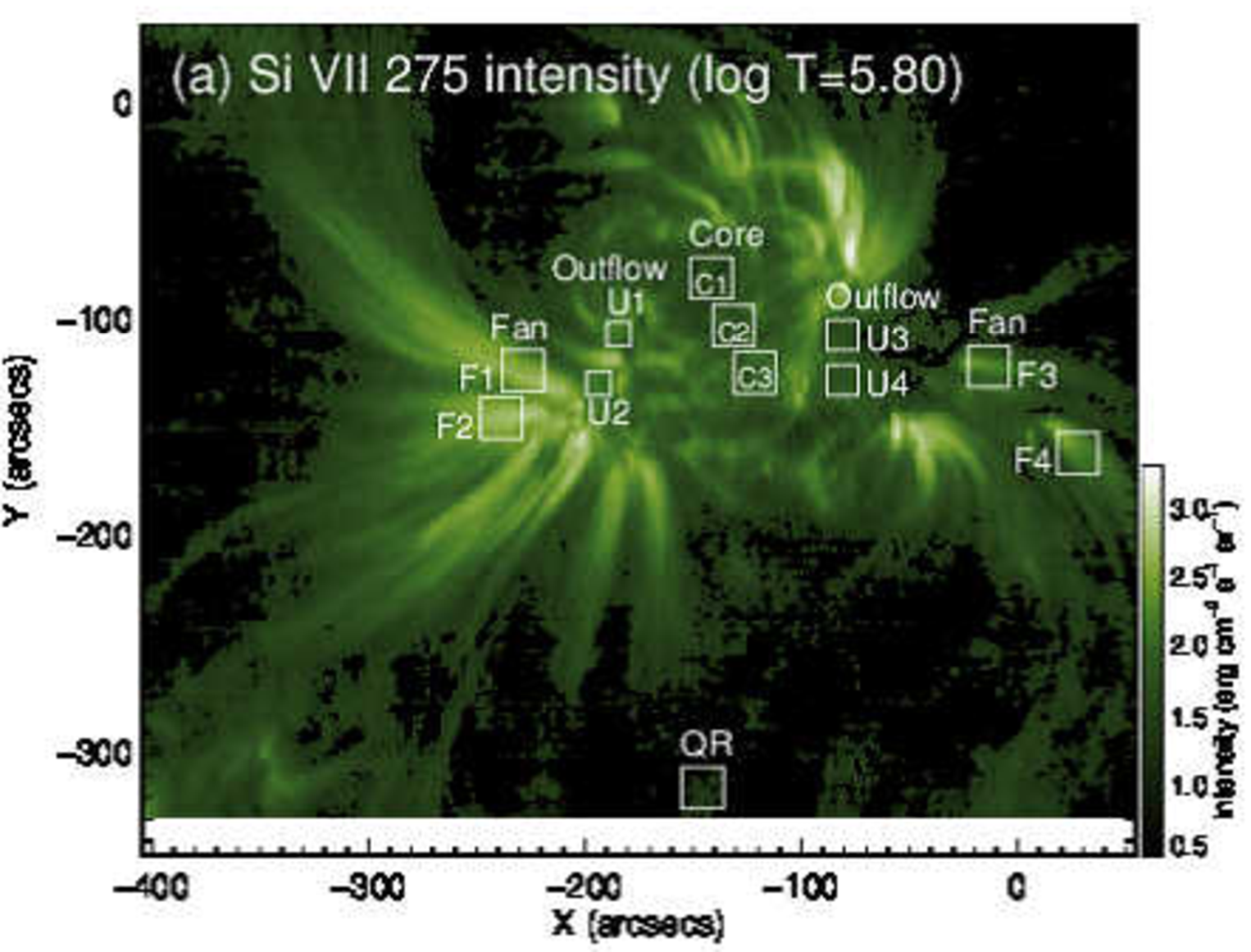}
  \includegraphics[width=8.4cm,clip]{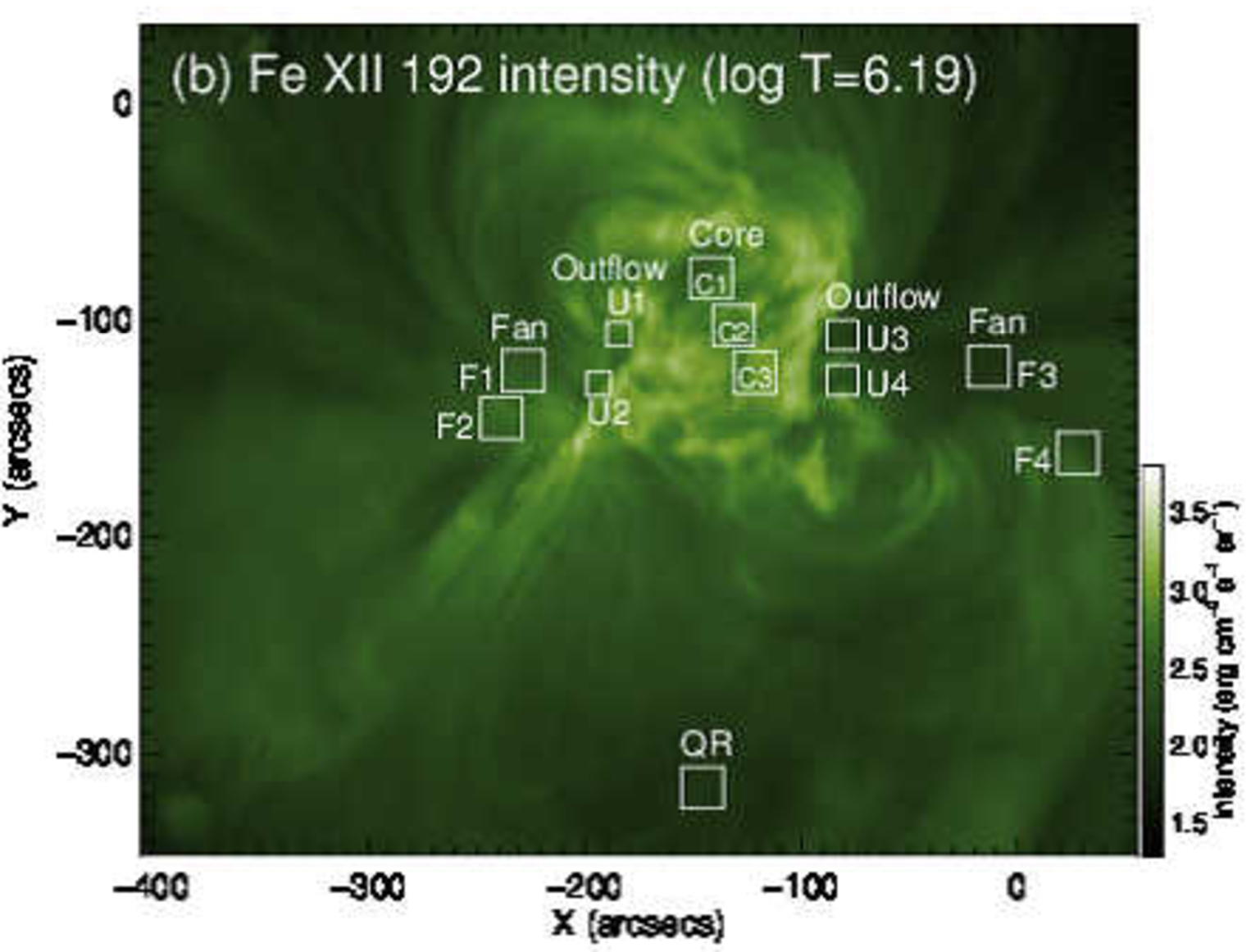}
  \includegraphics[width=8.4cm,clip]{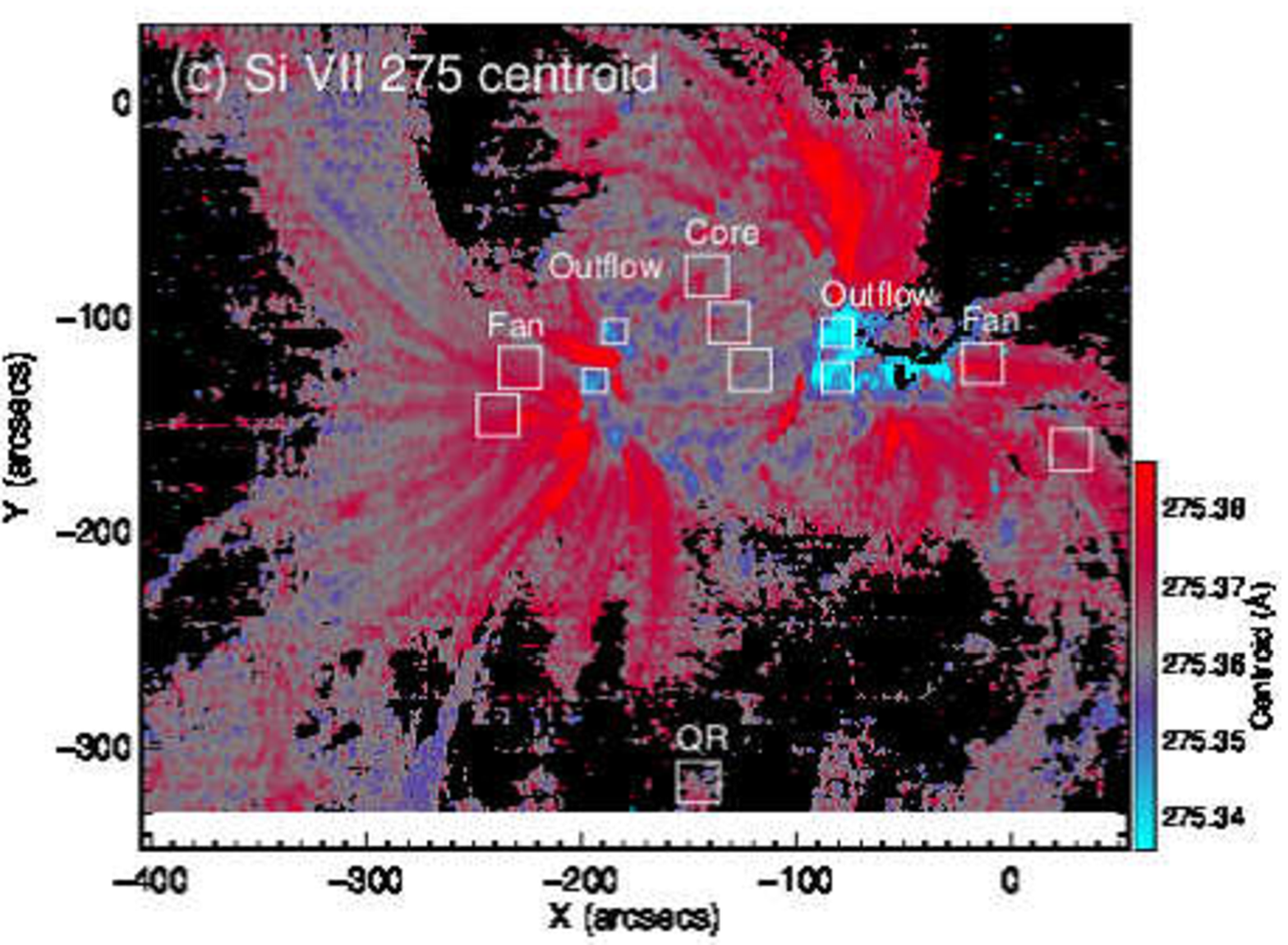}
  \includegraphics[width=8.4cm,clip]{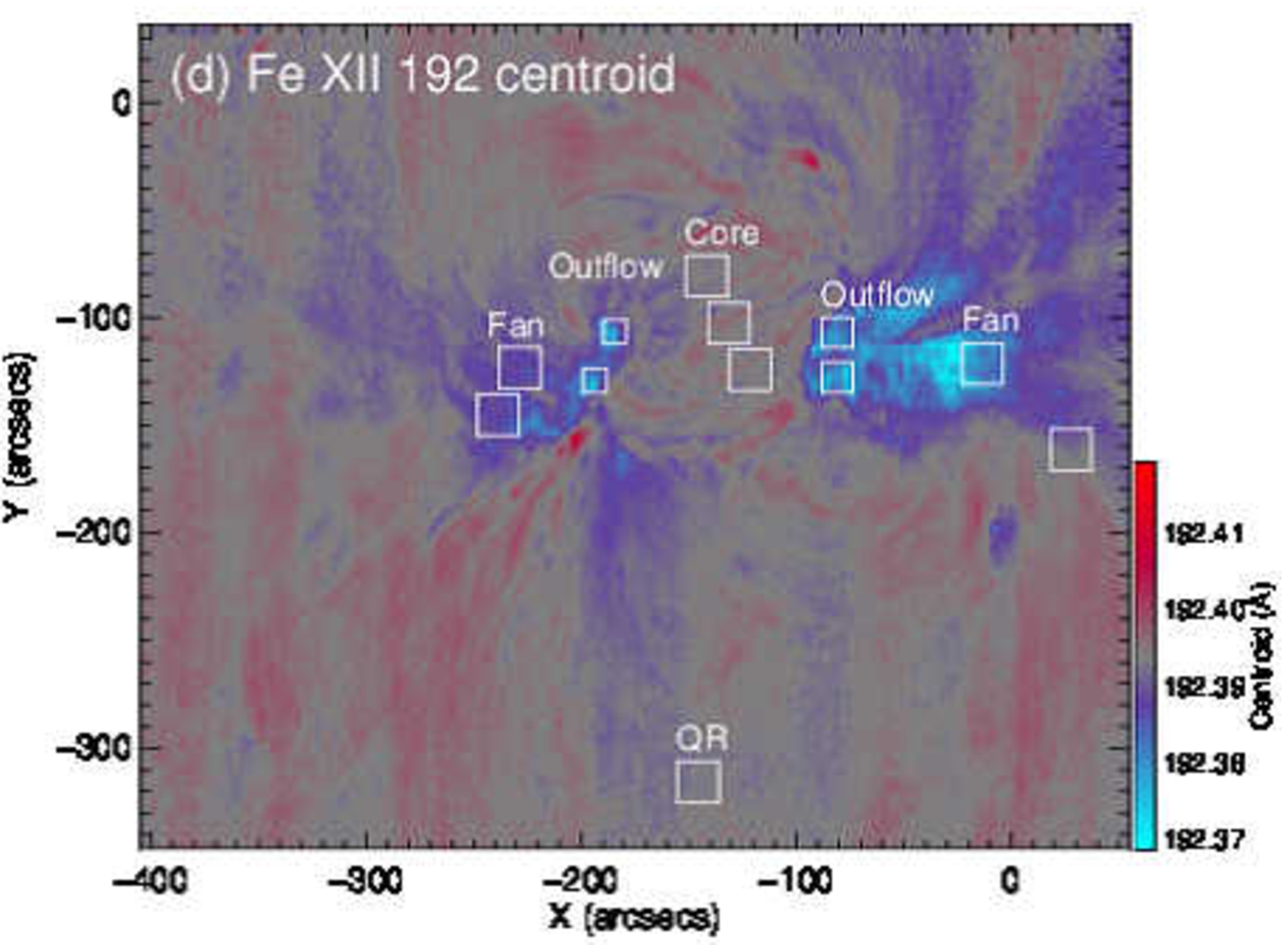}
  \includegraphics[width=8.4cm,clip]{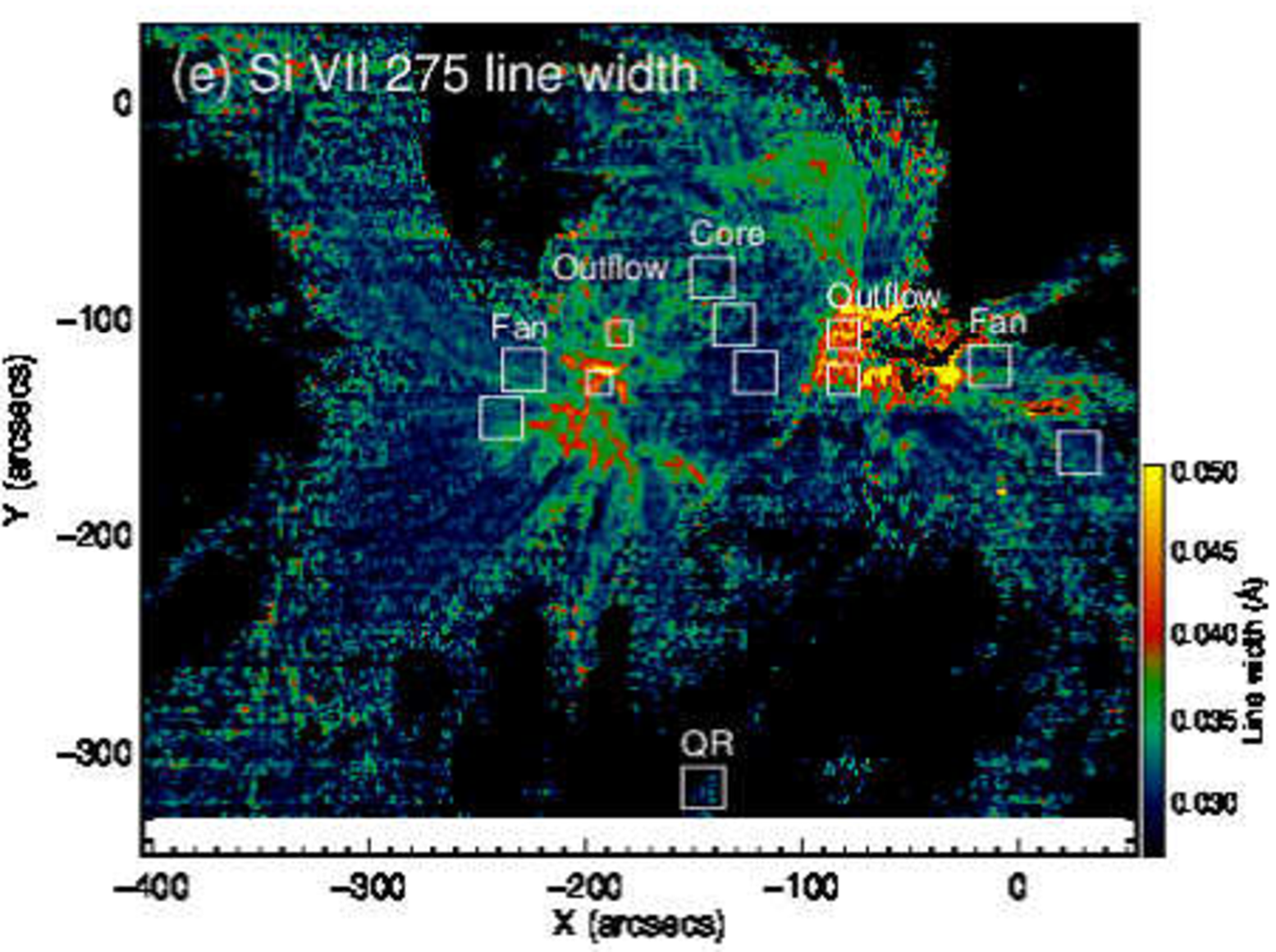}
  \includegraphics[width=8.4cm,clip]{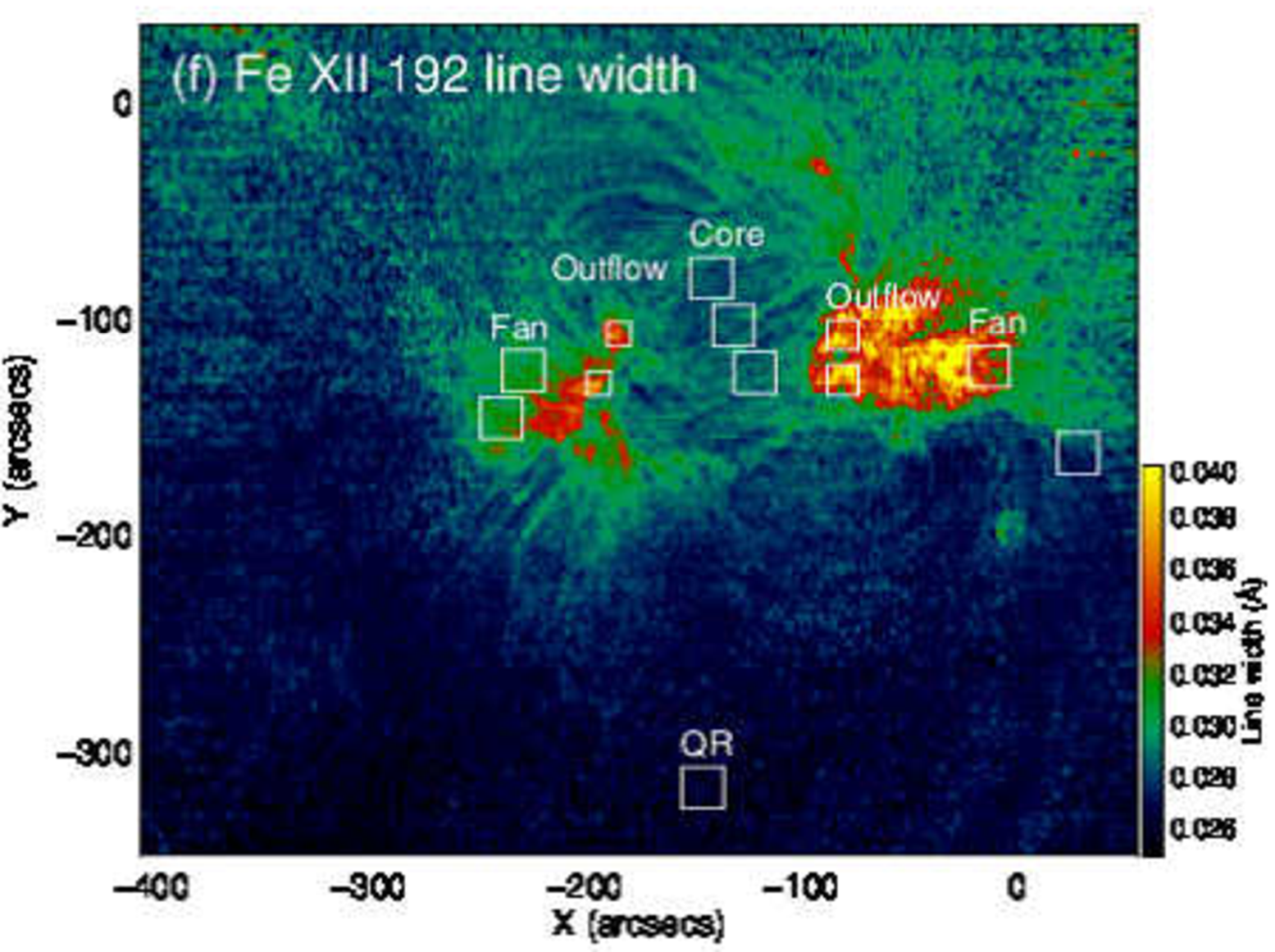}
  \caption{Maps of intensity (\textit{upper}), line centroid (\textit{middle}), and line width (\textit{lower}). \textit{Left} panels show those for (a) Si \textsc{vii} $275.35${\AA} and \textit{right} panels show those for (b) Fe \textsc{xii} $192.39${\AA}.  Boxes in the maps indicate the regions where Doppler velocities were measured in detail: C1--C3: active region core, F1--F4: fan loops, U1--U4: outflow regions, and QR: the quiet region.}
  \label{fig:vel_map_box}
\end{figure}

We focused on four kinds of coronal structure in this chapter: active region core, fan loop, outflow region, and the quiet region.  The active region core was defined by the loop system at the center of the active region which is bright in the high temperature emission (\textit{cf.} panel f in Fig.~\ref{fig:vel_context_ar10978}), which has the size of $\sim 100''$ in the projected plane.  In Fe \textsc{xii} and Si \textsc{vii} intensity map (panel d and e in Fig.~\ref{fig:vel_context_ar10978}), there are no clear loops at the active region core, while patchy structures are seen.  Fan loops are extracted from Si \textsc{vii} image in which they are most distinct.  We define outflow region as the location (1) where the line width of Fe \textsc{xii} $192.39${\AA} is enhanced, and (2) which can be spatially separated from fan loops.  Several regions were extracted from the raster scan which are indicated by white boxes in Fig.~\ref{fig:vel_map_box}.  Names written beside the boxes respectively means C1--C3: active region core, F1--F4: fan loop, and U1--U4: outflow region.  As the quiet region, we selected the region far from the active region as long as possible, which is indicated by the box named QR.  The size of the boxes was chosen so that they fit to the spatial size of target structures (C1--C3, F1--F4, and QR: $20'' \times 20''$, U1 and U2: $12'' \times 12''$, U3--U4: $15'' \times 15''$).  Since fan loops were well developed at the east edge, so we carefully select the location of the outflow regions U1 and U2 in order to avoid the influence of neighboring fan loops.  

% --- End of TeX ---

%% file: tex/tab_vel_emission_lines.tex
\begin{table}
  \centering
  \caption{Emission lines used in this chapter.  A symbol ($^\mathrm{b}$) after wavelength indicates that the emission line is blended by another ion at the active region core. The numbers after temperature represent the full width of half maximum of each contribution function $G (T)$. }
  \label{tab:vel_emission_lines}
  \begin{tabular}{llr}
    \toprule
    \multicolumn{1}{c}{Ion} & 
    \multicolumn{1}{c}{Wavelength} &
    \multicolumn{1}{c}{Temperature} \\
     & 
    \multicolumn{1}{c}{(\AA)} & 
    \multicolumn{1}{c}{($\log_{10} \, \mathrm{K}$)} \\
    \midrule
    Mg \textsc{v}   & $276.58^{\mathrm{b}}$ & $5.50_{-0.14}^{+0.14}$ \\
    Mg \textsc{vi}  & $268.99$ & $5.66_{-0.13}^{+0.13}$ \\
    Mg \textsc{vii} & $278.40^{\mathrm{b}}$ & $5.80_{-0.13}^{+0.13}$ \\
    Si \textsc{vii}  & $275.35$ & $5.80_{-0.13}^{+0.14}$ \\
    Si \textsc{x}  & $258.37$ & $6.15_{-0.11}^{+0.11}$ \\
    S  \textsc{x}  & $264.23$ & $6.18_{-0.13}^{+0.13}$ \\
    S \textsc{xiii}  & $256.69$ & $6.42_{-0.10}^{+0.11}$ \\
    Fe \textsc{viii} & $185.21^{\mathrm{b}}$, $186.60^{\mathrm{b}}$, $194.66$ 
    & $5.69_{-0.17}^{+0.21}$ \\
    Fe \textsc{ix} & $188.49$, $197.86$ & $5.92_{-0.17}^{+0.15}$ \\
    Fe \textsc{x} & $184.54$, $257.26$ & $6.04_{-0.13}^{+0.12}$ \\
    Fe \textsc{xi} & $182.17$, $198.54$ & $6.13_{-0.11}^{+0.11}$ \\
    Fe \textsc{xii} & $192.39$, $193.51$, $195.12^{\mathrm{b}}$, $196.64$ 
    & $6.19_{-0.10}^{+0.10}$ \\
    Fe \textsc{xiii} & $196.54$, $202.04$ & $6.25_{-0.09}^{+0.09}$ \\
    Fe \textsc{xiv} & $264.789$, $274.20$ & $6.30_{-0.09}^{+0.09}$ \\
    Fe \textsc{xv} & $284.163$ & $6.35_{-0.10}^{+0.11}$ \\
    Fe \textsc{xvi} & $262.976$ & $6.45_{-0.13}^{+0.20}$ \\
    \midrule
    Total & $26$ lines & \\
    \bottomrule
  \end{tabular}
\end{table}
%\footnotetext{Blended by Ni XVI $185.230${\AA} at the active region core.}
%\footnotetext{Blended by Fe XII $196.640${\AA}.}

%% file: tex/vel_ar10978_lp.tex
% ===========================
%   Project:
%     T vs. V
%   Contents:
%     Velocity of AR 10978
% ===========================

Before measuring Doppler velocities, we look at each emission line profile.  Even visual inspection of line profiles gives us an insight sometimes better than a fitting result.  We compare line profiles from four locations: the quiet region (QR), fan loops (F1), an active region core (C2), and the outflow region (U3) indicated by white boxes in Fig.~\ref{fig:vel_map_box}.  Maps in the figure show intensity, line centroid, and line width from \textit{upper} to \textit{lower} panels. \textit{Left} and \textit{right} columns respectively show those for Si \textsc{vii} $275.35${\AA} ($\log T \, [\mathrm{K}] = 5.80$) and Fe \textsc{xii} $192.39${\AA} ($\log T \, [\mathrm{K}] = 6.19$).  In the following sections, we show the emission line profiles in an order of the formation temperature and note their characteristics.  The spectra from four regions (C2, F1, U3, and QR) were averaged within each region.  

% --- End of Tex ---

%% file: tex/vel_ar10978_lp_tr.tex
% ===========================
%   Project:
%     T vs. V
%   Contents:
%     Velocity of AR 10978
% ===========================

\begin{figure}
  \centering
  \includegraphics[width=8.4cm,clip]{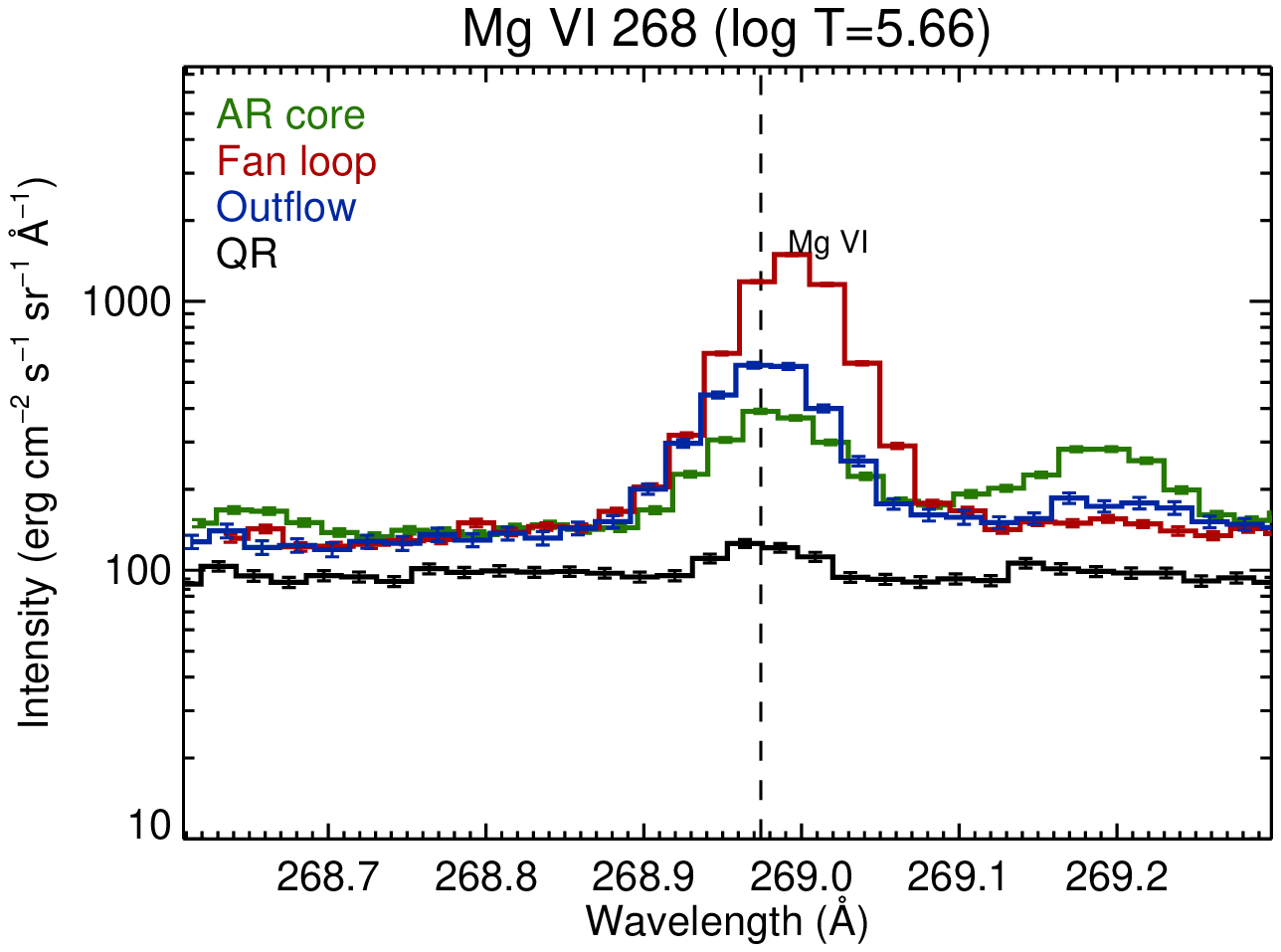}
  \includegraphics[width=8.4cm,clip]{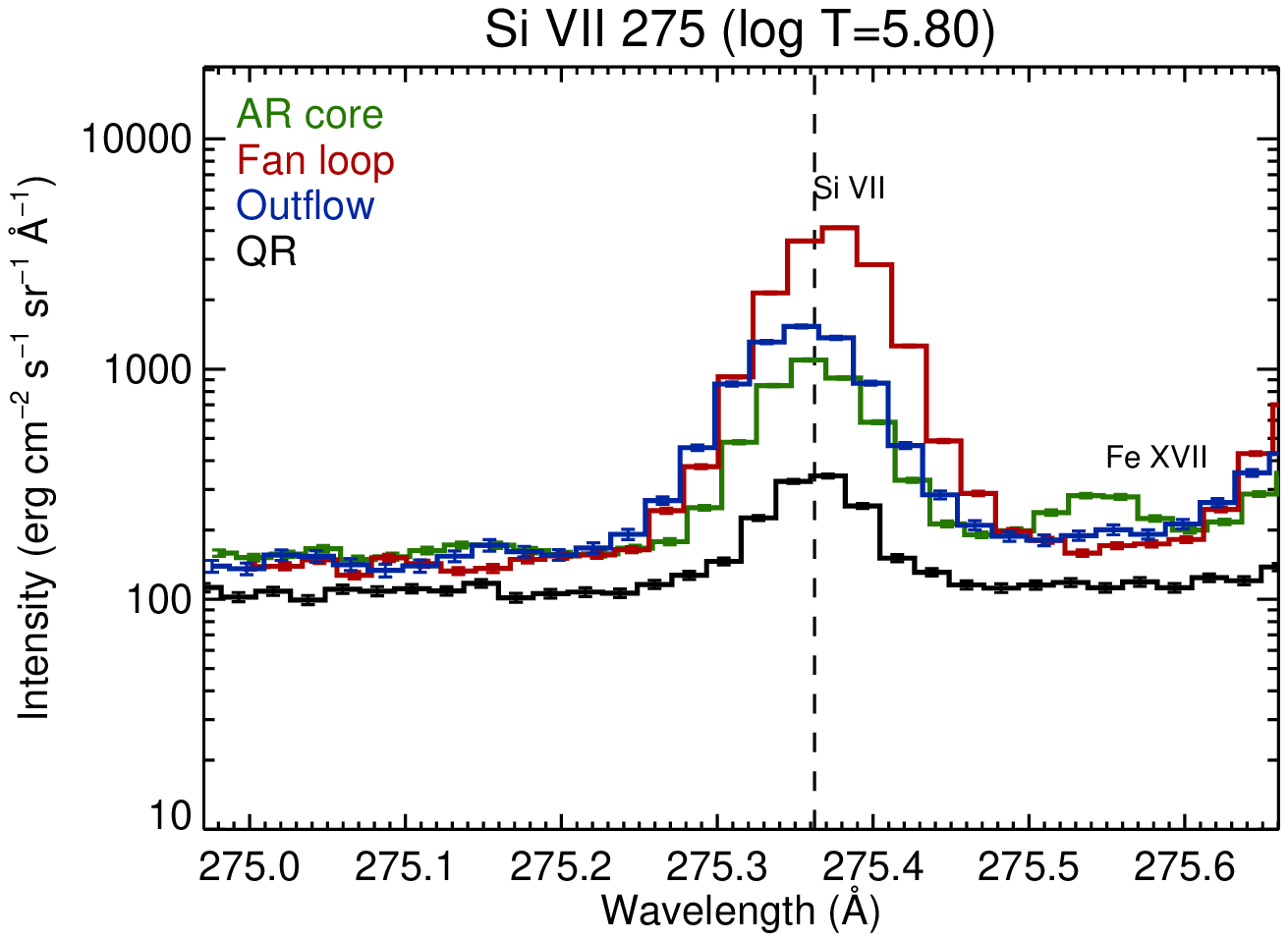}
  \caption{Line profiles at the quiet region (\textit{black}), the outflow region (\textit{blue}), a fan loop (\textit{red}), and the core region (\textit{green}). \textit{Left}: Mg \textsc{vi} $268.99${\AA}. \textit{Right}: Si \textsc{vii} $275.35${\AA}. A vertical dashed line in each panel indicates a line centroid of the emission line at the quiet region.}
  \label{fig:vel_lp_cmprsn_tr}
\end{figure}

Line profiles of two transition region emission lines Mg \textsc{vi} $268.99${\AA} ($\log T \, [\mathrm{K}] = 5.66$) and Si \textsc{vii} $275.35${\AA} ($\log T \, [\mathrm{K}] = 5.80$) are shown in Fig.~\ref{fig:vel_lp_cmprsn_tr}.  \textit{Black}, \textit{blue}, \textit{red}, and \textit{green} histograms respectively indicate the line profile in the quiet region, the outflow region, a fan loop, and the core region which were indicated by white boxes in Fig.~\ref{fig:vel_map_box}. A vertical dashed line in each panel indicates a line centroid of the emission line at the quiet region. The line profiles for fan loops are clearly shifted toward longer wavelength (\textit{i.e.,} redshift) by around $1 \, \mathrm{pix}$ (corresponding to $\simeq 0.022${\AA}) which corresponds to $\sim 20 \, \mathrm{km} \, \mathrm{s}^{-1}$ downward to the solar surface at this wavelength. On the other hand, the line profiles in the outflow region are blueshifted by $\simeq 0.5 \, \mathrm{pix}$. The line profiles look like almost symmetric in all regions. It should be noted that those in the outflow region are wider than those in other regions. 

% --- End of Tex ---

%% file: tex/vel_ar10978_lp_ct.tex
% ===========================
%   Project:
%     T vs. V
%   Contents:
%     Velocity of AR 10978
% ===========================

\begin{figure}
  \centering
  \includegraphics[width=8.4cm,clip]{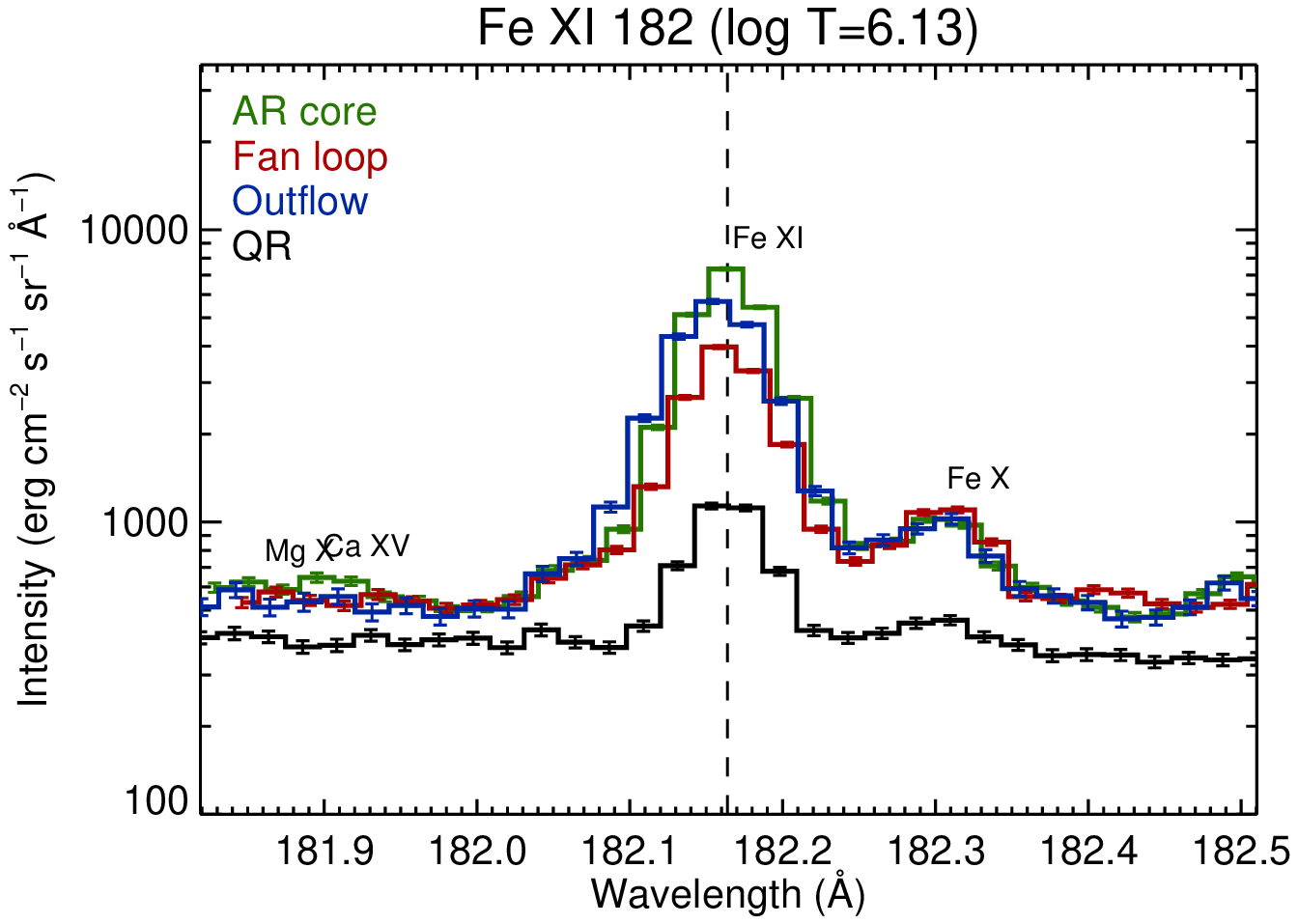}
  \includegraphics[width=8.4cm,clip]{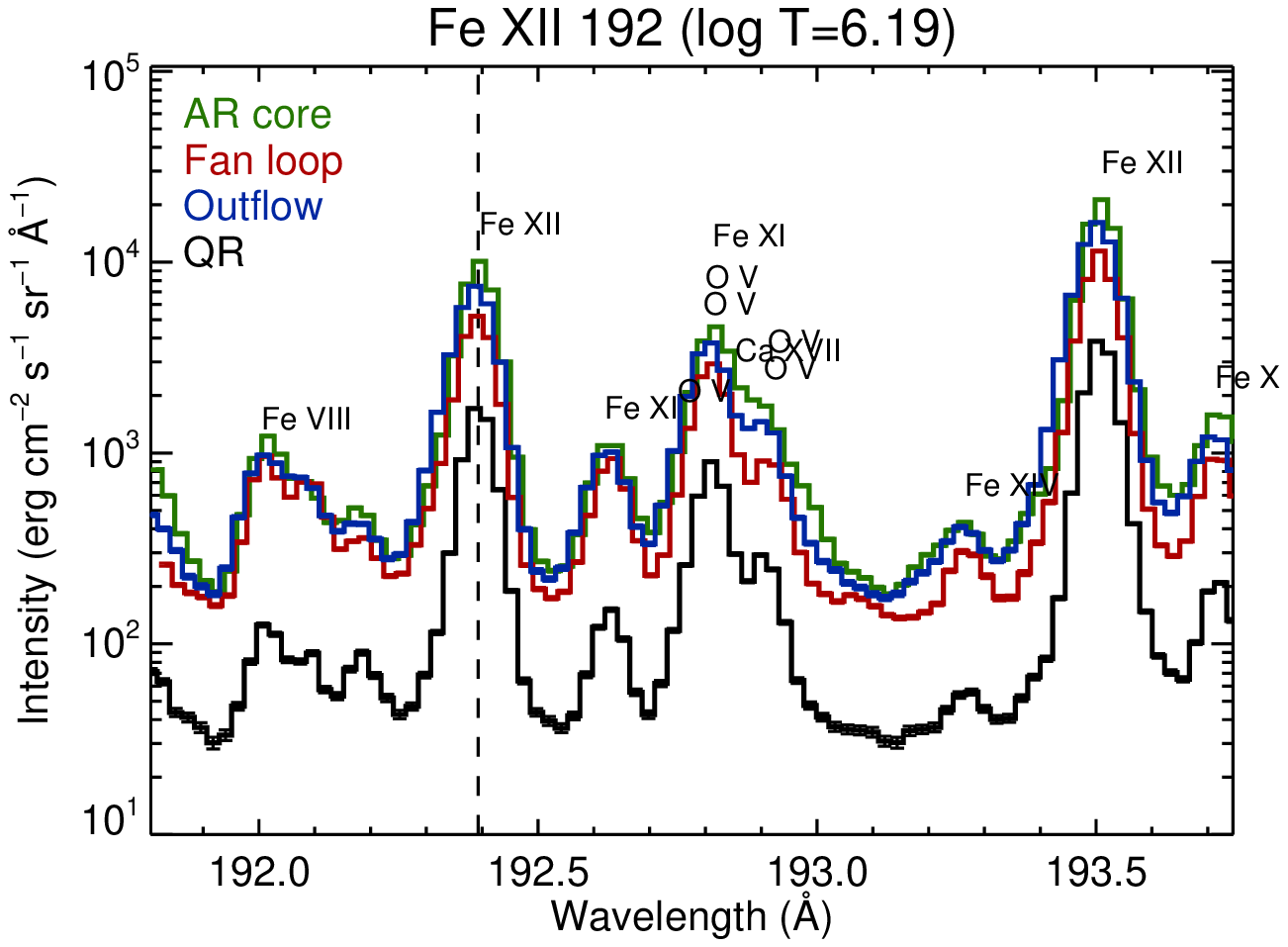}
  \caption{Line profiles at the quiet region (\textit{black}), the outflow region (\textit{blue}), fan loop (\textit{red}), and the core region (\textit{green}).  \textit{Left}: Fe \textsc{xi} $182.17${\AA}.  \textit{Right}: Fe \textsc{xii} $192.39${\AA}.  A vertical dashed line in each panel indicates a line centroid of the emission line at the quiet region.}
  \label{fig:vel_lp_cmprsn_ct}
\end{figure}

Line profiles of two coronal emission lines Fe \textsc{xi} $182.17${\AA} ($\log T \, [\mathrm{K}]=6.13$) and Fe \textsc{xii} $192.39${\AA} ($\log T \, [\mathrm{K}]=6.19$) are shown in Fig.~\ref{fig:vel_lp_cmprsn_ct}.  Different from the transition region emission lines, they exhibit an enhanced component at their blue wing in the outflow region.  This enhancement can be seen up to $4$--$5$ pixels far from the line centroid, which indicates that the upflow with a speed up to $100 \, \mathrm{km} \, \mathrm{s}^{-1}$ exists, while major portion of the emission is located near the same centroid as the quiet region where the plasma moves only by $\lesssim 10 \, \mathrm{km} \, \mathrm{s}$.  Note that the line profiles look symmetric in the quiet region. 

The enhanced blue wing is more clearly seen in Fe \textsc{xii} than in Fe \textsc{xi}, which means that the upflow dominantly consists of plasma with a temperature higher than the formation temperature of Fe \textsc{xi}. Another noticeable feature is that the enhanced component in Fe \textsc{xii} has even the same magnitude as the intensity in the core region at around $\lambda = 192.30${\AA}. We can see the similar behavior also for Fe \textsc{xii} $193.51${\AA} included in this spectral window.

% --- End of Tex ---

%% file: tex/vel_ar10978_lp_im.tex
% ===========================
%   Project:
%     T vs. V
%   Contents:
%     Velocity of AR 10978
% ===========================

\begin{figure}
  \centering
  \includegraphics[width=8.4cm,clip]{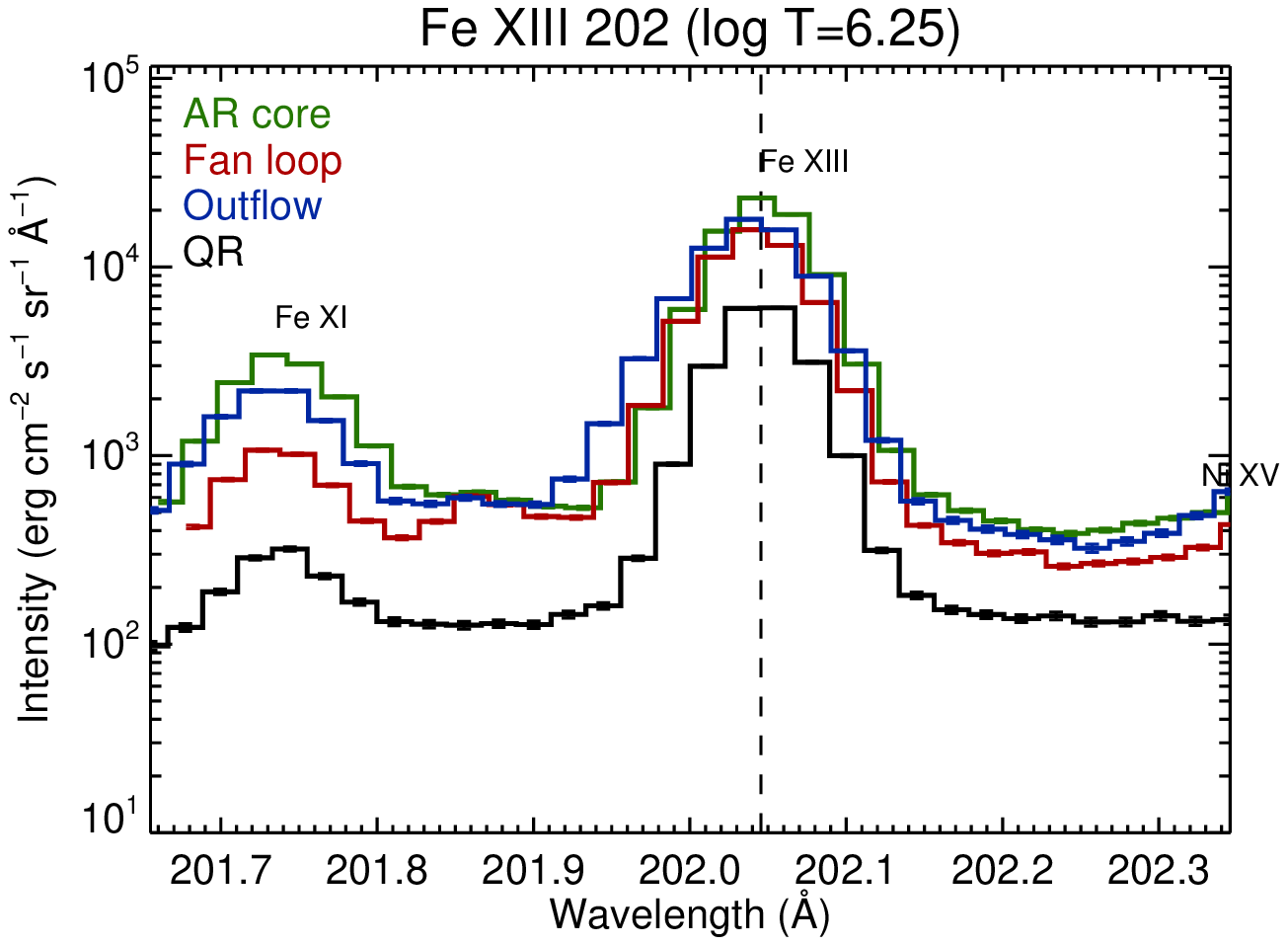}
  \includegraphics[width=8.4cm,clip]{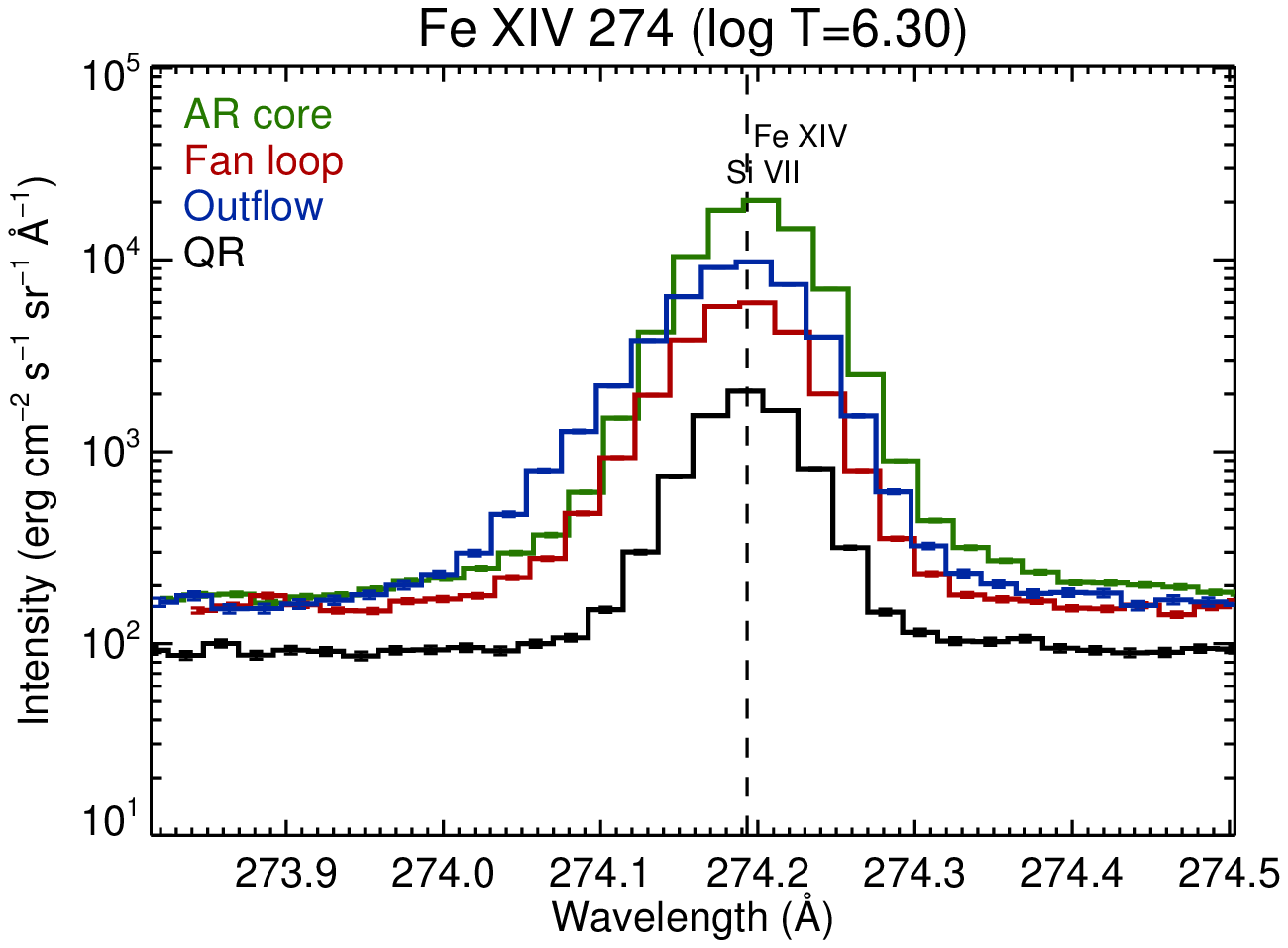}
  \caption{Line profiles at the quiet region (\textit{black}), the outflow region (\textit{blue}), fan loop (\textit{red}), and the core region (\textit{green}).  \textit{Left}: Fe \textsc{xiii} $202.04${\AA}.  \textit{Right}: Fe \textsc{xiv} $274.20${\AA}.  A vertical dashed line in each panel indicates a line centroid of the emission line at the quiet region.}
  \label{fig:vel_lp_cmprsn_im}
\end{figure}

Fig.~\ref{fig:vel_lp_cmprsn_im} shows line profiles for Fe \textsc{xiii} $202.04${\AA} ($\log T \, [\mathrm{K}]=6.25$) and Fe \textsc{xiv} $274.20${\AA} ($\log T \, [\mathrm{K}] = 6.30$).  While the line profiles in the quiet region (\textit{black}), a fan loop (\textit{red}), and the core region (\textit{green}) all look symmetric and their centroid position do not deviate significantly from the value of the quiet region indicated by a vertical dashed line, those in the outflow region (\textit{blue}) exhibit a significant enhancement in their blue wing (at $\lambda=201.85$--$201.95${\AA} for Fe \textsc{xiii} and at $\lambda=273.95$--$274.05${\AA} for Fe \textsc{xiv}) similar to Fe \textsc{xi}--\textsc{xii}.  Note that major portion of the emission is again located near the line centroid in the quiet region (\textit{i.e.,} at the vertical dashed line). 

% --- End of Tex ---

%% file: tex/vel_ar10978_lp_ht.tex
% ===========================
%   Project:
%     T vs. V
%   Contents:
%     Velocity of AR 10978
% ===========================

\begin{figure}
  \centering
  \includegraphics[width=8.4cm,clip]{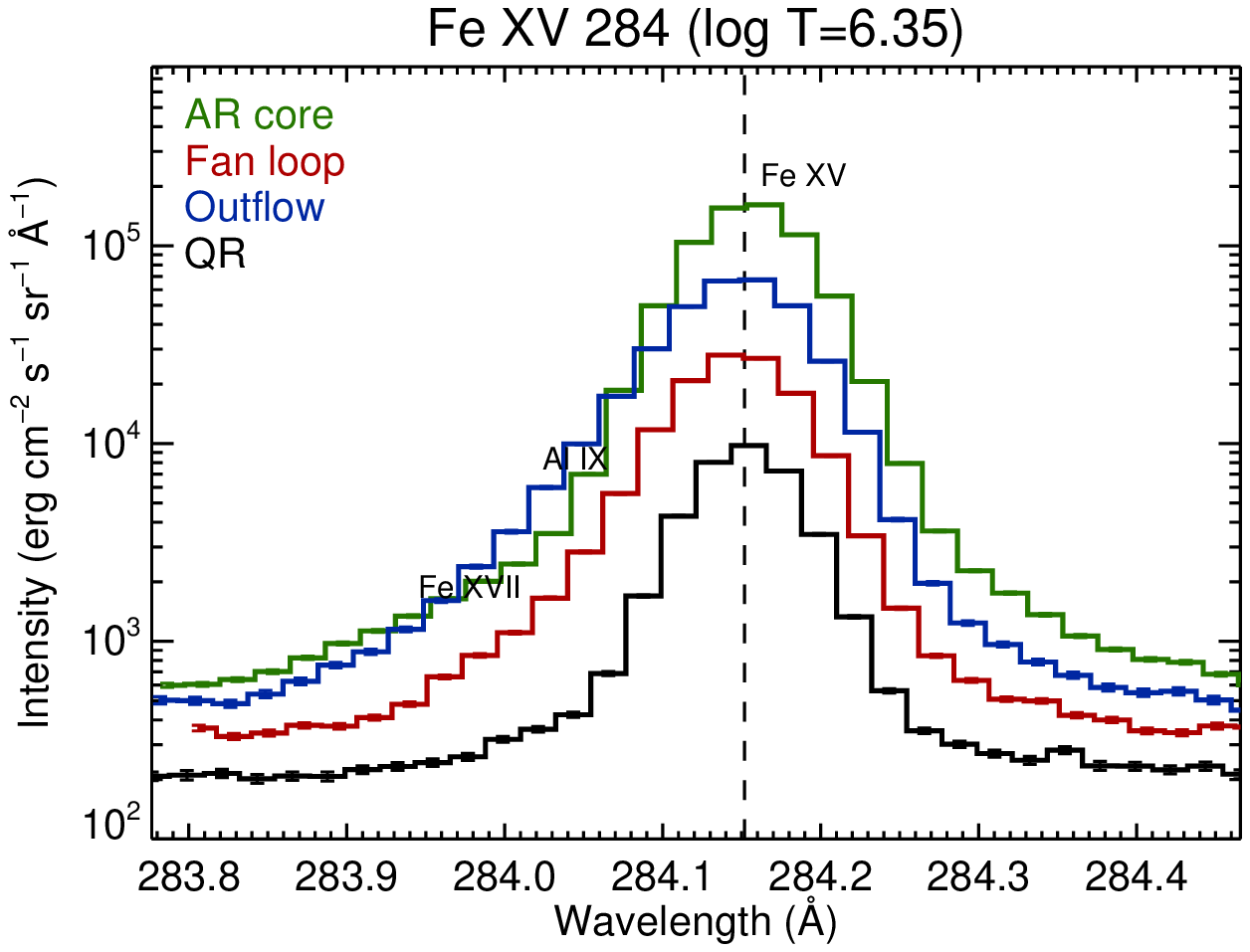}
  \includegraphics[width=8.4cm,clip]{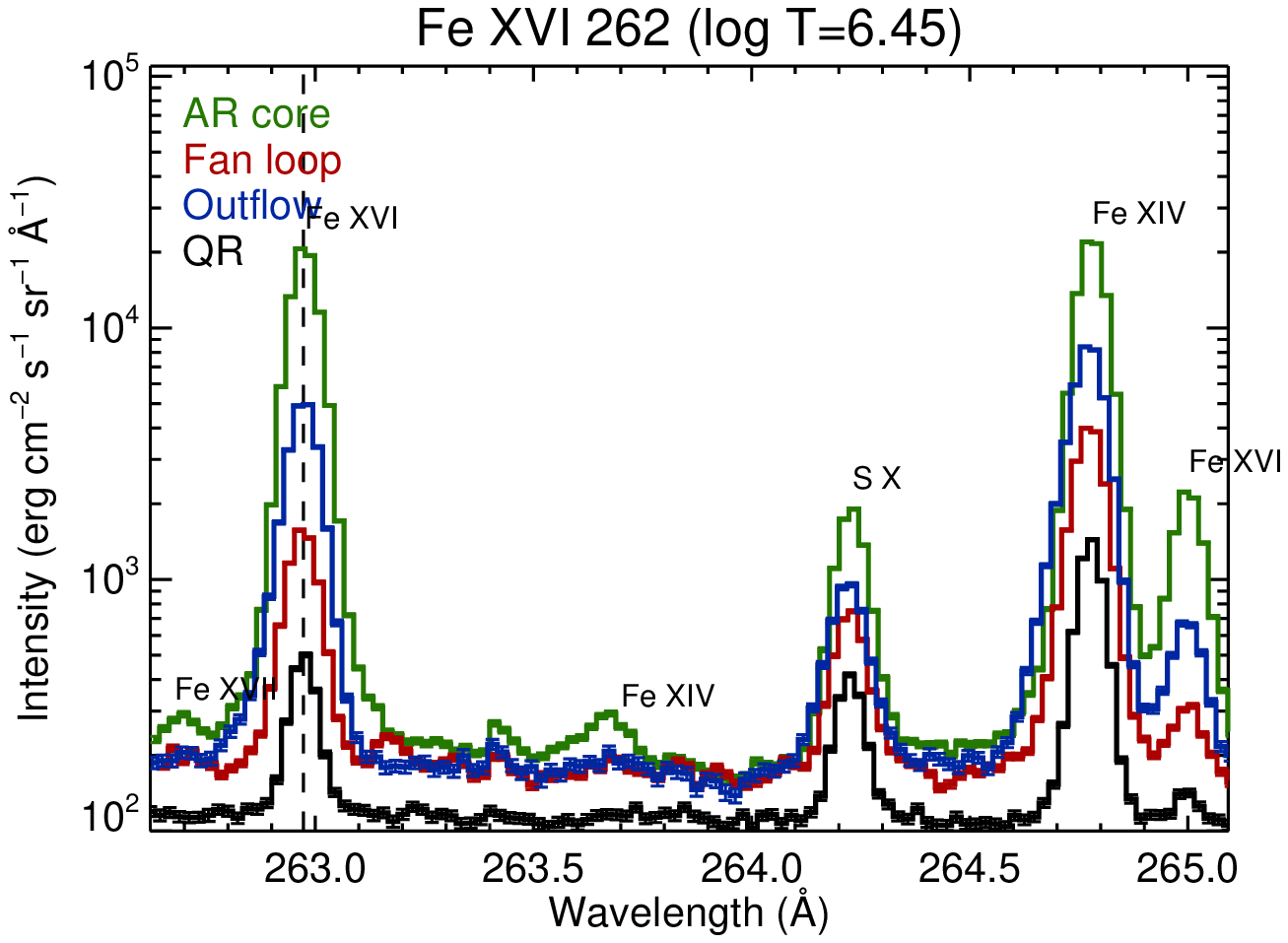}
  \caption{Line profiles at the quiet region (\textit{black}), the outflow region (\textit{blue}), fan loop (\textit{red}), and the core region (\textit{green}). \textit{Left}: Fe \textsc{xv} $284.16${\AA}. \textit{Right}: Fe \textsc{xvi} $262.98${\AA}. A vertical dashed line in each panel indicates a line centroid of the emission line at the quiet region.}
  \label{fig:vel_lp_cmprsn_ht}
\end{figure}

The hottest coronal plasma seen in the non-flare condition reaches around the temperature of $\log T \, [\mathrm{K}] \simeq 6.5$, which can be observed by Fe \textsc{xv} $284.16${\AA} ($\log T \, [\mathrm{K}]=6.35$) and Fe \textsc{xvi} $262.98${\AA} ($\log T \, [\mathrm{K}]=6.45$). Their line profiles are shown in Fig.~\ref{fig:vel_lp_cmprsn_ht}. The line profiles in the quiet region, a fan loop, and the core region respectively indicated by \textit{black, red,} and \textit{green} spectrum seem to be symmetric as same as those from lower temperature. Though an enhancement at the blue wing indeed exists around $\lambda =283.85$--$284.00${\AA} in the Fe \textsc{xv} emission line in the outflow region (\textit{blue}), that does not exceed the intensity of the core region (\textit{green}), which differs from the case of Fe \textsc{xii}--\textsc{xiv} ($\log T \, [\mathrm{K}]=6.2$--$6.3$). 

Another important feature to be mentioned here is that the enhancement at the blue wing does not exist in the line profile of Fe \textsc{xvi} in the outflow region (\textit{blue}). These indicates that the plasma producing the enhancement at blue wings in the line profiles have a temperature lower than $\log T \, [\mathrm{K}] \simeq 6.4$. Though not shown in the figure, S \textsc{xiii} $256.69${\AA} ($\log T \, [\mathrm{K}]=6.42$) also indicates the same characteristic as Fe \textsc{xvi} (\textit{i.e.}, no enhancement at its blue wing). 

% --- End of Tex ---

%% file: tex/vel_ar10978_vel_map.tex
% ===================================
%   Project:
%     T vs. V
%   Description:
%     Velocity maps.
% ===================================

\begin{figure}
  \centering
  \includegraphics[width=8.4cm,clip]{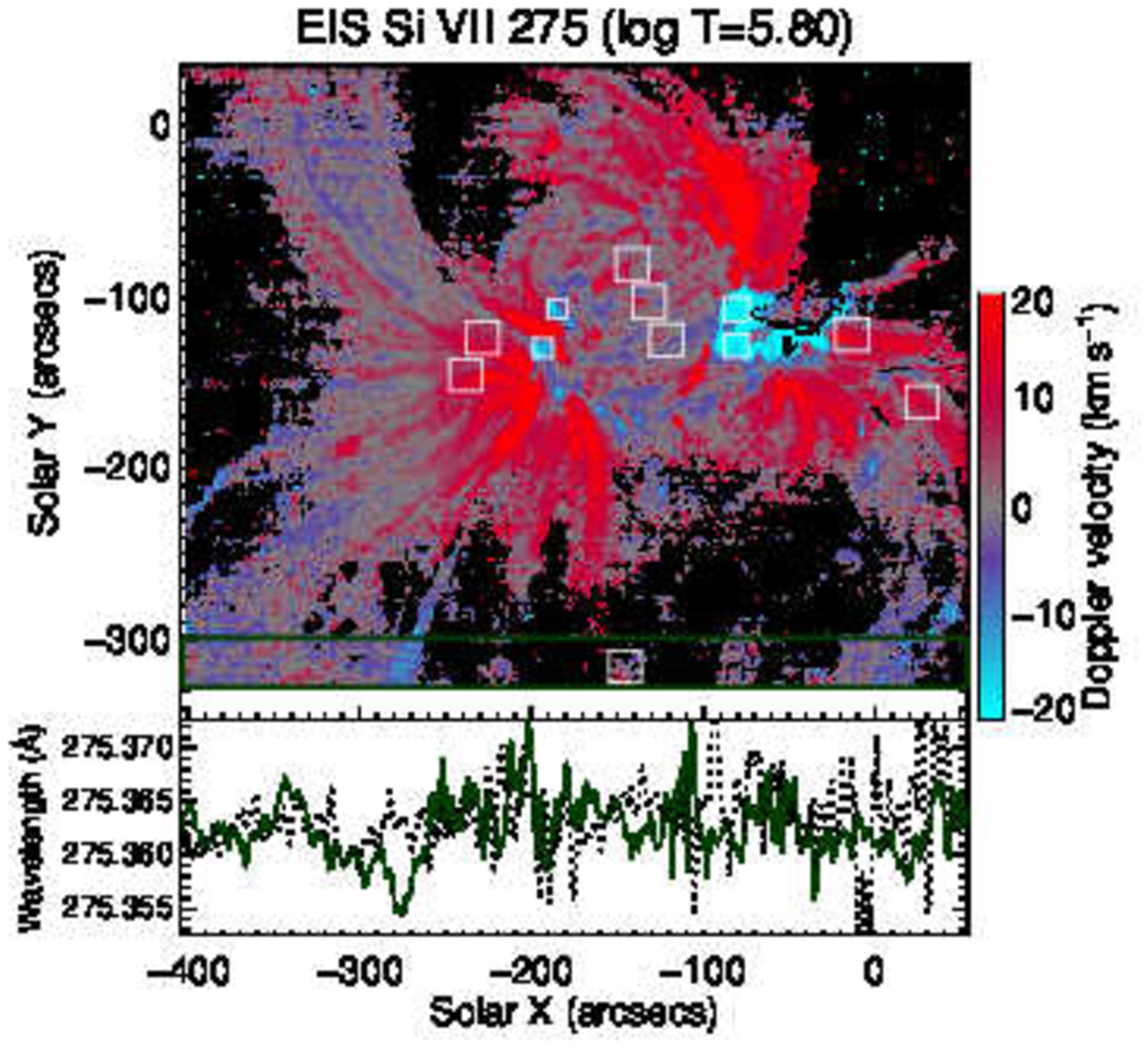}
  \includegraphics[width=8.4cm,clip]{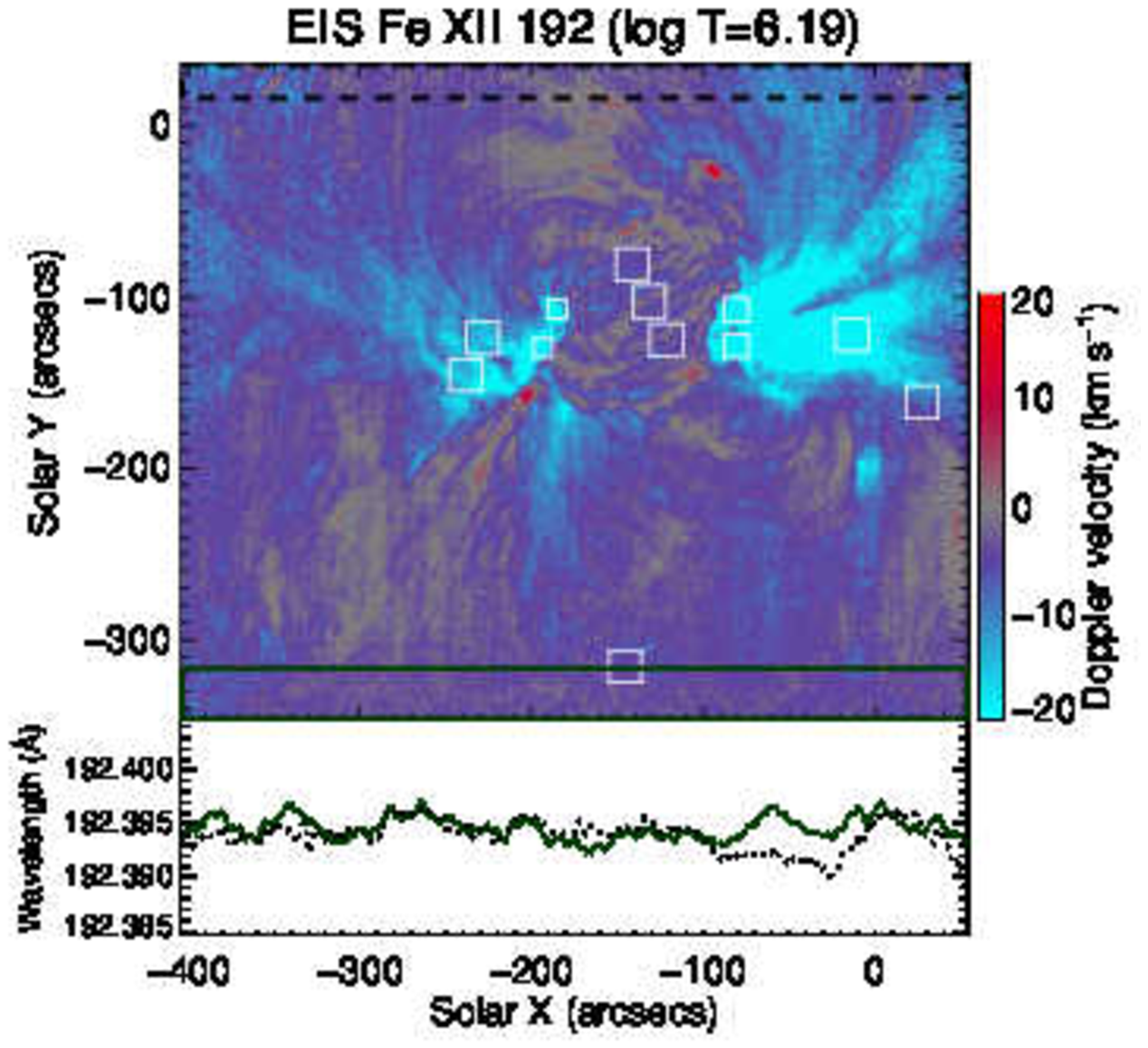}
  \includegraphics[width=8.4cm,clip]{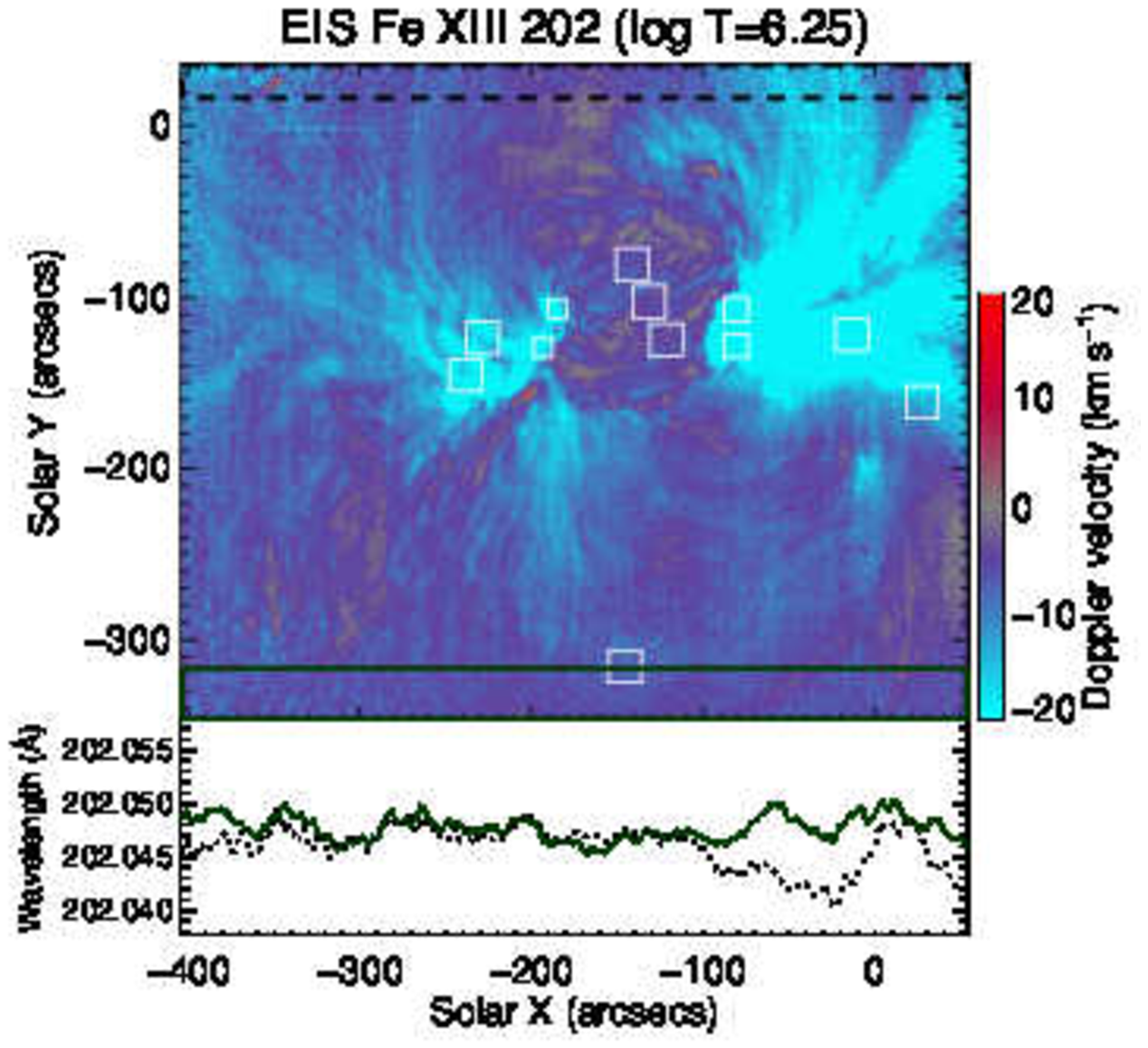}
  \includegraphics[width=8.4cm,clip]{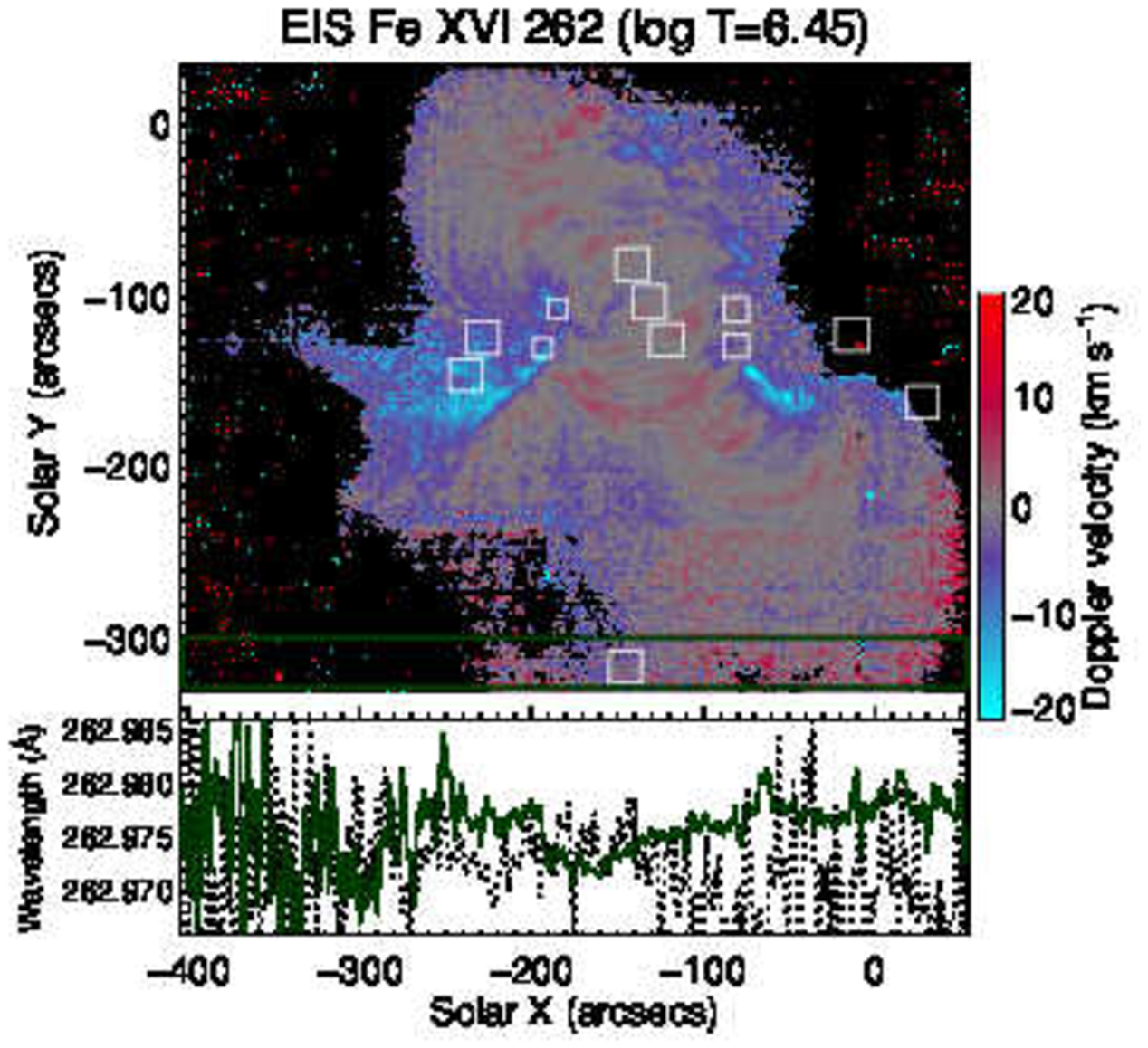}
  \caption{Doppler velocity maps of AR10978. 
    \textit{Left upper}: Si \textsc{vii} $275.35${\AA}. 
    \textit{Right upper}: Fe \textsc{xii} $192.39${\AA}. 
    \textit{Left lower}: Fe \textsc{xiii} $202.04${\AA}. 
    \textit{Right lower}: Fe \textsc{xvi} $262.98${\AA}.
  }
  \label{fig:vel_ar10978_vel_map}
\end{figure}

We calculated the Doppler velocity $v_{\mathrm{Dop}}$ by using the equation
\begin{equation}
  v_{\mathrm{Dop}} = \dfrac{\lambda_{\mathrm{c}} - \lambda_{0}}{\lambda_{0}} c \, \text{,}
\end{equation}
where $\lambda_{\mathrm{c}}$ is the line centroid, $\lambda_{0}$ is a rest wavelength, and $c$ is the speed of the light.  Here the fitting result by a single Gaussian was used for $\lambda_{\mathrm{c}}$.  This means that the Doppler velocity derived for the outflow region could be a weighted average of multiple components in the line profile, which leads to the speed of the outflow itself underestimated in the absolute value because there is a major component with much little shift.  The deviation from the single Gaussian fitting for a coronal emission line is described in Appendix \ref{sect:vel_res}.

The Doppler velocity map for Si \textsc{vii} $275.35${\AA} ($\log T \, [\mathrm{K}]=5.80$), Fe \textsc{xii} $192.39${\AA} ($\log T \, [\mathrm{K}] = 6.19$), Fe \textsc{xiii} $202.04${\AA} ($\log T \, [\mathrm{K}]=6.25$), and Fe \textsc{xvi} $262.98${\AA} ($\log T \, [\mathrm{K}]=6.45$) are shown in Fig.~\ref{fig:vel_ar10978_vel_map}.  The line profiles were spatially integrated by $3 \times 3 \, \mathrm{pix}^{2}$ before the single Gaussian fitting for Fe emission lines.  The regions where the statistical error originated in the photon noise exceeds $5 \, \mathrm{km} \, \mathrm{s}^{-1}$ are painted by \textit{black}. Note that we do not refer to Fe \textsc{xii} $195.12${\AA}, which is commonly used in the literature, due to the fact that the emission line is contributed by the neighboring Fe \textsc{xii} $195.18${\AA} at the region where the electron density becomes high (\textit{e.g.}, active region core), and it produces a fake redshift. 

The plot below each map in Fig.~\ref{fig:vel_ar10978_vel_map} shows the line centroids averaged in $y$ direction within squares indicated by \textit{green thick} line and \textit{black dashed} line in Doppler velocity maps.  At some locations, the line centroids are shifted coherently by several $\mathrm{m}${\AA} in these two regions which may be a residual of the orbital variation.  In order to compensate those shifts, we subtracted the 5-exposures running average (\textit{i.e.}, in the solar $x$ direction) of the centroid of Fe \textsc{xii} $192.39${\AA} indicated by the \textit{green thick} line in the lower plot.  This emission line was strong enough to obtain precise centroid ($\sigma \leq 1 \, \mathrm{km} \, \mathrm{s}^{-1}$) in the quiet region with spatial average of $3 \times 3 \, \mathrm{pix}^{2}$.  As far as the orbital variation dominantly comes from the spatial displacement of the grating component in the EIS instrument, all emission lines should be shifted by the same amounts.  It might be too much subtraction because this process could also remove the real fluctuation in the quiet region, so we set the window size of $5$ exposures in order not to compensate the statistical fluctuation of that size at least.

The references of the Doppler velocity were set to $v_{\mathrm{QR}}=0.2 \, \mathrm{km} \, \mathrm{s}^{-1}$ for Si \textsc{vii}, $v_{\mathrm{QR}}= - 4.3 \, \mathrm{km} \, \mathrm{s}^{-1}$ for Fe \textsc{xii}, $v_{\mathrm{QR}} = - 6.3 \,\mathrm{km} \, \mathrm{s}^{-1}$ for Fe \textsc{xiii} which were obtained by the procedures in Chapter \ref{chap:cal}.  By adjusting the average Doppler velocity within the region indicated by a green dashed box in the map to be $v_{\mathrm{QR}}$, we have obtained the Doppler velocity map.  The map for Fe \textsc{xii} is obviously different from those derived by setting the average Doppler velocity within in the map (or the quiet region) to be zero as in some previous studies, which of course become a mixture pattern of red and blue.  Note that since we could not measure the average Doppler velocity of the quiet region for $\log T \, [\mathrm{K}] \geq 6.3$ in Chapter \ref{chap:cal}, the average Doppler velocity in the green dashed box was set to zero for Fe \textsc{xvi}, which means that the Doppler velocity for Fe \textsc{xvi} is a relative quantity to that of the quiet region.  

Doppler velocity maps in Fig.~\ref{fig:vel_ar10978_vel_map} show some characteristic patterns.  Fan loops extended from both edges of the active region core are clearly redshifted by around $20 \, \mathrm{km} \, \mathrm{s}^{-1}$ as seen in the Si \textsc{vii} map.  We can see several filamentary structures at around $(-250'', -180'')$ and $(30'', -150'')$.  On the other hand, the outflow regions at $(-80'', -120'')$ and around $(-200'', -150'')$ are blueshifted by $v = - 20 \, \mathrm{km} \, \mathrm{s}^{-1}$.  Doppler velocity maps for coronal emission lines (\textit{i.e.}, Fe \textsc{xii}, \textsc{xiii}, and \textsc{xvi}) show quite different appearance from that for Si \textsc{vii}.  The locations corresponding to fan loops are blueshifted by larger than $20 \, \mathrm{km} \, \mathrm{s}^{-1}$.  This is common properties within coronal emission lines.  In contrast to the map for Si \textsc{vii}, the boundary between fan loops and the outflow region can not distinguished from the Doppler velocity. 

% --- End of TeX ---

%% file: tex/vel_ar10978_hist_vel.tex
% ===================================================
%   Project:
%     T vs. V
%   Contents:
%     Distribution of Doppler velocity of AR 10978
% ===================================================

The histograms of the Doppler velocities will be shown here in order to study their characteristics more quantitatively. We show the histograms for the same emission lines as described in Section \ref{sect:vel_lp} which covers wide temperature range of $\log T \, [\mathrm{K}]=5.7$--$6.5$. 

% --- End of Tex ---

%% file: tex/vel_ar10978_hist_vel_tr.tex
% ===================================================
%   Project:
%     T vs. V
%   Contents:
%     Distribution of Doppler velocity of AR 10978
% ===================================================

\begin{figure}
  \centering
  \includegraphics[width=8.4cm,clip]{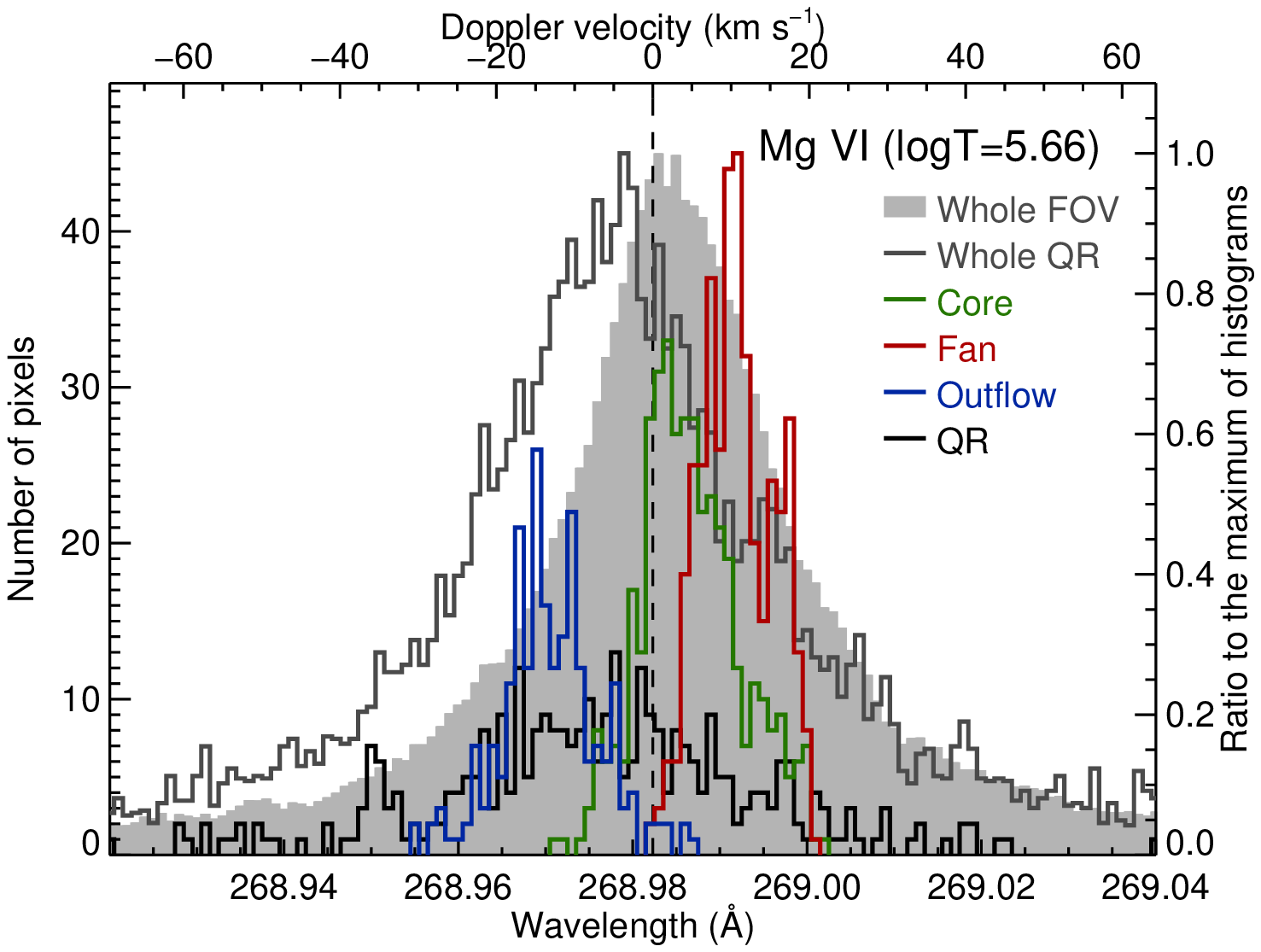}
  \includegraphics[width=8.4cm,clip]{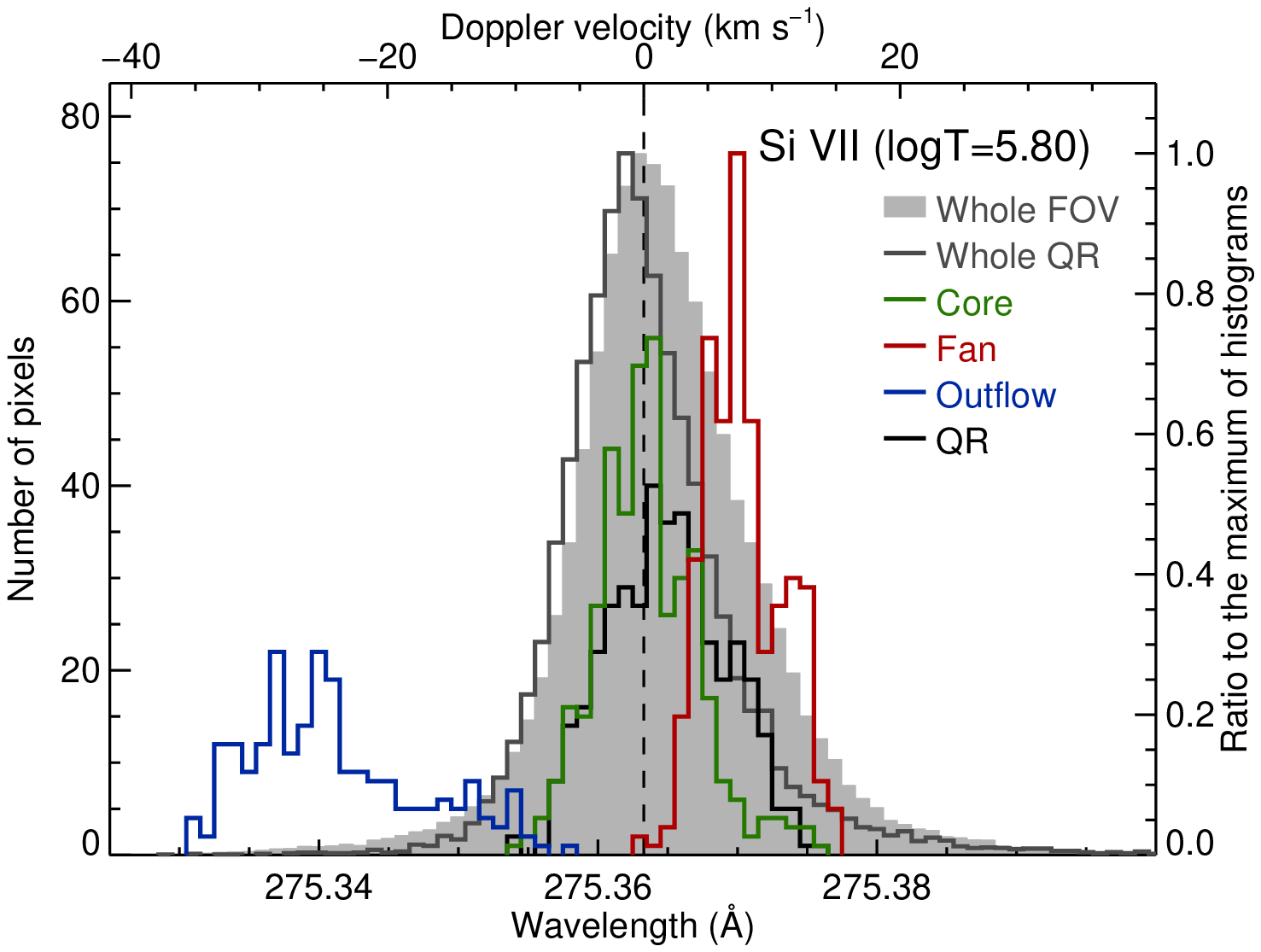}
  \caption{Histogram of line centroid/Doppler velocity at the quiet region (\textit{black}), the outflow region (\textit{blue}), fan loop (\textit{red}), and the core region (\textit{green}).  A \textit{gray solid} histogram is for the entire quiet region (defined as lower $30$ rows $[\mathrm{pixels}]$ in the map).  The histogram filled with \textit{gray} indicates that for the whole map.  \textit{Left}: Mg \textsc{vi} $268.99${\AA}.  \textit{Right}: Si \textsc{vii} $275.35${\AA}.  A vertical dashed line in each panel indicates zero point of Doppler velocity which was calculated so as the quiet region has the average Doppler velocity determined in Chapter \ref{chap:cal}.}
  \label{fig:vel_ar10978_hist_vel_tr}
\end{figure}

Histograms for Mg \textsc{vi} $268.99${\AA} ($\log T \, [\mathrm{K}]=5.66$) and Si \textsc{vii} $275.35${\AA} ($\log T \, [\mathrm{K}]=5.80$) are shown in Fig.~\ref{fig:vel_ar10978_hist_vel_tr}.  \textit{Black}, \textit{blue}, \textit{red}, and \textit{green} histograms respectively indicate the result from the quiet region, the outflow region, the fan loop, and the core region within white boxes C2, F1, U3, and QR in Fig.~\ref{fig:vel_map_box}.  A \textit{Gray solid} histogram is for the entire quiet region defined as lower $30$ rows $[\mathrm{pixel}]$ in the map.  The histogram filled with \textit{gray} indicates that for the whole map.  These two histograms were normalized by the maximum value of other four histograms.  A vertical dashed line in each panel indicates zero point of Doppler velocity which was determined so that the average Doppler velocity of the quiet region has the average Doppler velocity obtained in Chapter \ref{chap:cal}. 

The histogram for Mg \textsc{vi} in the \textit{left} panel of Fig.~\ref{fig:vel_ar10978_hist_vel_tr} shows that the Doppler velocity in the outflow region reaches $-20 \, \text{--} \, {-10} \, \mathrm{km} \, \mathrm{s}^{-1}$, while that in the fan loop indicates $10 \text{--} 20 \, \mathrm{km} \, \mathrm{s}^{-1}$ in reversal.  The core region shows almost no velocity.  The \textit{right} panel shows the histogram for Si \textsc{vii}.  It basically has the same behavior for the selected region as that for Mg \textsc{vi}, but more clearly.  The outflow region exhibits the upward speed of $\gtrsim 20 \, \mathrm{km} \, \mathrm{s}^{-1}$ and the histogram for this region significantly deviates from that of other regions.  On the other hand, the fan loop shows the downward speed around $10 \, \mathrm{km} \, \mathrm{s}^{-1}$.  The histogram for the core region is located at almost the same position as that for the quiet region.

% --- End of Tex ---

%% file: tex/vel_ar10978_hist_vel_ct.tex
% ===================================================
%   Project:
%     T vs. V
%   Contents:
%     Distribution of Doppler velocity of AR 10978
% ===================================================

\begin{figure}
  \centering
  \includegraphics[width=8.4cm,clip]{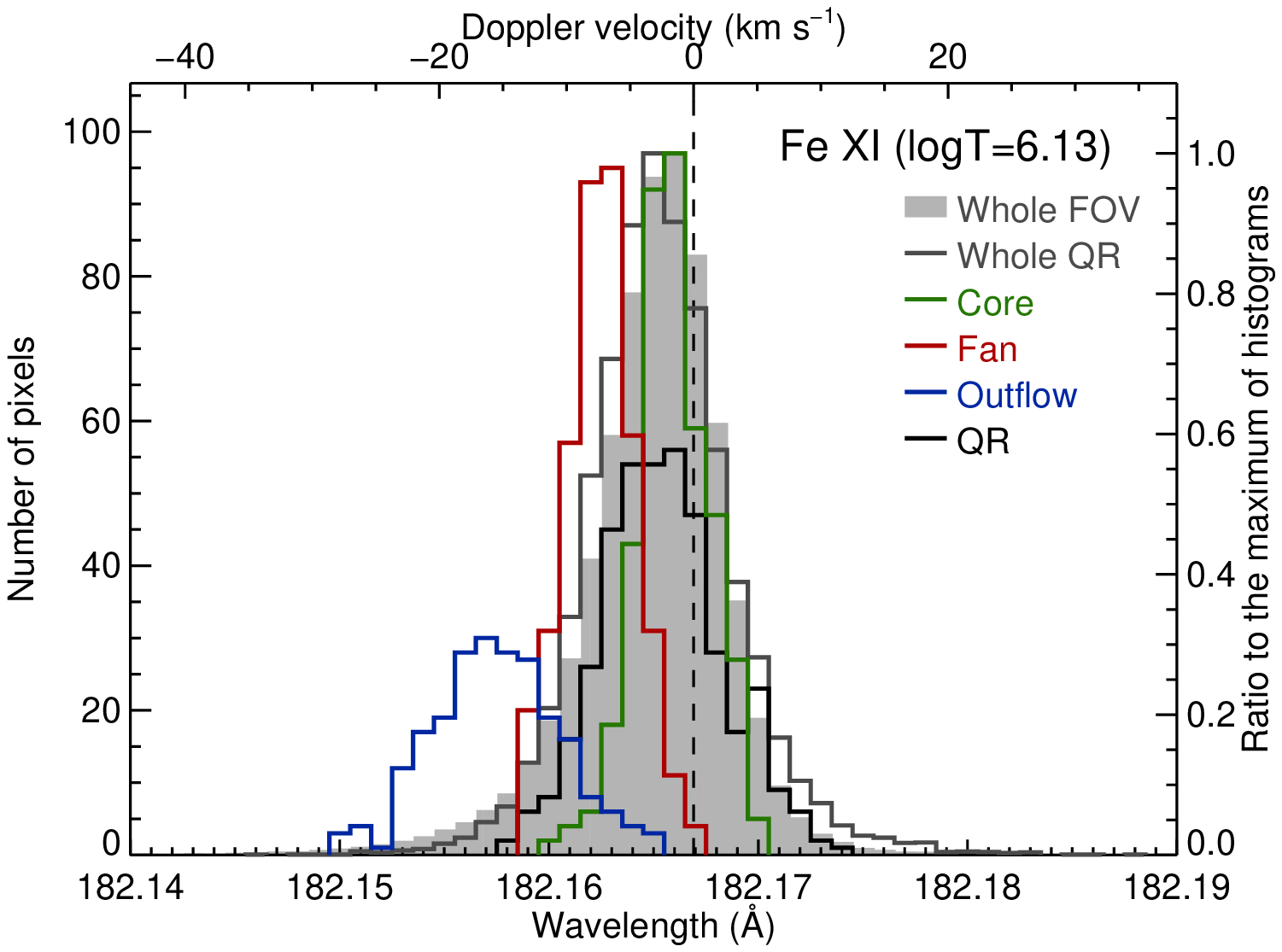}
  \includegraphics[width=8.4cm,clip]{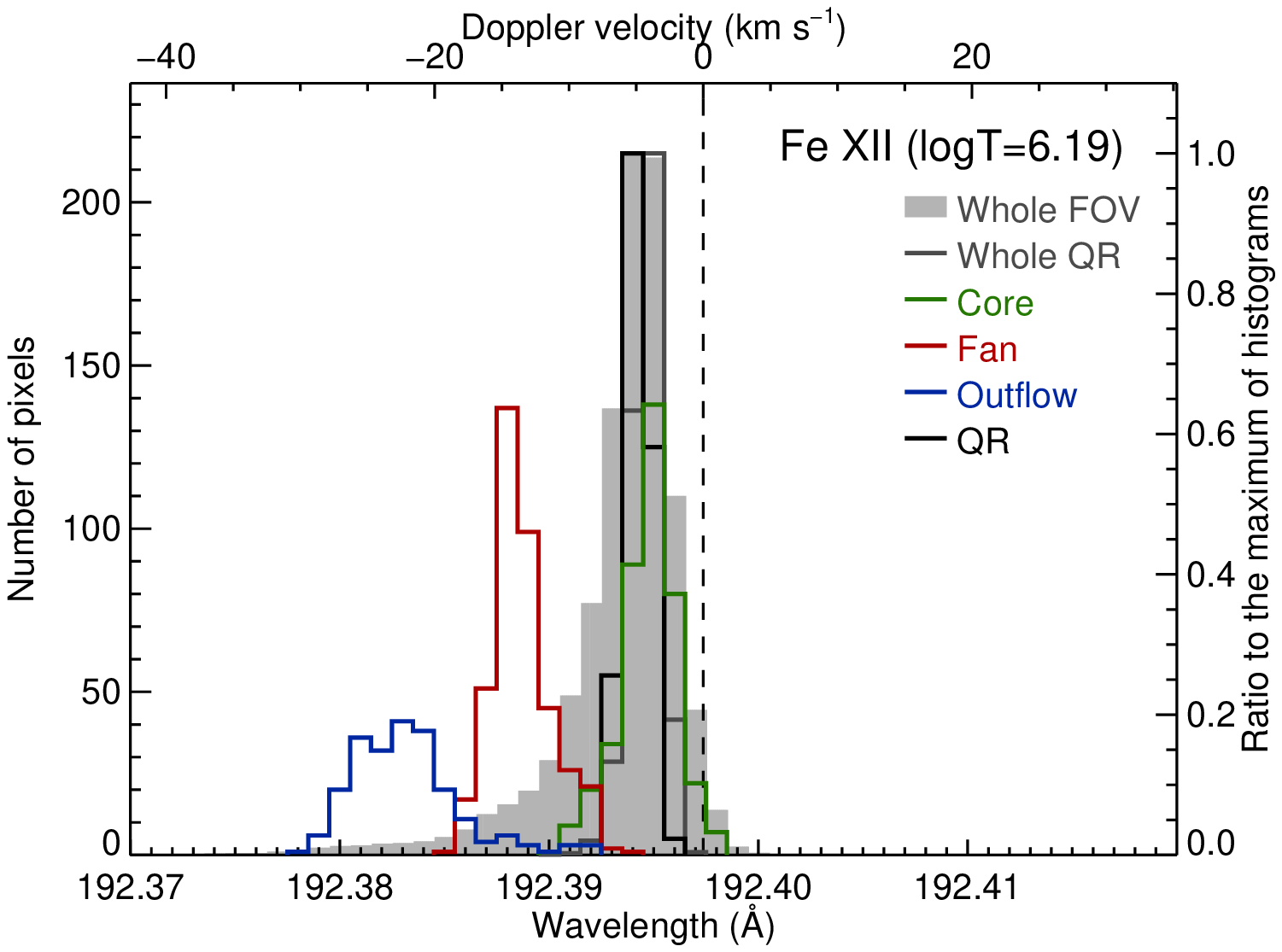}
  \caption{Histogram of line centroid/Doppler velocity at the quiet region (\textit{black}), the outflow region (\textit{blue}), fan loop (\textit{red}), and the core region (\textit{green}).  A \textit{Gray solid} histogram is for the quiet region (defined as lower $30$ pixels in the map).  The histogram filled with \textit{gray} indicates that for the whole map.  \textit{Left}: Fe \textsc{xi} $182.17${\AA}.  \textit{Right}: Fe \textsc{xii} $192.39${\AA}.  A vertical dashed line in each panel indicates zero point of Doppler velocity which was calculated so as the quiet region has the average Doppler velocity determined in Chapter \ref{chap:cal}.}
  \label{fig:vel_ar10978_hist_vel_ct}
\end{figure}

Fig.~\ref{fig:vel_ar10978_hist_vel_ct} shows histograms for Fe \textsc{xi} $182.17${\AA} and Fe \textsc{xii} $192.39${\AA} whose formation temperature is around $\log T \, [\mathrm{K}]=6.1$--$6.2$.  While the Doppler velocity for Fe \textsc{xi} is $-10 \, \text{--} \, {-5} \, \mathrm{km} \, \mathrm{s}^{-1}$ in the fan loops, that for Fe \textsc{xii} exceeds $-10 \, \mathrm{km} \, \mathrm{s}^{-1}$.  Though not shown in the figure, a neighbor emission line Fe \textsc{xii} $193.51${\AA} also shows the same behavior.  

Both emission lines exhibit the Doppler velocity of $\simeq - 20 \, \mathrm{km} \, \mathrm{s}^{-1}$ (\textit{i.e.}, upward) in the outflow region, and the speed is larger for Fe \textsc{xii} than for Fe \textsc{xi} (\textit{i.e.,} higher formation temperature) by roughly $5 \, \mathrm{km} \, \mathrm{s}^{-1}$.  

% --- End of Tex ---

%% file: tex/vel_ar10978_hist_vel_im.tex
% ===================================================
%   Project:
%     T vs. V
%   Contents:
%     Distribution of Doppler velocity of AR 10978
% ===================================================

\begin{figure}
  \centering
  \includegraphics[width=8.4cm,clip]{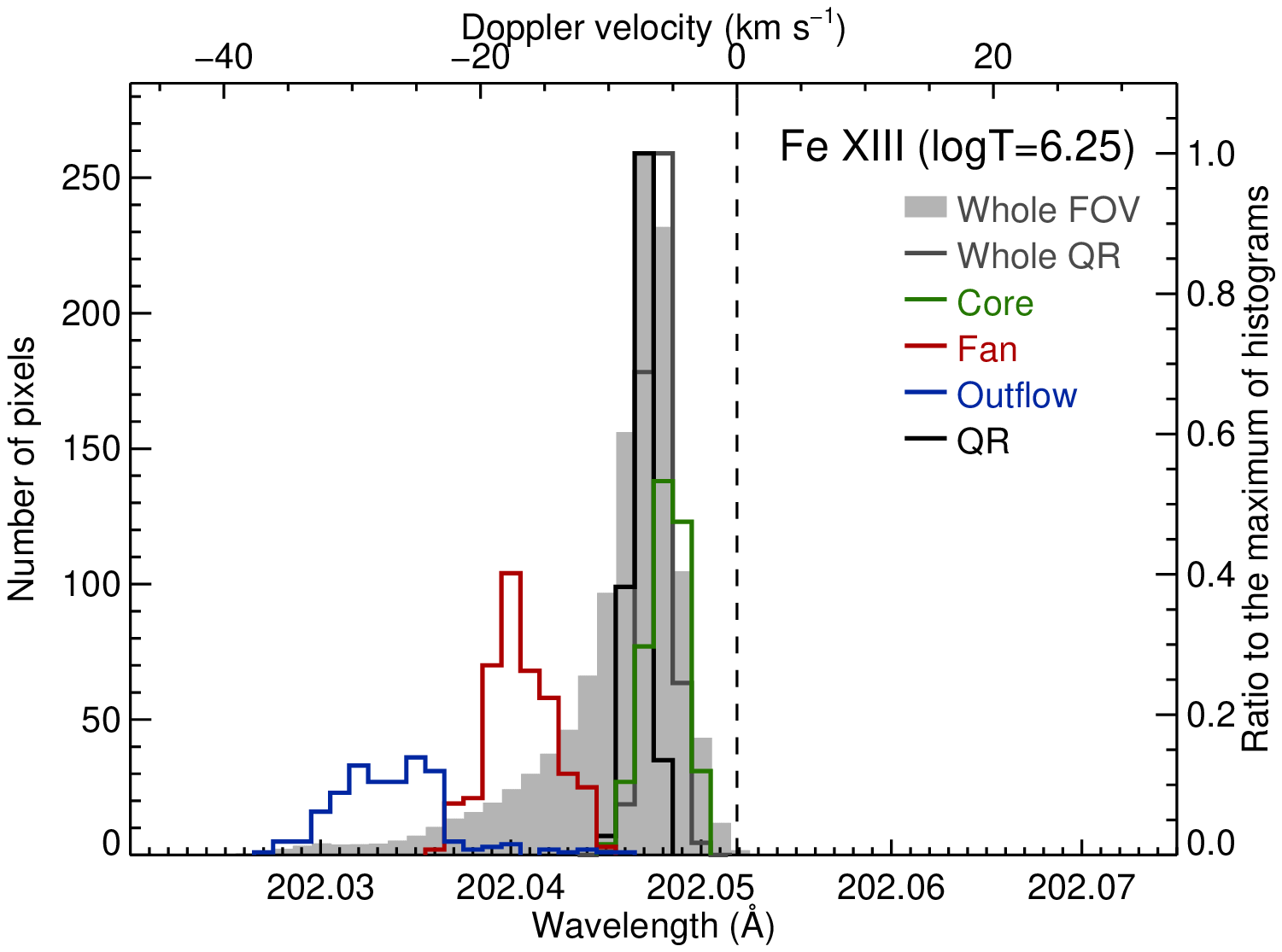}
  \includegraphics[width=8.4cm,clip]{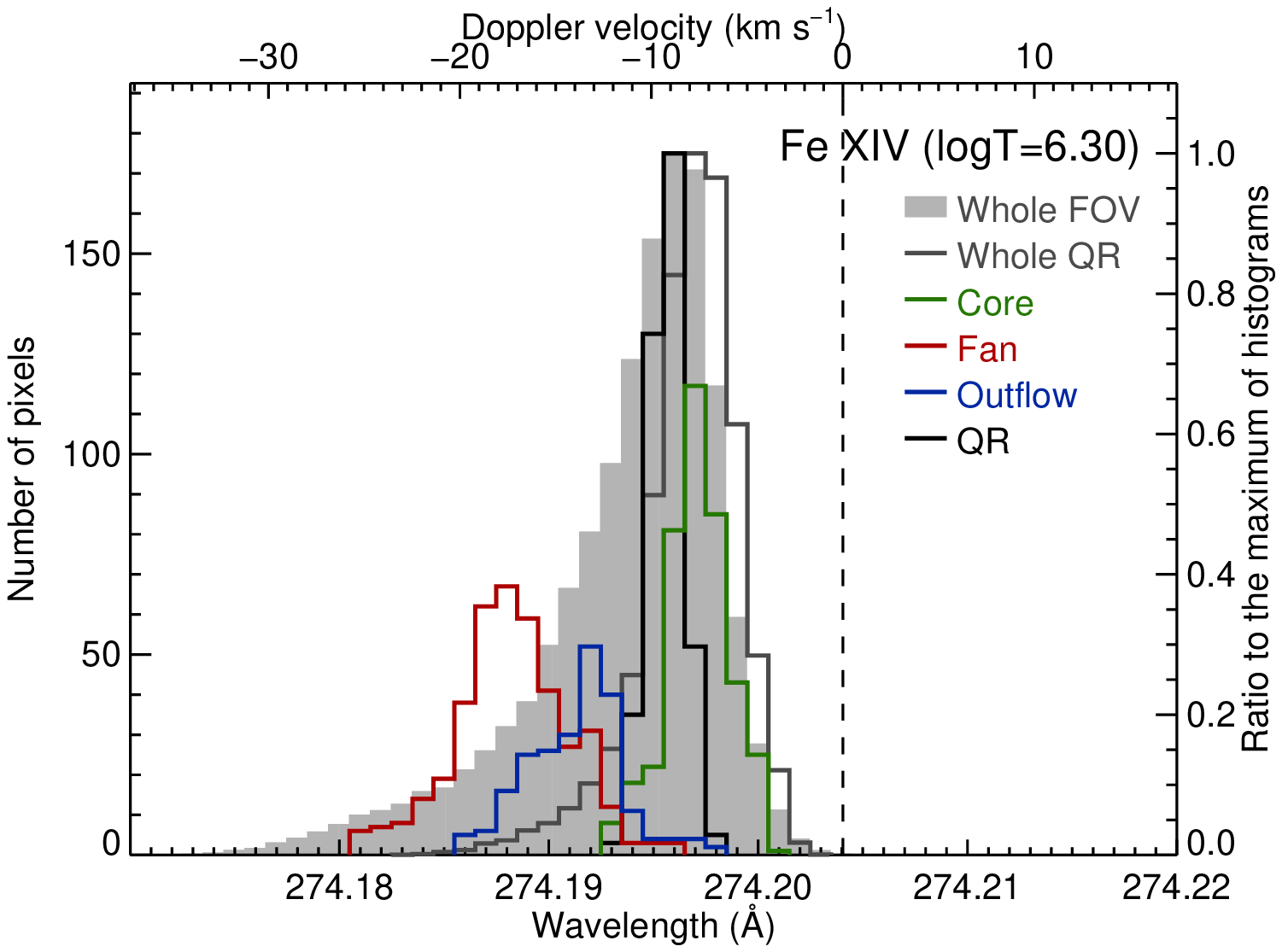}
  \caption{Histogram of line centroid/Doppler velocity at the quiet region (\textit{black}), the outflow region (\textit{blue}), fan loop (\textit{red}), and the core region (\textit{green}).  A \textit{Gray solid} histogram is for the quiet region (defined as lower $30$ pixels in the map).  The histogram filled with \textit{gray} indicates that for the whole map.  \textit{Left}: Fe \textsc{xiii} $202.04${\AA}.  \textit{Right}: Fe \textsc{xiv} $274.20${\AA}.  A vertical dashed line in each panel indicates zero point of Doppler velocity which was calculated so as the quiet region has the average Doppler velocity determined in Chapter \ref{chap:cal}. }
  \label{fig:vel_ar10978_hist_vel_im}
\end{figure}

The emission lines with the formation temperature around $\log T \, [\mathrm{K}] = 6.3$ have the largest enhancement in their blue wing compared to the major component in the line profiles as described in Section \ref{sect:vel_lp}.  As shown in Fig.~\ref{fig:vel_ar10978_hist_vel_im}, the Doppler velocity in the outflow region reaches $\simeq - 30 \, \mathrm{km} \, \mathrm{s}^{-1}$ for Fe \textsc{xiii}.  For Fe \textsc{xiv}, the Doppler velocity there becomes smaller, which is around $-20 \, \text{--} \, {-10} \, \mathrm{km} \, \mathrm{s}^{-1}$.  The Doppler velocity in the fan loop becomes larger than lower formation temperatures, and indicates blueshift of $v \simeq -20 \, \mathrm{km} \, \mathrm{s}^{-1}$ for both emission lines.  

Note that the histograms for whole FOV (\textit{filled with gray}) of the scan have an extended tail toward shorter wavelength for both emission lines, which was previously reported for Fe \textsc{xiv} $274.20${\AA} \citep{hara2008}.  This property clearly differs from the histograms for the whole quiet region (\textit{gray solid}) which look symmetric especially for Fe \textsc{xiii}.  

% --- End of Tex ---

%% file: tex/vel_ar10978_hist_vel_ht.tex
% ===================================================
%   Project:
%     T vs. V
%   Contents:
%     Distribution of Doppler velocity of AR 10978
% ===================================================

\begin{figure}
  \centering
  \includegraphics[width=8.4cm,clip]{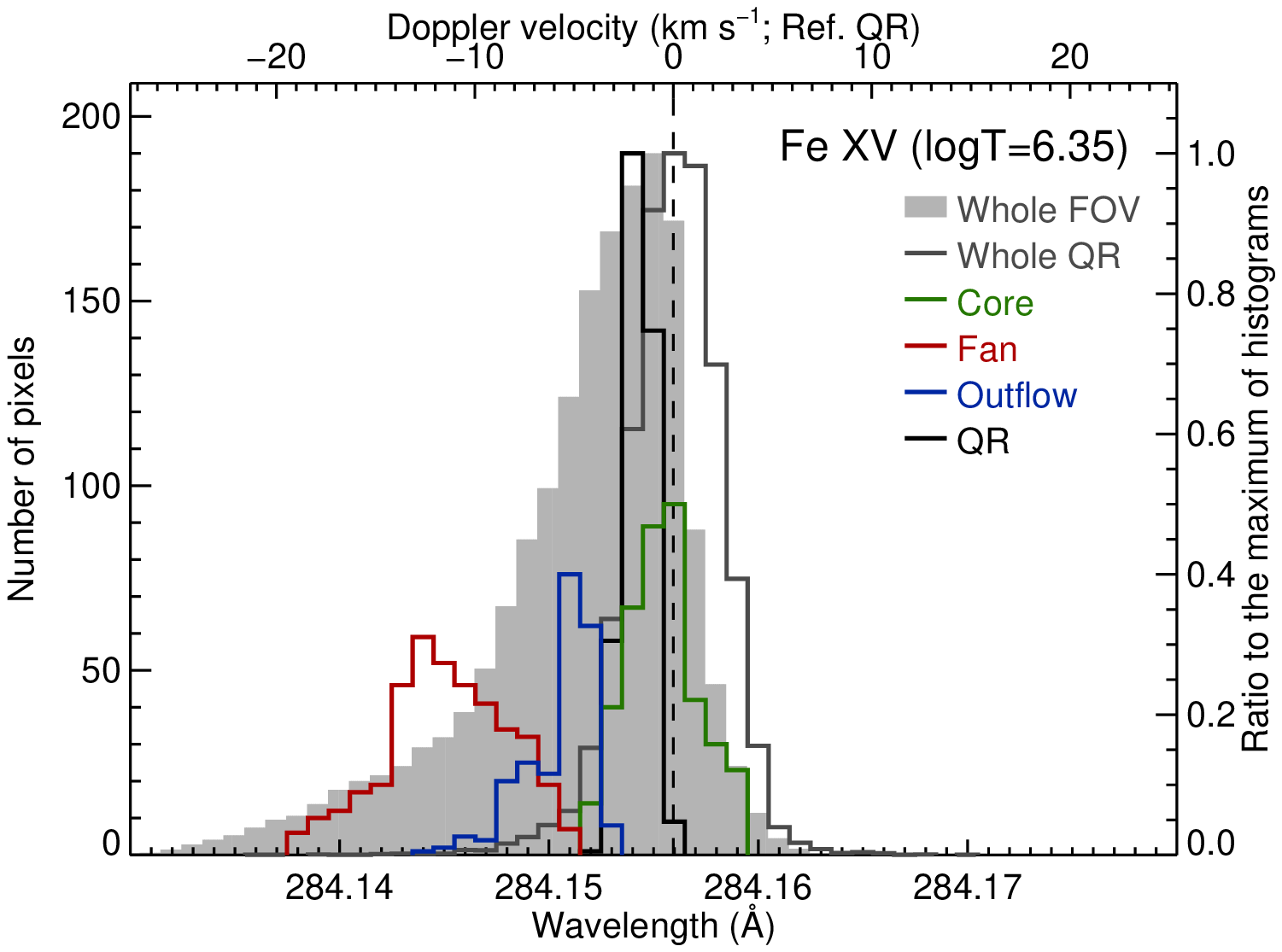}
  \includegraphics[width=8.4cm,clip]{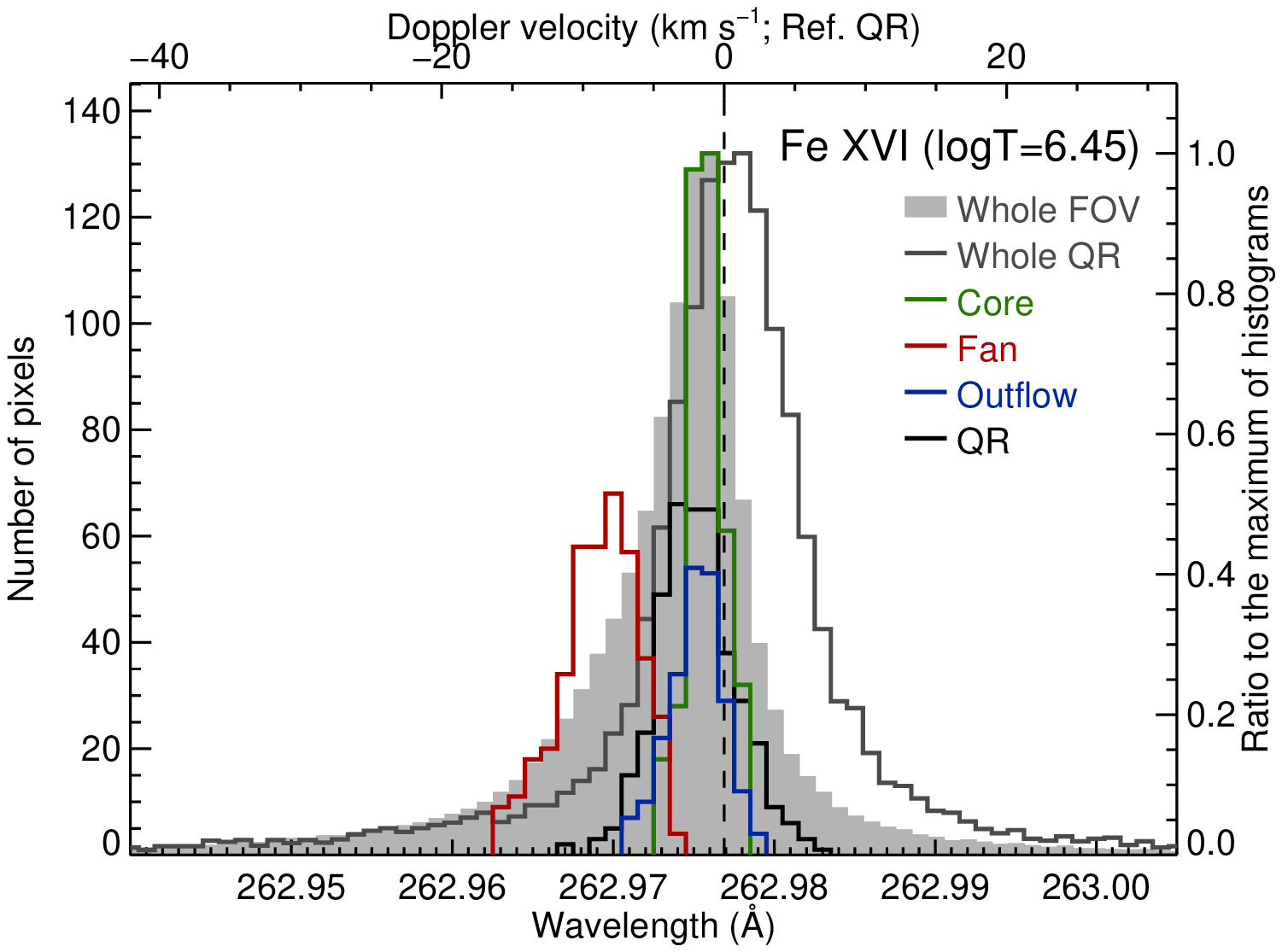}
  \caption{Histogram of line centroid/Doppler velocity at the quiet region (\textit{black}), the outflow region (\textit{blue}), fan loop (\textit{red}), and the core region (\textit{green}).  A \textit{Gray solid} histogram is for the quiet region (defined as lower $30$ pixels in the map).  The histogram filled with \textit{gray} indicates that for the whole map.  \textit{Left}: Fe \textsc{xv} $284.16${\AA}.  \textit{Right}: Fe \textsc{xvi} $262.98${\AA}.  The average value in the quiet region was used for zero point of Doppler velocity because we could not determine the reference at this temperature in Chapter \ref{chap:cal}.}
  \label{fig:vel_ar10978_hist_vel_ht}
\end{figure}

Fig.~\ref{fig:vel_ar10978_hist_vel_ht} shows the histograms of the Doppler velocity for Fe \textsc{xv} $284.16${\AA} and Fe \textsc{xvi} $262.98${\AA} ($\log T \, [\mathrm{K}] \geq 6.4$). While the quiet region, the fan loop, and the core region all have similar value within $\simeq 2$--$3 \, \mathrm{km} \, \mathrm{s}^{-1}$, the outflow region exhibits the Doppler velocity around $- 15 \, \mathrm{km} \, \mathrm{s}^{-1}$ for Fe \textsc{xv}, which is smaller than that of Fe \textsc{xiii} in the magnitude. The Doppler velocity in the fan loop is around $-10 \, \mathrm{km} \, \mathrm{s}^{-1}$, and slightly decreases from the temperature of $\log T \, [\mathrm{K}]=6.3$.  Note that the histogram obtained from the whole map for Fe \textsc{xv} again has an extended tail as same as for Fe \textsc{xiv}. 

One striking fact here is that the outflow ceases for Fe \textsc{xvi}.  The histogram for S \textsc{xiii} $256.69${\AA} (though not shown), which has a similar formation temperature ($\log T \, [\mathrm{K}] = 6.42$), exhibited the same behavior as Fe \textsc{xvi}. This reinforces the result that the outflow was significantly reduced at the temperature higher than $\log T \, [\mathrm{K}]=6.4$.

% --- End of Tex ---

%% file: tex/vel_ar10978_tvsv.tex
% ===========================
%   Project:
%     T vs. V
%   Contents:
%     Distribution of Doppler velocity of AR 10978
% ===========================

\begin{figure}
  \centering
  \includegraphics[width=14cm,clip]{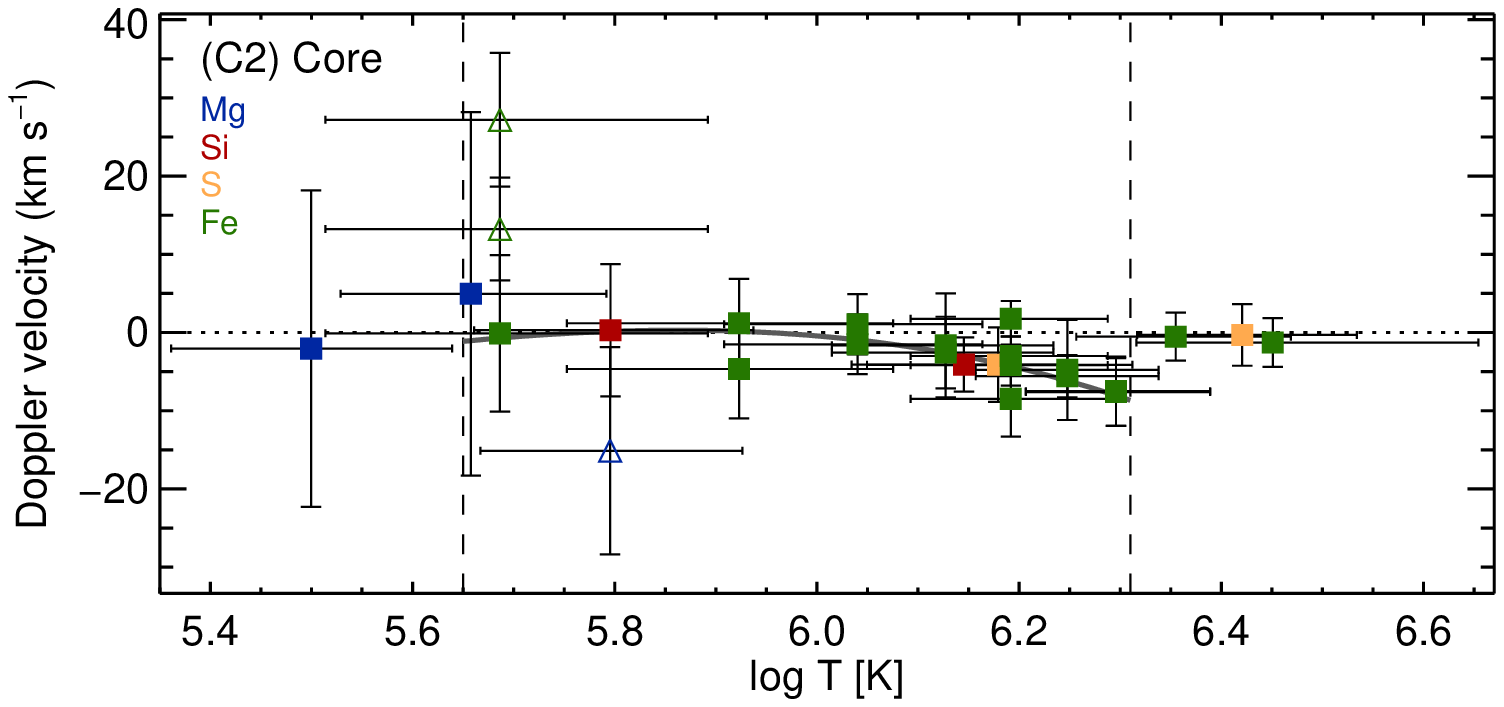}
  \includegraphics[width=14cm,clip]{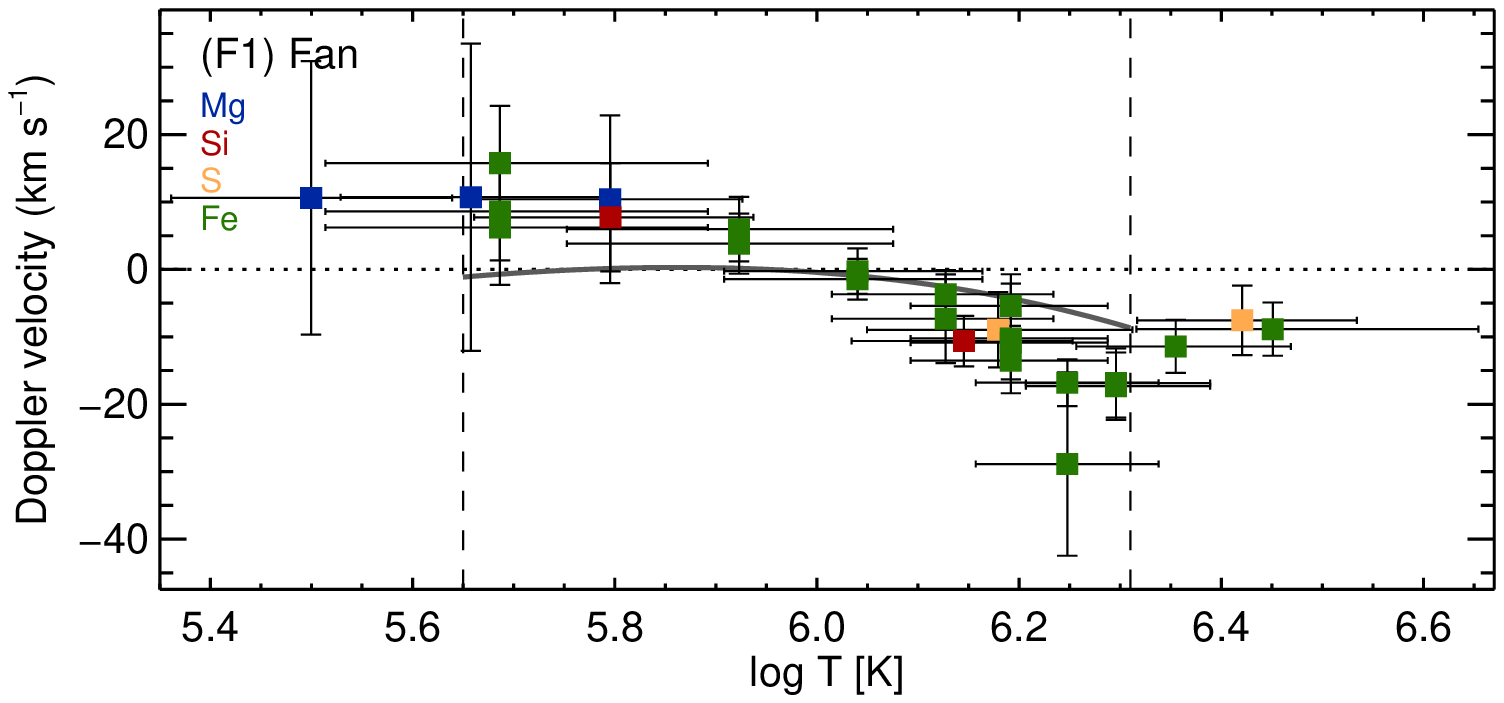}
  \includegraphics[width=14cm,clip]{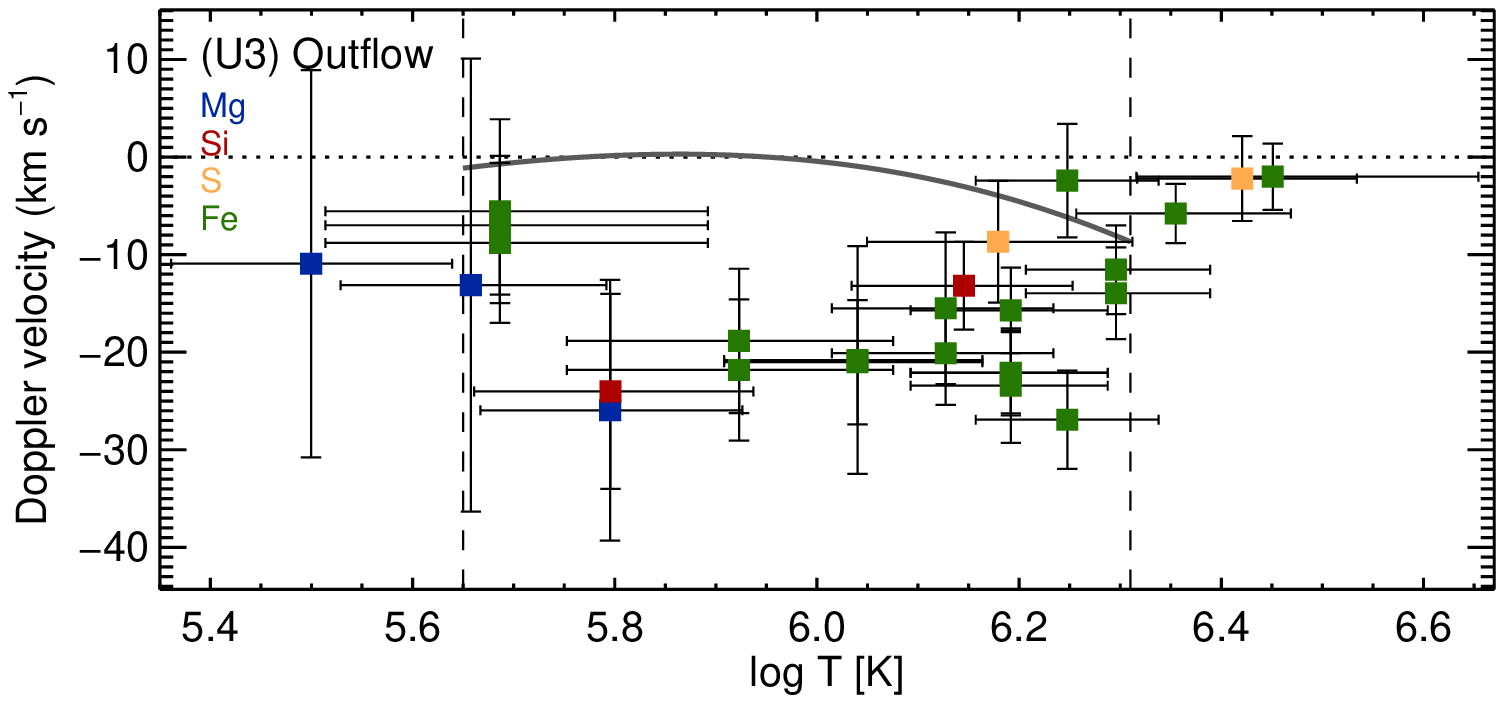}
  \caption{Temperature dependence of the average Doppler velocities.
           \textit{Upper}: active region core. 
           \textit{Middle}: fan loop.
           \textit{Lower}: outflow region.
           \textit{Vertical} error bars indicate the standard deviation of Doppler velocities in each region including the error in the reference Doppler velocity.  \textit{Horizontal} error bars indicate the full width of half maximum of the contribution functions.  Two \textit{vertical dashed} lines show the range where the reference Doppler velocity was measured in Chapter \ref{chap:cal}. \textit{Triangles} in \textit{upper} panel indicate emission lines blended by a high temperature coronal emission line.}
  \label{fig:vel_ar10978_tvsv}
\end{figure}

Fig.~\ref{fig:vel_ar10978_tvsv} shows the temperature dependence of the Doppler velocities in (a) the core region (C2), (b) fan loop (F1), and (c) the outflow region (U3).  The Doppler velocities here are the values averaged within the white boxes indicated in Fig.~\ref{fig:vel_map_box}.  Colors of data points indicate the ion species as denoted in legends.  Vertical error bars indicate the standard deviation of Doppler velocities in each region.  Those are not calculated from errors originated in Poisson noise of photons which is typically in the order of $1.5 \, \mathrm{km} \, \mathrm{s}^{-1}$ (the core region), $2.2 \, \mathrm{km} \, \mathrm{s}^{-1}$ (fan loop), and $4.7 \, \mathrm{km} \, \mathrm{s}^{-1}$ (the outflow region) for Fe \textsc{xi} $182.17${\AA} (\textit{i.e.}, medium strength in the studied emission lines).  These values are slightly less than the standard deviation shown in each panel, but they are in almost the same magnitude, from which it can be considered that the standard deviation of measured Doppler velocities includes Poisson noise and the physical fluctuation of Doppler velocity in the studied region.  Two \textit{vertical dashed} lines show the range where the reference Doppler velocity was measured in Chapter \ref{chap:cal}.  Since the Doppler velocity above the temperature of $\log T \, [\mathrm{K}]=6.25$ could not  determined, we plotted the difference from the quiet region value for Fe \textsc{xv}, S \textsc{xiii} and Fe \textsc{xvi} (\textit{i.e.}, rightmost three data points).  The formation temperature of Fe \textsc{xiv} ($\log T \, [\mathrm{K}]=6.30$) is located within the full width of half maximum of the contribution function of Fe \textsc{xiii}, and the Doppler velocity of the quiet region obtained in Chapter \ref{chap:cal} was extrapolated for Fe \textsc{xiv}.  For Mg \textsc{v} $276.58${\AA} ($\log T \, [\mathrm{K}]=5.50$) at the low temperature side, we adopted the result from SUMER observation: $v_{\mathrm{QR}}=6.7 \, \mathrm{km} \, \mathrm{s}^{-1}$ at $\log T \, [\mathrm{K}]=5.50$ \citep{teriaca1999}. 

In the active region core shown in panel (a), almost all of the emission lines exhibited the Doppler velocities within $-5 \, \mathrm{km} \, \mathrm{s}^{-1} \leq v_{\mathrm{Dop}} \leq 5 \, \mathrm{km} \, \mathrm{s}^{-1}$.  Only exceptions are Mg \textsc{vii} ($\log T \, [\mathrm{K}]=5.80$) which has the velocity of around $-15 \, \mathrm{km} \, \mathrm{s}^{-1}$, and Fe \textsc{viii} $185.21${\AA} and $186.60${\AA} ($\log T \, [\mathrm{K}] = 5.69$) which have the Doppler velocity larger than $10 \, \mathrm{km} \, \mathrm{s}^{-1}$.  Judging from the spectra of Mg \textsc{vii} at the core region (shown in Appendix \ref{sect:vel_app_mg_lines}), there seems to exist a P \textsc{xii} emission line ($\log T \, [\mathrm{K}] = 6.3$) exists at the blue wing of Mg \textsc{vii}.  Note that the spectra other than in the core region do not significantly suffer from those blends as seen in Fig.~\ref{fig:vel_app_mg_lines}.  Fe \textsc{viii} $185.21${\AA} and $186.60${\AA} are respectively blended by Ni \textsc{xvi} $185.23${\AA} and Ca \textsc{xiv} $186.61${\AA}, which are both located near Fe \textsc{viii} lines.  Since the separation between Fe \textsc{viii} and blending line is smaller than EIS spectral pixels ($\simeq 0.0223 \, \text{\AA} \, \mathrm{pix}^{-1}$), the spectra were not obviously distorted.  

The fan loop shows a characteristic dependence as shown in panel (b).  We can see clearly downward velocity in lower temperature and upward velocity in higher temperature within the temperature range $\log T \, [\mathrm{K}]=5.7$--$6.3$.  This result is consistent with that obtained by \citet{warren2011}.  The Doppler velocities above the formation temperature $\log T \, [\mathrm{K}] = 6.4$ includes systematic uncertainties due to the fact that we could not determine the zero point of Doppler velocities for emission lines with the formation temperature above $\log T \, [\mathrm{K}] = 6.4$ in Chapter \ref{chap:cal} because those emission lines were too weak in the quiet region.  The Doppler velocities above $\log T \, [\mathrm{K}] = 6.4$ represent the relative velocities from those of the quiet region.  
%However, the gradient of the decreasing velocities is reduced above the temperature of $\log T \, [\mathrm{K}]=6.3$. 

Panel (c) shows the temperature dependence of the outflow region, which is our main topic in this thesis.  All emission lines (\textit{i.e.}, all temperature range) analyzed here exhibit upward velocity.  The most critical result here is the upward velocity of $\simeq 20 \, \mathrm{km} \, \mathrm{s}^{-1}$ even at the transition region temperature (\textit{i.e.}, $\log T \, [\mathrm{K}] \leq 6.0$), which has not been revealed previously in the literature.  In addition to this, emission lines above the formation temperature of $\log T \, [\mathrm{K}]=6.4$ indicated almost the same Doppler velocity as in the quiet region, though we could not measure the absolute Doppler velocity for those emission lines.  
%The measurement of the line centroid of the emission line Fe \textsc{xii} $196.53${\AA} was possibly affected by a line blending at its red wing due to a neighboring Fe \textsc{xiii} $196.64${\AA}.  

%The implication from this fact will be discussed later. 
% --- End of Tex ---

%% file: tex/vel_sum.tex
% ================================================
%   Chapter:
%     Doppler velocities of the outflow region.
%   Description:
%     Summary.
% ================================================

% Purpose
In order to investigate the temperature dependence of Doppler velocities in the outflow region, we analyzed an active region NOAA AR10978 when it passed near the disk center.  The scan observation with \textit{Hinode}/EIS analyzed here was carried out on 2007 December 11.  Its FOV included the entire active region.  The data contains twenty four spectral windows and a number of emission lines within those.  We selected twenty six emission lines in total which are strong so that the line centroid could be measured accurately.  The formation temperatures of those emission lines range from $\logt=5.50$ (Mg \textsc{v}) up to $6.45$ (Fe \textsc{xvi}), which enabled us to derive the temperature dependence of Doppler velocities with a temperature range covering that of the typical corona. 

% Line profiles
The spectra were inspected in the active region core, fan loops, the outflow regions, and the quiet region.  The line profiles appeared to be symmetric and well fitted by a single Gaussian except for the outflow regions where an obvious enhancement at the blue wing was observed.  This enhancement was most prominent in the emission lines with a formation temperature of $\logt=6.1$--$6.3$ (Fe \textsc{xi}-\textsc{xv}).  At the formation temperature above $\logt=6.4$ (S \textsc{xiii} and Fe \textsc{xvi}), no such enhancement at the blue wing was observed.  
%{\color{red}
This might imply that the emission from S \textsc{xiii} and Fe \textsc{xvi} observed in the outflow region was actually a scattered spectrum in the spectrometer which came from the active region core (\textit{i.e.}, bright in those emission lines).  It is also implicated by the result that the line centroid position in the outflow region does not differ from that in the active region core. 
%}

% Analysis
After correcting the spectrum tilt, line centroids of the emission lines were derived through fitting by a single Gaussian.  We obtained reasonable Doppler velocities by adjusting an average Doppler velocity in the quiet region included in the FOV to the result in Section \ref{chap:cal}.  Our analysis has an advantage in this point over previous measurements of Doppler velocities in outflow regions and nearby locations.  Those measurements were based on a simple comparison of the obtained line centroid with the independent limb observation \citep{warren2011} processed through the correction of the orbital variation modeled by \citet{kamio2010} which potentially includes the error of $\simeq 10 \, \kmpers$ in total.  Others just assumed the average Doppler velocity along the EIS spectroscopic slit would be zero, which seems to be too strong assumption in the corona, and that is not the case according to our result derived in Section \ref{chap:cal}. 

% Results for active region core
We studied the temperature dependence of obtained Doppler velocity in several core regions, fan loops, outflow regions.  The most of the emission lines analyzed exhibit Doppler velocity of $-5 \, \kmpers \leq v_{\mathrm{Dop}} \leq 5 \, \kmpers$ at the core regions.  The interpretation may be rather complex because the corona is optically thin, and that we observe patchy structures at $\logt \leq 6.2$ while we observe multiple loops more clearly with increasing temperature above.  The patchy structures are considered to be the footpoints of the overlying multiple loops \citep{berger1999}.  The obtained temperature dependence has a weak negative slope, and Doppler velocity becomes negative at around $\logt=6.25$.  This leads to possible interpretations that (1) loops at the core moves upwardly with several $\kmpers$, or (2) there are plasma upflows with that temperature at footpoints of the multiple loops.  It is difficult to give a decisive implication here because the temperature around $\logt=6.25$ is a transition from the patchy structures to the multiple loops. 

% Results for the outflow regions and fan loops
As a significant result, we found that not only the coronal emission lines ($\logt \geq 6.0$) show the velocity of $\simeq -20 \, \kmpers$ (\textit{i.e.}, upward) but also the transition region lines ($\logt \leq 6.0$) in the outflow region did, which has not been revealed so far.  This tendency differs from that in fan loops where the Doppler velocities decreased with increasing formation temperature from $10\text{--}20 \, \kmpers$ at $\logt=5.7$ to $-20 \, \kmpers$ at $\logt=6.3$.  The definitive difference between fan loops and the outflow regions was found in the temperature dependence of the Doppler velocities, which indicates that they are actually composed of different sturctures although sometimes taken as identical.  The result that the plasma with wide temperature range ($\logt=5.5\text{--}6.3$) flows up in the outflow regions may be negative evidence for the scenario that EBWs seen in the outflow region indicates the flow induced by an impulsive heating with long time interval than the cooling timescale, since such situation produces a redshift of the transition region lines \citep{patsourakos2006}.  Our result implies that (1) the outflows in the western outflow region are induced by steady heating compared to the cooling timescale, and (2) they never return. 

% --- End of TeX ---

%% file: tex/vel_app_spec_tilt.tex
\begin{figure}
  \centering
  \includegraphics[width=11.5cm,clip]{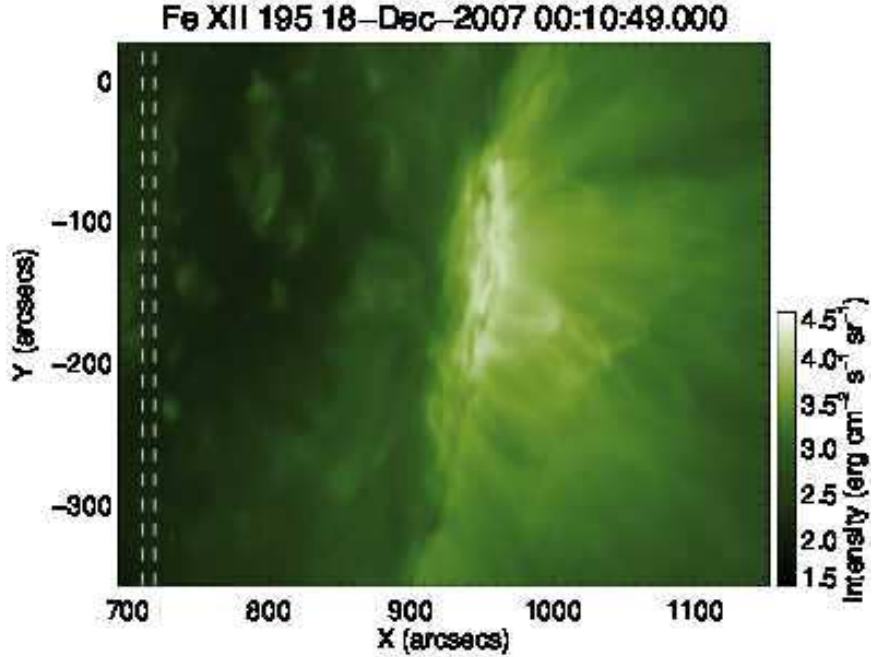}
  \caption{Intensity image of Fe \textsc{xii} $195.12${\AA} taken by a raster scan at the west limb on 2007 December 18. The region between two vertical dashed lines indicates the location where the spectrum tilts were investigated.}
  \label{fig:vel_app_wlimb}
\end{figure}

% ===========================================
%   Chapter:
%     Doppler velocity of the AR outflows.
%   Section:
%     Spectrum tilts.
% ===========================================
The spectrum tilts (described in Section \ref{sect:cal_append_itdn}) were investigated to check the validity of the standard EIS software and the calibration done by \citet{kamio2010}. Different from the analysis in Section \ref{sect:cal_append_itdn}, we could not find the data with full CCD in $y$ direction around the time period when NOAA AR10978 was observed. The scans during that time period, EIS FOV basically included the active region which may affect the line centroid significantly more than tha quiet region. We found the most preferable data which was done at the west limb as shown in Fig.~\ref{fig:vel_app_wlimb}. This raster scan relatively wide quiet region inside the limb. 

\begin{figure}
  \centering
  \includegraphics[width=7.9cm,clip]{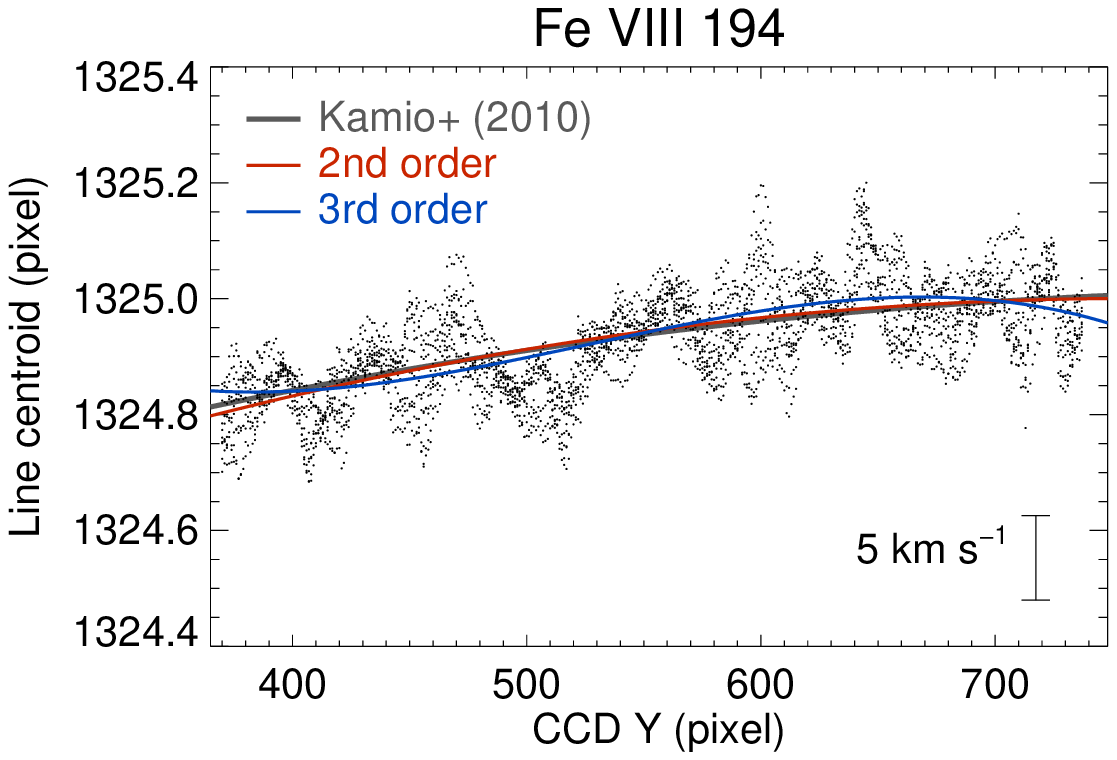}
  \includegraphics[width=7.9cm,clip]{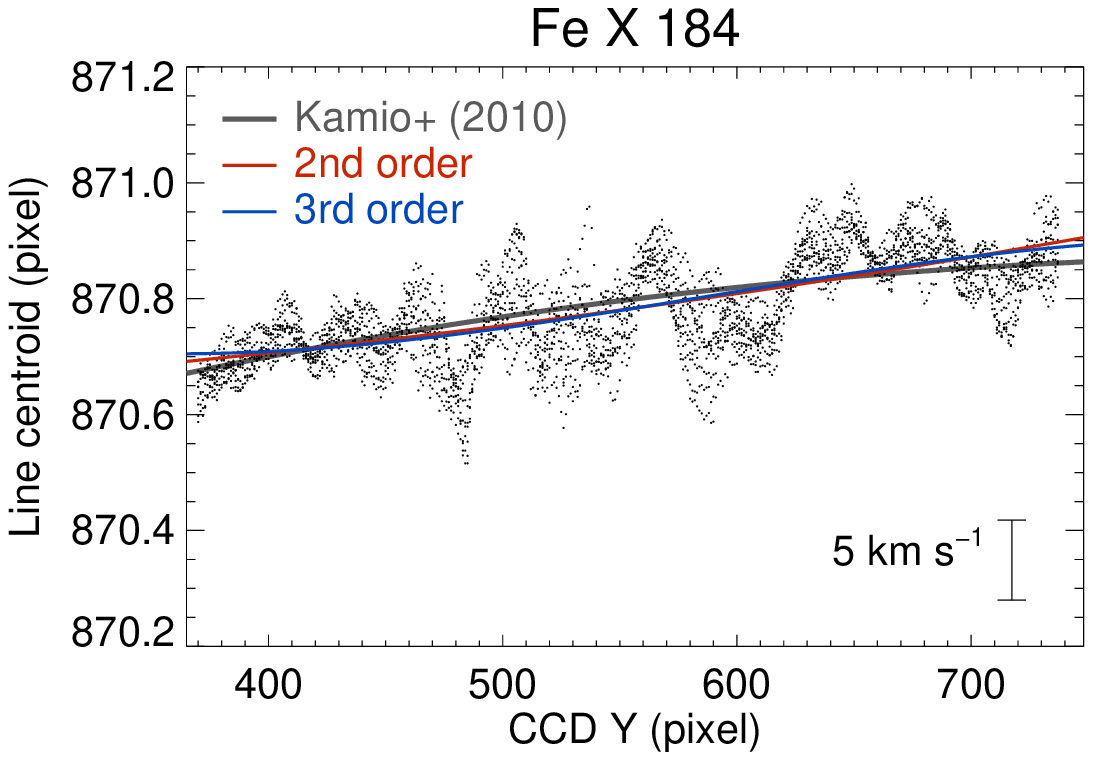}
  \includegraphics[width=7.9cm,clip]{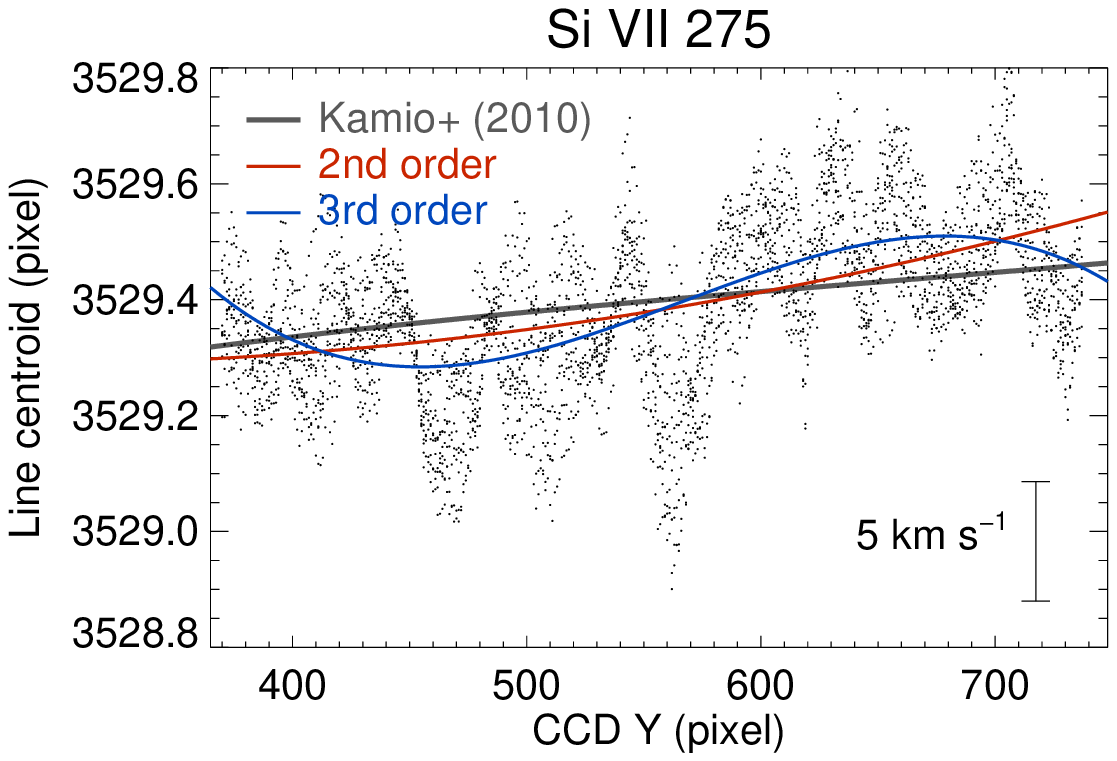}
  \includegraphics[width=7.9cm,clip]{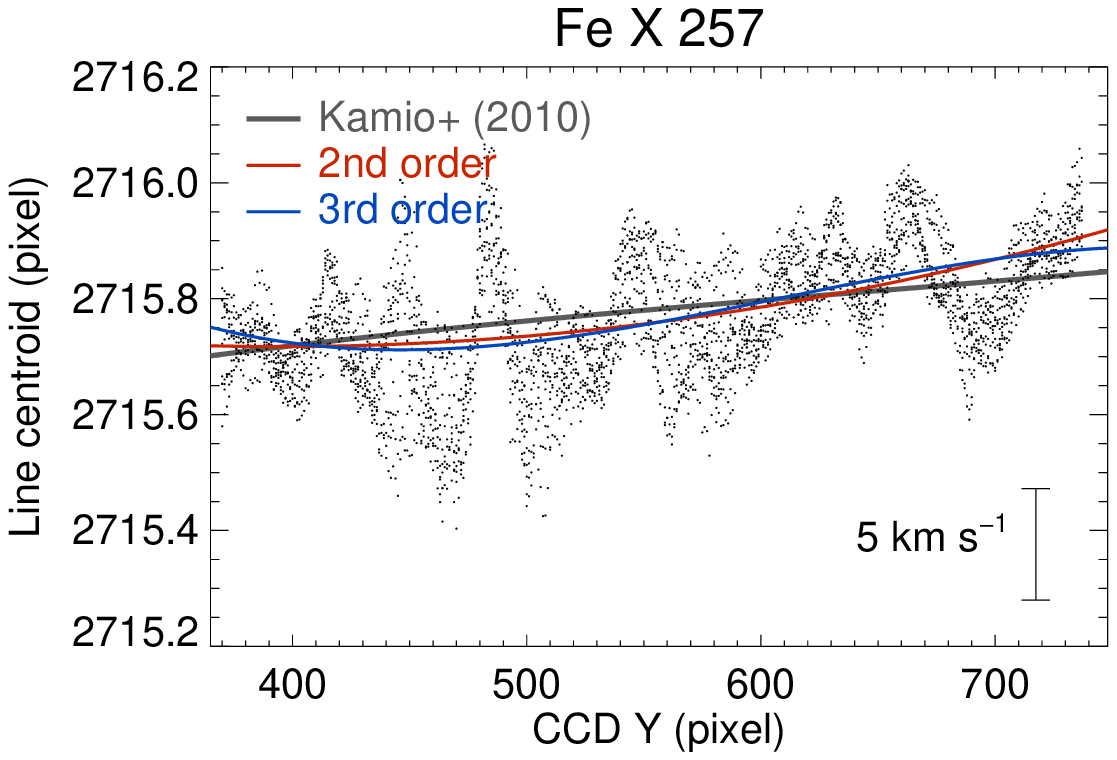}
  \caption{The spectrum tilts for Fe \textsc{viii} $194.66${\AA}, Si \textsc{vii} $275.35${\AA}, 
  and Fe \textsc{x} $184.54${\AA}/$257.26${\AA}. \textit{Gray,} \textit{red,} and \textit{blue} 
  solid line respectively indicate the tilt given by \citet{kamio2010}, second order 
  polynomial fitting, and third order polynomial fitting. A vertical bar in the right lower 
  corner in each panel shows the velocity scale of $5 \, \mathrm{km} \, \mathrm{s}^{-1}$.
  }
  \label{fig:vel_app_tilt}
\end{figure}

We chose the region to be analyzed where (1) away from the active region on the limb as far as possible, and (2) less coronal bright points along all $y$ positions in the FOV. The best region satisfying those criteria was between two vertical dashed lines in Fig.~\ref{fig:vel_app_wlimb}. Here we show the variation of four line centroids from both SW and LW CCDs in Fig.~\ref{fig:vel_app_tilt}: Fe \textsc{viii} $194.66${\AA} (\textit{upper left}) and Fe \textsc{x} $184.54${\AA} (\textit{upper right}) from the SW CCD, Si \textsc{vii} $275.35${\AA} (\textit{lower left}) and Fe \textsc{x} $257.26${\AA} (\textit{lower right}) from the LW CCD. \textit{Gray,} \textit{red,} and \textit{blue} solid line respectively indicate the tilt given by \citet{kamio2010}, second order polynomial fitting, and third order polynomial fitting. The calibration given by \citet{kamio2010} is shifted in each panel so that the value at the left vertical axis coincides with the second order polynomial fitting. A vertical bar in the right lower corner in each panel shows the velocity scale of $5 \, \mathrm{km} \, \mathrm{s}^{-1}$. Note that Fe \textsc{viii} and Si \textsc{vii} emission lines have similar formation temperature of $\log T \, [\mathrm{K}] \simeq 5.7$--$5.8$. 

The curve given by \citet{kamio2010} represents the behavior of line centroids quite well in the SW CCD as seen in two upper panel in Fig.~\ref{fig:vel_app_tilt}. Data points are a little more dispersed in lower panels because of the less number of photons of those emission lines, nevertheless, the curve given by \citet{kamio2010} again represents their behavior. Therefore we adopt the calibration given by \citet{kamio2010} in this chapter. 

% --- End of TeX ---

%% file: tex/vel_app_mg_lines.tex
% ================================================
%   Chapter:
%     Doppler velocities of the outflow region.
%   Description:
%     Mg lines as an appendix.
% ================================================

\begin{figure}[b]
  \centering
  \includegraphics[width=8.3cm,clip]{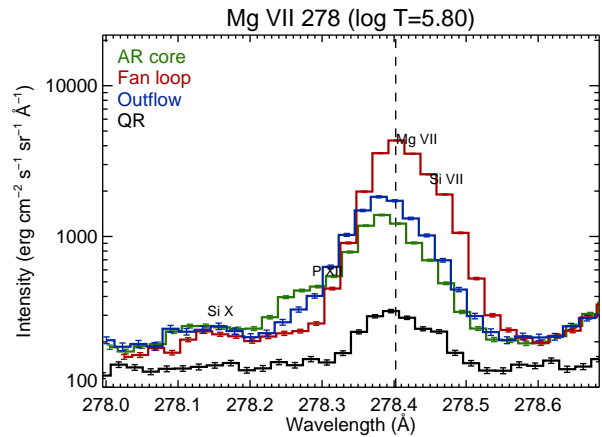}
  \caption{Line profiles of Mg \textsc{vii} $278.40${\AA} at the quiet region (\textit{black}), the outflow region (\textit{blue}), fan loop (\textit{red}), and the core region (\textit{green}).  A vertical dashed line in each panel indicates a line centroid of the emission line at the quiet region.}
  \label{fig:vel_app_mg_lines}
\end{figure}

At the active region core, Mg \textsc{vii} $278.40${\AA} exhibited Doppler velocity of $\simeq - 15 \, \mathrm{km} \, \mathrm{s}^{-1}$.  Here Fig.~\ref{fig:vel_app_mg_lines} shows line profiles of the emission line.  The line profile at the core region (\textit{green}) has a distinct uplift of the spectrum at $\lambda=278.20$--$278.30${\AA} which is considered to be P \textsc{xii} $278.29${\AA} ($\log T \, [\mathrm{K}]=6.3$). This could cause the apparent blueshift of Mg \textsc{vii}.

% --- End of TeX ---

%% file: tex/vel_app_tvsv.tex
% =======================================
%   Chapter:
%     Doppler velocity of outflow.
%   Section:
%     Appendix.
%     T vs. V including other samples.
% =======================================

\begin{figure}
  \centering
  \includegraphics[width=8.4cm,clip]{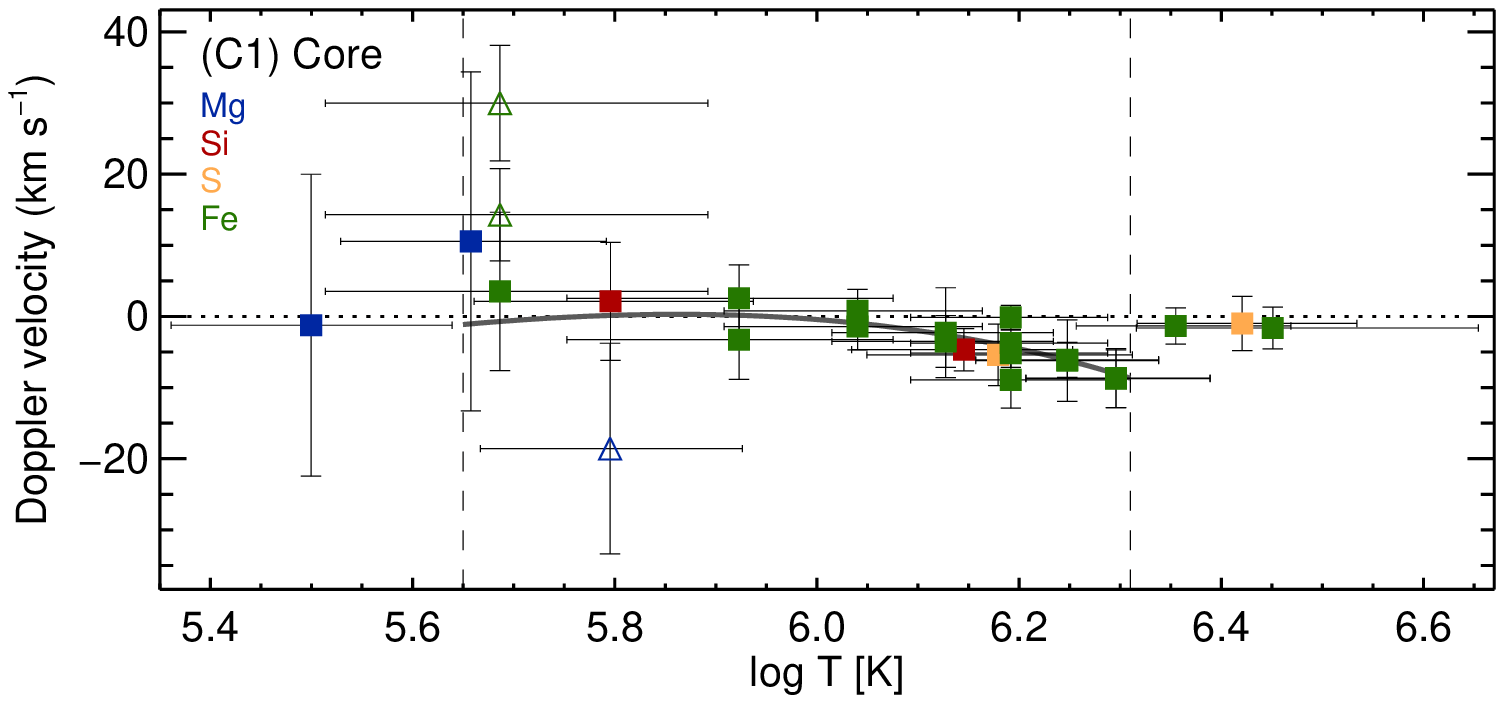}
  \includegraphics[width=8.4cm,clip]{images/arout_tvsv/eis/tvsv/tvsv_core_2.eps}
  \includegraphics[width=8.4cm,clip]{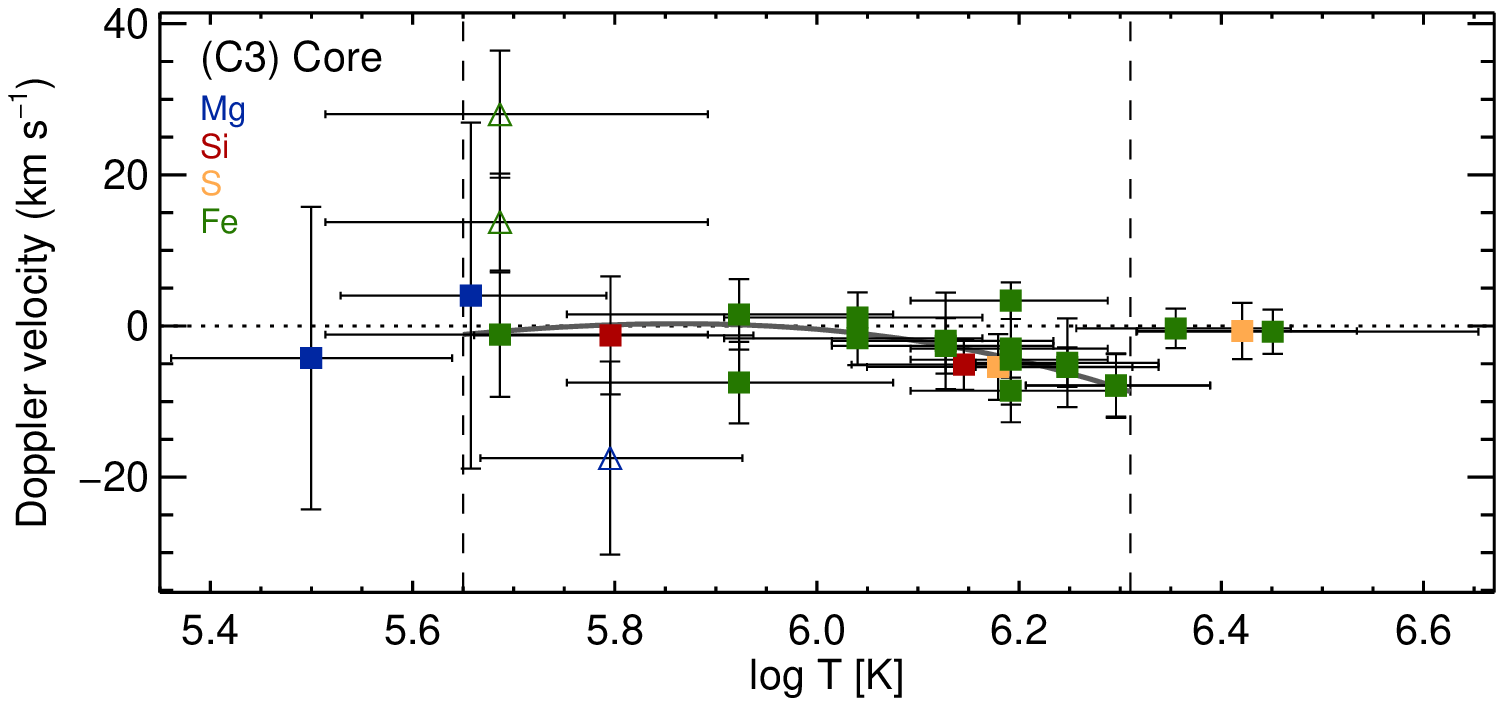}
  \caption{Temperature dependence of the average Doppler velocities in C1--C3 indicated by white boxes in Fig.~\ref{fig:vel_map_box}.}
  \label{fig:vel_app_tvsv_core}
\end{figure}

\begin{figure}
  \centering
  \includegraphics[width=8.4cm,clip]{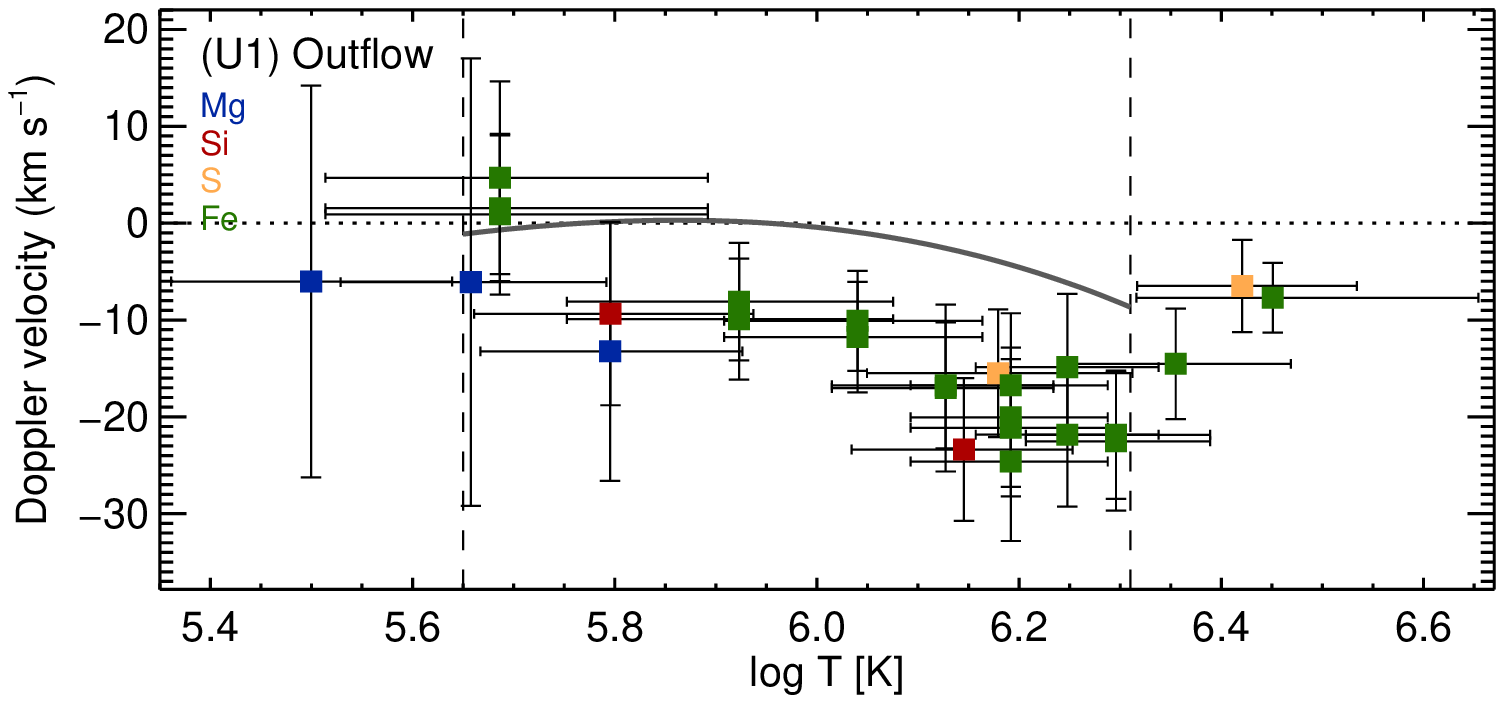}
  \includegraphics[width=8.4cm,clip]{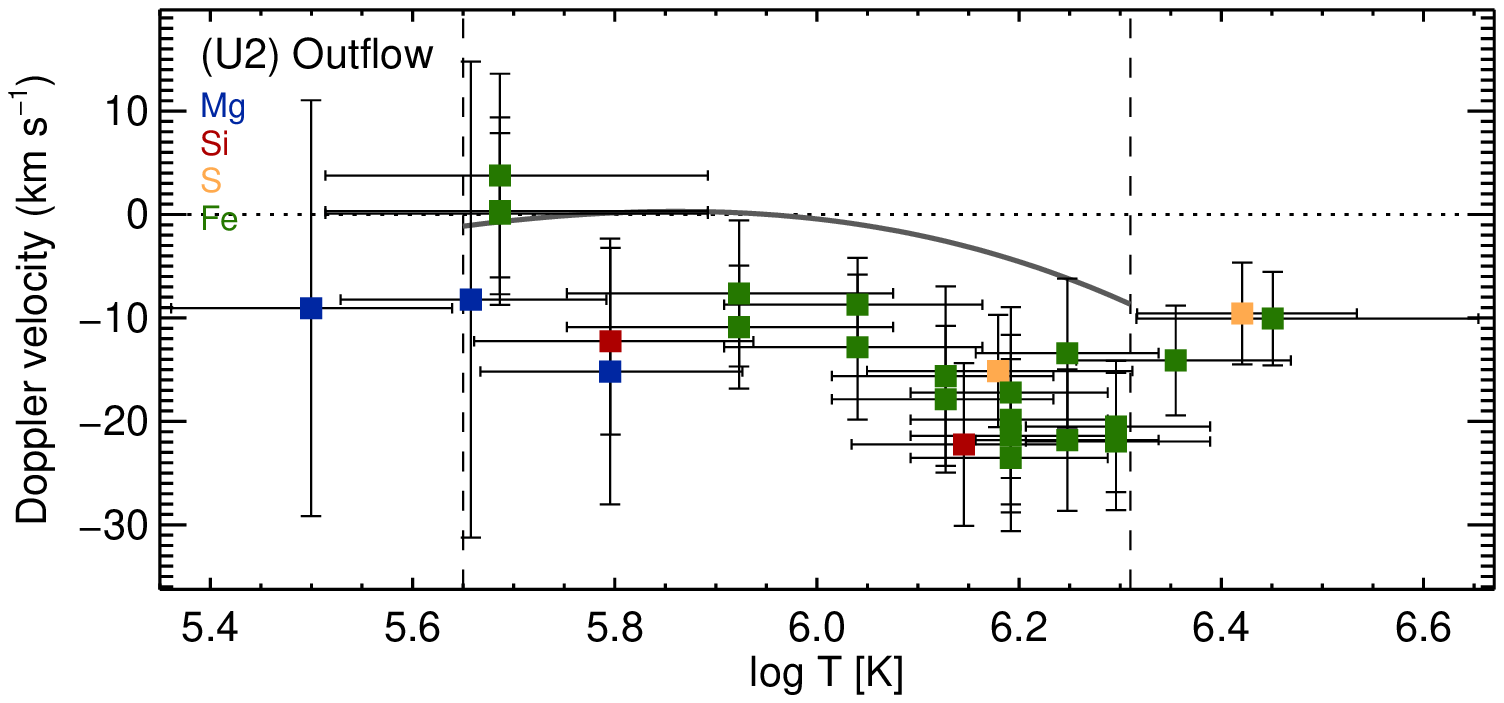}
  \includegraphics[width=8.4cm,clip]{images/arout_tvsv/eis/tvsv/tvsv_outflow_3.eps}
  \includegraphics[width=8.4cm,clip]{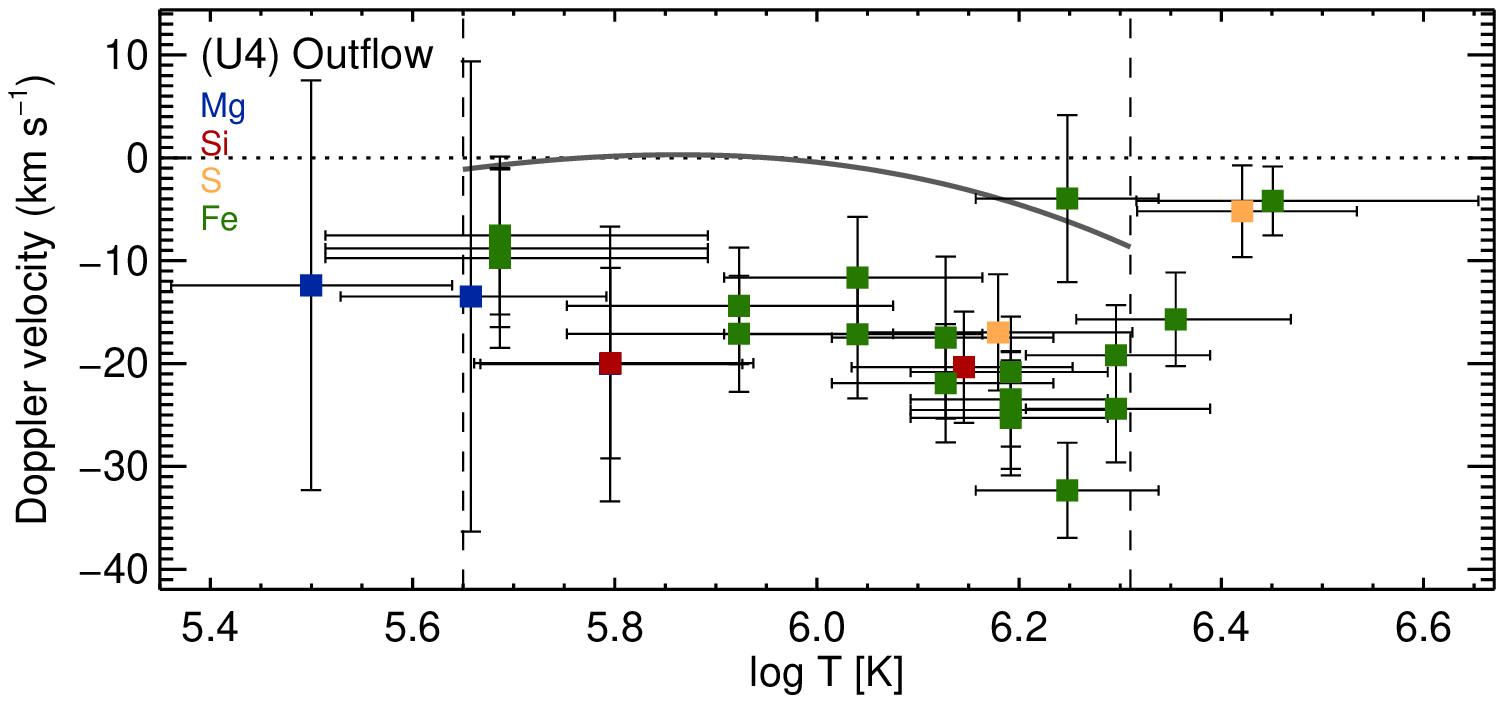}
  \caption{Temperature dependence of the average Doppler velocities in U1--U4 indicated by white boxes in Fig.~\ref{fig:vel_map_box}.}
  \label{fig:vel_app_tvsv_outflow}
\end{figure}

\begin{figure}
  \centering
  \includegraphics[width=8.4cm,clip]{images/arout_tvsv/eis/tvsv/tvsv_fan_1.eps}
  \includegraphics[width=8.4cm,clip]{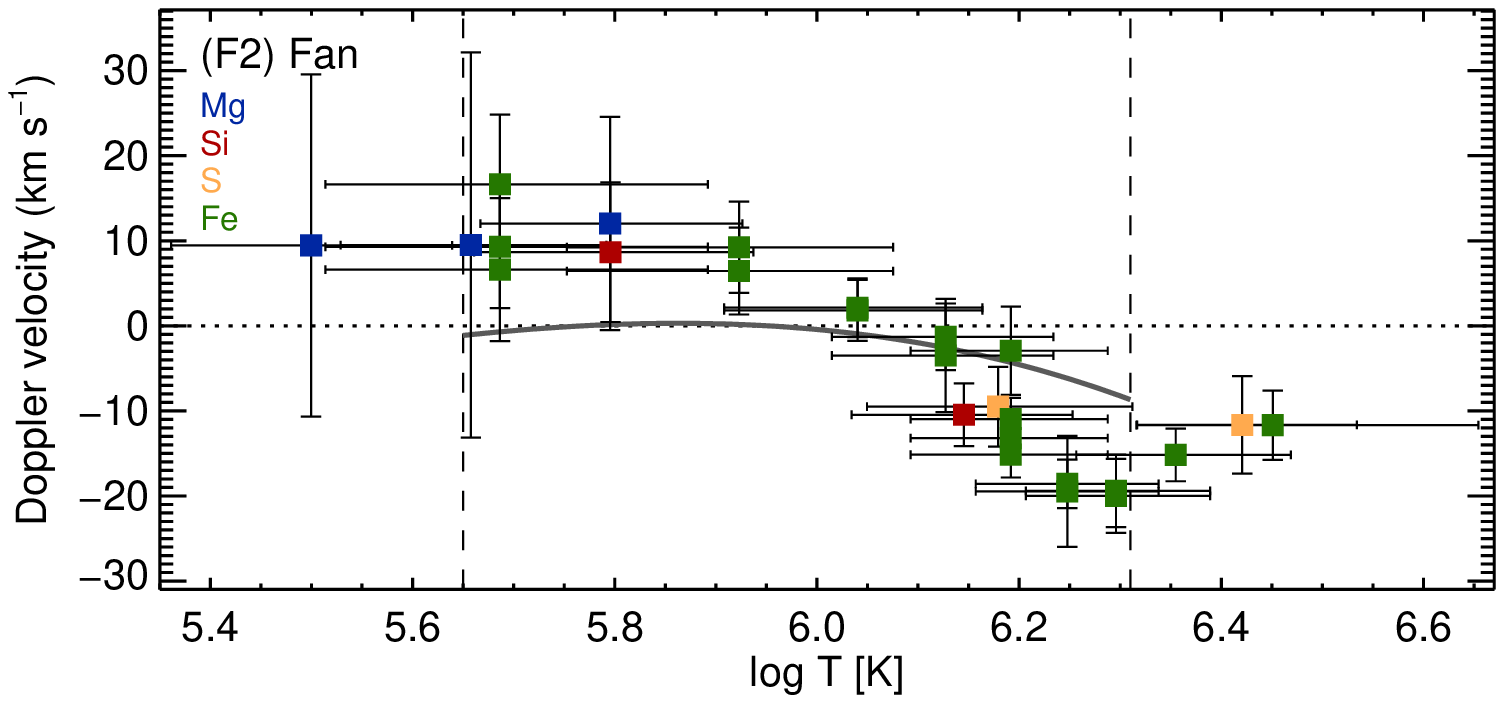}
  \includegraphics[width=8.4cm,clip]{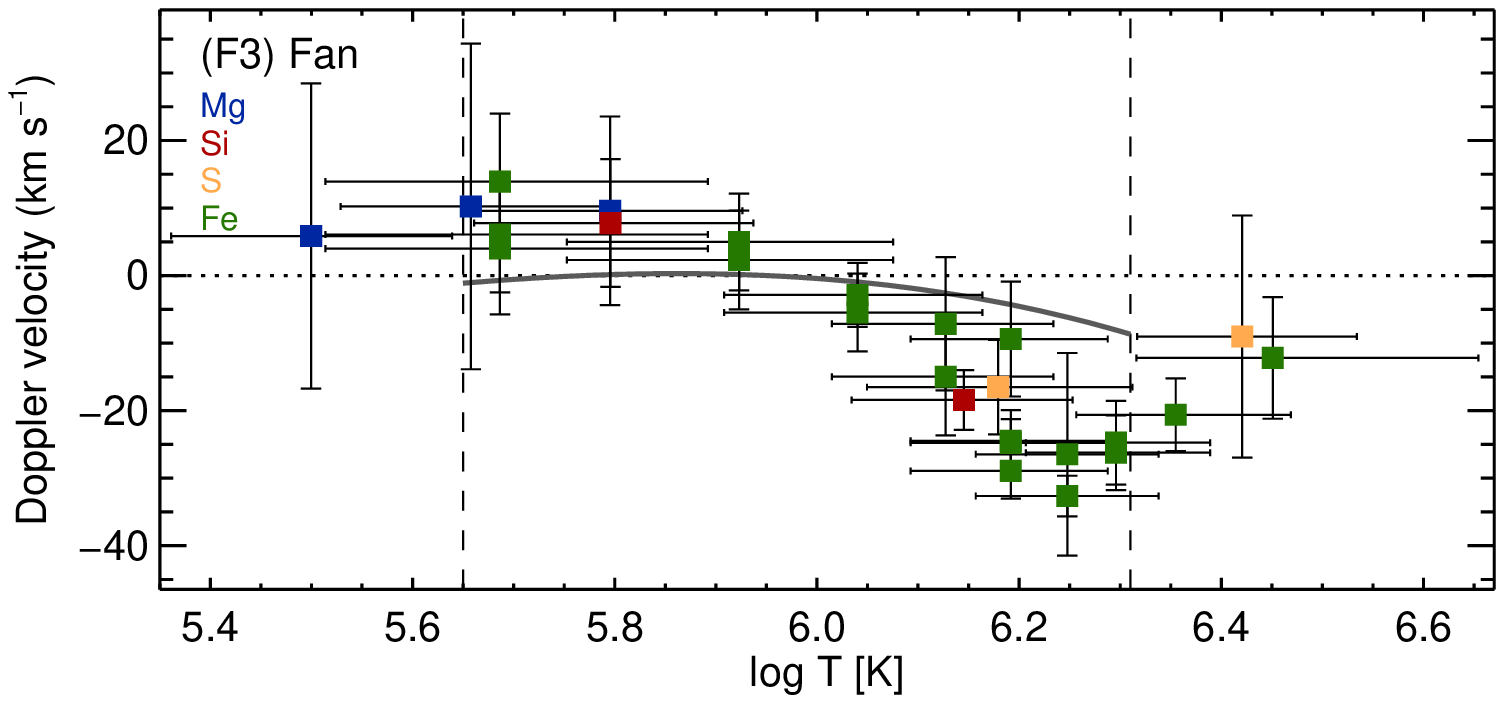}
  \includegraphics[width=8.4cm,clip]{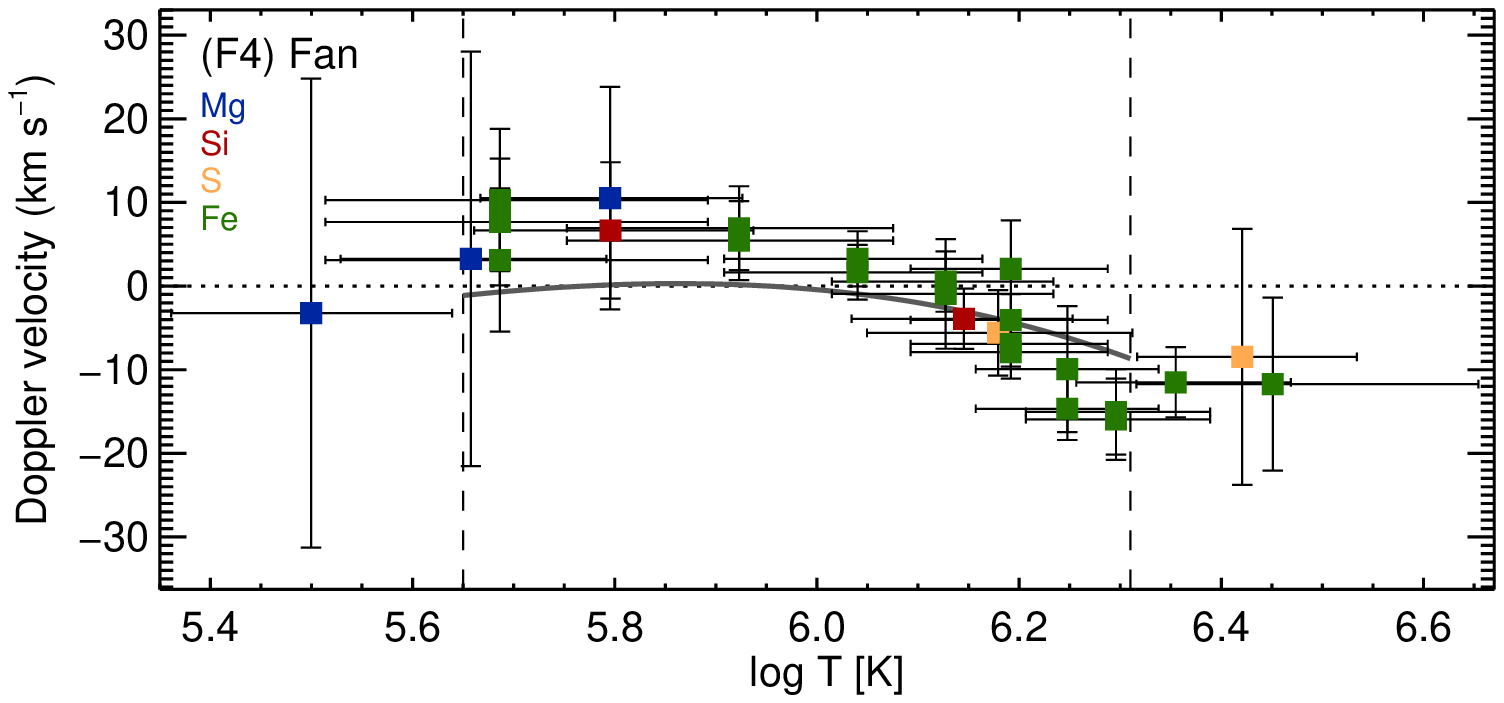}
  \caption{Temperature dependence of the average Doppler velocities in F1--F4 indicated by white boxes in Fig.~\ref{fig:vel_map_box}.}
  \label{fig:vel_app_tvsv_fan}
\end{figure}

The temperature dependence of the Doppler velocities for all \textit{white} boxes indicated in Fig.~\ref{fig:vel_map_box} is shown here. Fig.~\ref{fig:vel_app_tvsv_core}--\ref{fig:vel_app_tvsv_fan} respectively show the result for core regions (C1--C3), outflow regions (U1--U4), and fan loops (F1--F4).  

% --- End of TeX ---

%% file: tex/vel_ar10978_lp_res.tex
% ===========================
%   Project:
%     T vs. V
%   Contents:
%     Velocity of AR 10978
% ===========================

\begin{figure}
  \centering
  \includegraphics[width=8.3cm,clip]{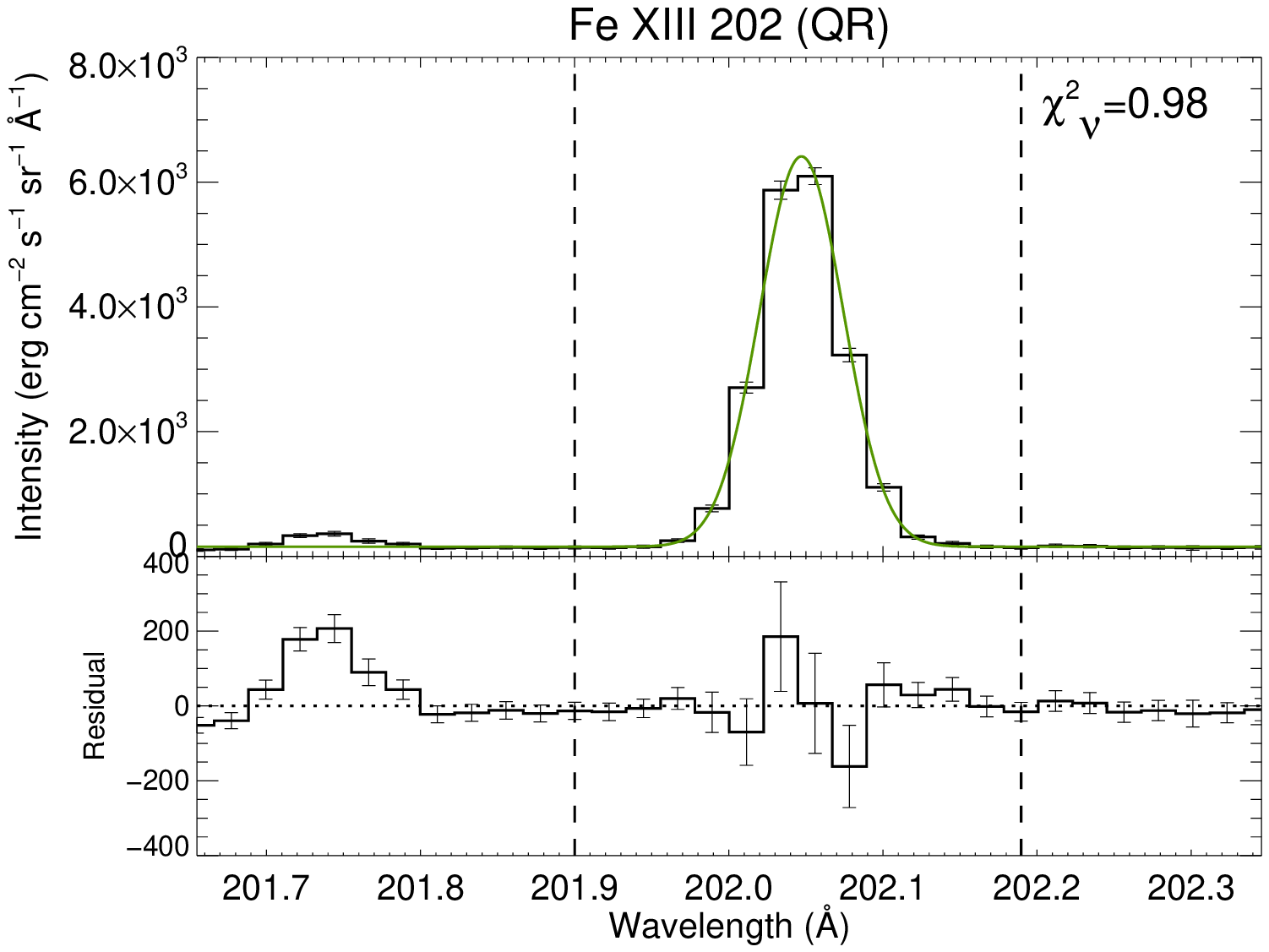}
  \includegraphics[width=8.3cm,clip]{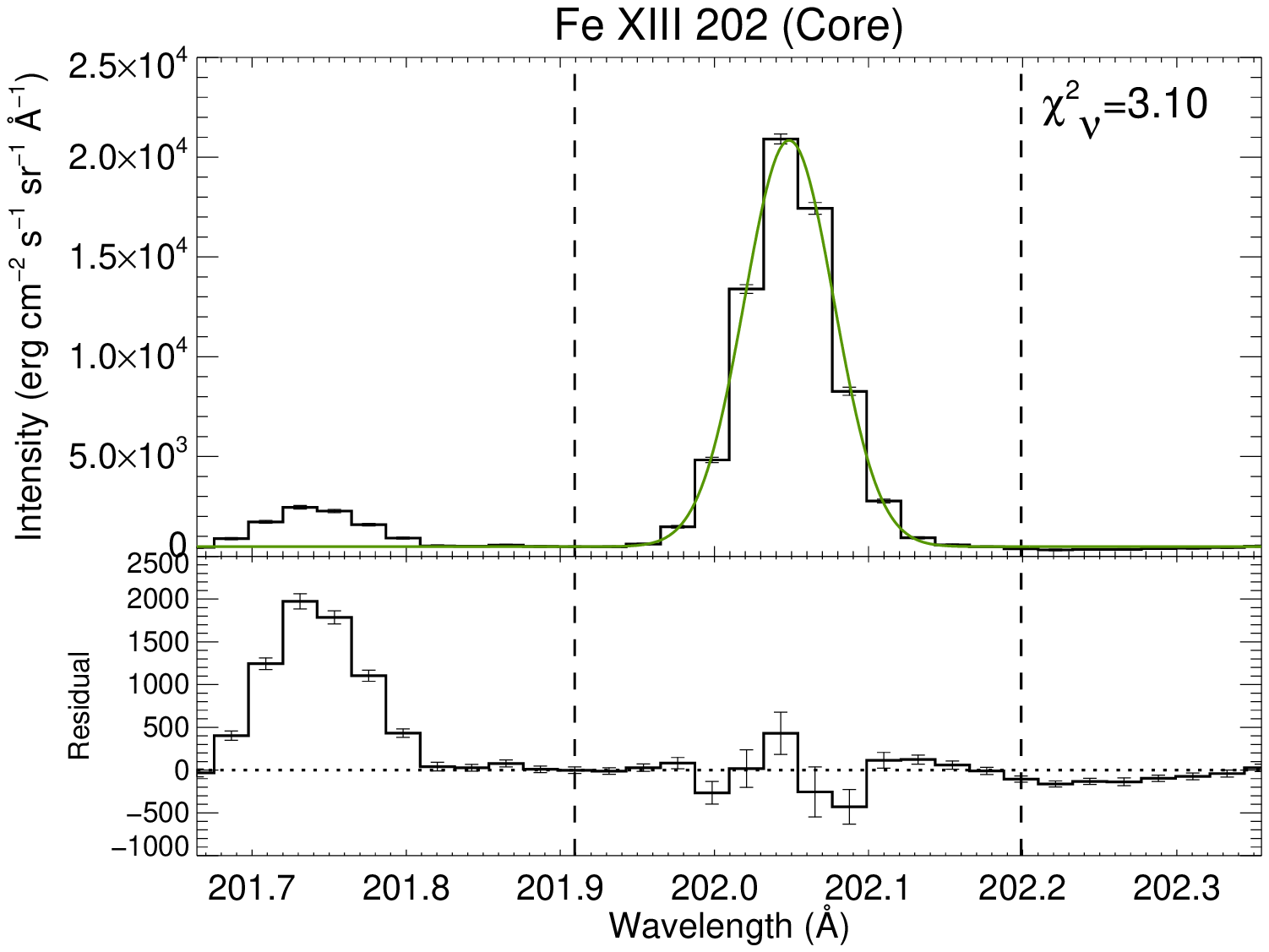}
  \includegraphics[width=8.3cm,clip]{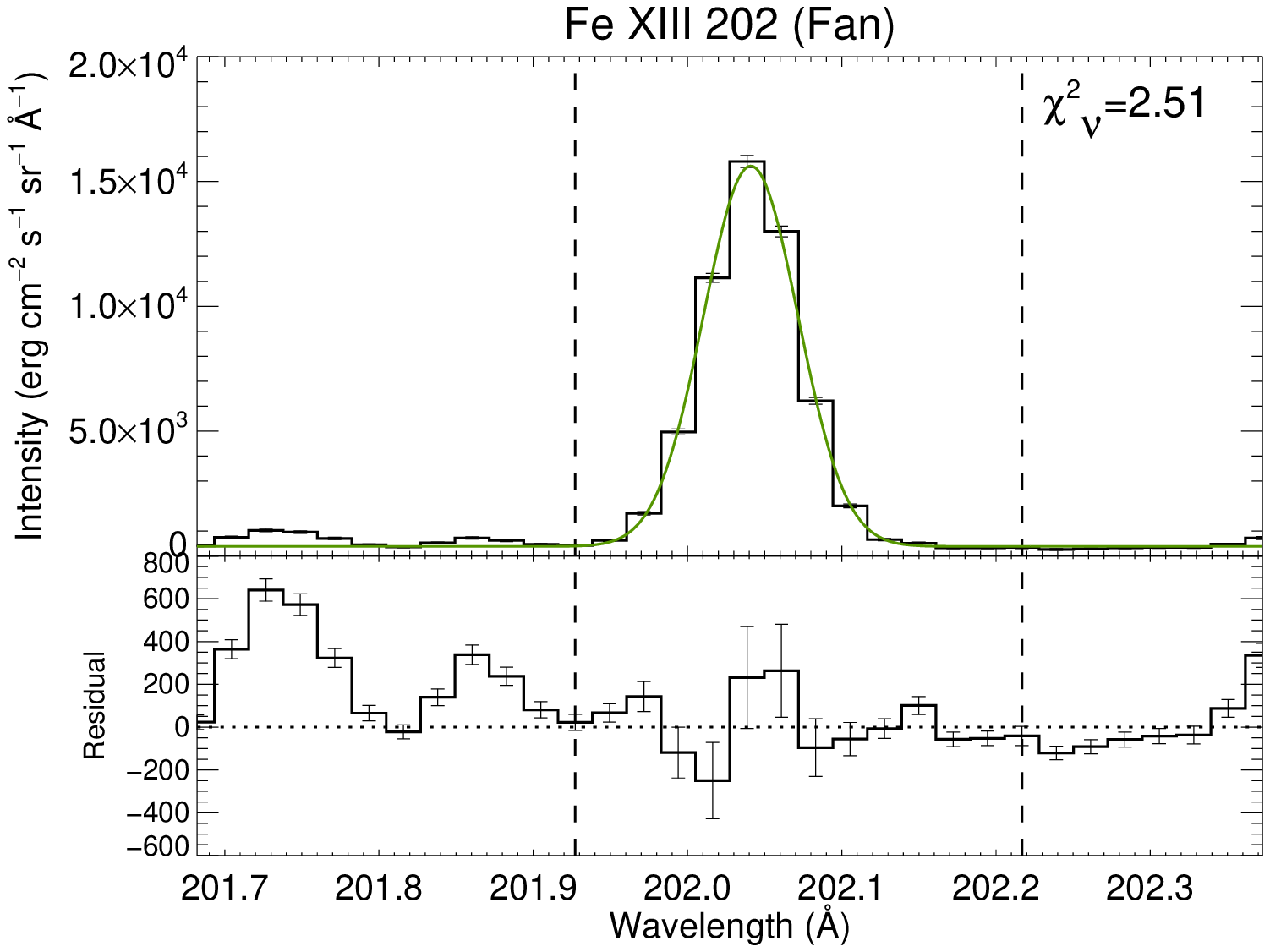}
  \includegraphics[width=8.3cm,clip]{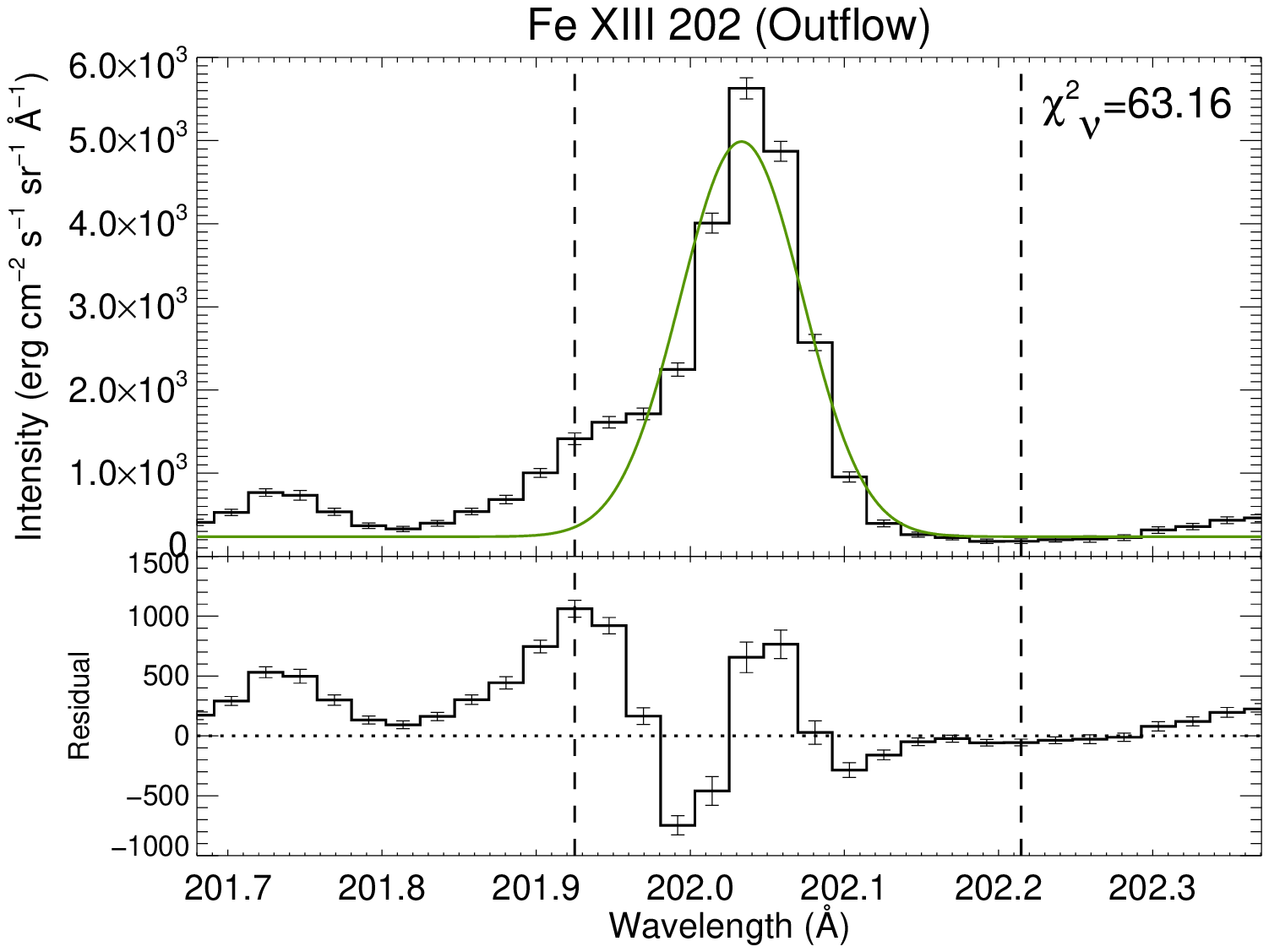}
  \caption{Line profiles of Fe \textsc{xiii} $202.04${\AA} and 
    their residual from a single Gaussian. 
    \textit{Upper left:} the quiet region. 
    \textit{Upper right:} the active region core.
    \textit{Lower left:} fan loop.
    \textit{Lower right:} the outflow region.}
  \label{fig:vel_lp_res_202}
\end{figure}

In order to look into the deviation from symmetric profile at the outflow region, we show four line profiles of Fe \textsc{xiii} $202.04${\AA} in Fig.~\ref{fig:vel_lp_res_202}.  This emission line exhibited the most significant enhancement of its blue wing in the outflow region. Fig.~\ref{fig:vel_lp_res_202} shows the line profile in the \textit{upper} part and the residual from a single Gaussian in the \textit{lower} part in each panel. A green line plotted over on the spectrum in each panel indicates a single Gaussian fitted to it. An emission line at around $\lambda=201.7$--$201.8${\AA} is Fe \textsc{xi} $201.73${\AA}, which is weaker than Fe \textsc{xiii} $202.04${\AA} in any regions. It is confirmed that the line profiles in the quiet region, the core region, and a fan loop are well represented by a single Gaussian as seen in the residuals. On the other hand, the line profile in the outflow region significantly deviates from a single Gaussian, and the deviation crudely reaches up to $1000 / 1.5 \times 10^{4} \simeq 0.07 = 7${\%} in the peak of major component.  The deviation from the single Gaussian can be evaluated quantitatively by the reduced chi square $\chi^2_{\nu}=\Sigma_{i} \left[\phi (\lambda_{i}) - y_{\mathrm{fit}} (\lambda_{i})\right]^2 / \nu$, where $\phi (\lambda_{i})$ is the spectral intensity at $\lambda = \lambda_{i}$, $y_{\mathrm{fit}} (\lambda)$ is the fitted Gaussian, and $\nu$ is the number of degrees of freedom.  The value of $\chi^2_{\nu}$ was written beside the fitted range indicated by two \textit{vertical dashed lines}. 

\begin{figure}
  \centering
  \includegraphics[width=8.3cm,clip]{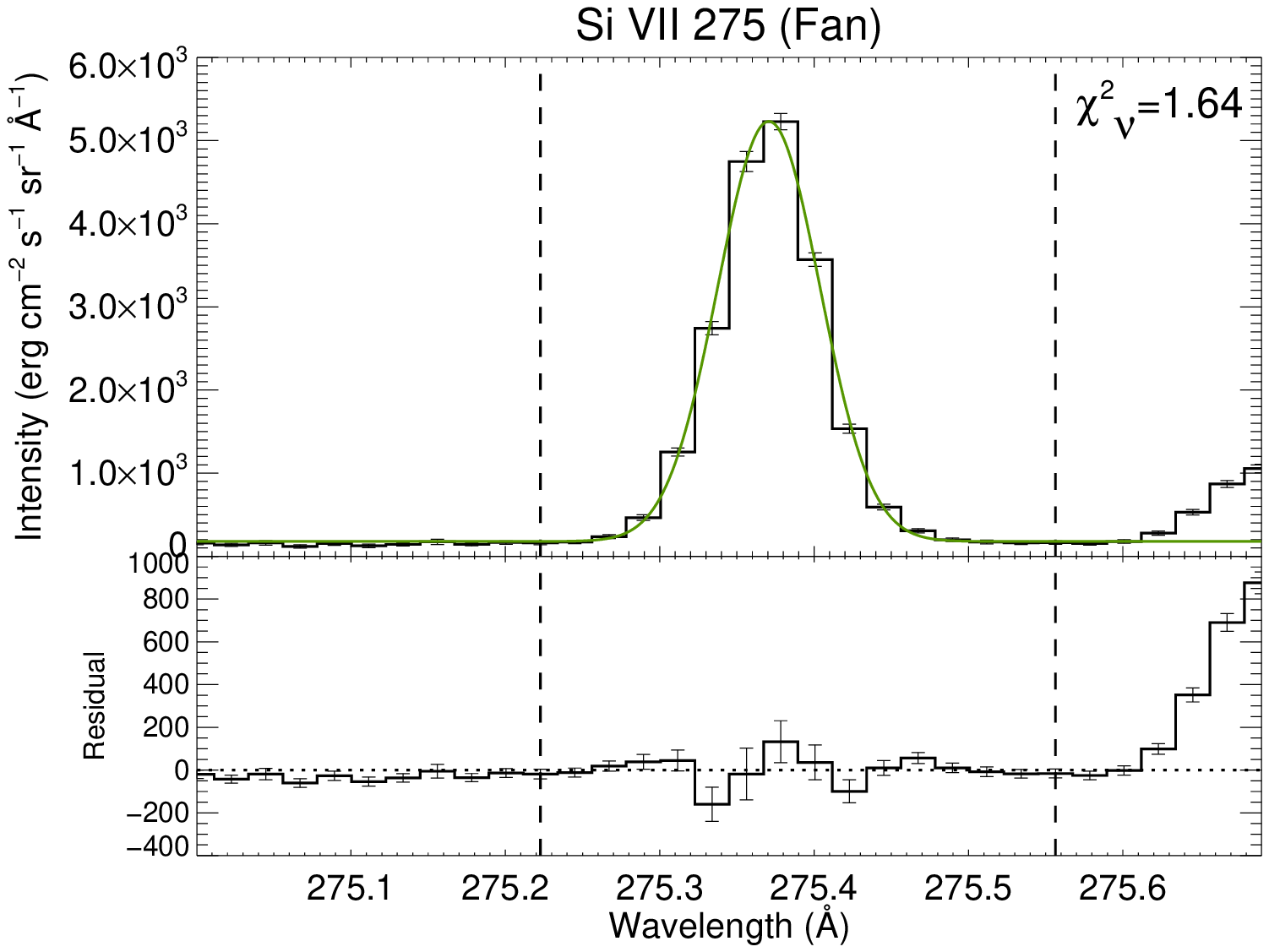}
  \includegraphics[width=8.3cm,clip]{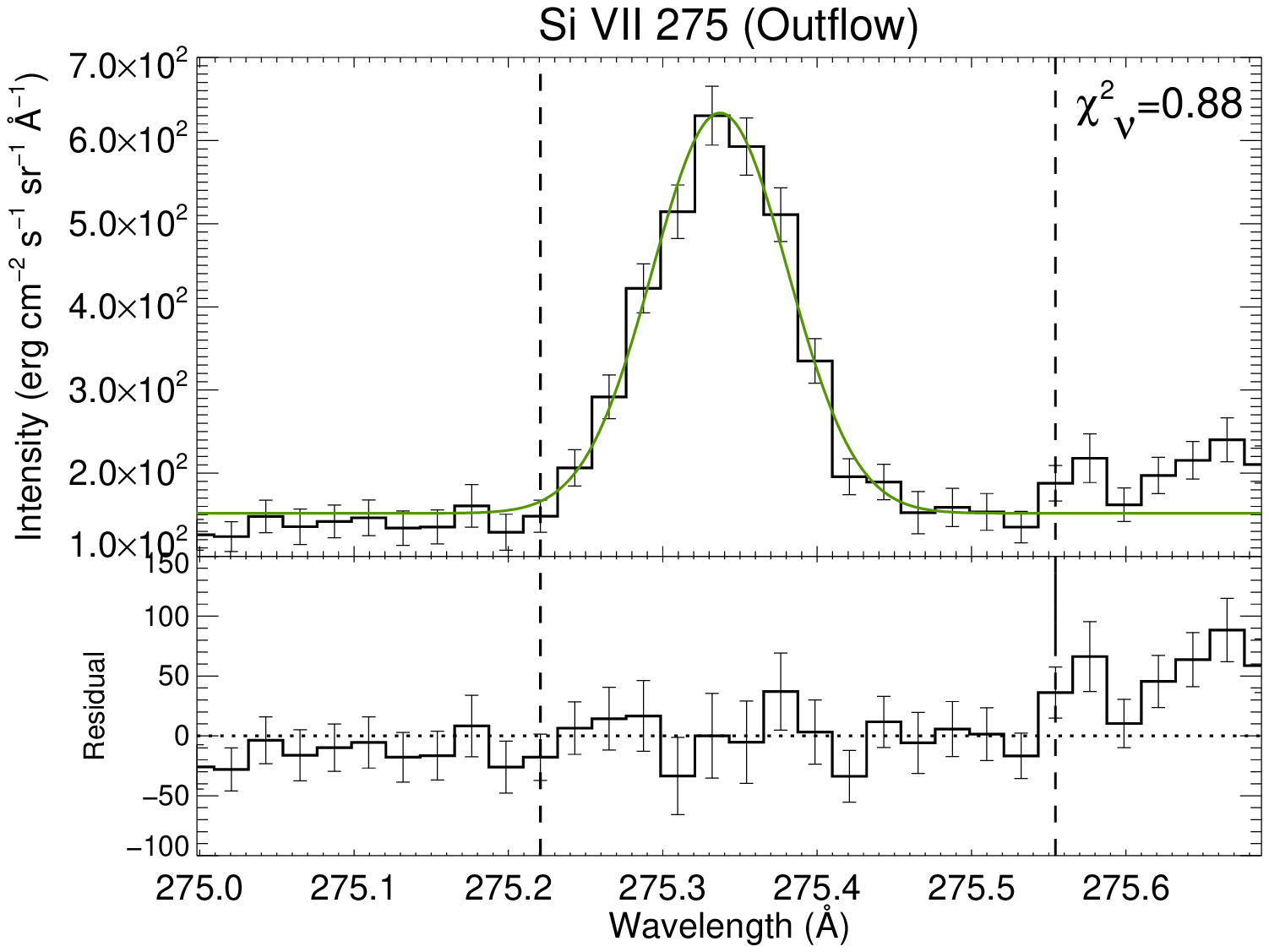}
  \caption{Line profiles of Si \textsc{vii} $275.35${\AA} ($\log T \, [\mathrm{K}]=5.80$) 
    and their residual from a single Gaussian.
    \textit{Left:} in a fan loop. 
    \textit{Right:} in the outflow region.}
  \label{fig:vel_lp_res_275}
\end{figure}

\begin{figure}
  \centering
  \includegraphics[width=8.3cm,clip]{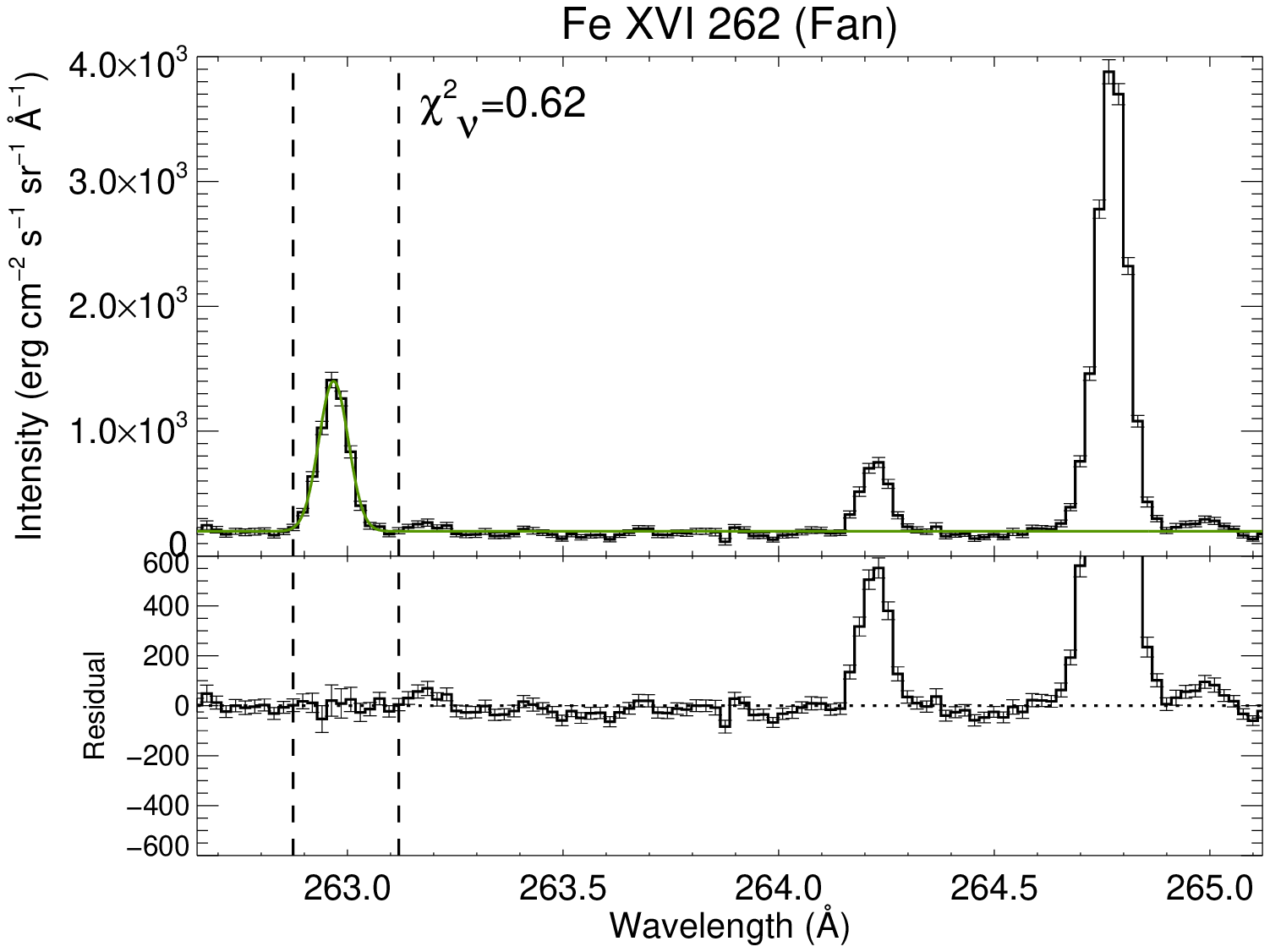}
  \includegraphics[width=8.3cm,clip]{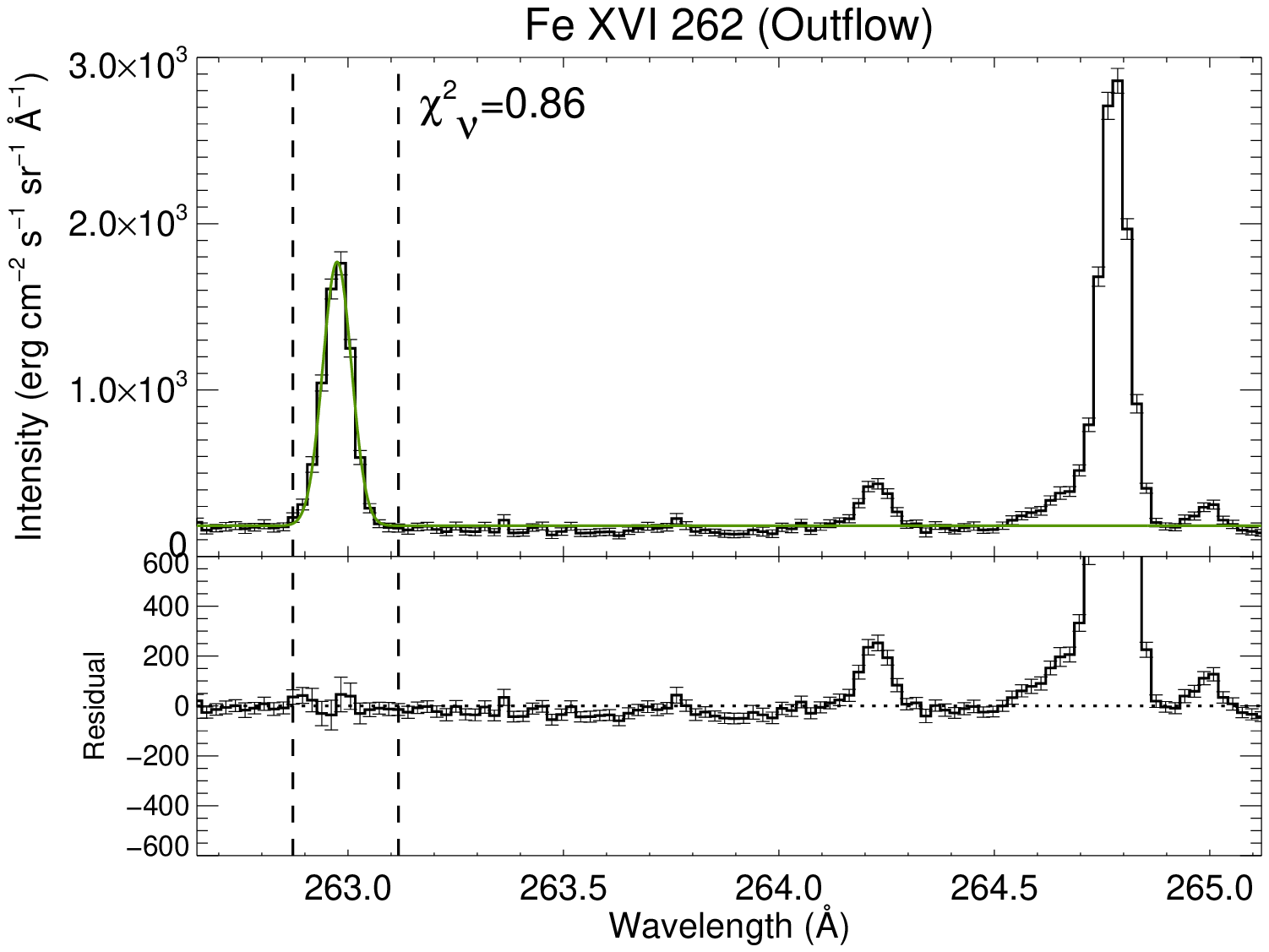}
  \caption{Line profiles of Fe \textsc{xvi} $262.98${\AA} ($\log T \, [\mathrm{K}]=6.45$)
    and their residual from a single Gaussian.
    \textit{Left:} in a fan loop.
    \textit{Right:} in the outflow region.}
  \label{fig:vel_lp_res_262}
\end{figure}

As an example of a line profile from the temperature below $\log T \, [\mathrm{K}]=6.0$, those of Si \text{vii} $275.35${\AA} ($\log T \, [\mathrm{K}]=5.80$) are shown in Fig.~\ref{fig:vel_lp_res_275}. In contrast to Fe \textsc{xiii}, this transition region emission line shows almost no significant deviations from a single Gaussian. There might be a little excess in the line profile in the outflow region around $275.27${\AA}, but since it is only a single pixel, it could be better not to think this excess as significant.  As inferred in Section \ref{sect:lp_ht}, the hottest emission line analyzed here, Fe \textsc{xvi} $262.98${\AA} ($\log T \, [\mathrm{K}]=6.45$) neither did not have significant deviations from a single Gaussian as shown in Fig.~\ref{fig:vel_lp_res_262}. 

\begin{figure}
  \centering
  \includegraphics[width=15cm,clip]{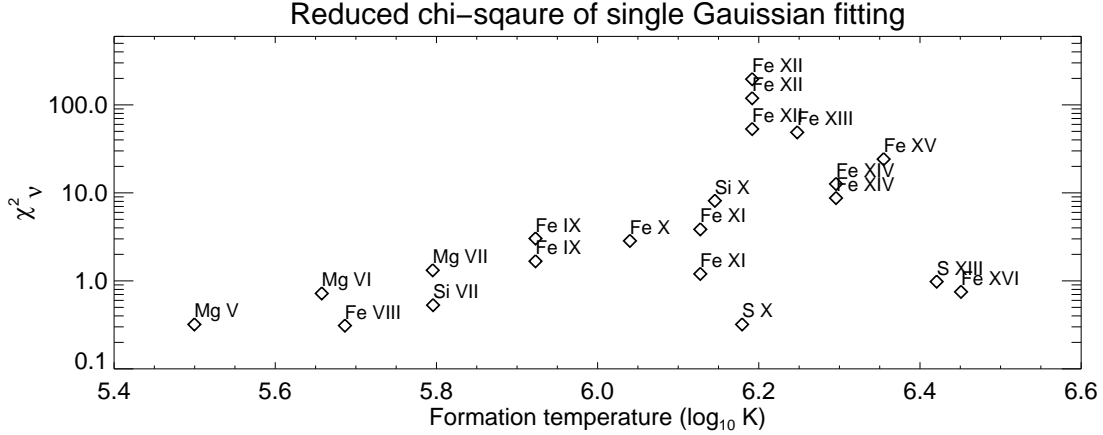}
  \caption{$\chi^2_{\nu}$ of the single-Gaussian fitting for emission lines within the temperature range of $\log T \, [\mathrm{K}]=5.5\text{--}6.5$. }
  \label{fig:vel_tvschi}
\end{figure}

The temperature dependence of $\chi^2_{\nu}$ for the temperature range of $\log T \, [\mathrm{K}]=5.5-6.5$ is shown in Fig.~\ref{fig:vel_tvschi}.  It is clearly shown that Fe \textsc{xii}--\textsc{xv} exhibit largest $\chi^2_{\nu}$.  The transition region lines (Mg \textsc{v}--\textsc{vii}, Si \textsc{vii}, etc.) and two hot emission lines (S \textsc{xiii} and Fe \textsc{xvi}) indicate much less $\chi^2_{\nu}$ of the order of unity as expected from Fig.~\ref{fig:vel_lp_res_275} and \ref{fig:vel_lp_res_262}.  The deviations for Fe \textsc{xii}--\textsc{xv} are caused by the existence of enhanced blue wing, and this plot shows that the enhanced blue wings are distributed around the temperature of $\log T \, [\mathrm{K}] = 6.2-6.3$.  Note that the number of fitted bins is not the same for the emission lines so that we could not directly interpret this plot as a significance probability.  

% --- End of Tex ---

%% file: tex/contents_dns.tex
\chapter{Density of the upflows}
  \label{chap:dns}
\section{Introduction}
\input{tex/dns_itdn.tex}
\section{Observation and calibration}
  \subsection{EIS raster scan}
    \input{tex/dns_raster_scans.tex}
  \subsection{Relative wavelength calibration}
    \label{sec:wavelength_adjust}
    \input{tex/dns_wvl_cal.tex}
\section{Density diagnostics of upflows}
    \label{subsect:dns_diag}
    \input{tex/dns_analysis_preface.tex}
  \subsection{Integration of observational pixels}
    \input{tex/dns_analysis_integration_pixels.tex}
  \subsection{De-blending of Si \textsc{vii} from Fe \textsc{xiv} 274.20{\AA}}
    \label{sec:de-blend}
    \input{tex/dns_analysis_de-blending.tex}
  \subsection{Simultaneous fitting of the two Fe \textsc{xiv} emission lines}
    \input{tex/dns_analysis_simul_fit.tex}
  \subsection{Density inversion}
    \label{sec:dns_inv}
    \input{tex/dns_analysis_dens_inv.tex}
\section{Density derived from Fe \textsc{xiv} 264.78{\AA}/274.20{\AA}}
  \label{sect:dns_results}
  \subsection{Results from single Gaussian fitting}
    \input{tex/dns_results_1G.tex}
  \subsection{Density of the upflows}
    \input{tex/dns_results_n.tex}
  \subsection{Column depth}
    \input{tex/dns_results_h.tex}
  \subsection{Uncertainty in Si \textsc{vii} density}
    \input{tex/dns_results_sivii.tex}
\section{Summary and discussion}
  \label{sect:dns_sum}
  \input{tex/dns_sum.tex}
\clearpage
\begin{subappendices}
\section{Fe \textsc{xvi} 262.98{\AA} and 265.01{\AA}}
  \label{sect:262n265}
  \input{tex/dns_app_262n265.tex}
\section{Fe \textsc{xiv} 264.78{\AA} intensity and electron density}
  \label{sect:dns_app_in}
  \input{tex/dns_app_in.tex}
  \section{Electron density at the footpoints of the outflow region}
    \label{sect:dns_app_mgvii}
    \input{tex/mph_mgvii_dens.tex}
\end{subappendices}

%% file: tex/dns_itdn.tex
% =========================
%   Project:
%     Density of upflows
% =========================

% Importance of density measurement in the upflow region and introduction to diagnostics
Although a number of observations have revealed the physical properties such as the source region and velocity of the outflows from the edge of active regions, there is one remained missing property: the electron density of the outflow itself.  The density of an outflow region derived by using the line ratio of Fe \textsc{xii} $186.88${\AA}/$195.12${\AA} was $\simeq 7 \times 10^{8} \, \mathrm{cm}^{-3}$ \citep{doschek2008}, which is slightly lower than the typical value in active region ($n_{\mathrm{e}} \ge 10^9 \, \mathrm{cm}$).  Recently, \citet{brooks2012} carried out differential emission measure (DEM) analysis at the outflow regions.  It was revealed that the properties of DEM and also the chemical abundance are rather close to those of active region, from which the authors concluded that the outflowing plasma originate in the active region loops.  The interchange reconnection was considered to be a candidate for accelerating the plasma into the outer atmosphere \citep{baker2009}. 

% Why density measurement of upflows?
The electron density of the outflow itself should help us to better understand the nature of the outflows, however, there has been no such challenge until present.  One point of view is that those outflows are directly linked to the coronal heating in a way which the outflowing plasma fills the outer atmosphere and form the corona \citep{depontieu2009,mcintosh2012}.  The impulsive heating in a coronal loop induces an upflow from its footpoint, which is a possible candidate what we see as the outflow \citep{delzanna2008,hara2008}.  Outflows can be also caused by the sudden change of the pressure environment in a coronal loop \citep{bradshaw2011}.  

% Klimchuk (2012)
The analytical approach was recently proposed in terms of the ratio of the electron density between major rest component ($n_{\mathrm{Major}}$) and minor outflow component (enhanced blue wing, $n_{\mathrm{EBW}}$) in coronal emission line profiles \citep{klimchuk2012}.  It was shown that if the tips of the chromospheric spicules supply the coronal plasma \citep{depontieu2011}, that ratio (here after denoted as $n_{\mathrm{EBW}}/n_{\mathrm{Major}}$) takes a value of an order of $10$--$100$, while tiny impulsive heating (\textit{i.e.}, nanoflare) creates the ratio of $0.4$--$1$ \citep{patsourakos2013}. Thus, it was suggested that the ratio $n_{\mathrm{EBW}}/n_{\mathrm{Major}}$ can be used as a diagnostic tool which enables us to discriminate these two mechanisms in the corona.

% What will this study reveal?
\begin{figure}
  \centering
  \includegraphics[width=8.4cm,clip]{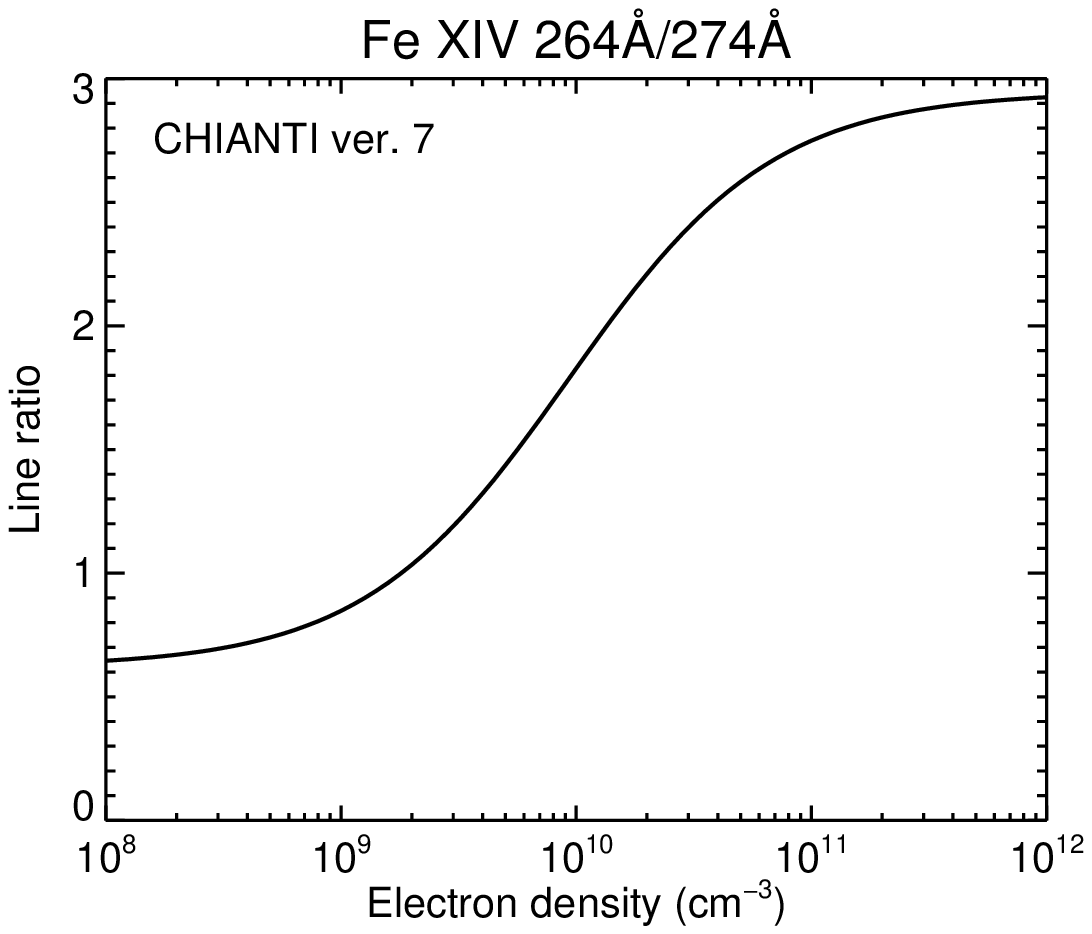}
  \includegraphics[width=8.4cm,clip]{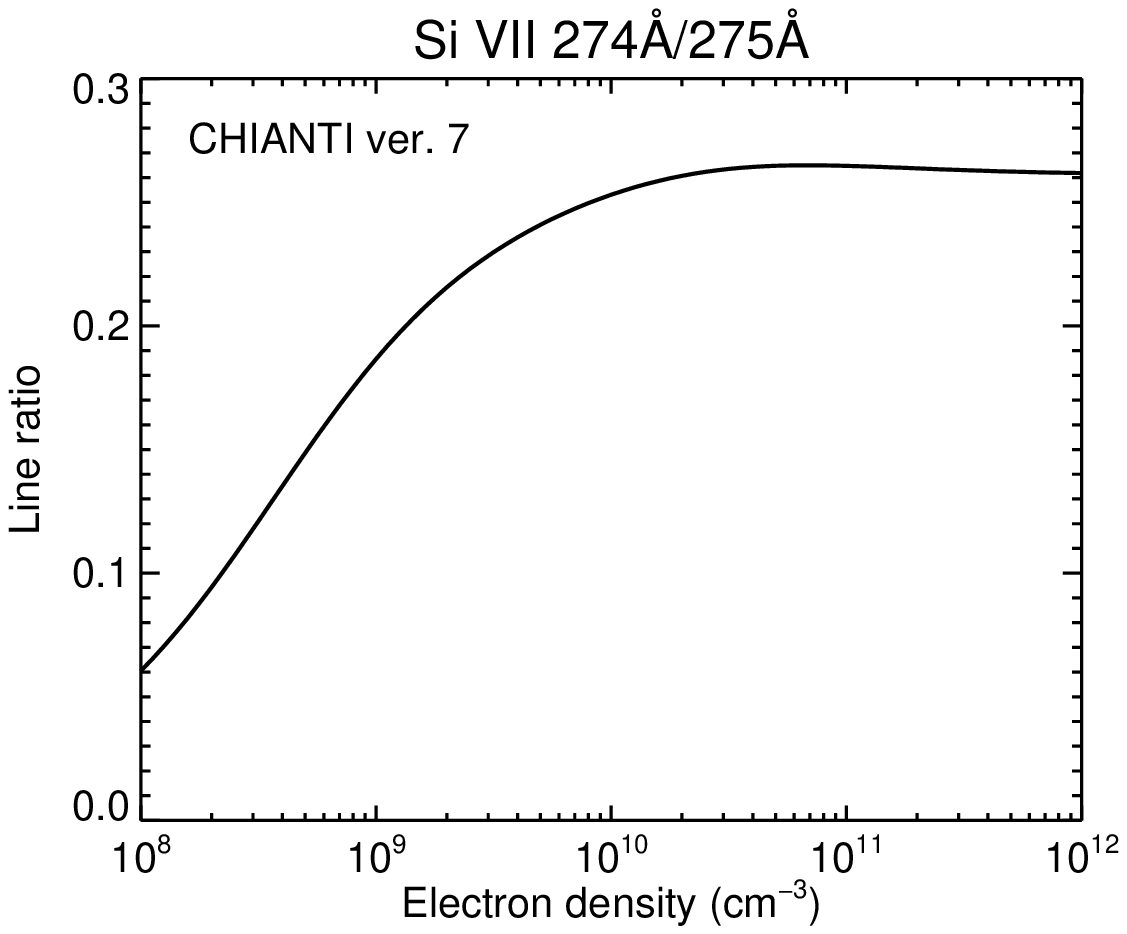}
  \caption{Theoretical line ratio calculated by CHIANTI database ver.~7 \citep{dere1997,landi2012}. (a) Fe \textsc{xiv} $264.78${\AA}/$274.20${\AA}. (b) Si \textsc{vii} $274.18${\AA}/$275.35${\AA}.}
  \label{fig:dns_itdn_rat_chianti}
\end{figure}

In this study, we used the spectroscopic data obtained with EIS onboard \textit{Hinode} in order to measure the electron density of the outflows for the first time.  As a line pair suitable for our purpose, Fe \textsc{xiv} $264.78${\AA} and $274.20${\AA} were chosen because (1) those emission lines have a distinct enhanced blue wing at the outflow region which leads to better signal-to-noise ratio, (2) they consist of relatively clean emission lines and their line wings in shorter wavelength side do not overlap with other emission lines, different from the cases for Fe \textsc{xii} $186.88${\AA}/$195.12${\AA} and Fe \textsc{xiii} $202.04${\AA}/$203.83${\AA}, and (3) the Fe \textsc{xiv} line pair is sensitive to the density range of $n_{\mathrm{e}} = 10^{8\text{--}12} \, \mathrm{cm}^{-3}$, which is wider than other line pairs.  

% --- End of Contents ---

%% file: tex/dns_raster_scans.tex
% ==================================
%   Project: 
%     DNS (density of upflows)
%   Description:
%     EIS raster scans of AR10978
% ==================================

In this study, we analyzed a raster scan obtained with \textit{Hinode}/EIS, which observed active region NOAA AR10978 (hereafter AR10978) at the center of the solar disk.  Overall properties of the active region was described in the previous chapter.  The scan with narrow $1''$ slit started on 2007 December 11 00:24:16UT and ended at 04:47:29UT. Field of view (FOV) was $256'' \times 256''$ and exposure time was $60 \, \mathrm{s}$.  The EIS data was processed through the standard software which detects the cosmic ray hits on the CCD pixels, subtracts the dark current bias, and corrects DN at warm pixels.  The DN is converted into the unit of intensity: $\mathrm{erg} \, \mathrm{cm}^{-2} \, \mathrm{s}^{-1} \, \mathrm{sr}^{-1} \, \text{\AA}^{-1}$.  This quantity should be called as \textit{spectral intensity} in the literature, however, we use a term \textit{intensity} for the simplicity.  One complicated point in the calibration is the thermal drift of the projected location on the CCD pixels due to the orbital motion of \textit{Hinode}.  Since the relative position of two emission lines Fe \textsc{xiv} $264.78${\AA} and $274.20${\AA} is the most important factor in this analysis, we roughly calibrated the wavelength through the method developed by \citet{kamio2010}.  

% --- End of Contents ---

%% file: tex/dns_wvl_cal.tex
% ========================================
%   Wavelength for simultaneous fitting
% ========================================

In order to make a precise fitting for Fe \textsc{xiv} $264.78${\AA}/$274.20${\AA} simultaneously, we have to know the wavelength positions of each emission line on the EIS spectrum which correspond to the same Doppler velocity.  Considering the Doppler effect, centroids of Fe \textsc{xiv} $264.78${\AA} and $274.20${\AA} shift as
\begin{align}
  & \lambda_{\mathrm{obs,}\, 274} 
  = \left( 1 + \frac{v_{\mathrm{LOS}}}{c} \right) \lambda_{0, \,274} 
  \label{eq:dns_wvl_cal_1}\\
  & \lambda_{\mathrm{obs,}\, 264} 
  = \left( 1 + \frac{v_{\mathrm{LOS}}}{c} \right) \lambda_{0, \,264}, 
  \label{eq:dns_wvl_cal_2}
\end{align}
where $\lambda_{\mathrm{obs,} 274}$ and $\lambda_{\mathrm{obs,} 264}$ are observed centroids, $v_{\mathrm{LOS}}$ is line-of-sight velocity (positive means that the plasma move away from an observer), and $\lambda_{0, 274}$ and $\lambda_{0, 264}$ are the rest wavelength that we do not know.  We can safely assume Doppler velocities in each equation have the same value because these two emission lines are radiated from the same degree of Fe ions. Dividing Eq.~(\ref{eq:dns_wvl_cal_1}) by Eq.~(\ref{eq:dns_wvl_cal_2}) leads to
\begin{equation}
  \frac{\lambda_{\mathrm{obs,} \, 274}}{\lambda_{\mathrm{obs,} \, 264}} 
    = \frac{\lambda_{0, \, 274}}{\lambda_{0, \, 264}}.
  \label{eq:dns_wvl_cal_3}
\end{equation}
This equation means that the ratio of the centroid wavelengths in two emission lines becomes constant independent of the Doppler velocity of the emitting plasma.  Once we obtain the constant $\lambda_{0, \, 274} / \lambda_{0, \, 264}$, it is possible to determine wavelength positions of the two emission lines corresponding to the same Doppler velocity without absolute calibration of the zero-velocity point.  

\begin{figure}
  \centering
  \includegraphics[width=8.4cm,clip]{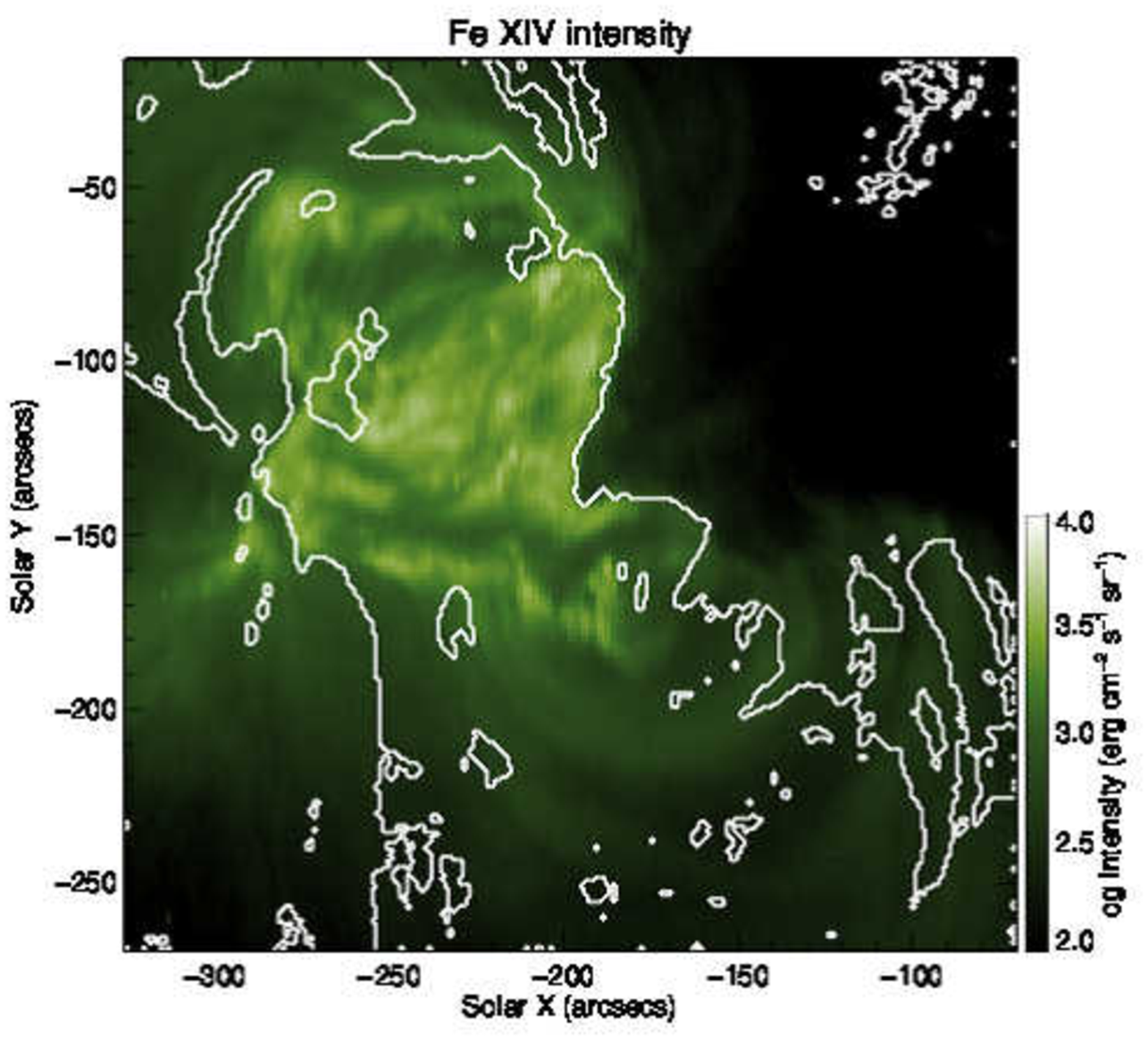}
  \includegraphics[width=8.4cm,clip]{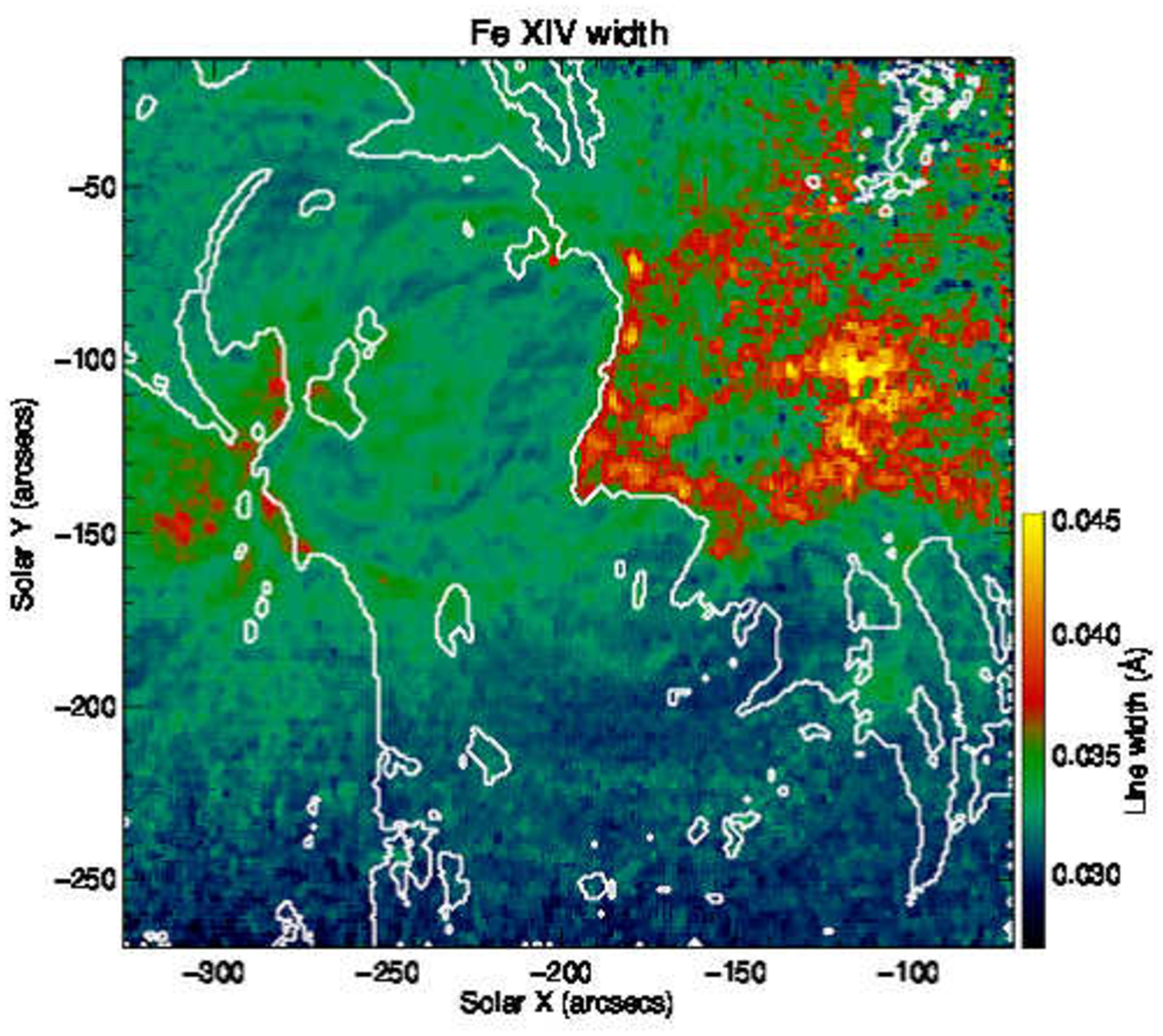}
  \caption{\textit{Left}: map of Fe \textsc{xiv} $264.78${\AA} intensity on 2007 December 11. \textit{Right}: map of Fe \textsc{xiv} $264.78${\AA} line width. White contours indicate the boundary of the area excluded when making the histogram of the ratio $\lambda_{\mathrm{obs,} 274} / \lambda_{\mathrm{obs,} 264}$.}
  \label{fig:map_wvl_cal}
\end{figure}

In order to determine the constant $\lambda_{0, \, 274}/\lambda_{0, \, 264}$ in Eq.~(\ref{eq:dns_wvl_cal_3}), we obtained the centroids $\lambda_{\mathrm{obs,} \, 274}$ and $\lambda_{\mathrm{obs,} \, 264}$ from the EIS scan.  We excluded the region where (1) line blend is significant, (2) the intensity is low, or (3) line profiles of Fe \textsc{xiv} obviously deviate from single Gaussian.  There is a weak Si \textsc{vii} 274.18{\AA} emission line near Fe \textsc{xiv} $274.20${\AA}, separated by roughly one EIS spectral pixel.  This may cause spurious blueshift of Fe \textsc{xiv} $274.20${\AA}. When considering the extreme example, if Si \textsc{vii} 274.18{\AA} has the same intensity as Fe \textsc{xiv} $274.20${\AA}, the fitted centroid becomes 274.19{\AA} (\textit{i.e.,} mean of the two emission lines), which means a blueshift of $\sim 10 \, \mathrm{km} \, \mathrm{s}^{-1}$. To avoid this spurious effect, we excluded the pixel for the intensity ratio Si \textsc{vii} $275.35${\AA}/Fe \textsc{xiv} $264.78${\AA} exceed 0.1, which might produce the spurious blueshift of $\sim 0.5 \, \mathrm{km} \, \mathrm{s}^{-1}$ in the maximum. Secondly, the region where the intensity of Fe \textsc{xiv} $264.78${\AA} is lower than $1500 \, \mathrm{erg} \, \mathrm{cm}^{-2} \, \mathrm{s}^{-1} \, \mathrm{sr}^{-1}$ was excluded to reduce the uncertainty in determining the value of Eq. (\ref{eq:dns_wvl_cal_3}).  Thirdly, we should exclude the area where the line profile significantly deviates from single Gaussian.  %This is naturally satisfied when the condition (2) was applied as later seen in Fig.~\ref{fig:map_wvl_cal}.

The boundary of excluded area is shown by white contours in Fig. \ref{fig:map_wvl_cal}. Only the active region core between the two white contours was used as seen from the map of Fe \textsc{xiv} intensity (the \textit{left} panel). The \textit{right} panel shows the map of line width, from which we can see that the outflow region was also excluded where the line profiles skews and have long tail in the shorter wavelength which produces enhanced line width when fitted by single Gaussian. 

\begin{figure}
  \centering
  \includegraphics[width=8.4cm,clip]{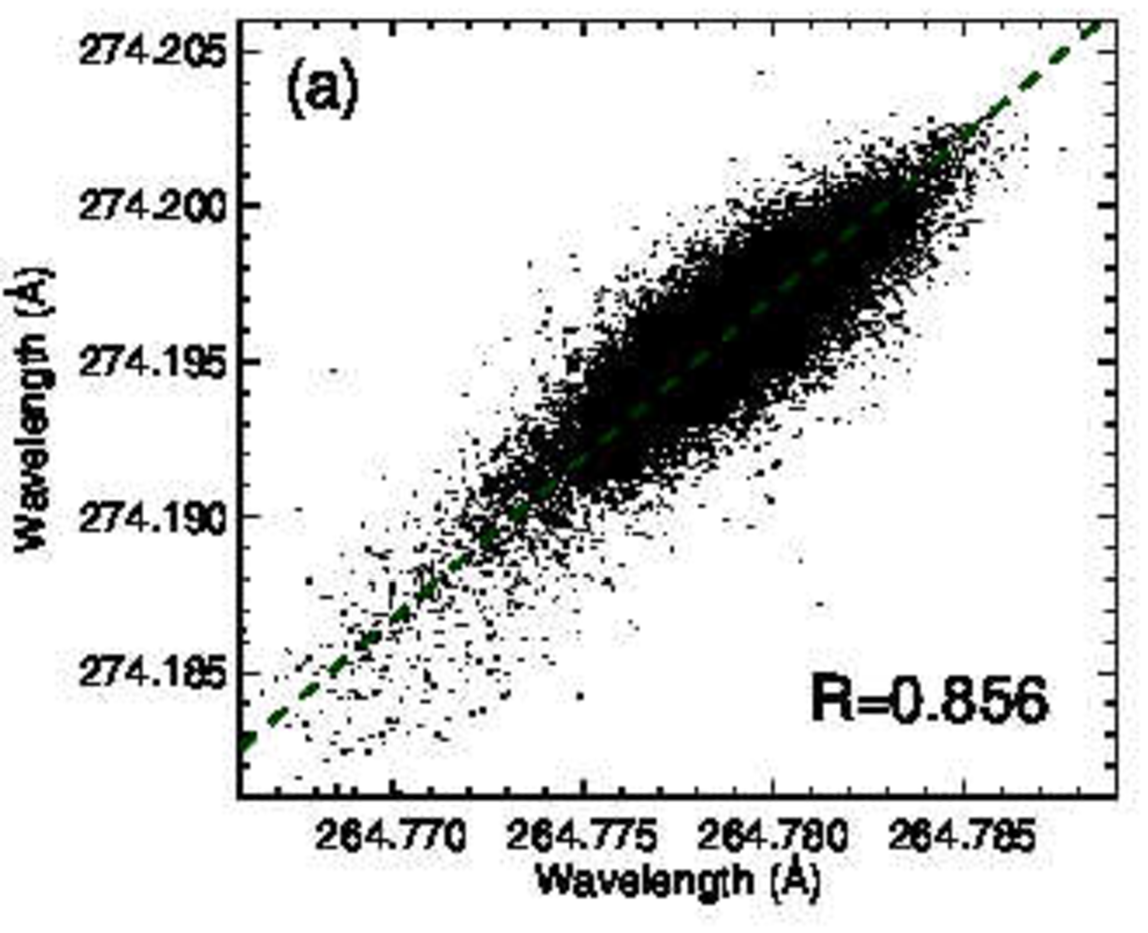}
  \includegraphics[width=8.4cm,clip]{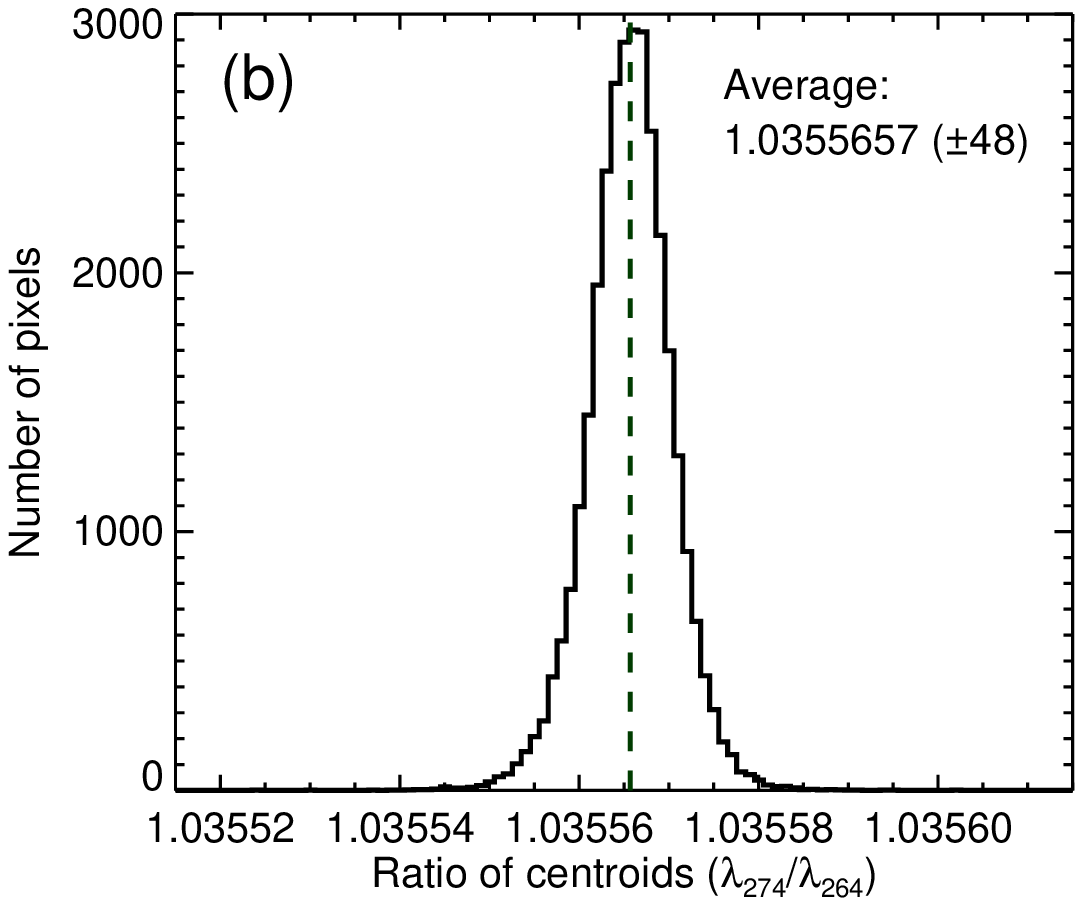}
  \caption{Histogram of the ratio of centroids for two emission lines Fe \textsc{xiv} 274.20/264.78 in the raster scan taken on 2007 December 11.  Vertical dashed line indicates the average value ($1.0355657 \pm 0.0000044$).}
  \label{fig:ratio_wvl_fexiv}
\end{figure}

Fig.~\ref{fig:ratio_wvl_fexiv} shows the histogram of the ratio $\lambda_{\mathrm{obs,} 274} / \lambda_{\mathrm{obs,} 264}$ after excluding the area describe above.  The average value ($\alpha$) was $\alpha=1.0355657$ and its standard deviation was $4.4 \times 10^{-6}$. We also derived the ratio $\lambda_{\mathrm{obs,} \,274} / \lambda_{\mathrm{obs,} \, 264}$ from another scan data in the same day (but different time; 10:25:42UT), which resulted in the exactly same average value but slightly larger standard deviation of $5.4 \times 10^{-6}$ which might came from the shorter exposure time ($40 \, \mathrm{s}$).  Note that the theoretical ratio predicted by CHIANTI is $1.0355559$ and it is within the deviation coming from the precision of the wavelength calibration for the EIS long wavelength band ($\simeq 0.002${\AA}; \citeauthor{brown2007}\ \citeyear{brown2007}).  We used the obtained $\alpha$ to fit the two emission lines from Fe \textsc{xiv} $264.78${\AA}/$274.20${\AA} simultaneously, and to confirm the results by exploiting a new method ($\lambda$-$n_{\mathrm{e}}$ diagram) in Chapter~\ref{chap:ndv}. 

% --- End of Tex ---

%% file: tex/dns_analysis_preface.tex
% ===================================
%   Density diagnostics of upflows
% ===================================

One of main achievements in this thesis is the density measurement of the outflow from the edge of an active region for the first time.  Previous observations have revealed that the density of the outflow region measured by using a line pair Fe \textsc{xii} $186.88${\AA}/$195.12${\AA} indicates $7 \times 10^{8} \, \mathrm{cm}^{-3}$ which is close to that of coronal holes rather than that of active regions \citep{doschek2008}.  However, density of the outflow itself, to be measured by separating its component from the major component in line profiles, has not been investigated so far. 

\begin{figure}
  \centering
  \begin{minipage}[c]{10.4cm}
    \includegraphics[width=10.4cm,clip]{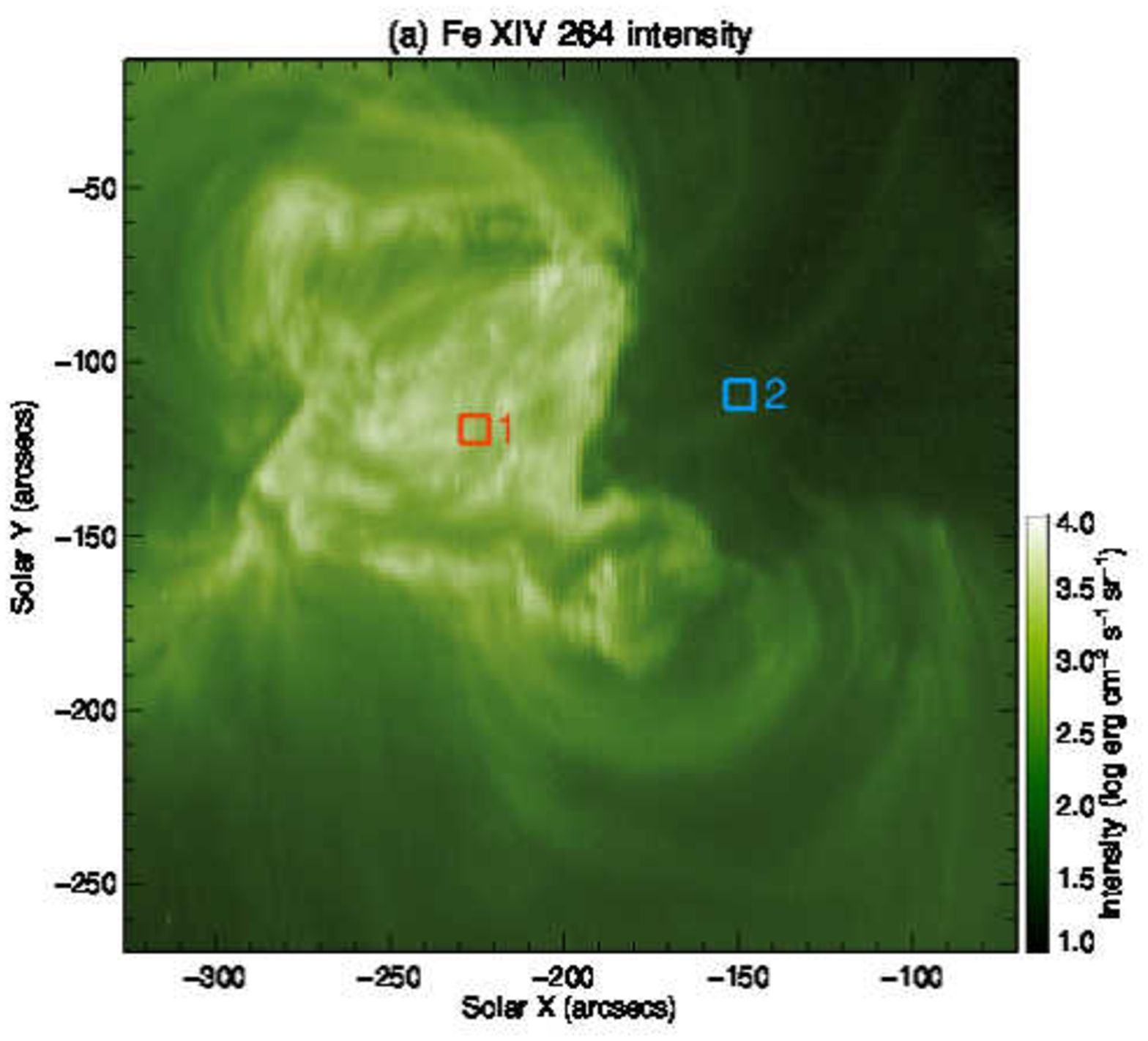}
  \end{minipage}  
  \begin{minipage}[c]{6.5cm}
    \includegraphics[width=6.5cm,clip]{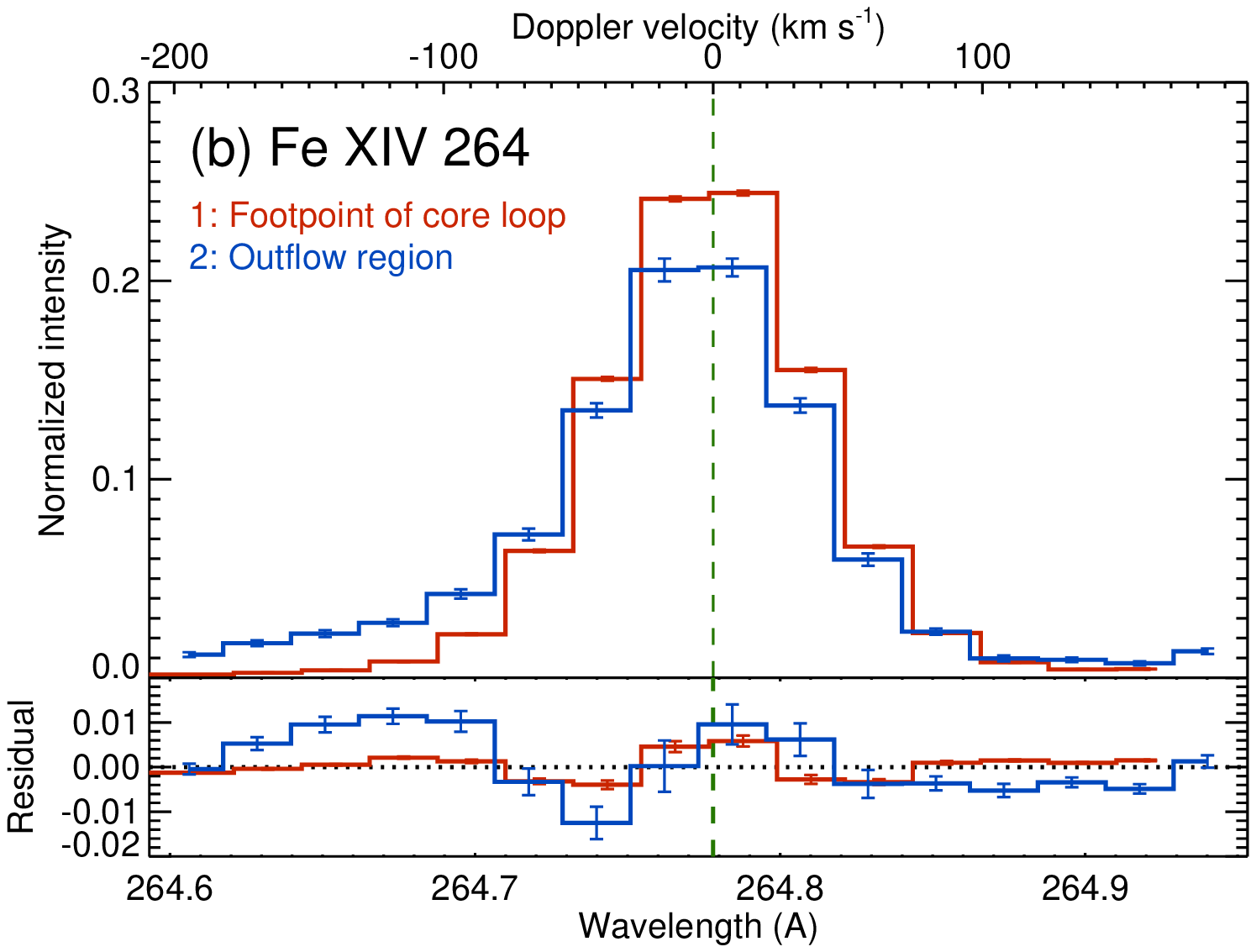}
    \includegraphics[width=6.5cm,clip]{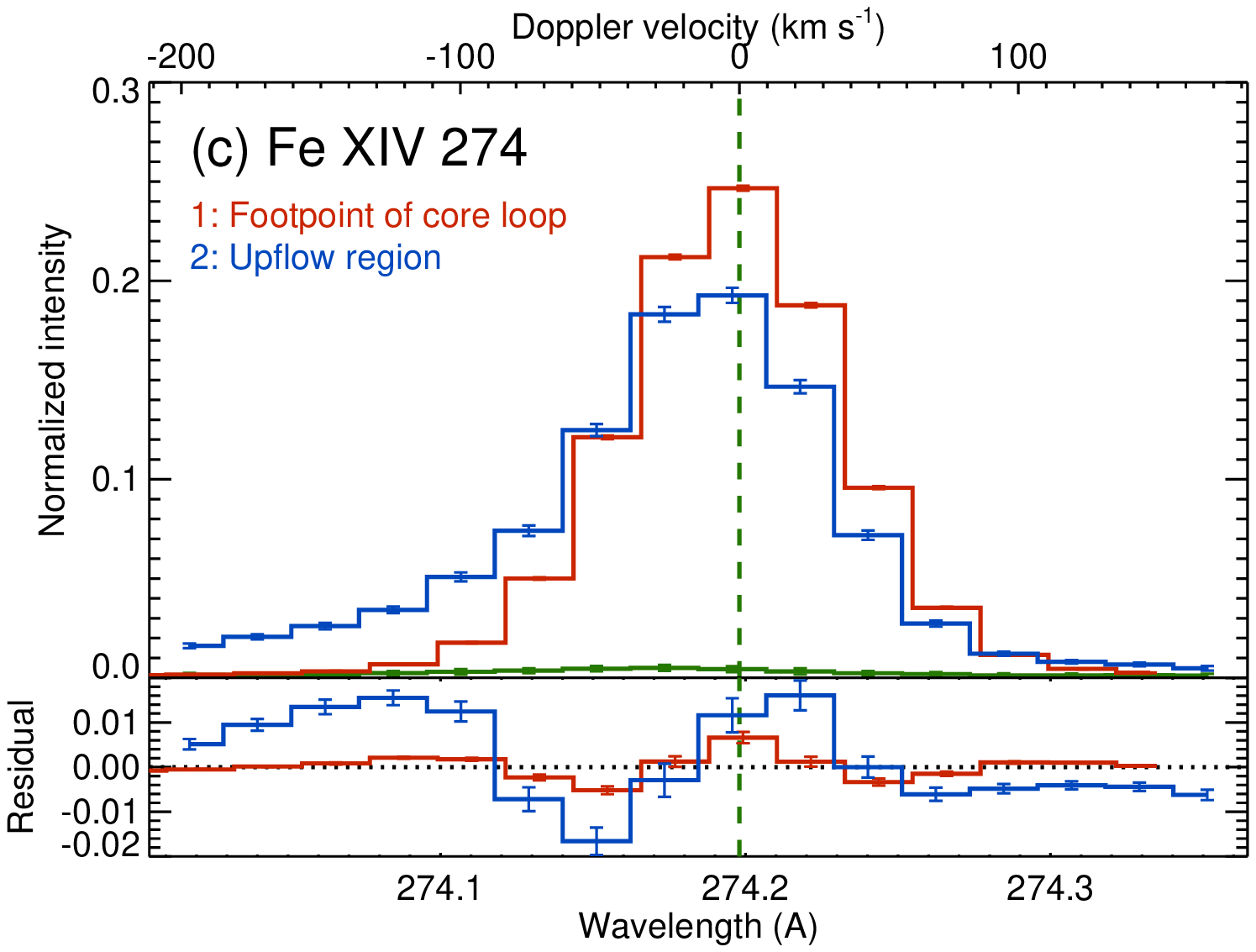}
  \end{minipage}
  \caption{Line profiles of the active region AR10978. (a) Context image of AR10978 obtained on 2007 December 11 00:24:16--04:47:29UT.  Intensity of Fe \textsc{xiv} $264.78${\AA} is shown.  Boxes numbered as 1 (\textit{red}) and 2 (\textit{blue}) respectively indicate a footpoint of core loops and the outflow region.  (b) Fe \textsc{xiv} $264.78${\AA} spectra.  (c) Fe \textsc{xiv} $274.20${\AA} spectra.  In each panel, line profiles at the footpoint of core loops (\textit{red} histogram) and at the outflow region analyzed here (\textit{blue} histogram) are shown in the \textit{upper} half.  The spectra were normalized by their integration.  Residuals from the single-Gaussian fitting of each histogram are shown in the \textit{lower} half.  \textit{Green} histogram in panel (c) shows estimated spectrum of Si \textsc{vii} $274.18${\AA}.}
  \label{fig:dns_lp_examples}
\end{figure}

There are three reasons of the difficulties in the analysis of spectroscopic data obtained by \textit{Hinode}/EIS.  Firstly, the signals from an upflow are detected as an enhanced blue wing (hereafter, EBW) component in emission line profiles.  Examples of them are shown in Fig.~\ref{fig:dns_lp_examples}.  In each panel, line profiles at the footpoint of a core loop (\textit{red} histogram) and at the outflow region analyzed here (\textit{blue} histogram) are shown in the \textit{upper} half.  Residuals from single-Gaussian fitting of each histogram are shown in the \textit{lower} half, which is quite useful in detecting weak signals in line wing \citep{hara2008}.  There is a significant enhancement at the blue wing ($\le -100 \, \mathrm{km} \, \mathrm{s}^{-1}$) both in Fe XIV $264.78${\AA} and $274.20${\AA} as shown by \textit{blue} histograms.  \textit{Green} histogram in panel (c) shows estimated spectrum of Si \textsc{vii} $274.18${\AA} which was subtracted in the density diagnostics described later.  The EBW component is weak in most cases as seen in spectra indicated by blue histograms shown in Fig.~\ref{fig:dns_lp_examples}.  In addition, EBW component is significantly dominated by the strong component almost at rest, which makes the analysis of upflows quite uncertain. 

Secondly, the density measurement of the outflow itself needs the accurate determination of the rest wavelengths of emission lines from which we fit the two emission lines simultaneously and deduce the intensity.  As described earlier in this thesis, this is often laborious because we do not have the absolute measure of the wavelength corresponding to each observational spectral pixels. 

Thirdly, density measurement needs at least two emission lines from the same ion (\textit{e.g.}, Fe \textsc{xiv} as used in this thesis).  This means that the two emission lines should be fitted simultaneously using same parameters such as Doppler velocity and line width.  No previous studies on the outflows from the edge of active region have dealt with such fitting. 

Procedure of density diagnostics in this thesis is as follows: (1) integration of neighboring multiple pixels in order to reduce the noise, (2) determination of the wavelength position corresponding to the same Doppler velocity, (3) removal of blending Si \textsc{vii} $274.18${\AA} from Fe \textsc{xiv} $274.20${\AA} using Si \textsc{vii} $275.35${\AA} as a reference, (4) simultaneous fitting of Fe \textsc{xiv} $264.78${\AA} and $274.20${\AA}, and (5) density inversion using a theoretical curve from CHIANTI as a function of the intensity ratio.  In following sections, each procedure will be described in detail. 

% --- End of Contents ---

%% file: tex/dns_analysis_integration_pixels.tex
% =======================================
%   Classification: analysis
%   Description: integration of pixels
% =======================================

The outflows from the edge of active regions are usually detected as an EBW in emission line profiles.  Its intensity does not exceed $\sim 25 \mathrm{\%}$ of that of the major component \citep{doschek2012}.  This makes the analysis difficult since the photon noise of the major component affects the emission from EBW.  In addition, the region where the outflows can be seen is usually dark (\textit{i.e.}, small signal-to-noise ratio).  In order to improve the signal-to-noise ratio, we integrated over multiple observational pixels in space using a square box with the size of $5'' \times 5''$.  Larger size of the integration box generally results in better signal-to-noise ratio, however, we chose the size of integration box so as not to lose the information of the outflow region.  In the integration, the pixels with instrumental problems (\textit{i.e.}, hot or bad pixels) were excluded. 

% --- End of contents ---

%% file: tex/dns_analysis_de-blending.tex
% =============================================================
%   Project: analysis
%   Description: De-blending of Si 274.18 from Fe XIV 274.20
% =============================================================
% General
%The spectroscopic analysis often becomes complicated when a neighbor emission line blends into the target emission line. 
%One of the famous blend is Fe XII $195.18 \, \mathrm{\AA}$ into Fe XII $195.12 \, \mathrm{\AA}$ (\textit{i.e.}, self blended) or Fe XI $192.81 \, \mathrm{\AA}$ and O V lines into Ca XVII $192.83 \, \mathrm{\AA}$. 
%In order to make robust analysis, we must subtract blended line(s) by using some references such like another emission line of the same ion as that to be subtracted. 
%But, for example, blended line Fe XII $195.18 \, \mathrm{\AA}$ is often ignored in the analysis of Fe XII $195.12 \, \mathrm{\AA}$ line when one studies the vicinity of active region or quiet Sun (\textit{i.e.}, relatively low-density region) because Fe XII 195.18 emission increases for higher density (especially for $\gtrsim 10^{10} \, \mathrm{cm}^{-3}$ in the active region core). 
%This thesis aims to deduce the density using a Fe \textsc{xiv} line pair, which needs accurate intensities. 

% Fe XIV 264.78
We should subtract blended lines from both Fe \textsc{xiv} lines.  Fe \textsc{xiv} $264.78${\AA} is isolated without significant contributions from other emission lines.  As for Fe \textsc{xvi} $265.01${\AA}, it is far enough from Fe \textsc{xiv} $264.78${\AA} in non-flare situation.  Moreover, estimated peak intensity of Fe \textsc{xvi} $265.01${\AA} was around $100 \, \mathrm{erg} \, \mathrm{cm}^{-2} \, \mathrm{s}^{-1} \, \mathrm{sr}^{-1} \, \text{\AA}^{-1}$ in the observed outflow region\footnote{We estimate the intensity by using Fe \textsc{xvi} $262.98${\AA} included in the data.  The line ratio Fe \textsc{xvi} $265.01${\AA}/$262.98${\AA} was determined in the raster scan used in Chapter \ref{chap:vel} and it resulted in the ratio of $0.083$.  See Appendix \ref{sect:262n265} for details.}, which is at most the background level of Fe \textsc{xiv} $264.78${\AA} as seen in Fig.~\ref{fig:fig_diff}. 

% Fe xiv 274.20 & Si vii 274.18
On the other hand, Fe \textsc{xiv} 274.20 potentially has a contribution from Si \textsc{vii} $274.18${\AA}, which may become significant in the vicinity of an active region because Si \textsc{vii} emission often comes from the footpoint of cool loops extending from the edge of the active region.  We should subtract this blend from Fe \textsc{xiv} $274.20${\AA}. 

In this study, the spectrum of Si \textsc{vii} $274.18${\AA} was calculated by using the observed line profile of Si \textsc{vii} $275.35${\AA} 
%{\color{red} 
which is known to be clean (\textit{i.e.}, without any significant blend).
%}  
The intensity ratio of Si \textsc{vii} $274.18${\AA}/$275.35${\AA} is at most $0.25$ as calculated from CHIANTI version 7 \citep{dere1997,landi2012}.  The value has a dependence in the density range $10^{8} \, \mathrm{cm}^{-3} \le n_{\mathrm{e}} \le 10^{10} \, \mathrm{cm}^{-3}$, and it varies $0.06$--$0.27$ (monotonically increasing) as shown in Fig.~\ref{fig:dns_itdn_rat_chianti}.  First we remove the blending Si \textsc{vii} $274.18${\AA} for the case $n_{\mathrm{e}} = 10^{9} \, \mathrm{cm}^{-3}$ (Si \textsc{vii} electron density), and after that we considered three cases of the ratio corresponding to the density of $10^{8}$, $10^{9}$, and $10^{10} \, \mathrm{cm}^{-3}$.  In order to make our analysis be more robust, we excluded the location where the estimated intensity of Si \textsc{vii} $274.18${\AA} exceeds $5${\%} of the Fe \textsc{xiv} intensity.  Using the theoretical ratio, the intensity of Si \textsc{vii} $275.35${\AA} was converted into that of Si \textsc{vii} $274.18${\AA}.  The spectrum of Si \textsc{vii} $275.35${\AA} was then placed at Si \textsc{vii} $274.18${\AA} taking into account the shift of Si \textsc{vii} $275.35${\AA} from the rest wavelength using the relative difference between wavelength of Si \textsc{vii} $274.18${\AA} and $275.35${\AA} (\textit{i.e.}, $1.1808${\AA}) given by CHIANTI database.  Note that since there were no locations where Si \textsc{vii} $274.18${\AA} dominates Fe \textsc{xiv} $274.20${\AA} in the data, we could not determine the relative wavelength position of the two Si \textsc{vii} lines, therefore we used the wavelength difference given by CHIANTI for the Si \textsc{vii} lines.  The data points of the estimated Si \textsc{vii} $274.18${\AA} in the wavelength direction were interpolated into the data points of Fe \textsc{xiv} $274.20${\AA} by cubic spline. Thus, we removed the blended Si \textsc{vii} $274.18${\AA} from Fe \textsc{xiv} $274.20${\AA}. 

% --- End of Contents ---

%% file: tex/dns_analysis_simul_fit.tex
% ======================================
%   Project: analysis
%   Description: simultaneous fitting
% ======================================

% Introduction
In order to make the fitting more robust, the two emission line profiles of Fe \textsc{xiv} $264.78${\AA}/$274.20${\AA} were fitted simultaneously.  It is based on the consideration that the emission line profiles coming from the same ion species must have the same Doppler shift and the same Doppler width. %The intensity ratio varies because that depends on the density. 
% Example of emission line profiles
As seen in Fig.~\ref{fig:dns_lp_examples}, emission line profiles of Fe \textsc{xiv} $264.78${\AA} and $274.20${\AA} from the active region core (\textit{red} histogram) are obviously symmetric, while those from the outflow region (\textit{blue} histogram) have an EBW.  This EBW did not exceed the major component at everywhere in the outflow region ($\le 30${\%}).  Previous observations have never shown such emission line profiles whose EBW dominates over the major component \citep{doschek2012}. 

\begin{figure}
  \centering
  \includegraphics[width=8.4cm,clip]{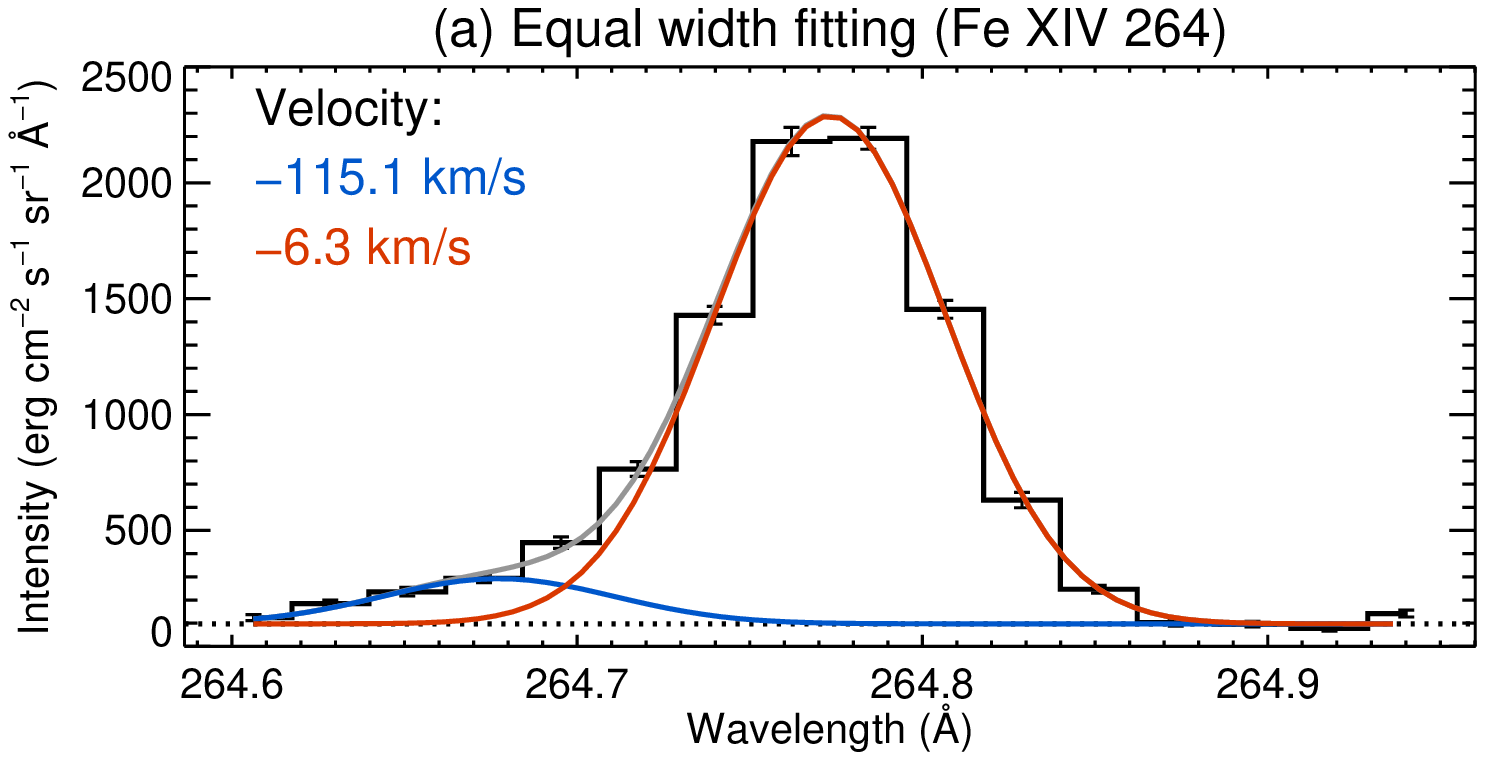}
  \includegraphics[width=8.4cm,clip]{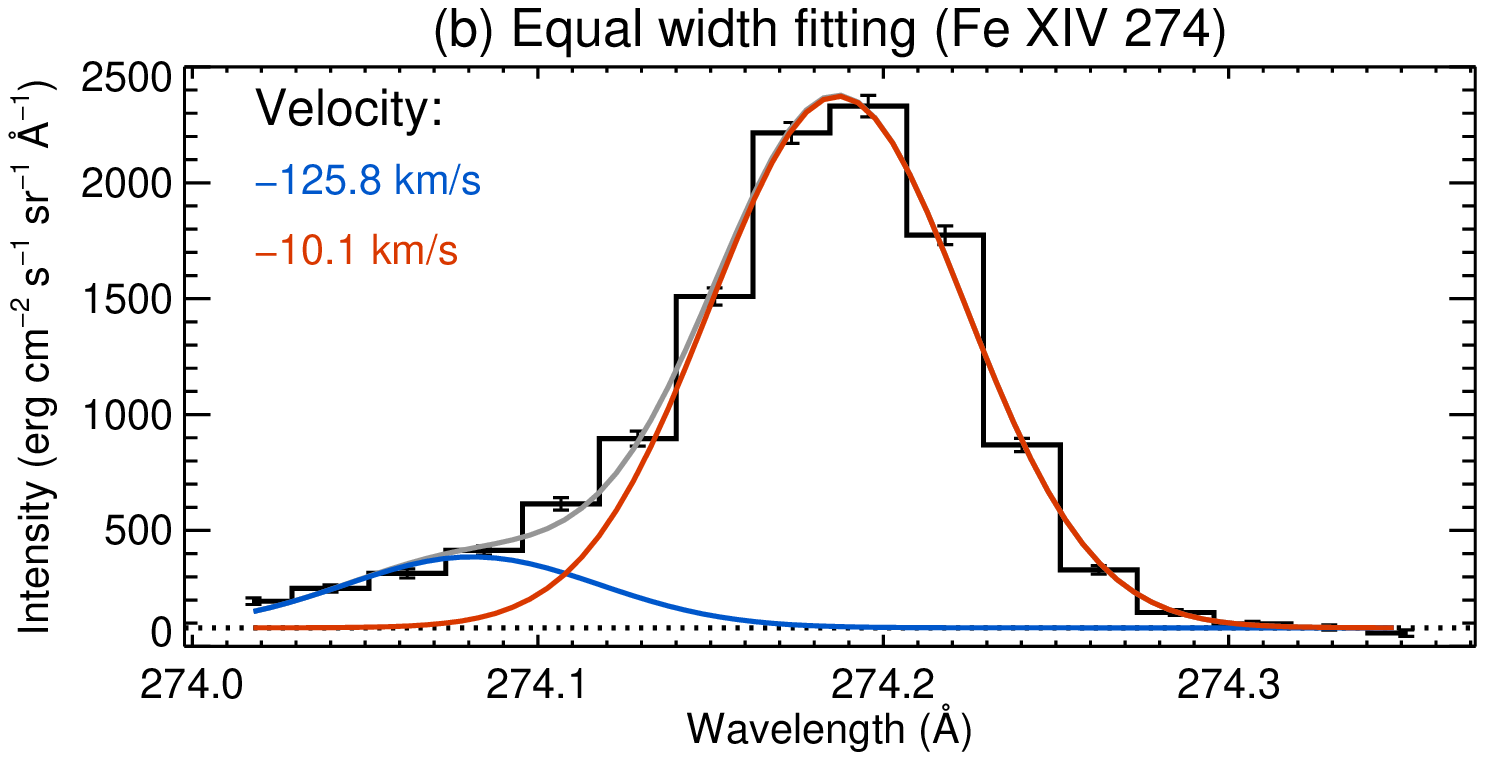}
  \includegraphics[width=8.4cm,clip]{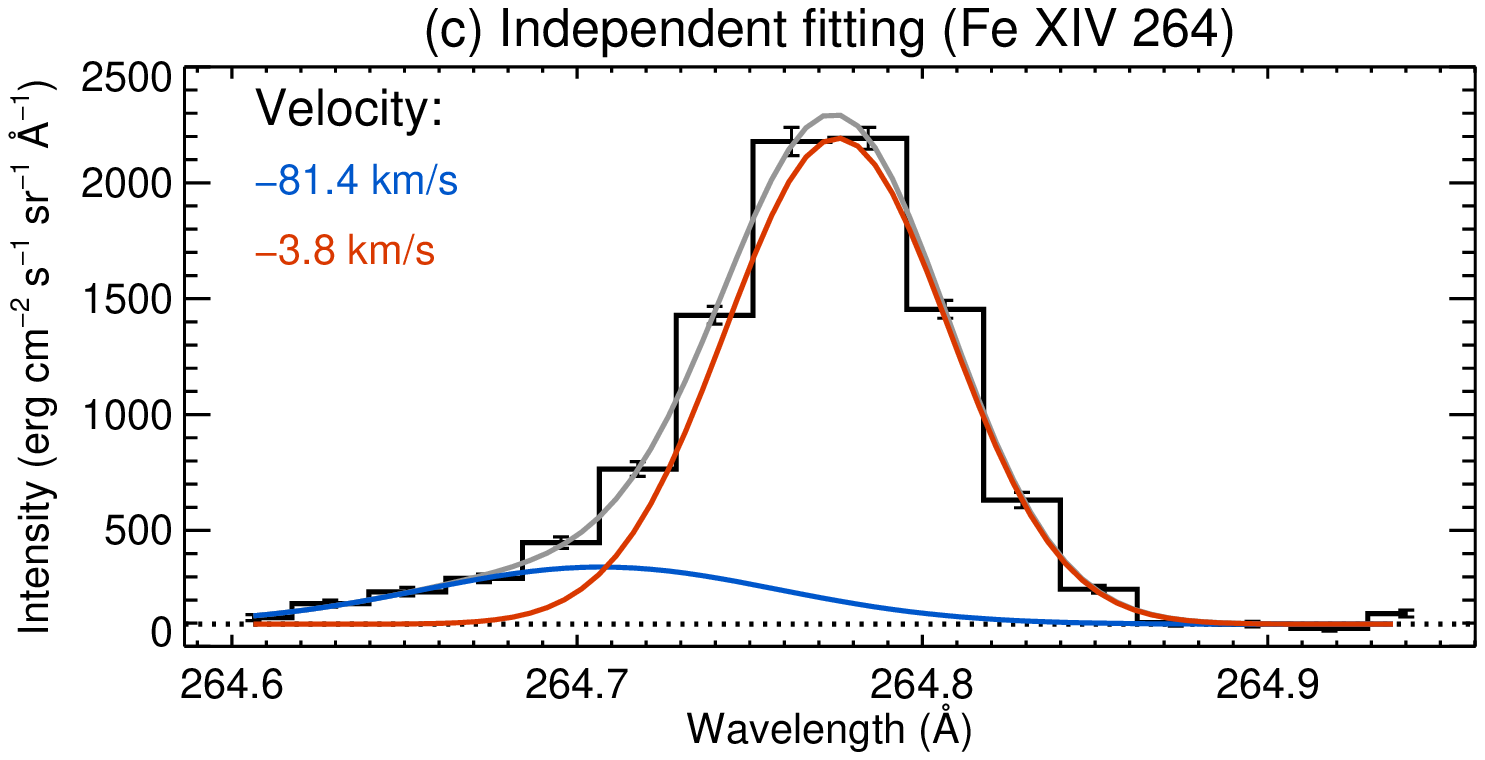}
  \includegraphics[width=8.4cm,clip]{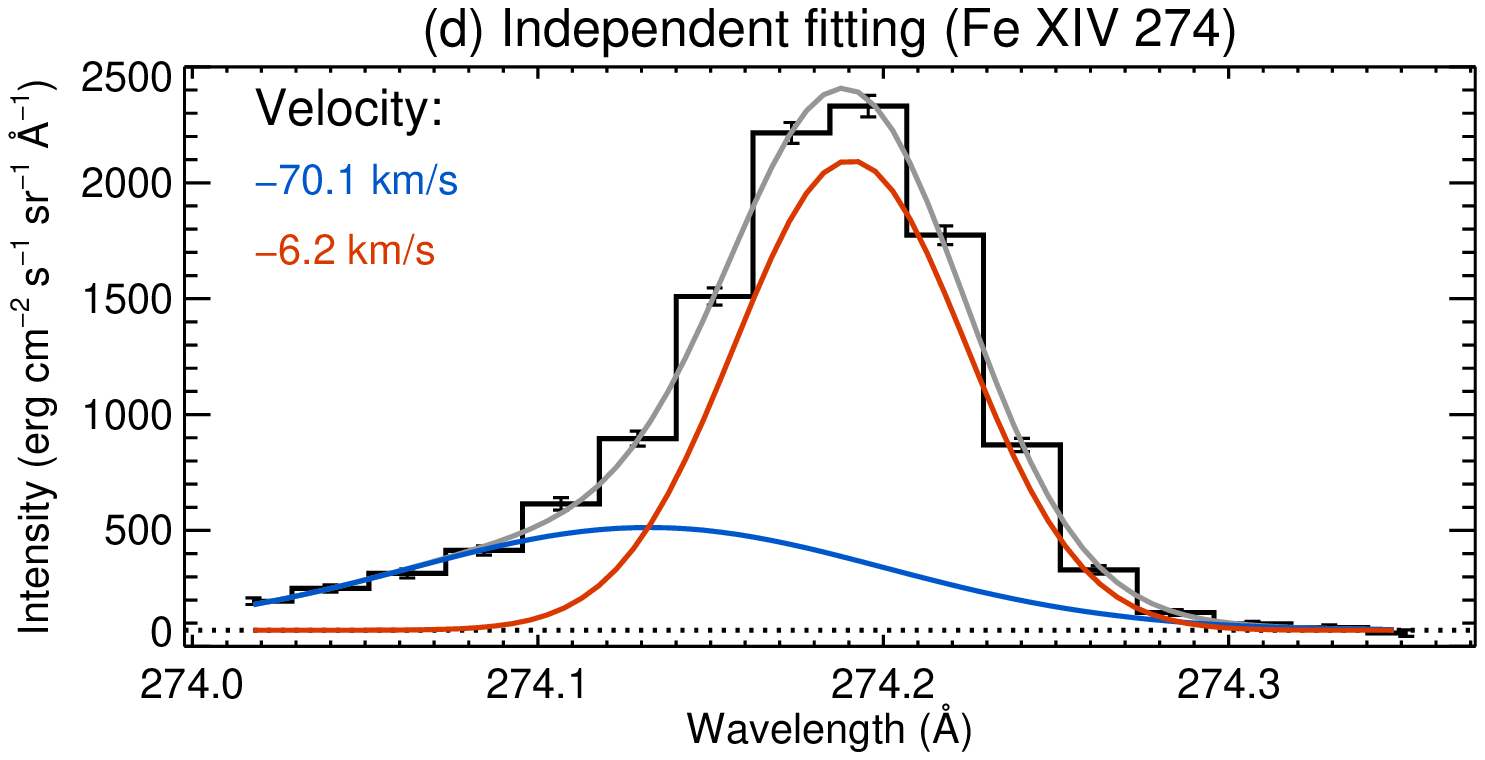}
  \includegraphics[width=8.4cm,clip]{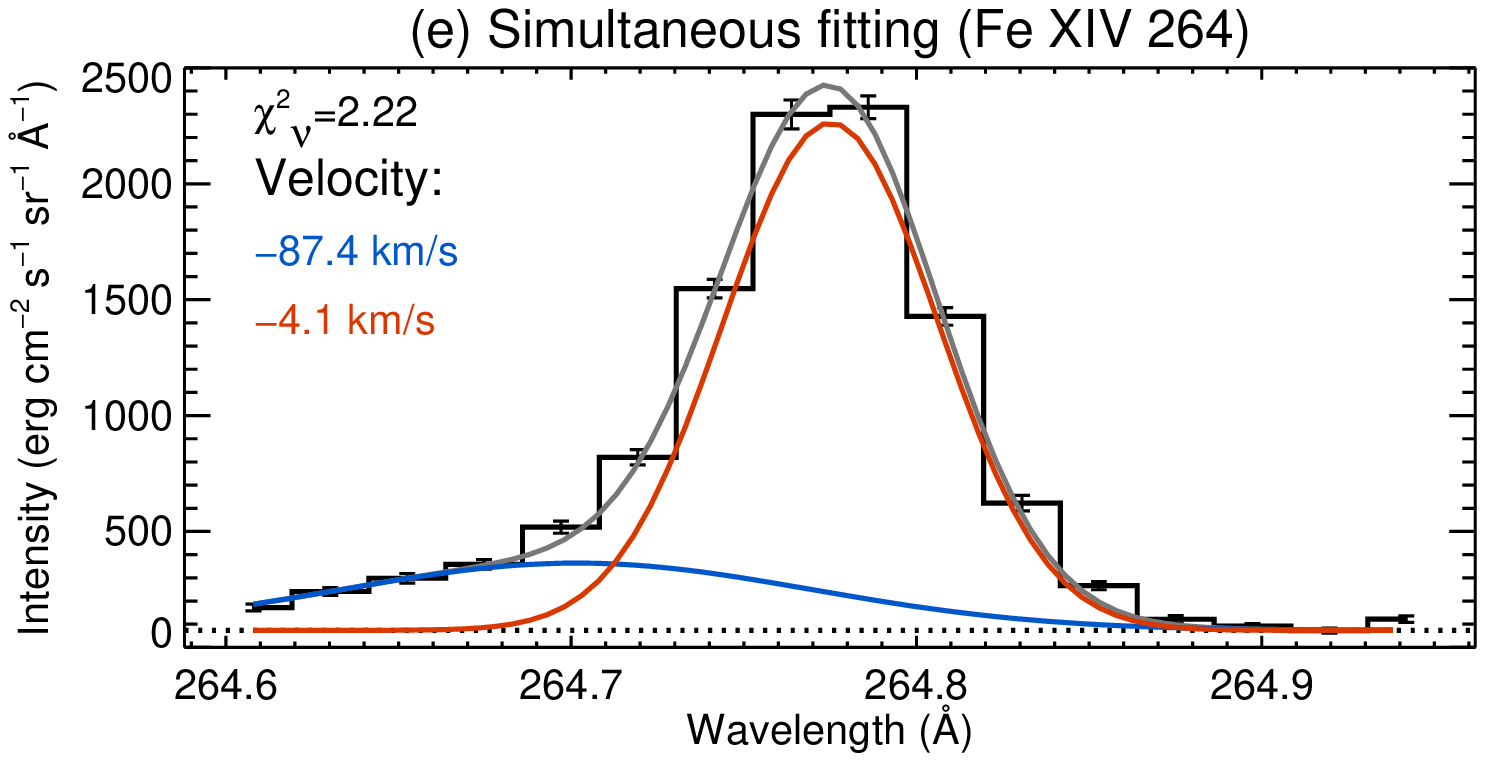}
  \includegraphics[width=8.4cm,clip]{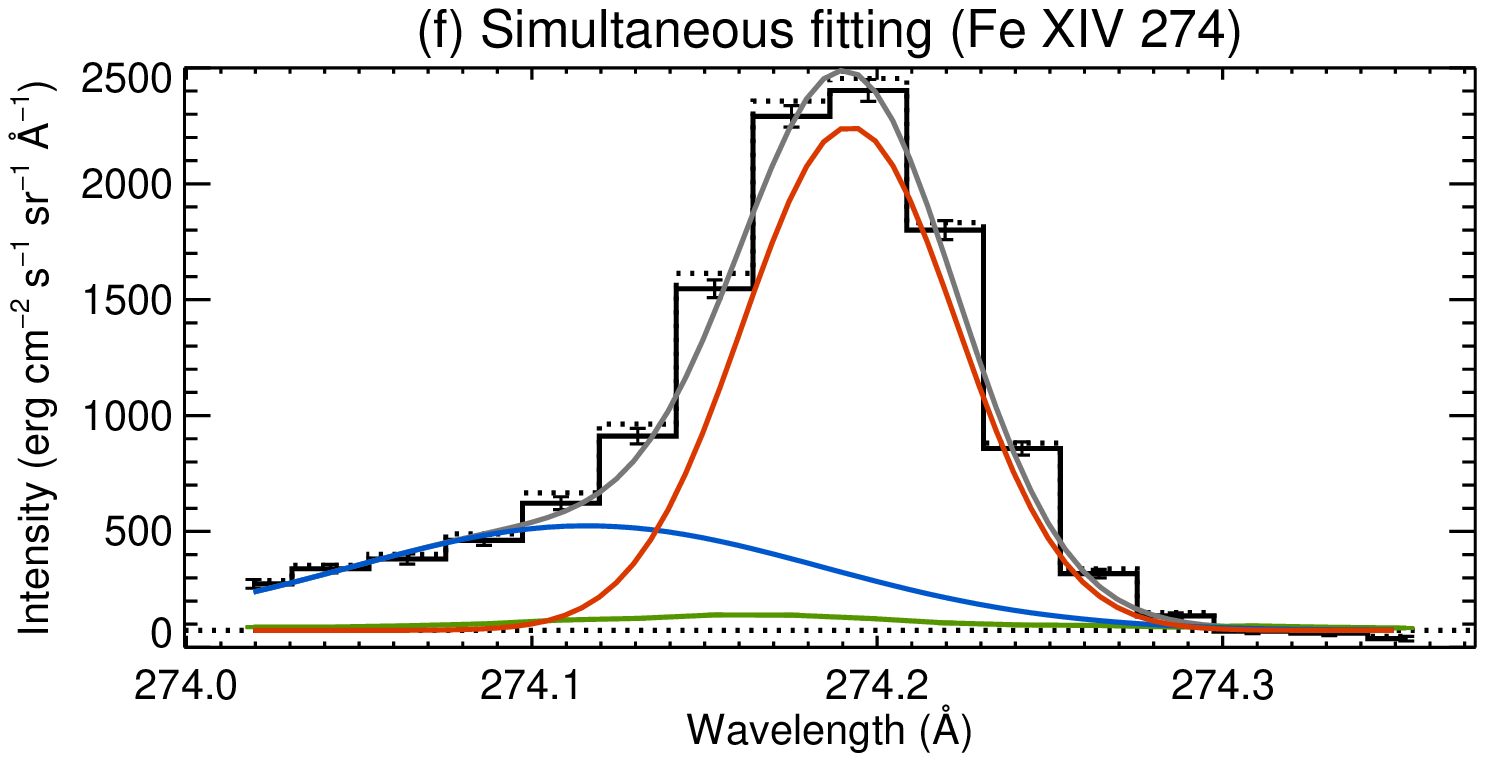}
  \caption{Fitting results for Fe \textsc{xiv} $264.78${\AA} and $274.20${\AA} obtained by three different models.  \textit{Upper row} (panels a and b): fitting with double Gaussians those have the same line width applied to each line profile independently (model 1).  \textit{Middle row} (panels c and d): fitting with double Gaussians those do not necessarily have the same line width applied to each line profile independently (model 2).  \textit{Lower row} (panels e and f): fitting with double Gaussians applied to two line profiles simultaneously without the assumption of the same line width of two components (model 3). }
  \label{fig:fig_diff}
\end{figure}

% The validity of double Gaussian fitting
In this study, the emission line profiles of the outflow region are assumed to be composed of two Gaussian components.  %\citet{bryans2010} implicated that the profiles have a possibility to be composed of three or more Gaussians, considering from their result that there is a negative correlation between Doppler shift and line width in the major component which was deduced from the double Gaussian fitting.  
Most of previous analysis on the outflows in the edge of an active region assumed that the main component and EBW have \textit{the same line width} in order to reduce the possibility of approaching an unreasonable solution in the fitting parameter space, but the assumption could strongly affect the fitting \citep{bryans2010,brooks2012}.  \citet{brooks2012} mentioned that this assumption may cause the underestimation of the intensity of EBW.  Line profile with EBW often shows rather longer tail in the line wing than that could be represented by a Gaussian which has the same line width as the major component. Moreover, the assumption that the major component and EBW have the same line width does not originate in the physical requirement. 

In order to examine the differences in the fitting result between different constraint on the fitting parameters, we applied three fitting models to a line profile pair of Fe \textsc{xiv} $264.78${\AA}.  Line centroid and line width are respectively denoted by $\lambda$ and $W$. The suffixes below represent: ``1'' for Fe \textsc{xiv} $264.78${\AA}, ``2'' for Fe \textsc{xiv} $274.20${\AA} followed by the component either ``Major'' or ``EBW''.  
% First fitting model
First model (model 1) assumes $W_{1, \mathrm{Major}} = W_{1, \mathrm{EBW}}$ and $W_{2, \mathrm{Major}} = W_{2, \mathrm{EBW}}$, and fits the line profiles of Fe \textsc{xiv} $264.78${\AA} and $274.20${\AA} separately with double Gaussians those have the same line width for each component.  
% Second fitting model
Second model (model 2) also fits the line profiles of the two Fe \textsc{xiv} separately, but with double Gaussians those do not necessarily have the same line width for each component.  
% Third fitting model
Third model (model 3) fits the two Fe \textsc{xiv} line profiles simultaneously by applying $\lambda_{2, \mathrm{Major}} = \alpha \lambda_{1, \mathrm{Major}}$, $\lambda_{2, \mathrm{EBW}} = \alpha \lambda_{1, \mathrm{EBW}}$ ($\alpha = 1.0355657$; determined in Section \ref{sec:wavelength_adjust}), $W_{1, \mathrm{Major}} = W_{2, \mathrm{Major}}$, and $W_{1, \mathrm{EBW}} = W_{2, \mathrm{EBW}}$.  We adopted model 3 for the electron density measurement in this study because it is physically most reasonable 
%{\color{red} 
in the sense that the model deals the parameters (line centroids and thermal widths) consistently for both emission lines and does not impose the artificial restriction on the line widths.
%}

The results for those three models are shown in Fig.~\ref{fig:fig_diff}.  We obtained smaller and more blueshifted second component (EBW) with the model 1 in panels (a) and (b), which confirms the suggestion in \citet{brooks2012}.  In contrast, larger and less blueshifted EBWs were obtained with the models 2 and 3 as clearly seen in panel (c)--(f).  In addition to this, the line widths of EBW component was much broader for the models 2 and 3 than for the model 1.  
%{\color{red} 
It is not clear at present whether the increased widths may indicate superposition of multiple upflow components, which will be another point to be revealed in the future.
%}
The comparison between those three models shows that the results in previous analysis should underestimate the intensity of EBW with an artificial assumption that two components in line profile have the same line width.  Moreover, independent fitting applied to two emission lines causes a discrepancy as seen in panel (c) and (d).  The Doppler velocity of EBW component was $-81.4 \, \kmpers$ for Fe \textsc{xiv} $264.78${\AA} while $-70.1 \, \kmpers$ for Fe \textsc{xiv} $274.20${\AA}.  Note that the rest wavelengths were determined from a limb observation on 2007 December 6, so these Doppler velocities have an uncertainty of $10 \, \kmpers$ at most.

% --- End of Contents ---

%% file: tex/dns_analysis_dens_inv.tex
% ======================
%   Density inversion
% ======================

Now the densities of EBW and the major component can be obtained by referring to the theoretical intensity ratio of Fe \textsc{xiv} $264.78${\AA}/$274.20${\AA} as a function of electron density shown in Fig.~\ref{fig:dns_itdn_rat_chianti}.  The intensity ratio monotonically increases within the density range of $10^{8} \, \mathrm{cm}^{-3} \le n_{\mathrm{e}} \le 10^{12} \, \mathrm{cm}^{-3}$.  The electron density in the solar corona generally falls between $10^{8} \, \mathrm{cm}^{-3}$ (for coronal holes) and $10^{11} \, \mathrm{cm}^{-3}$ (for flare loops), so the intensity ratio of Fe \textsc{xiv} $264.78${\AA}/$274.20${\AA} is quite useful.  The error in the density was calculated by using the 1-$\sigma$ error in the intensity ratio.  The electron density is obtained from
\begin{equation}
  n_{\mathrm{e}} = F^{-1} \left( \frac{I_{264}}{I_{274}} \right) \, \mathrm{,}
\end{equation}
where $F^{-1}$ is the inverse function of the theoretical intensity ratio, and $I_{264}$ and $I_{274}$ are respectively the observed intensity of Fe \textsc{xiv} $264.78${\AA} and $274.20${\AA}.  Using $\sigma_{I_{264} / I_{274}}$ as the error of observed intensity ratio, we estimate the error of the density $\sigma_{n_{\mathrm{e}}}$ as
\begin{equation}
  n_{\mathrm{e}} \pm \sigma_{n_{\mathrm{e}}} 
  = F^{-1} \left( \frac{I_{264}}{I_{274}} \pm \sigma_{I_{264} / I_{274}} \right) \, \mathrm{.} 
\end{equation}
The error $\sigma_{n_{\mathrm{e}}}$ was not dealt symmetrically in this definition, which comes from the fact the function $F$ has a curvature which can not be negligible compared to $\sigma_{I_{264} / I_{274}}$. 

% --- End of Contents ---

%% file: tex/dns_results_1G.tex
% ===============================
%   Project:
%     DNS (Density of upflows)
%   Description:
%     Results from 1-Gaussian
% ===============================

\begin{figure}
  \centering
  \includegraphics[width=8.4cm,clip]{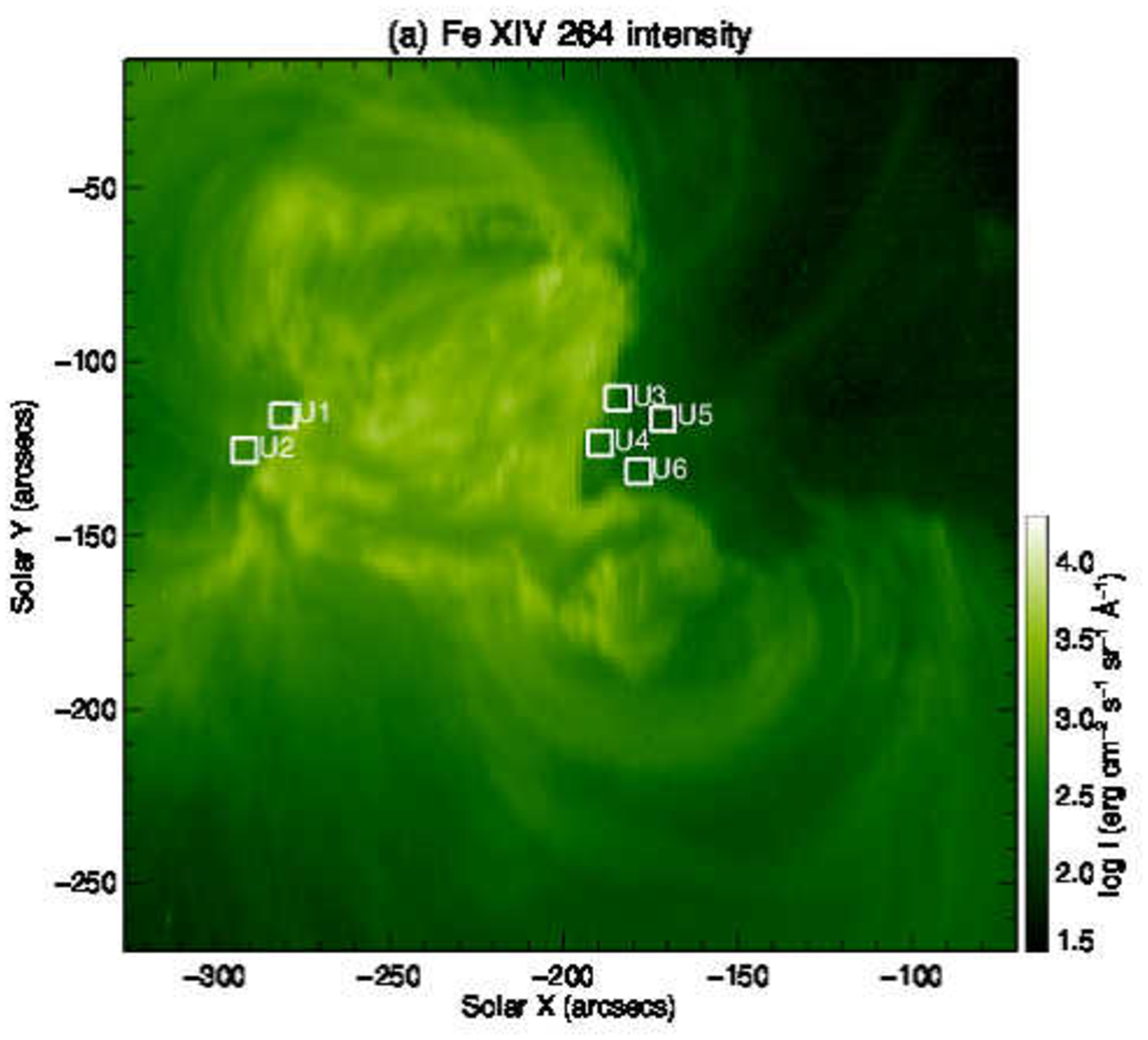}
  \includegraphics[width=8.4cm,clip]{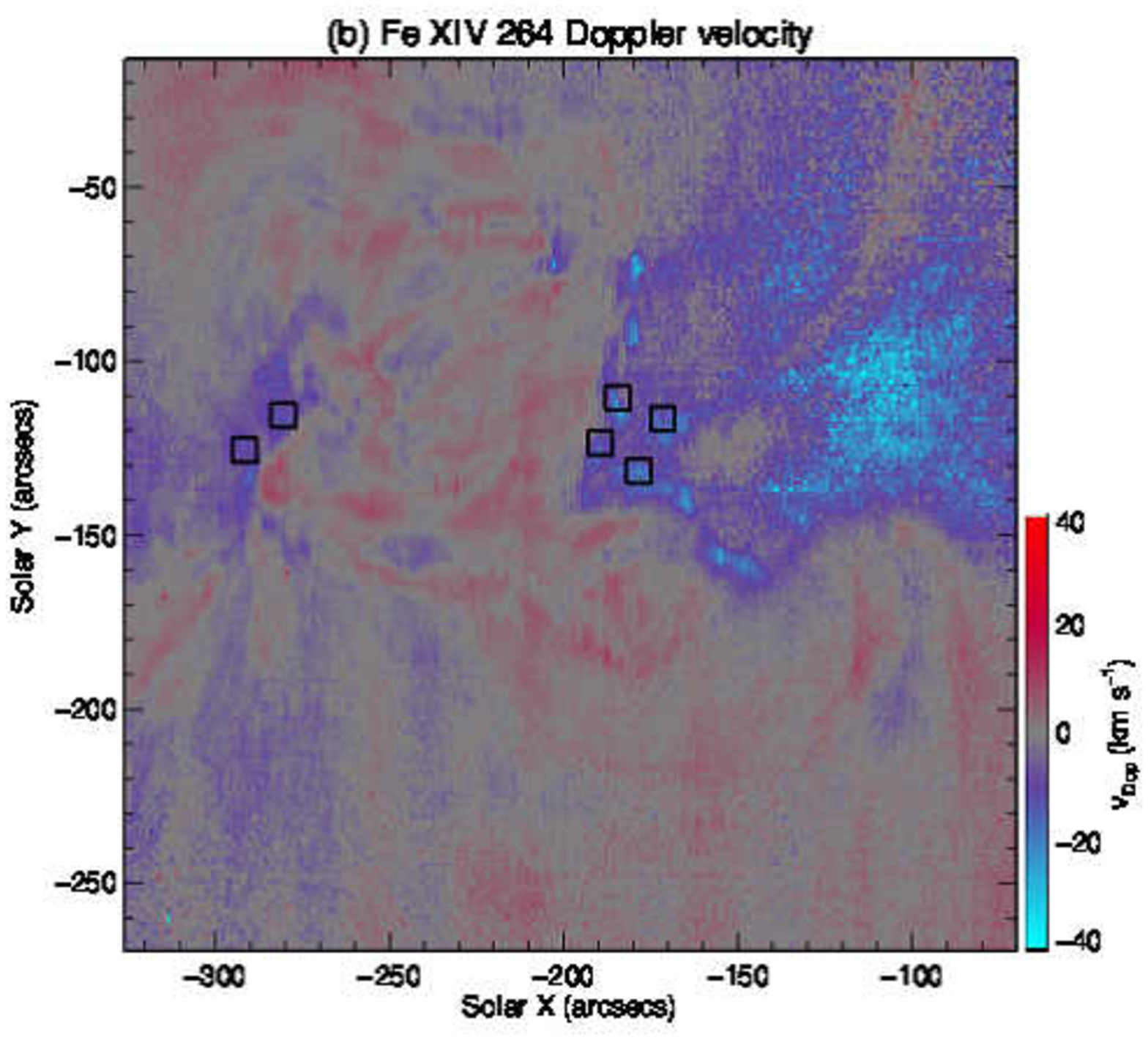}
  \includegraphics[width=8.4cm,clip]{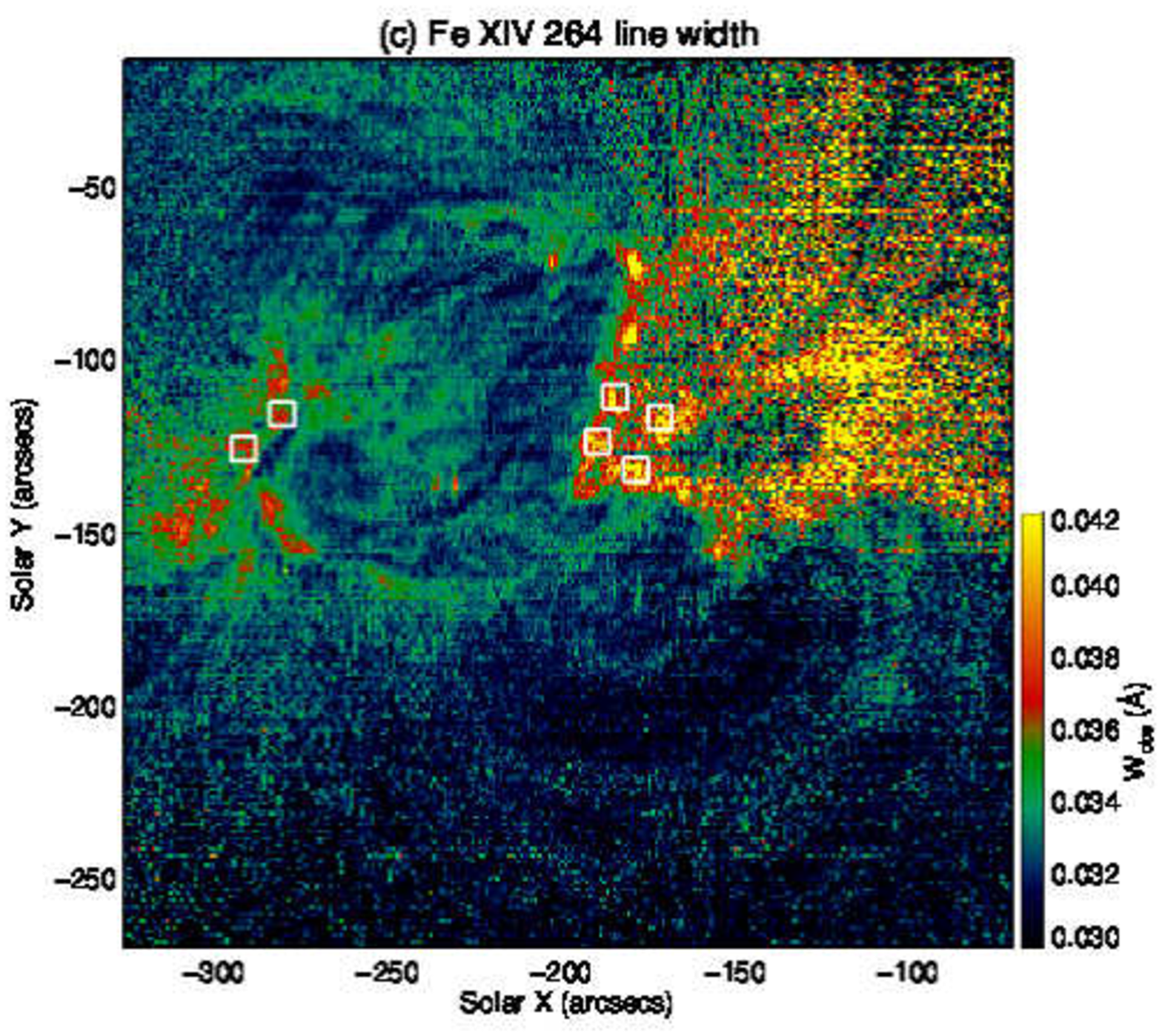}
  \includegraphics[width=8.4cm,clip]{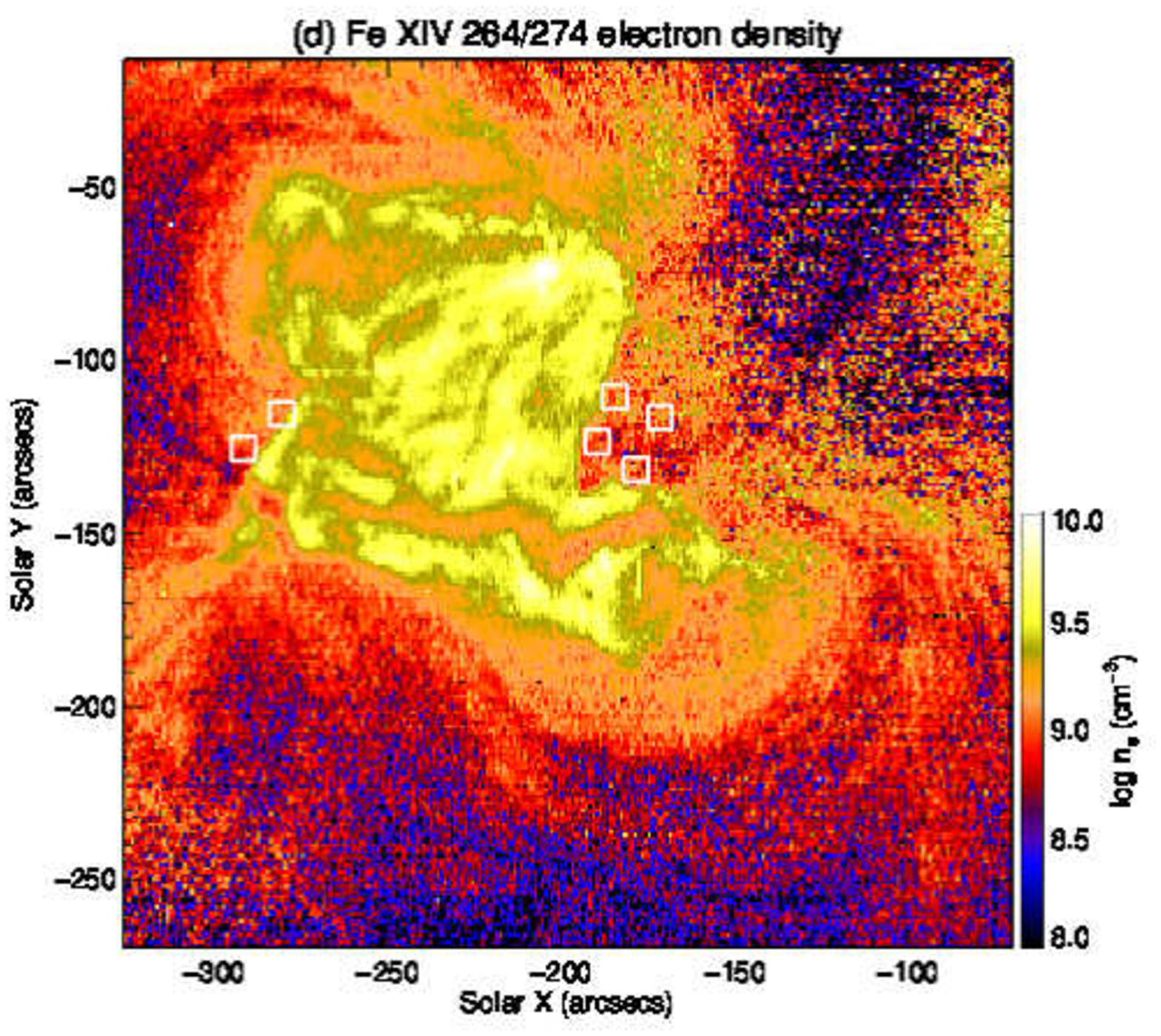}
  \caption{Physical quantities deduced from single Gaussian fitting for Fe \textsc{xiv} $264.78${\AA} obtained on 2007 December 11 00:24:16--04:47:29UT. (a) Intensity of Fe \textsc{xiv} $264.78${\AA}. (b) Doppler velocity of Fe \textsc{xiv} $264.78${\AA}. (c) Line width of Fe \textsc{xiv} $264.78${\AA}. (d) Electron density derived from the line ratio Fe \textsc{xiv} $264.78${\AA}/$274.20${\AA}.}
  \label{fig:fexiv_1G_map}
\end{figure}

First we describe the results deduced from single-Gaussian fitting.  As described above, line profiles at the outflow regions are known to have distorted shape which cannot be well represented by single Gaussian, nonetheless, the results deduced from single-Gaussian fitting may be useful because the fitting is much more robust in terms of the freedom of variables (\textit{e.g}, 4 parameters for single Gaussian with constant background and 7 parameters for double Gaussians).  Fig.~\ref{fig:fexiv_1G_map} shows the map of intensity, Doppler velocity, line width of Fe \textsc{xiv} $264.78${\AA}, and electron density derived from the line ratio Fe \textsc{xiv} $264.78${\AA}/$274.20${\AA}.  The blending Si \textsc{vii} $274.18${\AA} was taken into account and subtracted by referring to Si \textsc{vii} $275.35${\AA}.  It is clear from panel (b) that there is the outflow regions (\textit{i.e.}, blueshift) at the east/west edge of the active region core around $(x, y) = (-260'', -120'')$ and $(-175'', -125'')$.  Panel (c) shows that the line width at those outflow regions is larger than other locations by $\Delta W = 0.020\text{--}0.027$\text{\AA} (square root of the difference of squared line width) equivalent to $\delta v = 20 \text{--} 30 \, \mathrm{km} \, \mathrm{s}^{-1}$, which is similar to a result reported previously \citep{doschek2008,hara2008}.  The electron density at the outflow regions is $n_{\mathrm{e}} = 10^{8.5\text{--}9.5} \, \mathrm{cm}^{-3}$, which is lower than that at the core ($n_{\mathrm{e}} \geq 10^{9.5} \, \mathrm{cm}^{-3}$).  

% Selection of the outflow regions studied.
We selected the regions to be studied as similar way as that in Chapter \ref{chap:vel}.  We defined the outflow regions as the locations (1) where the line width of Fe \textsc{xiv} $264.78${\AA} is enhanced, and (2) which can be separated from fan loops seen in Si \textsc{vii} intensity map though not shown here.  The selected six regions are indicated by \textit{white} boxes in each map (numbered by U1--U6 as written in panel (a), whose size is $8'' \times 8''$.  Those regions are located beside the bright core as seen in the intensity map (panel a).  We hereafter refer U1--U2 as the eastern outflow region and U3--U6 as the western outflow region. 

% --- End of Contents ---

%% file: tex/dns_results_n.tex
% ===============================
%   Project:
%     DNS (Density of upflows)
%   Description:
%     Results for density
% ===============================

\begin{figure}
  \centering
  \includegraphics[width=8.4cm,clip]{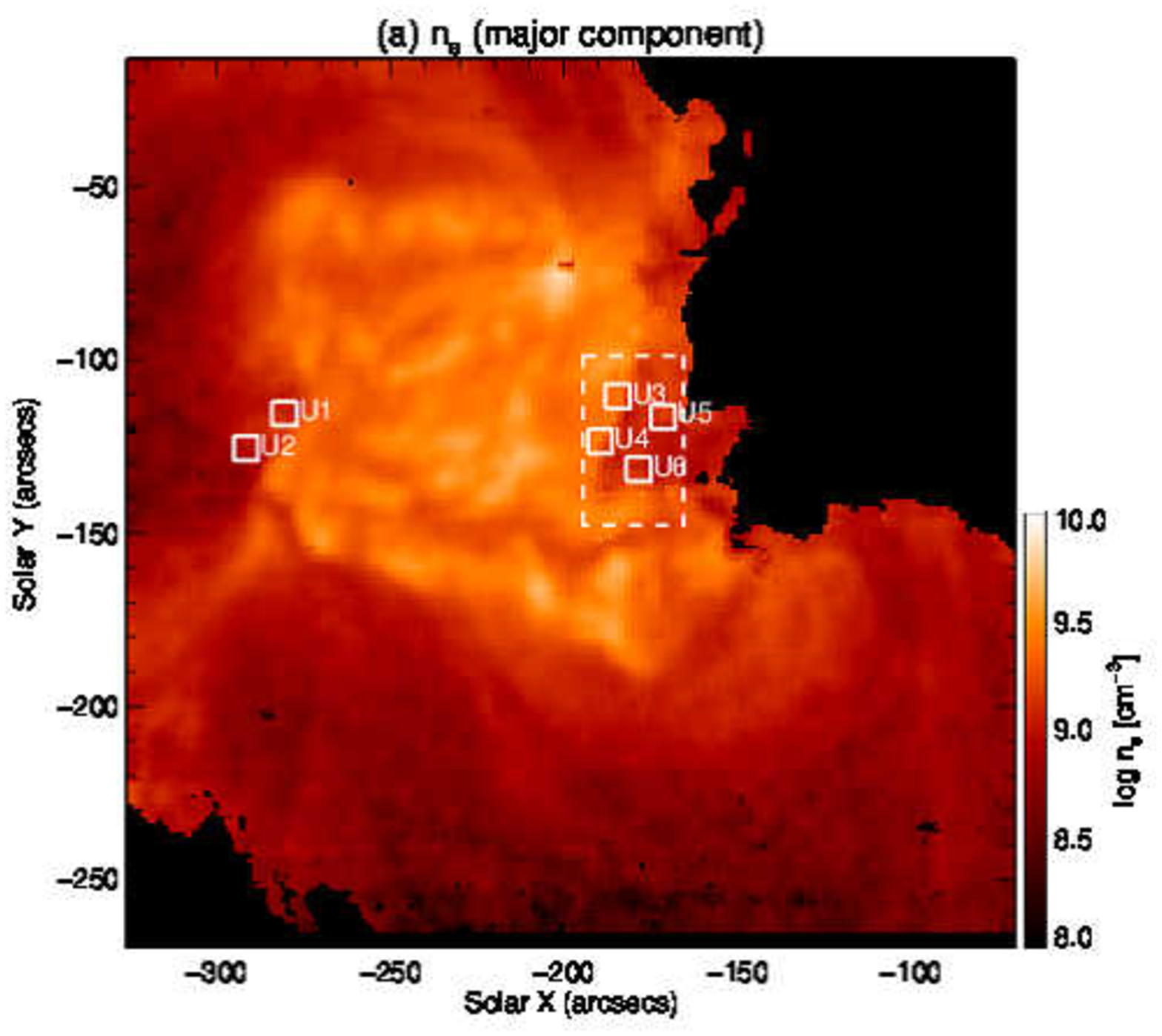}
  \includegraphics[width=8.4cm,clip]{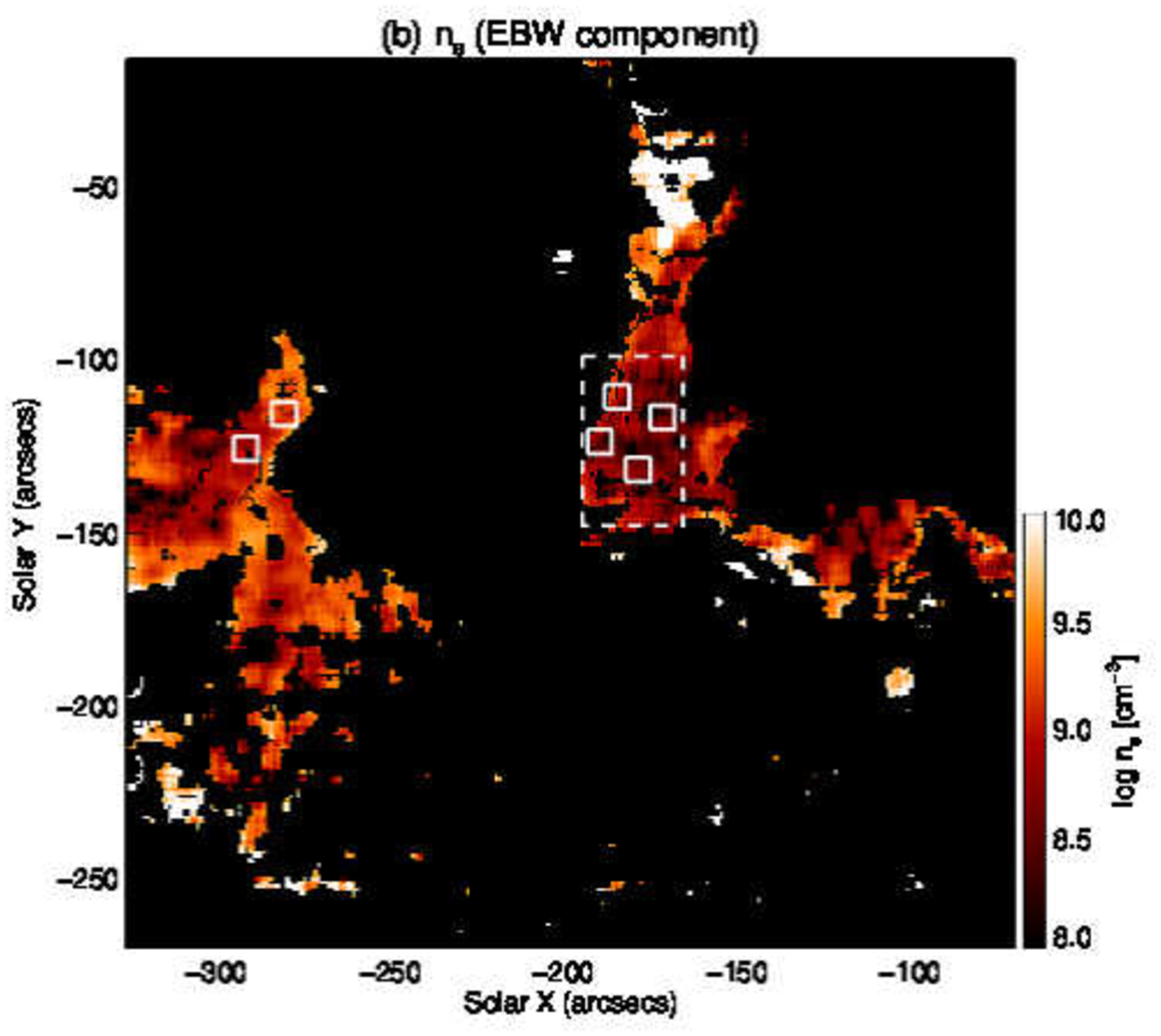}
  \caption{Electron density map deduced from two Gaussian fitting of an emission line pair Fe \textsc{xiv} $264.78${\AA}/$274.20${\AA} obtained by the raster scan on 2007 December 11 00:24:16--04:47:29UT.  (a) Electron density of the major component.  (b) Electron density of EBW component.  Same color contour are used in the two panels.  Pixels where the peak intensity of the major component ($I_{\mathrm{Major}}$) did not exceed $2.0 \times 10^{3} \, \mathrm{erg} \, \mathrm{cm}^{-2} \, \mathrm{s}^{-1} \, \mathrm{sr}^{-1} \, \text{\AA}^{-1}$ were masked by \textit{black}.  \textit{White} boxes numbered U1--U6 are the same as those in Fig.~\ref{fig:fexiv_1G_map}.  The \textit{white dashed} box indicate the entire western outflow region.}
  \label{fig:dens_map}
\end{figure}

The electron density of EBW component was measured through the analysis described in Section \ref{subsect:dns_diag}.  Fig.~\ref{fig:dens_map} shows the distributions of electron density for the major component ($n_{\mathrm{Major}}$) in panel (a) and EBW component ($n_{\mathrm{EBW}}$) in panel (b).  Pixels where the peak intensity of the major component ($I_{\mathrm{Major}}$) did not exceed $2.0 \times 10^{3} \, \mathrm{erg} \, \mathrm{cm}^{-2} \, \mathrm{s}^{-1} \, \mathrm{sr}^{-1} \, \text{\AA}^{-1}$ were masked by \textit{black}.  This threshold was determined by using the scatter plot of intensity and electron density of the major component shown in Appendix \ref{sect:dns_app_in}.  Pixels falling into the next three conditions were displayed, and others were masked by \textit{black}.  (1) $I_{\mathrm{Major}} \ge 2.0 \times 10^3 \, \mathrm{erg} \, \mathrm{cm}^{-2} \, \mathrm{s}^{-1} \, \mathrm{sr}^{-1} \, \text{\AA}^{-1}$.  (2) The intensity of EBW component ($I_{\mathrm{EBW}}$) exceeds $3${\%} of that of the major component ($I_{\mathrm{EBW}}/I_{\mathrm{Major}} \ge 0.03$).  (3) The difference between the Doppler velocity of EBW component ($v_{\mathrm{EBW}}$) and that of the major component ($v_{\mathrm{Major}}$) satisfies $v_{\mathrm{EBW}} - v_{\mathrm{Major}} < -30 \, \mathrm{km} \, \mathrm{s}^{-1}$ (\textit{i.e.}, the two components are well separated).

\begin{figure}
  \centering
  \begin{minipage}[c]{8.4cm}
    \includegraphics[width=8.4cm,clip]{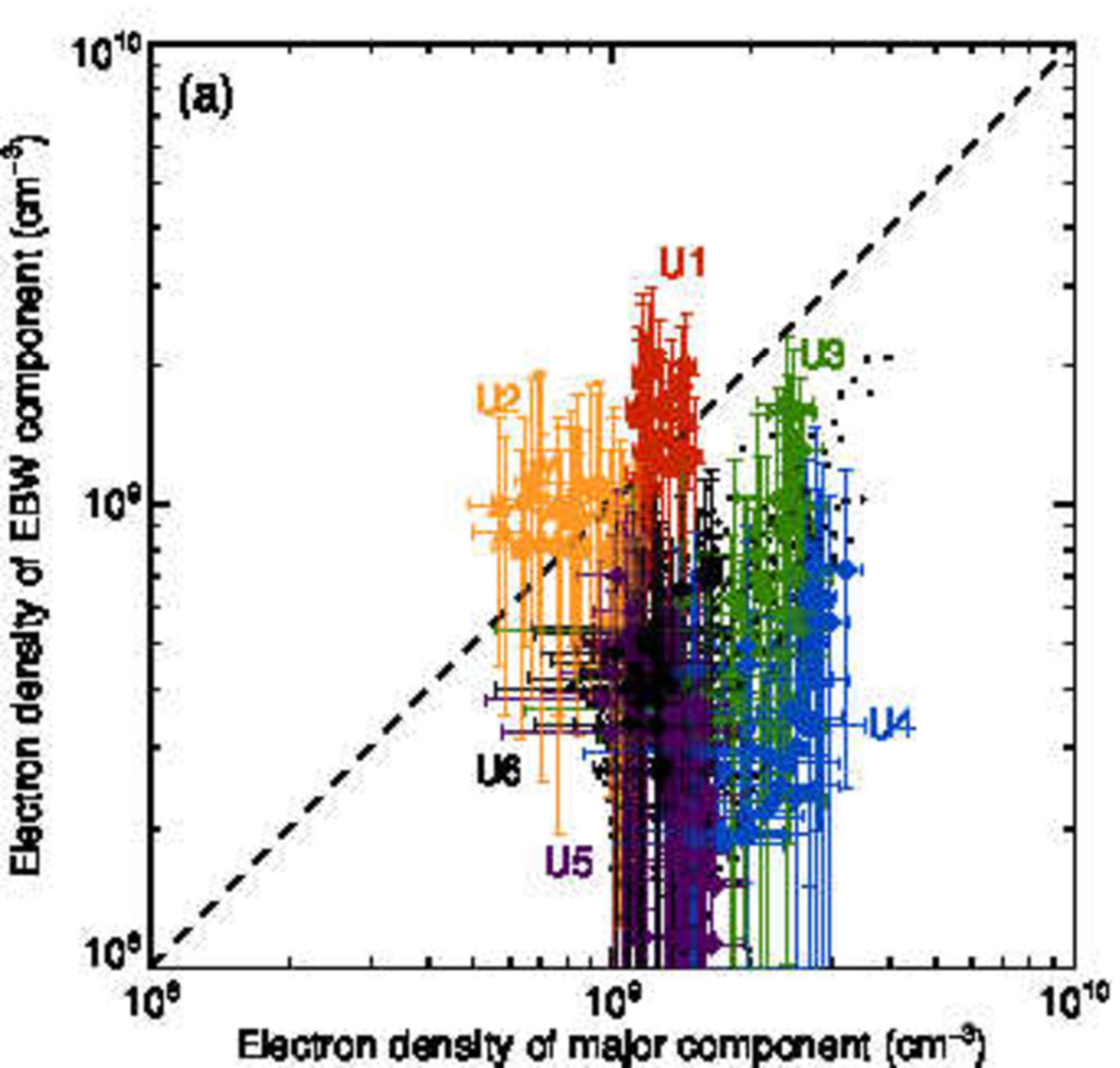}
  \end{minipage}
  \begin{minipage}[c]{8.4cm}
    \includegraphics[width=8.4cm,clip]{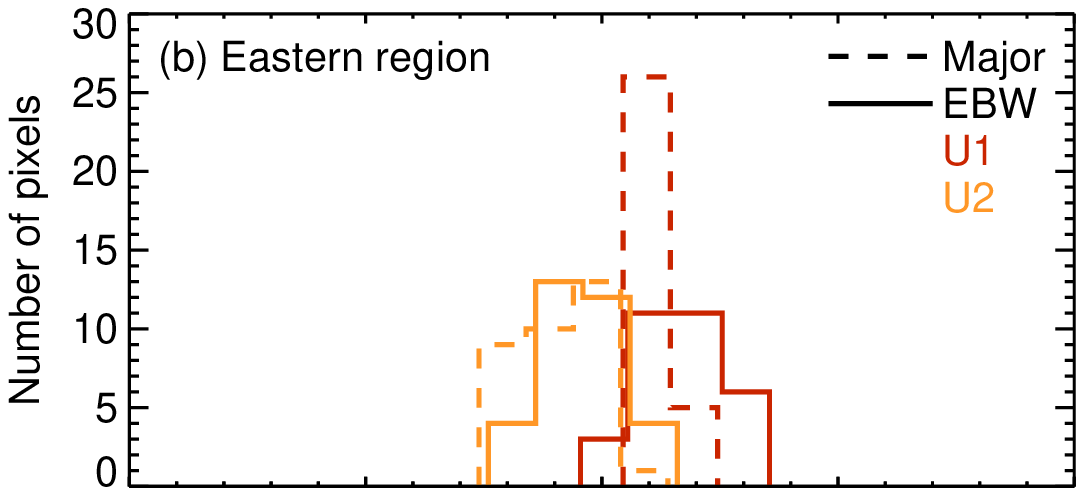}
    \includegraphics[width=8.4cm,clip]{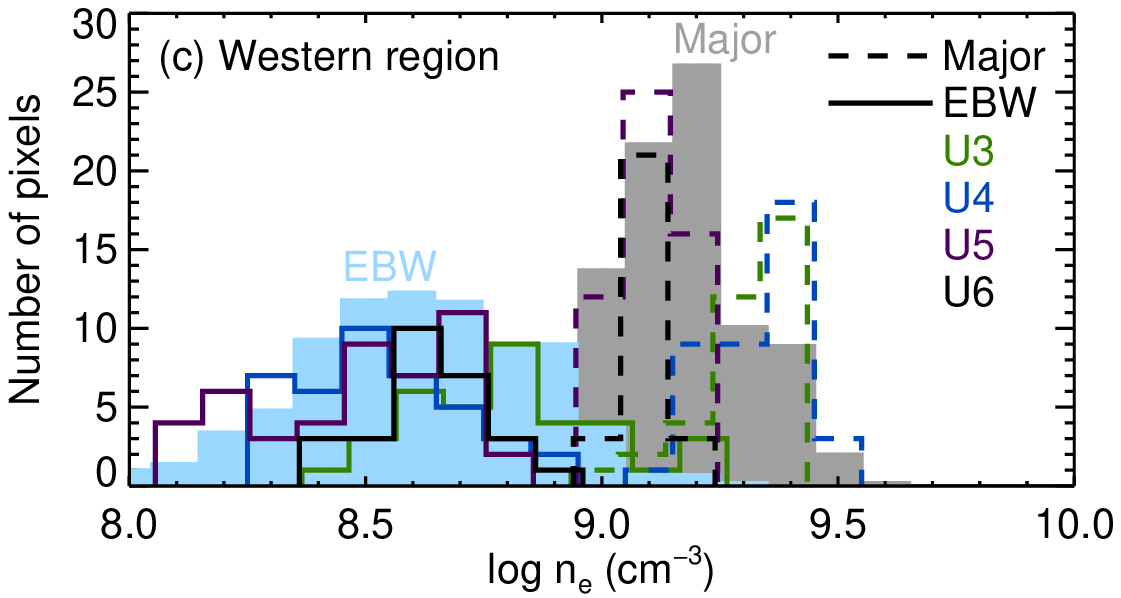}
  \end{minipage}
  \caption{(a) Scatter plot for Fe \textsc{xiv} electron density of the major component vs.\ that of EBW component.  \textit{Colors} indicate the selected region indicated by \textit{white} boxes in Fig.~\ref{fig:fexiv_1G_map}.  \textit{Triangles} (\textit{Diamonds}) represent the data points in the eastern (western) outflow regions.  Numbers beside data points correspond to the name of the \textit{white} boxes.  \textit{Black dots} show the electron density for the western outflow region indicated by a \textit{white dashed} box in Fig.~\ref{fig:dens_map}.  The dashed line indicates point where two densities equal each other.  (b) Histograms for the electron density of the major component (\textit{dotted}) and EBW component (\textit{solid}) in the eastern outflow region.  (c) Histograms for the electron density of the major component (\textit{dotted}) and EBW component (\textit{solid}) in the western outflow region.  The \textit{gray} (the major component) and \textit{turquoise} (EBW component) histograms in the background are made for the entire western outflow region.  Those two histograms are multiplied by $0.1$.}
  \label{fig:dns_sct}
\end{figure}

The relationship of electron density between the major component and EBW component are shown in Fig.~\ref{fig:dns_sct}.   Scatter plot in panel (a) shows the electron density for the outflow regions U1--U6 (\textit{colored symbols}) and for the entire western outflow region indicated by the \textit{white dashed} box in Fig.~\ref{fig:dens_map} (\textit{black dots}).  The eastern outflow regions (U1--U2) and west ones (U3--U6) exhibit different characteristics.  The scatter plots for U1--U2 indicate $n_{\mathrm{Major}} \leq n_{\mathrm{EBW}}$, while those for U3--U6 indicate $n_{\mathrm{Major}} \geq n_{\mathrm{EBW}}$.  Panels (b) and (c) show the same data but in histograms for which colors again indicate the selected outflow regions.  The \textit{gray} (the major component) and \textit{turquoise} (EBW component) histograms in the background of panel (c) are made for the entire western outflow region.  Those two histograms clearly indicate that $n_{\mathrm{EBW}}$ ($10^{8.61 \pm 0.24} \, \mathrm{cm}^{-3}$) is smaller than $n_{\mathrm{Major}}$ ($10^{9.18 \pm 0.13} \, \mathrm{cm}^{-3}$) at the entire western outflow region, which confirms that our selection of the studied regions was not arbitrary.  The average electron densities in each studied region are listed in Table \ref{tab:dns_cmpl}.  %The reason why the magnitude relationship between $n_{\mathrm{Major}}$ and $n_{\mathrm{EBW}}$ is different in the east and west outflow region will be mentioned in Chapter \ref{chap:dis}.  

% --- End of Contents ---

%% file: tex/dns_results_h.tex
% ===============================
%   Chapter:
%     Density of upflows.
%   Description:
%     Column depth.
% ===============================

\begin{figure}
  \centering
  \begin{minipage}[c]{8.4cm}
    \includegraphics[width=8.4cm,clip]{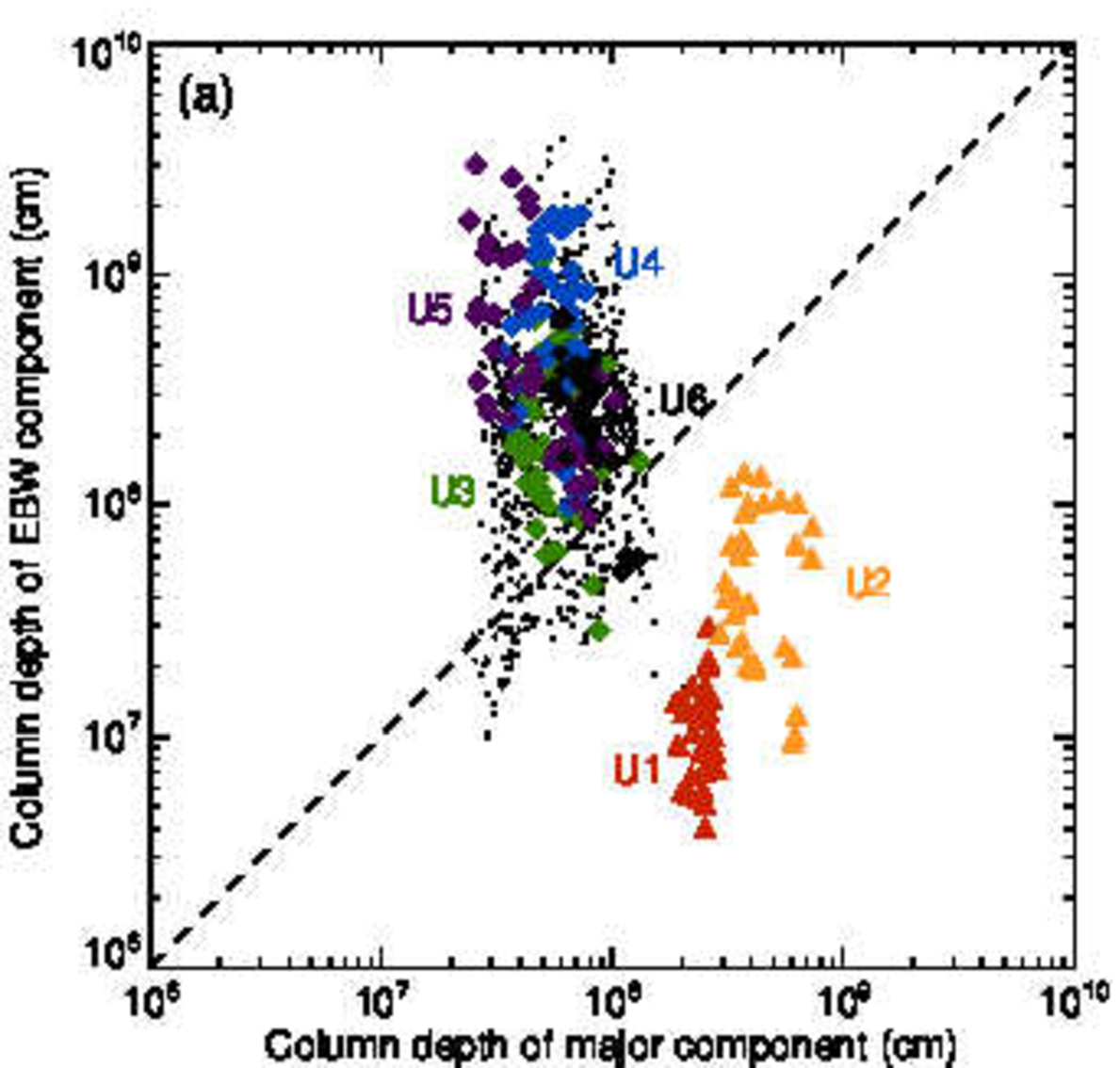}
  \end{minipage}
  \begin{minipage}[c]{8.4cm}
    \includegraphics[width=8.4cm,clip]{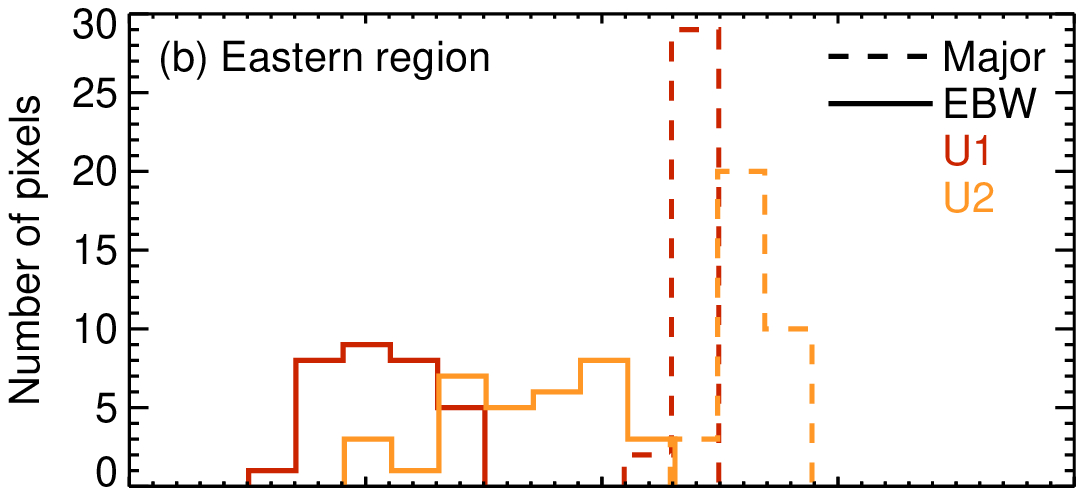}
    \includegraphics[width=8.4cm,clip]{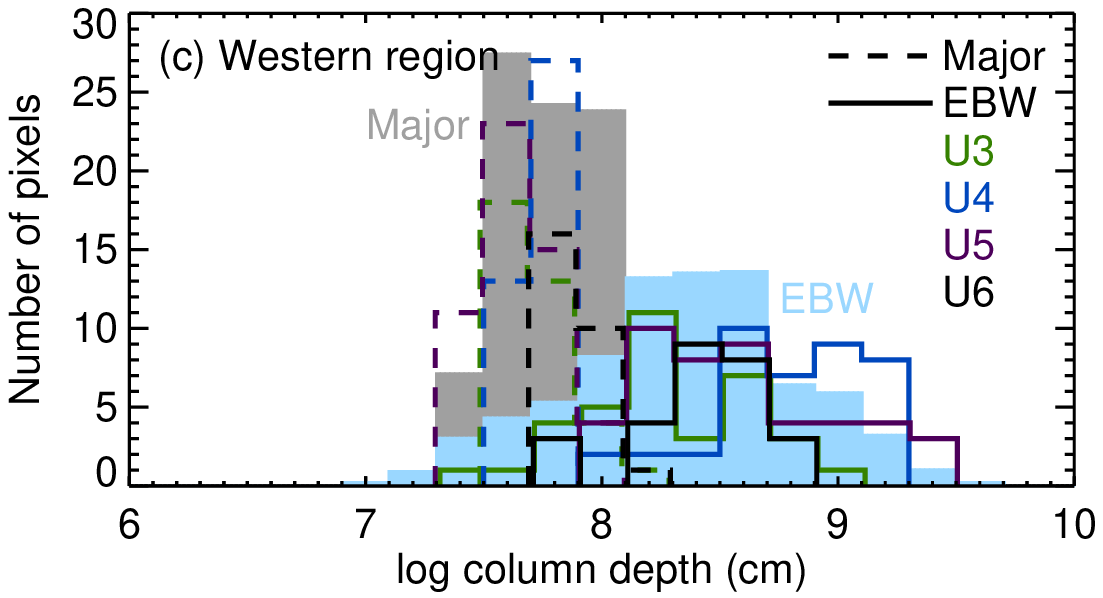}
  \end{minipage}
  \caption{%
    %\color{red} 
    (a) Scatter plot for column depth of the major component vs.\ that of EBW component.  \textit{Colors} indicate the selected region indicated by \textit{white} boxes in Fig.~\ref{fig:fexiv_1G_map}.  \textit{Triangles} (\textit{Diamonds}) represent the data points in the eastern (western) outflow regions.  Numbers beside data points correspond to the name of the \textit{white} boxes.  \textit{Black dots} show the column depth for the entire western region indicated by a \textit{white dashed} box in Fig.~\ref{fig:dens_map}.  The \textit{dashed} line indicates point where two quantities equal each other.  (b) Histograms for column depth of the major component (\textit{dotted}) and EBW component (\textit{solid}) in the eastern outflow region.  (c) Histograms for column depth of the major component (\textit{dotted}) and EBW component (\textit{solid}) in the western outflow region.  The \textit{gray} (the major component) and \textit{turquoise} (EBW component) histograms in the background are made for the entire western outflow region.  Those two histograms are multiplied by $0.1$.
  }
  \label{fig:column_depth}
\end{figure}

Using the obtained electron density for each component in Fe \textsc{xiv} line profiles, the column depth of each component can be respectively calculated.  We use the equation for the column depth including the filling factor, 
\begin{equation}
  h^{\ast} = hf = \dfrac{I}{n_{\mathrm{e}}^2 G (n_{\mathrm{e}}, \, T)} \, \text{,}
  \label{eq:column_depth}
\end{equation}
where $f$ is the filling factor, $I$ is the intensity of an emission line, $n_{\mathrm{e}}$ is the electron density, and $G (n_{\mathrm{e}}, \, T)$ is the contribution function of an emission line.  The quantity $h^{\ast}$ has a physical meaning as the plasma volume per unit area along the line of sight.  Here the temperature substituted to Eq.~(\ref{eq:column_depth}) was simply assumed to take a single value $T_{\mathrm{f}}$ at which the contribution function $G (n_{\mathrm{e}}, \, T)$ becomes maximum ($\log T_{\mathrm{f}} \, [\mathrm{K}]=6.30$ for the Fe \textsc{xiv} lines used here).  
%{\color{red}
Panel (a) in Fig.~\ref{fig:column_depth} shows a scatter plot for the column depth of the major component ($h_{\mathrm{Major}}$) and that of EBW component ($h_{\mathrm{EBW}}$).  Colored symbols respectively indicate the studied regions (U1--U2 for the eastern outflow region, and U3--U6 for the western outflow region).  As similar to the result for the electron density, the eastern and western outflow regions exhibit different characteristics: $h_{\mathrm{Major}} \ge h_{\mathrm{EBW}}$ in the eastern region, and $h_{\mathrm{Major}} \le h_{\mathrm{EBW}}$ in the western region.  Panels (b) and (c) display the same data in the form of histogram respectively for the eastern and western outflow region.  The \textit{gray} and \textit{turquoise} histograms in the background of panel (c) show the results for the entire western outflow region indicated by a white dashed box in Fig.~\ref{fig:dens_map}.  Table \ref{tab:dns_cmpl} shows the column depths averaged in each studied region.  

The result $h_{\mathrm{Major}} \le h_{\mathrm{EBW}}$ in the western outflow regions (U3--U6) means that the upflow dominates over the rest component in terms of the volume, opposite to the composition ratio of emission line profile itself.  The value of $h_{\mathrm{EBW}} \simeq 10^{8.0\text{--}9.0} \, \mathrm{cm}$ can be understood by considering that the inclination of the magnetic field lines in the western outflow region was $30^{\circ} \text{--} 50^{\circ}$ and the horizontal spatial scale of the region was the order of $10'' (\sim 10^{9} \, \mathrm{cm})$, which leads to the vertical height of nearly the same amount.  On the other hand, it is clearly indicated that $h_{\mathrm{EBW}}$ is smaller than $h_{\mathrm{Major}}$ by up to one order of magnitude in the eastern outflow region (U1--U2).  This means that the upflows possess only a small fraction compared to the rest plasma.
%}

% --- End of Contents ---

%% file: tex/dns_results_sivii.tex
% =======================================
%   Project:
%     Density of upflows
%   Section:
%     Estimation of error from Si VII.
% =======================================

\begin{figure}
  \centering
  \includegraphics[height=6.5cm,clip]{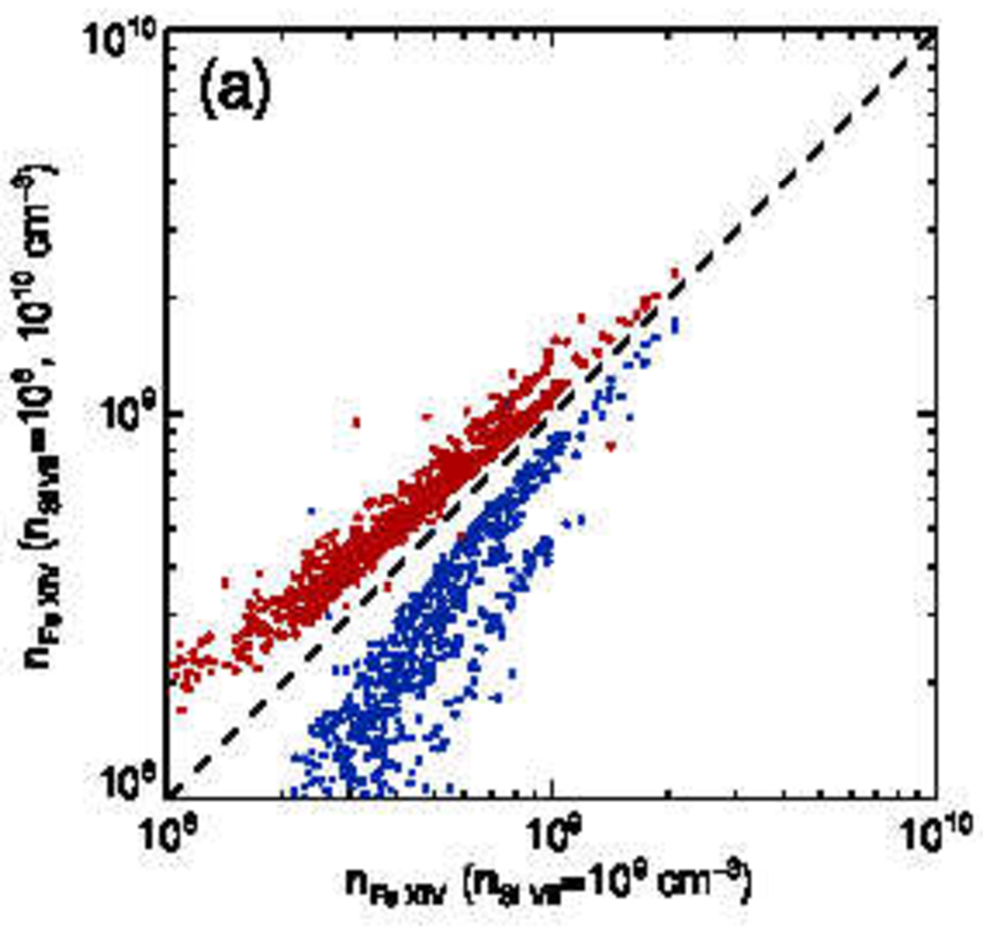}
  \includegraphics[height=6.5cm,clip]{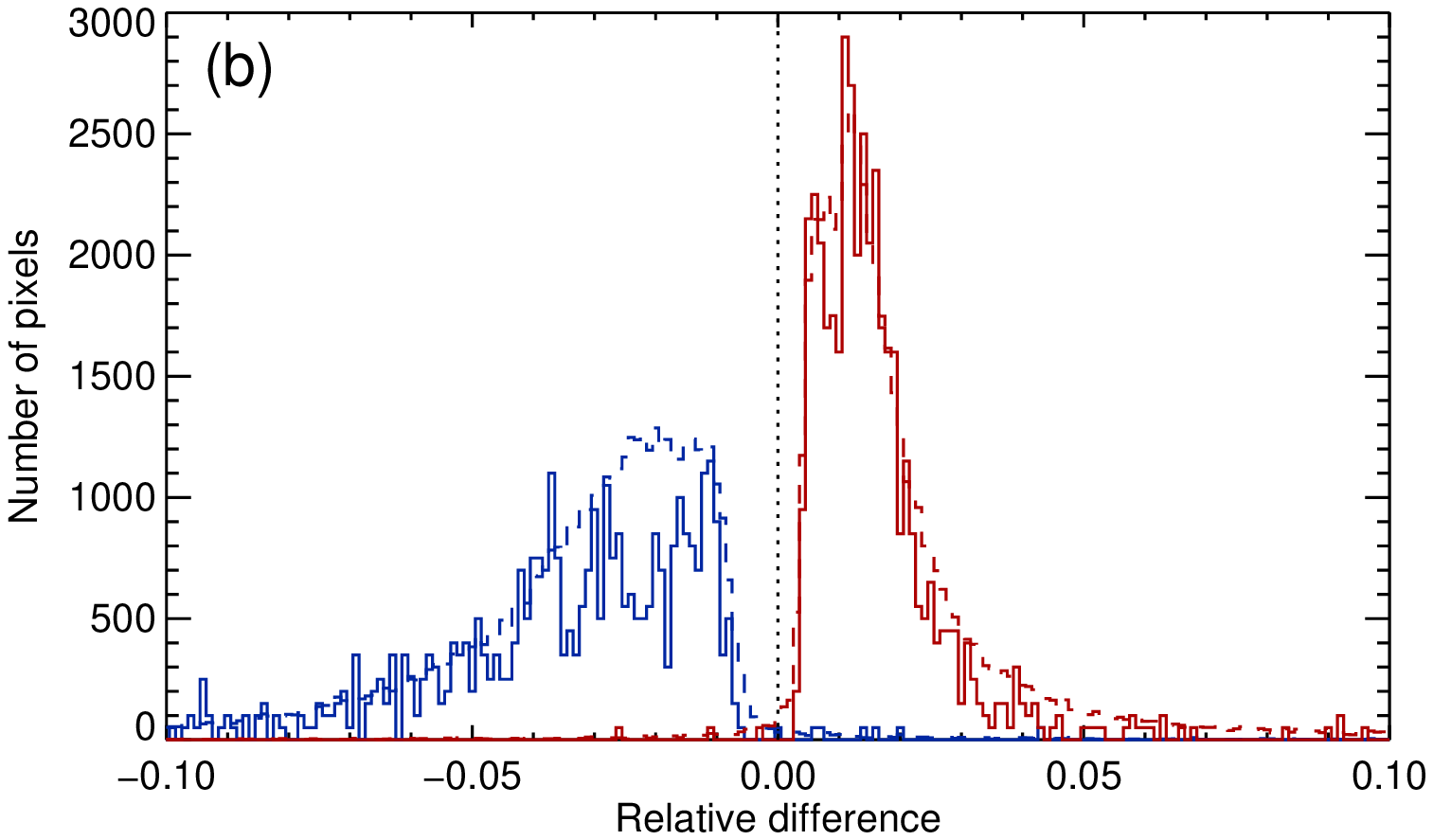}
  \caption{(a) Scatter plots of derived Fe \textsc{xiv} density ($n_{\mathrm{Fe \textsc{xiv}}}$) for different electron density of Si \textsc{vii} ($n_{\mathrm{Si \textsc{vii}}}$).  Horizontal axis indicates $n_{\mathrm{Fe \textsc{xiv}}}$ of EBW component derived by assuming $n_{\mathrm{Si \textsc{vii}}}=10^9 \, \mathrm{cm}^{-3}$.  Vertical axis indicates $n_{\mathrm{Fe \textsc{xiv}}}$ of EBW component derived by assuming $n_{\mathrm{Si \textsc{vii}}}=10^8 \, \mathrm{cm}^{-3}$ (\textit{blue}) and $10^{10} \, \mathrm{cm}^{-3}$ (\textit{red}). (b) The same data as in panel (a) but the \textit{horizontal} axis indicates a relative difference $\Delta n_{\mathrm{Fe \textsc{xiv}}} / n_{\mathrm{Fe \textsc{xiv}}}$, where $\Delta n_{\mathrm{Fe \textsc{xiv}}}$ is a difference of $n_{\mathrm{Fe \textsc{xiv}}}$ for different $n_{\mathrm{Si \textsc{vii}}}$ ($10^{8}$ and $10^{10} \, \mathrm{cm}^{-1}$) measured from the case for $n_{\mathrm{Si \textsc{vii}}}=10^{9} \, \mathrm{cm}^{-3}$.}
  \label{fig:err_sivii}
\end{figure}

In the line profile analysis, we assumed that the electron density corresponding to the temperature of Si \textsc{vii} (\textit{i.e.}, the transition region; hereafter $n_{\mathrm{Si \textsc{vii}}}$) was $10^{9} \, \mathrm{cm}^{-3}$.  Since the electron density is not the same for emission lines with different formation temperature, there is an uncertainty in $n_{\mathrm{Si \textsc{vii}}}$ which cannot be determined from the data used in this analysis.  In order to evaluate the error in the electron density derived for Fe \textsc{xiv} ($n_{\mathrm{Fe \textsc{xiv}}}$) coming from this uncertainty, we remove the blending Si \textsc{vii} at Fe \textsc{xiv} $274.20${\AA} in three cases for $n_{\mathrm{Si \textsc{vii}}}$: $10^{8}$, $10^{9}$, and $10^{10} \, \mathrm{cm}^{-3}$, and derived $n_{\mathrm{Fe \textsc{xiv}}}$ for each case.  Panel (a) in Fig.~\ref{fig:err_sivii} shows scatter plots for the electron density of EBW component within the entire western outflow region derived for the case $n_{\mathrm{Si \textsc{vii}}}=10^9 \, \mathrm{cm}^{-3}$ vs.\ $10^8 \, \mathrm{cm}^{-3}$ $(10^{10} \, \mathrm{cm}^{-3})$ by \textit{blue} (\textit{red}).  The $n_{\mathrm{Fe \textsc{xiv}}}$ of EBW component derived by assuming $n_{\mathrm{Si \textsc{vii}}}=10^8$ $(10^{10}) \, \mathrm{cm}^{-3}$ becomes smaller (larger).  Panel (b) in Fig.~\ref{fig:err_sivii} shows those relative differences $\Delta n_{\mathrm{Fe \textsc{xiv}}} / n_{\mathrm{Fe \textsc{xiv}}}$, where $\Delta n_{\mathrm{Fe \textsc{xiv}}}$ is a difference of $n_{\mathrm{Fe \textsc{xiv}}}$ for different $n_{\mathrm{Si \textsc{vii}}}$ ($10^{8}$ and $10^{10} \, \mathrm{cm}^{-1}$) measured from the case for $n_{\mathrm{Si \textsc{vii}}}=10^{9} \, \mathrm{cm}^{-3}$.  Colors (\textit{red} and \textit{blue}) indicate the same meaning as in panel (a).  \textit{Solid} and \textit{Dashed} histograms respectively indicate that for the western outflow region (the \textit{white dashed} box in Fig.~\ref{fig:dens_map}) and for the entire field of view.  These relative differences were calculated in log scale.  The histograms show that the error coming from the difference of $n_{\mathrm{Si \textsc{vii}}}$ does not exceed $5${\%}.  It means that the error is around $10^{0.4\text{--}0.5}$ at most for the density range $10^{8} \, \mathrm{cm}^{-3} \leq n_{\mathrm{e}} \leq 10^{10} \, \mathrm{cm}^{-3}$, roughly becomes a factor of $3$ (\textit{i.e.}, comparable to the error originated in the photon noise). 

% --- End of Tex ---

%% file: tex/dns_sum.tex
% =================================
%   Chapter:
%     Density.
%   Section:
%     Summary.
% =================================

\input{tex/tab_dns_cmpl.tex}

The electron density of the outflow from the edges of NOAA AR10978 was measured by using an emission line pair Fe \textsc{xiv} $264.78${\AA}/$274.20${\AA}.  The upflow component was extracted from an enhanced blue wing (EBW) in Fe \textsc{xiv} line profiles through double-Gaussian fitting.  We fitted those two Fe \textsc{xiv} emission lines simultaneously with a physical restriction that corresponding components in two emission lines must have the same Doppler velocity and thermal width, which previous EIS analysis on the density diagnostics have not been challenged.  
%{\color{red} 
The Doppler velocities, derived electron densities, and the column depths for the studied outflow regions are listed in Table \ref{tab:dns_cmpl}.
  
The derived electron density for the major component ($n_{\mathrm{Major}}$) and that for EBW component ($n_{\mathrm{EBW}}$) had opposite relationship in their magnitudes at the eastern and western outflow regions.  There are several possibilities which cause the difference in the magnitude relationship between the east and west outflow region as follows.  (1) The major component and EBW in Fe \textsc{xiv} line profiles are not directly related (\textit{e.g.}, superposition of structures along the line of sight).  The electron density of EBW component just reflects the energy input amount.  (2) The eastern outflow regions consist of the footpoints of corona loops extending to the north and connected to the opposite magnetic polarity around $(x,y)=(-170'', -70'')$, while longer coronal loops emanate in the western outflow regions and extend to the north west considering from the appearance in Fig.~\ref{fig:fexiv_1G_map}.  The difference in length may influence the plasma density even by the same driving mechanism for the outflow, since it is more easily for the upflows in a open structure to flow without condensation than for those in a closed loop.  

We also calculated the column depth for each component ($h_{\mathrm{Major}}$ and $h_{\mathrm{EBW}}$).  In the eastern region, $h_{\mathrm{EBW}}$ was smaller than $h_{\mathrm{Major}}$ by roughly one order of magnitude, which implies that the upflows possess only a small fraction ($\sim 0.1$) where the major rest plasma dominates in terms of the volume amount.  Considering this implication with the result for the electron density ($n_{\mathrm{EBW}} \ge n_{\mathrm{Major}}$), it leads to a picture that the upflows may play a role in supplying hot plasma ($\logt = 6.2 \text{--} 6.3$) into coronal loops.  On the other hand, in the western outflow region, the upflows have a larger volume by a factor of $5$--$6$ than the rest plasma, from which we consider the western outflow region as a structure composed of extending tubes with unidirectional upflows.%}

% --- End of TeX ---

%% file: tex/tab_dns_cmpl.tex
% ===================================
%   Chapter:
%     Density.
%   Section:
%     Summary.
% ===================================
\begin{table}
  \centering
  \caption{%
%{\color{red} 
  Doppler velocities, electron densities, and column depths of EBW component and the major component derived through the double-Gaussian fitting %}
%{\color{dartmouthgreen} \textbf{
applied to Fe \textsc{xiv} $264${\AA}/$274${\AA}.  Caution that the Doppler velocities listed in the table are calculated by using limb spectra observed independently on 2007 December 6 as a reference of zero velocity, which leads to the error up to $10 \, \kmpers$ at most (\textit{cf.} Chapter \ref{chap:cal}).%}}
}
  \begin{tabular}{lrrrrrr}
    \toprule
    & \multicolumn{3}{c}{EBW component} 
    & \multicolumn{3}{c}{The major component} 
    \\
    \cmidrule(lr){2-4}
    \cmidrule(lr){5-7}
    \rule[-2pt]{0pt}{13pt}
    & \multicolumn{1}{c}{$v_{\mathrm{Dop}} \, (\kmpers)$} 
    & \multicolumn{1}{c}{$\log n_{\mathrm{e}} \, [\mathrm{cm}^{-3}]$}
    & \multicolumn{1}{c}{$\log h \, [\mathrm{cm}]$}
    & \multicolumn{1}{c}{$v_{\mathrm{Dop}} \, (\kmpers)$} 
    & \multicolumn{1}{c}{$\log n_{\mathrm{e}} \, [\mathrm{cm}^{-3}]$}
    & \multicolumn{1}{c}{$\log h \, [\mathrm{cm}]$}
    \\
    \midrule
    \multicolumn{2}{l}{Eastern outflow region} & \rule[-2pt]{0pt}{15pt} & & & & \\
    U1     & $-92.4 \pm \mspace{9mu} 2.4$ & $9.17 \pm 0.09$ & $7.03 \pm 0.22$ & $-4.7 \pm 0.9$ & $9.10 \pm 0.04$ & $8.38 \pm 0.05$ \\
    U2     & $-84.8 \pm 21.4$ & $8.95 \pm 0.09$ & $7.67 \pm 0.34$ & $-3.6 \pm 1.7$ & $8.93 \pm 0.09$ & $8.64 \pm 0.12$ \\
    \midrule
    Ave.{} & $-88.8 \pm 15.2$ & $9.06 \pm 0.14$ & $7.36 \pm 0.43$ & $-4.2 \pm 1.4$ & $9.01 \pm 0.11$ & $8.51 \pm 0.16$ \\
    \multicolumn{2}{l}{Western outflow region} & \rule[-2pt]{0pt}{18pt} & & & & \\
    U3     & $-61.4 \pm 15.7$ & $8.79 \pm 0.21$ & $8.25 \pm 0.35$ & $-0.6 \pm 2.6$ & $9.31 \pm 0.09$ & $7.74 \pm 0.12$ \\
    U4     & $-56.3 \pm 15.2$ & $8.53 \pm 0.17$ & $8.80 \pm 0.34$ & $ 3.4 \pm 2.5$ & $9.34 \pm 0.10$ & $7.74 \pm 0.10$ \\
    U5     & $-73.2 \pm 12.2$ & $8.48 \pm 0.21$ & $8.59 \pm 0.41$ & $-1.3 \pm 1.0$ & $9.11 \pm 0.07$ & $7.67 \pm 0.18$ \\
    U6     & $-54.3 \pm 13.3$ & $8.64 \pm 0.12$ & $8.40 \pm 0.29$ & $-0.8 \pm 1.1$ & $9.10 \pm 0.05$ & $7.89 \pm 0.10$ \\
    \midrule
    Ave.{} & $-62.0 \pm 16.0$ & $8.60 \pm 0.22$ & $8.53 \pm 0.41$ & $ 0.1 \pm 2.7$ & $9.22 \pm 0.14$ & $7.74 \pm 0.15$ \\
    \bottomrule
  \end{tabular}
  \label{tab:dns_cmpl}
\end{table}

% --- End of TeX ---

%% file: tex/dns_app_262n265.tex
% ===========================================
%   Chapter:
%     Density.
%   Section:
%     Fe XVI 262 and 265.
% ===========================================

\begin{figure}
  \centering
  \includegraphics[width=8.4cm,clip]{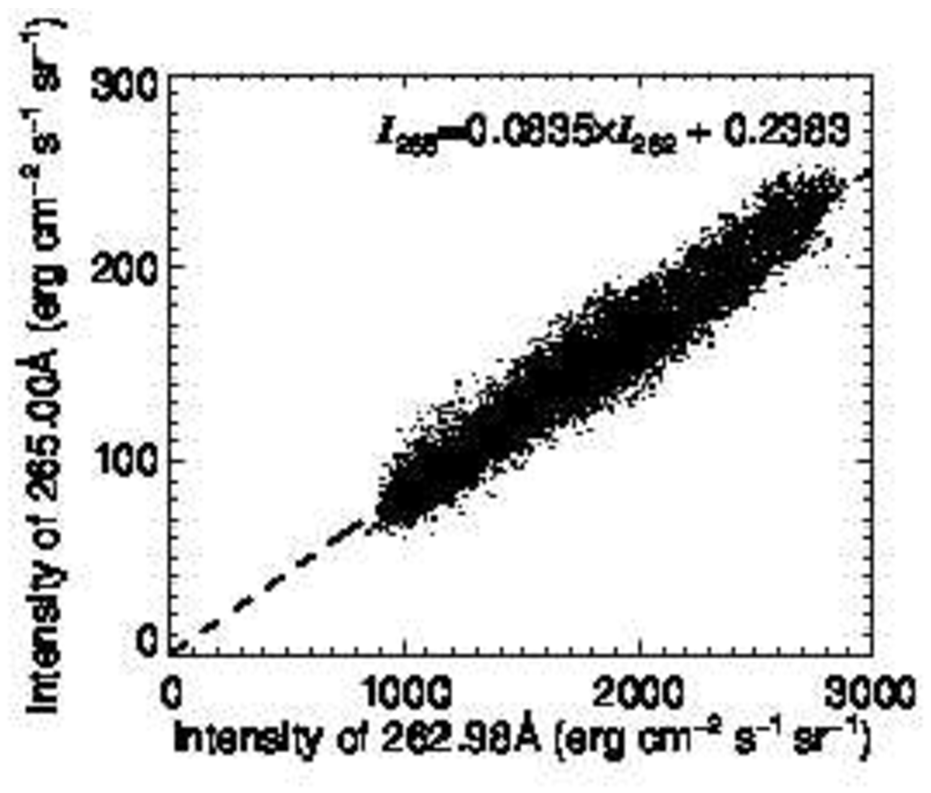}
  \includegraphics[width=8.4cm,clip]{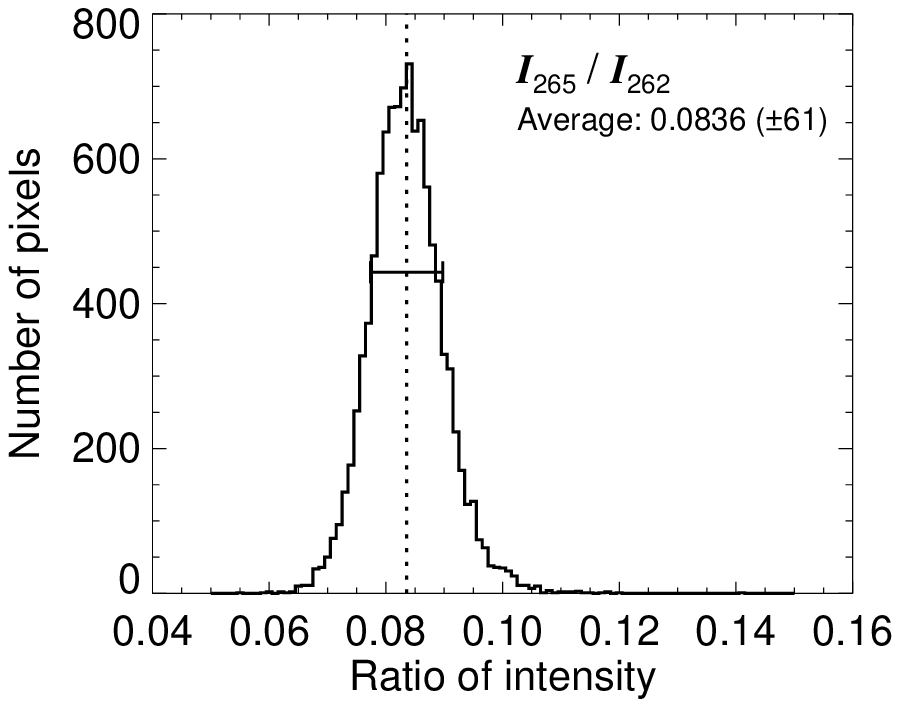}
  \caption{\textit{Left}: scatter plot for intensities of Fe \textsc{xvi} $262.98${\AA} and $265.01${\AA} derived from the scan data used in Chapter \ref{chap:vel}.  A \textit{dashed} line indicates a fitted linear function whose representation is written in the upper part. \textit{Right}: the ratio of intensities between those two emission lines. The average value of the ratio is written in the panel.}
  \label{fig:262n265}
\end{figure}

We investigated the intensity ratio between Fe \textsc{xvi} $262.98${\AA} and $265.01${\AA} here.  Since the scan used in this chapter did not include a line core of Fe \textsc{xvi} $265.01${\AA}, we exploited that used in Chapter \ref{chap:vel} which includes both emission lines.  \textit{Left} panel in Fig.~\ref{fig:262n265} shows the scatter plot for intensities of Fe \textsc{xvi} $262.98${\AA} ($I_{262}$) and $265.01${\AA} ($I_{265}$).  A \textit{dashed} line indicates the fitted linear function, whose slope is $0.0835$ (\textit{i.e.}, the intensity ratio).  \textit{Right} panel shows the histogram for the ratio between the two emission lines ($I_{265}/I_{262}$). The average value of $I_{265}/I_{262}$ is $0.0836 \, (\pm 6.1 \times 10^{-3})$, which is almost the same as the value obtained from the linear fitting in the \textit{left} panel.  

% --- End of TeX ---

%% file: tex/dns_app_in.tex
% ==============================================
%   Chapter:
%     Density.
%   Section:
%     Appendix for intensity-density scatter.
% ==============================================

We can see a clear positive correlation between peak intensity and electron density in the intensity range larger than $I_{\mathrm{Major}} = 2.0 \times 10^{3} \, \mathrm{erg} \, \mathrm{cm}^{-2} \, \mathrm{s}^{-1} \, \mathrm{sr}^{-1} \, \text{\AA}^{-1}$ (indicated by a \textit{vertical dashed} line) while the plot is more scattered below that intensity.  Not only the photon noise contributes to this large uncertainty, but also unidentified blended emission lines could do.  Therefore we analyzed the data points with $I_{\mathrm{Major}}$ larger than the value which the \textit{vertical dashed} line indicates. 

\begin{figure}
  \centering
  \includegraphics[width=9.6cm,clip]{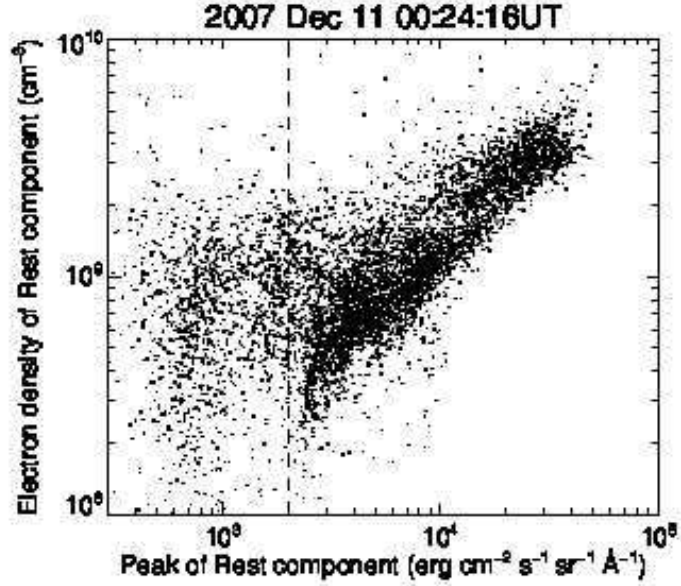}
  \caption{Peak intensity of Fe \textsc{xiv} $264.78${\AA} vs.\ electron density deduced from the major component of Fe \textsc{xiv} $264.78${\AA} and $274.20${\AA} in the double Gaussian fitting.}
  \label{fig:in}
\end{figure}

% --- End of TeX ---

%% file: tex/mph_mgvii_dens.tex
% ========================================
%   Chapter:
%     Morphology of the outflow region.
%   Section:
%     Mg VII density at the footpoints.
% ========================================

\begin{figure}[h]
  \centering
  \includegraphics[width=7.8cm,clip]{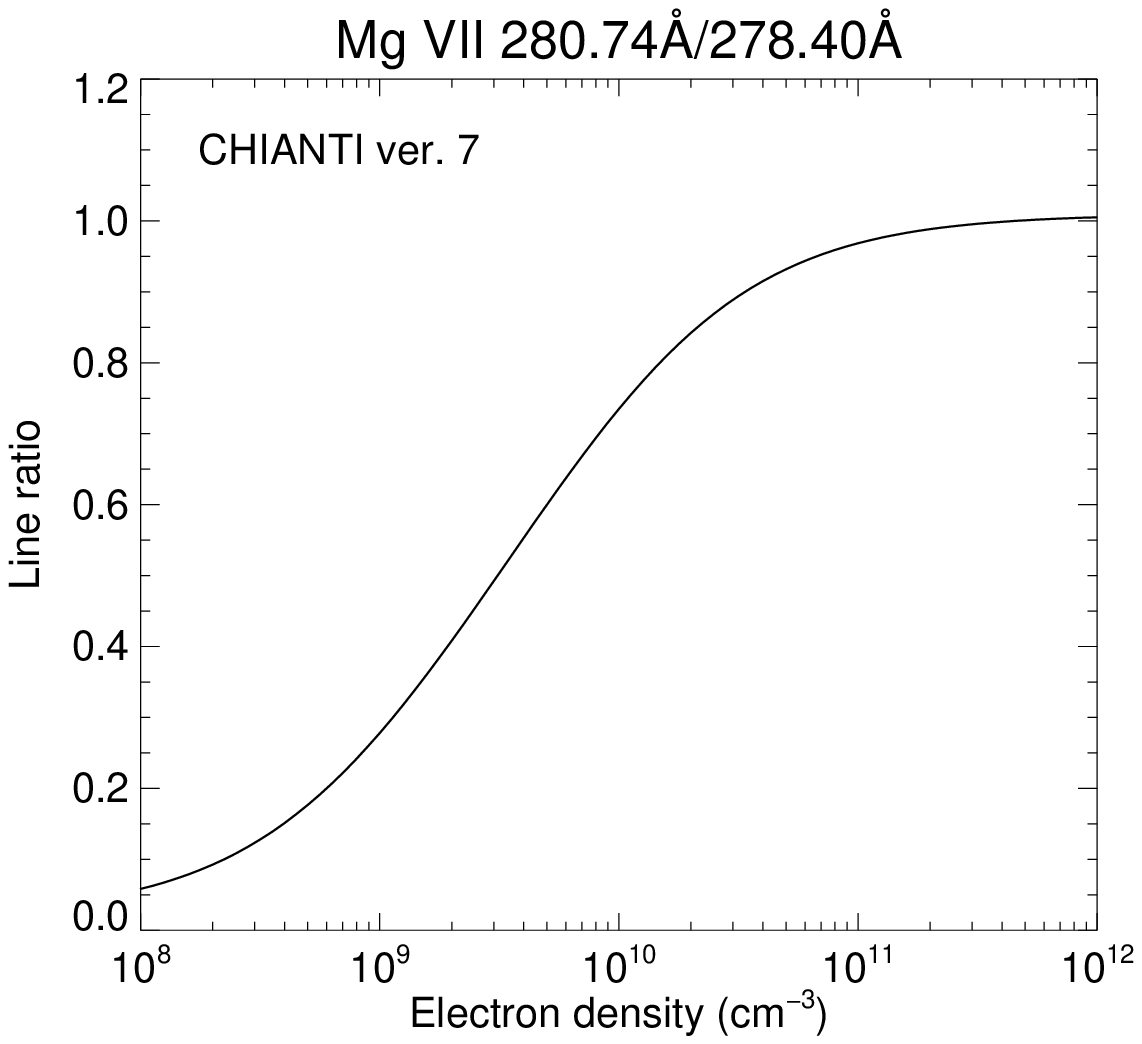}
  \includegraphics[width=9.1cm,clip]{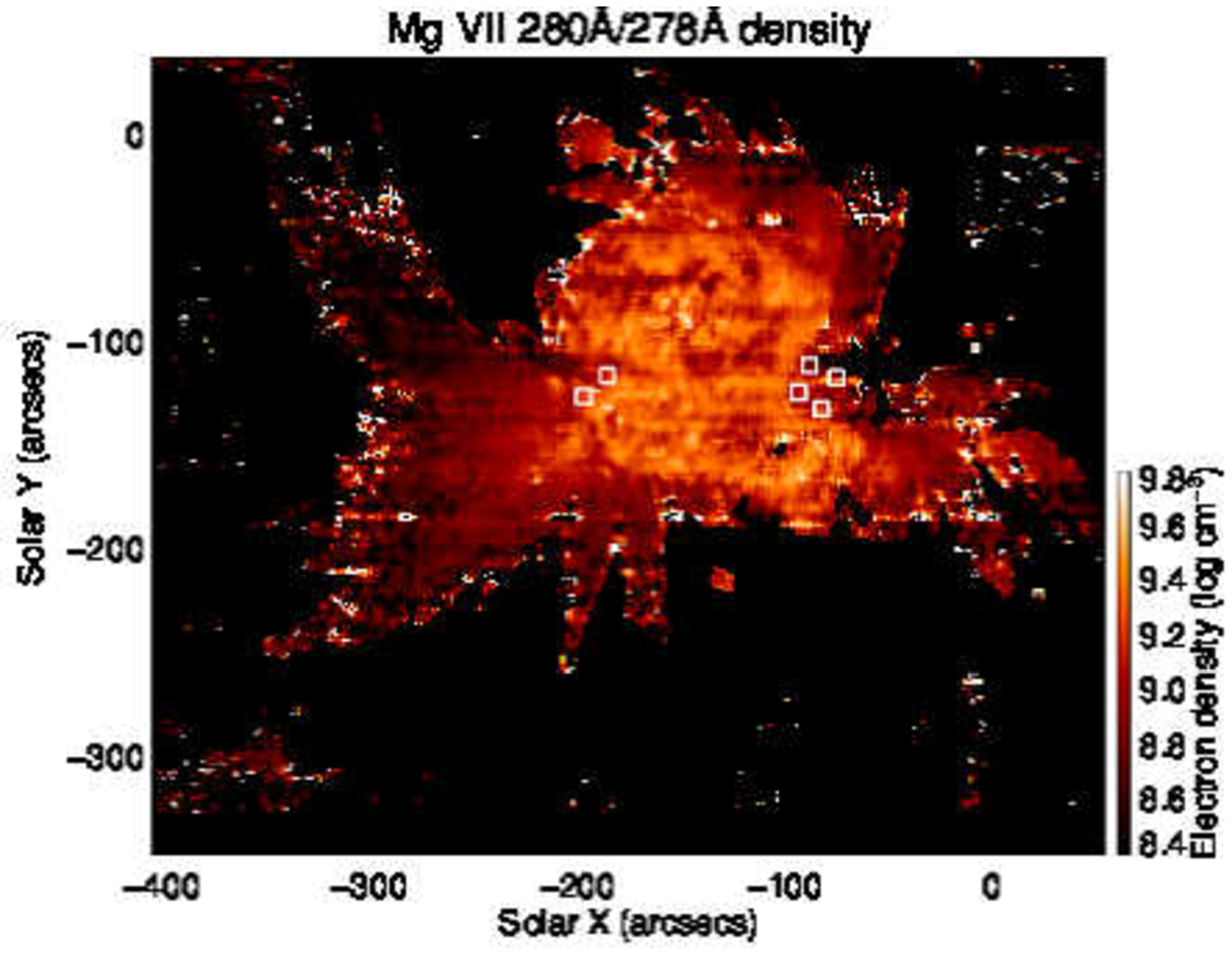}
  \caption{\textit{Left}: Intensity ratio of Mg \textsc{vii} $280.74${\AA}/$278.40${\AA} as a function of electron density.  CHIANTI ver.~7 \citep{dere1997,landi2012} was used in the calculation.  \textit{Right}: electron density map derived by the Mg \textsc{vii} line pair.}
  \label{fig:mph_mgvii_dens}
\end{figure}

The electron density in NOAA AR10978 was also measured by using a Mg \textsc{vii} emission line pair $280.74${\AA}/$278.40${\AA}.  This line pair was analyzed since the formation temperature of Mg \textsc{vii} is $\log T \, [\mathrm{K}]=5.8$, which is almost the same as that of Si \textsc{vii}.  We can check the validity of the assumption of $n_{\mathrm{Si \textsc{vii}}}=10^{9} \, \mathrm{cm}^{-3}$ in our line profile analysis.  The scan used in this chapter did not include these emission lines, so we exploited that used in Chapter \ref{chap:vel}.  The theoretical curve for the intensity ratio of Mg \textsc{vii} $280.74$/$278.40${\AA} is shown by \textit{left} panel in Fig.~\ref{fig:mph_mgvii_dens}.  This ratio has a good sensitivity to the electron density within the range of $n_{\mathrm{e}}=10^{8\text{--}11} \, \mathrm{cm}^{-3}$ and is suitable for the density diagnostics in the corona as the typical electron density falls within that range. 

The spectra were spatially averaged by $5'' \times 5''$ and the intensity of each emission line was calculated by single-Gaussian fitting.  Then we derived the electron density at each location referring to the theoretical curve.  \textit{Right} panel in Fig.~\ref{fig:mph_mgvii_dens} shows the spatial map for the electron density derived from the Mg \textsc{vii} line pair.  The color contour was scaled within $n_{\mathrm{e}} = 10^{8.4\text{--}9.8} \, \mathrm{cm}^{-3}$.  The region painted by \textit{black} means that the peak intensity of Mg \textsc{vii} $278.40${\AA} did not exceed $4.0 \times 10^{2} \, \mathrm{erg} \, \mathrm{cm}^{-2} \, \mathrm{s}^{-1} \, \mathrm{sr}^{-1} \text{\AA}^{-1}$ below which the spectra were too weak for reliable measurement.  \textit{White} boxes in the map are drawn by taking into account the solar rotation so as to be located at the same position as in Fig.~\ref{fig:fexiv_1G_map}.  The electron density around the eastern and western outflow regions is around $n_{\mathrm{e}}=10^{8.8\text{--}9.2} \, \mathrm{cm}^{-3}$, from which we can conclude that our line profile analysis was valid. 

% --- End of TeX ---

%% file: tex/contents_ndv.tex
\chapter{$\lambda$-$n_{\mathrm{e}}$ diagram}
  \label{chap:ndv}
  \input{tex/ndv_itdn.tex}
\section{Method}
  \input{tex/ndv_method.tex}
    \input{tex/ndv_method_2.tex}
    \input{tex/ndv_method_3.tex}
\section{Verification of the method}
  \input{tex/ndv_test.tex}
  \subsection{Dependence on electron density}
    \input{tex/ndv_test_density.tex}
  \subsection{Dependence on intensity}
    \input{tex/ndv_test_intensity.tex}
  \subsection{Dependence on velocity}
    \input{tex/ndv_test_velocity.tex}
  \subsection{Dependence on thermal width}
    \input{tex/ndv_test_width.tex}
  \subsection{Binning effect}
    \input{tex/ndv_test_binning.tex}
\section{$\lambda$-$n_\mathrm{e}$ diagram in AR10978}
  \input{tex/ndv_sv.tex}
\section{Summary and discussion}
  \input{tex/ndv_sum.tex}

%% file: tex/ndv_itdn.tex
% !TEX root = main.tex
% =========================
%   Lambda-N diagram
%   Introduction
% =========================

We modeled the spectra by the composition of two Gaussians in the above analysis.  However, it is difficult to prove whether this assumption is suitable for the outflow regions.  There are two alternative approaches to deal such a spectrum consist of more than two Gaussians.  One way is to adopt multiple-Gaussian functions (more than two components) and resolve multiple flows existing in a emission line.  More number of free parameters we use, the spectra would be fitted with less $\chi^2$.  But this does not mean that we extracted much more useful physical information from the spectra.  The number of local minima increase with complexity of fitting model, and the fitting process becomes an ill-posed problem.  

The other way is our new type of plot without assuming any fitting model.  Each spectral bin in a pair of a spectrum is used to derive electron density at those each bin, which we refer to as ``$\lambda$-$n_{\mathrm{e}}$ diagram'' hereafter.  In this method, we obtain the measure of electron density of the plasma which have the speed of $v_{\mathrm{Dop}}=c \, (\lambda - \lambda_0)/\lambda_0$ ($\lambda_0$: rest wavelength), which is a function of wavelength.  Consider a density-sensitive pair of spectra $\phi_1 (\lambda)$ and $\phi_2 (\lambda)$ emitted from the same degree of an ion.  These emission lines must have the same Doppler velocity because they came from the same degree of the ion, so after converting the variable $\lambda$ into Doppler velocity $v_{\mathrm{Dop}}$ as denoted by $\phi_i^* (v_{\mathrm{Dop}}) = \phi_i (\lambda)$ ($i=1,2$), we can calculate the electron density as a function of the Doppler velocity
\begin{equation}
  n_{\mathrm{e}}^{*} (v_{\mathrm{Dop}}) = 
    R^{-1}
    \left[
      \frac{\phi_2^* (v_{\mathrm{Dop}})}{\phi_1^* (v_{\mathrm{Dop}})} 
    \right] \, \text{.}
    \label{eq:ndv_rprsnt}
\end{equation}
The derived $n_{\mathrm{e}}^{*} (v_{\mathrm{Dop}})$ can be converted into a function of wavelength in either spectrum, $n_{\mathrm{e}} (\lambda)$, by the equation of Doppler effect.  Function $R (n_{\mathrm{e}})$ is the ratio of intensities from two emission lines which is a function of electron density, so when we know the intensities of two emission lines which are represented as
\begin{align}
  I_1 = \int \phi_1 (\lambda) d \lambda \, \text{,} \\
  I_2 = \int \phi_2 (\lambda) d \lambda \, \text{,} 
\end{align}
electron densities can be usually derived by
\begin{equation}
  n_{\mathrm{e}} = 
    R^{-1}
    \left( 
      \frac{I_2}{I_1} 
    \right) \, \text{.}
\end{equation}

As shown in above equations, $\lambda$-$n_{\mathrm{e}}$ diagram represents that of the particles which move with that speed, in other words, we does not obtain the electron density of the whole plasma as an ensemble of Maxwellian distribution.  We emphasize the advantage of our method using Eq.~(\ref{eq:ndv_rprsnt}) that even if we do not know the precise functional form of spectra, it gives us the electron density as a function of Doppler velocity without any modeling.  

% --- End of Tex ---

%% file: tex/ndv_method.tex
% =========================
%   Lambda-N diagram
%   Method
% =========================
Making $\lambda$-$n_\mathrm{e}$ diagram contains processes: (1) subtraction of blending emission line, (2) adjusting wavelength scale of Fe \textsc{XIV} $264.78${\AA} to $274.20${\AA} by interpolation, and (3) density inversion at each spectral pixel. Since the blend of an emission line Si \textsc{vii} $274.18${\AA} into Fe \textsc{xiv} $274.20${\AA} was already described in Section \ref{sec:de-blend}, here we explain the processes (2) and (3). 

% --- End of Tex ---

%% file: tex/ndv_method_2.tex
% ==================================================
%   Lambda-N diagram
%   Method: adjusting wavelength scale
% ==================================================

Since the EIS instrument does not have absolute wavelength scale, the corresponding wavelength location of the same velocity in Fe \textsc{xiv} $264.78${\AA} and $274.20${\AA} must be determined from data itself as described in Section \ref{sec:wavelength_adjust}.  Using obtained relation $\lambda_{\mathrm{obs,}274}/\lambda_{\mathrm{obs,}264} = 1.0355657$ $(\pm 0.0000044)$, each wavelength value imposed on the spectral window of Fe \textsc{xiv} $264.78${\AA} was projected onto the values on the spectral window of Fe \textsc{xiv} $274.20${\AA} by the scaling 
\begin{equation}
  \tilde{\lambda_{i}} = \alpha \lambda_{264,i} \, \, (\alpha = 1.0355657) \, \text{,}
  \label{eq:ndv_scaling}
\end{equation}
where a number $i$ indicates the $i$th spectral pixel in a spectrum of $264.78${\AA}.

At this point, we projected the spectrum of Fe \textsc{xiv} $264.78${\AA} into another location in the wavelength direction, which is shifted by as same Doppler shift as the spectrum of Fe \textsc{xiv} $274.20${\AA}.  However, since the spectral pixels on the CCD of EIS are positioned almost linearly as a function of wavelength, the wavelength values of the projected spectrum do not coincide to those of Fe \textsc{xiv} $274.20${\AA}.  In order to align them in a exactly same wavelength values, the projected spectrum of Fe \textsc{xiv} $264.78${\AA} was interpolated by a cubic spline.  

% --- End of Tex ---

%% file: tex/ndv_method_3.tex
% ==================================================
%   Lambda-N diagram
%   Method: density inversion
% ==================================================

We can calculate the ratio of spectral intensity Fe \textsc{xiv} $264.78${\AA}/$274.20${\AA} at each spectral bin.  Now we are able to derive the electron density as the same way described in the section \ref{sec:dns_inv}.  Because intensity at each spectral bin has larger errors compared to the integrated intensity (\textit{e.g., } double-Gaussian fitting), the estimated errors for the electron density in the $\lambda$-$n_{\mathrm{e}}$ diagram become large especially for the line wing.

% --- End of Tex ---

%% file: tex/ndv_test.tex
% =========================
%   Density of upflow
%   Lambda-N method
% =========================

In order to test the validity of $\lambda$-$n_{\mathrm{e}}$ method, we synthesized spectra of Fe \textsc{xiv} $264.78${\AA} and $274.20${\AA} taking into account the spectral resolution of EIS and instrumental broadening.  The spectra were composed of two components which represent plasma at the rest and an upflow.  While the physical parameters for the major rest component (peak, Doppler velocity, and width) were fixed, those for a minor blueshifted component (\textit{i.e.}, upflow) were taken as variables.  We made $\lambda$-$n_{\mathrm{e}}$ diagrams for the minor component with
\vspace{-5pt}
\begin{itemize}
\item electron density of $8.50$, $8.75$, $9.00$, $9.25$, and $9.50$ 
  in the unit of $\log \, \mathrm{cm}^{-3}$, 
  \vspace{-5pt}
\item intensity of $1$, $5$, $10$, $15$, and $20 \, \mathrm{\%}$ 
  (ratio to the major component in Fe \textsc{xiv} $274.20${\AA}), 
  \vspace{-5pt}
\item Doppler velocity of $0$, $-50$, $-100$, $-150$, 
  and $-200\,\mathrm{km}\,\mathrm{s}^{-1}$, 
  \vspace{-5pt}
\item thermal width of $2.0$, $2.5$, $3.0$, $3.5$, and $4.0\,\mathrm{MK}$.
\end{itemize}
\vspace{-5pt}
The nonthermal width was not considered in this test because it does not produce any differences essentially. The test for each variable will be given below. 

% --- End of Tex ---

%% file: tex/ndv_test_density.tex
% ==========================
%   Lambda-N method
%   Dependence on density
% ==========================

\begin{figure}
  \centering
  \begin{minipage}[c]{6.0cm}
    \includegraphics[width=6.0cm,clip]{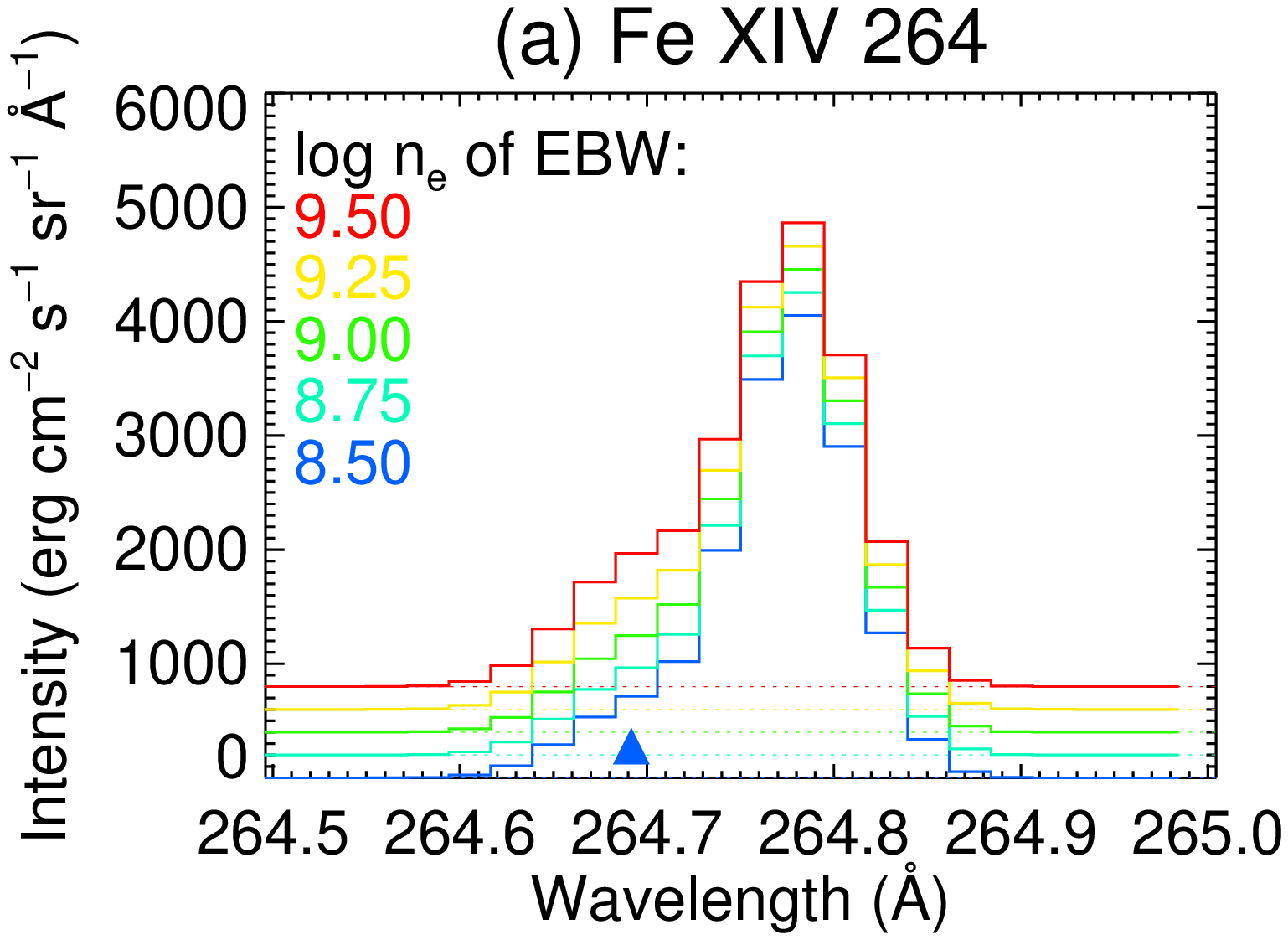}
    \includegraphics[width=6.0cm,clip]{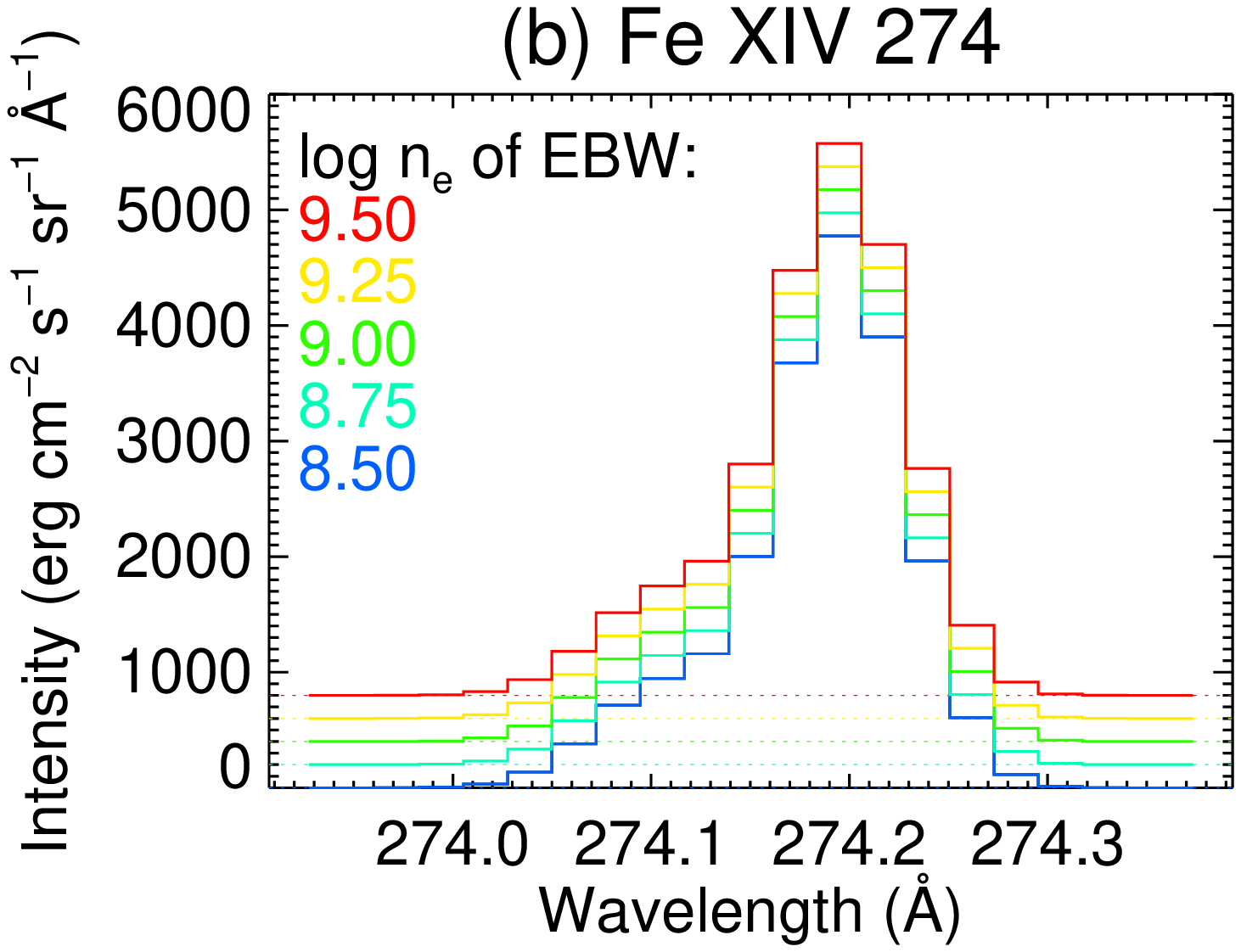}
  \end{minipage}
  \begin{minipage}[c]{10.6cm}
    \includegraphics[width=10.6cm,clip]{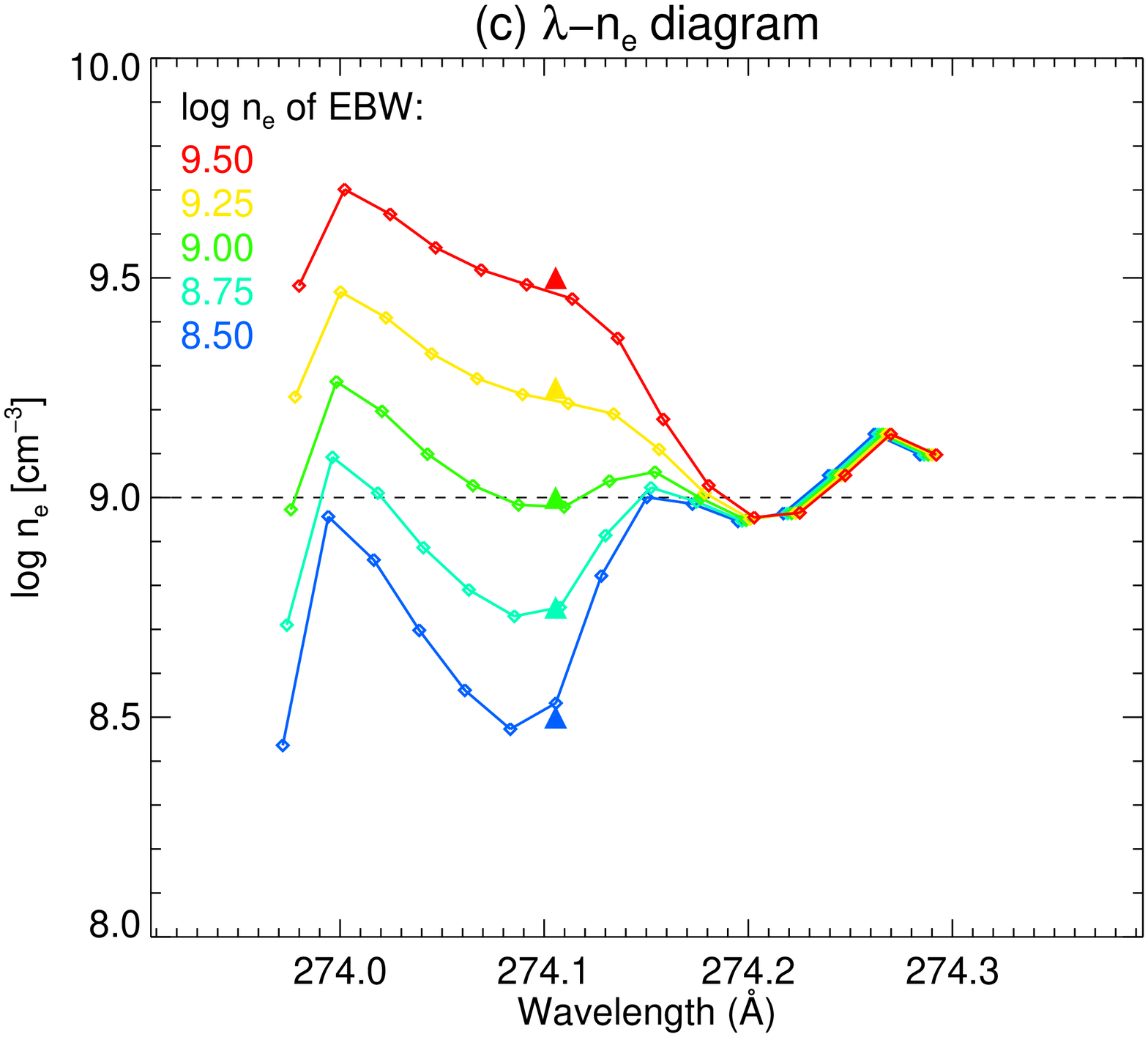}
  \end{minipage}
  \caption{
    (a) Line profiles of Fe \textsc{xiv} $264.78${\AA},
    (b) Line profiles of Fe \textsc{xiv} $274.20${\AA}, and 
    (c) $\lambda$-$n_\mathrm{e}$ diagrams.
    Each color indicates different electron density of minor blueshifted component (\textit{blue}: $8.50$, \textit{turquoise}: $8.75$, \textit{green}: $9.00$, \textit{yellow}: $9.25$, and \textit{red}: $9.25$ in the unit of $\log \, \mathrm{cm}^{-3}$). Electron density of major component was fixed to $\log n_{\mathrm{e}} \, [\mathrm{cm}^{-3}] = 9.00$. The triangles in panel (c) indicate centroid and electron density of the given minor component. 
  }
  \label{fig:test_density}
\end{figure}

The most important point on $\lambda$-$n_{\mathrm{e}}$ diagram is whether it reflects the electron density of the components which compose spectrum properly or not.  In order to test that, we synthesized the spectra which are composed of major component at the rest which has the fixed electron density of $\log n_{\mathrm{e}} [\mathrm{cm}^{-3}]=9.0$ and minor component which has the variable electron density. Five cases ($\log n_{\mathrm{e}} \, [\mathrm{cm}^{-3}] = 8.50$, $8.75$, $9.00$, $9.25$, and $9.50$) were analyzed, where the peak ratio of minor/major component was $15\mathrm{\%}$ with fixed upflow speed $v=-100 \, \mathrm{km} \, \mathrm{s}^{-1}$.  In panel (a) and (b) of Fig.~\ref{fig:test_density}, the spectra of Fe \textsc{xiv} $264.78${\AA} and $274.20${\AA} are respectively shown.  Colors (\textit{blue}, \textit{turquoise}, \textit{yellow}, \textit{green}, and \textit{red}) indicate the five cases calculated here.  After converting the wavelength scale of $264.78${\AA} to $274.20${\AA}, $\lambda$-$n_{\mathrm{e}}$ were obtained as shown in panel (c) of Fig.~\ref{fig:test_density}.  The triangles in panel (c) indicate centroid and electron density of the given minor component.  It is clear that those $\lambda$-$n_{\mathrm{e}}$ diagrams well reflect the change of the electron density from $\log n_{\mathrm{e}} \, [\mathrm{cm}^{-3}] = 8.50\text{--}9.50$.  Despite the spectra being composed of only two components, $\lambda$-$n_{\mathrm{e}}$ diagrams do not become a step function but a smooth function.  This is natural because the two Gaussians in the spectra contribute each other by their overlapping wings.  We claim that the method proposed here ($\lambda$-$n_{\mathrm{e}}$) is a good indicative of the electron density of components in the spectrum. 

% --- End of Tex ---

%% file: tex/ndv_test_intensity.tex
% ============================
%   Lambda-N method
%   Dependence on intensity
% ============================

\begin{figure}
  \centering
  \begin{minipage}[c]{6.0cm}
    \includegraphics[width=6.0cm,clip]{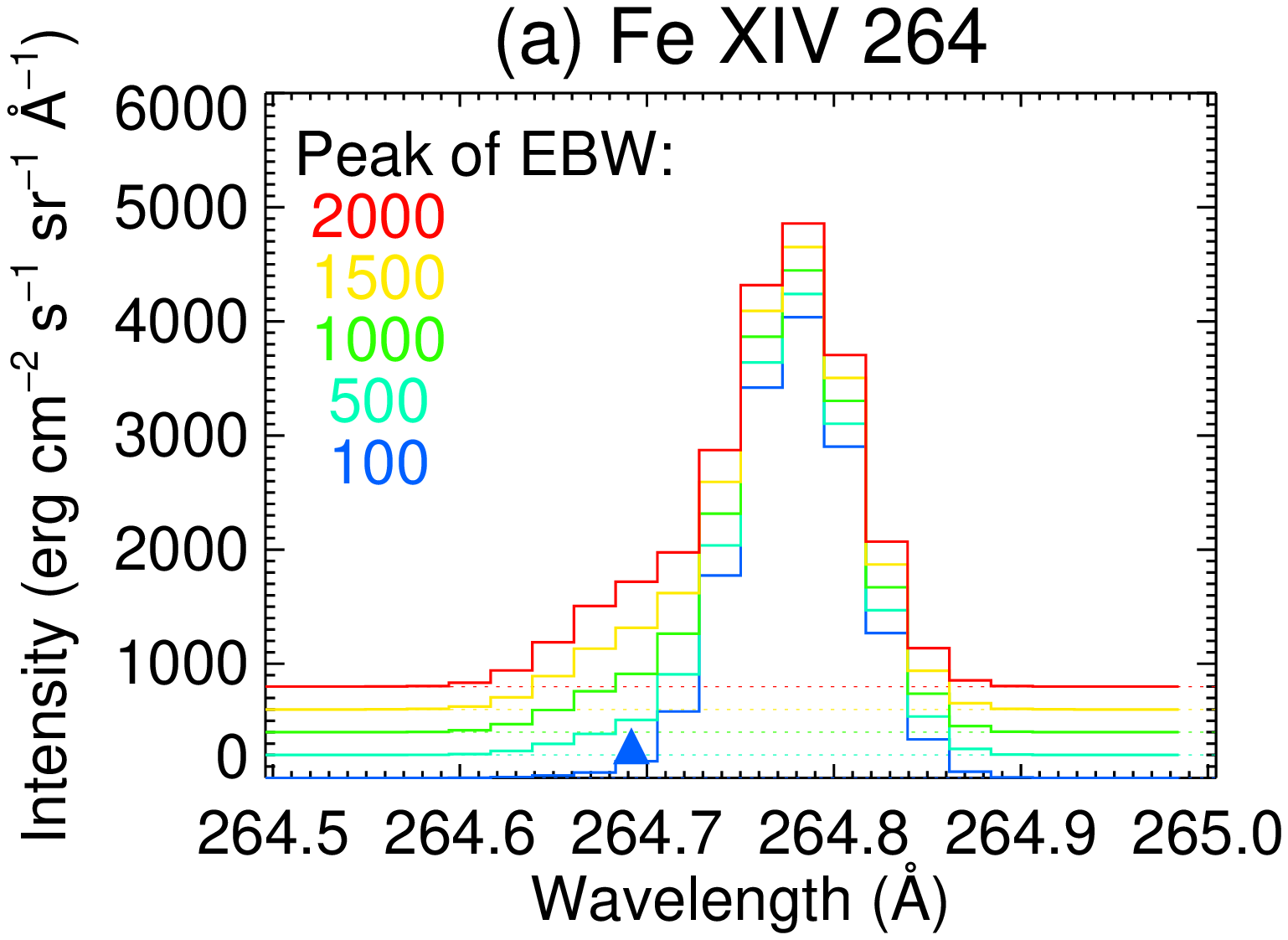}
    \includegraphics[width=6.0cm,clip]{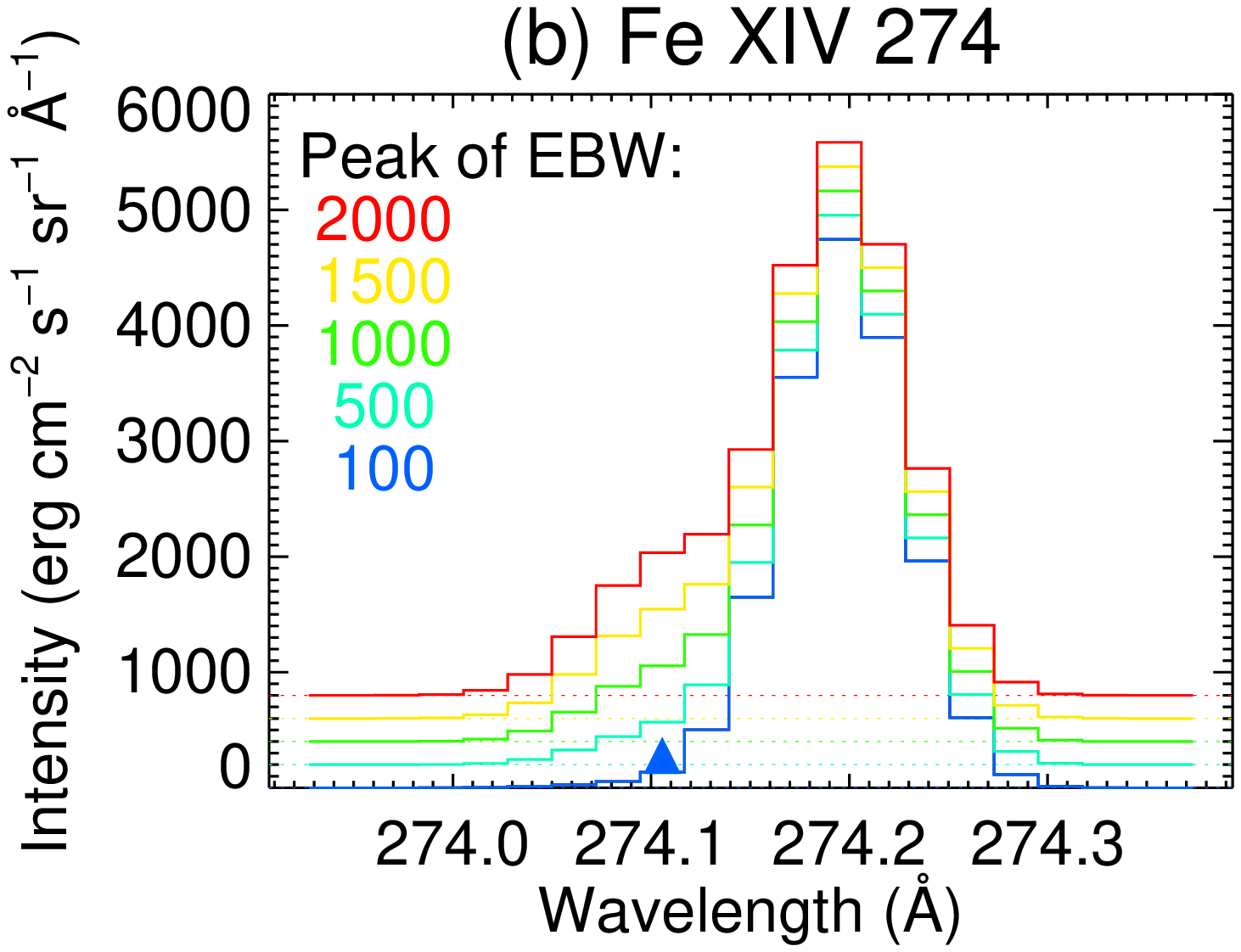}
  \end{minipage}
  \begin{minipage}[c]{10.6cm}
    \includegraphics[width=10.6cm,clip]{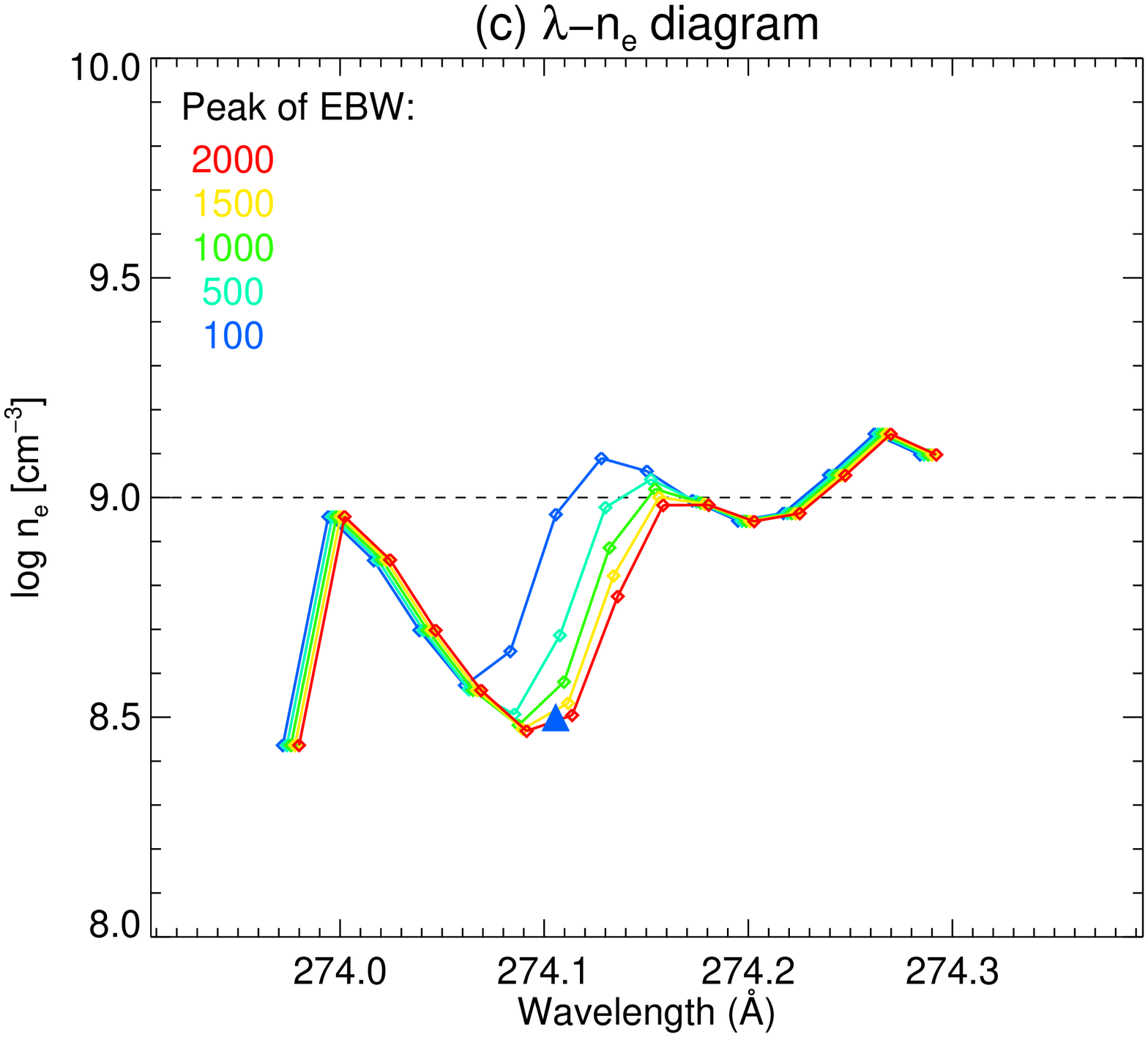}
  \end{minipage}
  \caption{
    (a) Synthetic line profiles of Fe \textsc{xiv} $264.78${\AA}, (b) Synthetic line profiles of Fe \textsc{xiv} $274.20${\AA}, and (c) $\lambda$-$n_\mathrm{e}$ diagrams.  Each color indicates different intensity of minor blueshifted component (\textit{blue}: $1${\%}, \textit{turquoise}: $5${\%}, \textit{green}: $10${\%}, \textit{yellow}: $15${\%}, and \textit{red}: $20${\%} of the major rest component). Intensity of the major rest component was fixed.}
  \label{fig:test_intensity}
\end{figure}

The dependence of $\lambda$-$n_{\mathrm{e}}$ diagram on the intensity of minor component is relatively trivial compared to previous section. As same as Fig.~\ref{fig:test_density}, the spectra of Fe \textsc{xiv} $264.78${\AA} and $274.20${\AA}, and $\lambda$-$n_{\mathrm{e}}$ diagrams are respectively shown in panel (a), (b), and (c) of Fig.~\ref{fig:test_intensity}.  Colors indicate the five cases for variable relative intensity calculated here(\textit{blue}: $1 \mathrm{\%}$, \textit{turquoise}: $5\,\mathrm{\%}$, \textit{green}: $10\,\mathrm{\%}$, \textit{yellow}: $15\,\mathrm{\%}$, and \textit{red}: $20\,\mathrm{\%}$ of the major rest component).  The blue triangle in panel (c) indicates centroid and electron density of given minor component.  Note that in this test of intensity, those two quantities were fixed ($v = -100 \, \mathrm{km} \, \mathrm{s}^{-1}$ and $\log n_{\mathrm{e}} \, [\mathrm{cm}^{-3}] = 8.5$).  As the intensity of minor blueshifted component increases (\textit{i.e.,} \textit{blue} to \textit{red}), dip around $274.07$--$274.10${\AA} becomes distinct.  This is simply because the region between two components is dominated by the major component when the minor component has relatively small peak.  On the other hand, the blue side (\textit{i.e.,} shorter wavelength) of the minor component around $274.0$--$274.04${\AA} does not change significantly for the five cases.  Although the location of the dip moves by a change in the relative intensity to a certain extent, the tendency of $\lambda$-$n_{\mathrm{e}}$ diagrams for relative intensity larger than $5\,\mathrm{\%}$ are very similar and well reflect the input electron density, in other words, they are useful for us to know the existence of tenuous upflow. 

% --- End of Tex ---

%% file: tex/ndv_test_velocity.tex
% ===========================
%   Lambda-N method
%   Dependence on velocity
% ===========================

\begin{figure}
  \centering
  \begin{minipage}[c]{6.0cm}
    \includegraphics[width=6.0cm,clip]{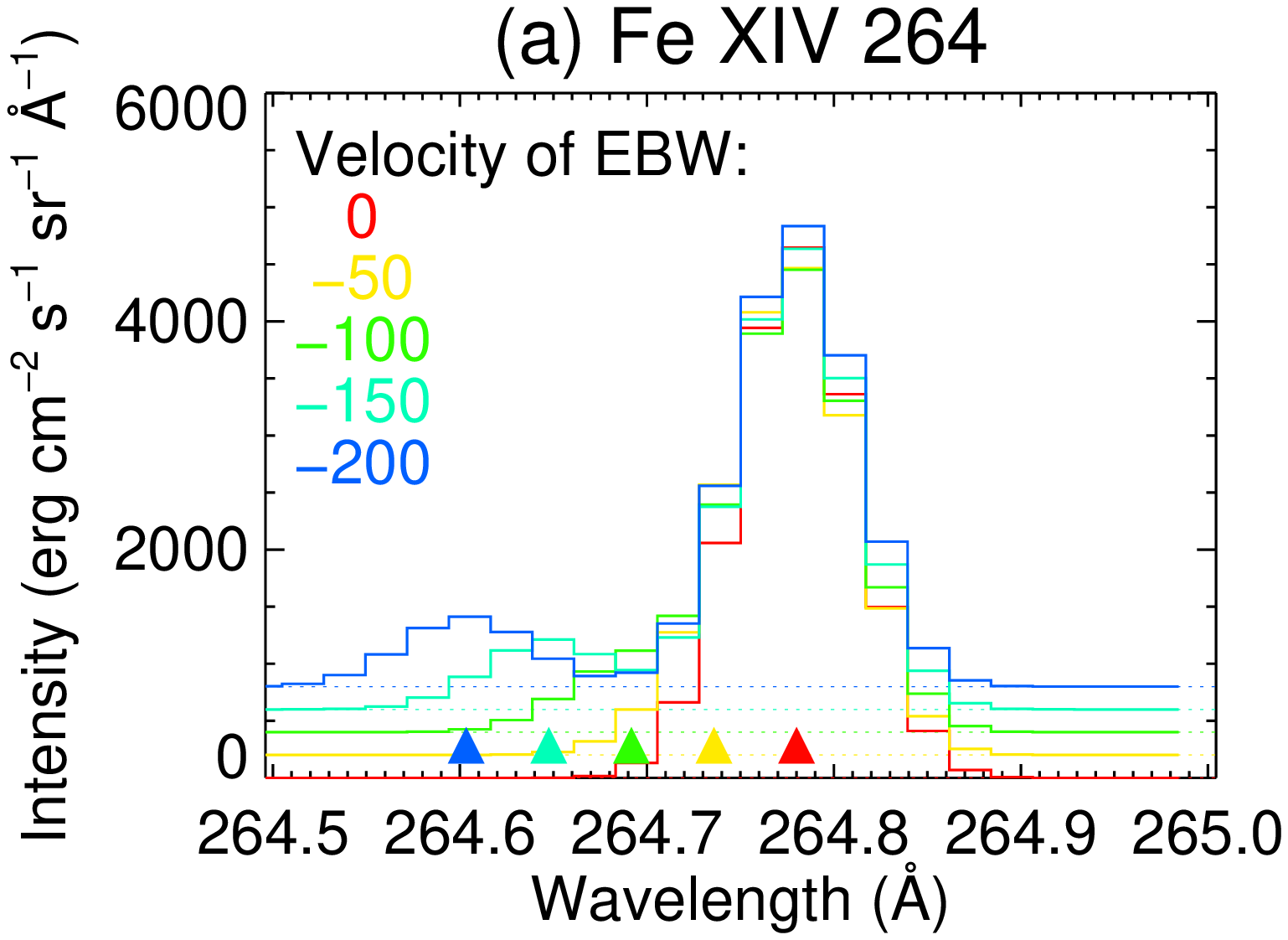}
    \includegraphics[width=6.0cm,clip]{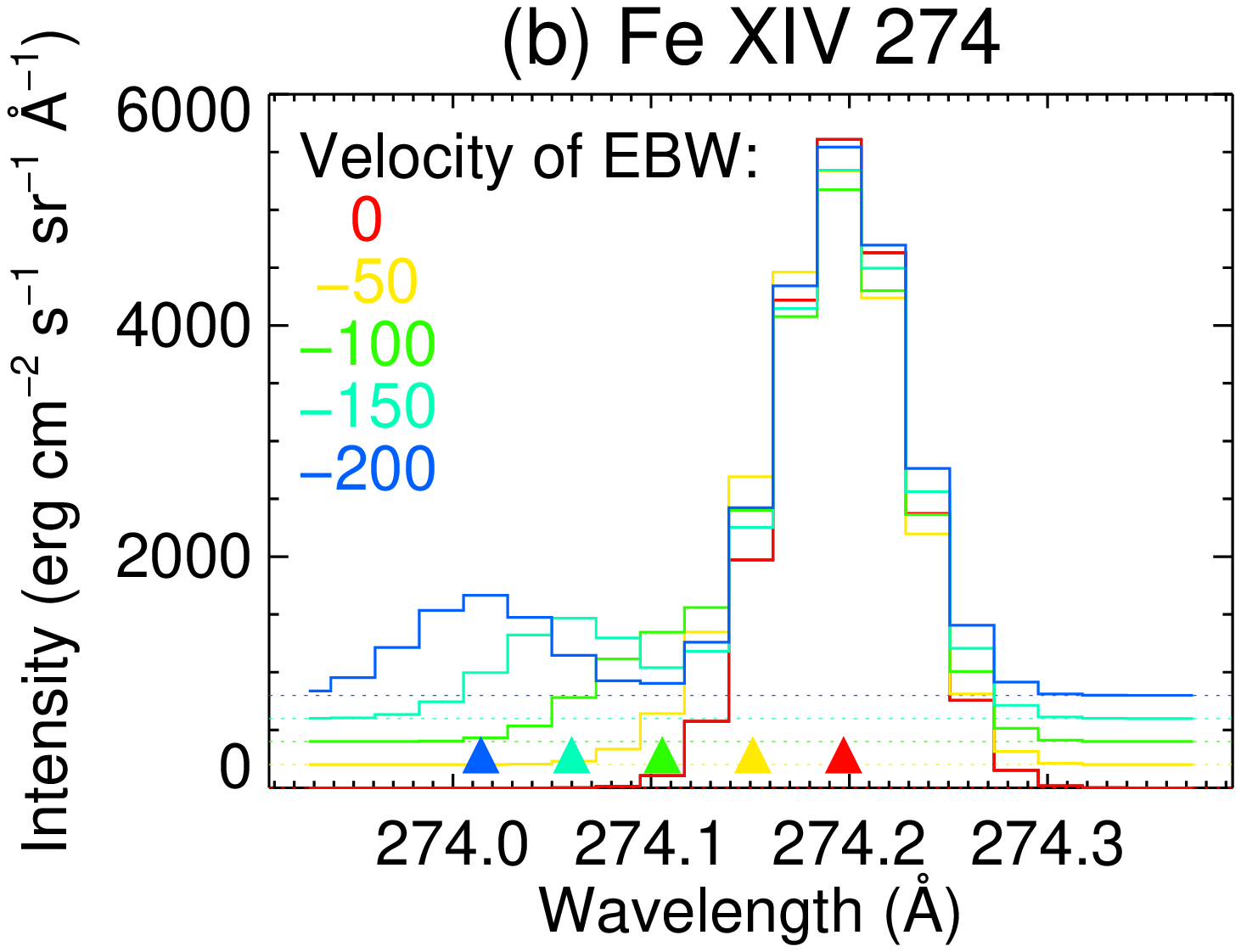}
    \end{minipage}
  \begin{minipage}[c]{10.6cm}
    \includegraphics[width=10.6cm,clip]{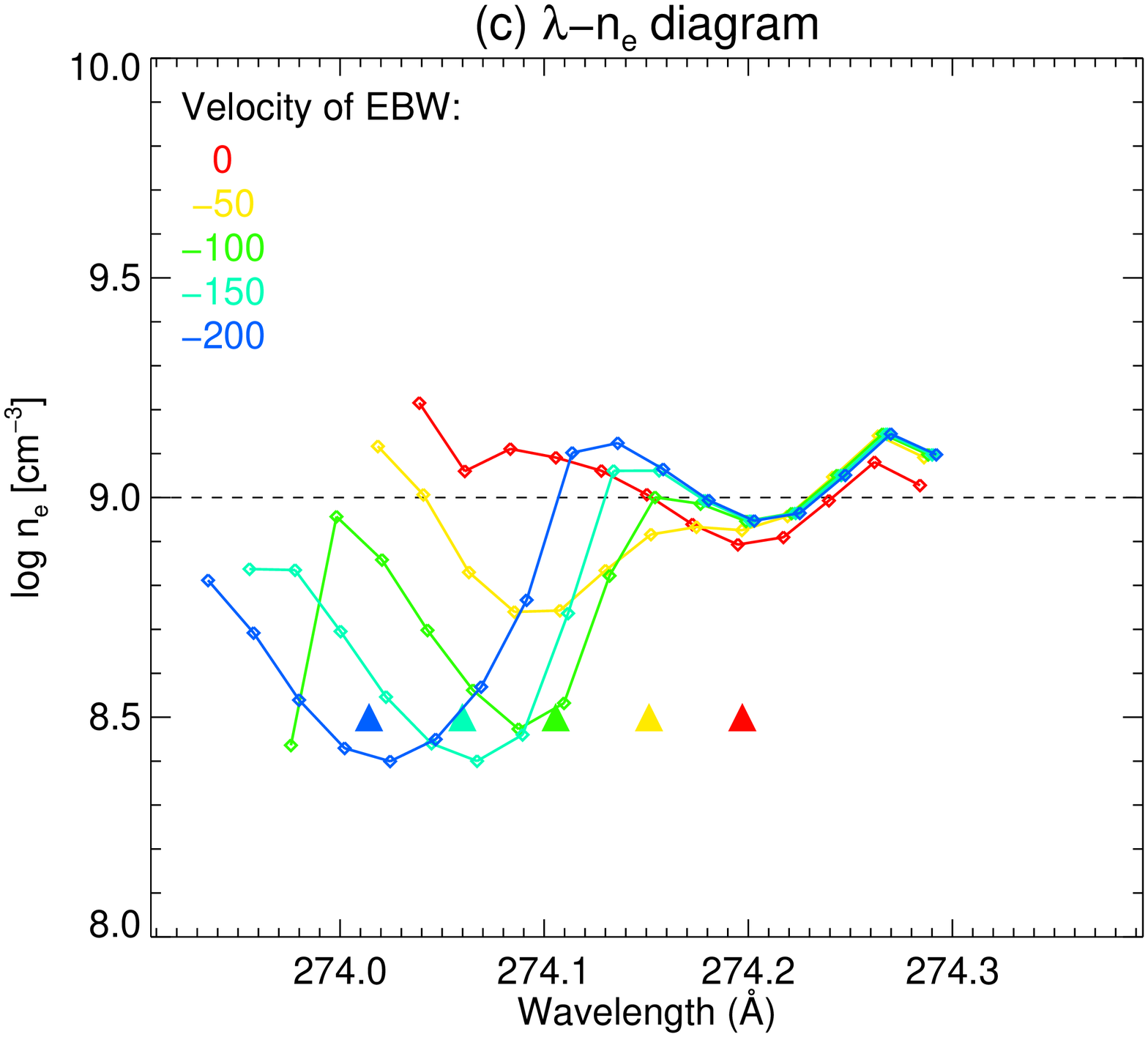}
  \end{minipage}
  \caption{
    (a) Synthetic line profiles of Fe \textsc{xiv} $264.78${\AA},
    (b) Synthetic line profiles of Fe \textsc{xiv} $274.20${\AA}, and 
    (c) $\lambda$-$n_\mathrm{e}$ diagrams.
    Each color indicates different velocity of minor blueshifted component (\textit{blue}: $0\,\mathrm{km}\,\mathrm{s}^{-1}$, \textit{turquoise}: $-50\,\mathrm{km}\,\mathrm{s}^{-1}$, \textit{green}: $-100\,\mathrm{km}\,\mathrm{s}^{-1}$, \textit{yellow}: $-150\,\mathrm{km}\,\mathrm{s}^{-1}$, and \textit{red}: $-200\,\mathrm{km}\,\mathrm{s}^{-1}$).  The major rest component was at rest ($0\,\mathrm{km}\,\mathrm{s}^{-1}$).
  }
  \label{fig:test_velocity}
\end{figure}

The dependence of $\lambda$-$n_{\mathrm{e}}$ diagram on the Doppler velocity of minor component is obvious.  The spectra of Fe \textsc{xiv} $264.78${\AA} and $274.20${\AA}, and $\lambda$-$n_{\mathrm{e}}$ diagrams are respectively shown in panel (a), (b), and (c) of Fig.~\ref{fig:test_velocity}.  Colors indicate the five cases for variable Doppler velocity calculated (\textit{blue}: $0\,\mathrm{km}\,\mathrm{s}^{-1}$, \textit{turquoise}: $-50\,\mathrm{km}\,\mathrm{s}^{-1}$, \textit{green}: $-100\,\mathrm{km}\,\mathrm{s}^{-1}$, \textit{yellow}: $-150\,\mathrm{km}\,\mathrm{s}^{-1}$, and \textit{red}: $-200\,\mathrm{km}\,\mathrm{s}^{-1}$).  Major rest component was at rest ($0\,\mathrm{km}\,\mathrm{s}^{-1}$) with the electron density of $\log n_{\mathrm{e}} \, [\mathrm{cm}^{-3}] = 9.0$.  The triangles in panel (c) indicate centroid and electron density of given minor component. The relative intensity of minor component is $15 \, \mathrm{\%}$ of that of major component and the electron density of minor component was set to $\log n_{\mathrm{e}} \, [\mathrm{cm}^{-3}] = 8.5$ in all five cases here.  The location of dips in $\lambda$-$n_{\mathrm{e}}$ diagram well represent the centroid position of the input minor component when two components are separated so that the spectrum is dominated by themselves near their centroids.  This is not the case for $v=-50 \, \mathrm{km}\,\mathrm{s}^{-1}$ (\textit{i.e.}, \textit{yellow}), where those two components are not separated so well.  In this case, $\lambda$-$n_{\mathrm{e}}$ diagram gradually decreases from longer to shorter wavelength.  One advantage of the method described here is that we are able to know the tendency of electron density of upflow/downflow without any fitting to the spectrum which might produce spurious result occasionally.

% --- End of Tex ---

%% file: tex/ndv_test_width.tex
% ========================
%   Lambda-N method
%   Dependence on width
% ========================

\begin{figure}
  \centering
  \begin{minipage}[c]{6.0cm}
    \includegraphics[width=6.0cm,clip]{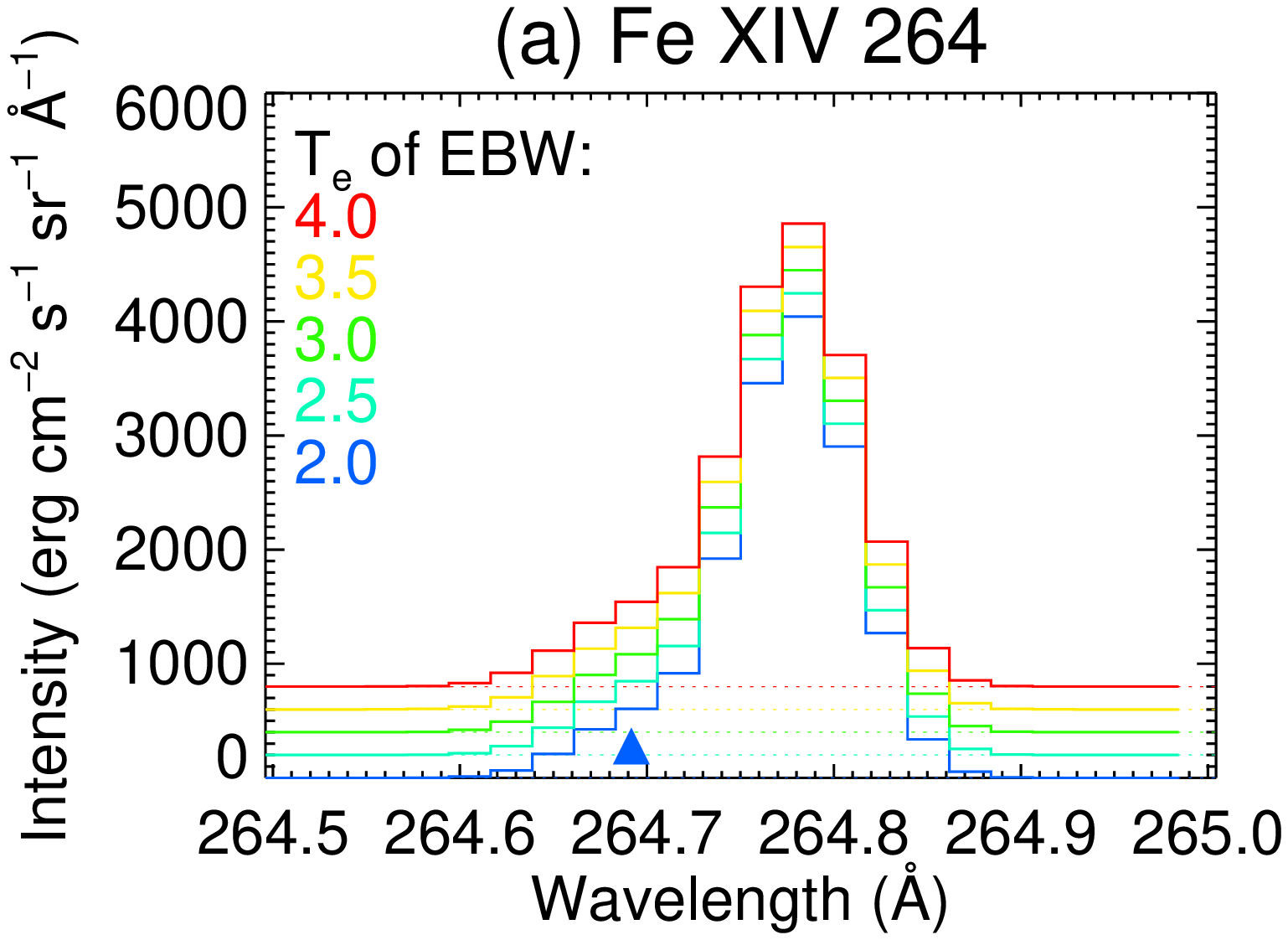}
    \includegraphics[width=6.0cm,clip]{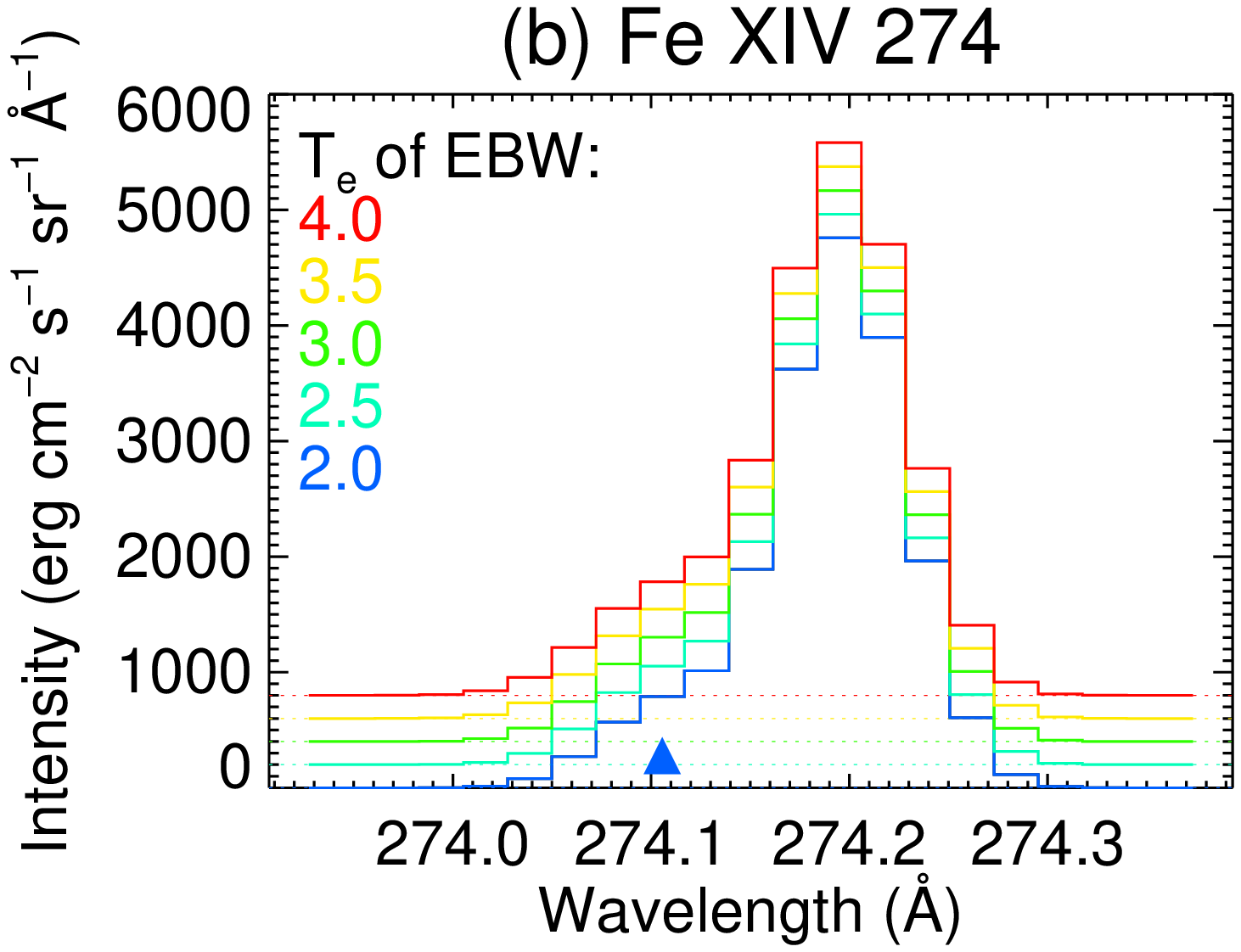}
  \end{minipage}
\begin{minipage}[c]{10.6cm}
    \includegraphics[width=10.6cm,clip]{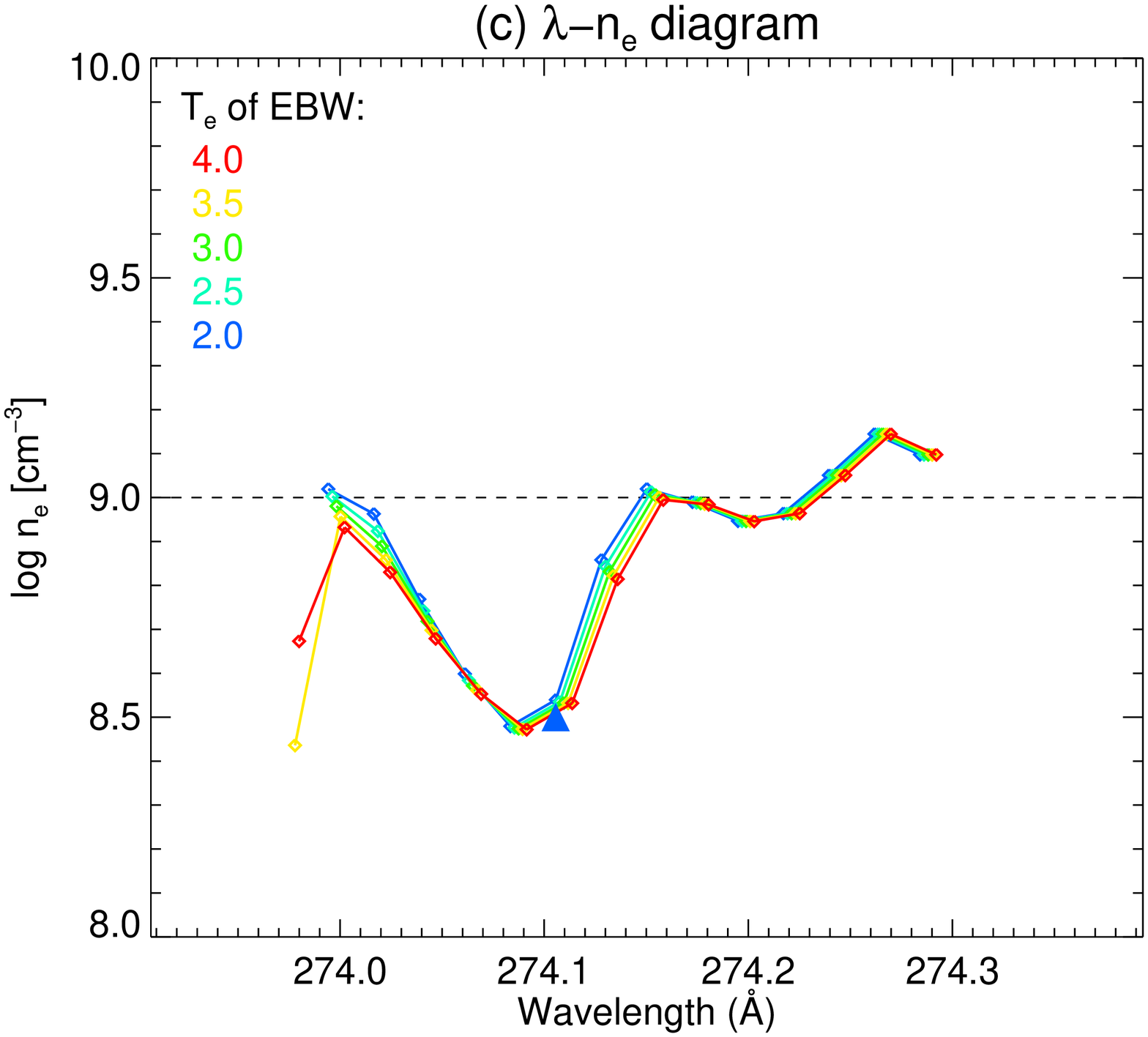}
  \end{minipage}
  \caption{
    (a) Synthetic line profiles of Fe \textsc{xiv} $264.78${\AA}, (b) Synthetic line profiles of Fe \textsc{xiv} $274.20${\AA}, and (c) $\lambda$-$N_{\mathrm{e}}$ diagrams.  Each color indicates different thermal width of minor blueshifted component (\textit{blue}: $2.0\,\mathrm{MK}$, \textit{turquoise}: $2.5\,\mathrm{MK}$, \textit{green}: $3.0\,\mathrm{MK}$, \textit{yellow}: $3.5\,\mathrm{MK}$, and \textit{red}: $4.0\,\mathrm{MK}$). Thermal width of major component was fixed to $2.0\,\mathrm{MK}$.}
  \label{fig:test_width}
\end{figure}

The thermal width of minor component can be considered to have less influence on $\lambda$-$n_{\mathrm{e}}$ diagram since the instrumental width of EIS is around $30 \, \kmpers$ ($0.027${\AA}), which corresponds to the thermal width of plasma with temperature of $5 \, \mathrm{MK}$. Taking into account the fact that the Differential emission measure analysis revealed that the temperature of the upflow is within $\logt = 6.1$--$6.5$ \citep{brooks2011}, we test the case for the temperature of minor component from $2$--$4 \, \mathrm{MK}$. The spectra of Fe \textsc{xiv} $264.78${\AA} and $274.20${\AA}, and $\lambda$-$n_{\mathrm{e}}$ diagrams are respectively shown in panel (a), (b), and (c) of Fig.~\ref{fig:test_width}.  Colors indicate the five cases for variable thermal width calculated (\textit{blue}: $2.0\,\mathrm{MK}$, \textit{turquoise}: $2.5\,\mathrm{MK}$, \textit{green}: $3.0\,\mathrm{MK}$, \textit{yellow}: $3.5\,\mathrm{MK}$, and \textit{red}: $4.0\,\mathrm{MK}$). The blue triangle in panel (c) indicates centroid and electron density of given minor component. As we expected, $\lambda$-$n_{\mathrm{e}}$ diagrams show much less change with increasing thermal width compared to the previous three sections.

The tests above for the four variables (\textit{i.e.}, density, intensity, velocity, and thermal width) indicate that the method proposed here ($\lambda$-$n_{\mathrm{e}}$ diagram) is a powerful diagnostic tool for coronal plasma which may constitute from several component along the line of sight and form non-single-Gaussian line profile. In the next section, we exploit this $\lambda$-$n_{\mathrm{e}}$ diagram so that the result obtained by double-Gaussian fitting would (\textit{i.e.}, upflows are more tenuous than the rest component) be confirmed. 

% --- End of Tex ---

%% file: tex/ndv_test_binning.tex
% =========================
%   Project:
%     Density of upflows
% =========================
\begin{figure}
  \centering
  \includegraphics[width=8.4cm,clip]{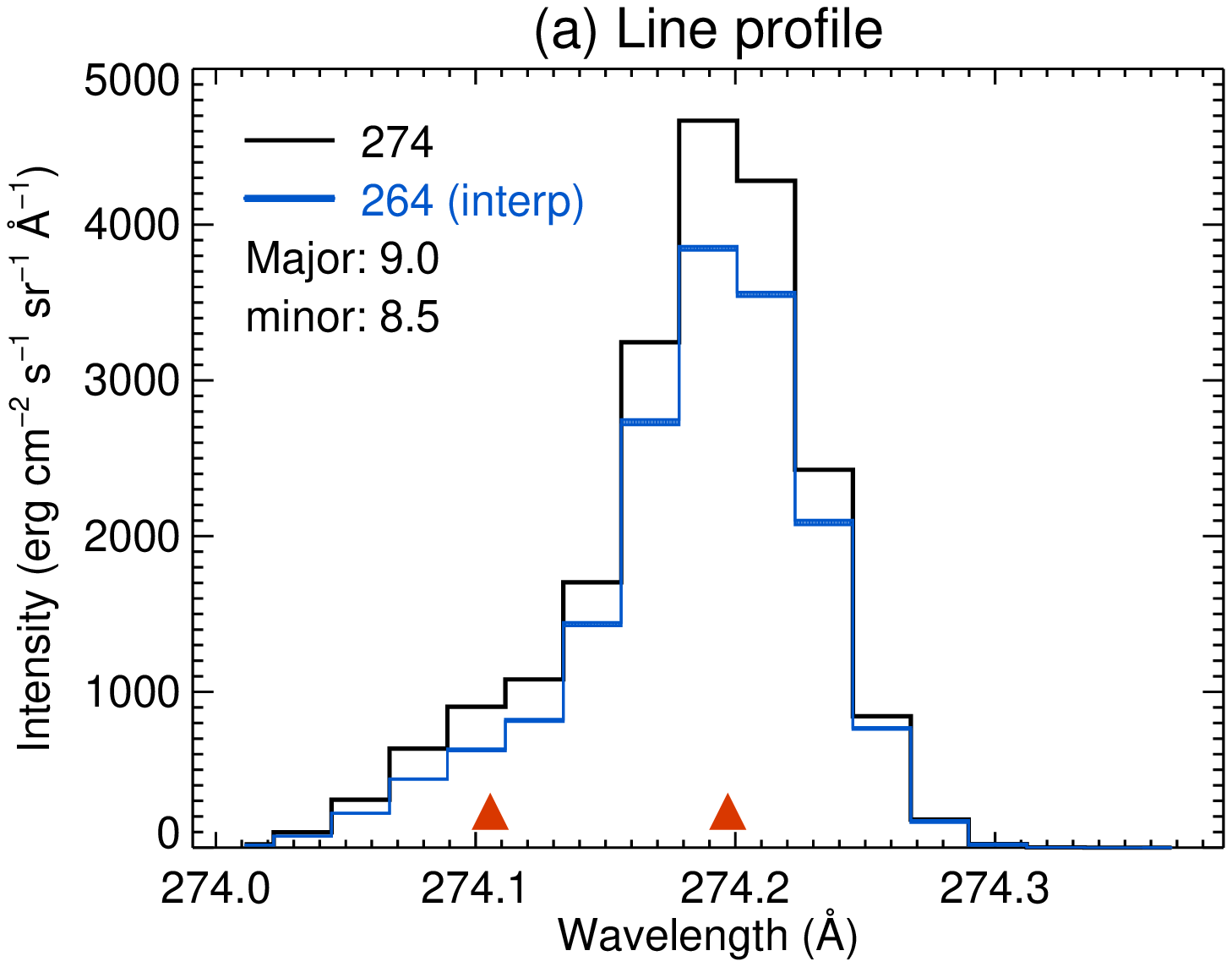}
  \includegraphics[width=8.4cm,clip]{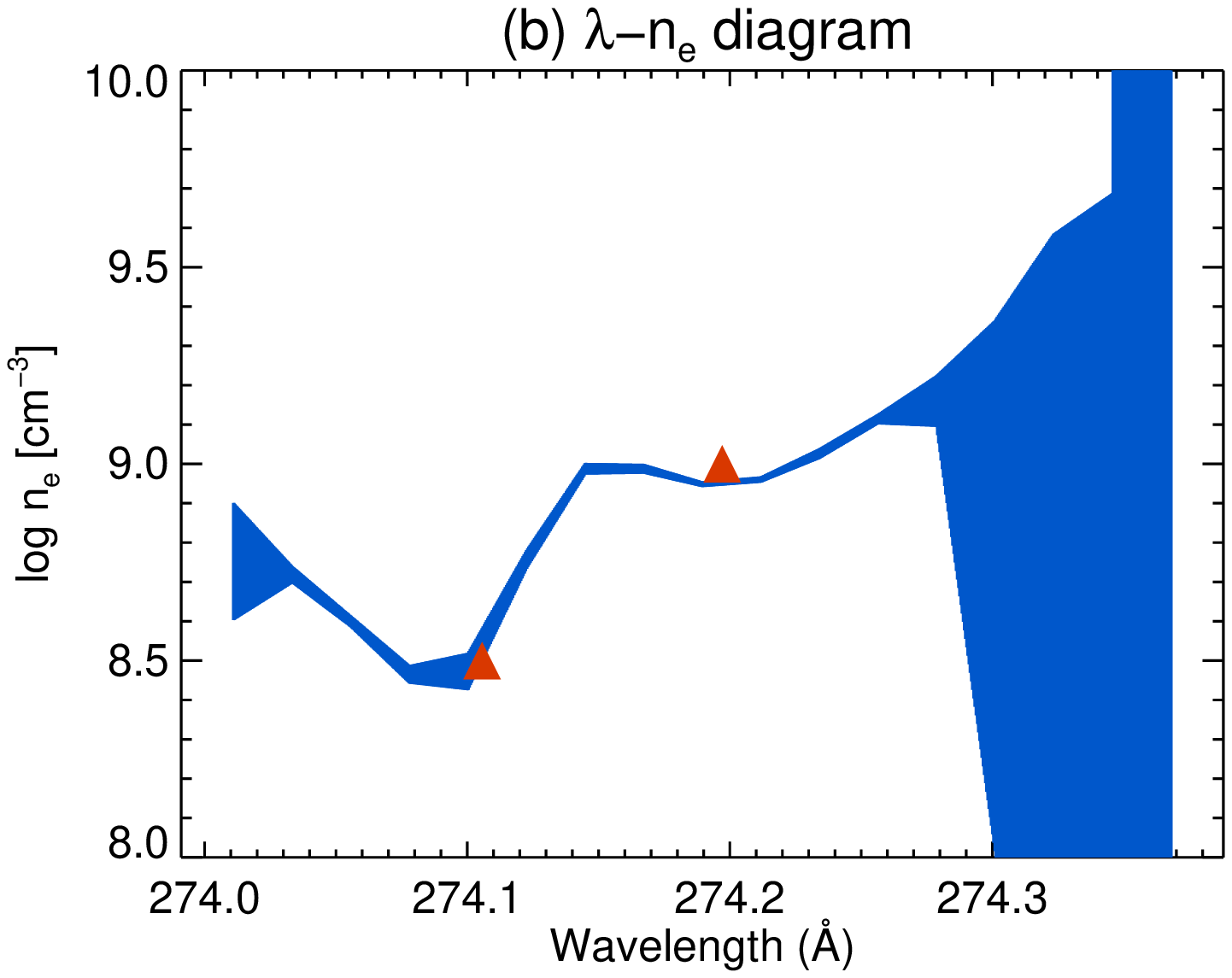}
  \caption{(a) Synthetic line profiles of $274.20${\AA} (\textit{black}) and $264.78${\AA} (\textit{blue}) with adjusted wavelength.  \textit{Red triangles} indicate the centroid of the two Gaussians given.  The electron density of major and minor component is respectively $10^{8.5}$ and $10^9 \, \mathrm{cm}^{-3}$.  The \textit{Blue} line profile has a finite width in the vertical axis (though it is narrow), which shows the range when changing the binning position at $264.78${\AA} spectral window.  (b) $\lambda$-$n_{\mathrm{e}}$ diagram with varying binning position at $264.78${\AA} spectral window. The vertical width of the diagram shows the range.  \textit{Red triangles} show the same meaning as in panel (a). }
  \label{fig:test_binning}
\end{figure}

As an instrumental effect, the binning of the wavelength direction may contribute to the uncertainty in $\lambda$-$n_{\mathrm{e}}$ diagram.  In order to test this possibility, we moved the position of binning at Fe \textsc{xiv} $264.78${\AA} spectral window consecutively by $0.001${\AA} up to $0.05${\AA} (\textit{cf.} spectral pixel of EIS $\simeq 0.0223${\AA}).  The results are shown in Fig. \ref{fig:test_binning}. Panel (a) shows synthetic line profiles of Fe \textsc{xiv} $274.20${\AA} (\textit{black} profile) and tested $264.78${\AA} (\textit{blue} profile).  \textit{Red triangles} indicate the centroid of the two Gaussians given.  The electron density of major and minor component is respectively $10^{8.5}$ and $10^9 \, \mathrm{cm}^{-3}$.  The \textit{Blue} line profile has a finite width in the vertical axis (though it is narrow and hard to see), which shows the range when changing the binning position at $264.78${\AA} spectral window.  Panel (b) shows $\lambda$-$n_{\mathrm{e}}$ diagram with varying binning position at $264.78${\AA} spectral window.  The vertical width of the diagram shows the range of values which derived $\lambda$-$n_{\mathrm{e}}$ diagrams took.  \textit{Red triangles} show the same meaning as in panel (a).  From panel (b), it can be seen that the uncertainty caused by the position of binning does not exceed $0.1 \, \log \, \mathrm{cm}^{-3}$ at the location where we are interested in, but becomes quite larger away from the emission line center because of the low intensity. 

% --- End of Tex ---

%% file: tex/ndv_sv.tex
% =======================
%   Lambda-N diagram
%   X-variation of NdV
% =======================

\begin{figure}
  \centering
  \includegraphics[width=12cm,clip]{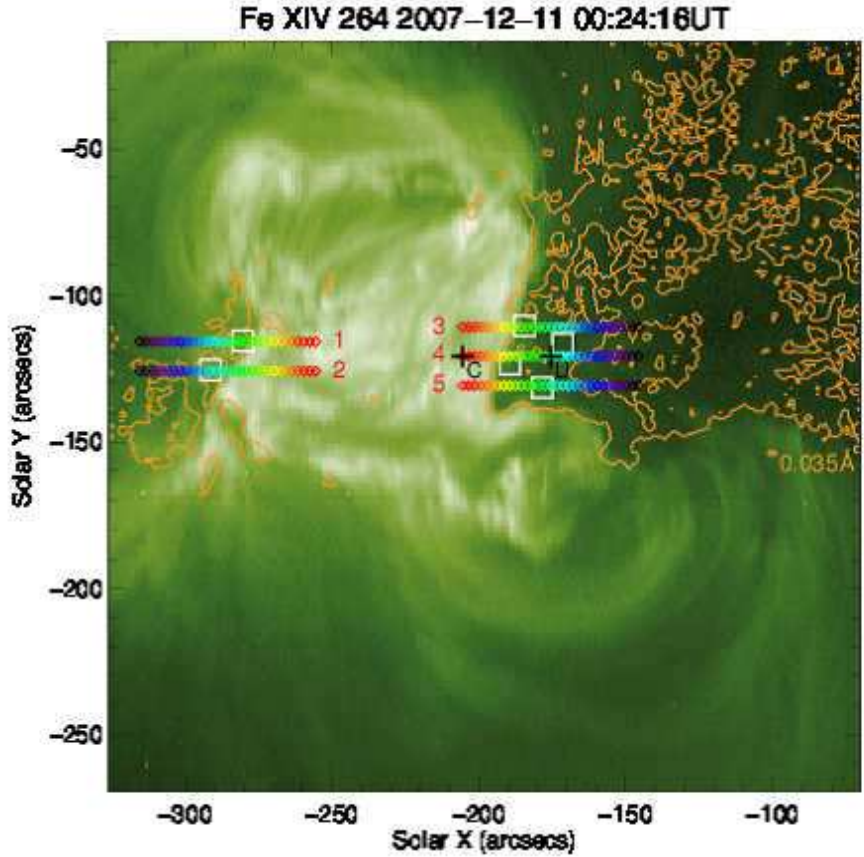}
  \caption{Intensity map of Fe \textsc{xiv} $264.78${\AA} obtained with EIS.  
%{\color{red}
Five arrays of colored diamonds (\textit{red}--\textit{violet}) are the locations where $\lambda$-$n_\mathrm{e}$ diagrams were made.
%}
The locations cut across the active region core and the outflow region.  \textit{Orange} contour indicates the line width of $0.035${\AA}.}
  \label{fig:eis_map_lp_sv}
\end{figure}

The electron density of the outflow region in AR 10978 is investigated through $\lambda$-$n_{\mathrm{e}}$ diagram here.  Fig.~\ref{fig:eis_map_lp_sv} shows intensity map of Fe \textsc{xiv} $264.78${\AA} obtained with EIS.  \textit{Orange} contours indicate the line width of $0.035${\AA}, which becomes an indication of the outflows.  Five horizontal arrays of colored diamonds (\textit{red}--\textit{violet}) which cut across the active region core and the outflow region are the locations where we made $\lambda$-$n_{\mathrm{e}}$ diagrams.  First, we look at the location indicated by \textit{black plus} signs named \texttt{C} (core) and \texttt{U} (outflow).  

\begin{figure}
  \centering
  \includegraphics[width=8.4cm,clip]{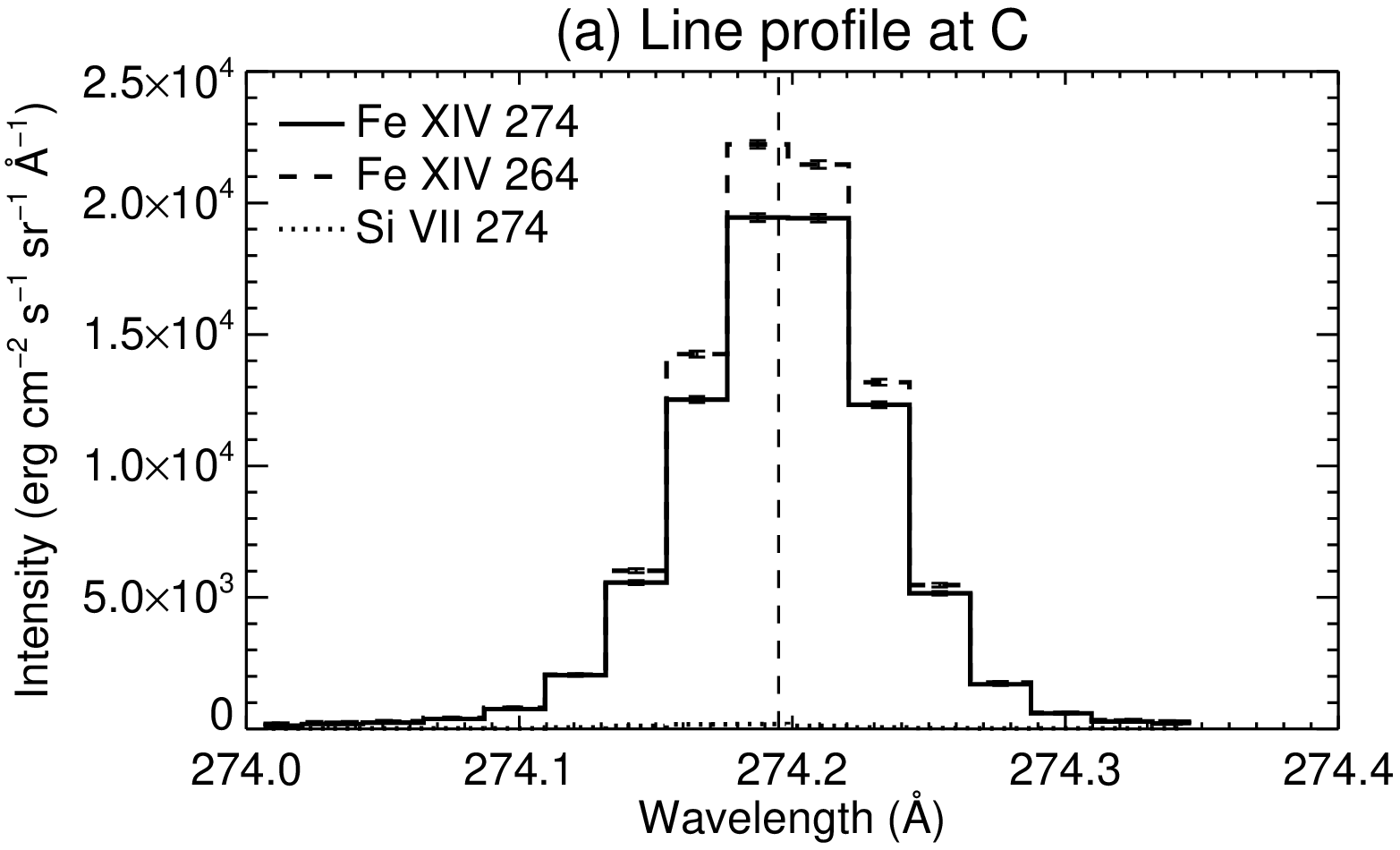}
  \includegraphics[width=8.4cm,clip]{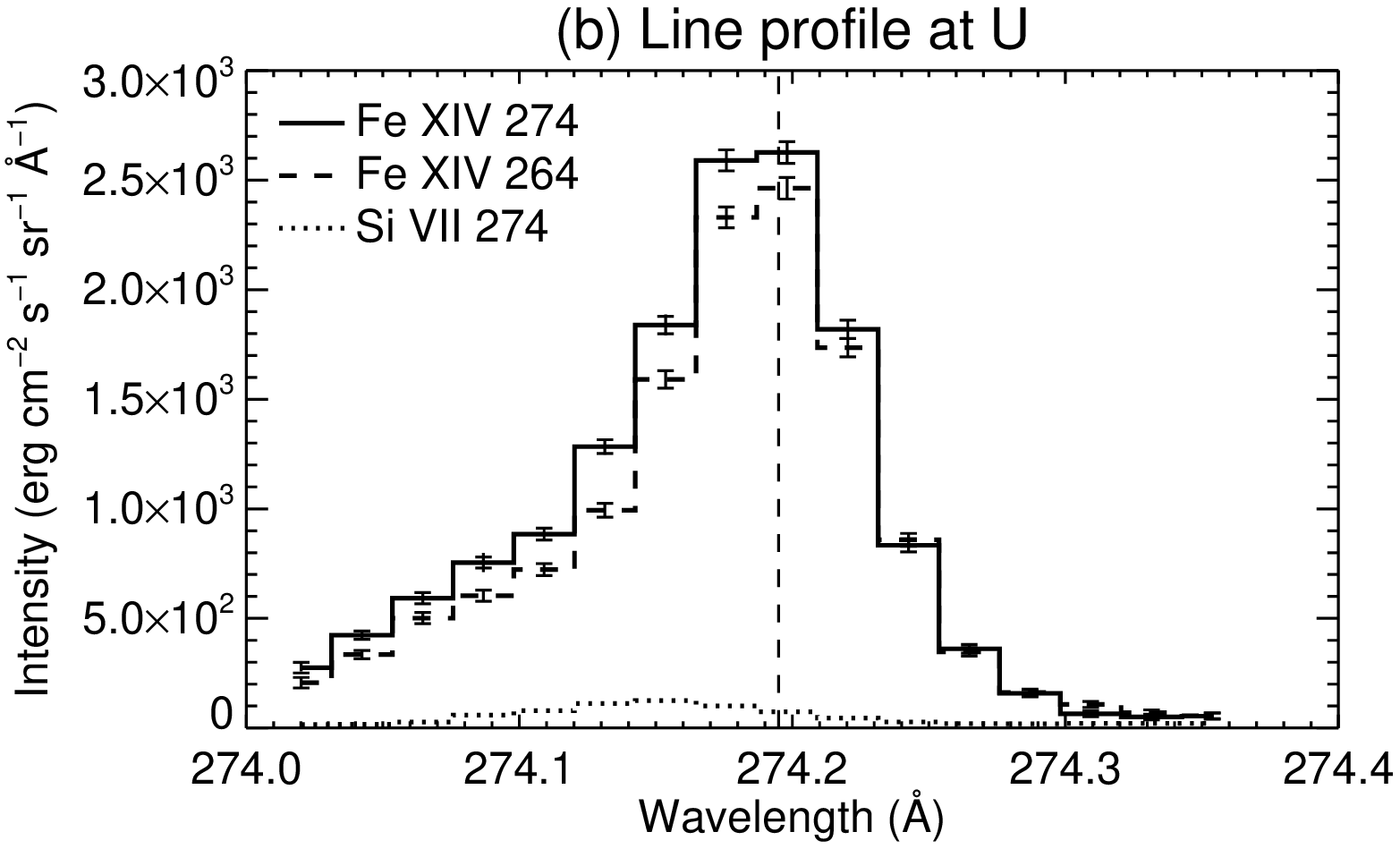}
  \includegraphics[width=8.4cm,clip]{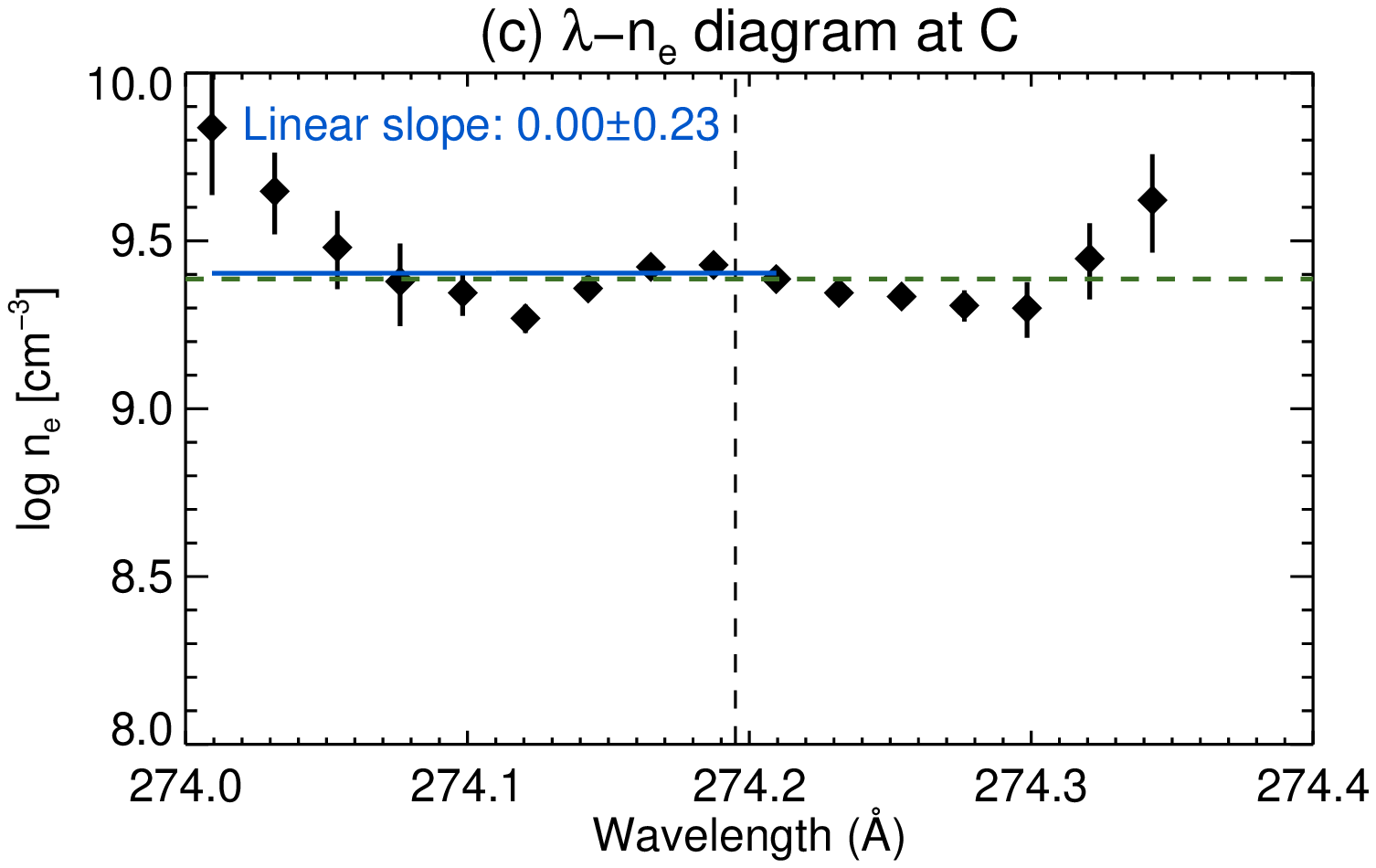}
  \includegraphics[width=8.4cm,clip]{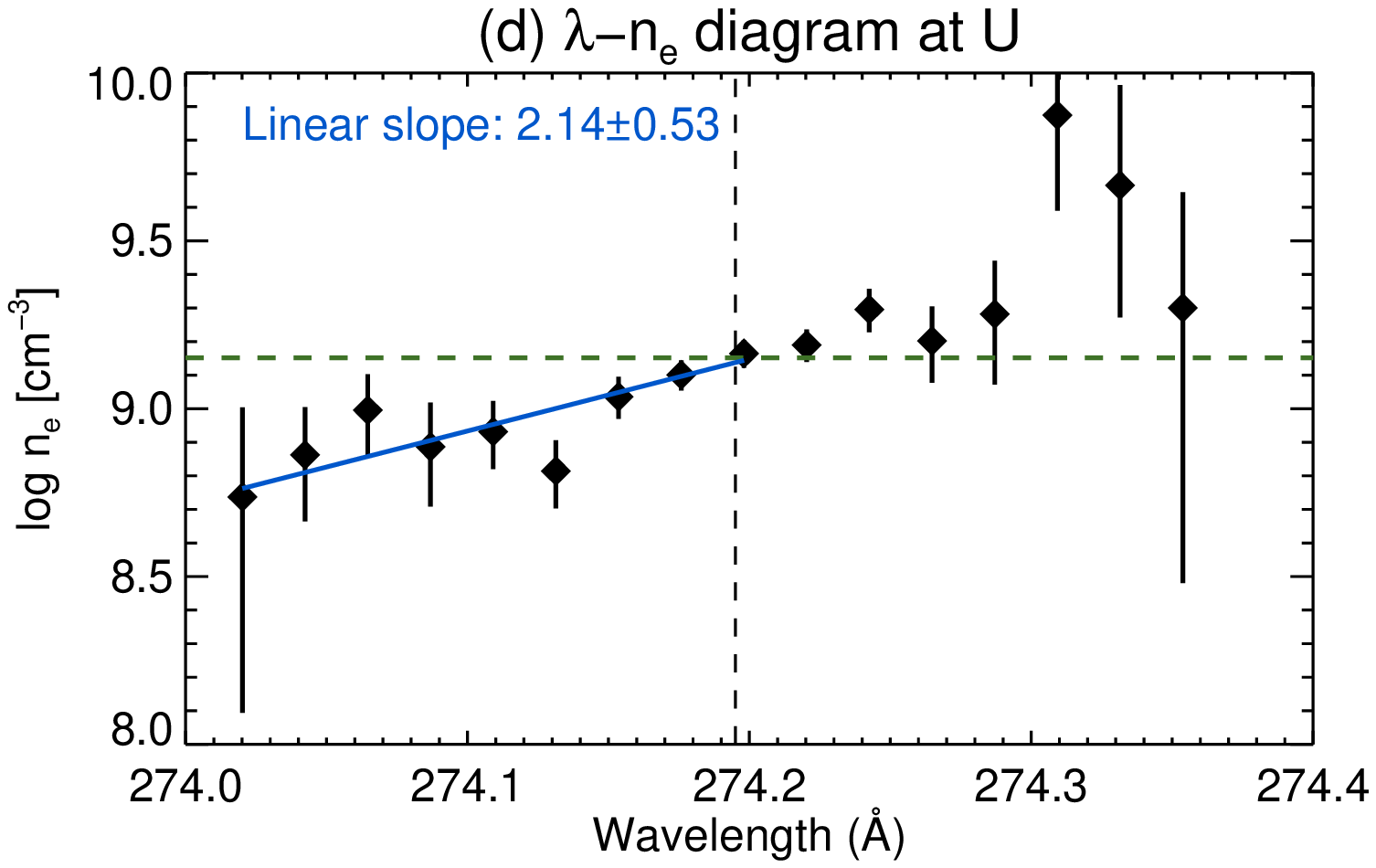}
  \caption{(a) Line profiles of Fe \textsc{xiv} $274.20${\AA} (\textit{solid})/$264.78${\AA} (\textit{dashed}; adjusted to the wavelength scale of $274.20${\AA}) and Si \textsc{vii} $274.18${\AA} (\textit{dotted}) estimated from $275.35${\AA} at the active region core.  (b) Line profiles at the outflow region.  (c) $\lambda$-$n_{\mathrm{e}}$ diagram at the active region core.  (d) $\lambda$-$n_{\mathrm{e}}$ diagram at the outflow region.}
  \label{fig:lp_and_ndv_274scale_example}
\end{figure}

In Fig.~\ref{fig:lp_and_ndv_274scale_example}, the line profiles of Fe \textsc{xiv} $274.20${\AA}, interpolated $264.78${\AA} and estimated Si \textsc{vii} $274.18${\AA} (see Section \ref{sec:de-blend}) are respectively shown by \textit{solid}, \textit{dashed}, and \textit{dotted} spectrum in panel (a) for the active region core and (b) for the outflow region.  We can see an enhanced blue wing in line profiles of Fe \textsc{xiv} in the outflow region.  The \textit{vertical dashed} lines indicate rough reference of the rest wavelength position $\lambda=274.195${\AA} which was the average line centroid above the limb in the 2007 December 18 data (possible error up to $0.01${\AA}). 

Panels (c) and (d) in Fig.~\ref{fig:lp_and_ndv_274scale_example} respectively show the $\lambda$-$n_{\mathrm{e}}$ diagram for the active region core and the western outflow region.  The \textit{horizontal green dotted} line in each plot indicate the electron density averaged in the neighbor three spectral bins which are nearest to $\lambda=274.20${\AA} (\textit{i.e.}, rest wavelength).  Those $\lambda$-$n_{\mathrm{e}}$ diagrams in the two locations exhibit a different behavior at shorter wavelength side around $\lambda=274.00\text{--}274.20${\AA}: the diagram in the active region core is roughly constant while that in the western outflow region slightly decreases at the shorter wavelength.  The number written in the upper left corner of each plot indicates the linear slope fitted within the wavelength range $\lambda \leq 274.20${\AA}.  This implies that the electron density of the outflows (\textit{i.e.}, shorter wavelength side) is smaller than that of the major rest component. 

In order to confirm the above implication more robust, we see the variation of $\lambda$-$n_{\mathrm{e}}$ diagram along $x$ direction from the active region core to the outflow regions.  The selected region spans from the active region core (\textit{red} diamond) to the outflow region (\textit{violet} diamond) as seen in Fig.~\ref{fig:eis_map_lp_sv}.  The boundary of the active region core corresponds to the color between \textit{yellow} and \textit{light green}.  
%{\color{red} 
The $\lambda$-$n_{\mathrm{e}}$ diagrams at each cut (1--5) are plotted in Fig.~\ref{fig:eis_ndv_sv}.  We can see clear change of the $\lambda$-$n_{\mathrm{e}}$ diagrams with colors.  The $\lambda$-$n_{\mathrm{e}}$ diagrams for cut 1 show a small hump around $274.00$--$274.10${\AA} representing that EBW component has larger electron density than the major component, though the hump at almost all locations (\textit{red}--\textit{black}) might mean that it was caused by an anomalous pixel (\textit{e.g.}, warm pixel).  Both for cut 1 and 2, the diagrams show flat or slightly decreasing behavior as a function of wavelength at all locations.  These behaviors are consistent with the result obtained in Chapter \ref{chap:dns} (region U1 and U2) which indicated that the electron density of the outflows in the eastern edge is almost the same or slightly larger.  On the other hand, in the western outflow region (cut 3--5), those for the outflow region show a dip around $274.10${\AA}.  This wavelength corresponds to $v = - 110 \, \kmpers$ for the emission line Fe \textsc{xiv} $274.20${\AA}, from which it is implied that the outflows in the western edge are composed of less dense plasma compared to the rest plasma (\textit{i.e.}, around $274.20${\AA}) existing along the line of sight.  Note that this velocity does not mean that of the upflows because no fitting was applied in $\lambda$-$n_{\mathrm{e}}$ diagram.
%}

The electron density of EBW component evaluated from $\lambda$-$n_{\mathrm{e}}$ diagrams around $\lambda=274.10${\AA} was $\log n_{\mathrm{e}} \, [\mathrm{cm}^{-3}]=9.0\text{--}9.2$ in the eastern outflow region, and $\log n_{\mathrm{e}} \, [\mathrm{cm}^{-3}] = 8.5\text{--}9.0$ in the western outflow region, which also coincides the result obtained in the previous chapter.  By exploiting $\lambda$-$n_{\mathrm{e}}$ diagram as a new diagnostic tool, we now support the result obtained by double-Gaussian fitting in Chapter \ref{chap:dns}.  

\begin{figure}
  \centering
  \includegraphics[width=10cm,clip]{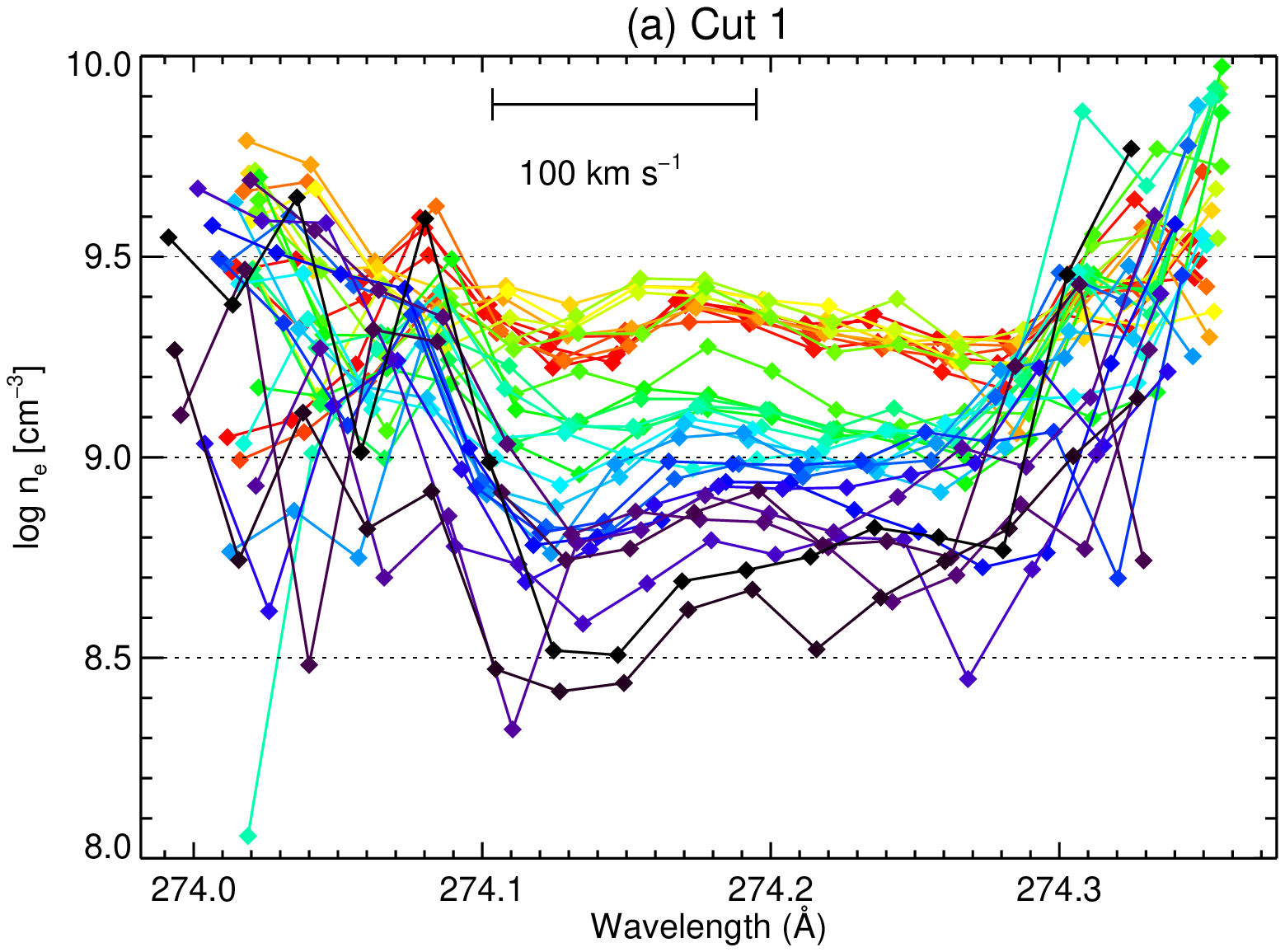} \\
  \includegraphics[width=10cm,clip]{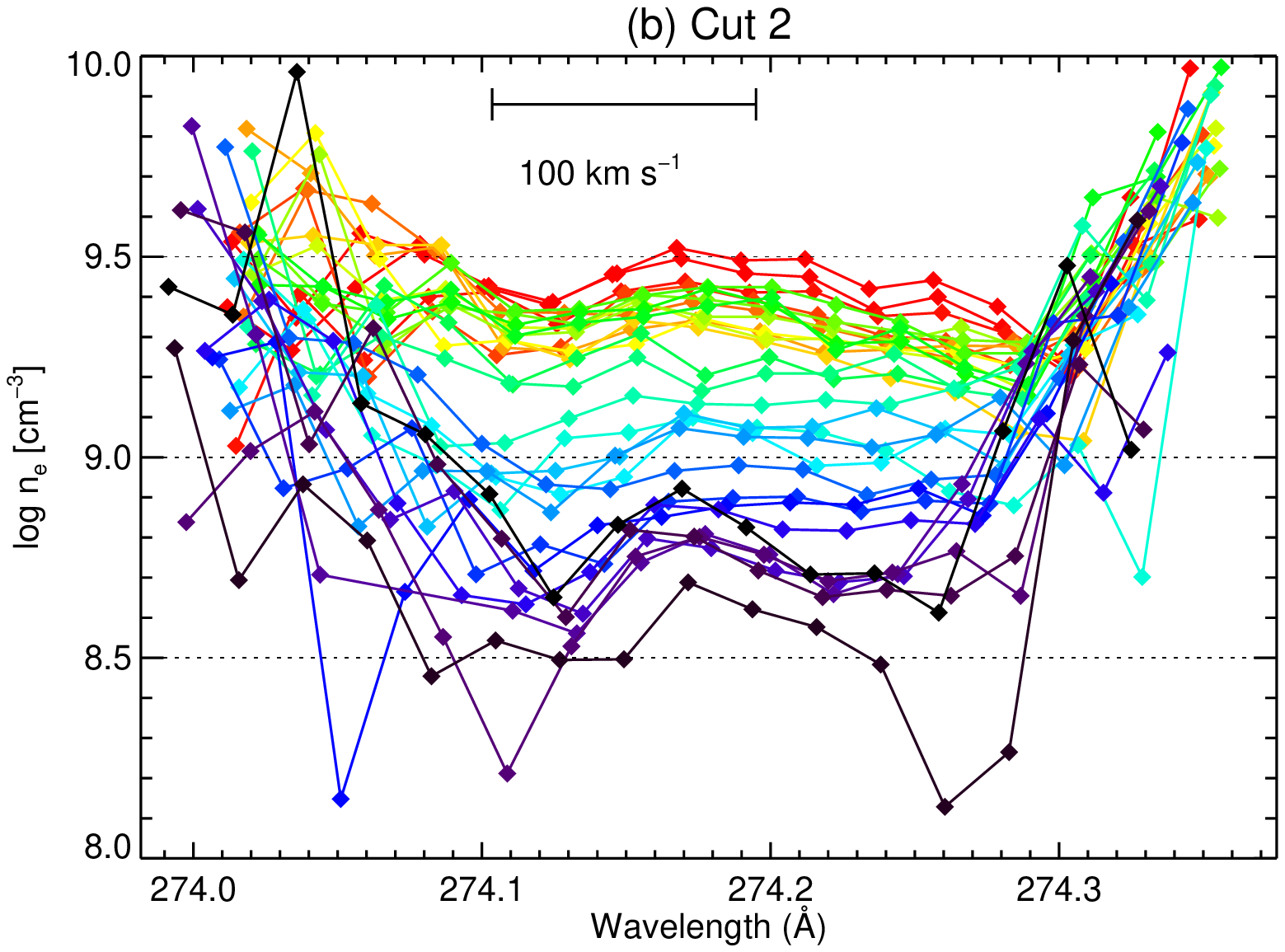}
  \caption{%\color{red} 
$\lambda$-$n_{\mathrm{e}}$ diagrams at the locations indicated by colored diamonds in Fig.~\ref{fig:eis_map_lp_sv} (Cut 1 and 2; including the eastern outflow region).}
  \label{fig:eis_ndv_sv}
\end{figure}

\addtocounter{figure}{-1}
\begin{figure}
  \centering
  \includegraphics[width=10cm,clip]{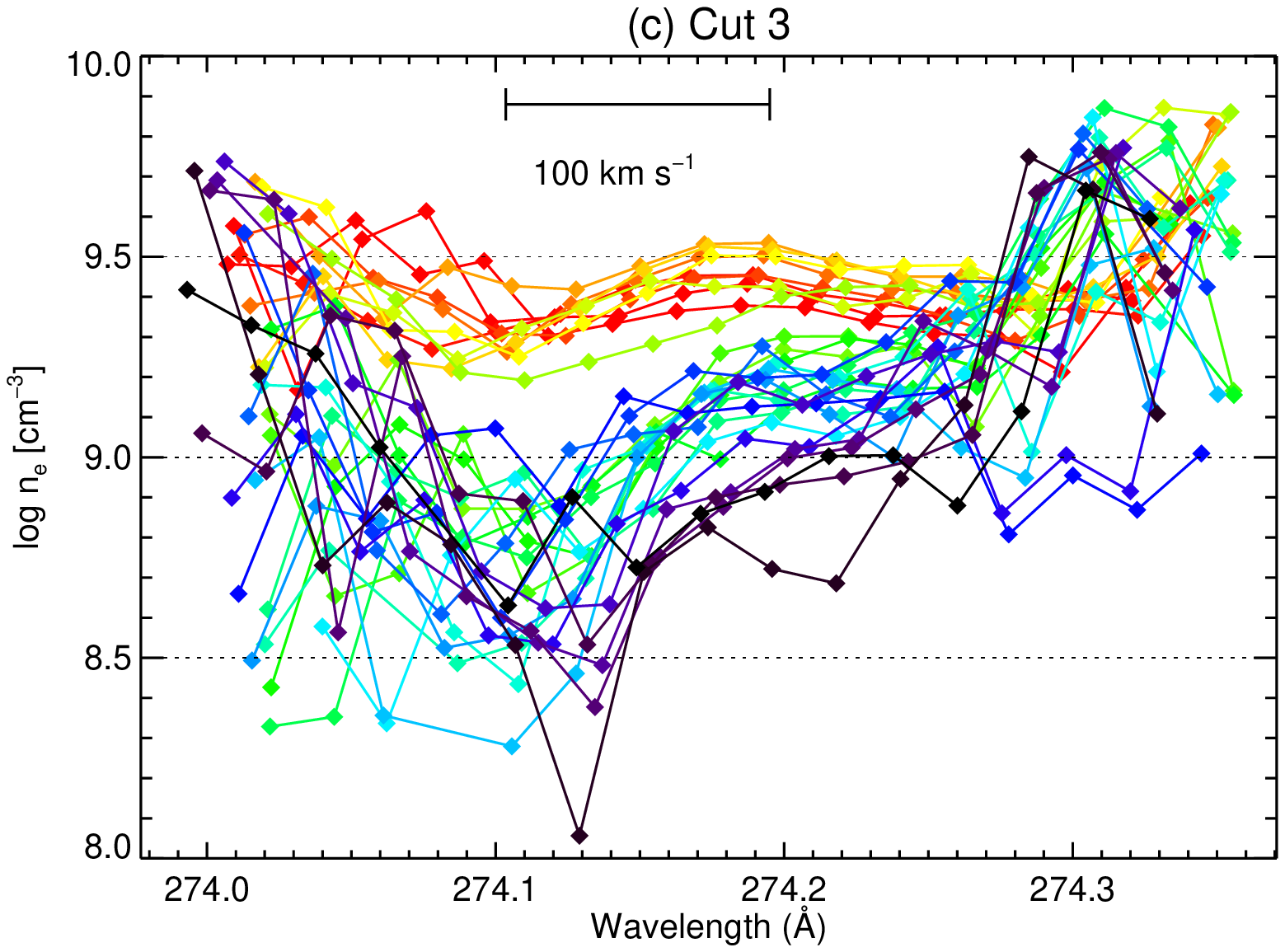} \\
  \includegraphics[width=10cm,clip]{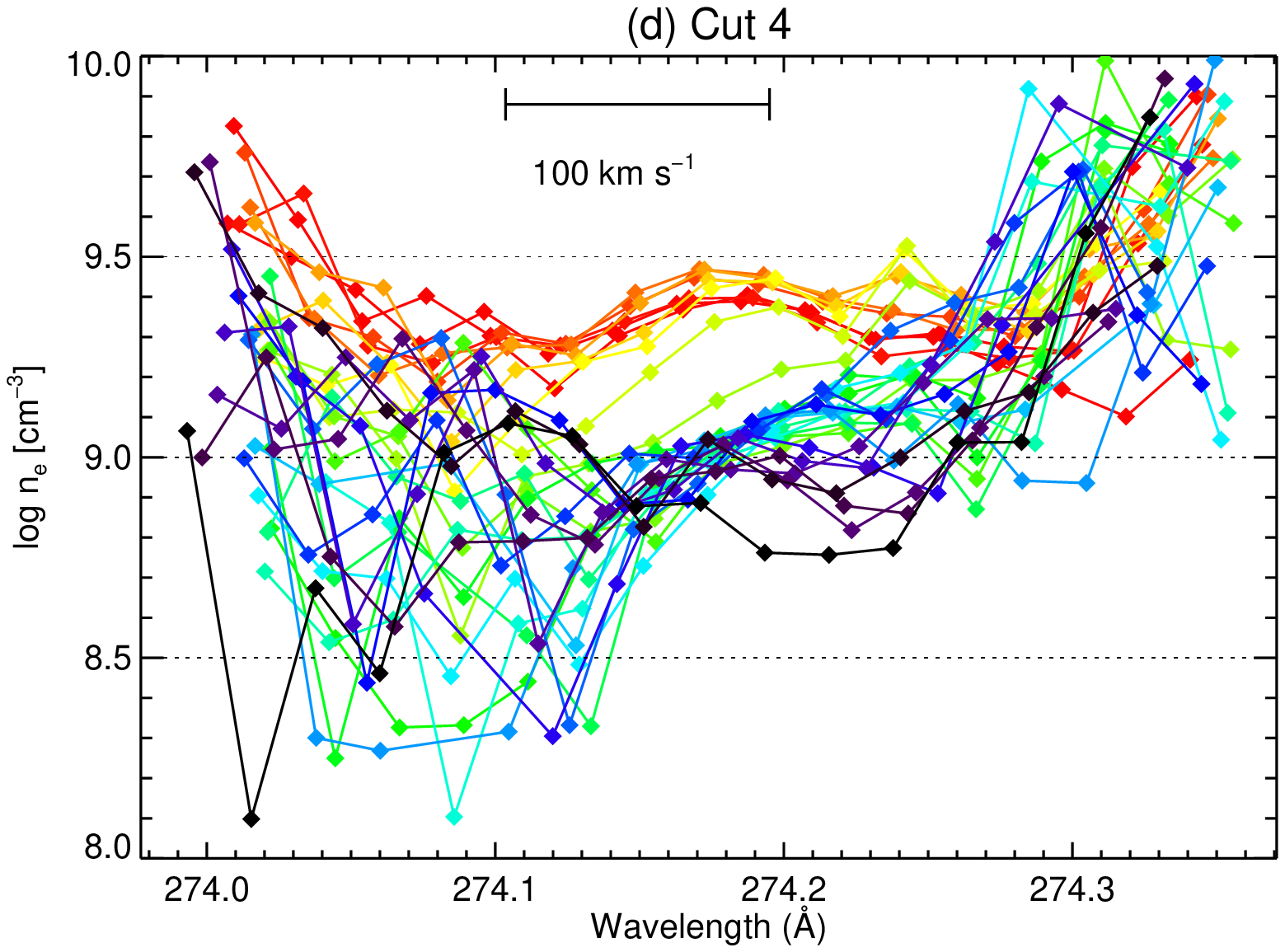} \\
  \includegraphics[width=10cm,clip]{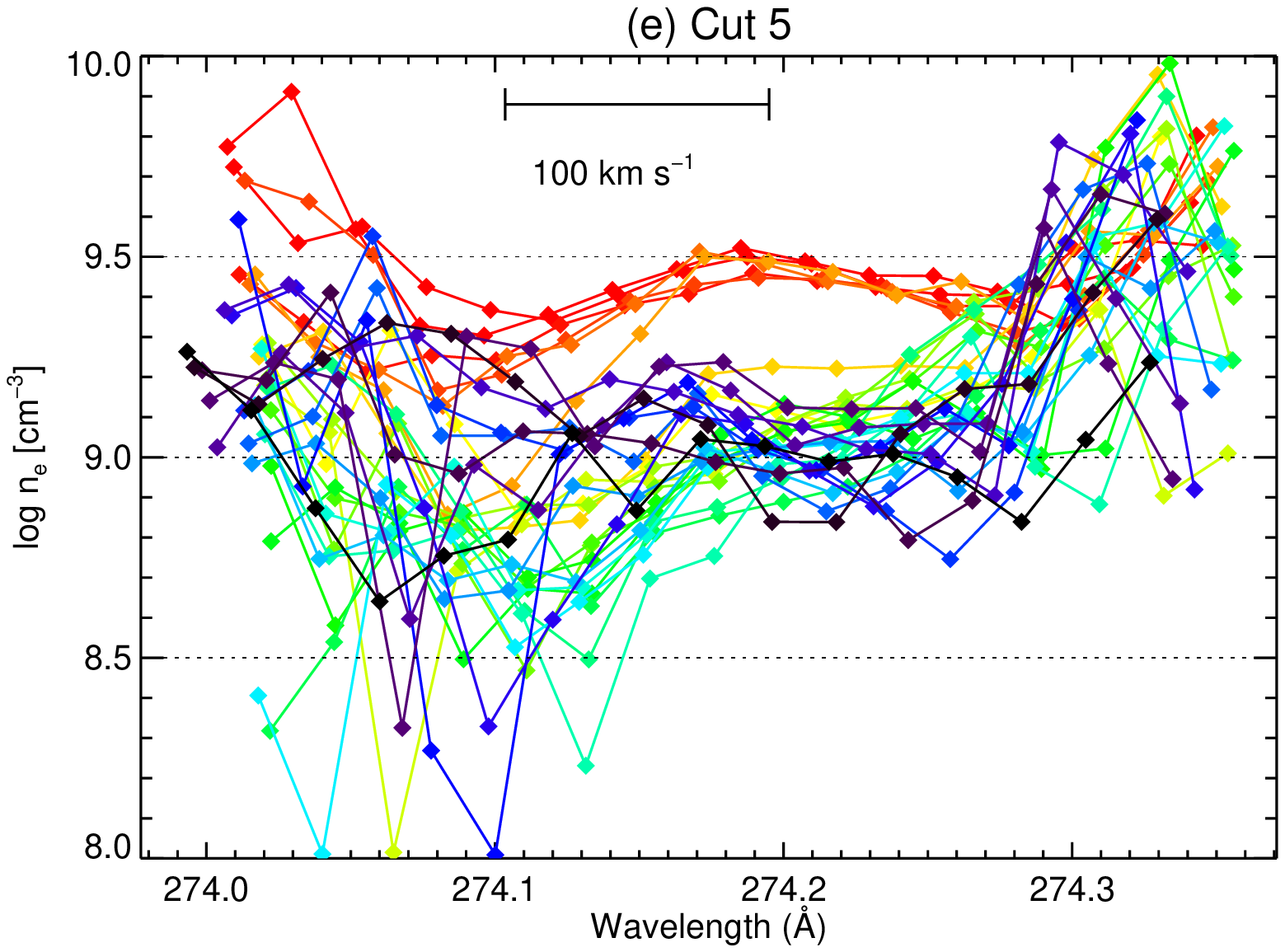}
  \caption{\textit{Continued.}  $\lambda$-$n_{\mathrm{e}}$ diagrams at the locations indicated by colored diamonds in Fig.~\ref{fig:eis_map_lp_sv} (Cut 3--5; including the western outflow region).}
\end{figure}

% --- End of Tex ---

%% file: tex/ndv_sum.tex
% =======================================
%   Chapter:
%     Lambda-N diagram.
%   Section:
%     Summary.
%   Contents:
%     - Purpose.
%     - Method and its advantage.
%     - Application to AR10978.
% =======================================

We introduced a density diagnostics from a new point of view in this chapter.  Electron density derived in our method is a function of Doppler velocity or wavelength (Eq.~\ref{eq:ndv_rprsnt}), referred to as $\lambda$-$n_{\mathrm{e}}$ diagram, which was found to be a good indication of the electron density of minor components in a line profile.  The method has an advantage that it does not depend any fitting model which might be ill-posed in some cases.  Our aim was to evaluate the electron density of the outflow seen at the edge of the active region, and reinforce the result obtained in Chapter \ref{chap:dns}.  

Using a density-sensitive emission line pair Fe \textsc{xiv} $264.78${\AA}/$274.20${\AA}, we studied $n_{\mathrm{e}} (\lambda)$ by making $\lambda$-$n_{\mathrm{e}}$ diagrams at the active region core and the outflow regions.  The increase in the diagrams was seen on the longer wavelength side for both structures, but we could not find whether that behavior actually implies the physical situation at present.  
%{\color{red}
The diagrams for the active region core were flat around $\log n_{\mathrm{e}} \, [\mathrm{cm}]^{-3} \simeq 9.5$, while those for the outflow regions exhibit some characteristic behaviors at shorter wavelength side.  They show a small hump around $v = -110 \, \kmpers$ in the eastern region (cut 1 and 2 in Fig.~\ref{fig:eis_map_lp_sv}), and a decrease trend from $\log n_{\mathrm{e}} \, [\mathrm{cm}^{-3}]=9.0$ to $\log n_{\mathrm{e}} \, [\mathrm{cm}^{-3}]=8.5$ in a velocity scale of $100 \, \kmpers$ in the western outflow region (cut 3--5 in Fig.~\ref{fig:eis_map_lp_sv}) as seen in Fig.~\ref{fig:eis_ndv_sv}.  Thus we confirmed the results obtained in Chapter \ref{chap:dns} through our new method independent of the double-Gaussian fitting. 
%}

% --- End of TeX ---

%% file: tex/contents_dis.tex
\chapter{Summary and discussion}
  \label{chap:dis}
\section{Summary of the results}
  \input{tex/dis_sum_result.tex}
\section{Temperature of the outflows}
  \label{sect:dis_te_outflow}
  \input{tex/dis_te_outflow.tex}
\section{Outflow region and fan loops}
  \label{sect:dis_outflow_fan}
  \input{tex/dis_outflow_fan.tex}
\section{Steady flow along coronal loops}
  \label{sect:dis_siphon}
  \input{tex/dis_siphon.tex}
\section{Interpretation in terms of impulsive heating}
  \label{sect:dis_impulsive}
  \subsection{Temperature dependence of the Doppler velocity}
    \label{sect:dis_intermittent}
    \input{tex/dis_intermittent.tex}
  \subsection{Analytical estimation of electron density}
    \input{tex/dns_discuss_n.tex}
    \label{sect:dis_n}
%\section{Electron density and column depth}
%  \label{sect:dis_depth}
%  \input{tex/dns_discuss_h.tex}
\section{Driving mechanisms of the outflow}
  \label{sect:dis_drive}
  \input{tex/dns_discuss_drive.tex}
\section{Mass transport by the outflow}
  \label{sect:dis_mass}
  \input{tex/dis_mass.tex}
\section{Eastern and western outflow region}
  \label{sect:dis_ew}
  \input{tex/dis_ew.tex}
\section{Future work}
  \label{sect:dis_future}
  \input{tex/dis_future.tex}

%% file: tex/dis_sum_result.tex
% ==============================
%   Chapter:
%     Discussion.
%   Section:
%     Future work.
% ==============================

Here we summarize the results obtained in Chapter \ref{chap:cal}--\ref{chap:ndv}. 

%{\color{blue}
%%%%%%%%%%%%%
\vspace{-5mm}
\subsection*{Average Doppler shifts of the quiet region (Chapter 3)}
\vspace{-2mm}

Two meridional scans were analyzed, from which we determined the Doppler velocity of the quiet region within $\logt = 5.7$--$6.3$ in the accuracy of $\simeq 3 \, \kmpers$ for the first time.  It was shown that emission lines below $\logt = 6.0$ have Doppler velocity of almost zero with an error of $1$--$3 \, \kmpers$, while those above that temperature are blueshifted with gradually increasing magnitude: $v=-6.3 \pm 2.1 \, \kmpers$ at $\logt = 6.25$ (Fe \textsc{xiii}).  These Doppler velocities were used as a reference in Chapter \ref{chap:vel}.  

%%%%%%%%%%%%%
\vspace{-5mm}
\subsection*{Doppler velocity measurement for AR outflows (Chapter 4)}
\vspace{-2mm}

Doppler velocities for emission lines with the formation temperature of $\logt = 5.5$--$6.5$ were measured by the single-Gaussian fitting in the active region NOAA AR10978.  The outflow regions and fan loops were distinguished in terms of the Doppler velocity in the transiton region temperature.  

\vspace{-3mm}
\begin{itemize}
  \item The line profiles with the formation temperature of $\logt = 6.1$--$6.3$ have significant EBW corresponding to $v \sim -100\,\kmpers$ in the outflow regions, while those below $\logt = 6.0$ and above $\logt = 6.4$ did not.  
  \vspace{-3mm}
  \item The Doppler velocities within $\logt=5.7$--$6.3$ were all blueshifted by $v=-20\,\kmpers$ (\textit{i.e.}, upflow) in the outflow regions.  The clear blueshift of the transition region lines 
were confirmed for the first time 
%{\color{dartmouthgreen} 
%\textbf{
precisely in the outflow regions.
%}} 
  \vspace{-3mm}
  \item Fan loops adjacent to the outflow regions exhibited decreasing Doppler velocity from $v=20\,\kmpers$ at the transition region temperature to $-10\,\kmpers$ at the coronal temperature, consistent with previous observations.
\end{itemize}
\vspace{-3mm}

%%%%%%%%%%%%%
\vspace{-5mm}
\subsection*{Density of the upflows (Chapter 5)}
\vspace{-2mm}

The electron density of the outflows was derived by using a density-sensitive line pair Fe \textsc{xiv} $264.78${\AA}/$274.20${\AA}.  The double-Gaussian fitting was simultaneously applied to the two emission line profiles from the physical requirement.  Obtained Doppler velocities, electron densities, and column depths are listed in Table \ref{tab:dis_ew} (excerpt from Table \ref{tab:dns_cmpl}).  
\vspace{-3mm}
\begin{itemize}
\item We found that the magnitude relationship between the electron density of the major component and that of EBW component is opposite in the eastern and western outflow regions.  
\vspace{-3mm}
\item The column depths calculated from the obtained electron density indicate that the volume amount of the upflows is much less than that of the major component in the eastern outflow region, while the former dominates the latter in the western outflow region.  
\end{itemize}
\vspace{-3mm}
%Obtained electron density was $n_{\mathrm{e}} = 10^{8.5\text{--}9.0} \, \mathrm{cm}^{-3}$ for EBW component (\textit{i.e.}, upflows), and $n_{\mathrm{e}}=10^{9.0\text{--}9.3} \, \mathrm{cm}^{-3}$ for the major component.

\input{tex/tab_dis_ew.tex}

%%%%%%%%%%%%%
\vspace{-5mm}
\subsection*{$\lambda$-$n_{\mathrm{e}}$ diagram (Chapter 6)}
\vspace{-2mm}

We developed a new method in line profile analysis to investigate the electron density of EBW component from another point of view.  The advantage of the method is that it does not depend on fitting models, from which we reinforced the results obtained in Chapter \ref{chap:dns}. 
%}
%%%%%%%%%%%%%
%\vspace{-5mm}
%\subsection*{Appendix (Morphology of the outflow region)}
%\vspace{-2mm}
%
%The potential magnetic field around the active region NOAA AR10978 was extrapolated from an MDI magnetogram.  The constructed magnetic field lines rooted in the outflow region were connected to the region slightly inside the opposite edge of the active region, and their lengths were $L=100\text{--}200 \, \mathrm{Mm}$. 
% --- End of TeX ---

%% file: tex/tab_dis_ew.tex
\begin{table}
  \centering
  \caption{%
    %{\color{red} 
      Derived physical quantities in the eastern and western outflow regions.  Topology indicates the magnetic connectivity which was inferred from the coronal magnetic field (constructed in Appendix \ref{chap:mph}) and the appearance of Doppler velocity maps: open means that a structure extends far from its footpoint (\textit{i.e.}, very long), and close means that a structure forms a corona loop.%}
  }
  \label{tab:dis_ew}
  \begin{tabular}{lrrrrrrc}
    \toprule
    & \multicolumn{2}{c}{$v_{\mathrm{Dop}} \, (\mathrm{km} \, \mathrm{s}^{-1})$} 
    & \multicolumn{2}{c}{$\log n_{\mathrm{e}} \, [\mathrm{cm}^{-3}]$} 
    & \multicolumn{2}{c}{$\log h \, [\mathrm{cm}]$}
    & Topology
    \\
    \cmidrule(lr){2-3}
    \cmidrule(lr){4-5}
    \cmidrule(lr){6-7}
    \rule[-2pt]{0pt}{13pt}
    & \multicolumn{1}{c}{EBW} & \multicolumn{1}{c}{Major}
    & \multicolumn{1}{c}{EBW} & \multicolumn{1}{c}{Major}
    & \multicolumn{1}{c}{EBW} & \multicolumn{1}{c}{Major}
    & 
    \\
    \midrule
    Eastern & $-88.8 \pm 15.2$ & $-4.2 \pm 1.4$ & $9.06 \pm 0.14$ & $9.01 \pm 0.11$ & $7.36 \pm 0.43$ & $8.51 \pm 0.16$ & closed
    \\
    Western & $-62.0 \pm 16.0$ & $0.1 \pm 2.7$ & $8.60 \pm 0.22$ & $9.22 \pm 0.14$ & $8.53 \pm 0.41$ & $7.74 \pm 0.15$ & open \\
    \bottomrule
  \end{tabular}
\end{table}

%% file: tex/dis_te_outflow.tex
% ===================================
%   Chapter:
%     Discussion.
%   Section:
%     Temperature of the outflows.
% ===================================
%{\color{blue}
% The main constituent of the outflows
The line profiles in the outflow regions are characterized by their EBW with a speed up to $100 \, \kmpers$ within the formation temperature of $\logt=6.1$--$6.3$.  This implies that the outflows are mainly composed of the plasma in that temperature range.  The differential emission measure (DEM) analysis of the outflows indicated that there is a peak at $1.4$--$1.8 \, \mathrm{MK}$ (\textit{i.e.}, around $\logt = 6.2$), which is consistent with our result.  
% The hot coronal lines
Though it has not been focused on in the literature, we found that the line profiles with the formation temperature of $\logt \ge 6.4$ (\textit{i.e.}, Fe \textsc{xvi} and S \textsc{xiii}) did not show the signature of EBW in the outflow region.  This means that the hot plasma does not exist in the outflows, which leads to the implication that the corona in the outflow regions are heated up to at most $\logt \le 6.3$.  Note that the lack of EBW in those hot coronal lines might indicate the scattered spectra from the active region core, since Fe \textsc{xvi} and S \textsc{xiii} are much brighter at the core as already discussed in section \ref{sect:vel_sum}.  

% The transition region temperatures
It was shown that the transition region lines (\textit{e.g.}, Si \textsc{vii} and Mg \textsc{vii}; $\logt = 5.8$) are symmetric in the outflow regions and did not have EBW, but they are significantly blueshifted by around $v = -20 \, \kmpers$.  It can be straightforwardly interpreted as the plasma with a temperature around $\logt = 5.8$ were all flowing upward with a speed of $v = -20 \, \kmpers$.  But since the line width was broadened as seen in Fig.~\ref{fig:vel_map_box}, we suppose two another possibilities: (1) the upflows in the transition region lines are actually hotter than their formation temperature, or (2) the line profiles from the transition region represent a superposition of several components 
%{\color{dartmouthgreen} 
%\textbf{
which have different line-of-sight velocities.
%}}
% First possibility
The interpretation of the line broadening in terms of the high temperature leads to the temperature of $T \simeq 1.5 \, \mathrm{MK}$ (\textit{i.e.}, $\logt = 6.2$) as calculated\footnote{%
%{\color{dartmouthgreen} \textbf{
In a case the difference of line width is attributed to the temperature, we can evaluate the temperature difference.  By using expressions $W_{1} = ( W_{\mathrm{inst}}^2 + \sigma T_{1} )^{1/2}$ and $W_{2} = ( W_{\mathrm{inst}}^2 + \sigma T_{2} )^{1/2}$ where $\sigma = (\lambda_{0}/c)^2 (2 k_{\mathrm{B}} / M_{\mathrm{i}})$ ($\lambda_{0}$: wavelength, $c$: the speed of the light, $k_{\mathrm{B}}$: the Boltzmann constant, and $M_{\mathrm{i}}$: mass of the ion), the temperature difference can be represented as 
$T_{2} - T_{1} = (W_{2}^2 - W_{1}^2) / \sigma$.  Taking $T_{1}$ as the formation temperature, we determined $T_{2}$ at the location where line widths are broadened.
%}}
}
 from the line width of $W \simeq 0.045${\AA} in the outflow regions and $W \simeq 0.030${\AA} in fan loops.  However, since the emission lines around $\logt = 6.2$ (\textit{i.e.}, Fe \textsc{xi}--\textsc{xiii}) clearly exhibit the existence of EBW, 
%{\color{dartmouthgreen}
%\textbf{
the line profiles at $\logt = 5.8$ must have EBW if the first possibility is the case.  However, the transition region lines showed symmetric, therefore the first possibility should be excluded.
% Second possibility
On the other hand, the second possibility remains reasonable when the emission measure of components in the line profiles is similar to each other.  In that case, the line profiles have a broadened and rather symmetric shape.  We do not know at present whether the components in the line profiles are all blueshifted or some of them are redshifted, which may become important to consider what is occurring at the transition region.  
% Test
One idea to test both possibilities is that we compare the line profiles both at the disk center and near at the limb.  The line profiles will keep their line width when the first possibility is the case, while they will become less broadened near at the limb than at the disk center when the second possibility is the case.
%}}
% Mention to the speed
The fact that the upflow speed in the transition region lines is much slower than that in the coronal lines may indicate the acceleration of the outflows from the transition region to the corona.  
%}
% --- End of TeX ---

%% file: tex/dis_outflow_fan.tex
% =============================================
%   Chapter:
%     Discussion
%   Section:
%     Outflow and fan loops.
% =============================================

Emission lines with the formation temperature $\log T \, [\mathrm{K}] \leq 6.4$ were all blueshifted by $v \leq -20 \, \mathrm{km} \, \mathrm{s}^{-1}$ (\textit{i.e.}, upflow) in the outflow region as shown in Chapter 4 (Fig.~\ref{fig:vel_ar10978_tvsv}).  This result is not consistent with \citet{warren2011}, one of few previous studies on the temperature dependence of outflow regions, which reported redshift of the transition region lines and blueshift of coronal lines.  This contradiction originates in the fact that they actually referred to the property of fan loops (Fig.~3 and 4 in their paper).  The outflows and fan loops are often located in neighboring region which leads to the confusing statement in the literature.  We found a clear difference in the temperature dependence of the Doppler velocity between the outflow region and surrounding fan loops.  \citet{baker2009} previously showed that an emission line from Si \textsc{vii} ($\log T \, [\mathrm{K}] = 5.8$) exhibits blueshift by several $\mathrm{km} \, \mathrm{s}^{-1}$.  However, their measurement included the uncertainty of $\sim 10 \, \mathrm{km} \, \mathrm{s}^{-1}$.  We have confirmed the blueshifts of emission lines with the formation temperature below $\log T \, [\mathrm{K}] = 6.0$ for the first time with much more careful procedures in determining the Doppler velocities (\textit{i.e.}, accuracy better than $5 \, \mathrm{km} \, \mathrm{s}^{-1}$; see Chapter \ref{chap:cal}). 

\input{tex/tab_dis_energy.tex}

\begin{figure}[b]
  \centering
  %bb=0 0 1267 644,
  %\includegraphics[width=10cm,clip]{images/dis/dis_outflow_fan.png}
  \includegraphics[width=10cm,clip]{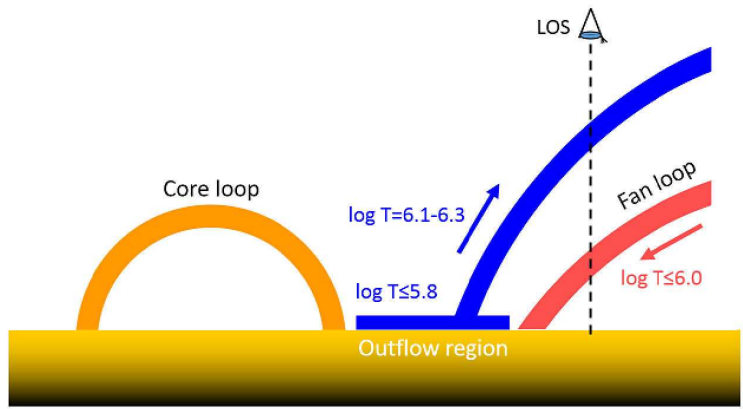}
  \caption{Schematic view of the outflow and a fan loop.}
  \label{fig:dis_outflow_fan}
\end{figure}

The fan loops extended from distinct footpoints beside the outflow region, while blueshifted region in Fe \textsc{xii}--\textsc{xiii} Doppler maps (see Section \ref{sect:vel_map}) was elongated from the outflow region.  No any distinct loops were seen in the outflow region observed with \textit{TRACE} $171${\AA} passband (see Section \ref{sect:mph_euv}).  Since EUV emission from the corona is optically thin, it is not obvious whether the temperature dependence of the Doppler velocity (\textit{i.e.}, $v \geq 0$ for the transition region lines and $v \leq 0$ for the coronal lines) represents the characteristics of fan loops.  We mention a possibility that our line of sight penetrates the fan loops and an elongated structure from the outflow region as shown in Fig.~\ref{fig:dis_outflow_fan}, which suggests that (1) the temperature dependence of fan loops would be a line-of-sight superposition of two structures, and (2) EBW component has been considered to be an indication of upwardly propagating disturbances in fan loops \citep{mcintosh2012}, but it may be originated from the outflow region.  

In order to consider the energetics of the outflow region and fan loops, we estimated the radiative flux ($Q$), thermal conductive flux ($F_{\mathrm{c}}$), and the kinetic energy flux ($K$) in both structures as listed in Table \ref{tab:dis_energy}.  
%{\color{red} 
The temperatures for both structures are taken by the fact that the deviation from single Gaussian has a peak around $\logt = 6.2$ (\textit{i.e.}, a large EBW component relative to the major component) for the outflows, and fan loops are most clearly seen in Fe \textsc{viii} and Si \textsc{vii}. 
%}
We used the radiative loss function $q = n_{\mathrm{e}}^2 \times 10^{-22} \, \mathrm{erg} \, \mathrm{cm}^3 \, \mathrm{s}^{-1}$ ($\log T \, [\mathrm{K}]=5.75$--$6.3$; \citeauthor{rosner1978}\ \citeyear{rosner1978}).  The relation of magnitude becomes $Q \approx F_{\mathrm{c}} \approx K$ for the outflow region, and $Q \gg F_{\mathrm{c}} > K$ for fan loops.  The total fluxes, a measure of the coronal heating term, were comparable for the outflow region and fan loops.  These relations may indicate that a significant part of the heating is converted into the kinetic energy (\textit{i.e.}, the outflow), while the radiation becomes the dominant loss and the thermal conduction plays less important role in fan loops.  The reason why they distribute the energy in different ways even with the same magnitude of the energy input could be important to better understand the formation process of those coronal structures, which should be investigated in the future.  
%{\color{blue}%
The observation which tracks the temporal variability both the outflow region and fan loops enables us to study the relationship between neighboring those structures and could shed the light to such a problem.  

%% file: tex/tab_dis_energy.tex
\begin{table}
  \centering
  \caption{Estimated energy loss from the outflow regions and fan loops.  The factor $q$ in the radiative flux $Q$ represents the radiative loss per volume noted in the main text.  We use the thermal diffusion coefficient of $\kappa_0 = 10^{-6} \, \mathrm{erg} \, \mathrm{cm}^{-1} \, \mathrm{s}^{-1} \, \mathrm{K}^{-7/2}$. 
  }
  \label{tab:dis_energy}
  \begin{tabular}{llllllll}
    \toprule
    & 
    & 
    & 
    & \multicolumn{4}{c}{\small Energy flux ($\mathrm{erg} \, \mathrm{cm}^{-2} \, \mathrm{s}^{-1}$)} \\
    \cline{5-8}
    \rule[-2pt]{0pt}{16pt}
    & {\small Length }
    & {\small $T$}
    & {\small Speed}
    & {\small Radiation}
    & {\small Conduction}
    & {\small Kinetic energy} 
    & \multicolumn{1}{c}{\small Total} \\
    & {\small ($\mathrm{Mm}$)}
    & {\small ($\mathrm{MK}$)}
    & {\small ($\mathrm{km} \, \mathrm{s}^{-1}$)}
    & $Q=q L$        
    & $F_{\mathrm{c}}\approx\kappa_0 \left| T \right|^{7/2} L^{-1}$ 
    & $K=\left[ (1/2) \rho v^2 \right] v$ 
    & \\
    \midrule
    Outflows
    & $\geq 100$   
    & $1$--$2$ 
    & $100$ 
    & $(1\text{--}10) \times 10^{5}$ 
    & $(1\text{--}11) \times 10^{5}$
    & $(2\text{--}5) \times 10^{5}$ 
    & $(4\text{--}20) \times 10^{5}$ \\
    Fan loops 
    & $100$--$200$ 
    & $0.8$    
    & $20$ 
    & $(1\text{--}13) \times 10^{5}$
    & $(3\text{--}5) \times 10^{4}$
    & $4 \times 10^{3}$ 
    & $(1\text{--}13) \times 10^{5}$ \\
    \bottomrule
  \end{tabular}
\end{table}

%% file: tex/dis_siphon.tex
% ==============================
%   Chapter:
%     Discussion.
%   Section:
%     Future work.
% ==============================

Taking the western outflow region as an example case, we consider the possibility whether the outflow can be the siphon flow, 
%{\color{dartmouthgreen}
%\textbf{
which is unidirectional flow along a coronal loop
%}} 
induced by the gas pressure difference between two footpoints of the coronal loop \citep{cargill1980}.  Here we check the conservation of mass flux and the difference of the gas pressure at both footpoints determined by the field lines constructed from an MDI magnetogram (see Appendix).  The conservation of mass flux can be represented as $n_{\mathrm{e}} \, \lvert v \rvert \, A = \mathrm{const.}$, where $A$ is a cross section of the magnetic flux tube ($\propto \lvert B \rvert^{-1}$).  The equation can be modified as $n_{\mathrm{e}} \, \lvert v_{\mathrm{Dop}} \rvert / \lvert B_z \rvert = \mathrm{const.}$.  

\begin{figure}
  \centering
  \includegraphics[width=13.8cm,clip]{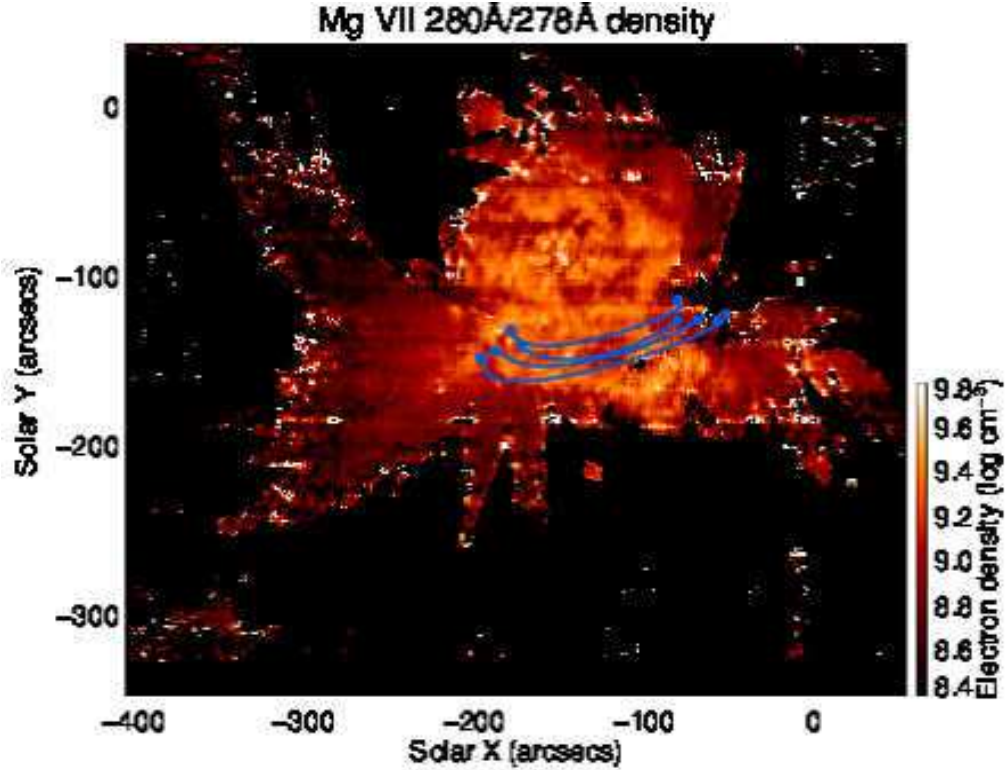}
  \caption{
    Electron density map derived by Mg \textsc{vii} $280.74${\AA}/$278.40${\AA} line ratio.  Four \textit{blue} lines are the field lines rooted at the western outflow region which were extrapolated from the magnetogram (see Appendix \ref{chap:mph}).  Their footpoints are indicated by \textit{circles}.}
  \label{fig:dis_mgvii_dens}
\end{figure}

Fig.~\ref{fig:dis_mgvii_dens} shows the electron density map derived by using a Mg line pair $280.74${\AA}/$278.40${\AA} described in Section \ref{sect:dns_app_mgvii}.  Four \textit{blue} lines indicate projected field lines.  The electron density and Doppler velocity of Mg \textsc{vii} in the western outflow region were respectively $n_{\mathrm{e}}=10^{8.7\text{--}8.9} \, \mathrm{cm}^{-3}$ and $v_{\mathrm{Dop}} = -29.3 \, \text{--} \, {-15.8} \, \kmpers$, while those at the opposite side was $n_{\mathrm{e}}=10^{9.3\text{--}9.4} \, \mathrm{cm}^{-3}$ and $v_{\mathrm{Dop}} = -0.21 \, \text{--} \, 10.4 \, \kmpers$.  The magnetic field strength in the outflow region was $B_z = -191 \, \text{--} \, {-305} \, \mathrm{G}$ while that at the opposite side was $B_z=285 \text{--} 810 \, \mathrm{G}$.  The ratio of the mass flux of the opposite side to the outflow region becomes $3.3 \times 10^{-1}$, $1.3$, $2.8 \times 10^{-2}$, $2.7 \times 10^{-1}$ for the four field lines.  Since the mass flux at both footpoints does not balance (\textit{i.e.}, too much supply for three out of four field lines checked), the steady flow along the coronal loops would not be realized.  Note that since the magnetic field strengths were measured at the photosphere while the Doppler velocities and electron densities were measured in the transition region, the above mass flux calculation may be a rough estimate.  

We also evaluate the pressure at the footpoints of loops connecting the western outflow region and the opposite side.  If we assume that the plasma emitting the emission line has the same temperature at both footpoints, the gas pressure in the western outflow region becomes $p_{\mathrm{gas}} = (4.4 \text{--} 6.9) \times 10^{-2} \, \mathrm{dyne} \, \mathrm{cm}^{-2}$, and that at the opposite side becomes $(1.7 \text{--} 2.2) \times 10^{-1} \, \mathrm{dyne} \, \mathrm{cm}^{-2}$.  The gas pressure at the western outflow region is smaller than that at the opposite side by an order of magnitude, which indicates that the outflow cannot be the siphon flow along the coronal loops.  
%{\color{red} 
This implies that the magnetic field extrapolation by potential calculation is not correct, or the field lines may continuously expand into higher location which can produce steady upflow from the footpoints.
%}

%{\color{darkmagenta}
%\textbf{
Some of previous observations interpreted the outflows as the source region of the slow solar wind \citep{sakao2007,harra2008,baker2009}.  Our results indicate that not only coronal lines but also the transition region lines are blueshifted, and clear EBW (\textit{i.e.}, fast component) exists only in the coronal line profiles, both of which are compatible with the Parker's solution for the solar wind \citep{parker1958} since it predicts strong acceleration near the bottom of the corona and monotonically increasing velocity into the interplanetary space.  However, the location where EBW becomes prominent (\textit{i.e.}, line broadening, \citeauthor{hara2008} \citeyear{hara2008}) is highly localized at the edges of the active region, and there is no EBW in the extended structure from the outflow regions.  It means that such fast component may be decelerated along the magnetic structure, which is not consistent with the solar wind solution.  We suggest two possibilities as follows.  (1) The deceleration occurs only due to the line of sight effect.  The angle between the line of sight and the magnetic field line is probably larger at the higher altitude.  (2) There exists some mechanism such as heating or a change in the magnetic topology, which induces the upflows from the transition region.  The direct measurement of the coronal magnetic field in the future will be helpful to test these possibilities.
%}}

% --- End of TeX ---

%% file: tex/dis_intermittent.tex
% ==============================
%   Chapter:
%     Discussion.
%   Section:
%     Intermittent flow.
% ==============================

When the impulsive heating occurs in a coronal loop, the energy will be transported dominantly through thermal conduction, which abruptly heat the transition region.  The transition region loses its energy increase by the radiation, but if the loss is not sufficient to consume the energy increase, the plasma starts to expand into the upper atmosphere.  One-dimensional hydrodynamic simulation on the impulsive heating showed that such upflow can be detected by hot emission lines and it has a speed of $\sim 100 \, \mathrm{km} \, \mathrm{s}^{-1}$ \citep{patsourakos2006}, which is a consistent value to the speed of the upflow component in line profiles of Fe \textsc{xii}--\textsc{xiv}.  Several observations linked the outflows to he upflow following an impulsive heating at the outflow regions \citep{delzanna2008,peter2010}.

There is a major discrepancy in the behavior of the transition region lines between our result and simulation by \citet{patsourakos2006}.  In their simulation, a line profile of Ne \textsc{viii} ($\log T \, [\mathrm{K}]=5.8$) was redshifted by $v \simeq 10 \, \mathrm{km} \, \mathrm{s}^{-1}$, which was explained by the plasma draining during the cooling phase.  Contrary to the simulation, our analysis indicated that even the transition region lines ($\log T \, [\mathrm{K}]=5.5\text{--}6.0$) are indeed blueshifted by $v \simeq -20 \, \mathrm{km} \, \mathrm{s}^{-1}$ as shown in Chapter 4.  Thus, we cannot apply the impulsive heating with long time intervals (\textit{i.e.}, leading to the plasma draining) directly to the outflow region.  

% --- End of TeX ---

%% file: tex/dns_discuss_n.tex
% =========================
%   Project:
%     Density of upflows
% =========================
%We have measured the electron density of the upflows from the edge of an active region. Before considering the nature of them, we note some equations which can be derived analytically in order to evaluate the electron density caused by different mechanisms.

% Spiclues
%Previous observations have linked the outflow from the edge of active regions to the coronal heating.  Here we discuss what the electron density obtained in Chapter 5 indicates.  
\citet{depontieu2011} proposed that the tip of the spicule is heated up to the coronal temperature (though the heating mechanism has not been revealed), and is injected to the higher atmosphere where the heated plasma form the corona.  The electron density of upflows from the tips of the spicules is estimated by Eq.~(10) in \citet{klimchuk2012} which considers the mass conservation,
\begin{equation}
  n_{\mathrm{UP,}\,s} \delta h_s = n_c h_c A,
  \label{eq:nup_spicule_1}
\end{equation}
where $n_{\mathrm{UP,}\,s}$ is the electron density of an upflow (a suffix $s$ denotes spicule), $\delta$ is the fraction of the spicule that is heated to coronal temperatures, $h_s$ is the height of the spicule, $n_c$ is the coronal density after the tip of the spicule expands into the corona, $h_c$ is the length of coronal loops, and $A$ is the expansion factor of the cross section of coronal loops from the chromosphere to the corona. Using typical coronal values: $n_c \simeq 10^9 \, \mathrm{cm}^{-3}$, $h_c \simeq 5 \times 10^9 \, \mathrm{cm}$, $\delta \simeq 10 \, \mathrm{\%}$ \citep{depontieu2011}, $h_s \simeq 10^9 \, \mathrm{cm}$ in the maximum height, and $A \sim 10$ (this factor has not been determined precisely yet, but is larger than unity), the electron density of upflows is estimated as
\begin{equation}
  n_{\mathrm{UP,}\,s} 
  \simeq 
  5 \times 10^{11} 
  \left( \frac{n_c}{10^9 \, \mathrm{cm}^{-3}} \right)
  \left( \frac{h_c}{5 \times 10^9 \, \mathrm{cm}} \right)
  \left( \frac{A}{10} \right)
  \left( \frac{\delta}{0.1} \right)^{-1}
  \left( \frac{h_s}{10^9 \, \mathrm{cm}} \right)^{-1}
    \, \mathrm{cm}^{-3} \, \text{.}
  \label{eq:nup_spicule_2}
\end{equation}

% Impulsive heating
For impulsive heating, giving the typical energy content of nanoflare (\textit{i.e.}, $10^{24} \, \mathrm{erg}$) and considering the enthalpy flux as a response of the transition region below the corona leads to
\begin{equation}
  \frac{5}{2} p v_{\mathrm{UP,}\, \mathrm{i}} 
  = 
  \frac{E_{\mathrm{i}}}{\pi r^2_{\mathrm{st}} \tau_{\mathrm{i}}} \, \text{,}
  \label{eq:nup_impulsive_1}
\end{equation}
where $p$ is the gas pressure of the upflow, $v_{\mathrm{UP,} \, \mathrm{i}}$ is the speed of the upflow, $E_{\mathrm{i}}$ is the released energy by the impulsive heating, $r^2_{\mathrm{st}}$ is the radius of the coronal strand (\textit{i.e.}, thin coronal loop as an elemental structure), and $\tau_{\mathrm{i}}$ is the duration of the impulsive heating.  Kinetic energy flux can be neglected because the upflow speed is around a half of the sound speed ($\simeq 200 \, \kmpers$ at $\logt = 6.3$), which means the ratio of the kinetic energy flux to the enthalpy flux is the order of $0.1$.  Typical parameters $E_{\mathrm{i}} \sim 10^{24} \, \mathrm{erg}$, $v_{\mathrm{UP, i}} \simeq 100 \, \kmpers$, $r^2_{\mathrm{st}} \sim 100 \, \mathrm{km}$ (moderate estimation considering the coronal filling factor of $0.01$--$0.1$), and $\tau_{\mathrm{i}} \sim 10\text{--}100 \, \mathrm{s}$ (this value contains a large uncertainty because of the lack of our knowledge in the present) imply 
\begin{align}
  n_{\mathrm{UP,}\, \mathrm{i}} 
  \simeq
  5 \times 10^{10} 
  \left( \frac{E_{\mathrm{i}}}{10^{24} \, \mathrm{erg}} \right)
  \left( \frac{r_{\mathrm{st}}}{10^7 \, \mathrm{cm}} \right)^{-2}
  \left( \frac{\tau_{\mathrm{i}}}{10 \, \mathrm{s}} \right)^{-1}
  \left( \frac{T_{\mathrm{i}}}{10^6 \, \mathrm{K}} \right)^{-1}
  \left( \frac{v_{\mathrm{UP,}\,\mathrm{i}}}{10^7 \, \mathrm{cm} \, \mathrm{s}^{-1}} \right)^{-1}
  \, \mathrm{cm}^{-3} \, \text{,}
  \label{eq:nup_impulsive_2}
\end{align}
for which we used $p = 2 n_{\mathrm{UP,}\,\mathrm{i}} k_{\mathrm{B}} T_{\mathrm{i}}$ where $n_{\mathrm{UP,}\,\mathrm{i}}$ is the electron density of the upflow and $T_{\mathrm{i}}$ is its temperature.
% Comparison with the observational results
It is clear that the predicted electron density estimated by adopting the typical coronal values from the spicule and impulsive heating significantly exceed the derived upflow density ($n_\mathrm{EBW} \leq 10^9 \, \mathrm{cm}^{-3}$ in our analysis).  %The upflow density for the spicule was estimated so that the prediction would be near the underestimation, which means that it is probably difficult to approach the observed value. 

Note that Eq.~(\ref{eq:nup_impulsive_2}) can be used to estimate the parameter range where the predicted upflow density becomes similar to the observed value since there is much uncertainty in the parameter $\tau_{\mathrm{i}}$.  If the heating continues for $\tau_{\mathrm{i}} = 500 \, \mathrm{s}$, Eq.~(\ref{eq:nup_impulsive_2}) leads to $n_{\mathrm{UP,}\,\mathrm{i}} \simeq 6 \times 10^8 \, \mathrm{cm}^{-3}$ (\textit{i.e.}, obtained upflow density) with other parameters kept to the typical value.  

% --- End of Tex ---

%% file: tex/dns_discuss_drive.tex
% ===========================================
%   Chapter:
%     Discussion
%   Description:
%     Discussion on driving mechanisms
% ===========================================

Driving mechanisms of upflows from the footpoint of the coronal structures proposed so far are classified into four categories: (1) impulsive heating at the footpoint \citep{hara2008,delzanna2008}, (2) the reconnection between open and closed fields \citep{harra2008,baker2009}, (3) active region expansion in the horizontal direction \citep{murray2010}, and (4) chromospheric spicules \citep{mcintosh2009a,depontieu2011}.  The mechanism (1) was discussed in Section \ref{sect:dis_intermittent} and \ref{sect:dis_n}.  The mechanism (3) is considered to work effectively during the initial phase of active region formation where the flux emergence occurs, which leads to the compression of pre-existing open (long) magnetic fields, and the upflows are induced.  The active region analyzed in this study was already mature and showed no significant expansion in SOT magnetograms during its disk passage.  We have already discussed the possibility of the mechanism (4) in Section \ref{sect:dis_n}.  Therefore, we concentrate discussion here on the mechanism (2). 

% QSL and predicted characteristics
The magnetic topology has been constructed from the photospheric magnetogram in previous observations, which suggested that the upflows are rooted at the boundary between closed region and open region \citep{harra2008,baker2009}.  The boundary is referred to as a quasi separatrix layer (QSL) where the magnetic reconnection between two field lines with different topology favorably occurs.  
%{\color{red}
It is suggested that the continuous reconnection at the QSL results in the persistent upflows as have been observed.  A numerical simulation by \citet{bradshaw2011} has shown that the propagation of a rarefaction wave excited at the reconnected point of closed (short; dense) and open (long; tenuous) loops indeed produces an upflow into the tenuous long loop (hereafter \textit{rarefaction wave scenario}).  One important point is that such reconnection induces the upflow with the electron density lower than that of the previously closed loop, in consideration of the nature of the rarefaction waves excited at the reconnected point.  It was clearly indicated that there is a negative correlation between the density and the upflow speed along the reconnected loop.  Another important point is that the speed of the upflow along the reconnected loop increases with a distance from the footpoint ($\partial v / \partial s < 0$; $s$ is a coordinate along the loop and negative $v$ indicates an upflow).  

We made the $\lambda$-$n_{\mathrm{e}}$ diagrams at the western outflow region, which show the electron density as a function of the wavelength decreases with the speed of the upflow (\textit{i.e.}, positive slope).  This result is consistent with the rarefaction wave scenario if the major component and EBW component came from the same field line.  However, EBW component in observed line profiles is prominent only near the footpoint of the outflow regions.  This clearly contradicts the second characteristic above because observation indicates that the fast component ($\sim 100 \, \mathrm{km} \, \mathrm{s}^{-1}$) concentrates in the root of the outflows (Fig.~\ref{fig:dens_map}), and the speed of the upflow becomes several tens of $\mathrm{km} \, \mathrm{s}^{-1}$ in the extending structure as shown in the Doppler velocity maps (Fig.~\ref{fig:vel_ar10978_vel_map}).  Thus, the rarefaction wave scenario could not be the driving mechanism of the outflows alone.  A heating process which creates the upflow from the bottom of the corona may exist.  
%}
% --- End of Contents ---

%% file: tex/dis_mass.tex
% ==============================
%   Chapter:
%     Discussion.
%   Section:
%     Mass supply.
% ==============================

Blueshifts within a wide temperature range ($\log T \, [\mathrm{K}] \leq 6.4$) in the outflow region indicate that the plasma flowing up into the outer atmosphere does not return at least in this temperature range.  In addition, the outflow region shows coherent pattern in the Doppler velocity maps, which means that there is a mass transport from that region.  We estimate the mass flux of the outflowing plasma $F_{\mathrm{out}}$ in the western outflow region by using Doppler velocity and electron density of EBW component obtained in Chapter \ref{chap:dns}.  The electron density was $n_{\mathrm{e}} \simeq 10^{8.7} \, \mathrm{cm}^{-3}$, and the Doppler velocity was ${-60} \, \mathrm{km} \, \mathrm{s}^{-1}$.  The total area ($S$) of the entire western outflow region was roughly $30'' \times 40''$ ($S \simeq 6 \times 10^{18} \, \mathrm{cm}^{2}$).  Considering the inclination angle of the magnetic field of $30^{\circ} \text{--} 50^{\circ}$, the speed of the outflow is roughly thought to be $v \sim 70\text{--}90 \, \mathrm{km} \, \mathrm{s}^{-1}$.  Thus, $F_{\mathrm{out}}$ can be estimated as $F_{\mathrm{out}}= 2 n_{\mathrm{e}} \mu v S = (4\text{--}5) \times 10^{10} \, \mathrm{g} \, \mathrm{s}^{-1}$ where $\mu$ is a mean mass of ions which was set to $1 \times 10^{-24} \, \mathrm{g}$.  For a comparison, we also evaluate the total mass contained in the active region.  Using volume of $V = (100'')^{3} = 4 \times 10^{29} \, \mathrm{cm}^{3}$ and typical density $n_{\mathrm{e}} = 10^{9\text{--}10} \, \mathrm{cm}^{-3}$, the total mass $M_{\mathrm{AR}}$ is evaluated as $M_{\mathrm{AR}} = 2 n_{\mathrm{e}} \mu V = 8 \times 10^{14\text{--}15} \, \mathrm{g}$. 

This implies that if the mass in the active region is actually lost by the outflow \citep{brooks2012}, the time scale of the mass drain becomes $\tau_{\mathrm{out}}=M_{\mathrm{AR}} / F_{\mathrm{out}} = 2 \times 10^{4\text{--}5} \, \mathrm{s}$ (\textit{i.e.}, several hours to a couple of days).  Since the lifetime of active regions is much longer (\textit{i.e.}, several weeks) than this time scale, the active region needs a certain mechanism to provide the plasma continuously.  We note that the outflow region is localized at the edge of the active region, which means that limited part of the active region is involved in the outflow.  In contrast to this mass drain scenario, the extrapolated magnetic field lines rooted in the outflow region were connected to near the opposite edge of the active region according to the potential field calculation described in Chapter 7.  The opposite side of the outflow region exhibit almost zero velocity, which indicates that the mass would accumulate from the outflow region.  This leads to the picture that the outflow actually provides the active region with the plasma.  However, Doppler velocity maps (Fig.~\ref{fig:vel_ar10978_vel_map}) show a blueshifted pattern extending to the north west from the western outflow region, which may indicate that it is connected to far higher atmosphere.  We must take into account the temporal evolution of the magnetic field in order to confirm the validity of these scenarios as mentioned in Section \ref{sect:dis_future}. 

% --- End of TeX ---

%% file: tex/dis_ew.tex
% =====================================================
%   Chapter:
%     Summary and discussion
%   Section:
%     Differences of eastern/western outflow region.
% =====================================================

\begin{figure}
  \centering
  \includegraphics[width=13.8cm,clip]{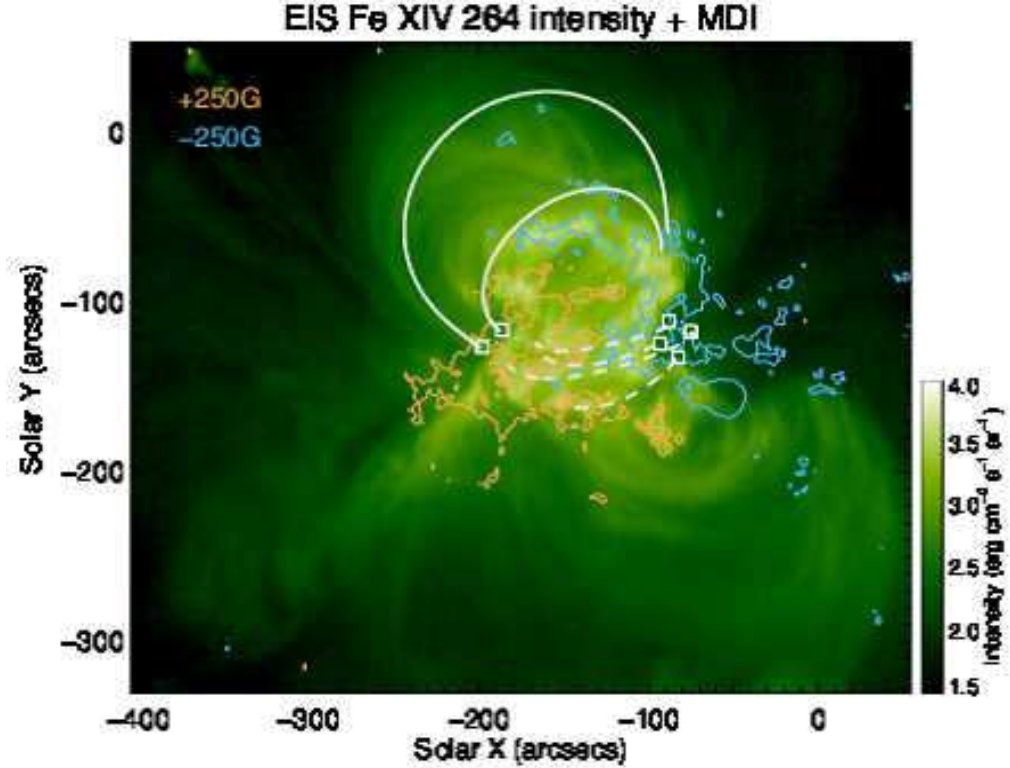}
  \caption{EIS Fe \textsc{xiv} $264.78${\AA} intensity map.  \textit{Orange} (\textit{Turquoise}) contours indicate a magnetic field strength of $+250 \, ({-250}) \, \mathrm{G}$ in the simultaneous MDI magnetogram.  Six \textit{white} boxes are located at the position corresponding to those studied in Chapter \ref{chap:dns}.  \textit{White} lines rooted at those boxes indicate the magnetic field lines extrapolated from the magnetogram (see Appendix \ref{chap:mph}).
%}
}
  \label{fig:dis_ew_map}
\end{figure} 

%{\color{red}
We have discussed about several physical properties of each outflow region in Sections \ref{sect:dis_siphon}--\ref{sect:dis_mass}, and here some implications will be put on the coronal formation (\textit{i.e.}, heating) from the viewpoint of the outflows.  The differences in those two outflow regions are listed in Table \ref{tab:dis_ew}.  The topology of magnetic field lines can be inferred from the extrapolated field lines (see Appendix \ref{chap:mph}) and Doppler velocity maps obtained in Chapter \ref{chap:vel} (Fig.~\ref{fig:vel_ar10978_vel_map}).  In order to confirm the connectivity of the magnetic field lines rooted at the studied outflow regions in Chapter \ref{chap:dns}, we drew projected field lines onto the intensity map of Fe \textsc{xiv} $264.78${\AA} as shown in Fig.\ref{fig:dis_ew_map}.  The outflow regions U1--U6 are indicated by \textit{white} boxes.  The contours with \textit{orange} (\textit{turquoise}) indicate a magnetic field strength of $+250 \, ({-250}) \, \mathrm{G}$ in the simultaneous MDI magnetogram.  

Two \textit{solid white} lines trace coronal loops, therefore we regarded the topology of the eastern outflow region as closed, which can be also seen as a coherent pattern tracing the coronal loops in the Doppler velocity maps.  Four \textit{dashed white} lines rooted at the western outflow region are connected to the opposite polarity around $(x,y)=(-160'', -150'')$, but the Doppler velocity maps clearly show that the blueshifted feature extends into the far west from which we suspected the topology of the western outflow region as open.  The closed loops rooted at the eastern outflow region are brighter than the open structures extending from the western outflow region by one order of magnitude.  This might reflect the length of each structure in the sense that the upflow easily fills a closed loop while it flows without obstacles in a open structure, which produces denser plasma in the closed loop.  

As a consequence, it leads to the implication that the upflow from the bottom of the corona becomes dense in the closed loop because of the pressure balance between the corona and the transition region, which is consistent with our result that the electron density of EBW component was larger in the eastern region than in the western region (see Table \ref{tab:dis_ew}).  Although the difference in the electron density of the major component would not be trivial, the relationship of the column depth (\textit{i.e.}, larger $h_{\mathrm{Major}}$ in the eastern outflow region than in the western outflow region) may represent that the eastern outflow region consists of more coronal loops than the western outflow region.  

\begin{figure}
  \centering
  \includegraphics[width=16.8cm,clip]{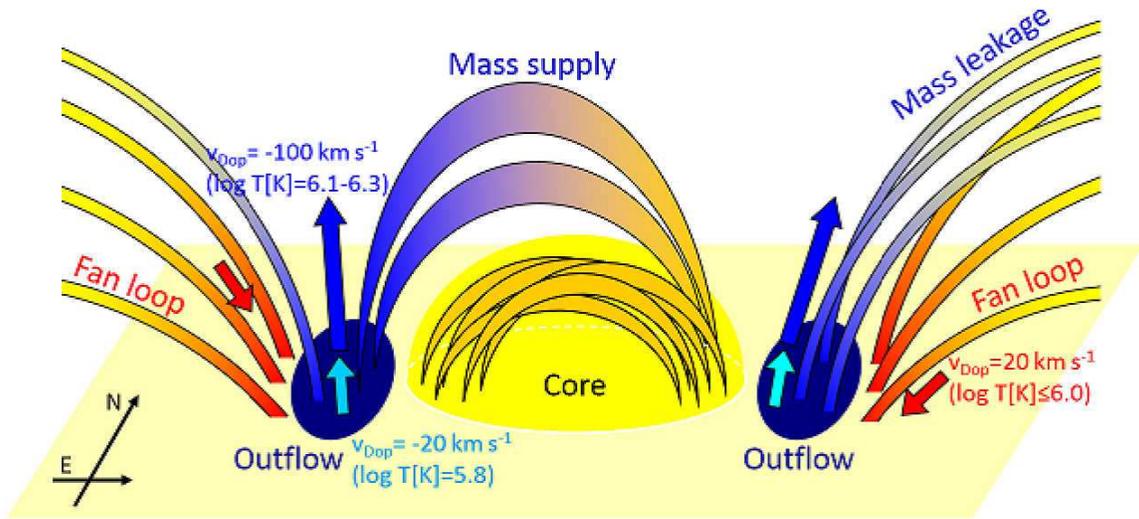}
  \caption{%\color{red} 
Schematic picture of active region outflows.}
  \label{fig:dis_mass_ew}
\end{figure}

We have evaluated mass leakage from the western outflow region in Section \ref{sect:dis_mass}, however, the closed topology of the eastern outflow region may actually imply mass supply to the active region.  If this is the case for a portion of the outflow region, it means that the outflow plays a crucial role in the coronal heating by supplying hot plasma into coronal loops.  We suggest a possible picture in Fig.~\ref{fig:dis_mass_ew} as a summary of discussion in this chapter. 
%}

% --- End of TeX ---

%% file: tex/dis_future.tex
% ==============================
%   Chapter:
%     Discussion.
%   Section:
%     Future work.
% ==============================

The temporal evolution of outflow regions will be a different point of view from this thesis.  Although a number of EIS scans revealed that the outflows are persistent for several days, it has not been clear how they evolve in the shorter timescale of $\sim \mathrm{min}$ because the scans usually take at least several hours to cover the whole active region.  EIS has an observation mode to study the temporal variation of spectra (sit-and-stare mode) which tracks the same position on the Sun in accordance with the solar rotation.  As examples of what are expected from such an observation, there are (1) identification of the origin in terms of the magnetic activity: if the outflows consist of elemental upflow events, we seek the signature of the driving magnetic activity at the photospheric level (\textit{e.g.}, shearing motion, coalescence, cancellation of magnetic patches), which is considered to be the underlying source of coronal dynamics, and (2) the evolution at different temperatures in response to the elemental upflows: this is useful to investigate the starting site (\textit{e.g.}, the upper transition region, higher up in the corona, etc.) which may be a clue to draw the geometry of the outflows. 

In association with the first example of our expectation above, the motion of magnetic patches at the photosphere within a whole active region scale should be investigated.  One approach is to study the flow field derived from the magnetograms through local correlation tracking (LCT) method which is feasible by using SOT and HMI magnetograms.  Since the outflows are localized at the edge of active region core, there is a possibility that the flow field at the edge shows particular characteristics related to the outflows.  It is also useful to construct the coronal magnetic field from consecutive magnetograms. The temporal evolution of coronal magnetic field lines connected to an outflow region can be studied from which we obtain the dynamical picture of the outflow region.  
 
We will also aim to detect a signature of outflows by EUV imaging observations.  A number of observations have been studied the propagating disturbances in fan loops observed by \textit{TRACE} and \textit{SDO}/AIA which are usually interpreted as waves or intermittent flows, however, the outflows themselves have not been observed in the images so far.  Since the outflows are dark compared to fan loops which often exist in the neighbor, we must carefully exclude the location where the influence of fan loops is significant.  The maximum of temperature response function of AIA $193${\AA} passband is $5 \times 10^{-25} \, \mathrm{DN} \, \mathrm{cm}^{5} \, \mathrm{s}^{-1} \, \mathrm{pix}^{-1}$, which leads to the DN of $4$--$10^{3}$ with exposure time of $2 \, \mathrm{s}$ (usual routine) and column emission measure of the outflow $n^2_{\mathrm{e}} h = 10^{24.6\text{--}27.0} \, \mathrm{cm}^{-5}$ which was obtained in Chapter 5.  That DN is enough for us to expect the detection of the outflows. % when they have fluctuations larger than $\simeq 4${\%} which is based on the photon noise.  

% --- End of TeX ---

%% file: tex/contents_rmk.tex
\chapter{Concluding remarks}
  \label{chap:rmk}
  \input{tex/conclusion_v1.tex}

%% file: tex/conclusion_v1.tex
% ==========================
%   Chapter:
%     Concluding remarks.
% ==========================
% What we have done
We investigated the Doppler velocity and electron density of the outflows in NOAA AR10978 observed with the EUV Imaging Spectrometer (EIS) onboard \textit{Hinode}.  

% Method (EIS, EUV emission lines, velocity reference)
In order to use the quiet region as a reference of the Doppler velocity, we analyzed the spectroscopic scans which cover the meridional line of the Sun and determined the Doppler velocity of emission lines with $5.7 \le \logt \le 6.3$ in the quiet region for the first time (Chapter \ref{chap:cal}).  It is shown that emission lines below $\logt = 6.0$ have Doppler velocity of almost zero with an error of $1\text{--}3 \, \kmpers$, while those above that temperature are blueshifted with gradually increasing magnitude: $v = - 6.3 \pm 2.1 \, \kmpers$ at $\logt=6.25$.  The Doppler velocity at $\logt = 5.8$ was consistent with previous observations with SUMER within the estimated error.  
%{\color{red}
This analysis enabled us to determine a Doppler velocity from EIS spectra in the accuracy of $\le 3 \, \kmpers$, which was greatly improved from previous works where the accuracy was considered to be $\ge 5 \, \kmpers$ (up to $10 \, \kmpers$).
%}

% Analysis & results (Chapter 4)
We measured the Doppler velocity of active region core, fan loops, and the outflow regions in NOAA AR10978 (Chapter \ref{chap:vel}) by using the velocity reference obtained in the previous chapter.  While the Doppler velocity at active region core did not deviate much from that of the quiet region, other two structures exhibited characteristic behavior.  Fan loops indicated $v \simeq 10 \, \kmpers$ at $\logt = 5.8$, declining to $v = - 20 \, \kmpers$ at $\logt=6.3$.  In contrast to those fan loops, the outflow regions exhibited a blueshift corresponding to $v \simeq - 20 \, \kmpers$ at all temperature range below $\logt = 6.3$, which implies that the plasma does not return to the solar surface at least in this temperature.  
%{\color{red}
The outflow regions have been sometimes regarded as identical to fan loops, however, we found the definitive difference of them in the Doppler velocity at the transition region temperature.
%}

% Analysis & results (Chapter 5)
In Chapter \ref{chap:dns}, the electron density of the outflowing plasma was derived for the first time by using a density-sensitive line pair Fe \textsc{xiv} $264.78${\AA}/$274.20${\AA}.  We extracted EBW component from the line profiles of Fe \textsc{xiv} through careful double-Gaussian fitting.  Since those two emission lines are emitted from the same ionization degree of Fe ion, they should be shifted by the same amount of Doppler velocity, which required the simultaneous fitting of those two line profiles.  We studied six locations selected from the eastern and western outflow regions.  The average electron density in those six locations was $n_{\mathrm{Major}} = 10^{9.16 \pm 0.16} \, \mathrm{cm}^{-3}$ and $n_{\mathrm{EBW}} = 10^{8.74 \pm 0.29} \, \mathrm{cm}^{-3}$.  The magnitude relationship between $n_{\mathrm{Major}}$ and $n_{\mathrm{EBW}}$ was different in the eastern and western outflow regions.  We also calculated the column depth of each component in the line profiles, which leads to the results that (1) the outflows possess only a small fraction ($\sim 0.1$) compared the major rest component in the eastern outflow region, while (2) the outflows dominate over the rest plasma by a factor of around five in the western outflow region.  The coronal structures beside those outflow regions show different appearance: loops connected to the opposite magnetic polarities at the eastern outflow regions, and only diffuse structures extending away from the active region, which might affect the density and column depth of the outflows. 

% Analysis & results (Chapter 6)
We developed a new method in line profile analysis to investigate the electron density of EBW component in Chapter \ref{chap:ndv}, which we refer to as \textit{$\lambda$-$n_{\mathrm{e}}$ diagram}.  This method has an advantage that it is independent from the fitting model.  By using $\lambda$-$n_{\mathrm{e}}$ diagram, we confirmed that EBW component in Fe \textsc{xiv} line profiles has smaller electron density than that of the major component at the western outflow region.  

% Discussion and conclusions
We discussed the implications from our results in Chapter \ref{chap:dis}.  The outflow regions and fan loops, which has been often discussed in the same context, exhibited different temperature dependence of Doppler velocity.  We concluded these structures are not identical (Section \ref{sect:dis_outflow_fan}).  We tried to interpret the outflows in terms of the siphon flow along coronal loops, but it turned out to be unreasonable (Section \ref{sect:dis_siphon}).  The temperature dependence of the Doppler velocity in the outflow regions obtained in Chapter \ref{chap:vel} were different from that was predicted by a numerical simulation \citep{patsourakos2006}, which dealt with impulsive heating with longer timescale than the cooling (Section \ref{sect:dis_intermittent}).  As for the case if intermittent heating is responsible for the outflows, the duration of heating was crudely estimated to be longer than $\tau = 500 \, \mathrm{s}$ for the energy input of $10^{24} \, \mathrm{erg}$ (\textit{i.e.}, nanoflare) so that the density of upflows from the footpoints becomes compatible with that of the observed outflows (Section \ref{sect:dis_n}).  
%{\color{red}
The electron density and column depth of the upflows in the eastern and western outflow regions were different, which was considered to be due to the magnetic structure above the outflow regions.  The mass leakage occurs at the western outflow region (small $n_{\mathrm{EBW}}$ and large $h_{\mathrm{EBW}}$), on the other hand, there is a possibility of the mass supply to active region loops at the eastern outflow region (large $n_{\mathrm{EBW}}$ and small $h_{\mathrm{EBW}}$), which may be related to the coronal heating process (Section \ref{sect:dis_ew}).  
%}

% Future work
In order to reveal the nature of active region outflows further, we proposed several targets: (1) temporal evolution of the outflows in a short timescale of $\sim \mathrm{min}$ for seeking the signature of drivers, (2) flow field in the photospheric magnetograms for investigating the dynamic topology above the outflow regions, and (3) moving feature in EUV images for detecting the outflows and related phenomena morphologically. 

% --- End of Contents ---

%% file: tex/contents_mph.tex
\chapter{Morphology of the outflow region}
  \label{chap:mph}
\section{Introduction}
  \input{tex/mph_itdn.tex}
\section{Potential field extrapolation from an MDI magnetogram}
  \input{tex/mph_pot_itdn.tex}
  \subsection{Calculation method}
    \input{tex/mph_pot_calc.tex}
  \subsection{Properties of calculated field lines}
    \input{tex/mph_pot_field_line.tex}
\section{EUV imaging observations}
  \label{sect:mph_euv}
  \input{tex/mph_trace.tex}
\section{Summary}
  \input{tex/mph_sum.tex}

%% file: tex/mph_itdn.tex
% ========================================
%   Chapter:
%     Morphology of the outflow region.
%   Section:
%     Observing TRACE images.
% ========================================

% Outflow regions have been emanated from the separatrix layers. But...
We study the magnetic field around the outflow region using a magnetogram taken by \textit{SoHO}/MDI in this chapter.  The configuration of the magnetic field helps us to infer the 3D structure of the plasma flow in the corona.  The outflow region has been said to be located at the region where magnetic topology switches and the lengths of coronal loops change drastically \citep{baker2009}.  %Their calculation of the magnetic field in the corona was based on the linear force free extrapolation.  
In this chapter, the potential magnetic field in the corona around NOAA AR10978 was constructed from an MDI magnetogram at the photosphere through Green's function method \citep{sakurai1982}.  The purpose is to study the connectivity of magnetic field lines rooted at the outflow region studied in this thesis. 

MDI is an instrument onboard \textit{SoHO} which measures the continuum intensity, Doppler velocity, and line-of-sight magnetic field strength of the whole Sun \citep{scherrer1995}.  Those observables are derived by processing the spectrum of an absorption line Ni \textsc{i} $6768${\AA}.  The magnetic field is measured by using the Zeeman splitting. The FOV of MDI is routinely set to the whole Sun, and MDI takes a magnetogram every $96 \, \mathrm{min}$ ($15$ images per day).  The spatial resolution of MDI is $\simeq 4''$ (pixel size of $1''.98$). In this chapter, we used a magnetogram around NOAA AR10978 when it passes near the center of the solar disk.

% Imaging observations  
We have measured Doppler velocity and electron density of the outflow region in previous chapters by analyzing the spectra obtained with \textit{Hinode}/EIS, which cannot be achieved by filter imagers.  However, a spectroscopic scan with EIS usually takes several hours to complete scanning a whole active region (size of $\sim 200''$) with normal exposure time of $\gtrsim 30 \, \mathrm{s}$.  It has been reported that the outflow continues for the timescale of several days \citep{bryans2010,demoulin2012}.  This means that the global structure of the outflow region should have been captured by current EIS data while some phenomena with short duration which might form the outflow were missed. 

% Outflow regions have been said to be just dark! But...
In order to seek such a signature which might be associated with the outflow, we analyze EUV images taken by \textit{SoHO}/EIT and \textit{TRACE}.  EIT is an EUV telescope onboard \textit{SoHO} which routinely takes a whole solar image every $12 \, \mathrm{min}$.  The spatial resolution of EIT is $\sim 5''$ (pixel size of $2''.6$), which is rather coarse.  However, it is useful when we align an MDI magnetogram with EUV images taken by other telescopes, since MDI and EIT are well coaligned by referring the solar limb.  \textit{TRACE} has a spatial resolution of $\sim 1''$ (pixel size of $0''.50$) which is higher than that of EIS (\textit{i.e.,} $2$--$3''$) so that it is useful to study the morphology of the outflow region.  The temporal cadence of \textit{TRACE} images used here was roughly $1 \, \mathrm{min}$, which is much better than that of the EIS scan.  Previous observations have revealed that the outflow emanated from dark region outside the active region core \citep{doschek2008}, but since it has not been revealed whether any activities are occurring in the site.  We aim to detect some signatures which might be linked to the outflow. 

% --- End of TeX ---

%% file: tex/mph_pot_itdn.tex
% ========================================
%   Chapter:
%     Morphology of the outflow region.
%   Section:
%     Introduction to potential field.
% ========================================

Coronal magnetic field plays a dominant role in the sense that the plasma is structured by the magnetic field (\textit{e.g.}, coronal loops), while motion across the field is hindered.  The coronal plasma is confined to magnetic field lines due to its small Larmor radius: $\simeq 2 \, \mathrm{cm}$ for electrons and $\simeq 1 \, \mathrm{m}$ for protons in the typical coronal environment ($T=10^6 \, [\mathrm{K}]$ and $B=10 \, [\mathrm{G}]$).  Therefore, the motion of the coronal plasma can be approximately considered to as one dimensional along the magnetic field, which helps us to implicate dynamical picture of the corona. 

However, direct measurement of the coronal magnetic field has been under developing and there are only a few measurements around an active region by an infrared emission line \citep{lin2004} and radio spectrum \citep{gary1994}.  As an alternative way, the magnetic field in the corona can be inferred through the extrapolation of a photospheric magnetogram into the corona.  Here we adopted the potential field calculation \citep{schmidt1964,sakurai1982} in order to extrapolate an MDI magnetogram at the photosphere around NOAA AR10978. 

% --- End of TeX ---

%% file: tex/mph_pot_calc.tex
% ========================================
%   Chapter:
%     Morphology of the outflow region.
%   Section:
%     Calculation of potential field.
% ========================================

The plasma beta in the corona is considered to be much less than unity ($\beta \ll 1$), which means that the Lorentz force dominates the gas pressure gradient, thus, 
\begin{equation}
  \left( \nabla \times \bm{B} \right) \times \bm{B} = \bm{0}
\end{equation}
holds in the static equilibrium, where $\bm{B}$ is the magnetic field. This equation is nonlinear in general, but it becomes rather simple in the lowest order approximation like
\begin{equation}
  \nabla \times \bm{B} = \bm{0} \, \text{,}
  \label{eq:rot_zero}
\end{equation}
which means that the electric current is zero everywhere in the corona. In this case, the magnetic field $\bm{B}$ can be represented as $\bm{B} = - \nabla \phi (\bm{r})$ where $\phi (\bm{r})$ is a potential function. Then Eq.~(\ref{eq:rot_zero}) becomes
\begin{equation}
  \nabla^2 \phi (\bm{r}) = 0 \mspace{18mu} (z>0) \, \text{,}
  \label{eq:laplace}
\end{equation}
which is the Laplace equation. A magnetogram at the photosphere plays a role as a boundary condition (\textit{e.g.,} Neumann problem), and it can be written as
\begin{equation}
  - \bm{n} \cdot \nabla \phi (\bm{r}) = B_\mathrm{ph} (x, y, 0) \mspace{18mu} (z=0) \, \text{,}
  \label{eq:bound_cnd}
\end{equation}
where $\bm{n}$ is a normal vector as to the solar surface, and $B_\mathrm{ph}$ is the vertical magnetic strength obtained at the photosphere ($z=0$). The $z$ direction was set to be toward the observer.

Here we used the Green's function method to solve Eq.~(\ref{eq:laplace}) with the boundary condition represented by Eq.~(\ref{eq:bound_cnd}) (\citeauthor{sakurai1982} \citeyear{sakurai1982}; classical Schmidt method). If we found a function $G (\bm{r}, \bm{r}^{\prime})$ which satisfies the following three conditions:
\begin{align}
  & \nabla^2 G (\bm{r}, \bm{r}^{\prime}) = 0 & (z > 0) \, \text{,} \\
  & G (\bm{r}, \bm{r}^{\prime}) \longrightarrow 0 
    & (\left| \bm{r} - \bm{r}^{\prime} \right| \rightarrow \infty; z>0) \, \text{,} \\
  & - \bm{n} \cdot \nabla G (\bm{r}, \bm{r}^{\prime}) 
  = \delta (x - x^{\prime}) \delta (y - y^{\prime}) & (z=0) \, \text{,}
\end{align}
the potential function $\phi$ can be represented by
\begin{equation}
  \phi (\bm{r}) = \int B_{\mathrm{ph}} (\bm{r}^{\prime}) G (\bm{r}, \bm{r}^{\prime}) d S^{\prime} 
  \, \text{,}
  \label{eq:phi_rprstn}
\end{equation}
where $\delta (x-x^{\prime})$ and $\delta (y-y^{\prime})$ are the Dirac's delta function. The term $d S^{\prime}$ indicates an area element in $x$--$y$ plane. It is easy to prove that Eq.~(\ref{eq:phi_rprstn}) satisfies the boundary condition Eq.~(\ref{eq:bound_cnd}). The function $G (\bm{r}, \bm{r}^{\prime})$ is called the Green's function and here has a functional form of 
\begin{equation}
  G (\bm{r}, \bm{r}^{\prime}) = \dfrac{1}{2 \pi \left| \bm{r} - \bm{r}^{\prime} \right|}
  \, \text{.}
\end{equation}
From this expression, the potential function can be calculated by Eq.~(\ref{eq:phi_rprstn}), then, magnetic field strength at any locations will be derived from $\bm{B} = - \nabla \phi (\bm{r})$. The integral in Eq.~(\ref{eq:phi_rprstn}) was replaced by the summation of discretized data points in the practical calculation. 

% --- End of TeX ---

%% file: tex/mph_pot_field_line.tex
% ========================================
%   Chapter:
%     Morphology of the outflow region.
%   Section:
%     Potential field.
% ========================================

\begin{figure}
  \centering
  \includegraphics[width=7.4cm,clip]{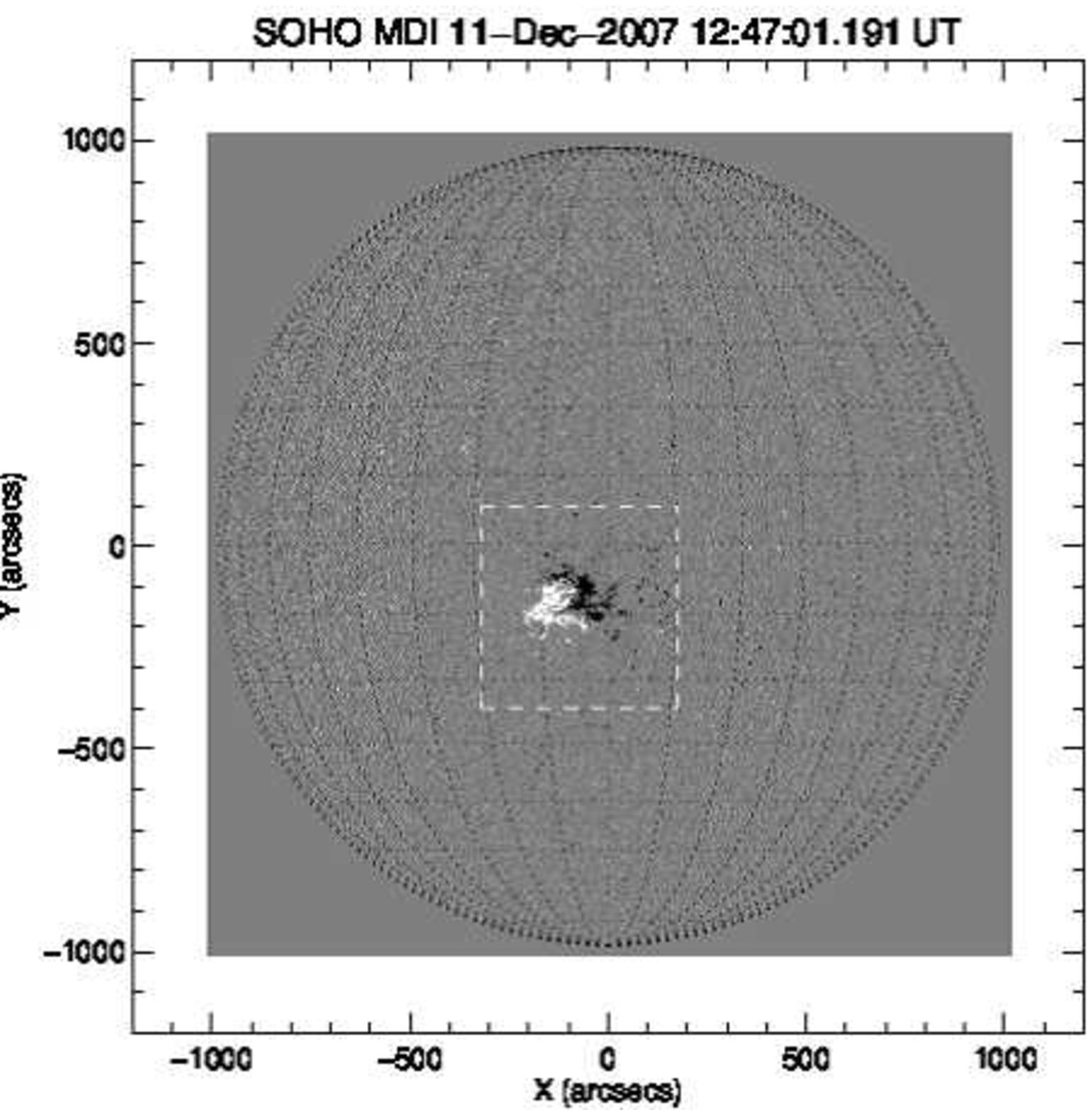}
  \includegraphics[width=0.925cm,clip]{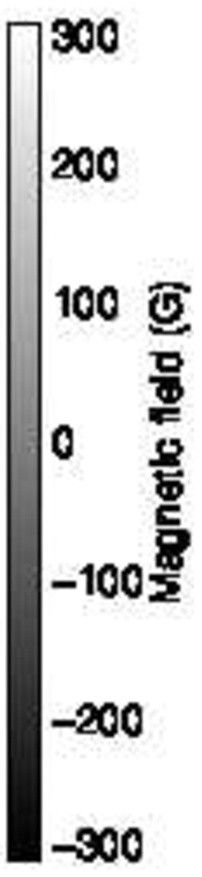}
  \includegraphics[width=7.4cm,clip]{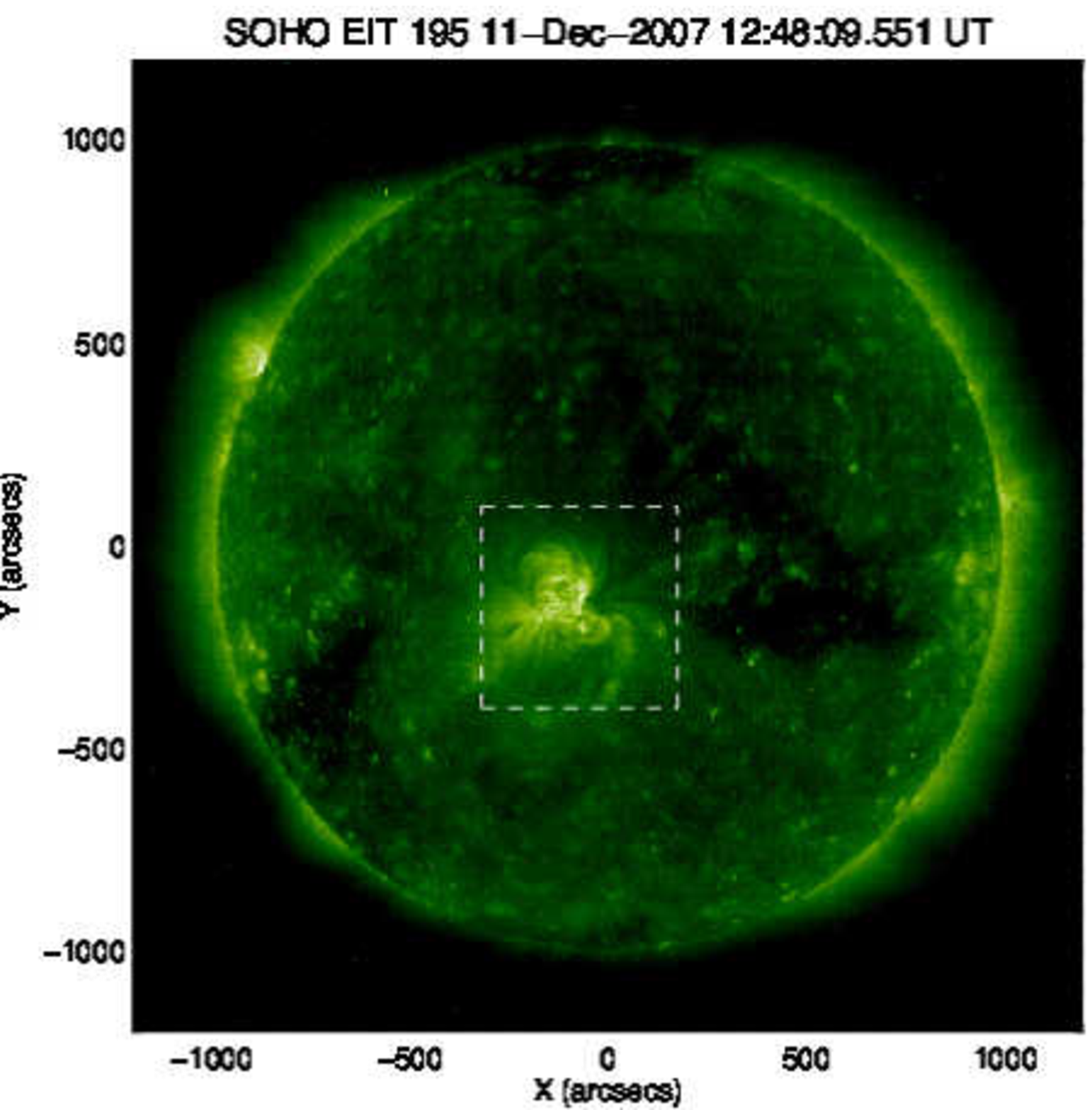}
  \includegraphics[width=0.925cm,clip]{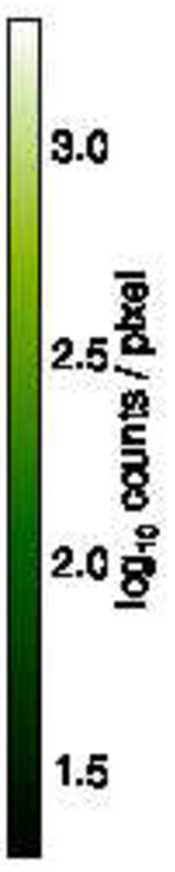}
  \caption{\textit{Left}: MDI magnetogram taken on 2007 December 11 12:47:01UT. 
    \textit{Right}: EIT image taken on 2007 December 11 12:48:09UT.
    \textit{White dashed} lines in each panel indicate the calculation box for potential field.
  }
  \label{fig:pot_whole_context}
\end{figure}

We applied the potential field extrapolation to an MDI magnetogram taken on 2007 Dec 11, which is shown in Fig.~\ref{fig:pot_whole_context}.  \textit{Left} and \textit{Right} panels respectively indicate an MDI magnetogram and an EIT image obtained almost simultaneously. The area including the active region NOAA AR10978 was extracted near the disk center around $(x,y)=(-150,-100)$ which is shown by \textit{white dashed} square, so that we could neglect the effect of the curvature of the solar surface ($\cos \theta \gtrsim 0.97$, $\theta$ is an angle between our line of sight and the normal vector at the solar surface).  The size of the calculated box was $250 \times 250 \times 200 \, \mathrm{pix}^{3}$ ($500'' \times 500'' \times 400''$), which is large enough to include whole active region, and the magnetic field was calculated in the orthogonal coordinate. 

\begin{figure}
  \centering
  \includegraphics[width=8.3cm,clip]{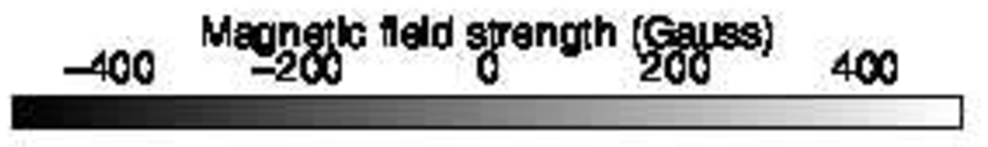}
  \includegraphics[width=8.3cm,clip]{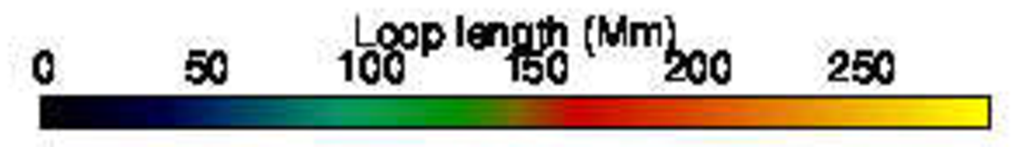}
  \includegraphics[width=8.3cm,clip]{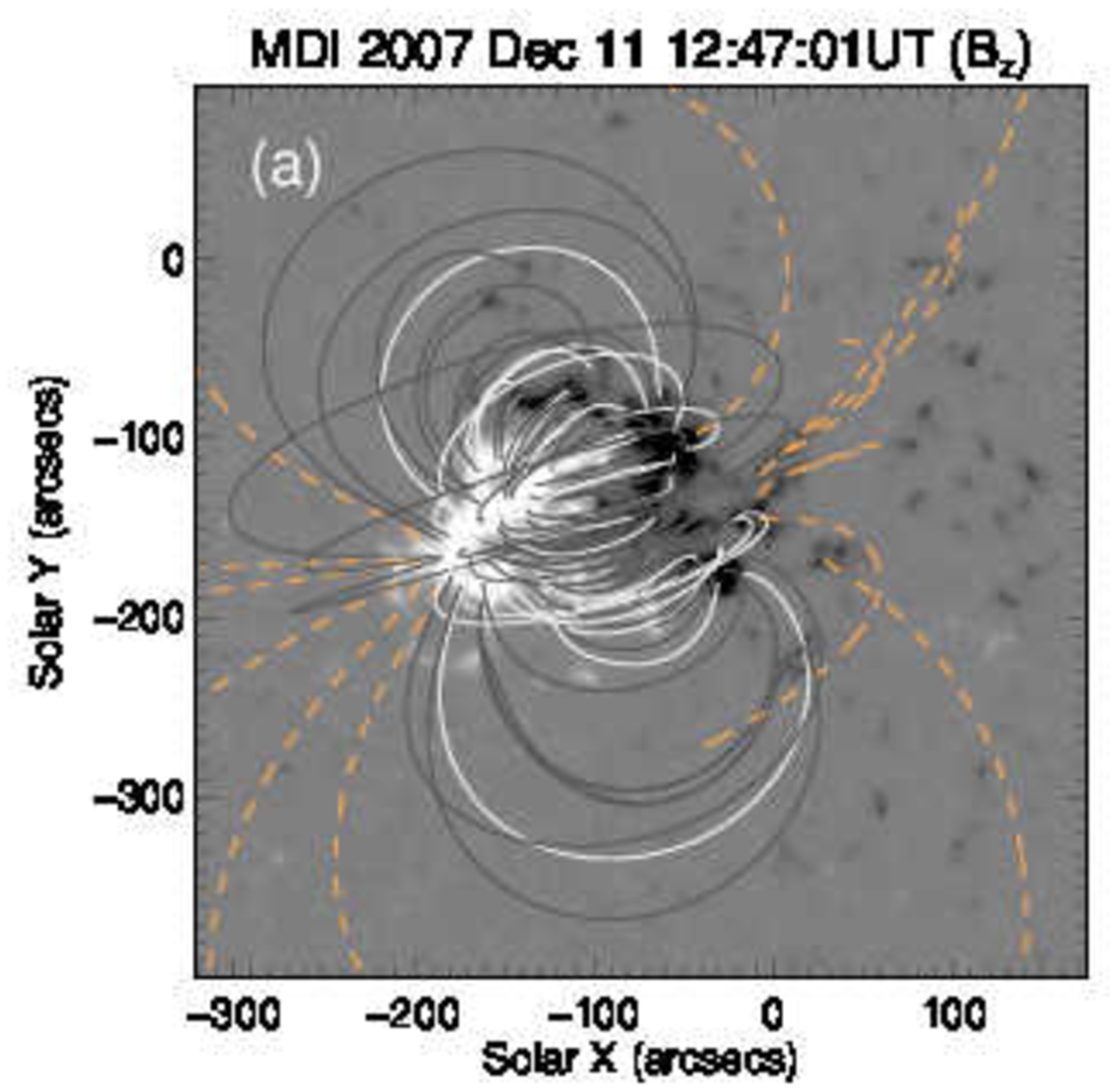}
  \includegraphics[width=8.3cm,clip]{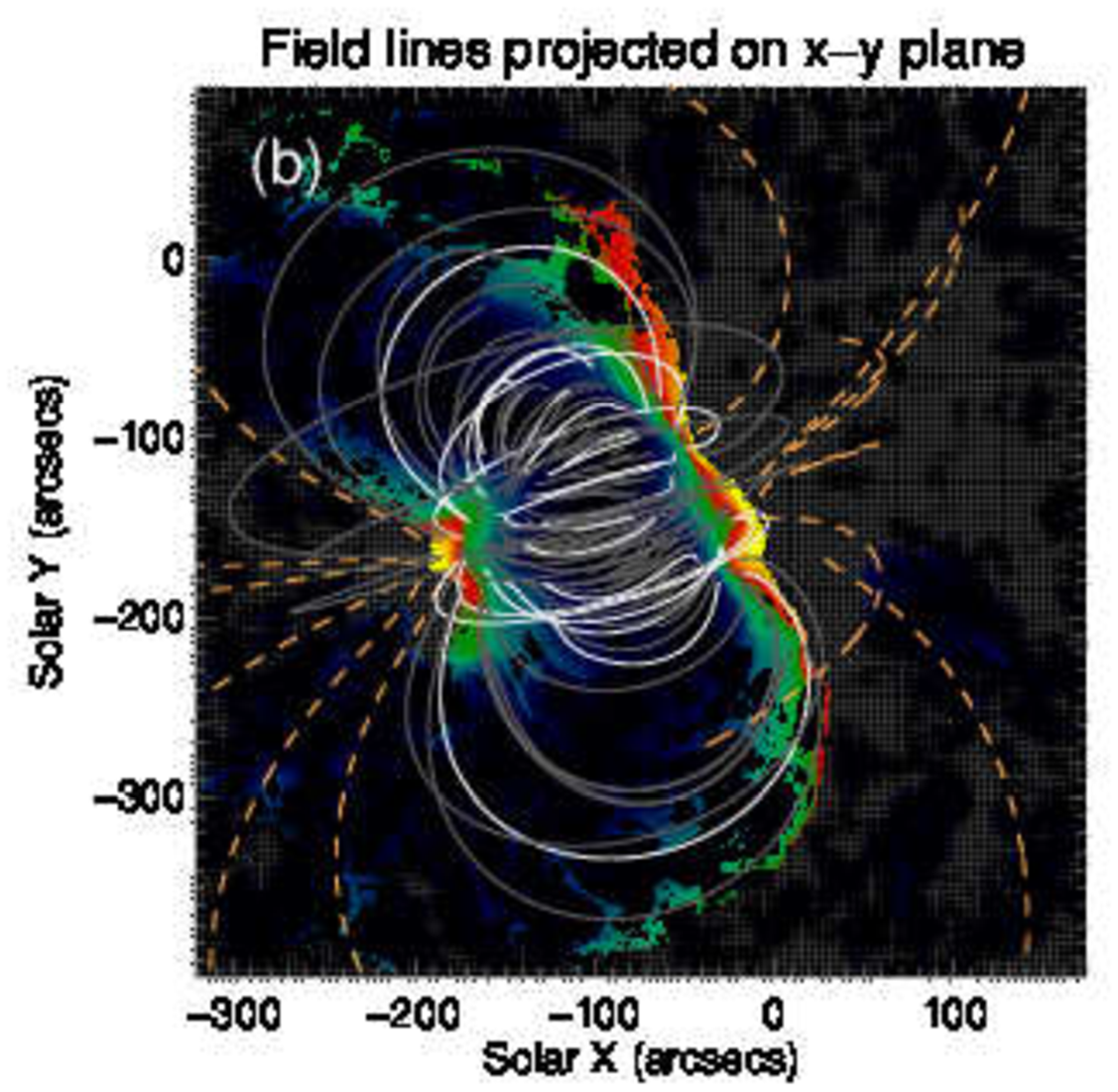}
  \includegraphics[width=8.3cm,clip]{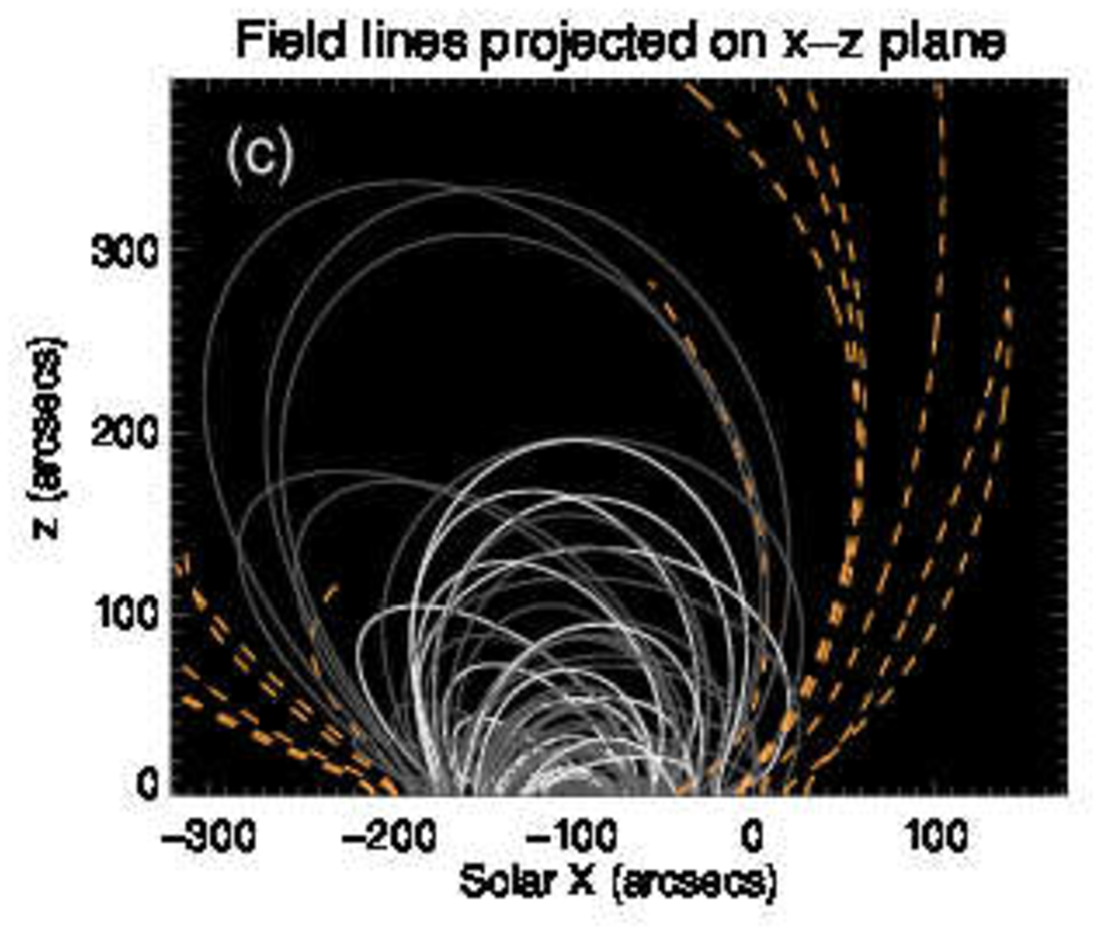}
  \includegraphics[width=8.3cm,clip]{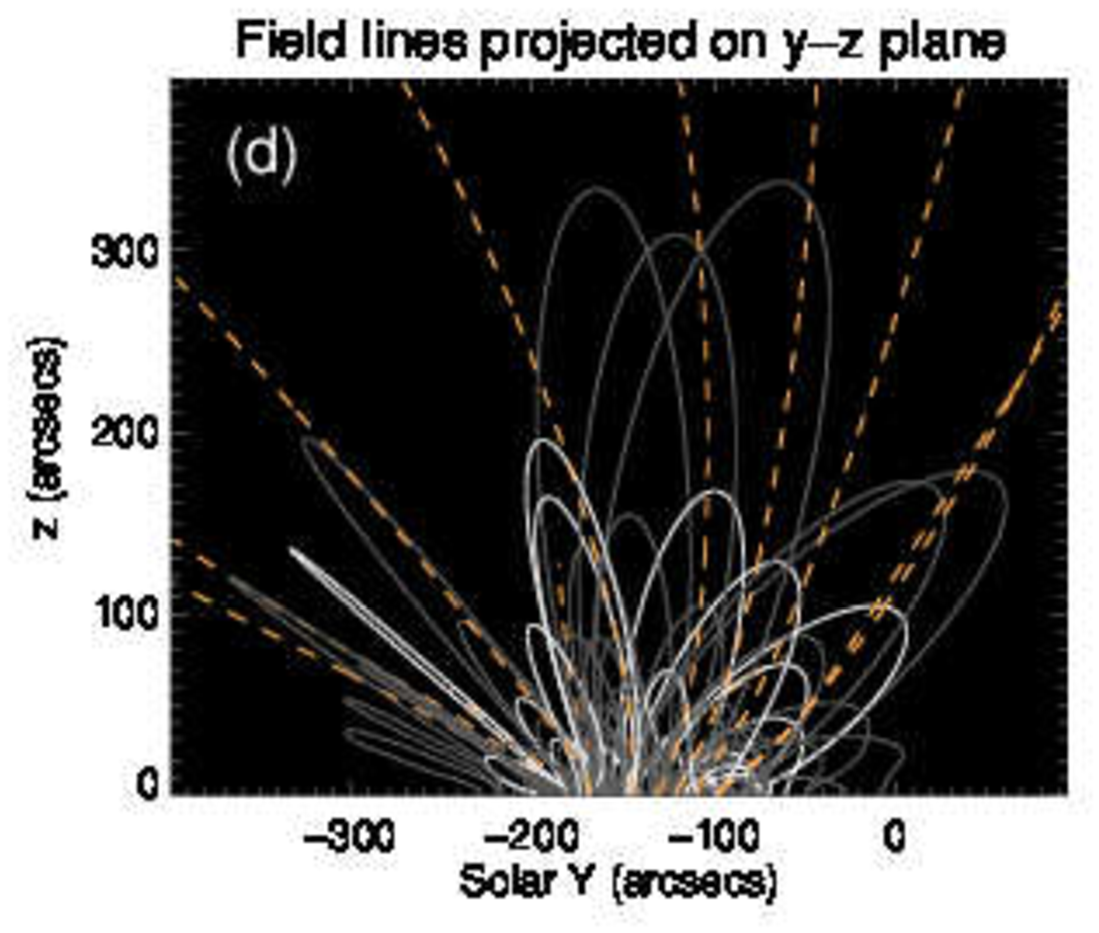}
  \caption{
    Magnetic field extrapolated by the potential calculation 
    using Green's function method described in \citet{sakurai1982}. 
    (a) Magnetogram at $z=0$ obtained by \textit{SoHO}/MDI. 
    (b) Calculated field lines projected onto $x$--$y$ plane. 
    \textit{White} (\textit{Gray}) lines indicated that the magnetic field strength 
    at their footpoint is larger than $500$ $(200) \, \mathrm{G}$.
    \textit{Orange dashed} lines indicate that the field line goes out from the side boundary. 
    Background color shows the length of field lines rooted at each pixel. 
    \textit{Gray hatched} region indicate that field lines rooted at the pixels 
    penetrate into the side boundary or the top of the calculation box 
    ($500'' \times 500'' \times 400''$).
    (c) Calculated field lines projected onto $x$--$z$ plane. 
    (d) Calculated field lines projected onto $y$--$z$ plane.
  }
  \label{fig:mph_mdi_pot}
\end{figure}

The result of the potential field extrapolation is shown in Fig.~\ref{fig:mph_mdi_pot}. The location $(x,y)$ are defined by heliocentric coordinates in which $y$ lies in the rotational axis of the Sun (positive value means north), and $z$ is in the direction vertical to the solar surface.  Panels respectively show calculated field lines projected onto (a) the magnetogram at $z=0$, (b) $x$--$y$ plane, (c) $x$--$z$ plane, and (d) $y$--$z$ plane.  For panel (a) and (b), \textit{white} (\textit{gray}) lines indicated that the magnetic field strength at their footpoint is larger than $500$ $(200) \, \mathrm{G}$.  \textit{Orange dashed} lines indicate that the field line goes out from the side boundary or the top of the calculation box.  Background color in panel (b) shows the length of field lines rooted at each pixel.  \textit{Gray hatched} region indicate that field lines rooted at the pixels penetrate into the side boundary or the top of the calculation box.  AR10978 had a leading negative sunspot around $(x, y) = (-30'', -180'')$ and following positive region less concentrated, which can be seen in panel (a).  The loop structures connecting those opposite polarities are prominent as shown by \textit{white}/\textit{gray} field lines in panels (a) and (b).  The field lines rooted at the region of strong magnetic field strength ($B_{z} \geq 500 \, \mathrm{G}$) reach the height up to $\sim 200 \, \mathrm{Mm}$ in the maximum as shown in panels (c) and (d).

\begin{figure}
  \centering
  \includegraphics[width=14cm,clip]{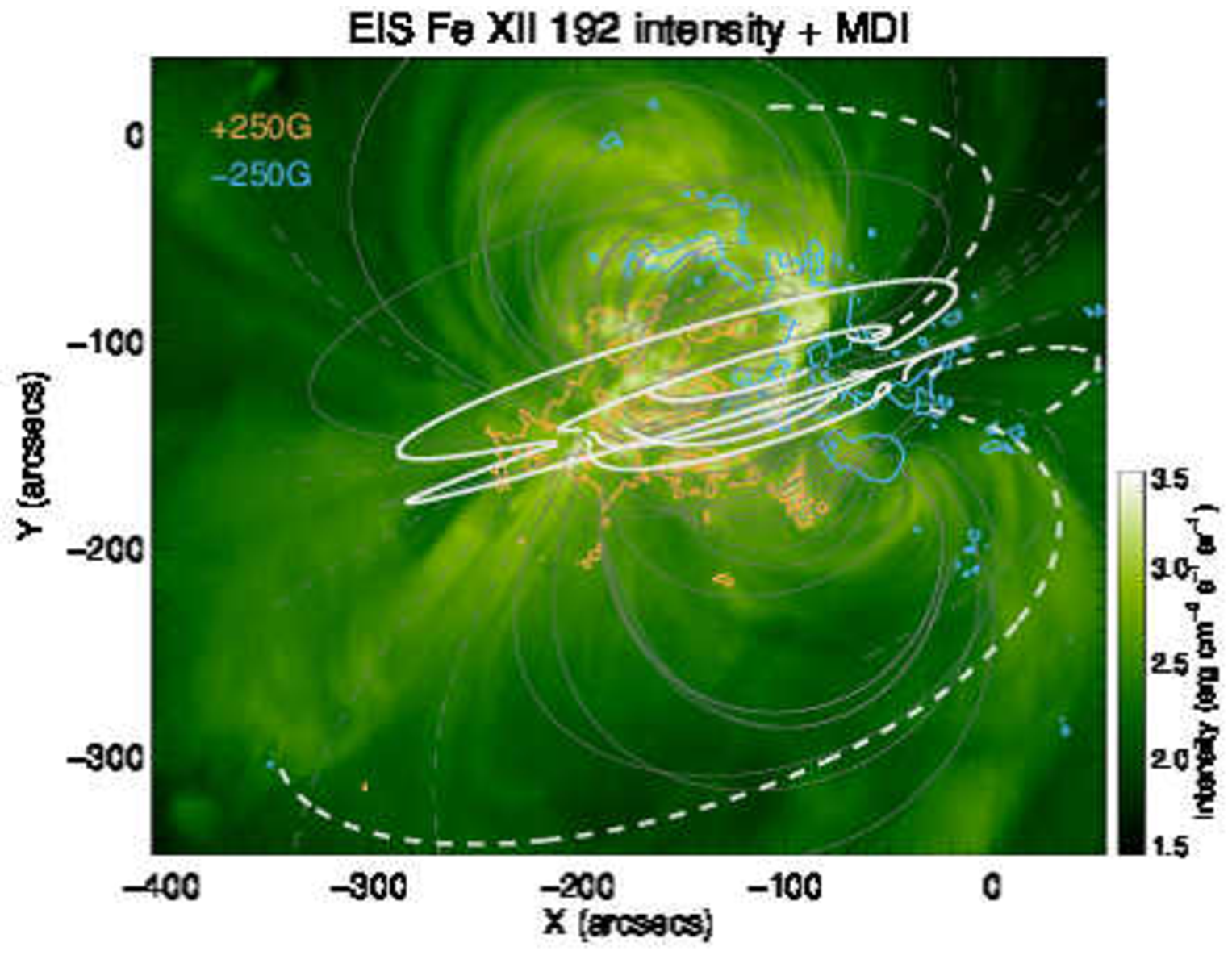}
  \includegraphics[width=14cm,clip]{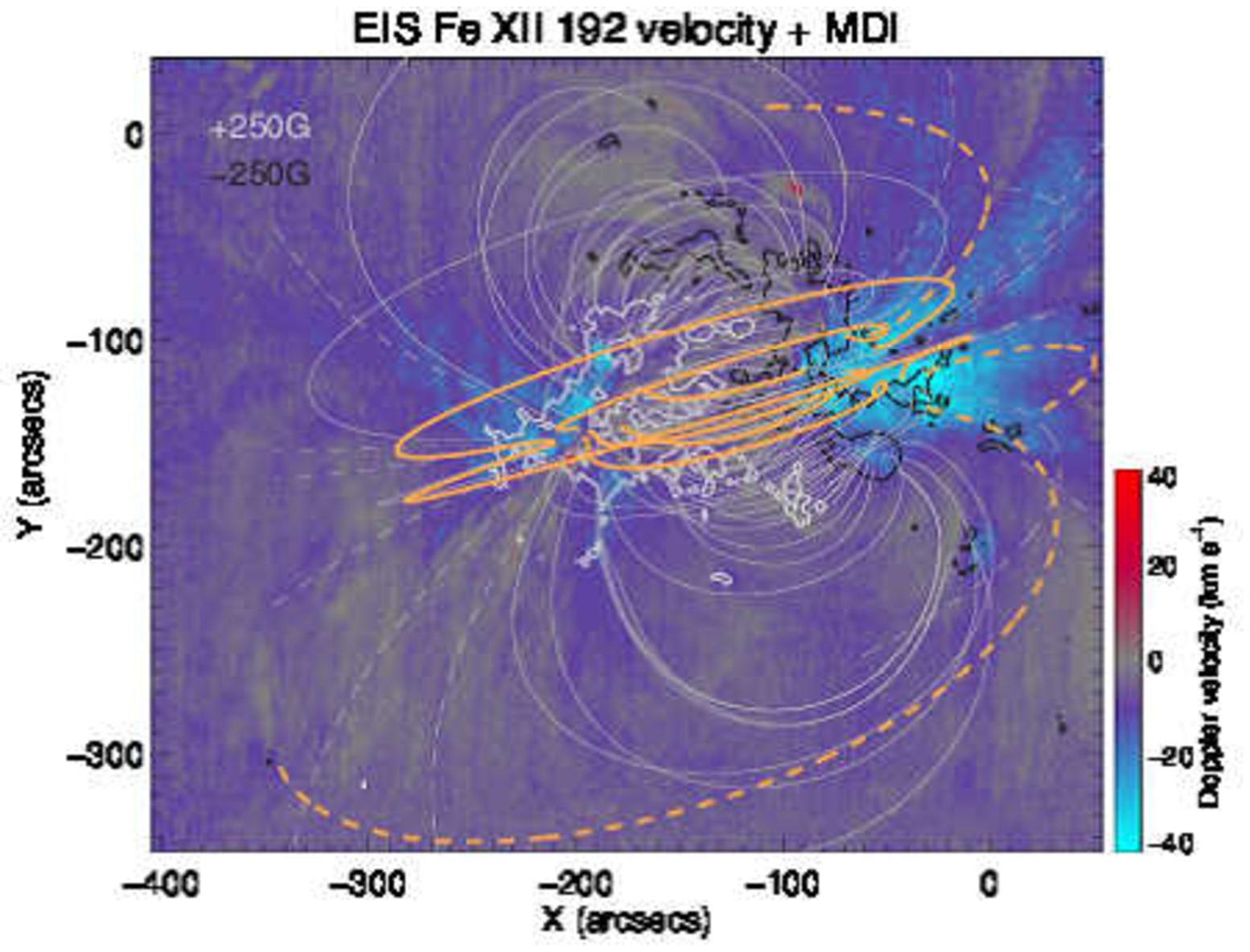}
  \caption{
    \textit{Left}: EIS Fe \textsc{xii} $192.39${\AA} intensity and 
    potential field calculated from MDI magnetogram. 
    \textit{Right}: EIS Fe \textsc{xii} $192.39${\AA} Doppler velocity and 
    potential field calculated from MDI magnetogram. 
    \textit{Dashed} lines indicate that the field line goes out 
    from the boundary of the calculation box of potential field. 
    \textit{White}/\textit{Orange} \textit{thick} lines show the field line
    rooted at the outflow region.
  }
  \label{fig:eis_field_line}
\end{figure}

In order to see the correspondence between magnetic field and coronal structures more clearly, the calculated field lines were projected on the EIS Fe \textsc{xii} $192.39${\AA} intensity (\textit{left}) and Doppler velocity (\textit{right}) map as shown in Fig.~\ref{fig:eis_field_line}. \textit{Gray solid} (\textit{dashed}) lines indicate the field line closed within (goes out from) the calculation box. In the intensity map, it is clearly seen that the legs of two fan loops extending to the south east well coincide with the projected \textit{gray dashed} lines. At the west region of the active region, the blueshifted area has a shape which is well traced by the projected \textit{gray dashed} lines as shown in the Doppler velocity map. The loop structures connecting the opposite polarities near the core are not clearly seen in the formation temperature of Fe \textsc{xii}, however, those are prominent in Fe \textsc{xvi} intensity map (\textit{cf.} panel f in Fig.~\ref{fig:vel_context_ar10978}).

\textit{White}/\textit{Orange} \textit{thick} lines are the loops rooted in the outflow region at their western footpoints.  The lengths of these loops were $\geq 100 \, \mathrm{Mm}$.  The inclination angle of those at their footpoint as to the solar surface was $30$--$50 \, \mathrm{deg}$. Potential field extrapolation clearly show that some of the field lines rooted in the outflow region were not long enough to reach the heliosphere but rather compact ($100$--$200 \, \mathrm{Mm}$ in the height).  The opposite footpoints of those field lines are located at the neighbor of the east edge of the core region (\textit{i.e.}, closed loop), and they look slightly being inside the edge.  Other field lines rooted in the outflow region are connected to the eastern outflow region, and the rest loops are so long enough to go out from the calculation box.  

\begin{figure}
  \centering
  \includegraphics[width=13cm,clip]{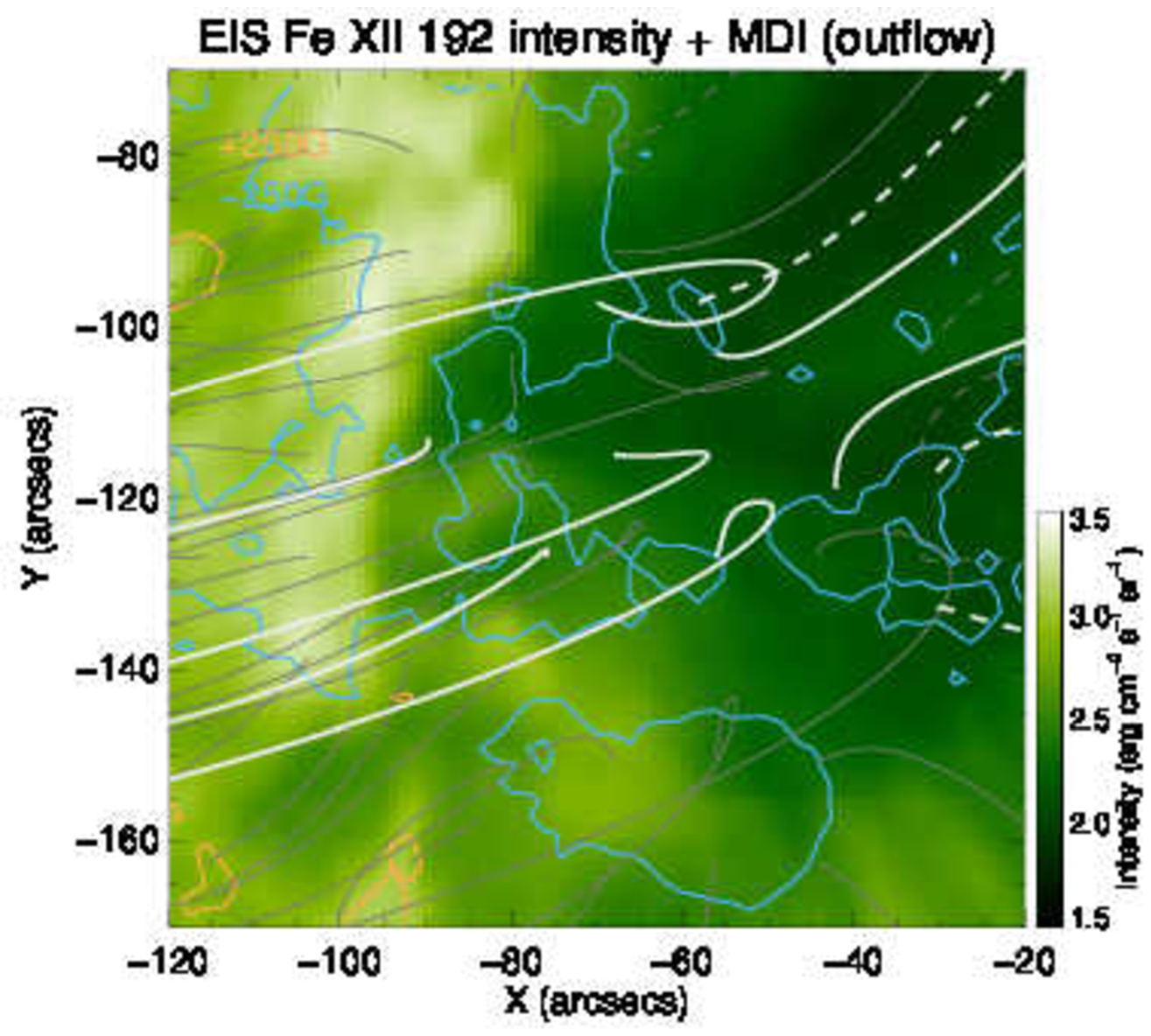}
  \includegraphics[width=13cm,clip]{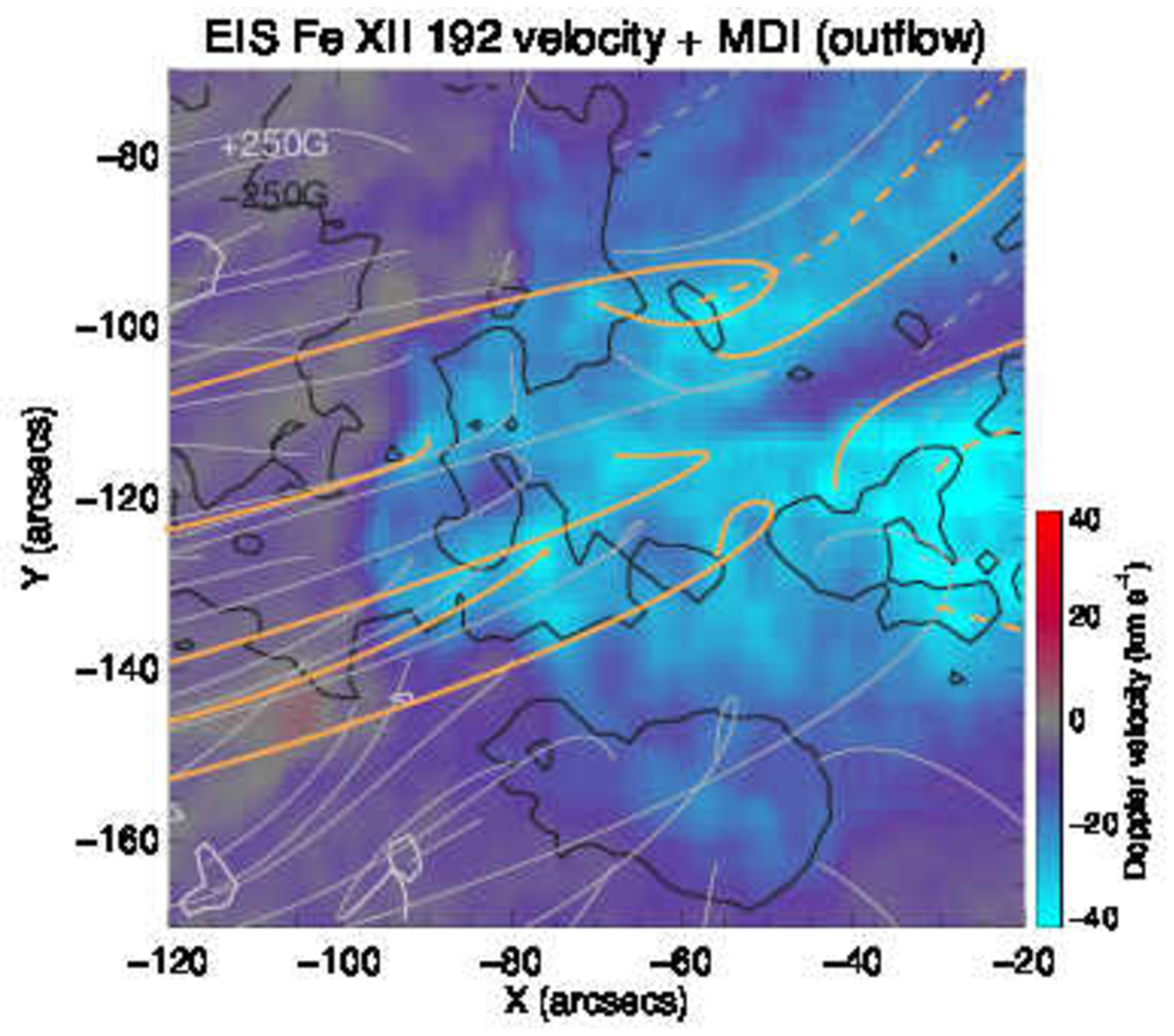}
  \caption{
    Zoomed outflow region as same format as Fig.~\ref{fig:eis_field_line}. 
  }
  \label{fig:eis_field_line_2}
\end{figure}

Fig.~\ref{fig:eis_field_line_2} shows the zoomed outflow region as same format as Fig.~\ref{fig:eis_field_line}.  We drew the \textit{orange}/\textit{white} \textit{thick} lines from the locations where the upflow speed is large as seen in \textit{lower} panel.  There are not spatial correspondence between the Doppler velocity and the magnetic field strength at the photosphere (\textit{black contour}: $-250 \, \mathrm{G}$).  The blueshifted region prevails more homogeneously than the magnetic field strength.  This might be caused by the line-of-sight integration.  Since almost all magnetic field lines have an inclination from the normal to the solar surface, the upflowing plasma at certain height in a coronal loop is not magnetically connected to the photospheric level along the same line of sight.  

% --- End of TeX ---

%% file: tex/mph_trace.tex
% ========================================
%   Chapter:
%     Morphology of the outflow region.
%   Section:
%     Observing TRACE images.
% ========================================

\begin{figure}
  \centering
  \includegraphics[width=12cm,clip]{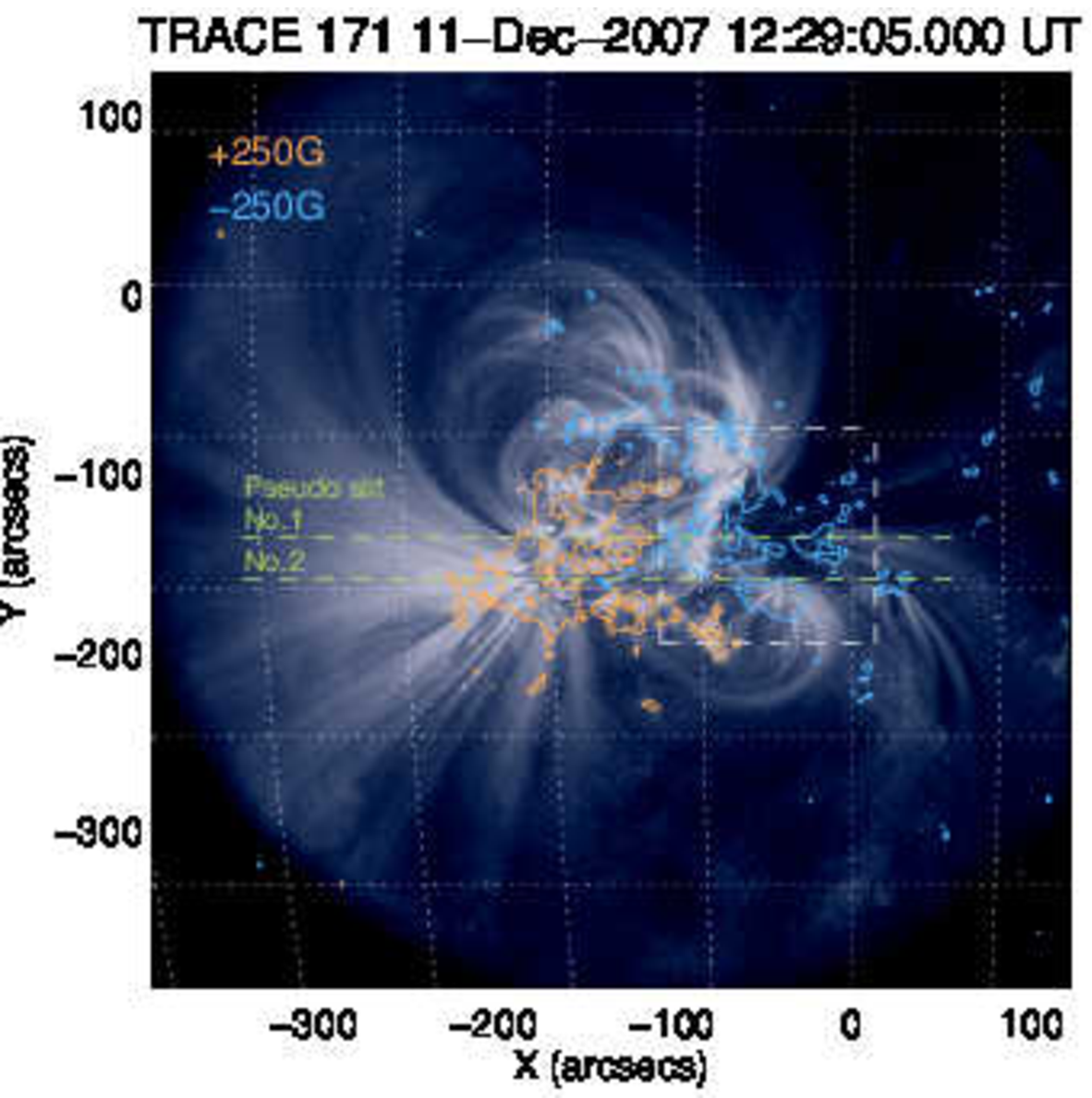}
  \caption{A \textit{TRACE} image taken on 2007 December 11 12:29:05UT. 
    \textit{Orange} (\textit{Blue}) contours indicate the magnetic field strength of $+250$ ($-250$) $\mathrm{G}$ in MDI magnetogram.  A \textit{white dashed} square shows the region zoomed in Fig.~\ref{fig:mph_mdi_euv}.  Two horizontal \textit{green dashed} lines indicate pseudo slits for $x$-$t$ diagrams shown in Fig.~\ref{fig:mph_tr_xt}.}
  \label{fig:mph_tr_context}
\end{figure}

\textit{TRACE} took EUV image of NOAA AR10978 mainly through the $171${\AA} passband on 2007 December 11. This passband includes dominant contribution from Fe \textsc{ix}/\textsc{x} emission lines and their formation temperature is $\log T \, [\mathrm{K}]=5.9$--$6.0$. Fig.~\ref{fig:mph_tr_context} shows a \textit{TRACE} $171${\AA} image taken at 12:29:05UT.  It is clearly seen that multiple loop system connects between positive polarity (\textit{orange} contour) and negative polarity (\textit{turquoise} contour), and fan loops extend out of the FOV at the both sides as more prominent in the east than in the west. Note that the outflow region focused in this thesis corresponds to the dark location within the \textit{white dashed} box where the magnetic field strength reaches several hundred Gauss. 

% Zoomed outflow region
\begin{figure}
  \centering
  \includegraphics[width=4.4cm,clip]{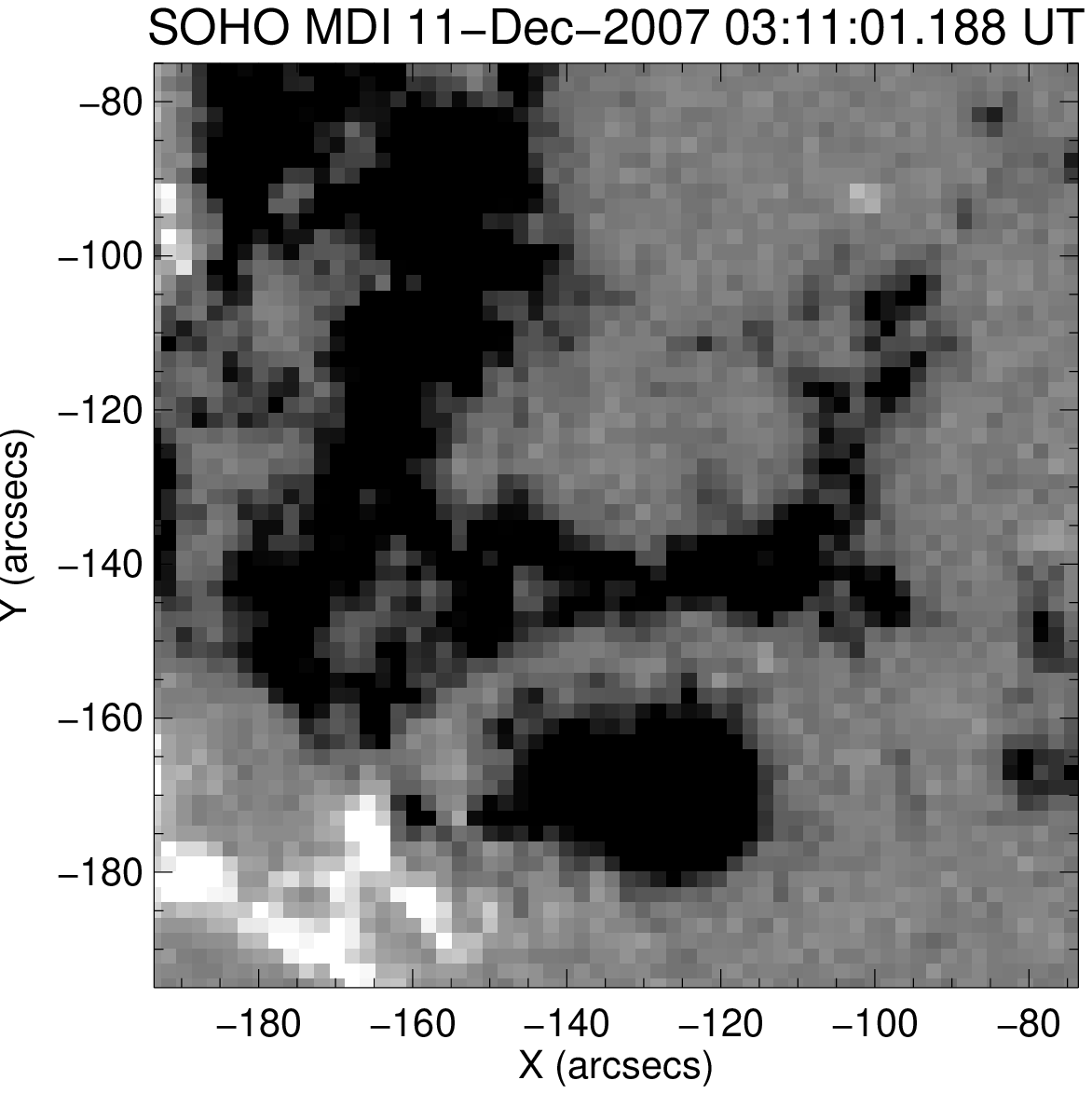}
  \includegraphics[width=4.4cm,clip]{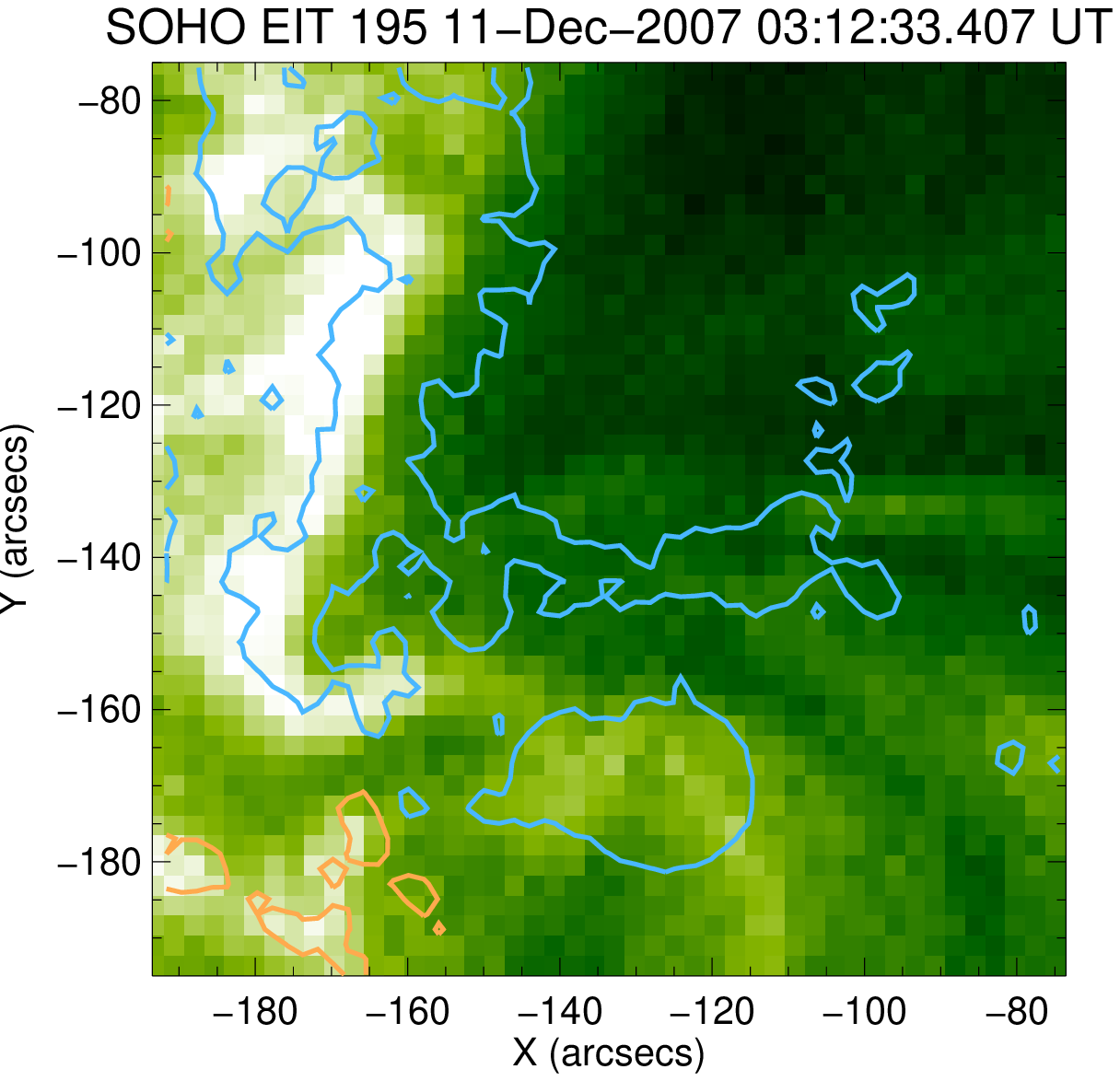}
  \includegraphics[width=4.4cm,clip]{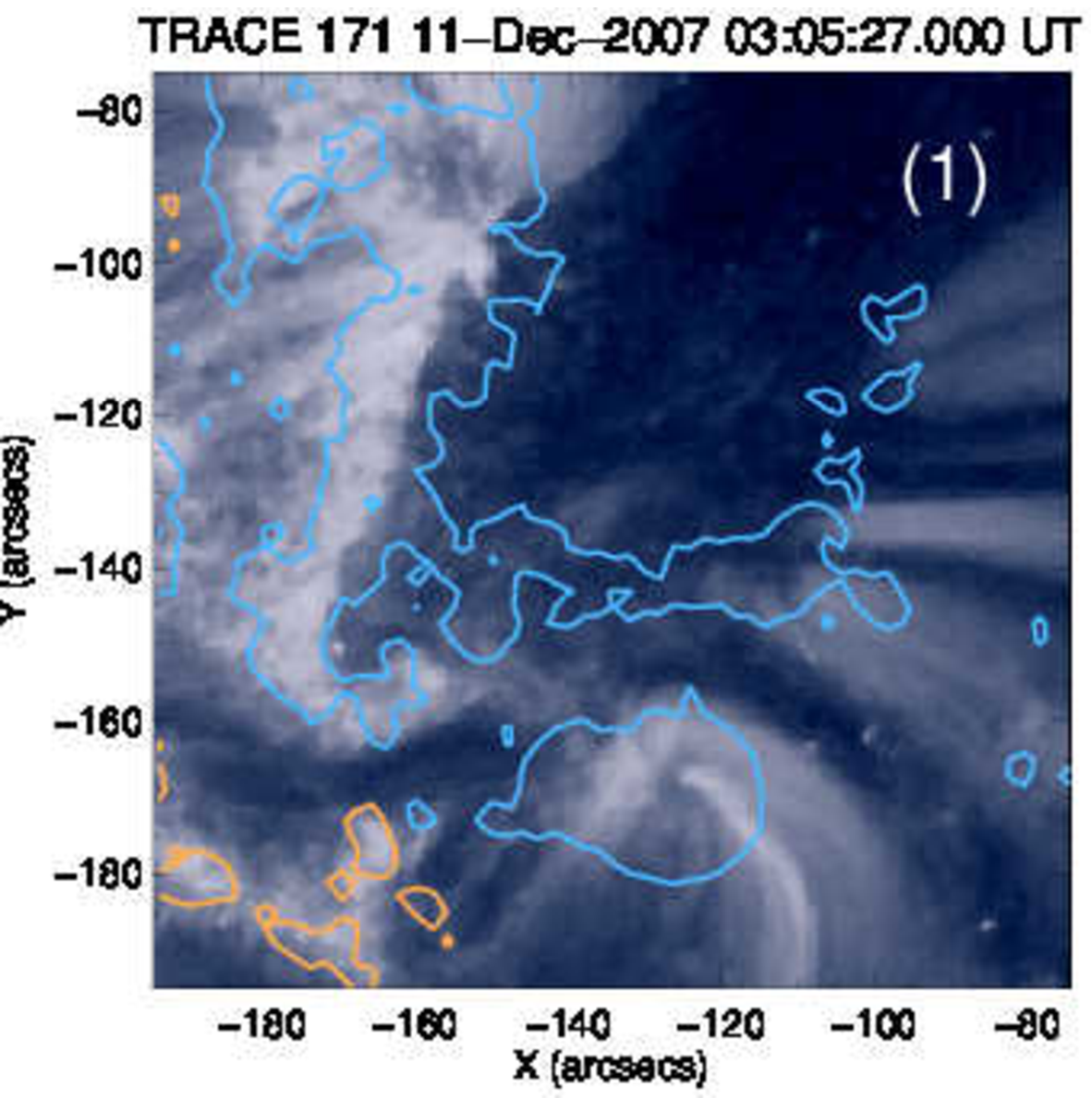}\\
  \includegraphics[width=4.4cm,clip]{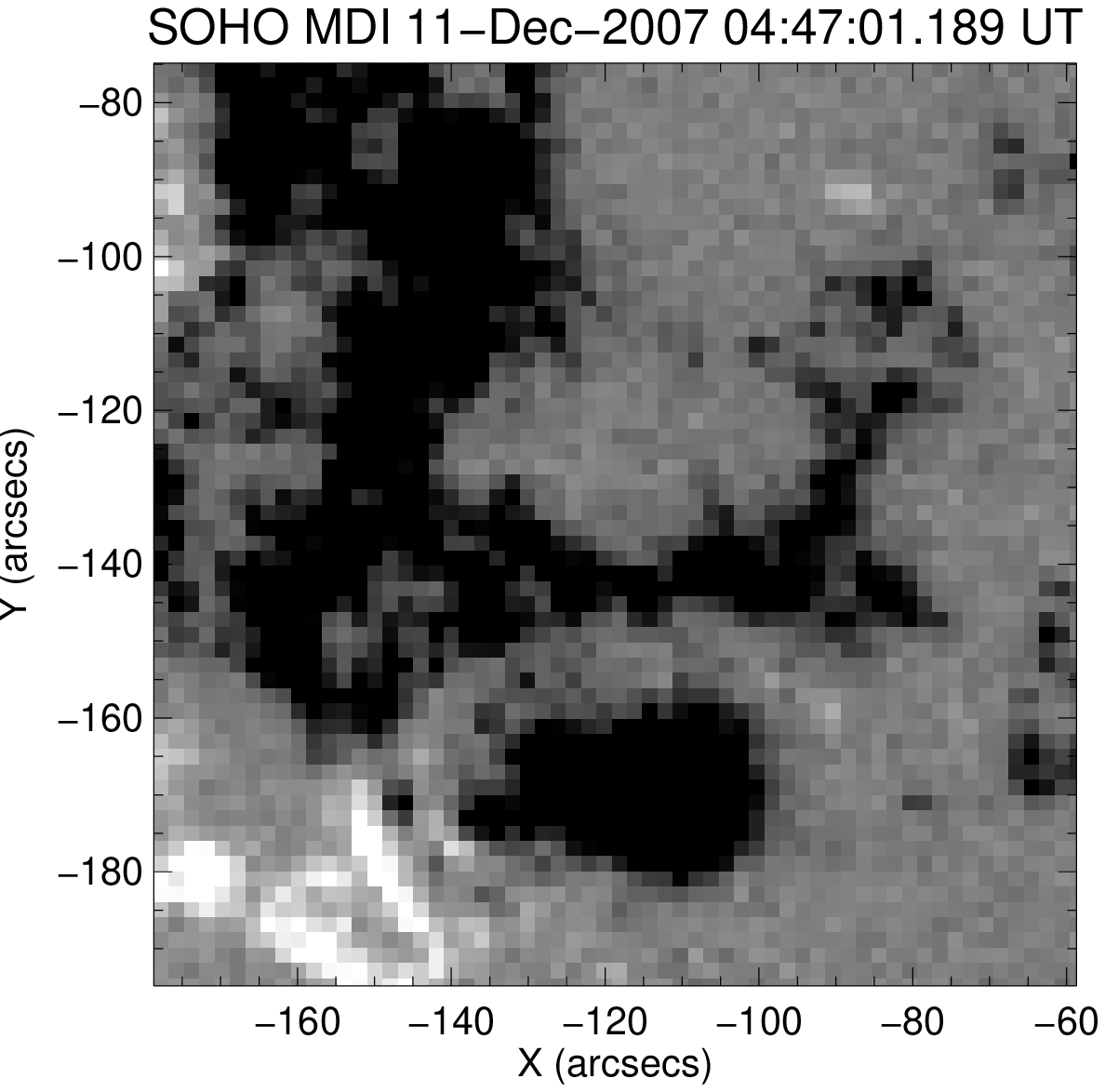}
  \includegraphics[width=4.4cm,clip]{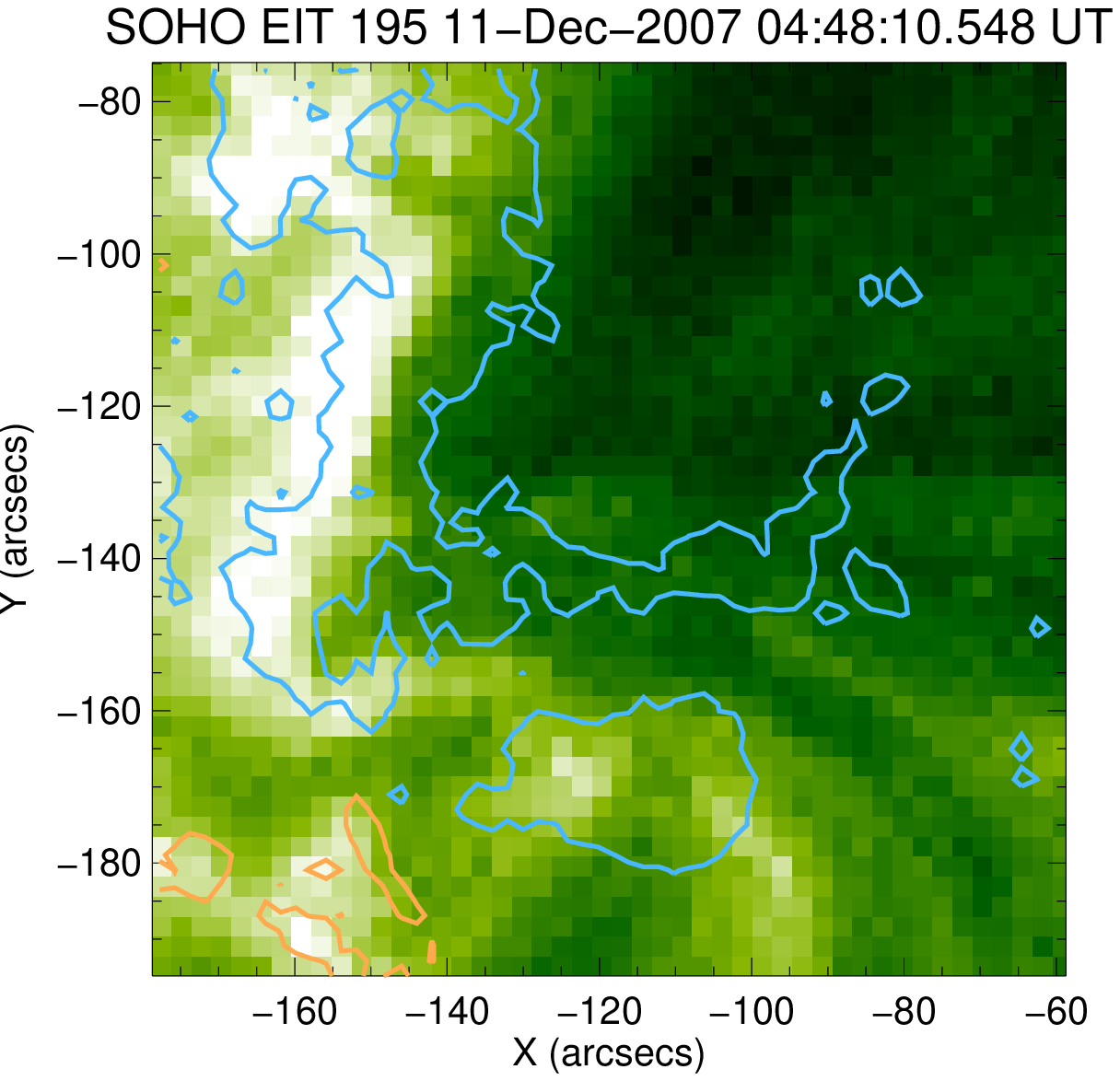}
  \includegraphics[width=4.4cm,clip]{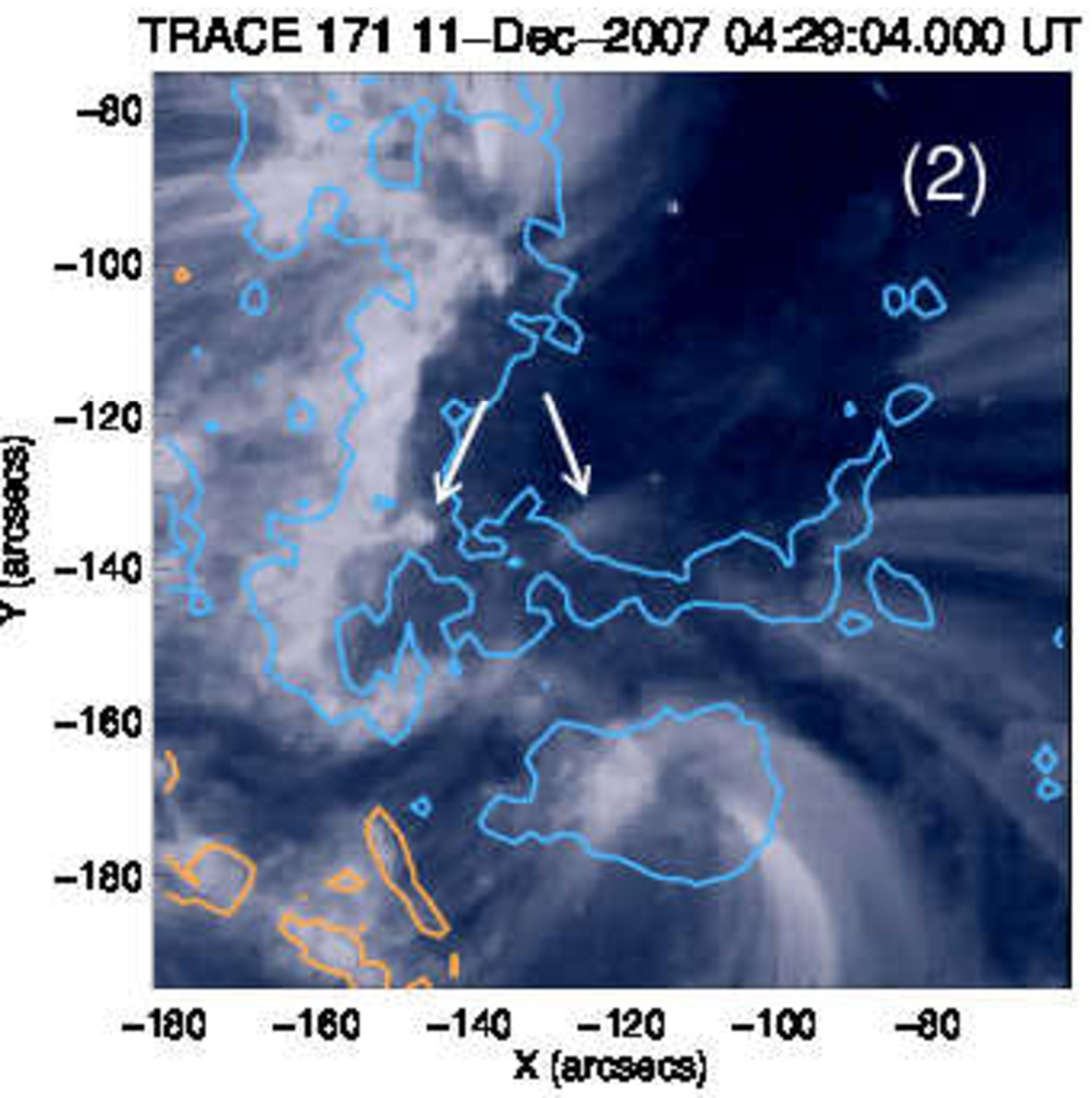}\\
  \includegraphics[width=4.4cm,clip]{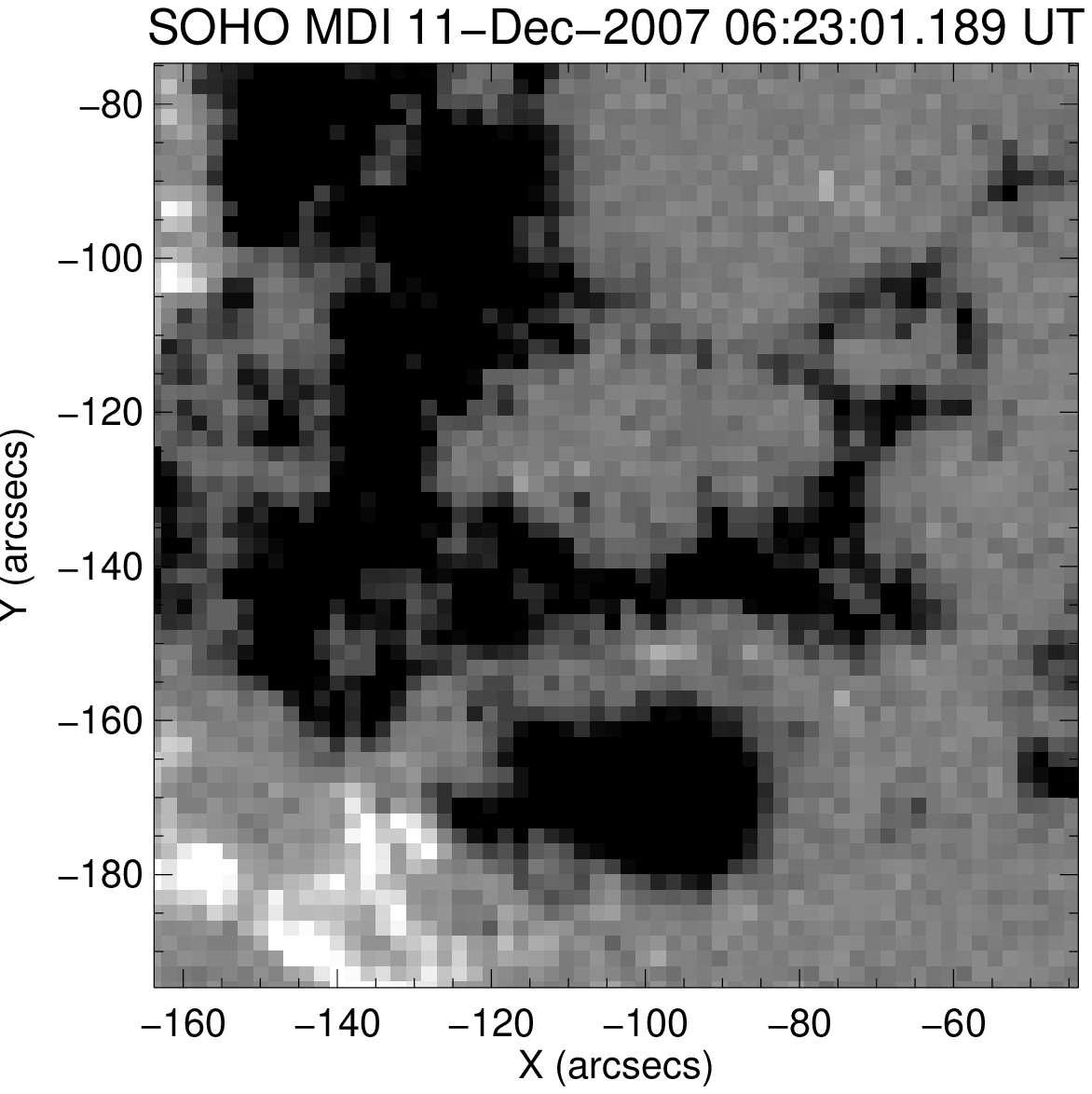}
  \includegraphics[width=4.4cm,clip]{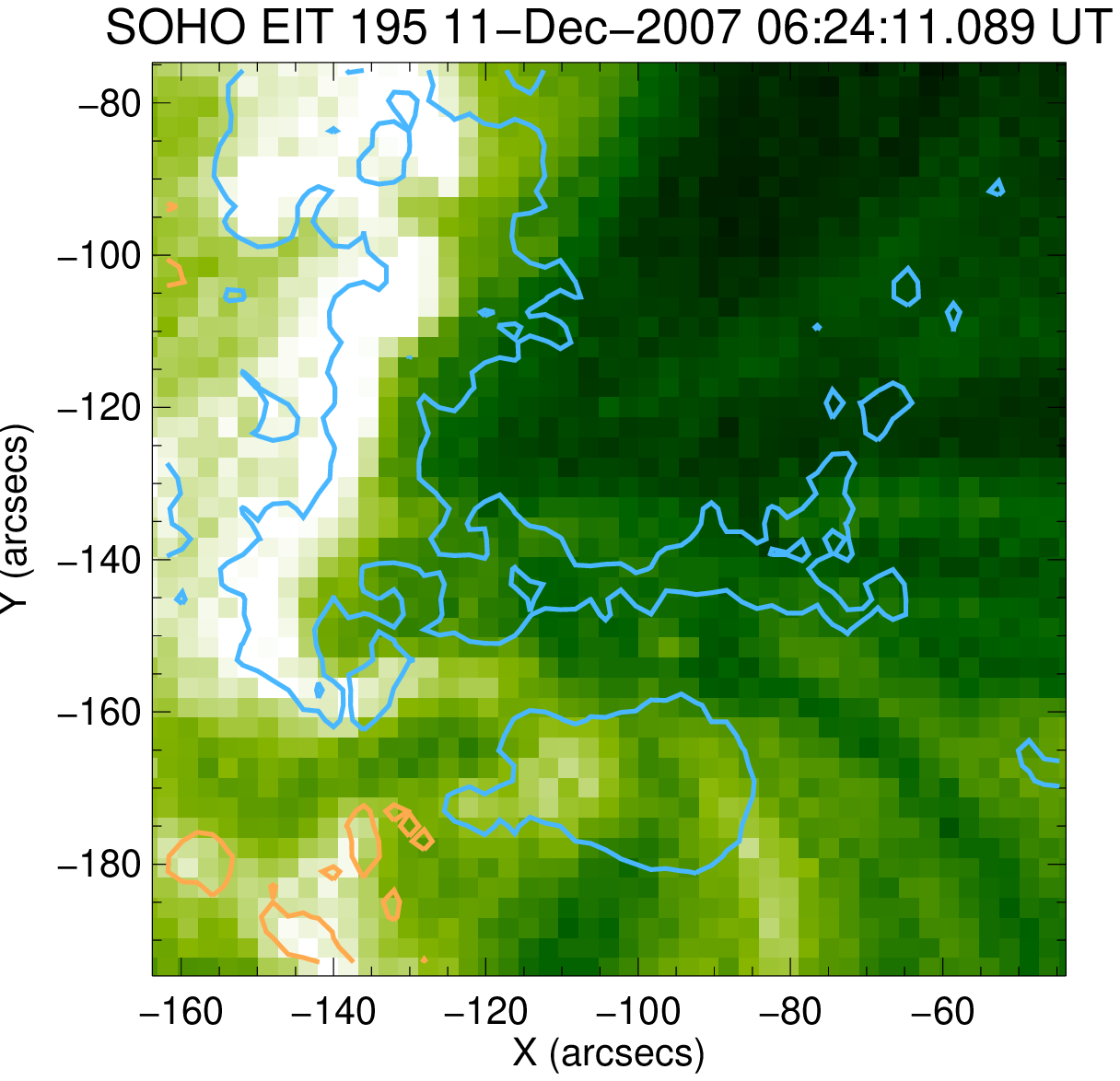}
  \includegraphics[width=4.4cm,clip]{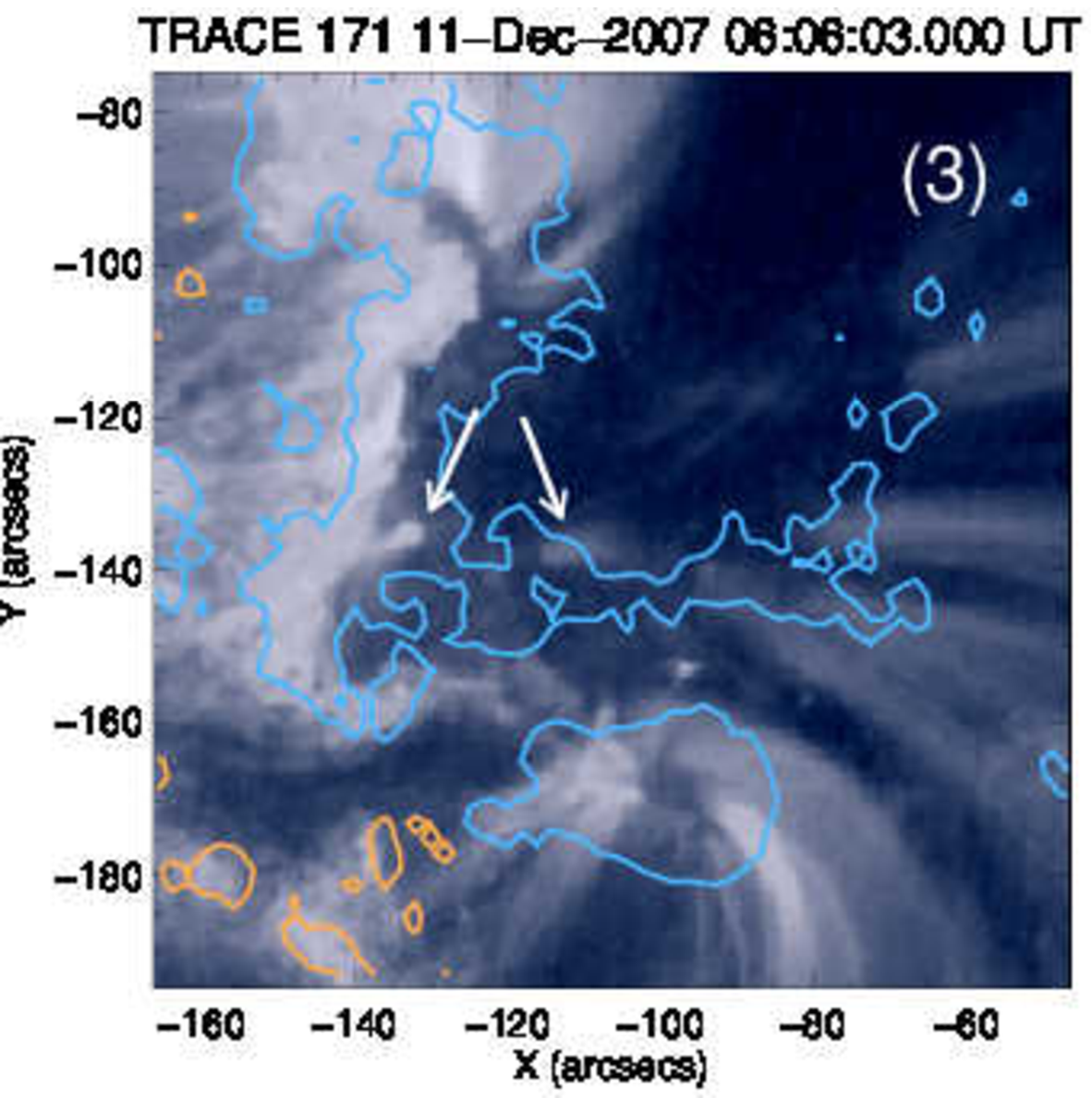}\\
  \includegraphics[width=4.4cm,clip]{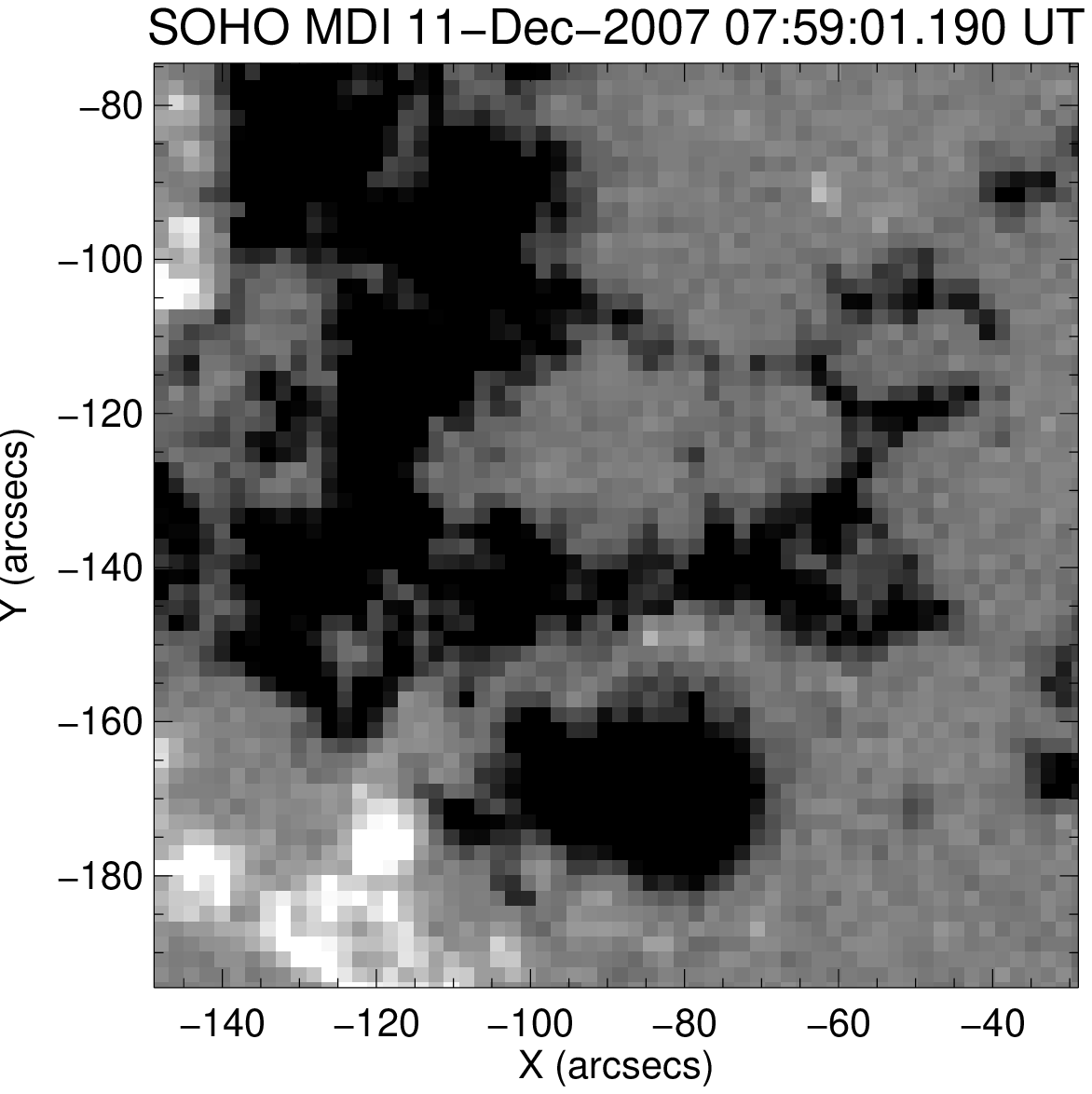}
  \includegraphics[width=4.4cm,clip]{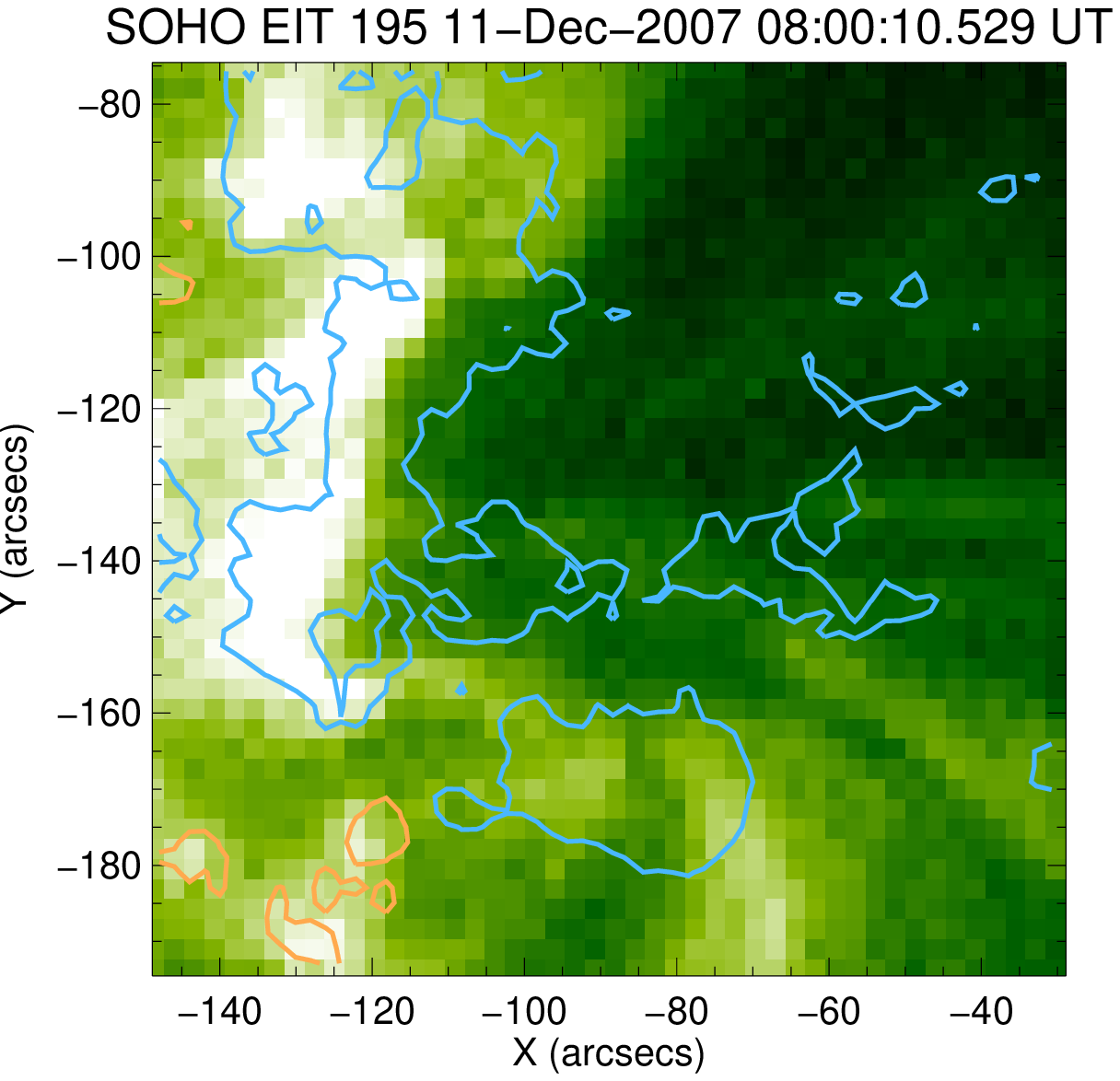}
  \includegraphics[width=4.4cm,clip]{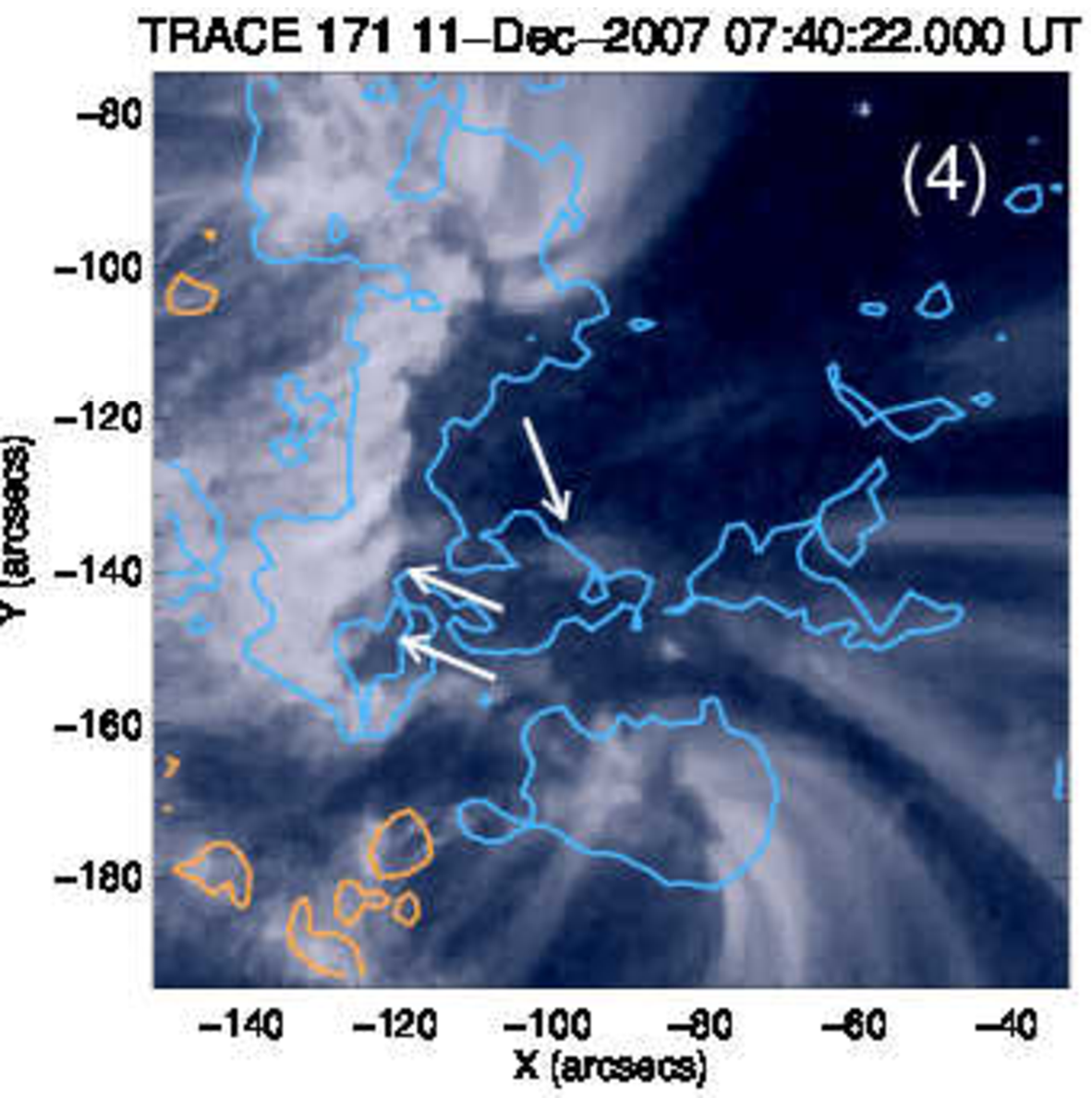}\\
  \includegraphics[width=4.4cm,clip]{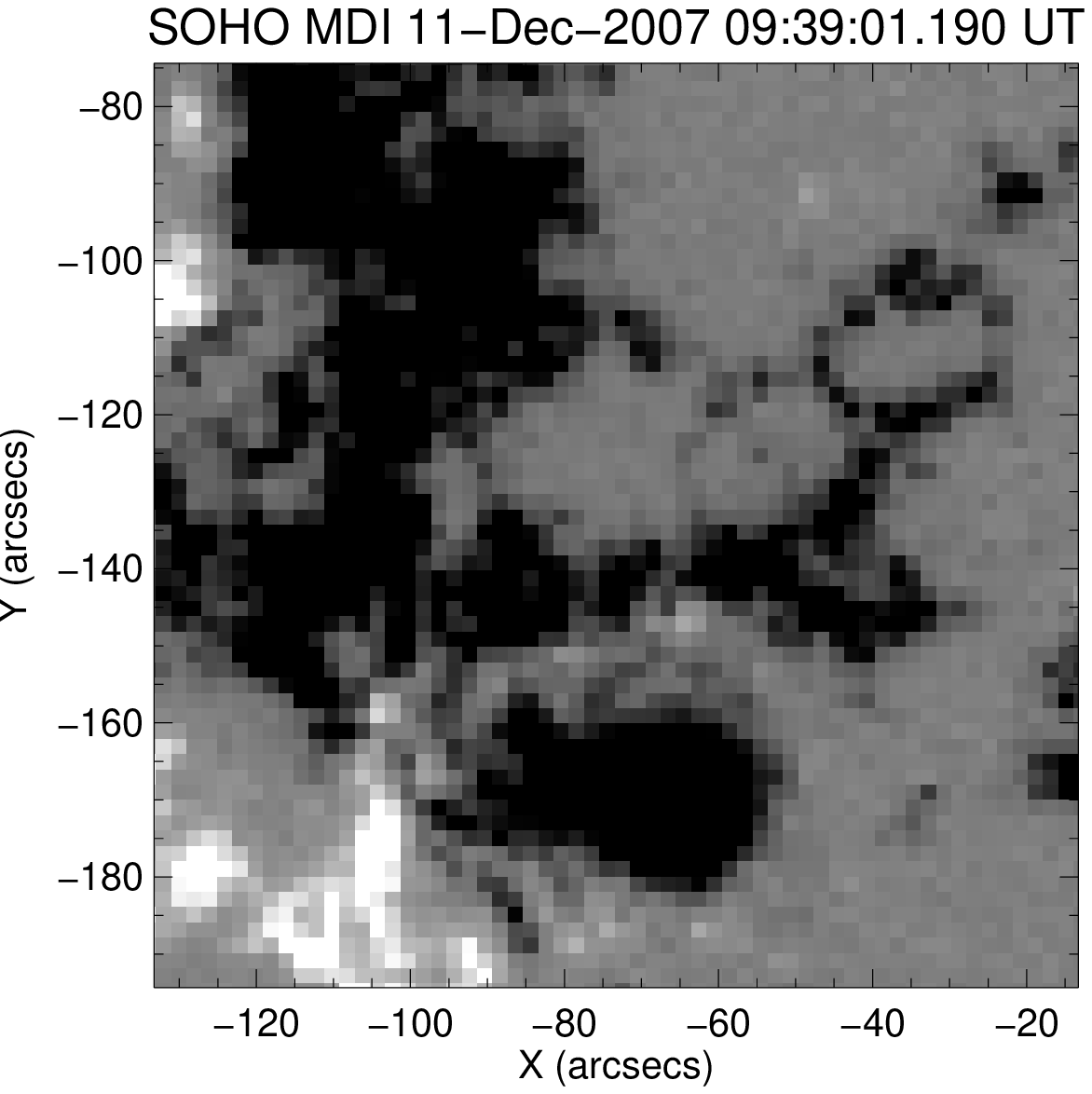}
  \includegraphics[width=4.4cm,clip]{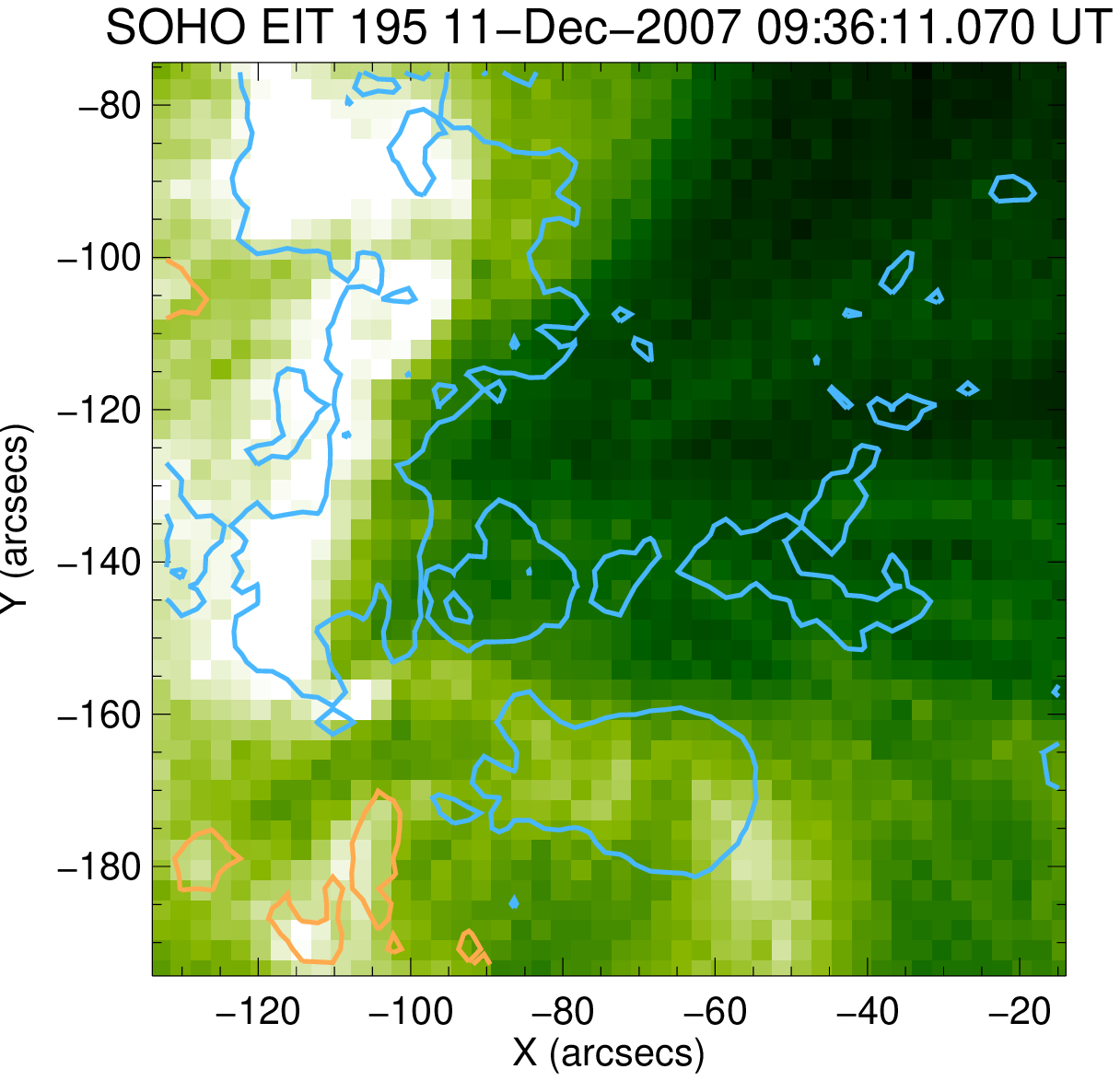}
  \includegraphics[width=4.4cm,clip]{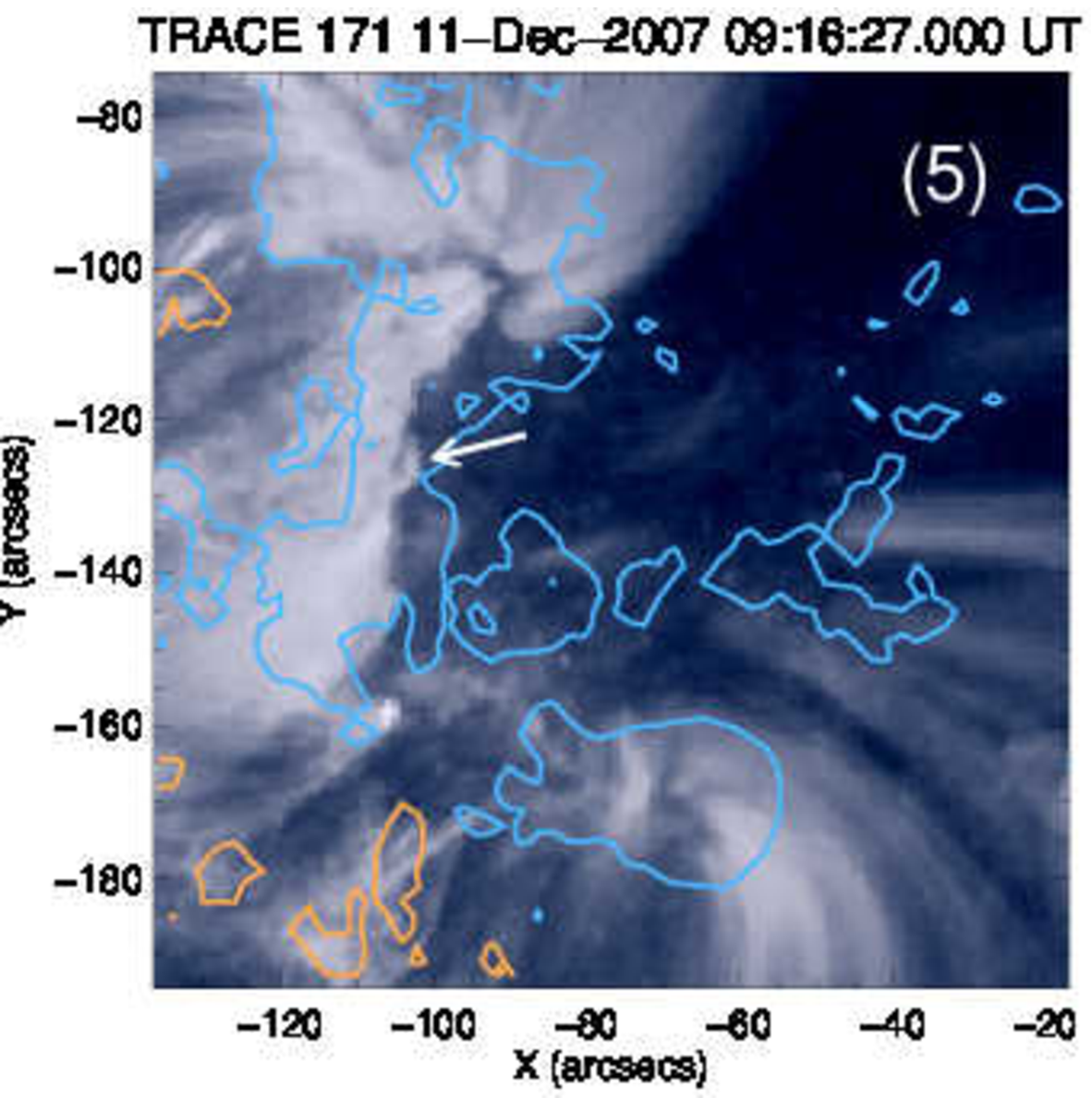}
  \caption{Magnetogram and EUV images of the outflow region on 2007 December 11 
    during around 3--20UT. 
    \textit{Left column}: MDI magnetograms.
    \textit{Middle column}: EIT $195${\AA} images.
    \textit{Right column}: \textit{TRACE} $171${\AA} images.}
    \label{fig:mph_mdi_euv}
\end{figure}
\addtocounter{figure}{-1}

\begin{figure}
  \centering
  \includegraphics[width=4.4cm,clip]{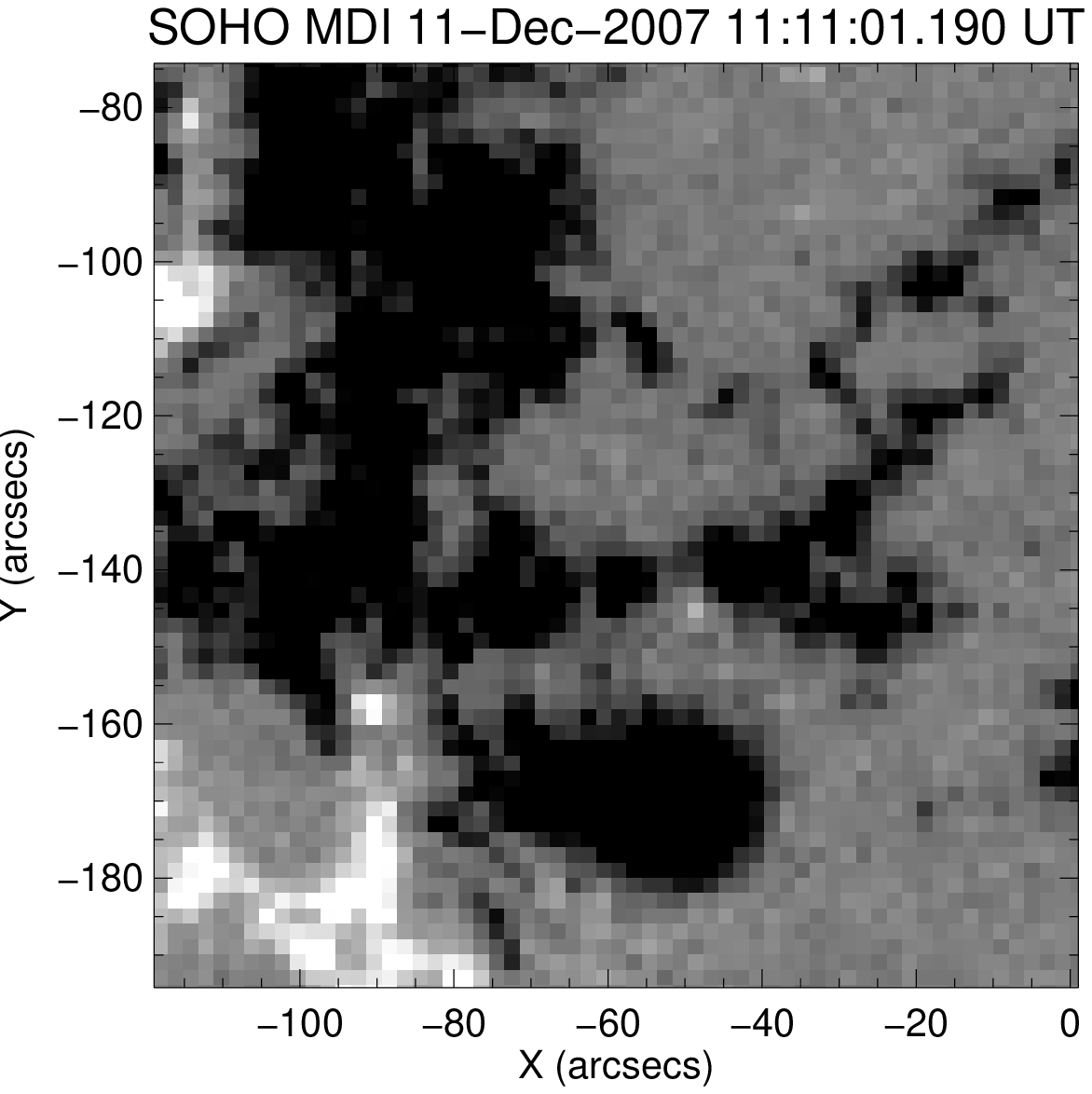}
  \includegraphics[width=4.4cm,clip]{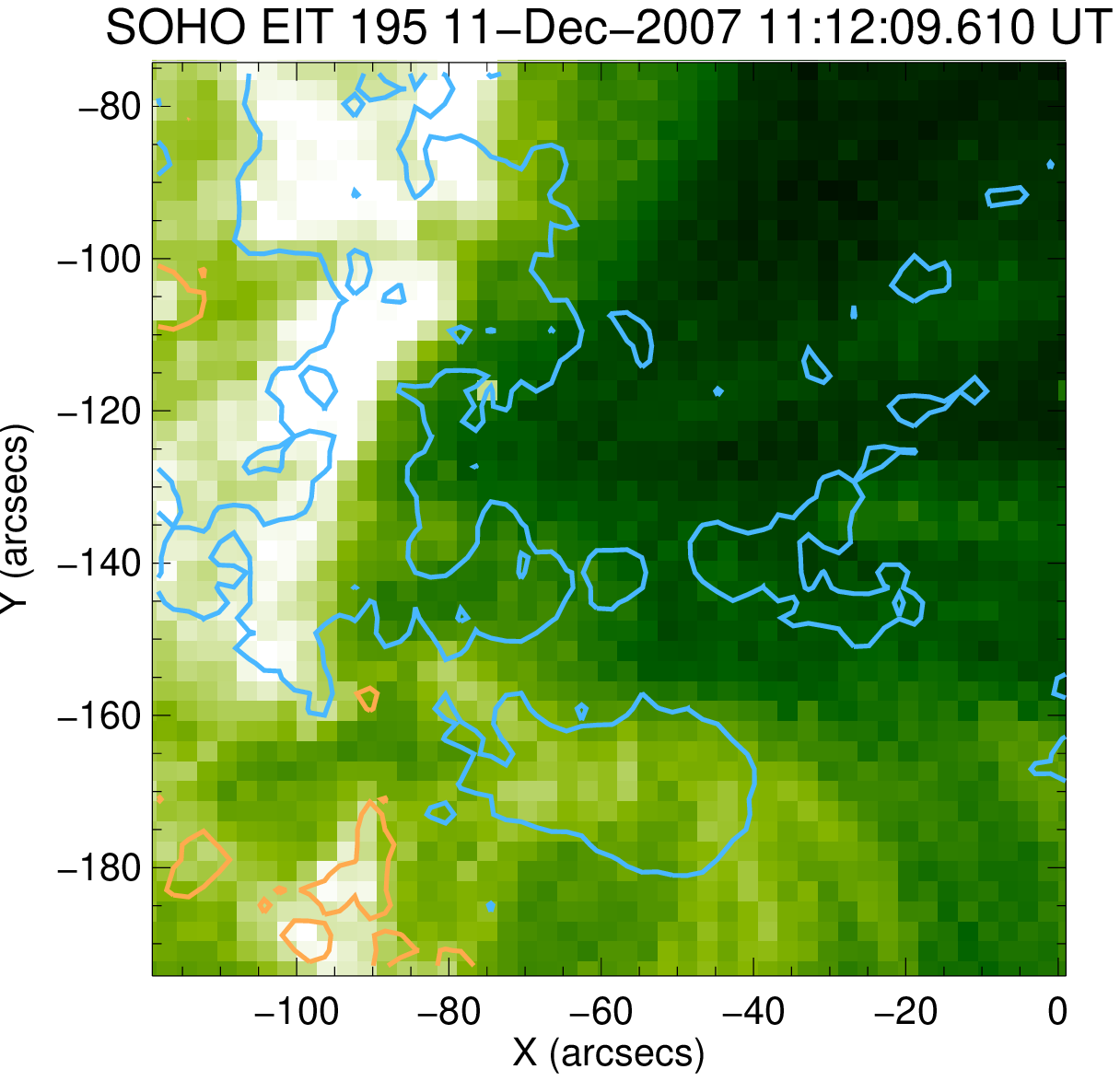}
  \includegraphics[width=4.4cm,clip]{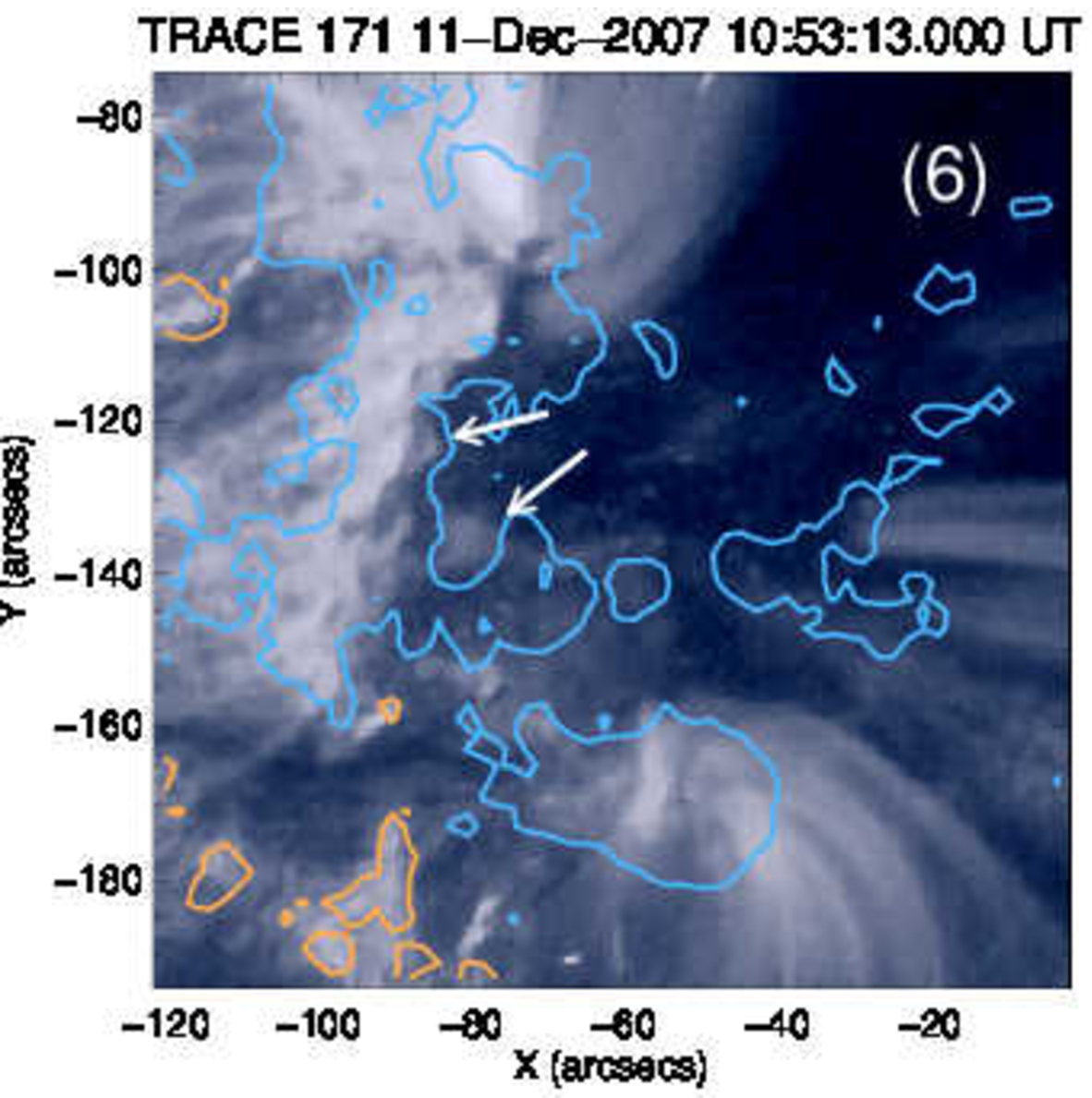}\\
  \includegraphics[width=4.4cm,clip]{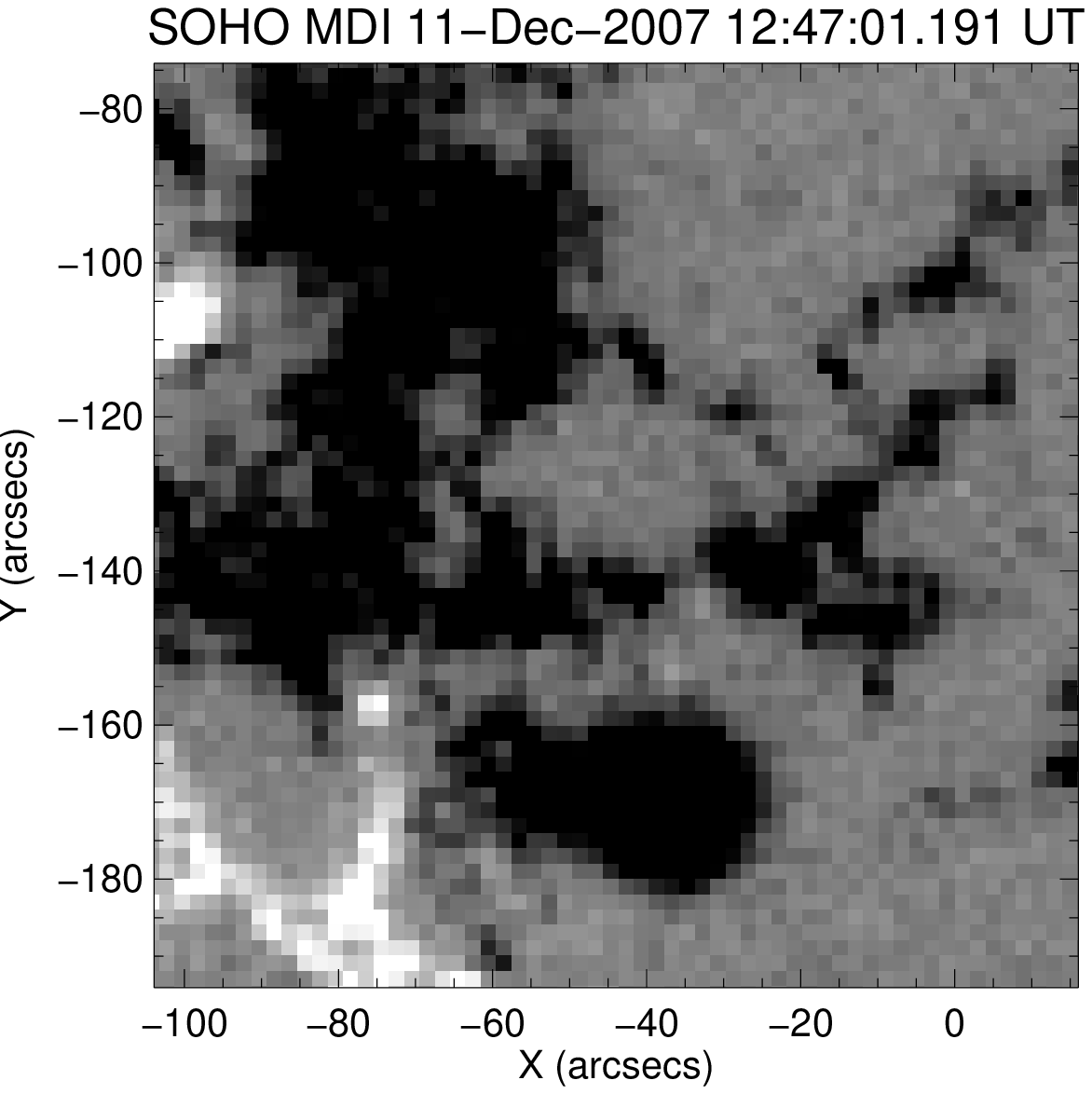}
  \includegraphics[width=4.4cm,clip]{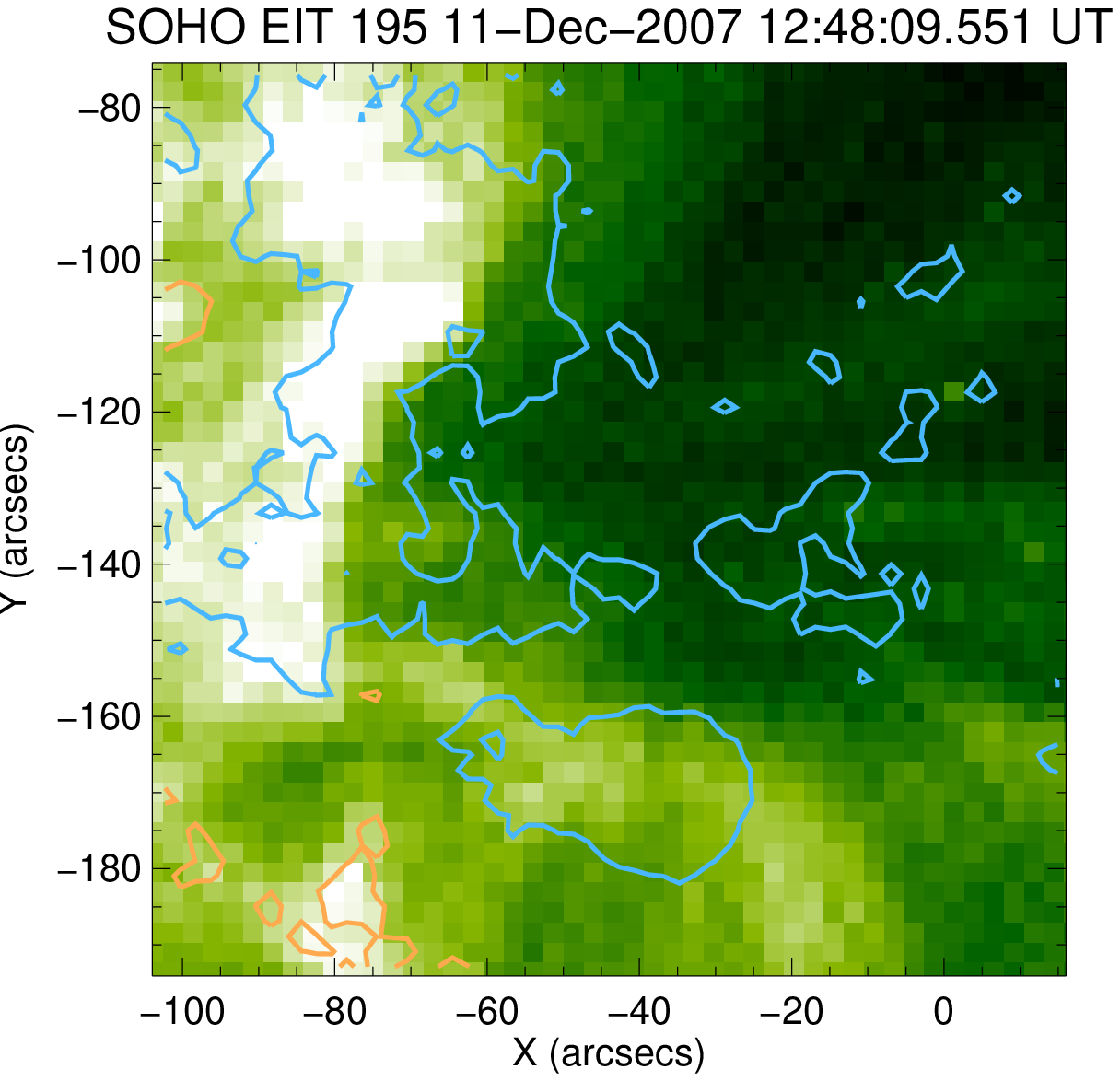}
  \includegraphics[width=4.4cm,clip]{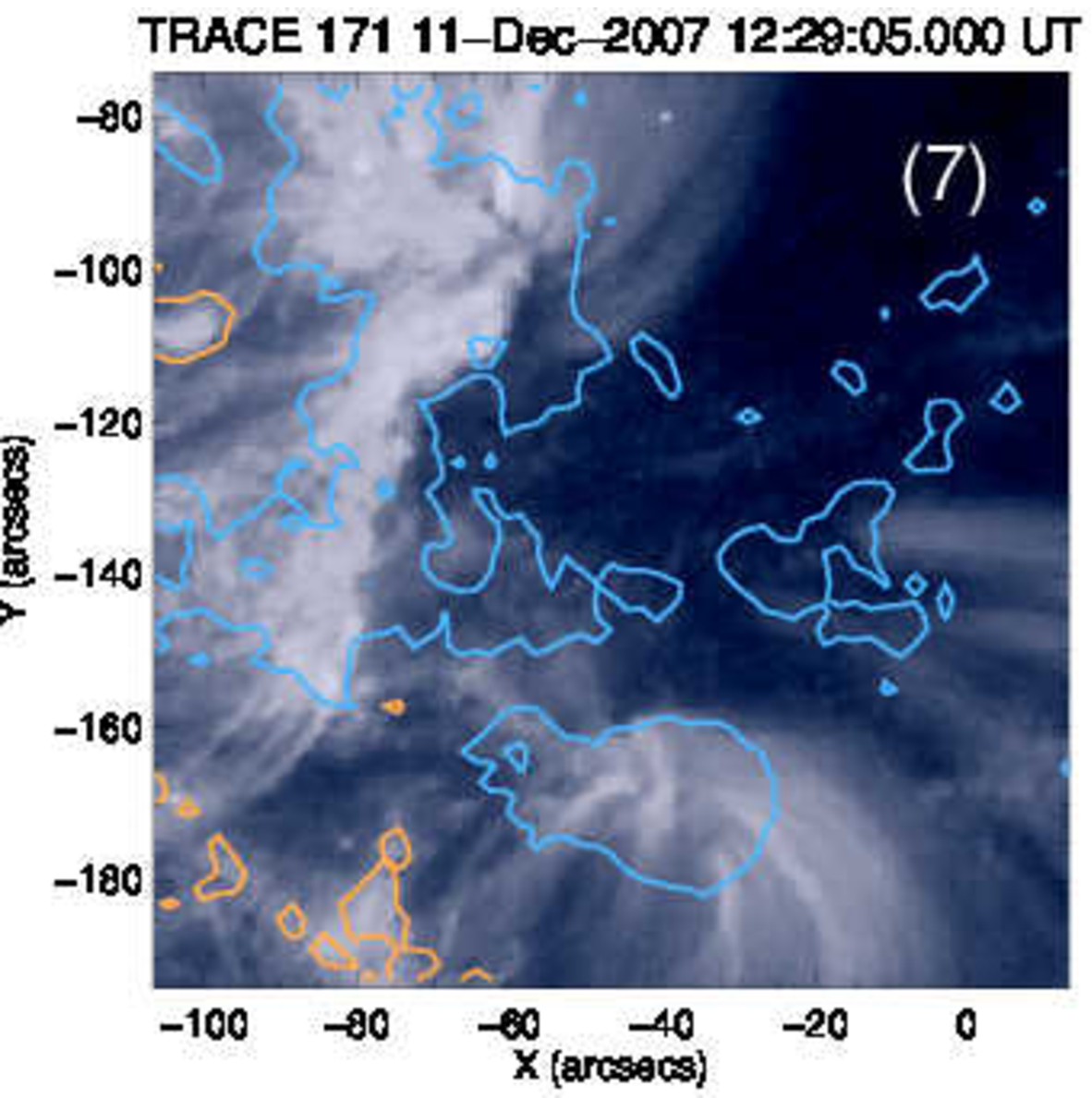}\\
  \includegraphics[width=4.4cm,clip]{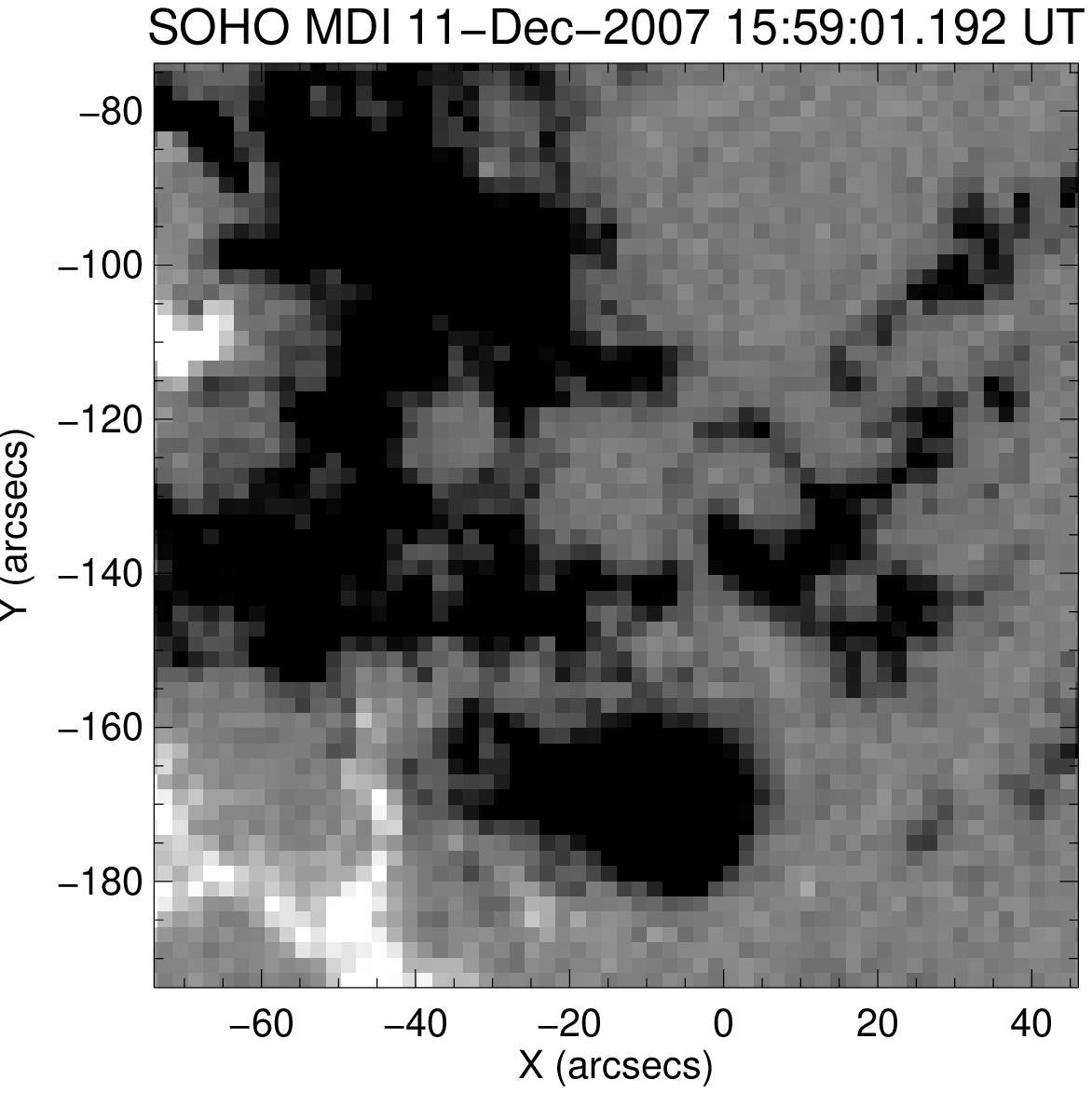}
  \includegraphics[width=4.4cm,clip]{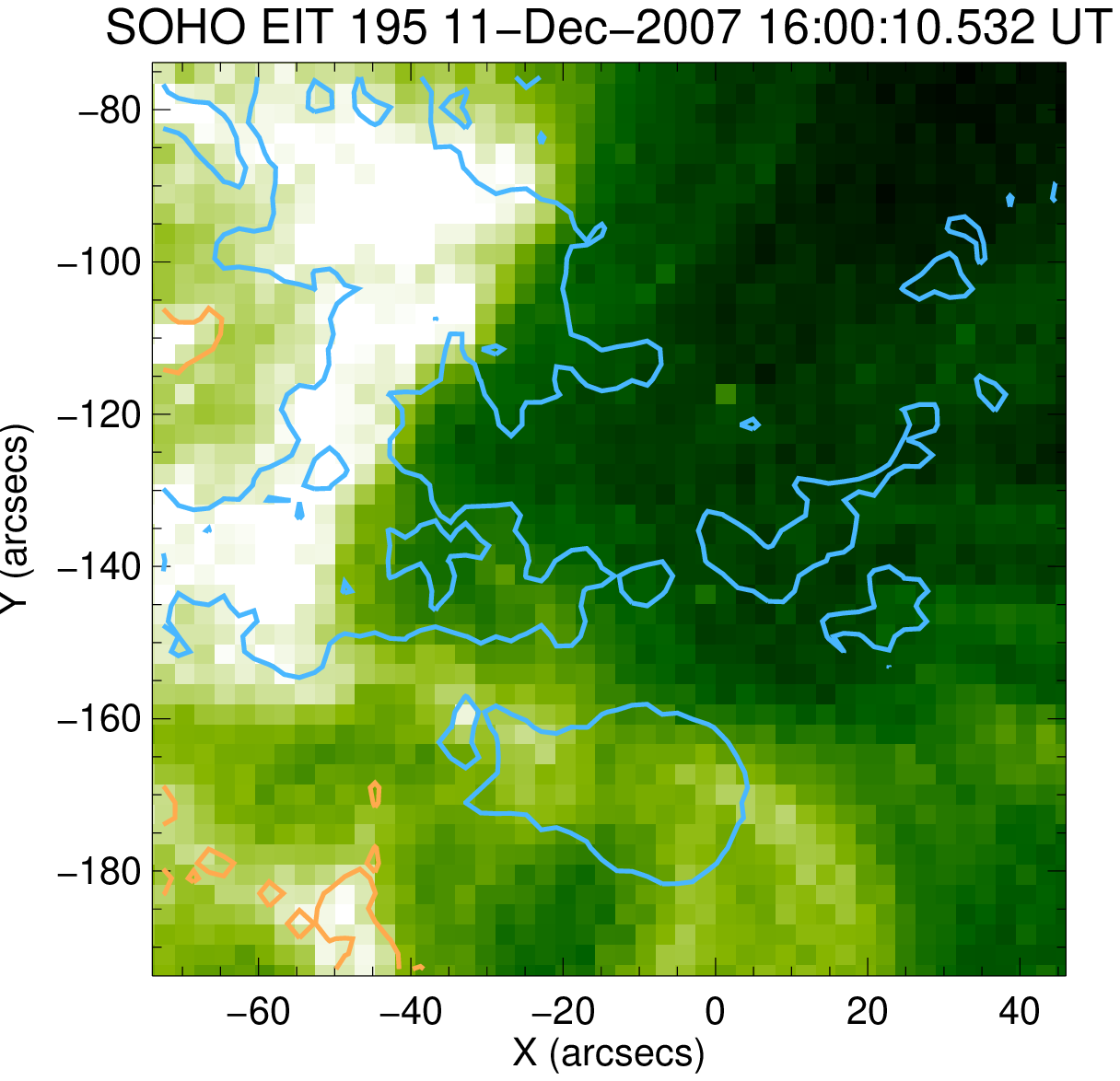}
  \includegraphics[width=4.4cm,clip]{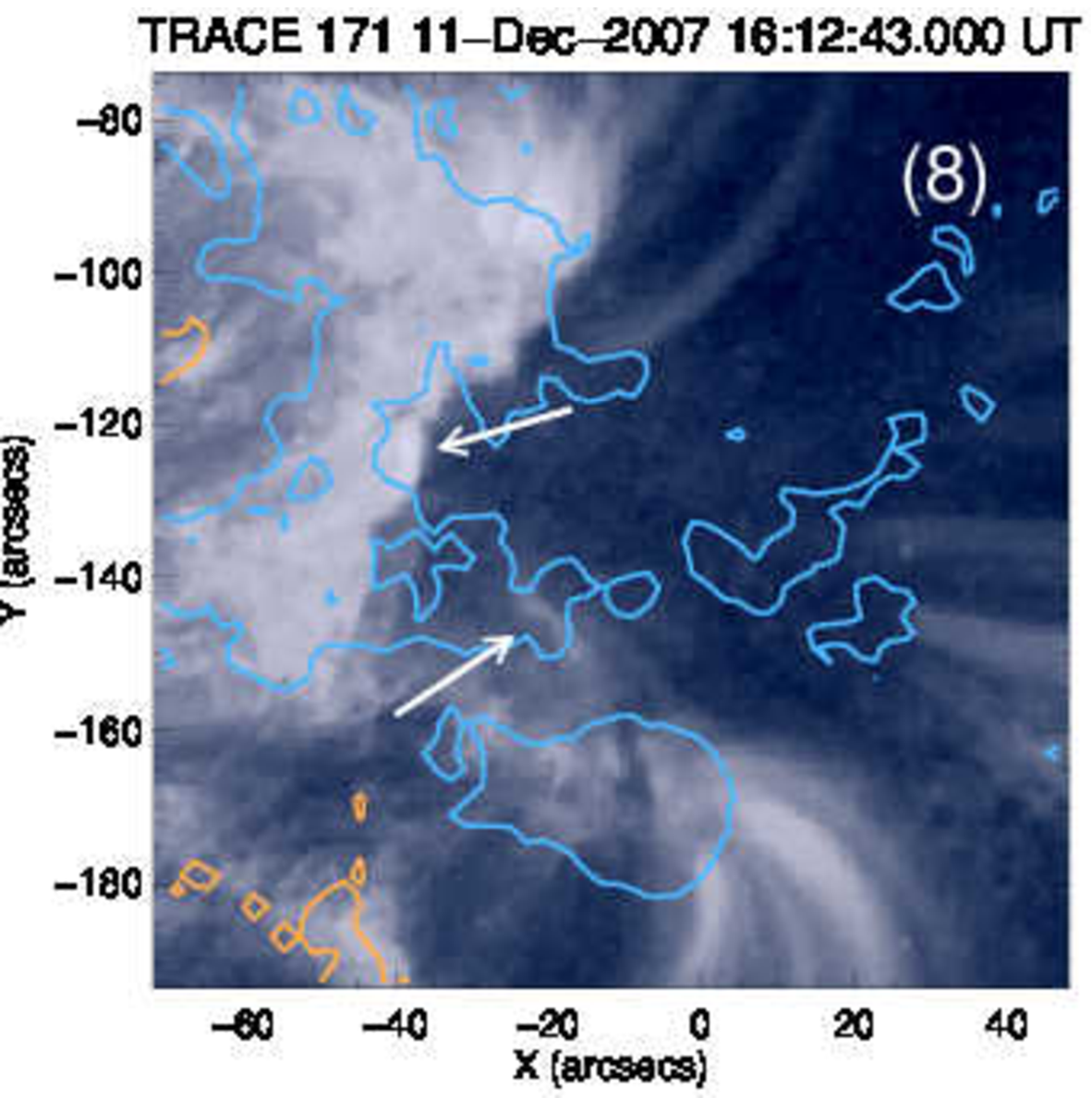}\\
  \includegraphics[width=4.4cm,clip]{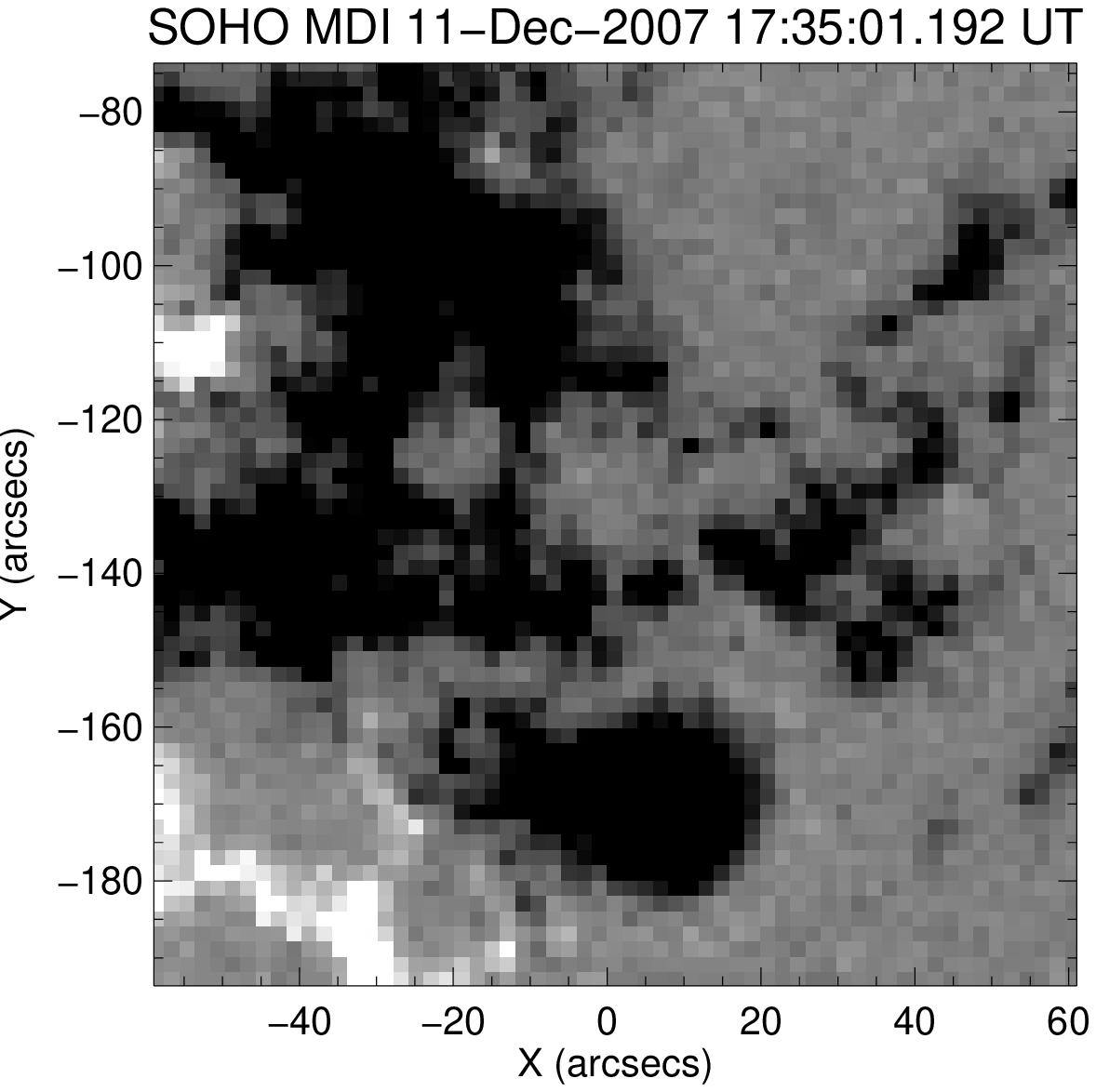}
  \includegraphics[width=4.4cm,clip]{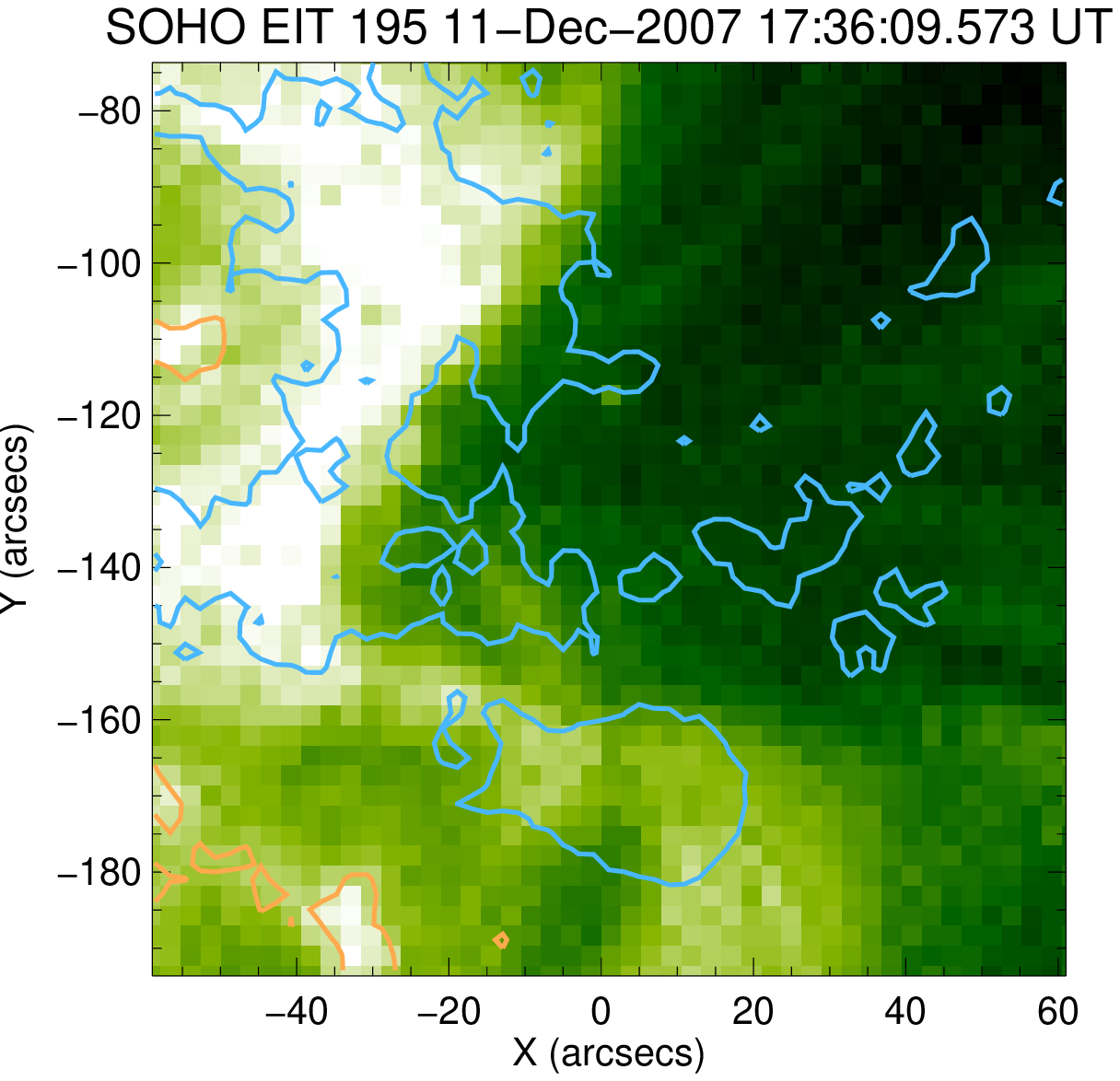}
  \includegraphics[width=4.4cm,clip]{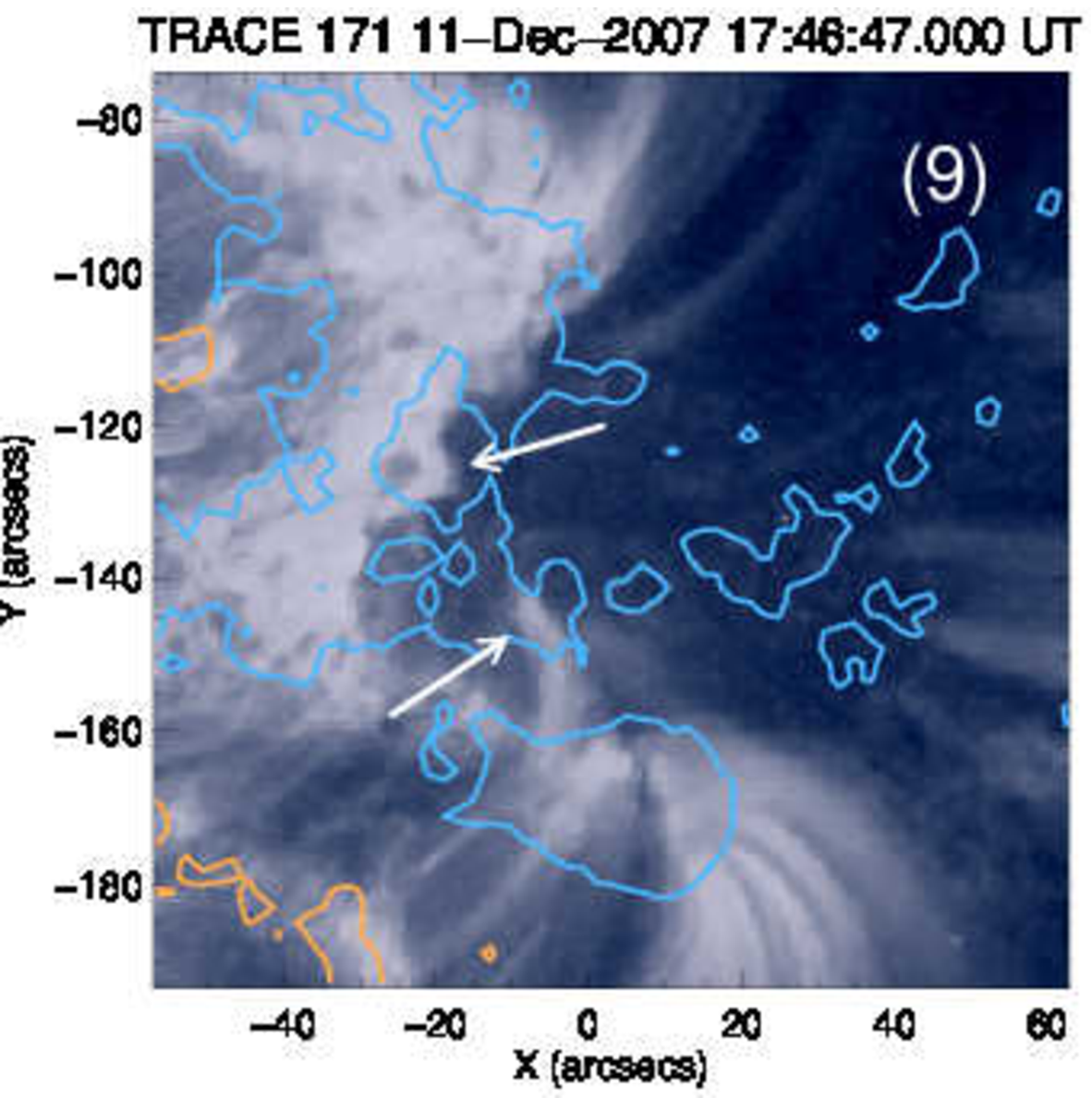}\\
  \includegraphics[width=4.4cm,clip]{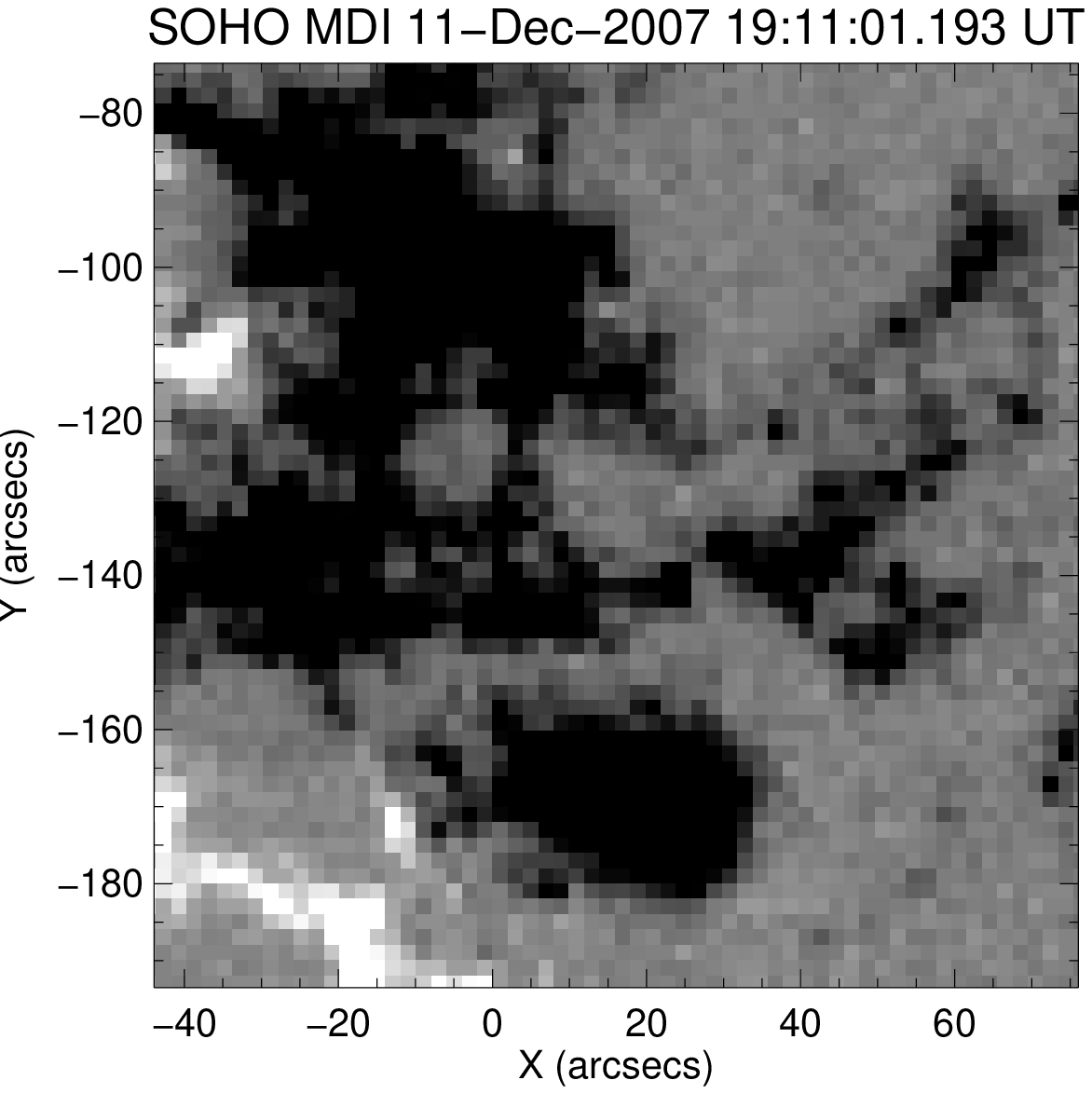}
  \includegraphics[width=4.4cm,clip]{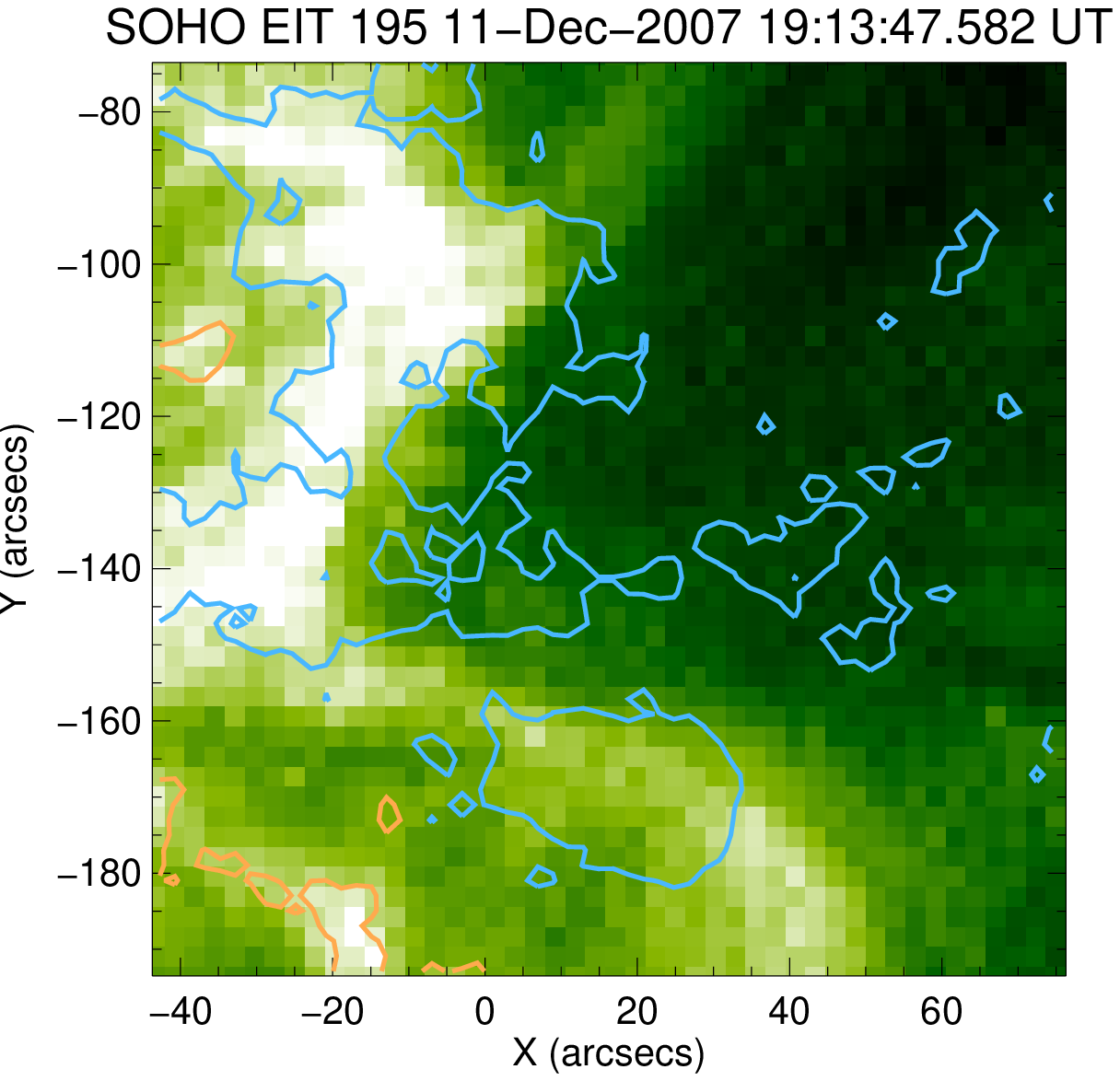}
  \includegraphics[width=4.4cm,clip]{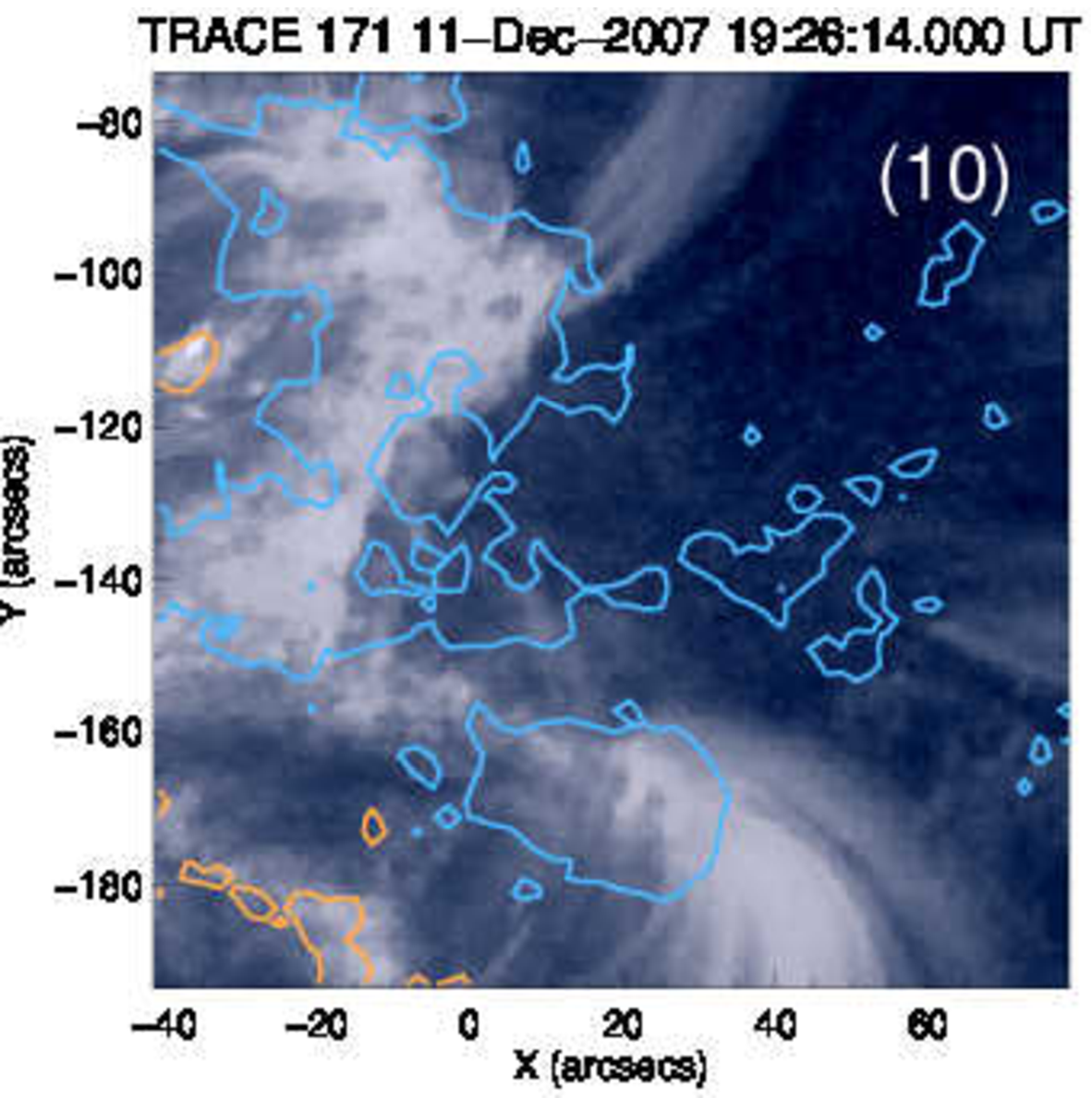}\\
  \caption{\textit{Continued.}}
\end{figure}

In Fig.~\ref{fig:mph_mdi_euv}, we show consecutive images taken by MDI (\textit{left column}), EIT $195${\AA} passband (\textit{middle column}), and \textit{TRACE} $171${\AA} passband (\textit{right column}) with the FOV indicated by the \textit{white dashed} box in Fig.~\ref{fig:mph_tr_context}.  Time intervals between each row are $\sim 96 \, \mathrm{min}$ (temporal cadence of MDI observations).  Colored contours in EIT images (\textit{middle column}) and \textit{TRACE} images (\textit{right column}) show $+250 \, \mathrm{G}$ (\textit{orange} contour) and $-250 \, \mathrm{G}$ (\textit{turquoise} contour) on each MDI magnetogram (\textit{left column}) which was taken at near timing. 
% MDI magnetograms and EIT images
As seen in the MDI magnetograms, there is a negative sunspot at the bottom of panels from which coronal loops extend to the south in EIT and \textit{TRACE} images.  An arc-like pattern lies at the north of the sunspot in MDI magnetograms (above $y=-150''$).  From left hand side of the arc-like structure, multiple loops system extends to the opposite polarity at the east, while there is no corresponding distinct loops at the bottom side of the arc-like structure.  The right hand side of the arc-like structure seems to be footpoints of fan loops extending to the west direction.  The multiple loop system exists only at the east half of the left hand side of that structure during $3$--$12$UT as especially seen in \textit{TRACE} images.  The overall structure in magnetograms did not change much.

% TRACE images
A careful inspection at the \textit{TRACE} images leads to the detection of intermittent appearance of some bright structures as indicated by \textit{white} arrows in Fig.~\ref{fig:mph_mdi_euv}.  There are two types of structures: one is a coronal loop which forms the multiple loop system connecting positive and negative polarities (\textit{e.g.}, a left arrow in the second and third rows from the upper), while the other is extended toward relatively outside (\textit{e.g.}, a right arrow in the second and third rows).  Note that the area possessed by these small bright structures are not dominant in the outflow region. 

% X-T diagrams
\begin{figure}
  \centering
  \includegraphics[width=16.5cm,clip]{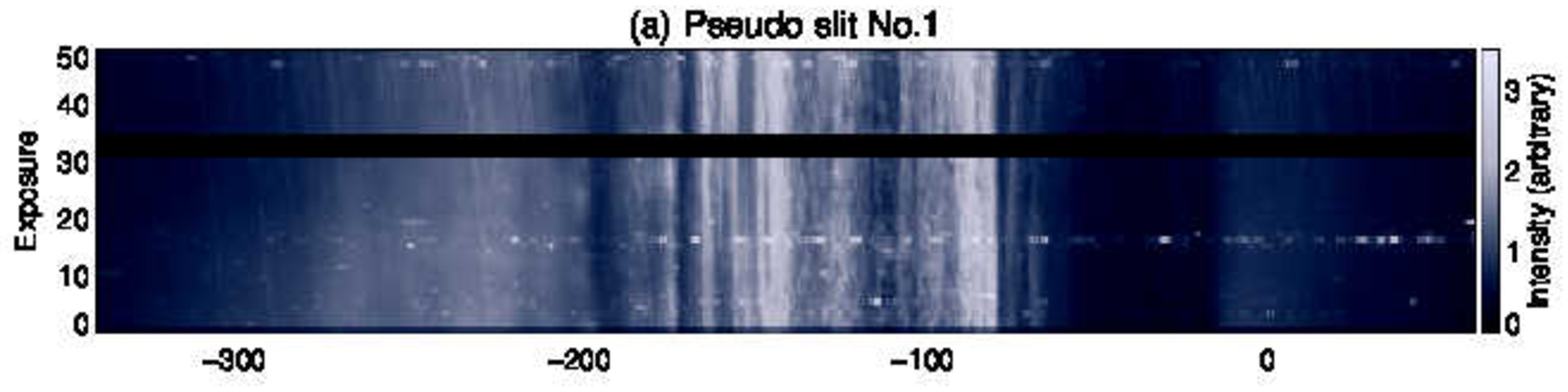}
  \includegraphics[width=16.5cm,clip]{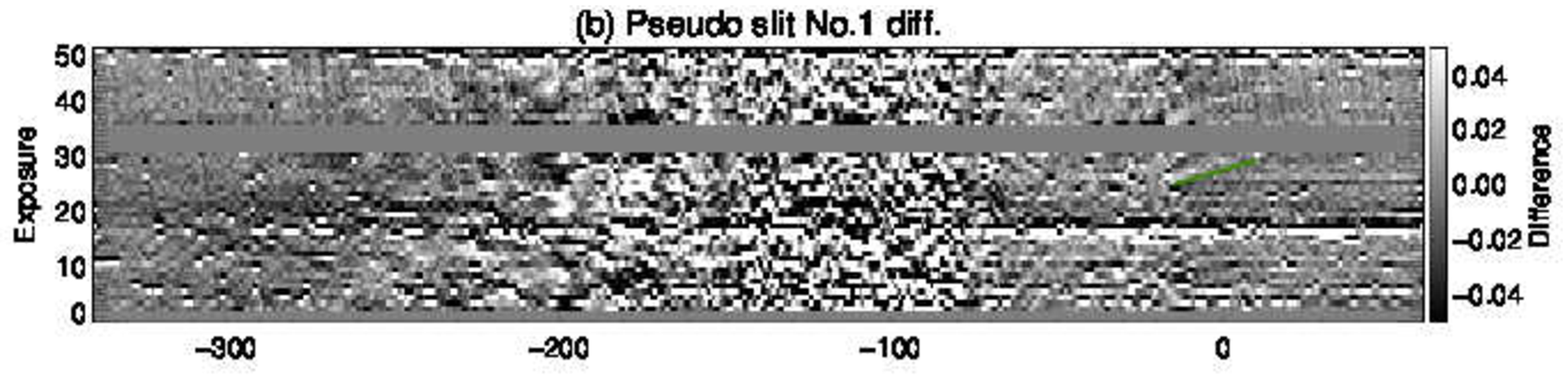}
  \includegraphics[width=16.5cm,clip]{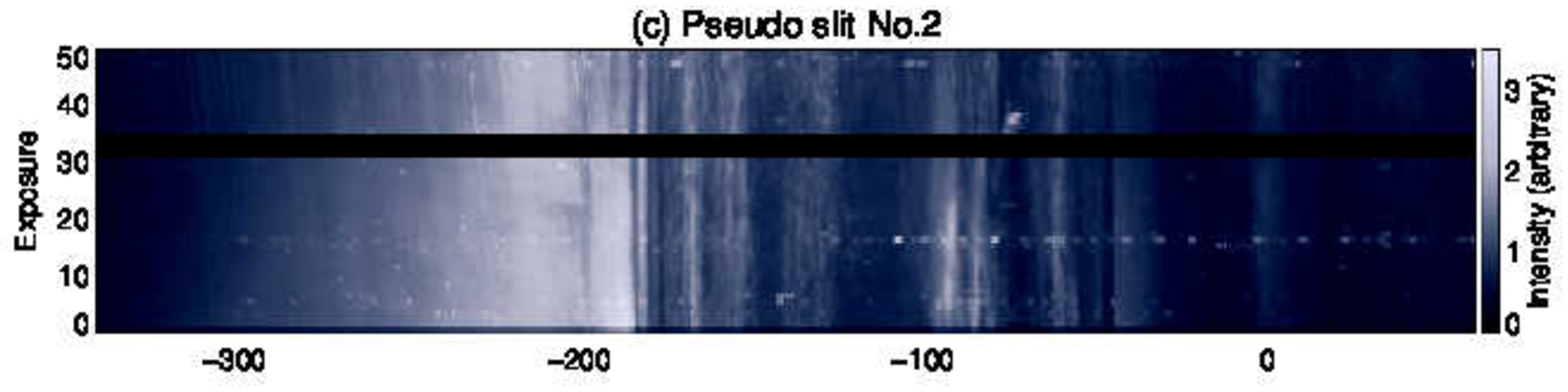}
  \includegraphics[width=16.5cm,clip]{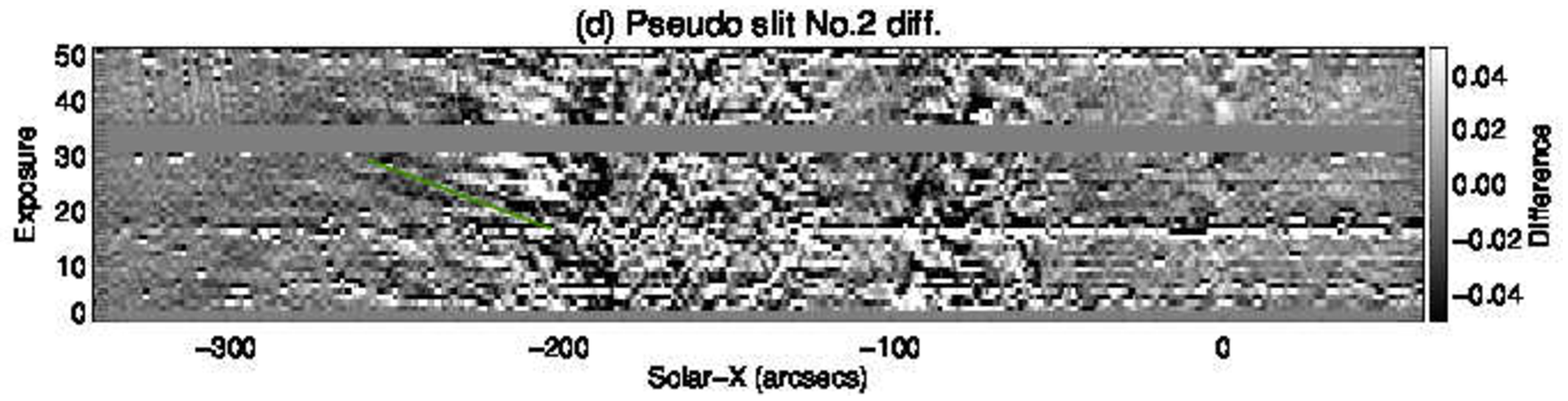}
  \caption{$x$-$t$ diagrams for pseudo slits indicated by horizontal \textit{green dashed} lines in Fig.~\ref{fig:mph_tr_context}. Each panel shows $x$-$t$ diagram for (a) intensity at pseudo slit No.1, (b) running difference at pseudo slit No.1, (c) intensity at pseudo slit No.2, and (d) running difference at pseudo slit No.2. A \textit{Green} thick line in each panel indicates the propagating disturbance traced visually.}
  \label{fig:mph_tr_xt}
\end{figure}

% Explain the figure.
Fig.~\ref{fig:mph_tr_xt} shows $x$-$t$ diagrams made at horizontal \textit{green dashed} lines in the \textit{TRACE} map (Fig.~\ref{fig:mph_tr_context}).  We selected the observational sequence during 2007 December 11 12:29--13:32UT when it observed the active region with a temporal cadence of $\sim 1 \, \mathrm{min}$ and the least gap in time.  The pseudo slits have a width of $5 \, \mathrm{pix}$ so that the signal-to-noise ratio would be improved.  The location of the pseudo slit was set in each image so that it continuously tracks the same region since the solar rotates about $15''$ ($30 \, \mathrm{pixels}$) during the \textit{TRACE} observation.  Panels (a) and (c) respectively show an $x$-$t$ diagram for the intensity at the pseudo slit No.1 and No.2.  The horizontal (vertical) axis shows the solar $X$ and exposure No.\ (intervals of roughly $1 \, \mathrm{min}$, but not exactly constant).  The unit of the intensity is normalized by the median at each exposure because we found that the mean intensity increases with time during the observation, which might be a result of the gradual change in the telescope environment.  Panels (b) and (d) show the running differences in temporal direction.  Note that sharp patterns slightly inclined to $-x$ direction (from right bottom to left upper), which are clearly seen around $x=-300''\,\text{--}\,{-200''}$, may be due to the instrumental effect (\textit{e.g.}, CCD characteristics).  We found these patterns were located at specific pixels on CCD.  The black region (exposure No.$31$--$34$) means that there is a time gap of a few minutes.

% Describe the feature seen in the figure.
The pseudo slit No.1 cuts across the center portion of the active region core, the outflow region, and goes through a fan loop extending toward the west.  The slit also cut fan loops at the east edge of the core, but it was not aligned along the fan loops.  The pseudo slit No.2 was set to cut along the fan loops at the east edge.  We can see several propagating features from the footpoints of fan loops: $x=-15''$ to $10''$ in panel (b) and $x=-200''$ to $-260''$ in panel (d) as indicated by \textit{green} lines.  These propagating disturbances occurred not only once but repetitively, and are considered to be identical to those observed by previous \textit{TRACE} observations \citep{demoortel2000,winebarger2001}.  In contrast to fan loops, the outflow region ($x=-80''\,\text{--}\,-20''$ in panel a and b) did not exhibit any clear propagating features in the analyzed data.  We also tried to seek signatures by looking into the \textit{TRACE} movie, however, we could not detect the prominent flow pattern. 

% --- End of TeX ---

%% file: tex/mph_sum.tex
% =================================
%   Chapter:
%     Morphology
%   Section:
%     Summary
% =================================

% Potential field extrapolation from an MDI magnetogram
The magnetic field above the active region NOAA AR10978 was extrapolated from an MDI magnetogram through solving the Laplace equation by Green's function method. The magnetogram was taken near at the disk center and we calculated the magnetic field in the orthogonal coordinate. The size of the calculation box was enough large to include the whole active region ($500'' \times 500'' \times 400''$). The extrapolated field lines are in well coincidence with the morphology of EUV loops seen in the EIS intensity and Doppler velocity maps. 

The field lines rooted at the outflow region had lengths of $\simeq 100$--$200 \, \mathrm{Mm}$ and their footpoints with the opposite polarity were located slightly inside the east edge of the active region. While a part of the calculated field lines corresponding to fan loops went out from the calculation box from its side or top boundary, those rooted at the outflow region were not so long enough to reach higher than $\simeq 200 \, \mathrm{Mm}$, which means rather ``closed'' field lines.  

% TRACE images. Intermittent phenomena and stack plots where I searched a signature of OF.
EUV images taken by \textit{SoHO}/EIT and \textit{TRACE} showed that the magnetic field strengths at the outflow region were relatively strong ($\lvert B_{z} \rvert \geq 200 \, \mathrm{G}$). A careful inspection at the \textit{TRACE} images helped us to find the intermittent appearance of some bright structures which looked like a leg of coronal loops emanated from the outflow region. However, these intermittent phenomena obviously did not possess dominant area in the outflow region, which may indicate that those are not a main contributor to the outflow. The $x$-$t$ diagrams clearly showed propagating disturbances in fan loops as reported by previous observations, however, we could not detect the prominent flow pattern in the outflow region. 

% --- End of TeX ---